\documentclass[letterpaper,12pt,titlepage,oneside,final]{book}
 
\newcommand{\href}[1]{#1} 

\newcommand{\ad}{\hat{a}}
\newcommand{\au}{\hat{a}^\dagger}
\newcommand{\bd}{\hat{b}}
\newcommand{\bu}{\hat{b}^\dagger}
\newcommand{\cd}{\hat{c}}

\newcommand{\td}{\hat{t}}
\newcommand{\tu}{\hat{t}^\dagger}
\newcommand{\sd}{\hat{s}}
\newcommand{\su}{\hat{s}^\dagger}
\newcommand{\sigu}{\hat{\sigma}^\dagger}
\newcommand{\sigd}{\hat{\sigma}}
\newcommand\mycom[2]{\genfrac{}{}{0pt}{}{#1}{#2}}
\newcommand{\ketbra}[2]{\ket{#1}\!\bra{#2}}
\newcommand{\tr}[1]{\text{Tr}\!\left(#1\right)}
\newcommand{\kett}[1]{\ket{#1}\!\rangle}
\newcommand{\prb}[1]{\text{Pr}\!\left(#1\right)}

\usepackage{ifthen}
\newboolean{PrintVersion}
\setboolean{PrintVersion}{false}

\usepackage{amsmath,amssymb,amstext} 
\usepackage{braket}
\usepackage{newtxtext}
\usepackage[backend=bibtex,citestyle=numeric-comp,bibstyle=phys,url=false,doi=false,isbn=false,biblabel=brackets,sorting=none,minbibnames=15,maxbibnames=20,giveninits=true]{biblatex}
\DeclareCaseLangs{}

\usepackage{fontawesome}

\DeclareMathAlphabet\urwscr{U}{urwchancal}{m}{n}%
\DeclareMathAlphabet\rsfscr{U}{rsfso}{m}{n}
\DeclareMathAlphabet\euscr{U}{eus}{m}{n}
\DeclareFontEncoding{LS2}{}{}
\DeclareFontSubstitution{LS2}{stix}{m}{n}
\DeclareMathAlphabet\stixcal{LS2}{stixcal}{m} {n}

\usepackage{setspace}
\usepackage{xcolor}
\usepackage{enumitem}

\usepackage{graphicx} 

\usepackage[pagebackref=false]{hyperref} 
\hypersetup{
    plainpages=false,       
    unicode=false,          
    pdftoolbar=true,        
    pdfmenubar=true,        
    pdffitwindow=false,     
    pdfstartview={FitH},    
    pdfnewwindow=true,      
    colorlinks=true,        
    filecolor=black,      
    linkcolor={blue!80!black},
    citecolor={blue!80!black},
    urlcolor={blue!80!black}
}

\usepackage[margin=1in]{geometry}

\let\origdoublepage\cleardoublepage
\newcommand{\clearemptydoublepage}{%
  \clearpage{\pagestyle{empty}\origdoublepage}}
\let\cleardoublepage\clearemptydoublepage

\begin{document}

\pagestyle{empty}
\pagenumbering{roman}

\begin{titlepage}
\doublespacing
        \begin{center}
        \vspace*{-0.5cm}
        \large
        { UNIVERSITY OF CALGARY }
        
        \vspace*{1.25cm}

        \large
        {Modelling Markovian light-matter interactions for quantum optical devices\\ in the solid state}

        \vspace*{1.25cm}

        \normalsize
        by \\

        \vspace*{1.25cm}

        \large
        Stephen Christopher Wein \\

        \vspace*{1.75cm}

        \normalsize
        A THESIS\\
        SUBMITTED TO THE FACULTY OF GRADUATE STUDIES\\
        IN PARTIAL FULFILLMENT OF THE REQUIREMENTS FOR THE\\
        DEGREE OF DOCTOR OF PHILOSOPHY
        \vspace*{1.5cm}
        
        GRADUATE PROGRAM IN PHYSICS AND ASTRONOMY
        
        \vspace*{1.5cm}

        CALGARY, ALBERTA\\
        MARCH, 2021 \\

        \vspace*{1.5cm}

        \copyright\ Stephen Christopher Wein 2021 \\
        \end{center}
\end{titlepage}

\pagestyle{plain}
\setcounter{page}{2}

\cleardoublepage 

\begin{center}\textbf{Abstract}\end{center}
\addcontentsline{toc}{chapter}{Abstract}
\doublespacing
The desire to understand the interaction between light and matter has stimulated centuries of research, leading to technological achievements that have shaped our world. One contemporary frontier of research into light-matter interaction considers regimes where quantum effects dominate. By understanding and manipulating these quantum effects, a vast array of new quantum-enhanced technologies become accessible. In this thesis, I explore and analyze fundamental components and processes for quantum optical devices with a focus on solid-state quantum systems. This includes indistinguishable single-photon sources, deterministic sources of entangled photonic states, photon-heralded entanglement generation between remote quantum systems, and deterministic optically-mediated entangling gates between local quantum systems. For this analysis, I make heavy use of an analytic quantum trajectories approach applied to a general Markovian master equation of an optically-active quantum system, which I introduce as a photon-number decomposition. This approach allows for many realistic system imperfections, such as emitter pure dephasing, spin decoherence, and measurement imperfections, to be taken into account in a straightforward and comprehensive way.

\cleardoublepage

\begin{center}\textbf{Preface}\end{center}
\onehalfspacing
\doublespacing

The content of this thesis was either directly developed for projects or indirectly inspired by projects that I was part of. Some of the content has already been published, some of it has not yet been published, and other parts may not be published other than in this thesis. In addition, some of the published material of which I was the primary author is presented verbatim, although it may be rearranged and modified as appropriate to suit the flow of this thesis. Other published material from works where I was a co-author but not the primary author is presented from my own perspective and cited appropriately. In every case, I have diligently detailed my contributions to all these original works to provide full transparency.

The following is a full list of published and submitted papers that I have co-authored during my graduate studies. The list is presented in chronological order. For each entry, the PDF symbol provides a link to the corresponding arXiv manuscript and the chain symbol links to the journal article page if published and the arXiv abstract page if submitted. The asterisk symbol ($^*$) indicates equal contribution.

\singlespacing
\vspace{5mm}
\begin{enumerate}[label={[\roman*]}]
\item \label{wein2016bellmeasurement} \textbf{S. Wein}, K. Heshami, C. A. Fuchs, H. Krovi, Z. Dutton, W. Tittel, and C. Simon. "Efficiency of an enhanced linear optical Bell-state measurement scheme with realistic imperfections." Phys. Rev. A \textbf{94}, 032332 (2016). \href{https://arxiv.org/pdf/1509.00088.pdf}{ \faFilePdfO } \href{https://journals.aps.org/pra/abstract/10.1103/PhysRevA.94.032332}
{\faLink}
\vspace{2mm}
\item \label{dhand2018mysteries} I. Dhand, A. D'Souza, V. Narasimhachar, N. Sinclair, \textbf{S. Wein}, P. Zarkeshian, A. Poostindouz, and C. Simon. "Understanding quantum physics through simple experiments: from wave-particle duality to Bell's theorem." arXiv:1806.09958 (2018). \href{https://arxiv.org/pdf/1806.09958.pdf}{\faFilePdfO }  \href{https://arxiv.org/abs/1806.09958}{\faLink }\\
Note: being developed into a textbook for Cambridge University Press.
\vspace{2mm}
\item \label{wein2018plasmonics} \textbf{S. Wein}, N. Lauk, R. Ghobadi, and C. Simon. ``Feasibility of efficient room-temperature solid-state sources of indistinguishable single photons using ultrasmall mode volume cavities.'' Phys. Rev. B \textbf{97}, 205418 (2018). \href{https://arxiv.org/pdf/1710.03742.pdf}{ \faFilePdfO } \href{https://journals.aps.org/prb/abstract/10.1103/PhysRevB.97.205418}{\faLink}
\vspace{2mm}
\item \label{asadi2018repeaters}
F. Kimiaee Asadi, N. Lauk, \textbf{S. Wein}, N. Sinclair, C. O`Brien, and C. Simon. ``Quantum repeaters with individual rare-earth ions at telecommunication wavelengths.'' Quantum \textbf{2}, 93 (2018). \href{https://arxiv.org/pdf/1712.05356.pdf}{\faFilePdfO } \href{https://quantum-journal.org/papers/q-2018-09-13-93}{ \faLink}
\vspace{2mm}
\item \label{ghobadi2019cryogenfree} R. Ghobadi$^*$, \textbf{S. Wein}$^*$, H. Kaviani, P. Barclay, and C. Simon. ``Progress toward cryogen-free spin-photon interfaces based on nitrogen-vacancy centers and optomechanics.'' Phys. Rev. A \textbf{99}, 053825 (2019). \href{https://arxiv.org/pdf/1711.02027.pdf}{\faFilePdfO } \href{https://journals.aps.org/pra/abstract/10.1103/PhysRevA.99.053825}{\faLink}
\vspace{2mm}
\item \label{ollivier2020reproducibility} H. Ollivier, I. Maillette de Buy Wenniger, S. Thomas, \textbf{S. C. Wein}, A. Harouri, G. Coppola, P. Hilaire, C. Millet, A. Lema\^{i}tre, I. Sagnes, O. Krebs, L. Lanco, J. C. Loredo, C. Ant\'{o}n, N. Somaschi, and P. Senellart. ``Reproducibility of high-performance quantum dot single-photon sources.'' ACS Photonics \textbf{7}, 1050 (2020). \href{https://arxiv.org/pdf/1910.08863.pdf}{\faFilePdfO } \href{https://pubs.acs.org/doi/abs/10.1021/acsphotonics.9b01805}{\faLink}
\vspace{2mm}
\item \label{asadi2020cavitygates} F. Kimiaee Asadi, \textbf{S. C. Wein}, and C. Simon. ``Cavity-assisted controlled phase-flip gates.'' Phys. Rev. A \textbf{102}, 013703 (2020). \href{https://arxiv.org/pdf/1911.02176.pdf}{\faFilePdfO } \href{https://journals.aps.org/pra/abstract/10.1103/PhysRevA.102.013703}{\faLink}
\vspace{2mm}
\item \label{asadi2020repeaters} F. Kimiaee Asadi, \textbf{S. C. Wein}, and C. Simon. ``Protocols for long-distance quantum communication with single $^{167}$Er ions.'' Quantum Sci. Technol. \textbf{5}, 045015 (2020). \href{https://arxiv.org/pdf/2004.02998.pdf}{\faFilePdfO } \href{https://iopscience.iop.org/article/10.1088/2058-9565/abae7c}{\faLink}
\vspace{-2mm}
\item \label{wein2020entanglement} \textbf{S. C. Wein}, J.-W. Ji, Y.-F. Wu, F. Kimiaee Asadi, R. Ghobadi, and C. Simon.``Analyzing photon-count heralded entanglement generation between solid-state spin qubits by decomposing the master-equation dynamics.'' Phys. Rev. A \textbf{102}, 033701 (2020). \href{https://arxiv.org/pdf/2004.04786.pdf}{\faFilePdfO } \href{https://journals.aps.org/pra/abstract/10.1103/PhysRevA.102.033701}{\faLink}
\vspace{2mm}
\item \label{thomas2020LAphonons} S. E. Thomas, M. Billard, N. Coste, \textbf{S. C. Wein}, Priya, H. Ollivier, O. Krebs, L. Taza\"{i}rt, A. Harouri, A. Lema\^{i}tre, I. Sagnes, C. Ant\'{o}n, L. Lanco, N. Somaschi, J. C. Loredo, and P. Senellart. "Bright polarized single-photon source based on a linear dipole." arXiv:2007.04330 (2020). \href{https://arxiv.org/pdf/2007.04330.pdf}{\faFilePdfO } \href{https://arxiv.org/abs/2007.04330}{\faLink}
\vspace{2mm}
\item \label{sharman2020repeaters} K. Sharman, F. Kimiaee Asadi, \textbf{S. C. Wein}, and C. Simon. "Quantum repeaters based on individual electron spins and nuclear-spin-ensemble memories in quantum dots." arXiv: 2010.13863 (2020). \href{https://arxiv.org/pdf/2010.13863.pdf}{\faFilePdfO } \href{https://arxiv.org/abs/2010.13863}{\faLink}
\vspace{2mm}
\item \label{ji2020roomtemperature} J.-W. Ji, Y.-F. Wu, \textbf{S. C. Wein}, F. Kimiaee Asadi, R. Ghobadi, and C. Simon. "Proposal for room-temperature quantum repeaters with nitrogen-vacancy centers and
optomechanics." arXiv:2012.06687 (2020). \href{https://arxiv.org/pdf/2012.06687.pdf}{\faFilePdfO } \href{https://arxiv.org/abs/2012.06687}{\faLink}
\vspace{2mm}
\item \label{ollivier2020g2hom} H. Ollivier$^*$, S. E. Thomas$^*$, \textbf{S. C. Wein}, I. Maillette de Buy Wenniger, N. Coste, J. C. Loredo, N. Somaschi, A. Harouri, A. Lema\^{i}tre, I. Sagnes, L. Lanco, C. Simon, C. Ant\'{o}n, O. Krebs, and P. Senellart. "Hong-Ou-Mandel interference with imperfect single photon sources."
Phys. Rev. Lett. \textbf{126}, 063602 (2021). \href{https://arxiv.org/pdf/2005.01743.pdf}{\faFilePdfO } \href{https://journals.aps.org/prl/abstract/10.1103/PhysRevLett.126.063602}{\faLink}
\vspace{2mm}
\item \label{loredo2020deterministic} \textbf{S. C. Wein}, J. C. Loredo, M. Maffei, P. Hilaire A. Harouri, N. Somaschi, A. Lema\^{i}tre, I. Sagnes, L. Lanco, O. Krebs, A. Auff{\`e}ves, C. Simon, P. Senellart, and C. Ant\'{o}n-Solanas. "Photon-number entanglement generated by sequential excitation of a two-level atom." In preparation (2021). 
\vspace{2mm}
\end{enumerate}

\onehalfspacing
\doublespacing
Throughout this thesis, I will be using the above roman numeral references whenever citing works where I am a co-author. However, not all of these papers are relevant to the present topic. The papers \ref{wein2018plasmonics} and \ref{wein2020entanglement} are the only two that are entirely included in this thesis, and the author contributions for these works are summarized below. Please also see appendix \ref{AppendixC} for copyright permissions.

Author contributions for \ref{wein2018plasmonics}---\textbf{SW} and CS conceived the idea. \textbf{SW} and RG developed the methods. \textbf{SW} performed the analysis and wrote the manuscript with guidance from NL and RG. NL and CS provided critical feedback. All authors contributed to editing the manuscript.

Author contributions for \ref{wein2020entanglement}---\textbf{SCW} and CS conceived the idea. \textbf{SCW} developed the methods. \textbf{SCW} performed the analysis and wrote the manuscript with help from JWJ, YFW, and FKA. RG and CS provided critical feedback. All authors contributed to editing the manuscript.

A portion of supplementary theory material from \ref{ollivier2020g2hom} and \ref{loredo2020deterministic} are also described in this thesis under fair dealing, as I was the primary author of that material within those experimental papers. Moreover, I will include unpublished research related to my contributions to \ref{asadi2020cavitygates} and \ref{asadi2020repeaters}. Both of these latter papers have already appeared in the thesis of the primary author, Dr. Faezeh Kimiaee Asadi. Because of this, I will make clear what material deviates from the published material and present any material from those published papers in my own original words with the appropriate citations. The remaining original material in this thesis is tangentially related to my other publications and I will detail those relationships as they become relevant.

The content of \ref{wein2018plasmonics} appears in section \ref{chapter3:roomtemperature} and appendix \ref{AppendixB}. In addition, \ref{wein2020entanglement} appears in chapter \ref{chapter2} and section \ref{chapter4:entanglementgeneration}. Some supplementary material related to \ref{ollivier2020g2hom} and \ref{loredo2020deterministic} is included in chapter \ref{chapter3}. The material related to \ref{asadi2020cavitygates} and \ref{asadi2020repeaters} appears in section \ref{chapter4:cavitygates}. Aside from the introduction, the remaining material that is not attributed to a paper is, to the best of my knowledge, original unpublished research that serves to aggregate my published works into a consistent framework.
\addcontentsline{toc}{chapter}{Preface}
\cleardoublepage
\phantomsection		


\begin{center}\textbf{Acknowledgements}\end{center}
\addcontentsline{toc}{chapter}{Acknowledgements}

I am genuinely grateful to everyone who has contributed to my personal and academic development, allowing for the completion of this milestone in my life. Although I cannot mention everyone, I hope to highlight those who have had the most significant impact.

I would like to express my heartfelt thanks to all the people who have mentored me through my studies. Most importantly, to my advisor Prof. Christoph Simon, whose wisdom and mindful guidance has undoubtedly shaped both my career and my life. To Dr. Dean Kokonas, whose words---\emph{``you just need to find your groove"} said to me when I was failing grade 10 math--- echo in my head whenever I struggle with a concept. To Mr. Bob Shoults, who never failed to pique my curiosity for physics. To Prof. Michael Wieser, whose lectures on quantum mechanics inspired me to switch undergraduate programs and study physics. And to Prof. David Feder, whose enthusiasm brought me into physics research.

This work would not be as it is without the influence from many of my collaborators, colleagues, and peers. I would like to thank all past and present members of Prof. Simon's theory group who have shaped my research environment; and also Prof. Pascale Senellart and all members of GOSS at the Centre for Nanosciences and Nanotechnologies for inspiring discussions and projects. In particular, I am very grateful to Prof. Khabat Heshami for guidance on my first research paper, to Prof. Sandeep Goyal for initiating my interest in open quantum systems, and Dr. Faezeh Kimiaee Asadi for the many productive discussions and successful collaborations.

The completion of my doctorate would not have been possible without my family and friends. I am foremost indebted to my parents, Beverly and Marcus Wein, who are the source of my confidence and continue to teach me to think critically. I cannot thank enough my partner in life and love, H\'{e}l\`{e}ne Ollivier, who inspires me everyday with her insight, fortitude, and wit. I am also grateful to my siblings, Carissa Murray and Michael Wein, whose encouragement and compassion give me the determination to pursue my dreams; to Florence and Yvonnick Ollivier for all their help and kindness; and to my closest friends, Levi Eisler and Emily Wein, for their moral support.

I am fortunate to have received financial help from many donors and institutions. Thank you to all the donors and estates thereof that contributed to scholarships and bursaries that I have received during my studies, including the TransAlta Corporation, the Don Mazankowski Scholarship Foundation, Gerald Roberts and Victor Emanuel Mortimer, Wilfred Archibald Walter, Alvin and Kathleen Bothner, William Robert Grainger, and the family, friends and colleagues of Dr. Robert J. Torrence. Thank you also to the University of Calgary's Faculty of Science, Faculty of Graduate Studies, Department of Physics and Astronomy, and Institute for Quantum Science and Technology for their support. My research would not have been possible without the early support from the University of Calgary's Program for Undergraduate Research Experience (PURE) and the Natural Sciences and Engineering Research Council of Canada (NSERC) Undergraduate Student Research Award. I am also thankful for the help from the Alberta Innovates-Technology Futures Graduate Student Scholarships, the NSERC Canadian Graduate Scholarships (CGS)-Master's, the NSERC Alexander Graham Bell CGS, and the SPIE Optics and Photonics Education Scholarships.

Lastly, I would like to thank all the members of my supervisory and exam committees during my graduate studies. Thank you to my supervisory committee members: Prof. Paul Barclay, Prof. Wolfgang Tittel, Prof. Daniel Oblak, and Prof. David Hobill, for their constructive feedback and guidance. Thank you to Prof. Belinda Heyne and to Prof. David Feder for serving on my candidacy exam committee, and to Prof. David Knudsen for serving on my thesis exam committee. Finally, thank you to Prof. Stephen Hughes who served as the external examiner for my thesis defense and provided critical feedback that helped me improve my work.

\cleardoublepage


\begin{center}\textbf{Dedication}\end{center}
\addcontentsline{toc}{chapter}{Dedication}

To Wolfgang and Sophie, my two wonderful children who fill my life with happiness.
\cleardoublepage

\setstretch{1.3} 
\renewcommand\contentsname{Table of Contents}
\addcontentsline{toc}{chapter}{Table of Contents}
\tableofcontents
\cleardoublepage
\phantomsection    
\doublespacing

\setstretch{1.4} 
\addcontentsline{toc}{chapter}{List of Figures}
\listoffigures
\cleardoublepage
\phantomsection		

\addcontentsline{toc}{chapter}{List of Tables}
\listoftables
\cleardoublepage
\phantomsection		

\addcontentsline{toc}{chapter}{List of Abbreviations}
\chapter*{List of Abbreviations}

\begin{tabular}{l l}
    Abbreviation\hspace{40mm}\vphantom{l} & Definition\hspace{75mm}\vphantom{l}\\\hline
    BD & Bin detector\\
    BS & Beam splitter\\
    BSM & Bell-state measurement\\
    CW & Continuous wave\\
    EPR & Einstein-Podolski-Rosen\\
    FWHM & Full width at half maximum\\
    HOM & Hong-Ou-Mandel \\
    LDOS & Local density of states\\
    LO & Local oscillator\\
    NV & Nitrogen-vacancy\\
    PNRD & Photon-number resolving detector\\ 
    POVM & Positive operator-valued measure\\
    PSB & Phonon sideband\\
    PVM & Projection-valued measure\\
    QD & Quantum dot\\
    QED & Quantum electrodynamics\\
    QIP & Quantum information processing\\
    QNM & Quasi-normal mode\\
    Qubit & Quantum bit\\
    SiV & Silicon-vacancy\\
    SPDC & Spontaneous parametric down-conversion\\
    SPS & Single-photon source\\
    ZPL & Zero-phonon line\\
\end{tabular}

\cleardoublepage
\phantomsection		

\pagenumbering{arabic}


\chapter{Introduction}
\setstretch{1}
\doublespacing

In this chapter, I will establish the context for this thesis. The first section will provide a brief overview of the history of research into light and matter without delving into the modern mathematical language. This will include a description of the contemporary applications of quantum light-matter interactions for fundamental research and technology and also a motivation for the systems of focus in this thesis. In the remaining of this chapter, I will then introduce the theoretical framework and basic mathematical background used throughout this thesis.

 \section{Motivation}

\subsection{History}

The conceptual relationship between light and matter has been debated for millennia. In the era of ancient Greek philosophers, it was postulated that light behaved like a ray while matter was composed of indivisible pieces or `atomos'. However, even in the ancient world, there were popular theories proposed that light rays were also composed of indivisible particles \cite{al2016optics}. During the Renaissance when the foundations of classical science were being established, Christiaan Huygens and Isaac Newton debated whether light was a wave-like phenomenon or if it was composed of discrete corpuscles, respectively. At the time, a particle-like model was generally accepted because it could explain the straight motion and polarization properties of light. This dominance shifted once the wave-like properties of interference and diffraction became well-understood by the likes of Thomas Young and Augustin-Jean Fresnel \cite{al2016optics}. On the other hand, some evidence for the particle nature of matter arose with the atomic theory of John Dalton in chemistry at the end of the 18th century \cite{thackray1966origin}. Regardless of their conceptual similarities and differences, the physics and theory of light and matter were largely segregated into very different fields of study.

The 19th century brought with it the industrial revolution instigated by discoveries in thermodynamics and statistical mechanics propelled by scientists such as Sadi Carnot and Ludwig Boltzmann, among many others. Alongside these advancements in understanding matter physics were the seminal electric and magnetic experimental observations by Michael Faraday that were later developed into the classical unified theory of electromagnetism and light founded by James Clerk Maxwell \cite{maxwell1865viii}. This fully established the concept that light was an electromagnetic wave. Furthermore, with the discovery of the electron by Joseph John Thomson \cite{falconer1997jj}, the particle nature of matter was gaining support. By the end of the 19th century, it may have appeared that physics was nearing completion.

Spawning from Max Planck's curiosity for black-body radiation \cite{al2016optics} that led him to revive the idea of discretized light, the definitive counter-intuitive connection between light and matter began to emerge. Soon after, this hypothesis was re-enforced by Albert Einstein's explanation of the photoelectric effect \cite{pais1979einstein}, which was observed by Heinrich Hertz. The behaviour of light could not be fully described as a wave when it interacted with matter. The existence of the fundamental particle of light---the photon---was discovered. Around the same time, Jean Perrin experimentally demonstrated the atomic nature of matter by verifying Einstein's theory of Brownian motion \cite{pais1979einstein}. Less than two decades after the particle nature of matter was unambiguously confirmed came the theoretical discoveries by Louis de Broglie that matter could also exhibit wave-like behaviour, which was itself confirmed shortly thereafter by observations of electron diffraction \cite{pais1979einstein}. Within a few decades, the conceptual line separating light and matter blurred into a perplexing duality that challenged the brightest minds of the 20th century.

The desire to understand and reconcile the intricate relationship between light and matter was the driving force behind the development of a new field of study---quantum physics. The pioneering advancements made by Einstein and Planck stimulated great debates and progress throughout the first few decades of the 20th century. For a while, physicists such as Niels Bohr and Arnold Sommerfeld sought to patch ever-growing inconsistencies between classical explanations and new observations in atomic physics. Then, in the fall of 1925, Werner Heisenberg along with Max Born and Pascual Jordan brought consistency by introducing the concept that properties of quantum particles of light and matter can be described by matrices \cite{al2016optics}. Just months after, Erwin Schr\"{o}dinger published his seminal paper on the differential wave equation for quantum states---the famous Schr\"{o}dinger equation---that reconciled wave-particle duality as a manifestation of quantum superposition \cite{pais1979einstein}. Shortly after that, Schr\"{o}dinger united his theory of wave mechanics with the matrix mechanics of Heisenberg, forming quantum mechanics---the foundations of modern quantum physics.

With a mathematically consistent framework, quantum mechanics was poised to stoke the fires of a new technological revolution paralleling that of the 19th century. However, quantum mechanics troubled many of its founding authors. Many of the steps advancing towards a quantum theory were taken reluctantly. Max Planck was unsatisfied to find that quantization explained black-body radiation. Louis de Broglie did not accept the mathematically abstract formulation of wave mechanics introduced by Schr\"{o}dinger. Most famously, Albert Einstein, Boris Podolski, and Nathan Rosen together published a paper in 1935 that challenged the extent to which quantum mechanics explained reality---the EPR paradox \cite{einstein1935can}. The paradox surrounded one peculiar consequence of quantum mechanics: the special case of \emph{entanglement} where two particles can seemingly influence each other's behaviour, even when separated by arbitrarily large distances. This ``spooky action at a distance" (``spukhafte Fernwirkung”), as Einstein called it, was completely counter-intuitive and it seemed to contradict his newly-developed theory of relativity. It was also somewhat ironic that Einstein had just resolved the problem of action at a distance in gravitational physics only to discover it appears in quantum physics. Unsurprisingly, this turn of events led Einstein and his colleagues to conclude that quantum physics must be incomplete.

The unanswered questions and confusing interpretations surrounding quantum mechanics were nonetheless impotent to hinder its influence. Regardless of its subtleties, quantum mechanics explained physical observations to such an accurate degree that it heralded an unstoppable technological revolution. It influenced developments of devices in fields such as nuclear physics, optics, and electronics; and medical technologies such as magnetic resonance imaging (MRI) and positron emission tomography (PET). In the 1960s, the laser was demonstrated after significant theoretical development following quantum theory. This invention alone allowed for uncountable applications in communications, sensing, and medical procedures. It also brought excimer laser lithography, modern transistors and, through these, semi-conductor microelectronics leading to modern computers and smart phones. 

Quantum theory itself continued to develop significantly through the mid 20th century. Notably, Paul Dirac developed relativistic quantum mechanics and introduced the contemporary Dirac notation for quantum physics. Mathematician and physicist John von Neumann developed the connection to functional analysis by pioneering the modern mathematical description in terms of Hilbert spaces of quantum states, quantum measurements, and operator theory. Still, the nagging question embodied by the EPR paradox lingered.

In 1964, almost 30 years after the EPR paradox was introduced, John Stewart Bell developed a theorem that brought with it a definitive proposal to test whether the quantum predictions arising from entanglement indeed explained reality in a way that no reasonable classical theory could \cite{bell1964einstein}. The proposal, now known as a Bell inequality, was a statistical test based on measuring entangled particles and comparing the average outcomes to the fundamental bounds allowed by classically correlated outcomes. Bell predicted that certain entangled states of quantum particles could indeed give stronger correlations than any classical analog. If proven correct, Bell's theorem would finally resolve the EPR paradox.

With Bell's theorem, entanglement went from being a peculiar consequence of quantum physics to a unique resource of quantum physics. It soon became a hard-driven goal to develop ways to prepare and manipulate quantum systems to generate entangled states, and then to measure those states with high accuracy. The concept of entanglement as a resource motivated a series of technological advances, setting the stage for a second quantum revolution.

Although Bell originally proposed to use electrons to test his theorem, it soon became apparent that photons were the better choice. In 1972, John Clauser and Stuart Jay Freedman provided strong evidence for the validity of Bell's theorem using entangled photons generated by cascaded emission from calcium atoms \cite{freedman1972experimental}. However, the important fundamental consequences of Bell's theorem required extreme care to ensure that the experiments leave no possible classical explanation, called loopholes \cite{larsson2014loopholes}. Hence, there were a series of experiments stretching from 1972 to 2013 that made important incremental steps towards the first loophole-free experiments published in 2015 \cite{hensen2015loophole,giustina2015significant,shalm2015strong}. 

Among the Bell test experiments were those of Alain Aspect \emph{et al.} \cite{aspect1982experimental}, Wolfgang Tittel \emph{et al.} led by Nicolas Gisin \cite{tittel1998violation}, and Gregor Weihs \emph{et al.} led by Anton Zeilinger \cite{weihs1998violation} that made use of spontaneous parametric down-conversion (SPDC) \cite{kwiat1995}, a process based on non-linear light-matter interaction used to generate entangled photon pairs. Notably, the first loophole-free Bell test was accomplished by Ronald Hanson's group in Delft by exploiting light-matter interaction in single atomic defects in diamond to generate entanglement between two solid-state quantum systems separated by 1.3 kilometers \cite{hensen2015loophole}. Shortly after, two other loophole-free Bell tests published by independent groups in Vienna and Boulder utilized entangled photons generated by SPDC, yielding impressive statistical significance of violation \cite{giustina2015significant,shalm2015strong}. All these Bell test experiments paralleled the development of quantum communication technology, which is one major category of second-generation quantum technology that can exploit entanglement.

Another major category of technology was spurred by the concept of entanglement as a resource. In the 1980s, it became apparent that quantum phenomena, such as entanglement, could allow for the implementation of computational algorithms that vastly exceeded classical limitations. Pioneered by physicists Paul Beinoff \cite{benioff1980computer} and Richard Feymann \cite{feynman1982simulating}, the field of quantum computation has seen huge growth over the past four decades, and a massive increase in commercial interest over the last decade. Just within the last two years, the first experimental demonstrations have been published showing that, for some specific tasks, quantum devices can outperform the best-known classical algorithms run on a supercomputer \cite{arute2019quantum,zhong2020quantum}.

\subsection{Applications}
\label{chapter1:applications}

To this day, light-matter interactions in quantum physics remains a key topic that motivates both fundamental and technological advances. The three pillars of quantum technology spearheading the second quantum revolution are (1) quantum communication, (2) quantum computation, and (3) quantum sensing. The first two pillars have already profoundly impacted the fundamental development of quantum physics, but all three open the door to devices and applications that would otherwise be impossible.

Quantum communication focuses on using the principles of superposition and entanglement to perform tasks between spatially separated parties. The most advanced application is quantum-secure key distribution \cite{gisin2002quantum}, where one uses the laws of quantum physics to ensure the security of a generated random key to be used for encrypting classical information. Current implementations rely on the transmission of photons through telecommunication fibre optic cables \cite{valivarthi2016quantum}, but long-distance communication will require free-space transmission using satellites \cite{aspelmeyer2003long,boone2015entanglement,liao2017satellite,yin2017satellite} and quantum repeater nodes along a network of fibre optics \cite{briegel1998quantum}. Quantum communication networks are currently being developed around the world \cite{liao2018satellite,valivarthi2020teleportation,pompili2021realization}. These networks may become the basis for a quantum internet \cite{kimble2008internet,simon2017internet,wehner2018internet}, where quantum resources can be distributed across the globe to enhance the operation of other novel devices and applications involving the other two pillars of quantum technology. Furthermore, quantum networks allow for a platform on which to test fundamental questions about the nature of quantum phenomena, such as Bell inequalities and quantum collapse models \cite{rideout2012fundamental}.

Quantum computation and simulation stand to provide the most technologically impressive results. Already, optical quantum systems are capable of performing information processing tasks in mere minutes that would take current classical algorithms on super computers billions of years \cite{zhong2020quantum}. Although the demonstrated tasks are system-specific and not yet useful, there are plenty of proposed algorithms and applications that quantum computers may be able to tackle in the future. The seminal example by Peter Shor promises an exponential speed-up of prime factorization \cite{shor1999polynomial}. Quantum simulation could solve complicated protein folding problems \cite{kassal2011simulating} that would lead to advances in medicine, shed light on the process of nitrogen fixation allowing for improved agriculture practices \cite{reiher2017elucidating}, accelerate materials engineering, and improve solar energy conversion and power transmission \cite{preskill2018quantum}. Quantum computation could also enhance artificial intelligence and machine learning \cite{biamonte2017quantum}, efficiently solve weather and climate models for accurate forecasting \cite{gaitan2020finding}, assist financial modelling \cite{orus2019quantum}, and optimize routing and traffic control \cite{stollenwerk2019quantum}. With the help of quantum networks, quantum computing could also be performed over spatially distributed systems and allow for blind quantum computing \cite{barz2012demonstration}.

Quantum sensing uses quantum effects to measure a system with accuracy and resolution exceeding the limits of classical technology \cite{giovannetti2006quantum,giovannetti2011advances}. This ranges from magnetic field sensing using atomic defects in nanoparticles \cite{mamin2013nanoscale} to superresolution of thermal light sources \cite{wang2020superresolution}. Quantum effects can also be exploited to improve interferometric phase measurements \cite{caves1981quantum}, which can increase the sensitivity of gravitational wave detectors \cite{aasi2013enhanced,mcculler2020frequency}. Quantum light may also improve methods for imaging biological tissues without damaging them \cite{genovese2016real}. With the help of quantum networks it may be possible to implement global synchronized access to a single quantum timekeeping device \cite{komar2014quantum} and improve the resolution of astronomical objects \cite{gottesman2012longer,stankus2020two}.

In all three of these pillars, one can find the extensive involvement of light-matter interactions \cite{northup2014quantum}. In particular, many technologies such as quantum networks, optical quantum computing, and biological imaging require the generation of non-classical light, such as single photons and entangled photonic states. These types of states are usually produced by the interaction of a laser with a solid-state material or atomic gas. For example, as mentioned before, SPDC can produce entangled photons. It is also one of the most common ways to generate high-quality single photons. This process occurs when laser light passes through a non-linear crystal. More recently, there has been significant motivation to study how laser light can manipulate optically-active single quantum defects in the solid-state to produce non-classical light and light-matter entanglement \cite{awschalom2018quantum,atature2018material}. These solid-state emitter systems have great potential to accelerate the development of many, if not all, of the technological applications mentioned above.

\subsection{Solid-state optical devices}
\label{chapter1:solidstateoptialdevices}
The technological evolution of classical computing and the internet has shown a clear trend towards favoring compact, scalable, and robust devices. For example, vacuum tube transistors were replaced by contemporary solid-state silicon-based technology while fiber optics and satellites have now replaced many electrical transmission lines for long-distance communication. A similar trend could be expected for the development of quantum technology whereby solid-state materials and optical light become more favored over time.

In the previous section, I touched on the multitude of applications of quantum technology. In particular, optically-active solid-state defects will continue to play a huge role in the future of the ongoing quantum technology revolution \cite{awschalom2018quantum,atature2018material}. These defects could provide many advantages over other platforms for quantum technology such as trapped ions \cite{haffner2008quantum} or superconducting circuits \cite{wendin2017quantum}. In addition, optically-active defects have the potential to satisfy any future demand for scalability and robustness. However, there are still many challenges to overcome. In this section, I will highlight some of the advantages and disadvantages of this platform.

The optical spectrum of light combines the infrared, visible, and ultraviolet ranges of the electromagnetic spectrum, comprising wavelengths near 100s of micrometers to 10s of nanometers. In terms of frequency, this can range from 100s of GHz to 1000s of THz. The range covers two prominent telecommunication bands at 1300~nm and 1500~nm, where transmission losses are low through fibre optics. It also covers the operational wavelengths of commercially available lasers such as titanium-sapphire and helium-neon lasers. Many semi-conductor materials, such as silicon, gallium arsenide, and diamond, have optical band-gaps that allow for addressable defects, such as quantum dots (QDs) \cite{woggon1997optical} and nitrogen-vacancy centers (NV centers) \cite{doherty2013nitrogen}. Optical light also transmits well in free space and can be manipulated with many off-the-shelf components such as silvered or dielectric mirrors, half-silvered beam splitters, objectives, and lenses. There are also plenty of commercially-available efficient detectors for optical light. 

Optical light (as with all light) is composed of photons, which generally interact only weakly with other particles. This makes it difficult to manipulate photons to perform logic gates, usually requiring probabilistic post-selection to accomplish \cite{pittman2001probabilistic}. They are also particles that travel quickly and are difficult to contain. This makes photons ideal for transmitting quantum information over long distances \cite{atature2018material} but a poor choice for storage of quantum information in one location. That said, there are very promising proposals for all-optical quantum computing \cite{o2007optical}. But even then, light-matter interactions are necessary to generate and detect the quantum states of light used to implement all-optical quantum information processing.

Solid-state quantum defects are microscopic or atomic-scale defects within a crystal lattice that can spatially confine electrons. This electron confinement is similar to how a nucleus can trap electrons to form an atom. These quantum systems are referred to as artificial atoms for this reason. Electrons trapped by a defect can exist in different quantum states, or electronic orbitals, dictated by the defect structure. When there are multiple electrons, they together occupy the molecular orbitals of the defect. Since electrons interact with light, laser pulses can be used to provide the right amount of energy to excite electrons from one orbital to another. Excited electrons may also spontaneously emit photons back into the environment. For this reason, I will refer to a single quantum defect, such as a QD or NV center, that is in an electronic configuration allowing for the emission of photons, as an emitter.

Each electron confined to a defect can itself exist in one of two spin states. Depending on the number of confined electrons and the defect symmetry, there may be optically-active spin states that can be manipulated by external static or dynamic electric and magnetic fields. This can provide a robust degree of freedom for manipulating and storing quantum information. Thus, optically active solid-state defects allow for a rich space of quantum states that are spatially localized, can be externally manipulated, may contain spin degrees of freedom, that can coherently interact with photons, and generate quantum states of light \cite{atature2018material}.

Solid-state defects have some prominent advantages over other stationary quantum systems, such as confined atomic gases or trapped atom/ion systems. Because the defects are integrated into a solid-state lattice, they do not require external manipulation to keep them localized in space. In addition, there are many well-developed methods for fabricating solid-state systems that can be adopted from semi-conductor microelectronics. This latter advantage also allows for the natural integration of quantum defects with classical information processing architectures, potentially providing a smooth and scalable transition to quantum-enhanced devices. 

By combining the benefits of optical light with those of optically-active solid-state defects that may contain spin degrees of freedom, we have a platform that can prepare, manipulate, store or transmit quantum information. This solid-state spin-photon platform satisfies all the main requirements for quantum technology to succeed \cite{divincenzo2000physical,awschalom2018quantum}. However, solid-state environments are hostile to delicate quantum phenomena. In the next section, I will discuss the biggest disadvantage of using solid-state materials for quantum information processing.

\subsection{Decoherence}
\label{chapter1:deoherence}

The primary challenge for developing quantum technology is decoherence. This occurs when a quantum system's environment perturbs the quantum system in an uncontrolled, probabilistic way \cite{fisher2003quantum}. The system then deviates from the desired quantum state, which can destroy the quantum properties such as superposition and entanglement that are required for quantum technology to operate. Decoherence occurs quickly for solid-state systems because there are many ways that the surrounding host environment can affect the state of the quantum defect. Thus, the main advantage of solid-state defects also gives rise to its biggest weakness. The three main culprits of decoherence in solid-state systems are thermal vibrations, electric noise, and magnetic noise \cite{chirolli2008decoherence}. 

Bulk thermal vibrations of atoms in the host lattice can displace the electronic orbitals of the quantum defect, causing decoherence. The quantized description of these thermal vibrations are called phonons. Decoherence becomes significantly worse as the lattice temperature increases. However, even at very low temperatures, energy contained in the defect can also spontaneously dissipate into the environment as phonons, which also causes decoherence. The interaction between the confined electrons and lattice phonons can dampen the light-matter interaction, putting limits on the speed and fidelity at which the electron can be manipulated with laser pulses \cite{mccutcheon2010quantum}.

Electronic noise is often induced by charge fluctuations \cite{somaschi2016}. This can occur if there are uncontrolled electrons moving into and out of other nearby defects or moving across the material surface. As a consequence, the local electric field experienced by the electrons within the defect can change. The rate of charge fluctuations is also temperature dependent. However, the severity depends a lot on the surface quality (if near a surface) and the density of the surrounding defects.

Magnetic noise is usually caused by random changes in the spin states of surrounding atomic nuclei. If a nearby nucleus flips from one spin state to another, this can change the local magnetic field experienced by electrons within the defect \cite{chirolli2008decoherence}. The many surrounding nuclei form what is called a spin bath. Random fluctuations in the state of this nuclear spin bath can cause decoherence of the electronic spin state of the defect.

In some cases, the environmental noise can be minimized. For example, low-frequency phonons may be suppressed by cutting a pattern into the material to forbid some vibration modes \cite{lutz2016modification}. However, this is not effective against high-frequency phonons. Charge noise can often be controlled by applying a static electric field to lock free charges in place \cite{somaschi2016}. Magnetic noise can sometimes be suppressed by isotopically engineering the material so that the nuclei do not have a spin degree of freedom at all \cite{balasubramanian2009}. Unfortunately, not all atoms have stable isotopes with zero nuclear spin. For some slower sources of noise, such as magnetic field fluctuations, rapid pulse sequences can also be used to dynamically decouple the quantum system from its environment \cite{viola1999dynamical}.

Besides minimizing noise, decoherence can be overcome by operating the device much faster than the timescale of decoherence. This means that the system should be prepared, manipulated, and measured before decoherence can occur. For interacting quantum subsystems, it is sometimes possible to enhance the rate of interaction to complete the protocol before decoherence becomes a major hindrance to the operation of the device. Unfortunately, often a portion of decoherence occurs on a timescale that cannot be realistically overcome \cite{iles2017phonon}. In some cases for large quantum systems, certain measurements and operations can be applied to correct for errors \cite{dur2007entanglement}. This may one day allow for protocols that can be implemented over arbitrary lengths of time. However, these error correction protocols require the system to already satisfy some threshold of quality to be successful \cite{dur1999quantum}. Hence, understanding and circumventing the effects of decoherence is still essential for the advancement of quantum technology.

 \subsection{Outline}

There are many different physical implementations of solid-state quantum optical devices for many different applications. Each implementation may be affected by its environment in a different way than the next. In this thesis, I analyze arguably the most simple type of model that can capture effects of decoherence on a quantum system: the phenomenological Markovian master equation. As I will discuss later in this chapter, this approach makes some simplifying assumptions about how a quantum system interacts with its environment. This allows for overarching and generalized results that capture the basic behaviour and limitations of many different devices. In addition, the methods I detail in chapter \ref{chapter2} and apply throughout the thesis can also be applied to devices with more specific types of Markovian master equations that take into account additional subtleties in the behaviour of a device. However, Markovian master equations usually cannot capture all subtle details of a particular system, especially those in solid-state environments, and may not capture limiting behaviour in some regimes of operation. For this reason, the Markovian master equation approach used in this thesis should be considered a first step in modelling decoherence that gives insight to further develop system-specific models that are more closely tailored for a particular device.

I will use the Markovian master equation approach to explore the effects of environment noise on two main types of applications of solid-state optical defects. The first type of application, discussed in chapter \ref{chapter3}, are those related to the generation of single photons (sections \ref{chapter3:roomtemperature} and \ref{chapter3:HOM}) and other pulsed photonic states (section \ref{chapter3:photonicstate}). The second type of application, discussed in chapter \ref{chapter4}, is related to using photons to mediate the interaction between two different defects. In section \ref{chapter4:entanglementgeneration}, I cover a few protocols that can be used to generate entangled spin states between remote defects. Section \ref{chapter4:cavitygates} follows up with a discussion on using photons to mediate deterministic local interactions between spins states of different defects. Finally, I summarize the main results and conclude the thesis in chapter \ref{chapter5}.

\section{Quantum systems}
\label{chapter1:quantumsystems}

In this section, I will introduce the basics of quantum mechanics that set the foundation for the analysis used in this thesis. In order to be concise and pertinent, I focus on using the Schr\"{o}dinger picture to introduce discrete variable systems that do not contain degenerate states nor utilize degenerate measurements of those states. The majority of content in this section and those that follow is drawn from the open quantum systems textbook by H.-P. Breuer and F. Petruccione \cite{breuer2002theory}.

A quantum system is a system described by a Hilbert space $\mathbb{H}$ of possible states. A state in this Hilbert space is represented by a complex-valued vector $\ket{\psi}$. This vector can be written as a linear superposition of orthonormal basis vectors $\ket{\phi_n}\in\mathbb{H}$ so that $\ket{\psi}=\sum_n c_n\ket{\phi_n}$. The complex amplitudes $c_n\in\mathbb{C}$ are given by the inner product $c_n=\braket{\phi_n|\psi}$, where $\bra{\phi_n}$ is the covector of $\ket{\phi_n}$ that belongs to the dual space of $\mathbb{H}$. The amplitudes $c_n$ have a particular physical interpretation as giving the probability $\prb{\phi_n}=|c_n|^2$ of finding the quantum system in the state $\ket{\phi_n}$. Hence, the state $\ket{\psi}$ must be restricted to satisfy the normalization condition $\sum_n\prb{\phi_n}=1$. The stochastic behaviour of the quantum system under measurement is entirely quantum in that it cannot be described by reasonable classical models \cite{bell1964einstein,kochen1975problem,pusey2012reality}. I will discuss measurements in more detail in the following section.

A quantum system can change from one state to another. Changes in the system state are described by linear operators denoted $\hat{U}:\mathbb{H}\rightarrow\mathbb{H}$, which have a particular property that they maintain the normalization condition of quantum states. That is, they are unitary transformations so that $\hat{U}^\dagger\hat{U}=\hat{U}\hat{U}^\dagger=\hat{I}$, where $\hat{I}$ is the identity operator. There are also linear operators $\hat{A}$ acting on the Hilbert space that are not unitary. These operators can describe the time dynamics, average physical quantities, and measurements of the system.

Just like a state, every linear operator can be written in terms of an orthonormal basis $\ket{\phi_n}$ of the Hilbert space as $\hat{A}=\sum_{n,m}A_{n,m}\ketbra{\phi_n}{\phi_m}$, where $A_{n,m}=\braket{\phi_n|\hat{A}|\phi_m}$ are the complex-valued matrix elements of $\hat{A}$ and $\ketbra{\phi_n}{\phi_m}$ is the outer product of states $\ket{\phi_n}$ and $\ket{\phi_m}$. The expectation value of a linear operator on the Hilbert space is determined by the quantum state $\ket{\psi}$ using $\braket{\hat{A}}=\braket{\psi|\hat{A}|\psi}=\sum_{n,m}c_n^*c_m A_{n,m}$. 

An operator $\hat{O}$ is said to be an observable if it is Hermitian $\hat{O}^\dagger=\hat{O}$. Observable operators correspond to physical quantities of the system that can be measured. By convention, we define the adjoint $\hat{O}^\dagger$ as the conjugate transpose of the operator's matrix representation. This implies that all observables have real eigenvalues and that they are diagonalizable. Hence, we can always find a new basis $\ket{\varphi_n}$ such that an observable $\hat{O}$ can be written as $\hat{O}=\sum_n\alpha_n\ketbra{\varphi_n}{\varphi_n}$, where $\alpha_n$ are the eigenvalues satisfying $\hat{O}\ket{\varphi_n}=\alpha_n\ket{\varphi_n}$.

In practice, our knowledge about the system state may be incomplete so that it is best described as a classical statistical mixture of possible quantum states. For example, it could be in the state $\ket{\psi_1}$ with probability $\prb{\psi_1}$ and state $\ket{\psi_2}$ with probability $\prb{\psi_2}=1-\prb{\psi_1}$. This type of classical uncertainty in the state of the system is physically very different from the quantum uncertainty described at the beginning of this section. Rather than tracking all possible statistical outcomes of a quantum system individually, we can describe the statistically mixed state of a quantum system using an observable operator, the density operator $\hat{\rho}$. Suppose we want the expectation value of $\hat{\rho}$ for a given state $\ket{\phi_n}$ to be $\prb{\phi_n}$, it is then intuitive that the operator $\ketbra{\psi}{\psi}$ is chosen so that $\prb{\phi_n}=\braket{\phi_n|\hat{\rho}|\phi_n}=\left|\braket{\phi_n|\psi}\right|^2$. In general, we have $\hat{\rho}\equiv\sum_n \prb{\psi_n}\ketbra{\psi_n}{\psi_n}$ where $\sum_n\prb{\psi_n}=1$. A quantum system that is in a state $\ket{\psi}$ where $\prb{\psi}=1$ is said to be in a pure quantum state. Otherwise, it is said to be in a mixed state.

Written in terms of the orthonormal basis $\ket{\phi_n}$, a general quantum state is
\begin{equation}
    \hat{\rho}=\sum_{n,m}\rho_{n,m}\ketbra{\phi_n}{\phi_m}
\end{equation}
where $\rho_{n,m}$ are called the density matrix elements. The diagonal elements $\rho_{n,n}=\braket{\phi_n|\hat{\rho}|\phi_n}=\prb{\phi_n}$ are equal to the probability of finding the quantum system in state $\ket{\phi_n}$ while the off-diagonal elements $\rho_{n,m}=\braket{\phi_n|\hat{\rho}|\phi_m}$ ($n\neq m$) quantify the amount of quantum coherence between states $\ket{\phi_n}$ and $\ket{\phi_m}$. Hence, with this formalism, both the quantum and classical statistical behaviours of a quantum system are captured.

The density operator is Hermitian $\hat{\rho}^\dagger=\hat{\rho}$ and by definition we have $\tr{\hat{\rho}}=\sum_n\rho_{n,n}=1$, where $\text{Tr}$ is the trace. As a consequence of these two properties, the magnitude of the off-diagonal elements of the density matrix are bounded by their corresponding diagonal elements by $|\rho_{n,m}|^2\leq \rho_{n,n}\rho_{m,m}$. In a related way, the trace of the density operator squared always satisfies $1/\text{dim}(\hat{\rho})\leq\tr{\hat{\rho}^2}\leq 1$, where $\text{dim}(\hat{\rho})$ is the dimension of $\hat{\rho}$ \cite{breuer2002theory}. This inequality is only saturated to 1 when $\hat{\rho}$ is a pure state and it is saturated at the lower bound when it is fully mixed. This quantity $\euscr{P}=\tr{\hat{\rho}^2}$ is called the trace purity, and it gives a natural way to characterize how pure a quantum state is.

\subsection{Measurements}
\label{chapter1:measurements}

In order to observe and exploit quantum properties of the system, it is necessary to measure the state of the system. At the beginning of the previous section, I alluded to the idea of `finding' the quantum system in a particular state, but gave little indication to how that is mathematically achieved. I will first introduce the idea of an ideal quantum measurement and then expand this concept to include imperfections.

The ideal von Neumann approach \cite{pechen2006quantum} used for instantaneous measurements associates every measurement with an observable acting on the Hilbert space. As discussed before, the average value of an observable operator is given by the state of the system $\ket{\psi}$ through the real-valued expectation $\braket{\hat{O}}=\braket{\psi|\hat{O}|\psi}$. This concept can be extended to a quantum system state represented by a density operator $\hat{\rho}$ by the relation $\braket{\hat{O}}=\tr{\hat{O}\hat{\rho}}$.

Since every observable is diagonalizable, it can be written in terms of a set of operators $\hat{\Pi}_n=\ketbra{\varphi_n}{\varphi_n}$ so that $\hat{O}=\sum_n\alpha_n\hat{\Pi}_n$. The operators $\hat{\Pi}_n$ are projection operators because $\ket{\varphi_n}$ form an orthonormal basis making $\hat{\Pi}_n^2=\hat{\Pi}_n$ idempotent. They are also complete: $\sum_n\hat{\Pi}_n=\hat{I}$. Thus, the action of $\hat{O}$ becomes clear when applied to the system state. It projects the state $\ket{\psi}$ onto state $\ket{\varphi_n}$ with probability $\prb{\alpha_n}=\left|\braket{\psi|\varphi_n}\right|^2$ and returns the associated eigenvalue $\alpha_n$. The expectation value of $\hat{O}$ is then interpreted as the average value of all possible measurement outcomes $\alpha_n$ weighted by the probability of projecting the quantum system onto the eigenstate $\ket{\varphi_n}$.

Once the quantum system has been ideally measured with the outcome $\alpha_n$, it is in the state $\ket{\varphi_n}$. This occurs with the probability $\tr{\hat{\Pi}_n\hat{\rho}\hat{\Pi}_n}=\prb{\alpha_n}$. Hence, after an ideal measurement, the system is in the pure state
\begin{equation}
    \ketbra{\varphi_n}{\varphi_n} = \frac{\hat{\Pi}_n\hat{\rho}\hat{\Pi}_n}{\tr{\hat{\Pi}_n\hat{\rho}\hat{\Pi}_n}}
\end{equation}
even if it was in a mixed state before the measurement. This type of idealized measurement directly implemented using projection operators is called a projection-valued measure (PVM).

To simplify notation, it is convenient to define an operator $\mathcal{P}_n$ so that $\mathcal{P}_n\hat{\rho}=\hat{\Pi}_n\hat{\rho}\hat{\Pi}_n$. This operator is our first example of a superoperator, which is an operator that takes an operator on the Hilbert space and returns another operator on the Hilbert space. For clarity and consistency, I will notate all superoperators in this thesis using the calligraphic font. In addition, I assume that they act on all operators situated to their right, unless otherwise specified. Using this projection superoperator, we can write $\ketbra{\varphi_n}{\varphi_n}=\mathcal{P}_n\hat{\rho}/\tr{\mathcal{P}_n\hat{\rho}}$. The projection superoperators are idempotent $\mathcal{P}_n^2=\mathcal{P}_n$. However, in general, they are not complete. That is, $\sum_n\mathcal{P}_n$ is not necessarily the identity superoperator $\mathcal{I}$ because the sum does not contain the superoperator projections onto $\ketbra{\varphi_n}{\varphi_m}$. This is because a projective measurement will remove quantum coherence between different eigenstates $\ket{\varphi_n}$ of the observable.

The power of measurements is now clear. They allow one to purify a quantum system by preparing it in a specific quantum state. In addition, a measurement gives a small piece of information about what state the system was in before the measurement. By repeatedly preparing and measuring a quantum system, it is possible to build up enough information to get a clear picture about how the quantum system behaves.

Not every measurement is ideal in that it can project the system onto a pure state. For example, in most realistic scenarios, there can be some classical noise in the device used to measure the quantum system. This noise may prevent the device from faithfully reporting the value $\alpha_n$ given that the quantum system was in the state $\ket{\varphi_n}$. Instead, it could erroneously indicate some other outcome $\alpha_m\neq\alpha_n$. In that case, it is not guaranteed that the state $\hat{\rho}$ is in the state $\ket{\varphi_m}$ if the device reports $\alpha_m$. There is some remaining chance that it is actually in the state $\ket{\varphi_n}$ and hence the quantum system is in a mixed state of $\ket{\varphi_m}$ and $\ket{\varphi_n}$ after the measurement.

We can describe the effect of classical noise of this non-ideal measurement in terms of conditional probabilities. I will first describe a simple scenario so that when I write the result it becomes clear what it physically represents. Suppose we have a device that gives two possible outcomes $0$ and $1$, and projects onto two possible orthogonal states $\ket{\downarrow}$ and $\ket{\uparrow}$. If we characterize the device by supplying it with many copies of $\ket{\uparrow}$ (ignoring the fact that we need a source of perfect $\ket{\uparrow}$ states), we can determine the probabilities $\prb{1|\!\uparrow}$ ($\prb{0|\!\uparrow}$) that it reports $1$ (0) given the input $\ket{\uparrow}$. Likewise, we can do the same for the input $\ket{\downarrow}$. Hence, after characterization, we have a set of four conditional probabilities $\prb{1|\!\uparrow}$, $\prb{1|\!\downarrow}$, $\prb{0|\!\uparrow}$, and $\prb{0|\!\downarrow}$.

Using Bayes' theorem we can write the probability of the state being in either $\ket{\uparrow}$ or $\ket{\downarrow}$ given that we measure $0$ or $1$. For example, $\prb{\uparrow\!|1}=\prb{1|\!\uparrow}\prb{\uparrow}/\prb{1}$. Then, if we measure outcome $1$, we know that the state after measurement is the mixed state $\hat{\varrho}_{1}=\prb{\uparrow\!|1}\ketbra{\uparrow}{\uparrow}+\prb{\downarrow\!|1}\ketbra{\downarrow}{\downarrow}$. Substituting in Bayes' theorem and applying our ideal measurement postulate gives us $\hat{\varrho}_{1}=\left(\prb{1|\!\uparrow}\mathcal{P}_\uparrow\hat{\rho}+\prb{1|\!\downarrow}\mathcal{P}_\downarrow\hat{\rho}\right)/\prb{1}$. We can now identify another important superoperator $\mathcal{F}_1=\prb{1|\!\uparrow}\mathcal{P}_\uparrow+\prb{1|\!\downarrow}\mathcal{P}_\downarrow$ so that $\hat{\varrho}_1=\mathcal{F}_1\hat{\rho}/\prb{1}$. Also, since $\tr{\hat{\varrho}_1}=1$, we can easily identify that $\tr{\mathcal{F}_1\hat{\rho}}=\prb{1}$. Now, we can do the same procedure for $\hat{\varrho}_0$ to find the superoperator $\mathcal{F}_0$. These two superoperators together form an implementation of what is called a positive operator valued measure (POVM).

Let us generalize this concept. Using a set of projection superoperators $\mathcal{P}_n$, we saw from above that we can write a POVM implementation as $\mathcal{F}_n=\sum_m \prb{\alpha_n|\varphi_m}\mathcal{P}_m$. Then, the probability of measuring outcome $\alpha_n$ is $\prb{\alpha_n}=\tr{\mathcal{F}_n\hat{\rho}}$ and the state after measurement is
\begin{equation}
    \hat{\varrho}_n=\frac{\mathcal{F}_n\hat{\rho}}{\tr{\mathcal{F}_n\hat{\rho}}}.
\end{equation}
Therefore, for the extension to imperfect measurements using a POVM implementation, the superoperator $\mathcal{F}_n$ plays the role analogous to $\mathcal{P}_n$ in the PVM implementation. However, in general, the set of operators $\mathcal{F}_n$ are neither complete $\sum_n\mathcal{F}_n\neq\mathcal{I}$ nor idempotent $\mathcal{F}_n\neq\mathcal{F}_n^2$.

In both the PVM and POVM implementations described above, the basis of projection operators $\hat{\Pi}_n$ (or projection superoperators $\mathcal{P}_n$) determines how the measurement affects the state. However, for a given set of measurement outcome probabilities $\prb{\alpha_n|\varphi_n}$, the set of superoperators $\mathcal{F}_n$ is not uniquely determined for a POVM. In the simplified scenario above, I assumed \emph{a priori} that the measurement performed projections onto the $\{\ket{\uparrow}$, $\ket{\downarrow}\}$ basis. However, this cannot be confirmed by only measuring the two quantum states $\ket{\uparrow}$ and $\ket{\downarrow}$. In practice, it is necessary to characterize the measurement device by measuring multiple sets of orthogonal states to fully determine the unique set of superoperators $\mathcal{F}_n$ that form the POVM implementation. Furthermore, I have neglected to describe the generalization of $\mathcal{F}_n$ where the comprising superoperators $\mathcal{P}_m$ are not necessarily projection superoperators. In fact, a general POVM implementation is built from any set of positive operators $\hat{F}_m=\hat{M}_m^\dagger\hat{M}_m$ known as POVM elements that satisfy the property $\sum_m\hat{F}_m=\hat{I}$. Then, the general form of $\mathcal{P}_m$ is the jump superoperator of $\hat{M}_m$: $\mathcal{P}_m\hat{\rho}=\hat{M}_m\hat{\rho}\hat{M}_m^\dagger$. The details of this generalization and the general properties of $\mathcal{F}_n$ are given in Ref.~\cite{breuer2002theory}.

\subsection{Closed quantum systems}

The time evolution of a quantum state is governed by the observable operator corresponding to the total system energy---the Hamiltonian $\hat{H}(t)$. This time evolution is described by the Schr\"{o}dinger equation
\begin{equation}
    i\hbar\frac{d}{dt}\ket{\psi(t)}=\hat{H}(t)\ket{\psi(t)},
\end{equation}
where $\hbar$ is Planck's constant. Here, I am using the total time derivative, as in Ref. \cite{breuer2002theory}, to illustrate that the equation is basis independent, as opposed to a partial derivative used in the common Schr\"{o}dinger wave equation that additionally includes position as a variable. The solution to the Schr\"{o}dinger equation is given by the unitary transformation acting on the Hilbert space $\hat{U}(t,t_0)$, where $\ket{\psi(t)}=\hat{U}(t,t_0)\ket{\psi(t_0)}$. This unitary operator also satisfies the Schr\"{o}dinger equation
\begin{equation}
\label{ch1eq:propagatorU}
    i\hbar\frac{\partial}{\partial t}\hat{U}(t,t_0)=\hat{H}(t)\hat{U}(t,t_0),
\end{equation}
with the initial condition $\hat{U}(t_0,t_0)=\hat{I}$. Here, I am using a partial derivative to emphasize that $t_0$ is held constant. Likewise, the evolution of the covector $\bra{\psi(t)}$ is given by the adjoint propagator $\hat{U}^\dagger(t,t_0)$ and so $\bra{\psi(t)}=\bra{\psi(t_0)}\hat{U}^\dagger(t,t_0)$.

For time-independent Hamiltonians, and like any matrix differential equation, the propagation operator is readily computed by exponentiation $\hat{U}(t,t_0)=e^{-i(t-t_0)\hat{H}/\hbar}$, which can be solved by diagonalizing the Hamiltonian $\hat{H}$. For time-dependent Hamiltonians, it is possible to solve the system using a time-ordered integration, analytic coarse graining \cite{fischer2018derivation}, or by various numerical integration techniques.

The evolution of the density operator is also governed by the propagator $\hat{U}$. Since each quantum state $\ket{\phi_n(t_0)}$ that form a basis evolves like $\ket{\phi_n(t)}=\hat{U}(t,t_0)\ket{\phi_n(t_0)}$ (and similarly for $\bra{\phi_n(t)}$), we have that $\hat{\rho}(t) = \hat{U}(t,t_0)\hat{\rho}(t_0)\hat{U}^\dagger(t,t_0)$. In fact, by taking the derivative of $\hat{\rho}(t)$, applying the product rule, and using Eq.~(\ref{ch1eq:propagatorU}) along with its adjoint, we arrive at the von Neumann equation of motion for the density operator
\begin{equation}
    i\hbar\frac{d}{dt}\hat{\rho}(t)=\left[\hat{H}(t),\hat{\rho}(t)\right]
\end{equation}
where $[\hat{A},\hat{B}]=\hat{A}\hat{B}-\hat{B}\hat{A}$ is the commutator of operators $\hat{A}$ and $\hat{B}$.

I will follow the convention of \cite{breuer2002theory} and define a closed quantum system as one whose evolution is described by the Schr\"{o}dinger equation or von Neumann equation for mixed states. It is also useful to distinguish a closed quantum system from an isolated quantum system, which is a closed system where the Hamiltonian is time-independent. This latter case implies that the average energy $\braket{\hat{H}}$ of the system remains constant with respect to time. However, a closed quantum system can still be externally manipulated, and hence experience a change in average energy.

One important property of a closed quantum system is that the trace purity $\euscr{P}$ remains constant. That is, an initial pure state will remain pure while an initial mixed state will stay just as mixed. The proof of this follows readily from the fact that $\hat{U}$ is unitary, which implies $\hat{U}^\dagger\hat{U}=\hat{I}$, and the cyclic property of trace. Because the evolution of a closed quantum system is governed by a unitary transformation, it is also said to be time reversible. This means that an original state $\ket{\psi(t_0)}$ can be recovered by applying the inverse transformation $\hat{U}^\dagger(t,t_0)=\hat{U}(t_0,t)$, which implies $\hat{U}(t_0,t)\ket{\psi(t)}=\ket{\psi(t_0)}$ for $t_0\leq t$. However, there are many quantum systems where reversible evolution seemingly does not occur because they are not closed systems---they inevitably interact with their environment.

\subsection{Open quantum systems}

In many cases, the size of the quantum system model would have to be very large to capture all the interacting systems affecting the device. This is particularly true for solid-state based quantum optical devices where there are many different interacting subsystems, such as confined electrons, charge noise, vibrations in the host lattice, a surrounding bath of nuclear spins, and electromagnetic cavity modes. This makes an exact mathematical description of the dynamics using the Schr\"{o}dinger equation very difficult to solve. Even if a solution is managed, the vast number of degrees of freedom also makes it difficult to distill that solution into relevant results that are easily applicable to experiments. Open quantum systems theory tackles this problem by first reducing the state space of the total system using a series of approximations. However, the dynamics of the resulting reduced state space is no longer governed by the Schr\"{o}dinger equation as irreversible state evolution becomes possible.

The first step towards obtaining the dynamics of the reduced system is to categorize the subsystems composing the total system. Some subsystems may interact only weakly with other subsystems such that they primarily introduce a quantum noise to the desired state evolution. Some other subsystems may interact strongly, contributing significantly to the quantum dynamics. Finally, some subsystems may be measured by an external system to gain information about the state.

\begin{figure}
    \centering
    \includegraphics[width=0.6\textwidth]{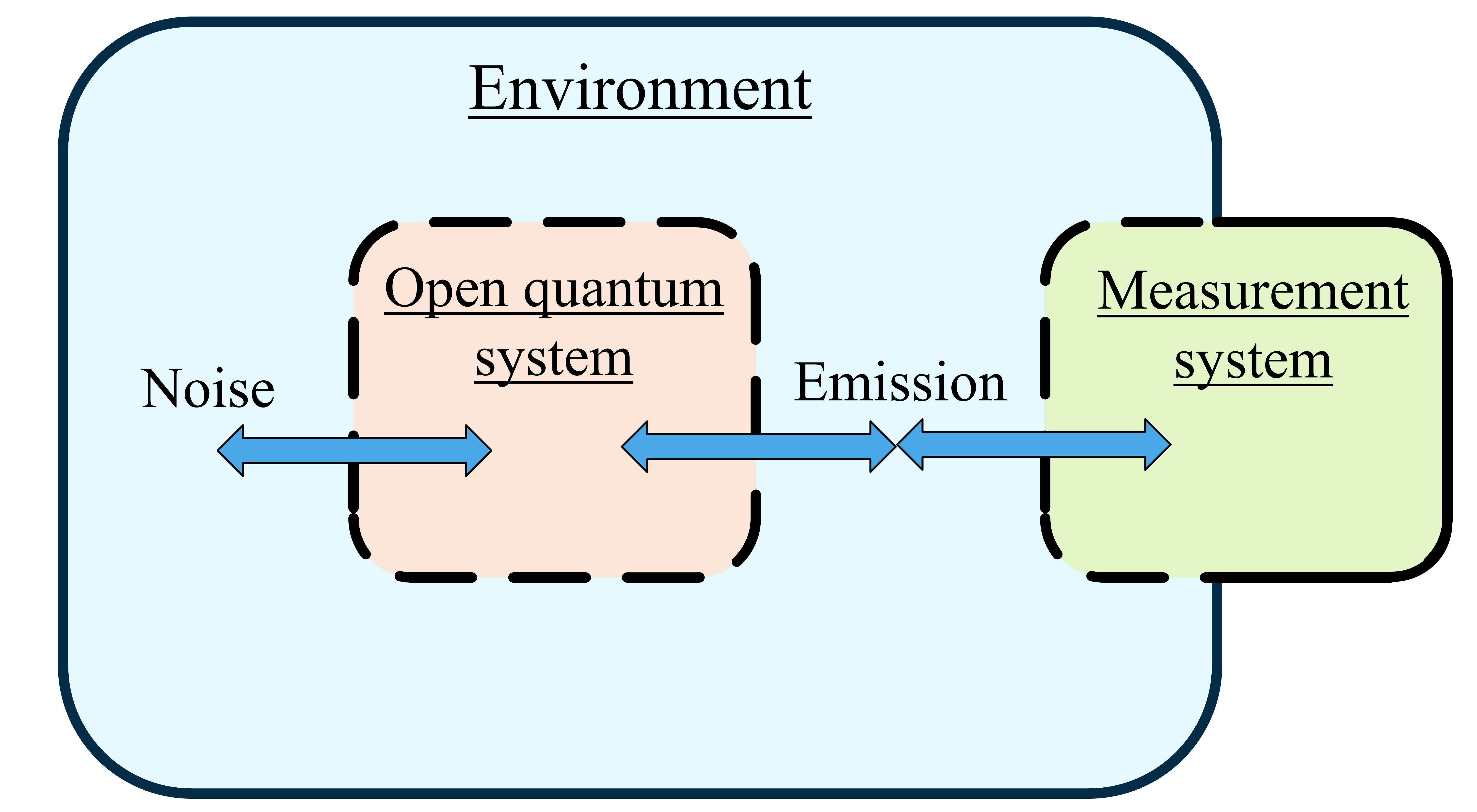}
    \caption[An illustration of an open quantum system within its environment.]{\small\textbf{An illustration of an open quantum system within its environment.} The open quantum system is affected by the noisy environment. It also leaks information into the environment by, for example, photon emission. This emission can be detected by an auxiliary measurement system to gain information about the state of the open quantum system.}
    \label{ch1fig:opensystems}
\end{figure}

I will proceed in the fashion outlined above by dividing subsystems into three categories (see Fig.~\ref{ch1fig:opensystems}). First, I call the \emph{open quantum system} the subsystem of interest combined with all subsystems that strongly interact with it. The open quantum system is also generally referred to as the reduced quantum system or, plainly, the quantum system. This open quantum system interacts with its \emph{environment}, which is composed of the remaining subsystems that interact only weakly with the open quantum system. Finally, we can have an auxiliary system which serves to measure one or more subsystems of the environment. This \emph{measurement system} may be quantum or classical. However, in this thesis, I will only ever consider a classical measurement system. Furthermore, if a subsystem in the environment has infinite degrees of freedom, such as a continuum of electromagnetic modes, it is called a \emph{reservoir}. If a reservoir is in a thermal equilibrium state, I will refer to it as a \emph{bath}.

For now, let us set aside the measurement system and consider only the open quantum system state $\hat{\rho}_\mathrm{S}=\text{Tr}_\mathrm{E}\!\left(\hat{\rho}\right)$ and the environment state $\hat{\rho}_\mathrm{E}=\text{Tr}_\mathrm{S}\!\left(\hat{\rho}\right)$, where $\text{Tr}_\mathrm{k}$ is the partial trace over subsystems within $k\in\{\mathrm{S},\mathrm{E}\}$. By applying the Schr\"{o}dinger equation, the total system density operator $\hat{\rho}=\ketbra{\psi}{\psi}$ was found to follow $i\hbar~ d\hat{\rho}(t)/dt=[\hat{H}(t),\hat{\rho}(t)]$. Then the reduced system state formally satisfies the Liouville-von Neumann first-order differential equation
\begin{equation}
\label{redliouville}
    i\hbar\frac{d}{dt}\hat{\rho}_\mathrm{S}(t) = \text{Tr}_\mathrm{E}\!\left[\hat{H}(t),\hat{\rho}(t)\right].
\end{equation}

\subsection{Markovian master equations}
\label{chapter1:markovian}

For any closed quantum system, there exists a dynamical map $\hat{U}(t_\mathrm{f},t_0)$ so that any future quantum state $\ket{\psi(t_\mathrm{f})}=\hat{U}(t_\mathrm{f},t_0)\ket{\psi(t_0)}$ depends only on the current quantum state $\ket{\psi(t_0)}$. In general, this is not true for an open quantum system because the reduced system can alter its environment and the environment can, in turn, affect how the reduced quantum system evolves. This environment back-action means that the evolution of a quantum state $\hat{\rho}_\mathrm{S}(t_0)$ of the reduced system depends on the system state before time $t_0$. However, if the environment is a bath that returns to its equilibrium state very quickly after a slight perturbation due to the evolution of the reduced quantum system, the characteristic timescale $\tau_\mathrm{B}$ of this back-action can be very short---meaning that the evolution of $\hat{\rho}_\mathrm{S}(t_0)$ only depends on its history up to $\sim \tau_\mathrm{B}$ backwards in time. This timescale is called the bath memory time \cite{breuer2002theory}.

If the dominant dynamics of the reduced quantum system occurs on a timescale $\tau_\mathrm{S}$ that is much longer than the bath memory time $\tau_\mathrm{B}$, the evolution of the state of the reduced quantum system approximately no longer depends on its history. In this case, there can exist a dynamical map such that any future state depends only on its current state. This property is known as Markovianity \cite{breuer2002theory}. All systems studied in this thesis are assumed to satisfy this very useful property of Markovianity. Formally, a Markovian system is one where there exists a dynamical map $\mathcal{U}(t_\mathrm{f},t_0)$ with the semigroup property $\mathcal{U}(t_\mathrm{f},t^\prime)\mathcal{U}(t^\prime,t_0)=\mathcal{U}(t_\mathrm{f},t_0)$ where $t_\mathrm{f}\geq t^\prime\geq t_0$ such that the state of the open quantum system $\hat{\rho}_\mathrm{S}$ at time $t_\mathrm{f}$ is fully described by $ \hat{\rho}_\mathrm{S}(t_\mathrm{f})=\mathcal{U}(t_\mathrm{f},t_0)\hat{\rho}_\mathrm{S}(t_0)$ for any $t_\mathrm{f}\geq t_0$.

The superoperator $\mathcal{U}$ is analogous to the unitary transformation $\hat{U}$ in that it serves as a propagator of the system state but with the very important difference that the propagtion superoperator $\mathcal{U}$ is almost never unitary, meaning that it captures irreversible processes such as the spontaneous emission of a photon from an excited emitter. It can also capture basic aspects of other decoherence processes discussed in section \ref{chapter1:deoherence}.

For a fixed time $t_0$, the superoperator $\mathcal{U}(t,t_0)$ forms a quantum dynamical semigroup \cite{breuer2002theory}. The generator superoperator $\mathcal{L}$ of this semigroup satisfies the partial differential equation
\begin{equation}
    \frac{\partial}{\partial t}\mathcal{U}(t,t_0) = \mathcal{L}(t)\mathcal{U}(t,t_0)
\end{equation}
for $t\geq t_0$. By multiplying both sides by $\hat{\rho}_\mathrm{S}(t_0)$, this gives rise to the most general form of the Markovian master equation for the reduced system dynamics
\begin{equation}
\label{markovianeq}
    \frac{d}{dt}\hat{\rho}_\mathrm{S}(t) = \mathcal{L}(t)\hat{\rho}_\mathrm{S}(t),
\end{equation}
where the generator $\mathcal{L}$ is called the Liouville superoperator. Comparing this result to that of Eq.~(\ref{redliouville}), one can see that we should have
\begin{equation}
    \mathcal{L}(t)\hat{\rho}_\mathrm{S}(t) = -\frac{i}{\hbar}\text{Tr}_\mathrm{E}\!\left[\hat{H}(t),\hat{\rho}(t)\right].
\end{equation}
However, the right-hand-side of this equation does not, in general, allow for a quantum dynamical semigroup. To derive $\mathcal{L}(t)$, it is necessary to apply approximations to satisfy Markovianity \cite{breuer2002theory}.

Not all physical systems can be successfully modeled using a Markovian master equation, and there are many interesting situations where Markovian behaviour breaks down \cite{alicki2002invitation}. One such case is when a reduced quantum system is manipulated faster than the bath memory time $\tau_\mathrm{B}$ to decouple it from its environment \cite{viola1999dynamical}. However, the phenomenological application of the Markovian master equation allows for computationally tractable solutions that often correctly predict the general behaviour of a device. It also allows for the derivation of simple bounds on figures of merit for devices or protocols, which can help guide experimental design and development.

The standard form of a Markovian master equation that preserves the properties of $\hat{\rho}_\mathrm{S}$ was derived by Lindblad, Gorini, Kossakowski and
Sudarshan \cite{manzano2020short} and is given by
\begin{equation}
\begin{aligned}
\label{diagonalmeq}
    \frac{d}{dt}\hat{\rho}_\mathrm{S}(t) &= -\frac{i}{\hbar}[\hat{H}_\mathrm{S}(t),\hat{\rho}_\mathrm{S}] + \sum_{i,j} u_{i,j}\left[\hat{L}_i\hat{\rho}_\mathrm{S}(t)\hat{L}_j^\dagger -\frac{1}{2}\left\{\hat{L}_j^\dagger\hat{L}_i,\hat{\rho}_\mathrm{S}(t)\right\}\right]\\
\end{aligned}
\end{equation}
where the Lindblad operators $\hat{L}_i$ are a unitless set of orthonormal operators acting on the reduced system space, $\{\hat{A},\hat{B}\}=\hat{A}\hat{B}+\hat{B}\hat{A}$ is the anti-commutator, the coefficients $u_{i,j}\geq 0$ correspond to the rate of system relaxation induced by the environment, and $\hat{H}_\mathrm{S}$ is the reduced system Hamiltonian. The exact form of these operators and corresponding coefficients depends on the approach taken to move from the Liouville-von Neumann form (Eq.~(\ref{redliouville})) to the Markovian form (Eq.~(\ref{markovianeq})).

One popular approach to achieve Markovianity and derive the Lindblad operators depends on the application of two main assumptions, which together are referred to as the Born-Markov approximations \cite{breuer2002theory}. The Born approximation assumes that the environment is a bath that is weakly coupled to the open quantum system and that the total system state is initially separable $\hat{\rho}(t_0)\simeq\hat{\rho}_\mathrm{S}(t_0)\otimes\hat{\rho}_\mathrm{E}(t_0)$. The Markov approximation then assumes the necessary requirement that $\tau_\mathrm{B}\ll\tau_\mathrm{S}$, so that the environment quickly returns to its initial equilibrium state. These two approximations reliably produce the effects of spontaneous emission and allow for the derivation of quantum optical master equations, as I will reference in the following section. 

There are other approaches to solving open quantum system dynamics that do not require the Born approximation, such as the derivation by a coarse graining of time \cite{fischer2018derivation}. This coarse graining approach can also reproduce the same Markovian master equation as the Born-Markov approach for cavity quantum electrodynamics, which is the topic of section \ref{chapter1:quantumED}. Another approach that can capture additional dynamics neglected by the Born-Markov approximations is obtained by applying a weak-coupling approximation in a more mathematically rigorous way \cite{alicki2002invitation}. This approach provides Lindblad operators and corresponding rate coefficients that are often time dependent to capture the nontrivial interaction dynamics between the reduced system and the bath. Although I do not explore these forms of the Markovian master equation in this thesis, many of the methods I use may be directly applied to more detailed Markovian models.

\subsection{Multi-time correlations}
\label{chapter1:correlations}

Using a Markovian master equation, we can compute the state of the reduced system at any time and also compute the expectation values of reduced system operators, which may be directly tied to measured quantities of the environment. The expectation value of an operator $\hat{A}$ in the reduced system space in the Schr\"{o}dinger picture is given by $\braket{\hat{A}(t)}\equiv\tr{\hat{A}\hat{\rho}(t)}$, where I will denote $\hat{\rho}$ as the reduced system density operator moving forward and take the trace to be over the reduced system space unless otherwise specified. Applying the solution to the master equation gives
\begin{equation}
    \braket{\hat{A}(t)}=\tr{\hat{A}\mathcal{U}(t,t_0)\hat{\rho}(t_0)}.
\end{equation}
It is convenient to define the adjoint propagation superoperator $\mathcal{U}^\dagger$, which is defined as the dynamical map satisfying $\tr{\left(\mathcal{U}^\dagger(t,t_0)\hat{A}\right)\hat{\rho}(t_0)}=\tr{\hat{A}\mathcal{U}(t,t_0)\hat{\rho}(t_0)}$. Then, we can define the operators $\hat{A}_\mathrm{H}(t)$ in the Heisenberg picture $\hat{A}_\mathrm{H}(t,t_0)=\mathcal{U}^\dagger(t,t_0)\hat{A}$. With this notion, it is straightforward to write out multi-time correlation functions in terms of the propagator $\mathcal{U}$. For example, consider the two-time correlation function:
\begin{equation}
\begin{aligned}
    \braket{\hat{A}(t^\prime)\hat{B}(t)}&\equiv\tr{\hat{A}_\mathrm{H}(t^\prime,t)\hat{B}_\mathrm{H}(t,t_0)\hat{\rho}(t_0)}=\tr{\left(\mathcal{U}^\dagger(t^\prime,t)\hat{A}\right)\left(\mathcal{U}^\dagger(t,t_0)\hat{B}\right)\hat{\rho}(t_0)}\\
    &=\tr{\hat{A}\mathcal{U}(t^\prime,t)\hat{B}\hat{\rho}(t)},
\end{aligned}
\end{equation}
where $t^\prime\geq t\geq t_0$. This result can be extended to higher order multi-time correlation functions. For more details on Heisenberg operators and the adjoint space, please again refer to Ref.~\cite{breuer2002theory}. For the solution to $\braket{\hat{A}(t)\hat{B}(t^\prime)}$, we can also use the useful Hermitian property that $\braket{\hat{A}(t+\tau)\hat{B}(t)}=\braket{\hat{A}(t-\tau)\hat{B}(t)}^*$ where $\tau>0$. Although I am using the Heisenberg picture to summarize the result above for multi-time correlations, all the derivations and computations in this thesis are performed in the Schr\"{o}dinger picture. The above result is intuitive because it implies that two-time correlation functions can be computed by propagating the initial state $\hat{\rho}(t_0)$ forward until time $t$ when the first operator $\hat{B}$ is applied, then propagating the result forward again until time $t^\prime$ when the second operator $\hat{A}$ is applied. It also leads to a useful theorem pertaining to the coupled differential equations of two-time correlation functions for Markovian systems called the quantum regression theorem, which I will present in section \ref{chapter1:quantumregression}.

We can also formulate multi-time correlation functions in terms of superoperators, which provides physical intuition and a simple approach to computation in the Fock-Liouville space (to be presented in section \ref{chapter1:fockliouville}). For example, consider the two-time second-order correlation function $G^{(2)}(t,t^\prime)=\braket{\hat{A}^\dagger(t)\hat{A}^\dagger(t^\prime)\hat{A}(t^\prime)\hat{A}(t)}$. Let us then define the jump superoperator $\mathcal{J}$, which is of the form $\mathcal{J}\hat{\rho}=\hat{A}\hat{\rho}\hat{A}^\dagger$ for an operator $\hat{A}$. Then the correlation is written in a transparent way:
\begin{equation}
    G^{(2)}(t,t^\prime)
    =\tr{\mathcal{J}\mathcal{U}(t^\prime,t)\mathcal{J}\mathcal{U}(t,t_0)\hat{\rho}(t_0)}.
\end{equation}
From this notation it is clear that the system propagates from time $t_0$ to time $t$ when it jumps by the instantaneous action of $\hat{A}$. Then it propagates to time $t^\prime$ when it jumps again. We can then interpret $G^{(2)}(t,t^\prime)$ as being an unnormalized probability density function of the system jumping once at time $t$ and again at time $t^\prime$. This superoperator notation can be extended to arbitrary time-ordered multi-time correlation functions \cite{breuer2002theory}.

\section{Quantum electrodynamics}
\label{chapter1:quantumED}

The work presented in this thesis is foremost an application of quantum electrodynamics (QED) to study and propose optical devices and protocols. QED develops from the quantization of the general solution to the classical electromagnetic wave equation \cite{griffiths2005introduction}. When subject to boundary conditions dictated by an arbitrary mode volume $V$, the classical free-space field solution to the wave equation is given by a summation of forward-propagating and backward-propagating plane waves of frequency $\omega_\mathbf{k}$ with discrete wave vectors $\mathbf{k}$ and polarization $\mathbf{u}_\epsilon(\mathbf{k})$, where $\epsilon$ is one of two possible polarizations of the plane wave. By quantizing the plane-wave amplitudes of the electric field solution and subjecting them to the canonical commutation relations \cite{scully1999quantum}, the quantized electric field in a homogeneous dielectric material is found to be
\begin{equation}
    \hat{\mathbf{E}}(\mathbf{r}) = -\sum_{\mathbf{k},\epsilon}\sqrt{\frac{\hbar\omega_\mathbf{k}}{2n^2\varepsilon_0 V}}\left(\mathbf{u}_\epsilon(\mathbf{k})\ad_\epsilon(\mathbf{k})e^{i\mathbf{k}\cdot\mathbf{r}}+\mathbf{u}^*_\epsilon(\mathbf{k})\au_\epsilon(\mathbf{k}) e^{-i\mathbf{k}\cdot\mathbf{r}}\right),
\end{equation}
where $\mathbf{r}$ is the position in space within $V$, $n$ is the index of refraction, and $\varepsilon_0$ is the vacuum permitivity. The quantized amplitudes $\au_\epsilon(\mathbf{k})$ ($\ad_\epsilon(\mathbf{k})$) are the photon creation (annihilation) operators that satisfy the canonical commutation relations $[\ad_\epsilon(\mathbf{k}),\au_{\epsilon^\prime}(\mathbf{k}^\prime)]=\delta_{\mathbf{k},\mathbf{k}^\prime}\delta_{\epsilon,\epsilon^\prime}$, where $\delta$ is the Kronecker delta function defined by $\delta_{ij}=1$ if $i=j$ and $\delta_{ij}=0$ if $i\neq j$. The creation and annihilation operators are ladder operators of a quantum harmonic oscillator. When these operators act on the state $\ket{n}$ of the quantized mode containing $n$ photons, the energy of the state is increased $\au\ket{n}=\sqrt{n+1}\ket{n+1}$ or decreased $\ad\ket{n}=\sqrt{n}\ket{n-1}$ by one quantum of energy $\hbar\omega_\mathbf{k}$. The creation and annihilation operators are not observables since $\ad\neq\au$; however, the electric field operator is an observable since $\hat{\mathbf{E}}^\dagger=\hat{\mathbf{E}}$.

Using the quantized electric field, we can obtain the Hamiltonian of the field as a summation of the total quantum harmonic oscillator energy \cite{scully1999quantum} for each mode
\begin{equation}
    \hat{H}_{E} = \sum_{\mathbf{k},\epsilon}\hbar\omega_\mathbf{k}\left(\frac{1}{2}+\au_\epsilon(\mathbf{k})\ad_\epsilon(\mathbf{k})\right),
\end{equation}
where often the divergent vacuum energy $\sum_\mathbf{k}\hbar\omega_\mathbf{k}/2$ is physically inconsequential and ignored by renormalization \cite{scully1999quantum,breuer2002theory}.

If the mode volume is physically confined in such a way to support only a single mode with wavenumber $k$ and polarization $\mathbf{u}(k)$ within $V$, we obtain the one-dimensional single-mode waveguide field
\begin{equation}
\label{waveguideEfieldeq}
    \mathbf{\hat{E}}(x) = -\sum_k\sqrt{\frac{\hbar\omega_k}{2n^2\varepsilon_0 V}}\left(\mathbf{u}(k)\bd(k)e^{ikx}+\mathbf{u}^*(k)\bu(k) e^{-ikx}\right),
\end{equation}
where $\bu$ and $\bd$ are the photon creation and annihilation operators of the waveguide.

\subsection{Spontaneous emission and pure dephasing}
\label{chapter1:spontaneousemission}

One of the most simple quantum system models with non-trivial dynamics is the two-state system, or two-level system. These two states could represent different electronic or molecular orbitals of a defect, two different spin states, or any two substates that are decoupled from a larger Hilbert space.

A two-level dipole emitter is a two-level system of different electronic states with a ground state $\ket{\mathrm{g}}$ and an excited state $\ket{\mathrm{e}}$ separated by an average energy $\hbar \omega_o$. The Hamiltonian is given by $\hat{H}_o=\hbar\omega_o\sigu\sigd$ where $\sigu=\ketbra{\mathrm{e}}{\mathrm{g}}$ is the raising operator and $\sigd=\ketbra{\mathrm{g}}{\mathrm{e}}$ is the lowering operator. The defining characteristic of a two-level dipole emitter is that it has a non-zero electric transition dipole moment between the two levels. This can arise if the electronic orbitals have different symmetry so that there is a shift in charge distribution when moving from one state to another. The dipole operator is a vector operator $\hat{\mathbf{d}}=q\hat{\mathbf{r}}$ analogous to the classical dipole moment, where $q$ is the electric charge and $\hat{\mathbf{r}}$ is the position operator. The transition dipole moment is then $\braket{\mathrm{g}|q\hat{\mathbf{r}}|\mathrm{e}}$, which is nonzero for an emitter. 

Under the dipole approximation, the interaction Hamiltonian for an emitter at position $\mathbf{r}_o$ in an electric field is given by $-\hat{\mathbf{d}}\cdot\hat{\mathbf{E}}(\mathbf{r}_o)$ \cite{scully1999quantum}. Using the quantized electric field from the previous section, this gives us the total emitter-field Hamiltonian
\begin{equation}
\label{spontaneousemissionhamiltonian}
    \hat{H} = \hbar\omega_o\sigu\sigd+\sum_{\mathbf{k},\epsilon}\hbar\omega_\mathbf{k}\au_\epsilon(\mathbf{k})\ad_\epsilon(\mathbf{k})+\sum_{\mathbf{k},\epsilon}\sqrt{\frac{\hbar\omega_\mathbf{k}}{2n^2\varepsilon_0 V}}\hat{\mathbf{d}}\cdot\left(\mathbf{u}_\epsilon(\mathbf{k})\ad_\epsilon(\mathbf{k})e^{i\mathbf{k}\cdot\mathbf{r}_o}+\mathbf{u}_\epsilon^*(\mathbf{k})\au_\epsilon(\mathbf{k})e^{-i\mathbf{k}\cdot\mathbf{r}_o}\right).
\end{equation}

To obtain a Markovian master equation for the dynamics of the two-level emitter, we can follow Ref.~\cite{breuer2002theory} by applying the Born-Markov approximations to the Liouville-von Neumann equation corresponding to Eq.~(\ref{spontaneousemissionhamiltonian}). If we also assume that the electric field is a bath with a temperature $T$, then we arrive at the quantum optical master equation for the reduced system $d\hat{\rho}/dt=\mathcal{L}\hat{\rho}$ given by the Liouville superoperator \cite{breuer2002theory}
\begin{equation}
\label{chapter1eq:spontemissionexcite}
    \mathcal{L}=-\frac{i}{\hbar}\mathcal{H}+\gamma^+\mathcal{D}(\sigu)+\gamma^-\mathcal{D}(\sigd),
\end{equation}
where $\mathcal{H}$ is the Liouville-von Neumann superoperator defined by $\mathcal{H}\hat{\rho}=[\hat{H}_\mathrm{S},\hat{\rho}]$ corresponding to the reduced system Hamiltonian $\hat{H}_\mathrm{S}=\hbar\omega_o\sigu\sigd$. The dissipation superoperator $\mathcal{D}$ is given by $\mathcal{D} = \mathcal{J}-\mathcal{A}/2$, where $\mathcal{J}$ is the jump superoperator defined by $\mathcal{J}(\hat{A})\hat{\rho}=\hat{A}\hat{\rho}\hat{A}^\dagger$ and $\mathcal{A}$ is the amplitude damping superoperator defined by $\mathcal{A}(\hat{A})\hat{\rho}=\{\hat{A}^\dagger\hat{A},\hat{\rho}\}$. The temperature-dependent rates associated with the dissipation superoperators are $\gamma^+=\gamma_\mathrm{r}\overline{n}(T)$ and $\gamma^-=\gamma_\mathrm{r}(1+\overline{n}(T))$, where
\begin{equation}
\begin{aligned}
    \gamma_\mathrm{r} &=\gamma^--\gamma^+= \frac{n\omega_o^3d^2}{3\pi \varepsilon_0\hbar c^3},
\end{aligned}
\end{equation}
is the zero-temperature radiative rate \cite{scully1999quantum}, $d=|\mathbf{d}|$ is the dipole magnitude, $\overline{n}(T)=(e^{\hbar\omega_o/k_\mathrm{B}T}-1)^{-1}$ is the average number of photons in the mode of the field that is resonant with the two-level system, and $k_\mathrm{B}$ is the Boltzmann constant. The reduced system resonance frequency $\omega_o$ is slightly different from the original frequency due to the Lamb shift induced by the vacuum fluctuations of the electromagnetic field and a possible thermal Stark shift if $\overline{n}(T)\neq 0$. However, since this only amounts to a shift in the zero-point energy, I will continue to write $\omega_o$ without specification.

The result of spontaneous emission is profound because it shows that an excited quantum dipole emitter will dissipate its energy into the environment by spontaneously emitting a single photon of frequency $\hbar\omega_o$ at the rate $\gamma_r$ even if the electromagnetic field is in the vacuum state ($T=0$). This passive coupling to the electromagnetic vacuum causes the emitter to experience decoherence. To see this, consider an arbitrary initial state of the two-level system $\ket{\psi(0)}=\alpha\ket{\mathrm{g}}+\beta\ket{\mathrm{e}}$, where $\alpha$ and $\beta$ are complex amplitudes satisfying $|\alpha|^2+|\beta|^2=1$. At low temperature where $\overline{n}(T)\simeq 0$, the Liouville superoperator in the frame rotating at the emitter resonance $\omega_o$ is $\mathcal{L}=\gamma_\mathrm{r}\mathcal{D}(\sigu)$ corresponding to the propagator $\mathcal{U}(t,t_0)=e^{\mathcal{L}(t-t_0)}$. Then the state of the reduced system at time $t$ is computed by $\hat{\rho}(t)=\mathcal{U}(t,0)\ketbra{\psi(0)}{\psi(0)}$ to give
\begin{equation}
\label{chapter1eq:spontemisrho}
    \hat{\rho}(t)=\left(|\alpha|^2+|\beta|^2(1-e^{-\gamma_\mathrm{r} t})\right)\ketbra{\mathrm{g}}{\mathrm{g}}+|\beta|^2e^{-\gamma_\mathrm{r} t}\ketbra{\mathrm{e}}{\mathrm{e}}+\left(\alpha^*\beta\ketbra{\mathrm{e}}{\mathrm{g}}+\alpha\beta^* \ketbra{\mathrm{g}}{\mathrm{e}}\right)e^{-\gamma_\mathrm{r} t/2}.
\end{equation}

From Eq.~(\ref{chapter1eq:spontemisrho}) for $\hat{\rho}(t)$, we can see that the coherence $\braket{\mathrm{e}|\hat{\rho}|\mathrm{g}}$ is degraded at the rate $\gamma_\mathrm{r}/2$ while the excited state population decays at the rate $\gamma_\mathrm{r}$. Another interesting observation is that the trace purity $\euscr{P}=\text{Tr}(\hat{\rho}^2(t))$ of the state is degraded during emission, meaning that the coherence is not always saturated to the upper bound dictated by the populations. For any initial pure state $\ket{\psi(0)}$ of the emitter, the trace purity begins at 1 and reaches a minimum value of $\text{min}(\euscr{P})=|\alpha|^4+2|\alpha|^2|\beta|^2+|\beta|^4/2$ at the time $t_\text{min}=\ln(2)/\gamma_\mathrm{r}$ before increasing back to unity as the emitter settles to its ground state $\ket{\mathrm{g}}$. This purity dip is related to the fact that the emitter briefly becomes entangled with the electromagnetic environment during spontaneous emission, which is the topic of section \ref{chapter3:photonicbellstates}.

In some systems, especially those in a solid state environment, the total dissipative rate of decay $\gamma$ can be larger than the radiative rate $\gamma_\mathrm{r}$. An increased decay rate can be caused by non-radiative decay pathways that can emit phonons instead of photons. This additional non-radiative decay rate $\gamma_\mathrm{nr}$ can be accounted for in a phenomenological way by replacing $\gamma_\mathrm{r}$ with $\gamma=\gamma_\mathrm{r}+\gamma_\mathrm{nr}$. 

In addition to an increased dissipative rate, solid-state environments can cause the rate of decoherence to be even larger than the $\gamma/2$ predicted by the total dissipation alone. This can occur when environmental fluctuations, such as charge noise, cause the emitter resonance $\omega_o$ to rapidly fluctuate in time on a timescale faster than its decay rate. Alternatively, this dephasing can be seen as random elastic scattering of particles off of the emitter, such as phonons. These particles can carry information about the emitter state into the environment. From either perspective, the result is that the phase of emitter superposition states becomes dependent on (or entangled with) the state of the environment. Looking in the reduced state picture, this emitter-environment entanglement manifests as random phase fluctuations, which on average decrease the emitter coherence. This dephasing of the emitter coherence is referred to as \emph{pure emitter dephasing}.

For some emitters, such as quantum dots, pure dephasing due to phonons can be modeled by rigorously including electron interactions with a continuum of longitudinal acoustic phonon modes \cite{manson2016polaron,nazir2016modelling}. For example, this can be done using a polaron transformation and a subsequent Markovian approximation to obtain a master equation \cite{manson2016polaron}. However, the exact quantum processes that account for all pure dephasing phenomena for any given solid-state system are often unidentified or very complicated to rigorously include. Regardless of its origin, we can still account for pure dephasing in a less-accurate phenomenological way \cite{grange2015cavity} by considering the effect of the superoperator $\mathcal{D}(\sigu\sigd)$. If we consider just the dynamics induced by the Liouville superoperator $\mathcal{L}=2\gamma^\star\mathcal{D}(\sigu\sigd)$, we can see that for an initial state $\ket{\psi(0)}$ as above, we get the solution $\hat{\rho}(t)=|\alpha|^2\ketbra{\mathrm{g}}{\mathrm{g}}+|\beta|^2\ketbra{\mathrm{e}}{\mathrm{e}}+e^{-t\gamma^\star}\left(\alpha\beta^*\ketbra{\mathrm{g}}{\mathrm{e}}+\alpha^*\beta\ketbra{\mathrm{e}}{\mathrm{g}}\right)$. Hence, the coherence $\braket{\mathrm{g}|\hat{\rho}|\mathrm{e}}$ will degrade exponentially at an additional rate $\gamma^\star$ while leaving the population untouched. By including the additional term $2\gamma^\star\mathcal{D}(\sigu\sigd)$ along with $\gamma\mathcal{D}(\sigd)$, the total decoherence rate of the emitter becomes $\Gamma/2$ where $\Gamma=\gamma+2\gamma^\star$. The rate $\Gamma$ is also the spectrally-broadened full width at half maximum (FWHM) of the Lorentzian emission line of the fluctuating emitter. Note that there are two conventions for defining the pure dephasing rate $\gamma^\star$ that differ by a factor of two: (1) as decay rate of the \emph{amplitude} of the coherence such that the FWHM is $\Gamma=\gamma+2\gamma^\star$ and (2) as the decay rate of the \emph{squared magnitude} of the coherence such that the FWHM is $\Gamma=\gamma+\gamma^\star$. This thesis uses the former convention with the exception of section \ref{chapter3:roomtemperature} where the latter convention is used.

In the literature, the total decay rate $\gamma$, pure dephasing rate $\gamma^\star$, and emitter linewidth $\Gamma$ are often expressed as a decay time rather than a rate, similar to the $T_1$ and $T_2$ notation in the field of nuclear magnetic resonance \cite{mlynarik2017introduction}. Drawing from the notation convention used in Ref.~\cite{grange2015cavity} for optical systems, and being careful to note the factor of 2 difference in choice of definition for $\gamma^\star$, we have that $T_1=1/\gamma$, $T_2^*=1/\gamma^\star$, and $T_2=2/\Gamma$. This implies that $1/T_2 = 2/T_1+1/T_2^*$ and $2T_2\leq T_1$. In addition, it is common to specify the radiative component of the decay time $T_{1,\text{rad}}=1/\gamma_\mathrm{r}$. When appropriate, e.g. for spin qubits, we can also distinguish between spontaneous decay $T_1^-=1/\gamma^-$ and incoherent (thermal) excitation $T_1^+=1/\gamma^+$ for a model utilizing the general form of Eq.~(\ref{chapter1eq:spontemissionexcite}).

The assumption that the coupling between the emitter and the bath is weak is implicit in the Born-Markov approximations used to derive spontaneous decay and thermal excitation. If $V$ is small enough for a particular mode that is resonant with the two-level system, it is possible to violate the Born-Markov weak-coupling condition. This scenario can be described by the Jaynes-Cummings Hamiltonian, which is the topic of the next section.

\subsection{Jaynes-Cummings Hamiltonian}
\label{chapter3:jaynescummings}

Suppose that we confine the electromagnetic field to one dimension so that the modes along this direction are described as in Eq.~(\ref{waveguideEfieldeq}). However, now let us assume that the modes are highly confined so that the quantized mode resonances are far separated in frequency ($\omega_k$ are far apart). This can occur if, for example, $V$ is on the order of the cubic wavelength of the mode resonance. This confinement of an electromagnetic mode is called a cavity.

If a two-level emitter with oscillation frequency $\omega_o$ is placed inside this cavity, it will now primarily couple to the cavity mode resonance $\omega_k$ that is closest to $\omega_o$, which I call $\omega_\mathrm{c}$. For simplicity, let us assume that the dipole is oriented parallel to the cavity mode polarization direction and that it is placed at the antinode of the electric field so that the coupling is maximized. Under these conditions, the cavity-emitter dipole interaction can be much stronger than the coupling between the emitter and the rest of the electromagnetic bath. This violates the weak-coupling assumption used in the previous section. Considering only this single cavity mode-emitter interaction, in the dipole gauge \cite{le2020theoretical}, we can write the quantum Rabi model Hamiltonian
\begin{equation}
\begin{aligned}
\hat{H} =\hbar\omega_o\sigu\sigd + \hbar\omega_\mathrm{c}\au\ad+\hbar g(\sigd+\sigu)(\ad+\au)
\end{aligned}
\end{equation}
where $\ad$ ($\au$) is the cavity mode photon annihilation (creation) operator and
\begin{equation}
    \hbar g=d\sqrt{\frac{\hbar\omega_\mathrm{c}}{2n^2\varepsilon_0 V}}
\end{equation}
is the vacuum cavity-emitter coupling rate and $d$ is the dipole magnitude of the emitter. Here, I am choosing a phase convention where $g$ is a positive real-valued rate. If the dipole is not oriented perfectly or placed at an antinode, the magnitude of the coupling coefficient $g$ will be reduced from its maximum value.

For optical systems, the cavity-emitter coupling rate $g$ is usually on the order of GHz at most. This is much smaller than $\omega_o\simeq\omega_\mathrm{c}$, which generally both fall into the THz regime. Thus, this thesis deals exclusively with interaction strengths where the quantum Rabi model satisfies gauge invariance \cite{le2020theoretical}. As an additional consequence, the eigenenergies of the system are only slightly perturbed in magnitude and are on the order of $\omega_o\pm g\simeq \omega_o$. This allows us to make an important simplification called the rotating wave approximation \cite{scully1999quantum,reiserer2015cavity}. To easily see how this approximation can be implemented, we must first move into the interaction picture \cite{breuer2002theory} by transforming $\hat{H}$ by the unitary transformation $\hat{U}=e^{-it\hat{H}_0/\hbar}$ given by the homogeneous part of the Hamiltonian $\hat{H}_0=\hbar\omega_o\sigu\sigd + \hbar\omega_\mathrm{c}\au\ad$. In this interaction picture, we have $\hat{H}_I=\hat{U}^\dagger\hat{H}\hat{U}-\hat{H}_0$, which gives
\begin{equation}
    \hat{H}_I=\hbar g\left(\sigd\ad e^{it(-\omega_o-\omega_\mathrm{c})}+\sigu\au e^{it(\omega_o+\omega_\mathrm{c})}+\sigd\au e^{it(-\omega_o+\omega_\mathrm{c})}+\sigu\ad e^{it(\omega_o-\omega_\mathrm{c})}\right).
\end{equation}
The first two terms show a phase rotation at the rates $\pm(\omega_o+\omega_\mathrm{c})$ while the second two terms show a phase rotation at the rates $\pm(\omega_o-\omega_\mathrm{c})$. Hence, the first two terms will only contribute to the dynamics of the system if the coupling rate $g$ is on the order of or bigger than the sum of $\omega_o$ and $\omega_\mathrm{c}$ whereas the second two terms simply require $g$ to be on the order of or bigger than the difference $\omega_o-\omega_\mathrm{c}$. For optical systems, the former case is very difficult to achieve (see more on this in section \ref{chapter1:cavityregimes}). Thus, we can eliminate the first two terms for having a much faster phase rotation.

By using the rotating wave approximation to eliminate the terms with a fast phase rotation, and transforming back into the original picture, we arrive at the Jaynes-Cummings Hamiltonian for a cavity-emitter system
\begin{equation}
\label{chapter1eq:jaynescummings}
    \hat{H}=\hbar\omega_o\sigu\sigd+\hbar\omega_\mathrm{c}\au\ad+\hbar g(\sigd\au+\sigu\ad).
\end{equation}
The physical interpretation of this Hamiltonian is clear. The emitter and the cavity mode both evolve with their respective homogeneous frequencies of $\omega_o$ and $\omega_\mathrm{c}$. The interaction term coherently converts a single excitation of the emitter into a single photon in the cavity or vice versa at the cavity-emitter coupling rate $g$.

The Jaynes-Cummings Hamiltonian has been very successful at explaining the quantum coherent interaction between an emitter and a cavity mode in the optical regime. However, its derivation makes a simplistic assumption about the spectrum of the cavity resonance that may be violated in some systems. Furthermore, the quantization of the electromagnetic field used here assumes that the entire field is confined perfectly to the mode volume $V$. Of course, this is never true because light inevitably leaks out of the cavity or becomes absorbed by the material forming the cavity. In section \ref{chapter1:qnmmastereq}, I will briefly introduce a more general approach that can account for both of these failings. However, in the sections preceding that discussion, I will introduce a simpler and more common approach that only accounts for the lossy nature of a nearly-ideal single-mode cavity with defined boundaries.

\subsection{Input-output theory}
\label{chapter1:inputoutput}

The cavity mode that gives rise to the Jaynes-Cummings interaction discussed in the previous section can inevitably lose energy into the environment. This cavity disspation can be modeled by considering that the electromagnetic field of the cavity mode must satisfy certain electromagnetic boundary conditions at the edges of the cavity mode volume $V$. This gives rise to an additional coupling between the confined cavity mode and the continuum of unconfined modes surrounding the cavity. These boundary conditions are reflected by the input-output relations derived by Gardiner and Collett \cite{gardiner1985input}.

The input-output relation between the confined mode $\ad$ and the remaining 1-dimensional continuum of modes (or waveguide mode) is, $\bd(t) - \bd_o(t) = \sqrt{\kappa\eta_\mathrm{c}}\ad(t)$, where $\bd$ is the forward propagating waveguide mode and $\bd_o$ is the reverse-propagating input waveguide mode. This holds in the limit that the cavity has a very low dissipative rate, or that it is highly reflective. For optical light, we can assume that the input waveguide mode is in the vacuum state since the thermal occupation $\overline{n}(T)$ is very small for frequencies on the order of THz, even at room temperature. The collection efficiency $\eta_\mathrm{c}$ can be introduced by assuming that the cavity has two possible decay channels, one into the waveguide at the rate $\kappa_\mathrm{c}$ and another $\kappa-\kappa_\mathrm{c}$ due to emission into other propagating modes or absorption losses. Then $\eta_\mathrm{c}=\kappa_\mathrm{c}/\kappa$ is the collection efficiency of the waveguide. We can further take into account the propagation phases of these fields to yield $\bd(t)e^{ikx}=\sqrt{\kappa\eta_\mathrm{c}}\ad(t)+\bd_o(t)e^{-ikx}$. 

This thesis deals exclusively with photon counting measurements of one or more waveguide modes. I will not explore measurements where it is necessary to consider the direct impact of vacuum fluctuations, such as quadrature measurements. As a result, I quite often write the direct equivalence between system operators and waveguide operators, such as $\bd(t)=\sqrt{\kappa\eta_\mathrm{c}}\ad(t)e^{-ikx}$, when computing quantities related to $\bd$. This is common practice in the literature \cite{carmichael,kiraz2004quantum,fischer2016dynamical}. However, this proportionality should be seen as a short-hand notation because it violates the necessary commutation relations for the propagating mode $\bd$.

\subsection{Dissipative cavity quantum electrodynamics}
\label{chapter1:cavityQED}

By accounting for the input-output coupling of the cavity mode and the waveguide, and subsequently tracing out the waveguide states, we arrive at the master equation for the reduced cavity-emitter system given by the Liouville superoperator $\mathcal{L}=-i\mathcal{H}/\hbar+\gamma\mathcal{D}(\sigd)+\kappa\mathcal{D}(\ad)$,
where the cavity-emitter Hamiltonian superoperator $\mathcal{H}\hat{\rho}=[\hat{H},\hat{\rho}]$ is given by Eq.~(\ref{chapter1eq:jaynescummings}) \cite{breuer2002theory}. Here, I have taken the temperatures of both environment baths of the emitter and cavity to be such that $\hbar\omega\gg k_\mathrm{B} T$ so that only spontaneous emission remains.

One last important ingredient that we need in order to arrive at a functional device model is the ability to externally control our two-level system within the cavity. This can be done by illuminating the two-level system with coherent laser pulses. If these pulses occupy electromagnetic modes that do not couple to the previously-described cavity mode or the collection waveguide mode, we can make a semi-classical approximation for the dipole interaction Hamiltonian and append it to our Jaynes-Cummings Hamiltonian with few negative consequences \cite{breuer2002theory,carmichael}. In the semi-classical approximation of the dipole interaction $-\hat{\mathbf{d}}\cdot\mathbf{E}$, the electromagnetic field is taken to be classical while the dipole remains quantized. This gives the semi-classical driving Hamiltonian in the rotating wave approximation 
\begin{equation}
    \hat{H}_D = -\hat{\mathbf{d}}\cdot\mathbf{E}(t) \simeq \frac{\hbar}{2}\left(\Omega^*(t)\sigd e^{i\omega_\mathrm{L}t}+
\Omega(t)\sigu e^{-i\omega_\mathrm{L}t}\right),
\end{equation}
where $\Omega(t)$ is the Rabi frequency induced by the coherent driving of frequency $\omega_\mathrm{L}$. This direct emitter driving scenario occurs in practice using a cross-polarization setup where excitation and emission are orthogonally polarized \cite{somaschi2016}\ref{ollivier2020reproducibility}. In addition, it is possible that the semi-classical drive is applied to the emitter via a cavity mode orthogonal to $\hat{a}$, which can increase $\Omega$ and may also distort pulses that are spectrally broad compared to the cavity linewidth. 

\begin{figure}[t]
    \centering
    \includegraphics[trim=0cm 2cm 0cm 1cm,width=0.9\textwidth,clip]{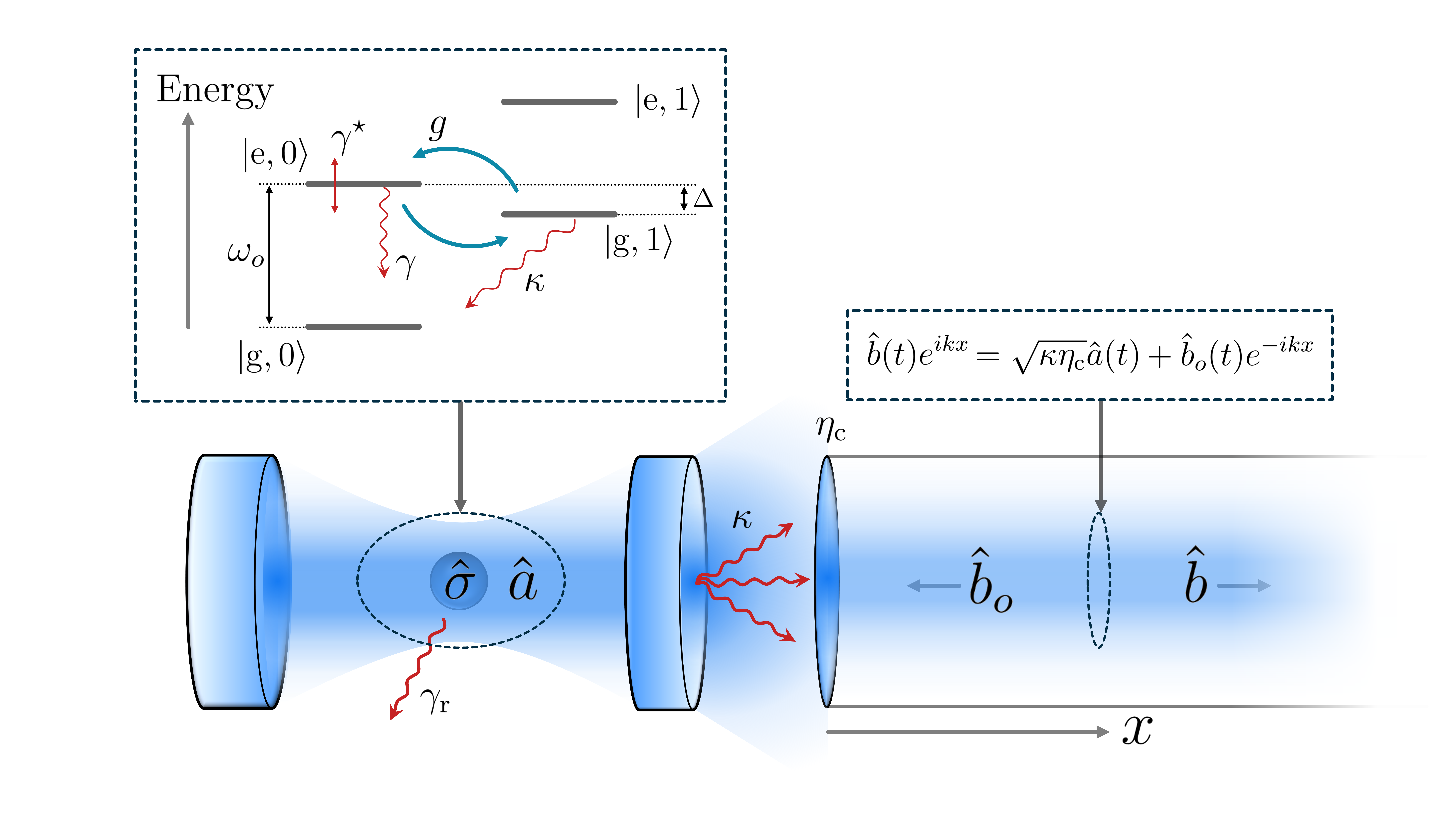}
    \caption[The emitter-cavity system coupled to a one-dimensional collection waveguide.]{\small\textbf{The emitter-cavity system coupled to a one-dimensional collection waveguide.} The emitter with optical frequency $\omega_o$, represented by the lowering operator $\hat{\sigma}=\ketbra{\mathrm{g}}{\mathrm{e}}$, experiences pure dephasing at the rate $\gamma^\star$ and decays from its excited state $\ket{\mathrm{e}}$ to its ground state $\ket{\mathrm{g}}$ at the rate $\gamma$. Part of the dissipated energy is radiated into the environment at the rate $\gamma_\mathrm{r}$. The cavity mode with frequency $\omega_\mathrm{c}$, represented by the bosonic annihilation operator $\hat{a}$ where $\hat{a}\ket{n}=\sqrt{n}\ket{n-1}$, decays exponentially into the environment at the rate $\kappa$. The interaction between the cavity mode and the optical transition of the emitter is described by the Jaynes-Cummings interaction with a coupling rate of $g$. The input-output relation describes the propagating mode $\hat{b}$ of the one-dimensional waveguide mode, which depends also on the collection efficiency $\eta_\mathrm{c}$, position along the waveguide $x$, and the wavenumber $k$. The vacuum input mode $\hat{b}_o$ of the collection waveguide must be included to satisfy the commutation relations, although it does not contribute to intensity measurements of the waveguide mode.}
    \label{ch1fig:emittercavitywaveguide}
\end{figure}

Taking into account the semi-classical driving Hamiltonian and the phenomenological pure dephasing term introduced in section \ref{chapter1:spontaneousemission}, we arrive at the driven dissipative cavity QED Markovian master equation model used in this thesis. This master equation is described by the Liouville superoperator
\begin{equation}
    \mathcal{L}(t) = -\frac{i}{\hbar}\mathcal{H}(t)+\kappa\mathcal{D}(\ad)+2\gamma^\star\mathcal{D}(\sigu\sigd)+\gamma\mathcal{D}(\sigd),
\end{equation}
where
\begin{equation}
    \hat{H}(t) = \hbar\Delta_o\sigu\sigd + \hbar\Delta_\mathrm{c}\au\ad + \hbar g\left(\au\sigd+\ad\sigu\right) +\frac{\hbar}{2}\left(\Omega^*(t)\sigd+\Omega(t)\sigu\right),
\end{equation}
and with a collection waveguide mode given by $\bd(t)=\sqrt{\kappa\eta_\mathrm{c}}\ad(t)e^{ikx}$. In writing the Hamiltonian, I have moved into the rotating frame defined by the driving laser frequency $\omega_\mathrm{L}$ to remove the explicit time-dependent phase. This is done by applying the unitary transformation $\hat{U}=e^{-i\omega_\mathrm{L} t(\sigu\sigd+\au\ad)}$ to the original Hamiltonian. Hence, $\Delta_o = \omega_o-\omega_\mathrm{L}$ is the detuning between the emitter resonance and the driving frequency and $\Delta_\mathrm{c}=\omega_\mathrm{c}-\omega_\mathrm{L}$ is the detuning between the cavity resonance and the driving frequency. Fig.~\ref{ch1fig:emittercavitywaveguide} shows a summary diagram of this quantum model for $\Omega(t)=0$ and emitter-cavity detuning $\Delta=\omega_o-\omega_\mathrm{c}$.

In many cases studied in this thesis, I assume that the emitter is initially perfectly prepared in the excited state $\ket{\mathrm{e}}$. For the cavity QED model above, this can be achieved by applying a square $\pi$ pulse with non-zero amplitude $\Omega=\pi/t_\mathrm{p}$ during a duration of time $t_\mathrm{p}$. Then, if $t_\mathrm{p}$ is much smaller than the timescale of the remaining system dynamics (so that the pulse is very fast), the initial emitter state $\ket{\mathrm{g}}$ is excited perfectly to $\ket{\mathrm{e}}$ after $t_\mathrm{p}$. Although this perfect $\pi$ pulse excitation is achievable using the Markovian master equation model, in practice there are limits to the regimes where it is valid. In particular, phonon damping of the solid-state emitter will prevent arbitrarily perfect excitation and the Markovian approximation breaks down when $t_\mathrm{p}$ is reduced to be on the order of the phonon bath memory time \cite{mccutcheon2010quantum}. In addition, emission can occur during the excitation pulse if $t_\mathrm{p}$ is not much faster than the total emitter decay rate. This may cause re-excitation to occur, leading to multi-photon emission. These additional details generally hinder the perfect operation of a device. Hence, in the spirit of this thesis, the perfect excitation approximation serves to derive general bounds on figures of merit and explore the general regimes of operation for a device. That said, I do also extensively explore multi-photon effects and re-excitation processes in section \ref{chapter3:HOM}.

To get insight into the emission dynamics in the idealized scenario where the emitter is perfectly excited, we can derive a set of  differential equations known as the optical Bloch equations for when $\Omega=0$. These describe the time dynamics of the expectation values of the operators $\sigd$ and $\ad$. In the Schr\"{o}dinger picture, we have $d\!\braket{\hat{A}(t)}\!/dt=\tr{\hat{A} \mathcal{L}\hat{\rho}(t)}$ for an operator $\hat{A}$ in the Hilbert space. Applying this to $\sigd$ and $\ad$, where for convenience I will drop the explicit dependence on time, gives
\begin{equation}
\label{chapter1eq:amplitudedifeq}
\begin{aligned}
    \frac{d}{dt}\braket{\sigd}&=-i\Delta\braket{\sigd}+ig\braket{\ad\sigd_z}-\frac{\Gamma}{2}\braket{\sigd}\\
    \frac{d}{dt}\braket{\ad}&=-ig\braket{\sigd}-\frac{\kappa}{2}\braket{\ad},\\
\end{aligned}
\end{equation}
where I have set $\Delta_\mathrm{c}=0$ so that the cavity frequency $\omega_\mathrm{c}$ defines the reference frame, $\Gamma=\gamma+2\gamma^\star$ is the FWHM of the homogeneously broadened spectrum of the emitter. Recall from section \ref{chapter1:spontaneousemission} that the quantity $\Gamma/2$ here describes the total decoherence of the emitter two-level system. The correlation function $\braket{\ad\sigd_z}$ couples these two equations to an infinite series of equations of motion for higher-order correlation functions. However, if the system begins with at most one excitation, either in the emitter or cavity, then the state subspace accessible to the system reduces to $\{\ket{\mathrm{g},0},\ket{\mathrm{g},1},\ket{\mathrm{e},0}\}$. In this subspace, the operator $\ad\sigd_z$ vanishes when acting on all states except $\ket{\mathrm{g},1}$ where it returns $-\ket{\mathrm{g},0}$. Hence, in this single-excitation approximation we have $\ad\sigd_z=-\ad$, which decouples the above coupled differential equations from the higher-order correlations.

Using the master equation, we can also derive the equations of motion for the populations $\braket{\sigu\sigd}$ and $\braket{\au\ad}$ as well as correlations $\braket{\au\sigd}$ and $\braket{\ad\sigu}$ in the single-excitation approximation, which are
\begin{equation}
\begin{aligned}
    \frac{d}{dt}\braket{\sigu\sigd}&=-ig\braket{\ad\sigu}+ig\braket{\au\sigd}-\gamma\braket{\sigu\sigd}\\
    \frac{d}{dt}\braket{\ad\sigu}&=i\Delta\braket{\ad\sigu}-ig\braket{\sigu\sigd}+ig\braket{\au\ad}-\frac{1}{2}\left(\kappa+\Gamma\right)\braket{\ad\sigu}\\
    \frac{d}{dt}\braket{\au\sigd}&=-i\Delta\braket{\au\sigd}+ig\braket{\sigu\sigd}-ig\braket{\au\ad}-\frac{1}{2}\left(\kappa+\Gamma\right)\braket{\au\sigd}\\
    \frac{d}{dt}\braket{\au\ad}&=ig\braket{\ad\sigu}-ig\braket{\au\sigd}-\kappa\braket{\au\ad}.
\end{aligned}
\end{equation}
If the emitter-cavity correlations decay very quickly compared to the coupling rate so that $2g<\kappa+\Gamma$, then the emitter-cavity system is said to be in the weak-coupling regime. I will present the different regimes of cavity QED in more detail in the following section. In the weak-coupling regime and when the initial system state does not give cavity-emitter correlations, the values of $\braket{\ad\sigu}$ and $\braket{\au\sigd}$ can be adiabatically eliminated \cite{auffeves2009pure}. This is done by setting their time derivatives to zero so that the correlations depend on time implicitly through the populations $\braket{\sigu\sigd}$ and $\braket{\au\ad}$. Solving the resulting linear system of equations for the correlations gives
\begin{equation}
    \braket{\ad\sigu}^*=\braket{\au\sigd}=\frac{2ig}{\Gamma+\kappa+2i\Delta}\left(\braket{\sigu\sigd}-\braket{\au\ad}\right).
\end{equation}
Substituting this solution into the set of differential equations results in
\begin{equation}
\label{chapter1eq:coupledRdifeq}
\begin{aligned}
    \frac{d}{dt}\braket{\sigu\sigd}&=-(\gamma+R)\braket{\sigu\sigd}+R\braket{\au\ad}\\
    \frac{d}{dt}\braket{\au\ad}&=-(\kappa+R)\braket{\au\ad}+R\braket{\sigu\sigd},\end{aligned}
\end{equation}
where
\begin{equation}
    R = \frac{4g^2(\kappa+\Gamma)}{(\kappa+\Gamma)^2+4\Delta^2}
\end{equation}
is the effective rate of population transfer between the emitter and the cavity. This rate is maximized under the resonant condition $\Delta=0$, giving $R=4g^2/(\kappa+\Gamma)$. Although this quantity $R$ only accurately approximates the rate of emitter-cavity population transfer when $2g<\kappa+\Gamma$, it is still a very useful quantity to discuss and explain various emitter-cavity behaviours. For this reason, the quantity $R$ will be discussed even when $\kappa+\Gamma\geq 2g$. Furthermore, from Eq.~(\ref{chapter1eq:coupledRdifeq}), we can see that the effective decay rate of the two-level system becomes $\gamma+R$. This decay rate enhancement is called the Purcell effect \cite{purcell1946}, and it is quantified by the unitless quantity known as the Purcell factor.

The Purcell factor $F_\mathrm{p}$ is defined as the ratio of the cavity-enhanced radiative rate emitted through the cavity mode to the bare radiative rate $\gamma_\mathrm{r}$ of the emitter. In the weak-coupling regime \cite{auffeves2009pure,auffeves2010controlling}, this cavity rate is $R$ and so
\begin{equation}
\label{ch1eq:purcellfactor}
    F_\mathrm{p} = \frac{R}{\gamma_\mathrm{r}} = \frac{4g^2(\kappa+\Gamma)}{\gamma_\mathrm{r}(\kappa+\Gamma)^2+4\Delta^2}.
\end{equation}
Recall from section \ref{chapter3:jaynescummings} that the cavity coupling rate can be written as $g=d\sqrt{\omega_\mathrm{c}/2n^2\varepsilon_0\hbar V}$. In addition, in section \ref{chapter1:spontaneousemission}, we found that $\gamma_\mathrm{r}=n\omega_o^3d^2/3\pi\varepsilon_0\hbar c^3$. Substituting these into $F_\mathrm{p}$, we have
\begin{equation}
\label{chapter1eq:purcellfactor}
    F_\mathrm{p}=\frac{4g^2}{\gamma_\mathrm{r}\kappa}F_\mathrm{inh} = \left(\frac{2\omega_\mathrm{c} d^2}{n^2\varepsilon_0\hbar V}\right)\left(\frac{3\pi \varepsilon_o\hbar c^3}{n\omega^3 d^2\kappa}\right)F_\mathrm{inh}=\frac{3}{4\pi^2}\left(\frac{\lambda}{n}\right)^3\left(\frac{Q}{V}\right)F_\mathrm{inh},
\end{equation}
where $Q=\omega_\mathrm{c}/\kappa$ is the cavity quality factor and $\lambda=2\pi c/\omega_o$ is the emission wavelength in the vacuum. The inhibition factor due to detuning and emitter dephasing is
\begin{equation}
    F_\text{inh}=\frac{\kappa(\kappa+\Gamma)}{(\kappa+\Gamma)^2+4\Delta^2}.
\end{equation}
Note that I am still assuming that the dipole is oriented and positioned to maximize the cavity-emitter coupling rate. The more general expression would have an additional inhibition factor $F_\text{geom}$ that depends on the geometry of the emitter-cavity system. The above result is valid for a phenomenological pure dephasing model. It is also possible to derive Purcell inhibition factors that rigorously include the electron-phonon interactions giving rise to emitter dephasing, which consequently affect the cavity-emitter interaction \cite{roy2015spontaneous}. These more detailed approaches are highly applicable to quantum dot systems, especially those that operate at warmer temperatures.

The expression given by Eq.~(\ref{chapter1eq:purcellfactor}) illustrates that the Purcell factor is proportional to the ratio of the cavity quality factor $Q$ to the cavity mode volume $V$. Thus, the decay rate enhancement experienced by the emitter is larger if the cavity is smaller for a given $Q$ or if the cavity dissipates energy into the environment more slowly for a given $V$. However, the validity of the Purcell factor is restricted to the weak-coupling regime of cavity QED. If a cavity approaches the bounds of this regime, the Purcell factor expression derived above can over-estimate the decay rate enhancement. In section \ref{chapter3:generalizedpurcell}, I will generalize the Purcell factor to include this additional inhibition effect induced by strong cavity coupling. In the following section, I will expand on the different regimes of cavity QED.

\subsection{Regimes of cavity quantum electrodynamics}
\label{chapter1:cavityregimes}

Depending on the relative magnitudes of the optical emitter-cavity QED system parameters, the dynamics of their interaction can be categorized into various regimes that display particular properties. The names of these regimes vary in the literature and so for clarity I will define my choice of vernacular following the convention taken in Ref.~\cite{auffeves2010controlling}, which is somewhat followed in Ref.~\cite{grange2015cavity} as well.

In this section, I will assume that the emitter is resonant with the cavity so that $\Delta=0$. I will also not consider regimes related to ultra-strong or deep-strong coupling \cite{kockum2019ultrastrong,forn2019ultrastrong}, where the cavity coupling $g$ is on the order of or greater than the frequency of the emitter $\omega_o$, which invalidates approximations made to obtain the Jaynes-Cummings model. This requires $g$ to be orders of magnitude larger than what has been demonstrated for a single optical emitter. However, effects of this regime can be observed in microwave systems where $\omega_o$ is much smaller or for ensembles of emitters where $g$ is effectively enhanced to $g\sqrt{N_o}$ for $N_o$ emitters and may one day be reached for a single optical emitter.

\begin{figure}[t]
    \centering
    \hspace{-35mm}(a)\hspace{51mm}(b)\hspace{51mm}(c)
    \includegraphics[width=\textwidth]{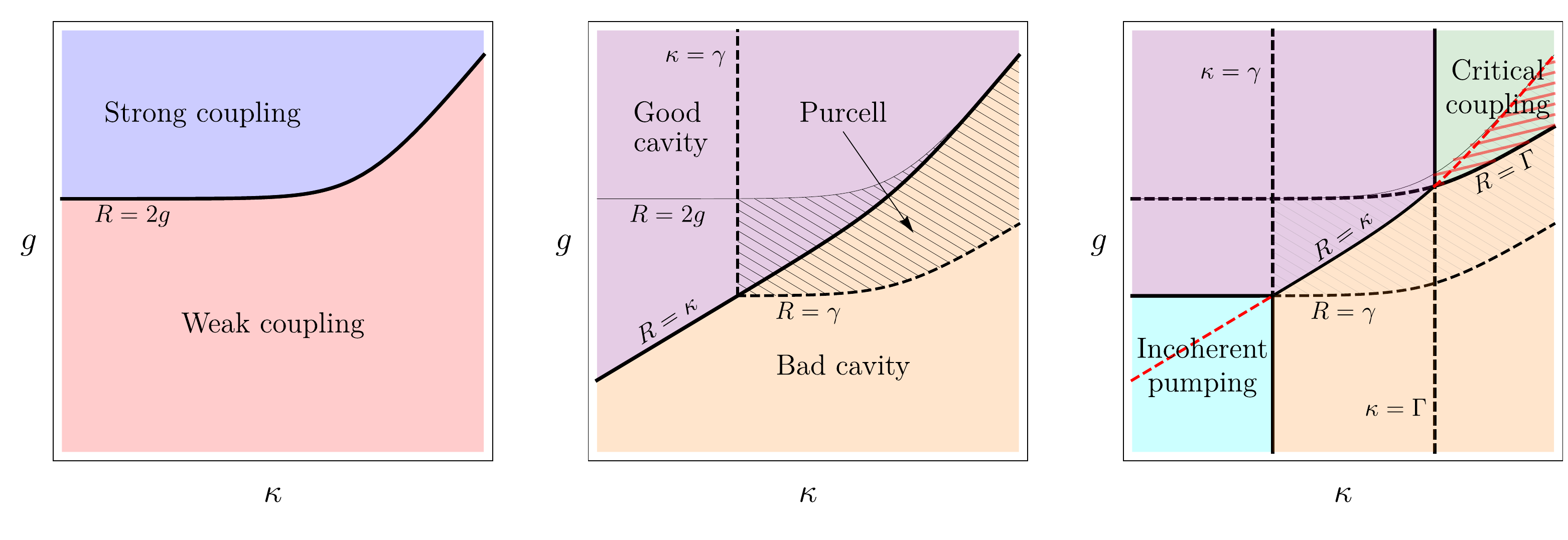}
    \caption[Log-scale illustrations of emitter-cavity QED regimes.]{\small\textbf{Log-scale illustrations of emitter-cavity QED regimes} as a function of the cavity decay rate $\kappa$ and cavity coupling rate $g$, inspired by plots from Ref.~\cite{grange2015cavity}. (a) The division of the strong-coupling (blue) and weak-coupling (red) regimes by the curve $R=2g$ drawn in black, where $R=4g^2/(\kappa+\Gamma)$, $\Gamma=\gamma+2\gamma^\star$, $\gamma$ is the bare emitter decay rate, and $\gamma^\star$ is the emitter pure dephasing rate. (b) The division of the good-cavity (purple) and bad-cavity (orange) regimes separated by the line $R=\kappa$. The hatched region indicates the Purcell regime ($R/\gamma\geq 1$) where the cavity can greatly enhance the emitter decay rate. Note that the entire region in the top right corner where $\kappa,R>\gamma$ (indicated by the dashed lines) can also allow for a Purcell enhancement, but the spectrum will show a vacuum Rabi splitting when $2g>R$ and then the standard Purcell factor is invalid. (c) The critical coupling (green) and incoherent-pumping (cyan) regimes that allow for spectrally/temporally pure emission from the cavity mode. The curve labeled by $R=\kappa$ shows the Goldilocks condition, which extends into both the green and cyan regions (red dashed lines). The faded gray hatched area shows the Purcell regime from panel (b) and the red hatched area shows the overlap between the critical regime and the Purcell regime. Panels (b) and (c) are drawn under the assumption that $\gamma\ll 2\gamma^\star$ so that $\Gamma\simeq 2\gamma^\star$.}
    \label{chapter1fig:cavityregimes}
\end{figure}

The cavity system can be first divided into the strong ($2g\geq \kappa+\Gamma$) and weak ($2g<\kappa+\Gamma$) coupling regimes (see Fig.~\ref{chapter1fig:cavityregimes}~(a)). Note that it is equivalent to define them as $2g\leq R$ for strong coupling and $2g> R$ for weak coupling. The strong-coupling regime is attained when the cavity coupling rate exceeds the decay rate of the cavity-emitter correlations, causing vacuum Rabi oscillations. This is also characterized by a visible vacuum Rabi splitting in the emission spectrum. The spectrum provided in the weak-coupling regime remains Lorentzian in shape, albeit with a width and quantum characteristics that further depend on the parameters.

The cavity system is said to be in the good-cavity regime when the emission rate from the cavity is slower than the effective rate of population transfer from the emitter ($\kappa<R$) and otherwise it is said to be in the bad-cavity regime ($\kappa\geq R$) (see Fig.~\ref{chapter1fig:cavityregimes}~(b)). It is also important to consider the regime where an enhancement of the emission rate is induced by the cavity. The enhancement occurs when the rates $R$ and $\kappa$ both exceed the emitter lifetime $\gamma$, shown by the upper right region bounded by the dashed lines in Fig.~\ref{chapter1fig:cavityregimes}~(b). Within this regime of enhancement, the area that is also within the weak-coupling regime is the classic Purcell regime, where the Purcell factor is valid and larger than 1.

In the case that $\gamma^\star$ is very small, all the regimes discussed so far will allow for emission that is spectrally pure, meaning that emission is in a pure quantum state. For pulsed systems, the emission will also be temporally pure and sequentially indistinguishable (I will discuss indistinguishability more in section \ref{chapter1:indistinguishability}). However, for a pulsed resonantly excited emitter that experiences dephasing, there are two regimes where highly indistinguishable emission can be achieved. I will refer to these two regimes as the incoherent pumping regime ($R,\kappa <\gamma$) and the critical regime ($R,\kappa>\Gamma$) (see Fig.~\ref{chapter1fig:cavityregimes}~(c)). The incoherent pumping regime is named so because the emitter inefficiently and incoherently excites the cavity mode so that the characteristics of emission are dominated by the cavity rather than the emitter. Since the cavity acts as the emitter, and it does not suffer from dephasing, indistinguishable emission can be achieved \cite{grange2015cavity}. I refer to the other regime as the critical regime because it allows for system dynamics that are reminiscent of a critically damped oscillator, although not all points within this regime satisfy that analogy. In this regime, the emission rate is so fast that it overcomes the dephasing rate, allowing for indistinguishable emission.

These latter two regimes contain the condition $R=\kappa$ that defines the boundary between the good- and bad-cavity regimes, which is a `Goldilocks condition' \cite{goldilocks} for single-photon sources. This condition implies that the cavity mode decays at the same rate that it is effectively populated by the emitter. Here, the perfect balance is struck between having a cavity that is good enough to enhance the light-matter coupling but still dissipative enough to not prolong the light-matter interaction. In the critical regime, this condition reduces to $\kappa=2g$, which is analogous to an impedance matching condition, allowing for an efficient transfer of emitter population into the environment. In the incoherent pumping regime, the condition reduces to $\kappa\Gamma=4g^2$. Of these two regimes, only the critical regime is within the region where a significant rate enhancement can be achieved. It also lies on the boundary between the strong-coupling and the weak-coupling regimes. This boundary is defined by $R=2g$, which incidentally converges to the Goldilocks condition in the critical regime. The small region between this boundary and the curve $R=\Gamma$ is the only region where the standard Purcell factor is valid (there are no vacuum Rabi oscillations) and the emission is highly indistinguishable. Interestingly, reasonably efficient emission that is partially indistinguishable can be achieved along the Goldilocks condition between the incoherent pumping and critical regimes. This bridge between the regions arises due to the cavity funnelling effect \cite{grange2015cavity}, where the broader dephased emitter linewidth is funneled into the narrower cavity linewidth.

Although these different regimes are useful to keep in mind, the remainder of this thesis will focus primarily on the critical regime and, in most cases, only the part of the critical regime that overlaps with the Purcell regime, as indicated by the red hatched area in Fig.~\ref{chapter1fig:cavityregimes}~(c). In the next section, I will describe how the cavity-emitter-waveguide system in the bad-cavity regime reduces to an effective `one-dimensional atom'.

\subsection{Cavity-enhanced one-dimensional atom}
\label{chapter1:1datom}

In section \ref{chapter1:cavityQED}, I introduced how the cavity-emitter dynamics in the weak-coupling regime can be described by the effective rate $R$ that couples the equation of motion for $\braket{\sigu\sigd}$ and $\braket{\au\ad}$ (see Eq.~(\ref{chapter1eq:coupledRdifeq})). We can push this idea one step further by eliminating the cavity mode from the dynamics entirely. In this case, the two-level emitter effectively couples directly to the collection waveguide. This simplified picture of a two-level system coupled to a waveguide is referred to as a one-dimensional atom \cite{auffeves2007giant}.

Instead of eliminating the correlation between the emitter and the cavity to get Eq.~(\ref{chapter1eq:coupledRdifeq}), we can directly eliminate the cavity amplitude in Eq.~(\ref{chapter1eq:amplitudedifeq}). By setting $d\!\braket{\ad}\!/dt=0$, we obtain $\braket{\ad} = -(2ig/\kappa)\braket{\sigd}$. Similarly, from Eq.~(\ref{chapter1eq:coupledRdifeq}) we can eliminate the cavity population to get $\braket{\au\ad}=R\braket{\sigu\sigd}/(\kappa+R)$. In this case, for $\kappa\gg R$, the cavity-emitter system reduces to an effective two-level system whose evolution can be described by the effective Liouville superoperator
\begin{equation}
    \mathcal{L} = -\frac{i}{\hbar}\mathcal{H} + \gamma^\prime\mathcal{D}(\sigd)+2\gamma^\star\mathcal{D}(\sigu\sigd),
\end{equation}
where $\hat{H}\simeq\hbar\omega_o^\prime\sigu\sigd+\hbar(\Omega^*(t)\sigd+\Omega(t)\sigu)/2$ is the effective Hamiltonian, $\gamma^\prime\simeq\gamma+R$ is the Purcell-enhanced decay rate, and $\omega_o^\prime\simeq\omega_o-4g^2\Delta/(\kappa^2+4\Delta^2)$ is the cavity Lamb-shifted resonance of the two-level system. Since this model is effectively identical to the model introduced in section \ref{chapter1:spontaneousemission}, I will write $\gamma^\prime$ as $\gamma$ without specifying the parameters of the cavity that give rise to the decay rate.

A special note must be made for including the semi-classical driving in the Hamiltonian. When performing adiabatic elimination, one will find that the effective decay rate $\gamma^\prime$ depends on $\Omega$ and that $\Omega$ reduces $\gamma^\prime$. Physically, this occurs when the semi-classical driving dresses the two-level system, effectively decoupling it from the cavity mode and reducing the Purcell factor. Furthermore, a time-dependent $\Omega$ will cause $\gamma^\prime$ to become time dependent. However, any transient behaviour of the cavity-emitter interaction due to a time-dependent $\Omega$ is inherently neglected when performing adiabatic elimination. Hence, in addition to requiring the bad-cavity condition $\kappa\gg R$ so that any transients are short compared $1/\gamma^\prime$, this effective two-level model is valid either when (1) $\Omega(t)$ is small compared to $g$ or (2) $\Omega(t)$ is nonzero for a time small compared to $1/\gamma^\prime$. This latter condition is often satisfied for state-of-the-art pulsed single-photon sources operating far into the bad-cavity regime, where the effective two-level model performs well \ref{ollivier2020reproducibility}. I will discuss this topic in more detail at the end of section \ref{chapter3:temporalcoherenceandstatistics}.

For the one-dimensional atom model, the cavity amplitude $\ad$ becomes effectively proportional to the amplitude of the two-level system $\sigd$. Strictly speaking, these two operators cannot be proportional because $\sigd$ is a finite dimensional operator whereas $\ad$ is an infinite dimensional operator. However, under the condition that $\braket{\ad},\braket{\au\ad}\ll 1$, which is satisfied in the bad-cavity regime, the cavity mode operator becomes an effective two-level operator itself. Furthermore, this effective proportionality extends to the waveguide mode $\bd$ provided that we only perform measurements on the waveguide that are not sensitive to the vacuum fluctuations of the continuum, such as photon-counting measurements.

\subsection{Quasi-normal mode master equation}
\label{chapter1:qnmmastereq}

The derivation of the Jaynes-Cummings model and the cavity QED master equation makes a key assumption that the modes of the cavity are far separated in frequency compared to their linewidth $\kappa$. That is, we must assume that the spectrum of the cavity is described by a sum of well-separated independent Lorentzian-shaped resonances. This allows for the light-matter interaction to be described by a single quantized mode $\ad$ interacting with the two-level system. However, this assumption is often violated for low-$Q$ devices that incorporate very lossy materials, such as metals or other plasmonic material, that can absorb and/or scatter light at a high rate.

If the spectrum of a cavity cannot be modeled by a summation of independent narrow Lorentzian modes, then it is necessary to revisit the light-matter interaction from first principles. Although there have been many approaches to generalizing the Jaynes-Cummings model for highly dissipative environments \cite{dorier2020critical}, one recent approach by Franke \emph{et al.} \cite{franke2019quantization} stands above the rest in that it provides a clear and computationally-friendly \cite{ren2020near} link between the classical properties of the electromagnetic field and a Markovian master equation predicting the quantum dynamics. This approach is based on a quasi-normal mode (QNM) expansion of the field where the cavity is described by a summation of complex Lorentzian modes that can interfere if they overlap in space and frequency. These QNMs are modes with complex eigenfrequencies that describe the mode resonance peak as the real part of the eigenfrequency and the spectral width, or damping rate, as the imaginary part of the eigenfrequency.

The QNM master equation has been successfully used to analyze Purcell enhancements of single emitters \cite{franke2019quantization} and study the quantum properties of light emitted by a cavity-emitter system \cite{hughes2019theory}. In this thesis, I will explore some properties predicted by the phenomenological application of the quantum model based on the generalized constraints of its parameters. Hence, I will not go into depth on the subtleties and fundamental considerations of its derivation, which can be found in Ref.~\cite{franke2020quantized}.

The form of the QNM master equation is similar to a multi-mode Jaynes-Cummings interaction with the key difference that the mode quantization accounts for possible interference and light-matter coupling renormalization. Let us begin by supposing that our cavity field is described by a set of QNMs $\tilde{a}_i$ \cite{franke2019quantization} with complex resonances $\tilde{\omega}_i=\omega_i -i\kappa_i/2$, where $\omega_i$ is the mode resonance and $\kappa_i$ is the FWHM spectral intensity linewidth. The modes $\tilde{a}_i$ contain all the spatial and spectral structure of the cavity field. However, in general, they do not immediately satisfy the canonical commutation relations for harmonic oscillators because they are not plane-wave modes---they have a finite width due to dissipation. To enforce these commutation relations, the modes can be symmetrized by defining new operators $\ad_i$ such that $\ad_i=\sum_{j}(\mathbf{S}^{-1/2})_{ij}\tilde{a}_j$ and $\au_i=\sum_{j}(\mathbf{S}^{-1/2})_{ji}\tilde{a}^\dagger_j$, where $\mathbf{S}$ is a positive semi-definite Hermitian matrix fully describing the spatial and spectral overlap of the QNMs \cite{franke2019quantization}.

After QNM symmetrization, the Liouville superoperator that governs the quantum evolution of an emitter-cavity interaction is given generally as \cite{franke2019quantization}
\begin{equation}
    \mathcal{L} = -\frac{i}{\hbar}\mathcal{H} + \gamma\mathcal{D}(\sigd,\sigd)+\sum_{ij}2\chi_{ij}^-\mathcal{D}(\ad_i,\ad_j)
\end{equation}
where $\mathcal{D}$ is given by $\mathcal{D}(\ad_i,\ad_j)\hat{\rho}=\ad_i\hat{\rho}\au_j-\{\au_j\ad_i,\hat{\rho}\}/2$ and the Hamiltonian corresponding to $\mathcal{H}\hat{\rho}=[\hat{H},\hat{\rho}]$ is
\begin{equation}
    \hat{H} = \hbar\omega_o \hat{\sigma}^\dagger\hat{\sigma} + \sum_{ij}\hbar\chi^{+}_{ij}\au_i\ad_j+\sum_{i}\hbar g_i\sigd\au_i+\hbar g_i^*\sigu\ad_i.
\end{equation}
The parameters $\chi^\pm_{ij}$ describing the QNM interference are given by $\chi_{ij}^+=(\chi_{ij}+\chi_{ji}^*)/2$ and $\chi_{ij}^-=i(\chi_{ij}-\chi_{ji}^*)/2$, where $\chi_{ij}=\sum_k(\mathbf{S}^{-1/2})_{ik}\tilde{\omega}_k(\mathbf{S}^{1/2})_{kj}$. The complex emitter-cavity coupling rate $g_i$ is given in terms of the QNM coupling rates before symmetrization $\tilde{g}_j$ by $g_i=\sum_j(\mathbf{S}^{1/2})_{ji}\tilde{g}_j$. The quantized mode $\ad_i$ has a bare resonance $\chi_{ii}^+$ and interacts with mode $\ad_j$ through the Hermitian coupling described by $\chi_{ij}^+$. It is important to note that the quantized mode operators $\ad_i$ describe the quasinormal modes after performing a symmetrization that maintains the required commutative properties. Hence, the resonances $\chi_{ii}^+$ and the decay rates $\chi_{ii}^-$ do not correspond to $\tilde{\omega}_{i}$ directly, but are already altered due to the interference of the QNMs. Furthermore, we have a coupling between the quantized modes. This coupling appears in the Hamiltonian described by parameters $\chi_{ij}^+$ and a coupling in the dissipative part of the master equation described by parameters $\chi_{ij}^-$.

Since the matrix $\mathbf{\chi}^-$ that is described by the elements $\chi_{ij}^-$ is Hermitian, there exists a unitary transformation of modes $\ad_i$ that diagonalizes the dissipative part of the master equation \cite{franke2020quantized}. I will denote these dissipative eigenmodes as $\cd_i=\sum_jv_j\ad_j$ for eigenvector elements $v_i$ of $\mathbf{\chi}^-$. The eigenvalues $\kappa_i^\prime$ of this transformation correspond to the exponential decay of modes $\cd_i$. This transformation allows for a straightforward derivation of input-output relations for the symmetrized QNM operators \cite{franke2020quantized}. It also provides a familiar Markovian master equation form where all of the mode coupling is described by the Hamiltonian and the dissipative part simply describes the exponential decay of those coupled modes. It is important to emphasize that, although this approach provides a model that is similar in form to a multi-mode Jaynes-Cummings model, the light-matter coupling and mode interference now become directly dependent on dissipation. This can lead to a significant reduction in the coupling magnitude compared to the standard Jaynes-Cummings approach when considering highly dissipative cavities \cite{franke2020quantized}.

In this thesis, I will not go into detail about the quantum properties of emission predicted by this QNM master equation, as this is already actively being explored \cite{hughes2019theory,ren2020near,franke2020quantized}. However, I do expect that many of the methods I use in this thesis for photonic state analysis can be applied to this generalized system to great effect. That said, in section \ref{chapter4:cavitygates}, I will apply the QNM master equation to explore the potential application of plasmonic cavities to mediate high-fidelity local spin-spin interactions.

\section{Figures of merit}

I will now introduce a handful of important quantities that appear multiple times throughout this thesis. First, I will discuss how the Purcell factor is related to the cavity cooperativity, which is a figure of merit used extensively in section \ref{chapter4:cavitygates}. Then I will introduce three important figures of merit for single-photon sources: the brightness, the single-photon purity, and the indistinguishability. These quantities arise often in chapter \ref{chapter3} and again in section \ref{chapter4:entanglementgeneration}. Finally, I will define state fidelity and concurrence. These will be used in section \ref{chapter3:photonicstate} to quantify the quality of photonic state entanglement, and again in chapter \ref{chapter4} to discuss spin-spin entanglement.

\subsection{Cavity cooperativity}
\label{chapter1:cavitycooperativity}

The rate $R$ not only describes the Purcell enhancement in the Purcell regime as we saw in section \ref{chapter1:cavityQED}, it is also a useful quantity to characterize the cavity-emitter system. As we found in section \ref{chapter1:cavityregimes}, it allows us to easily define many interesting regimes of a dephased cavity-emitter system. However, since $R$ is not a unitless quantity, it is not useful when comparing different physical systems. Instead, we can talk about the rate $R$ normalized by the emitter lifetime $\gamma$: the cavity cooperativity.

Let us consider a resonant system ($\Delta=0$) without dephasing in the regime where $\gamma\ll \kappa$. Then the cavity cooperativity is
\begin{equation}
\label{ch1eq:cooperativity}
    C = \frac{4g^2}{\kappa\gamma},
\end{equation}
which is a quantity that appears very often in the literature and will arise often in section \ref{chapter4:cavitygates}.
Similar to the Purcell factor, the cavity cooperativity quantifies how quickly the system decays via the cavity mode as opposed to direct decay via the emitter. Note that, if $\gamma\ll \kappa$, $C>1$ does not imply that there is a vacuum Rabi splitting. Hence, I do not consider a system with $C>1$ as necessarily being in the strong-coupling regime. We can also discuss the effective cavity cooperativity, which I denote $C^\star$, that is inhibited by emitter broadening and detuning
\begin{equation}
    C^\star=CF_\text{inh}.
\end{equation}
The effective cavity cooperativity is related to the effective Purcell factor by $C^\star=\eta_\mathrm{r}F_\mathrm{P}$, where $\eta_\mathrm{r}=\gamma_\mathrm{r}/\gamma$ is the radiative efficiency of the emitter.

\subsection{Brightness}
\label{chapter1:brightness}

\begin{figure}
    \centering
    \hspace{-61mm}(a)\hspace{31mm}(b)\hspace{45mm}(c)
    \includegraphics[width=\textwidth]{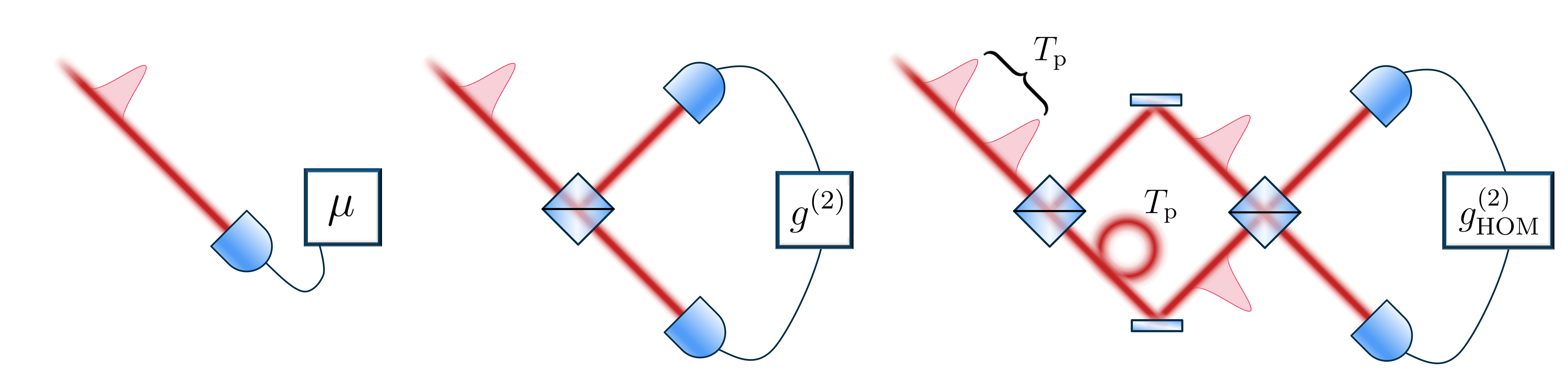}
    \caption[Characterization of a pulsed single-photon source.]{\textbf{Characterization of a pulsed single-photon source.} (a) Integrated intensity $\mu$ measured by direct detection. By knowing the losses prior to detection, this can be used to estimate the source brightness $\beta$, which is $\beta=1$ in the ideal case. (b) Single-photon purity measured by the integrated intensity correlation $g^{(2)}$ using a balanced beam splitter and normalizing by the intensity $\mu$. In the ideal case where the photonic state contains at most one photon, we have that $g^{(2)}=0$. (c) Characterization of the purity in the time/frequency domain. Subsequent photonic states are interfered in a Mach-Zehnder interferometer with one arm delayed by the pulse separation $T_\mathrm{p}$. The normalized intensity correlation after Hong-Ou-Mandel (HOM) interference $g^{(2)}_\text{HOM}$ should be zero in the ideal case due to perfect photon bunching.}
    \label{ch1fig:mug2hom}
\end{figure}

For many applications of single-photons, it is necessary that the photons are produced very efficiently. This is usually characterized by the source `first lens' brightness $\euscr{B}$, which is the ratio of the photon rate collected at position $x=0$ of the waveguide divided by rate of pulsed excitation of the device. In the case that at most one photon can be produced by a single pulse, the first lens brightness becomes equal to the probability of generating a single photon.

Since it is not easy to place a perfectly efficient detector at the first lens, the measurement of brightness is usually dependent on the characterization of additional losses and inefficiencies of the setup, such as transmission and detection efficiency. It is also affected by the probability that the emitter is in the correct `bright state' configuration to emit a photon, which I will denote $\eta_\text{b}$.

In this thesis, I will discuss only the contribution to the first lens brightness due to the dynamics of the light-matter interaction when the emitter is in the correct configuration. This is described by the average photon number in the waveguide $N(t)=\braket{\bu(t)\bd(t)}$ integrated over time
\begin{equation}
\label{ch1eq:brightness}
    \mu = \int N(t)dt=\kappa\eta_\mathrm{c}\int\braket{\au(t)\ad(t)}dt,
\end{equation}
where the integration is taken over all dynamics induced by a single pulsed excitation or state preparation protocol. It is also convenient to introduce the quantity $\overline{\mu}=\mu/\eta_\mathrm{c}=\sum_n np_n$, where $p_n=\prb{n}$ is the probability of having $n$ photons leaving the cavity mode. This quantity has important implications for the photon statistics of the photonic state produced by the cavity mode. Since I am dividing out the collection efficiency $\eta_\mathrm{c}$, some of these photons may be directly absorbed by the cavity material, or leave the cavity but never make it to the waveguide. Using $\overline{\mu}$, we can write the total first-lens brightness, as $\euscr{B}=\eta_\text{b}\eta_\mathrm{c}\overline{\mu}$.

For a given emitter system, it is often enlightening to look at what $\overline{\mu}$ is expected for a perfect state preparation of the emitter. This gives insight into the upper bound on the brightness under perfect single-photon generation conditions. I will denote this idealized definition of brightness as $\beta$ and refer to it either as the intrinsic cavity efficiency or brightness without further specification. For the cavity-emitter system introduced so far, a perfect state preparation corresponds to $\ket{\mathrm{e}}$, which can give at most one photon. In the Purcell regime, the value of $\beta$ then has an upper bound that is given by the ratio of decay from the emitter at the rate $\gamma$ and the effective rate $R$. Using the effective rate $R$, we can write this upper bound in terms of the effective cavity cooperativity as
\begin{equation}
    \beta\leq \frac{R}{\gamma+R}=\frac{C^\star}{1+C^\star}\leq 1.
\end{equation}

For single-photon sources, one ultimate goal is to achieve $p_1=1$. From the above definition of $\euscr{B}$, we can see that $\euscr{B}=1$ does not necessarily imply $p_1=1$ since $\overline{\mu}=1$ can imply an infinite number of sets of $p_n$ values. For example, $p_0=p_2=0.5$ can also give $\overline{\mu}=1$. Therefore, it is necessary to characterize the quality of a single-photon source using more than just the brightness.

\subsection{Single-photon purity}

An ideal single-photon source must provide at most one photon per pulse. This can be characterized by the antibunching of the photonic state at the output of a beam splitter (see Fig.~\ref{ch1fig:mug2hom}~(b)). If the photonic state does not contain more than one photon, the intensity correlation between output modes $\bd_3$ and $\bd_4$ measured by the two detectors will vanish. For a balanced beam splitter, the modes $\bd_3$ and $\bd_4$ will be in identical states proportional to the state of the input waveguide mode $\bd=\bd_1$, however the detection times may be different. Integrating the two-time intensity correlation of the waveguide mode $G^{(2)}(t_1,t_2)=\braket{\bu(t_1)\bu(t_2)\bd(t_2)\bd(t_1)}$ over all times and normalizing by the integrated intensity of each mode gives the normalized integrated intensity correlation characterizing the single-photon purity
\begin{equation}
\label{ch1eq:g2}
    g^{(2)}=\frac{1}{\mu^2}\iint G^{(2)}(t_1,t_2)dt_1dt_2,
\end{equation}
where the integration is taken over the dynamics as for $\mu$. This quantity is often denoted $g^{(2)}(0)$ to emphasize that it is the intensity correlation around zero delay, which is a relic of continuous wave (CW) source characterization. However, the true equivalent to $g^{(2)}(0)$ for pulsed sources would be given by the time-integrated intensity correlation $g^{(2)}(\tau)$ as a function of detection delay $\tau$ given by $g^{(2)}(\tau)=\mu^{-1}\int G^{(2)}(t,t+\tau)dt$, which is not equal to the full time-integrated quantity if $\tau=0$. Hence, for pulsed sources where subsequently produced photon states are uncorrelated, I find the emphasis on $0$ unnecessary when discussing the full time-integrated quantity and so I will use the notation $g^{(2)}$ and not $g^{(2)}(0)$. 

Furthermore, it is important to note that the intensity normalized $g^{(2)}$ is independent of photon losses, such as non-unity collection efficiency $\eta_\mathrm{c}$ and transmission efficiency $\eta_\mathrm{t}$. It is also independent of detector efficiency $\eta_\mathrm{d}$ provided that the beam splitter is balanced so that each detector receives the same average intensity. This means that $g^{(2)}$ measured after significant losses is identical to the value at the device.

If the source experiences irreversible evolution besides that which is caused by the collected emission, it is not straightforward to compute two-time correlations such as $\braket{\bu(t_1)\bu(t_2)\bd(t_2)\bd(t_1)}$ in the Schr\"{o}dinger picture when $t_2<t_1$. However, due to the time symmetry of the correlation function and normalization intensity, we can restrict the integration to a time ordering such that $t_2\geq t_1$ to compute $g^{(2)}/2$ and multiply by the factor of 2 to obtain the full time-integrated value. As a consequence, I will often switch between the form given in Eq.~(\ref{ch1eq:g2}) and the equivalent expression
\begin{equation}
    g^{(2)}=\frac{2}{\mu^2}\int_{-\infty}^\infty\int_{t_1}^\infty G^{(2)}(t_1,t_2)dt_2dt_1,
\end{equation}
depending on which is more convenient. I will discuss more about time ordering conventions in chapter \ref{chapter2}.

\subsection{Indistinguishability}
\label{chapter1:indistinguishability}

Some applications of single-photon sources, such as quantum state teleportation \cite{valivarthi2020teleportation}, quantum repeaters \cite{dur1999quantum}, and optical quantum computing \cite{o2007optical}, require that the photons are indistinguishable. This is characterized by how well two photons interfere at a beam splitter to give rise to two-photon bunching, which is known as the Hong-Ou-Mandel (HOM) effect \cite{hong1987measurement}. The topic of HOM interference is covered in significant detail in section \ref{chapter3:HOM}. However, in this introductory section, I will give the basics in order to define the quantities of mean wavepacket overlap and indistinguishability.

For perfectly indistinguishable photons at the input of a balanced beam splitter $\ket{\psi}=\bu_1\bu_2\ket{0}$, the output after the beam splitter transformation $\bu_1=(\bu_3+\bu_4)/\sqrt{2}$ and $\bu_2=(\bu_4-\bu_3)/\sqrt{2}$ is
\begin{equation}
    \ket{\psi} = \frac{1}{2}\left(\bu_3+\bu_4\right)\left(\bu_4-\bu_3\right)\ket{0} = \frac{1}{2}\left(\hat{b}_4^{\dagger 2}-\hat{b}_3^{\dagger 2}\right)\ket{0},
\end{equation}
which implies that the intensity correlation between detectors monitoring modes $\bd_3$ and $\bd_4$ vanishes. Here, I am using $\bd$ to represent a plane-wave mode of the waveguide(s) to demonstrate the basic nature of the HOM effect. In practice, finite wavepacket length plays an important role in determining the output state of the beam splitter.

To quantify the indistinguishability of photons emitted by the same source, a Mach-Zehnder interferometer is often used where one arm is delayed by the separation between pulses $T_\mathrm{p}$ (see Fig.~\ref{ch1fig:mug2hom}~(c)). By following the methods similar to Ref.~\cite{kiraz2004quantum}, we can take into account the time dynamics of the identical input photonic states. The intensity correlation at the output of a beam splitter is given as in the previous section $G^{(2)}_\text{HOM}(t_1,t_2)=\braket{\bu_3(t_1)\bu_4(t_2)\bd_4(t_2)\bu_3(t_1)}$. Applying the inverse balanced beam splitter relations $\bu_3=(\bu_1-\bu_2)/\sqrt{2}$ and $\bu_4=(\bu_1+\bu_2)/\sqrt{2}$ we can expand this intensity correlation in terms of the input modes $1$ and $2$. By making the assumption that these input modes are not entangled or classically correlated, the correlations such as $\braket{\bu_1\bu_2\bd_2\bd_1}$ can be separated into $\braket{\bu_1\bd_1}\braket{\bu_2\bd_2}$. Then, assuming that the states of each mode are identical, we can take $\bd_1(t)=\bd(t)$ and $\bd_2(t)=\bd(t)$. Finally, it is often the case that the input states carry no one- and two-photon coherence so that $\braket{\bd(t)}=\braket{\bd(t)\bd(t)}=0$. This is true when the photonic state leaving the cavity gives $p_1\simeq 1$, because any photon losses from collection or transmission will not generate photon number coherence. This is also true when the Mach-Zehnder interferometer used to measure indistinguishability has a highly unstable phase on the timescale of the measurement. In this case, any effect of nonzero $\braket{\bd(t)}$ or $\braket{\bd(t)\bd(t)}$ will average to zero. Under either condition, we obtain
\begin{equation}
    G^{(2)}_\text{HOM}(t_1,t_2)= \frac{1}{2}\left[N(t_1)N(t_2)+G^{(2)}(t_1,t_2)-\left|G^{(1)}(t_1,t_2)\right|^2\right],
\end{equation}
where $G^{(1)}(t_1,t_2)=\braket{\bu(t_2)\bd(t_1)}$ is the two-time first order amplitude correlation function of the mode $\bd$.

By integrating the intensity correlation over all arrival times for a single pulse or state-preparation protocol, and normalizing by the total integrated intensity, we obtain
\begin{equation}
\label{ch1eq:g2hom}
    g^{(2)}_\text{HOM} = \frac{1}{\mu^2}\iint G^{(2)}_\text{HOM}(t_1,t_2)dt_1dt_2=\frac{1}{2}\left(1+g^{(2)}-M\right),
\end{equation}
where I have defined the mean wavepacket overlap as
\begin{equation}
\label{ch1eq:mwpo}
    M = \frac{1}{\mu^2}\iint \left|G^{(1)}(t_1,t_2)\right|^2dt_1d_2.
\end{equation}
It is also common to talk about the HOM visibility $V_\text{HOM}$, which is further defined to be 
\begin{equation}
\label{ch1eq:vhom}
    V_\text{HOM} =  1 - 2g^{(2)}_\text{HOM}=M-g^{(2)}.
\end{equation}
Like $g^{(2)}$, all of these quantities are normalized by intensity and so they are independent of photon losses. Unlike $g^{(2)}$, good HOM interference requires that the intensity of each input is equal at the beam splitter.

In practice, the HOM visibility is normalized with respect to the total uncorrelated coincident counts of the interferometer \cite{kiraz2004quantum}\ref{ollivier2020reproducibility}\ref{ollivier2020g2hom}. This is because the uncorrelated counts provide a very good estimate for the squared intensity $\mu^2$ when the measured photonic state contains little one-photon coherence. If this is not the case, the uncorrelated coincident counts can underestimate $\mu^2$ \cite{gustin2018pulsed}, leading to an over-estimation of $V_\text{HOM}$ as defined above. This subtlety is not an issue for good single-photon sources where $\braket{\bd(t)}\simeq 0$, but it is very important to consider when using HOM visibility to characterize more general photonic states. I will discuss this more in section \ref{chapter3:selfhomodyne}.

The quantities $M$ and $V_\text{HOM}$ are both often referred to as the indistinguishability of photons. It is also common to use $1-g^{(2)}_\text{HOM}$ \cite{gustin2018pulsed}, which for a single photon is the HOM bunching probability. Indeed, all these quantities characterize the quality of interference between two photonic states. However, the term indistinguishability refers to a property of two particles. Hence, to discuss indistinguishability, we should be clear about which two particles we are considering. In the context of single-photon source characterization, we are concerned with measuring the indistinguishability of two successive uncorrelated photons emitted by the same source. That is, the two particles are in identical states that may not be pure. 

Recently, there has also been discussion of another quantity: the trace purity of the single-photon component of a photonic density matrix \cite{fischer2018particle,trivedi2020generation}. Let $\hat{\varrho}$ be the photonic state where $\hat{\varrho}_1$ is the unnormalized state after being projected onto the one-photon subspace with probability $p_1$, then the single-photon trace purity is $\euscr{P}_1\equiv \text{Tr}\left(\hat{\varrho}_1^2\right)/p_1^2$. In my opinion, $0\leq \euscr{P}_1\leq 1$ is the most appropriate quantity to call indistinguishability, as it is by definition the purity of two photons described by the same mixed photonic state and gives their maximum HOM bunching probability by $(1+\euscr{P}_1)/2$.

The mean wavepacket overlap $M$, the HOM visibility $V_\text{HOM}$, and $g^{(2)}_\text{HOM}$ are all dependent on the \emph{entire} photonic state measured at the interferometer, including multi-photon components, but are independent of photon losses prior to measurement. On the other hand, $\euscr{P}_1$ is given by the single-photon component of a photonic state and hence is generally dependent on losses. For example, a two-photon component will contribute to a measurement of $\euscr{P}_1$ if one photon was lost prior to the measurement. In the limit of large losses where only the single-photon component remains, $\euscr{P}_1$ of the photonic state \emph{at the interferometer} is equal to $M=V_\text{HOM}+g^{(2)}$. This follows from the fact that $G^{(1)}$ in this limit is proportional to temporal density function of the single-photon component. Temporal density functions will be the subject of section~\ref{chapter2:temporaldensityfunctions}. If $g^{(2)}=0$, then $\euscr{P}_1$ is independent of losses and so $\euscr{P}_1=M=V_\text{HOM}$. However, in general, it is not possible to exactly obtain $\euscr{P}_1$ of the photonic state \emph{at the source} by performing HOM measurements after losses \cite{trivedi2020generation}.

The single-photon trace purity $\euscr{P}_1$ of the photonic state at the source is an important quantity that characterizes the quantum purity of the source. It can also be used to make predictions about the behaviour of the device in various other applications, hence it is useful to try to estimate it. For these reasons, I will define the \emph{single-photon} indistinguishability $I$ as the value of $\euscr{P}_1=M=V_\text{HOM}$ when $g^{(2)}\rightarrow 0$. In this case, $I$ serves to quantify the upper bound on the interference visibility for a particular device, similar to $\beta$ for brightness. In section \ref{chapter3:roomtemperature}, I will compute $I$ and $\beta$ in the critical regime. The subtleties of the relationship between $M$, $V_\text{HOM}$, $g^{(2)}$, $\euscr{P}_1$, and $I$, was explored in Ref.~\ref{ollivier2020g2hom} and is a main topic of sections~\ref{chapter3:HOM} and \ref{chapter3:photonicstate}. In section \ref{chapter4:entanglementgeneration}, the mean wavepacket overlap becomes the focus as photons from two different sources are interfered after significant losses.

\subsection{Fidelity}
\label{chapter1:fidelity}

When discussing the general quality of a particular quantum state, it is often convenient to use the measure of state fidelity. However, this requires knowledge about the desired state, which may not be available. If the model is well-understood, then it may be the case that a desired state is apparent. The state fidelity then allows for a convenient way to quantify how close the measured state is to the desired state.

The fidelity between the desired state $\hat{\rho}$ and the measured state $\hat{\varrho}$ is defined generally by \cite{jozsa1994fidelity}
\begin{equation}
\label{ch1eq:fidelity}
    \euscr{F}=\text{Tr}^2\!\left(\sqrt{\sqrt{\hat{\rho}}\hat{\varrho}\sqrt{\hat{\rho}}}\right).
\end{equation}
Note that for $\hat{\rho}=\hat{\varrho}$ the fidelity is $\euscr{F}=1$, even if $\hat{\rho}$ is not a pure state $\tr{\hat{\rho}^2}< 1$. Very often, the desired state is a pure state $\hat{\rho}=\ketbra{\psi}{\psi}$. In this case, the fidelity reduces to the expectation value of the density operator with respect to the pure state $\euscr{F} = \braket{\psi|\hat{\varrho}|\psi}$. If both states are pure, the fidelity reduces to $\euscr{F}=\left|\braket{\phi|\psi}\right|^2$ where $\hat{\varrho}=\ketbra{\phi}{\phi}$ and $\hat{\rho}=\ketbra{\psi}{\psi}$.

In some protocols, such as those studied in section \ref{chapter4:entanglementgeneration}, it is possible to post-select different (unnormalized) measured states $\hat{\varrho}_m$ based on the measurement result $m$ in the space of measurements $\mathbb{M}$ such that $\hat{\varrho}=\sum_{m\in\mathbb{M}}\hat{\varrho}_m$ where $\prb{m}=\tr{\hat{\varrho}_m}$ is the measurement probability. In this case, I define the protocol fidelity as a weighted average of all the post-selected measurements $m$ in the set of accepted measurements $\mathbb{A}$ so that $\euscr{F}=\sum_{m\in\mathbb{A}}\prb{m}\euscr{F}_m/\eta$ where $\eta=\sum_{m\in\mathbb{A}}\prb{m}$ is the total protocol efficiency, $\euscr{F}_m=\text{Tr}^2\!\left(\sqrt{\sqrt{\hat{\rho}_m}\hat{\varrho}_m\sqrt{\hat{\rho}_m}}\right)/\prb{m}$, and $\hat{\rho}_m$ is the desired state for measurement $m$. In this way, the individual measurement probabilities $\prb{m}$ cancel and the weight is carried by the unnormalized density operator $\hat{\varrho}_m$.

\subsection{Entanglement and concurrence}
\label{chapter1:concurrence}

Although the fidelity is a good measure of the distance between two states, it is quite influenced by the classical statistics of the states and it requires knowledge about a particular desired state. This makes it difficult to use the fidelity to determine, for example, how entangled a state is. Formally, a quantum state $\ket{\psi}$ of two or more subsystems is said to be entangled if it cannot be written as a product of individual subsystem states. That is, if $\ket{\psi}\neq \ket{\psi_1}_1\otimes\ket{\psi_2}_2\otimes\cdots\otimes\ket{\psi_N}_N$ for any set of states $\ket{\psi_i}$ of $N$ subsystems. Entangled states give rise to quantum correlations that can be very useful in quantum information processing and fundamental experiments such as Bell inequality violations. However, it is generally a difficult task to quantify ``how much" entanglement a state has if there are more than a few quantum systems involved.

In this thesis, I only explore the simplest case: entanglement between qubits of two quantum systems. Suppose we have two systems each with states $\ket{0}$ and $\ket{1}$. A very useful set of entangled states that spans the total system of two qubits are the Bell states. These are defined by $\ket{\psi^\pm}=(\ket{10}\pm\ket{01})/\sqrt{2}$ and $\ket{\phi^\pm}=(\ket{00}\pm\ket{11})/\sqrt{2}$. These Bell states give maximum possible quantum correlations and so they are said to be maximally entangled. I will discuss Bell states more in sections \ref{chapter3:photonicstate} and \ref{chapter4:entanglementgeneration}.

Entanglement is straightforward to quantify for two-qubit systems, and there are multiple ways to do so. For this purpose, I will sometimes compute the fidelity with respect to a maximally entangled state. But occasionally, I will also compute the Wootters concurrence \cite{wootters2001entanglement}, which is independent of a reference state. If the subsystems have more than two states, I will compute the concurrence after projecting the total system onto the two-qubit subsystem. This measure of entanglement is defined by
\begin{equation}
\label{ch1eq:concurrence}
    \euscr{C} = \text{max}\left(0,\sqrt{\alpha_1}-\sqrt{\alpha_2}-\sqrt{\alpha_3}-\sqrt{\alpha_4}\right),
\end{equation}
where $\alpha_i\geq\alpha_{i+1}$ is the $i^\text{th}$ eigenvalue of $(\mathcal{I}_\text{s}\hat{\varrho})(\mathcal{Y}_\text{s}\hat{\varrho}^*)$. Here, $\mathcal{I}_\text{s}\hat{\rho} = (\hat{I}_\text{s}\otimes\hat{I}_\text{s})\hat{\rho}(\hat{I}_\text{s}\otimes\hat{I}_\text{s})$ is the density operator projected onto the two-qubit subsystem and $\hat{I}_\text{s}=\hat{\sigma}^\dagger\hat{\sigma}+\hat{\sigma}\hat{\sigma}^\dagger$ is the identity operator of the two-qubit subsystem, where $\hat{\sigma}$ is the spin qubit lowering operator; and $\mathcal{Y}_\text{s}\hat{\rho}^* =(\hat{\sigma}_y\otimes\hat{\sigma}_y)\hat{\rho}^*(\hat{\sigma}_y\otimes\hat{\sigma}_y)$, where $\hat{\sigma}_y = i(\hat{\sigma}-\hat{\sigma}^\dagger)$ is the Pauli $y$ operator. 

A state with a concurrence greater than 0 is an entangled state. A maximally entangled state of two qubits, such as a Bell state, has a concurrence of 1. The concurrence is also related to other measures of entanglement \cite{bartkiewicz2013entanglement}. For example, a state with a concurrence bigger than $1/\sqrt{2}$ is guaranteed to violate the Clauser-Horne-Shimony-Holt (CHSH) Bell inequality \cite{clauser1969proposed}. However, this does not mean that a concurrence less than $1/\sqrt{2}$ will not violate the CHSH inequality.

\section{Analytic tools}

In this section, I will summarize some important analytic methods used throughout this thesis. Each subsection describes a `tool' that I use at least once, and some more often than others. The first tool is the Fock-Liouville space, which is a formalism that I use in all of my numerical simulations. So, although it will not appear again directly in the lines of this thesis, nearly all of the figures are taking advantage of it in some way. The second tool is the quantum regression theorem, which is used to compute some analytic results shown in section \ref{chapter3:roomtemperature} and the details of its application are in appendix \ref{AppendixB}. The third tool is the non-Hermitian Hamiltonian approximation, which I apply to derive results presented in section \ref{chapter4:cavitygates}. It is also closely related to material in chapter \ref{chapter2}. The fourth tool, adiabatic elimination, has already been used in section \ref{chapter1:cavityQED}, and will appear again in section \ref{chapter4:cavitygates}. Lastly, I will discuss time-dependent perturbation theory in the context of superoperators, which is the foundation for material in chapter \ref{chapter2}.

\subsection{The Fock-Liouville space}
\label{chapter1:fockliouville}

To solve the dynamics of a Markovian master equation, it is very convenient to use of the Fock-Liouville space representation \cite{manzano2020short}. This transforms the square matrix form of the density operator $\hat{\rho}$ into a vector representation notated by $\kett{\rho}$ and linear superoperators of the form $\mathcal{S}\hat{\rho}=\sum_{i}\hat{A}_i\hat{\rho}\hat{B}_i$ into a square matrix notated by $\tilde{\mathcal{S}}$ acting on the density state vector. In this representation, the Markovian master equation is $d\ket{\rho}\!\rangle/dt=\tilde{\mathcal{L}}\ket{\rho}\!\rangle$, where $\tilde{\mathcal{L}}$ is the matrix representation of the linear superoperator $\mathcal{L}$.

Given that $\hat{\rho}=\sum_{i,j}\rho_{i,j}\ketbra{\phi_i}{\phi_j}$, the vector form of $\hat{\rho}$ becomes $\kett{\rho}=\sum_{i,j}\rho_{i,j}\ket{\phi_i}\otimes\ket{\phi_j}$. This can be viewed as taking each column of the matrix $\hat{\rho}$ and stacking them into a single column vector. Likewise, for a superoperator of the form $\mathcal{S}\hat{\rho}=\sum_i\hat{A}_i\hat{\rho}\hat{B}_i$, where $\hat{A}_i$ and $\hat{B}_i$ are operators acting on the Hilbert space, the matrix representation is given by the relation $\tilde{\mathcal{S}} = \sum_i\hat{A}_i\otimes\hat{B}_i^\text{T}$, where $\hat{B}_i^\text{T}$ is the transpose of $\hat{B}_i$. For a finite Hilbert space with dimension $N$, the Fock-Liouville space has dimension $N^2$.

The Fock-Liouville representation can be used to solve the evolution of the master equation using either time-dependent or time-independent techniques. If $\mathcal{L}$ is time dependent, the state evolution can be solved using one of many numerical algorithms, such as Runge-Kutta. If $\mathcal{L}$ is time independent, then the propagation superoperator $\mathcal{U}$ can be solved by standard matrix exponentiation in the Fock-Liouville space: $\tilde{\mathcal{U}}(t,t_0)=e^{(t-t_0)\tilde{\mathcal{L}}}$. This propagator then acts on $\kett{\rho}$ to solve for $\kett{\rho(t)}$ given $\kett{\rho(t_0)}$. The original density operator $\hat{\rho}(t)$ can then be recovered by applying the reverse transformation back into its matrix form.

In some cases, $\tilde{\mathcal{L}}$ can be analytically diagonalized, providing an analytic form for the propagation superoperator. This often allows for analytic derivations of useful figures of merit for a quantum device, such as efficiency, fidelity, or emission indistinguishability. Otherwise, it is also often possible to numerically diagonalize $\tilde{\mathcal{L}}$ for a fixed set of parameters resulting in $\tilde{\mathcal{U}}(t_\text{f},t_0)$ that is still analytic with respect to time. This approach can drastically decrease the time needed to simulate the time dynamics and two-time correlations.

\subsection{Quantum regression theorem}
\label{chapter1:quantumregression}

For Markovian systems, there is a useful theorem that sometimes allows for a simplified computation of multi-time correlations. The so-called quantum regression theorem \cite{breuer2002theory} implies that the statistics of multi-time correlation functions are determined by the statistics of the one-time expectation values along with a proper application of the dynamical map. That is, if we know the set of operator expectations $\braket{\hat{A}_i(t)\hat{B}_j(t)}$, and the equations of motion for $\braket{\hat{A}_i(\tau)}$, we can obtain $\braket{\hat{A}_i(t+\tau)\hat{B}_j(t)}$. Formally, the theorem states that if
\begin{equation}
    \frac{d}{d\tau}\braket{\hat{A}_i(\tau)}=\sum_{j}X_{i,j}\braket{\hat{A}_j(\tau)}
\end{equation}
then
\begin{equation}
    \frac{\partial}{\partial\tau}\braket{\hat{A}_i(t+\tau)\hat{A}_k(t)}=\sum_{j}X_{i,j}\braket{\hat{A}_j(t+\tau)\hat{A}_k(t)},
\end{equation}
where $X_{i,j}$ are the coefficients of the coupled set of first-order linear differential equations.

Suppose we know
\begin{equation}
    \frac{d}{dt}\braket{\hat{A}_i(t)\hat{A}_j(t)} = \sum_{k,l}Y_{i,k}^{j,l}\braket{\hat{A}_k(t)\hat{A}_l(t)}.
\end{equation}
Let $\mathbf{x}(t)$ be the vector of expectation values $\braket{\hat{A}_i(t)}$ and let $\mathbf{y}(t)$ be the vector of expectation values $\braket{\hat{A}_i(t)\hat{A}_j(t)}$. Then we can write $\dot{\mathbf{x}}=X\mathbf{x}$ and $\dot{\mathbf{y}}=Y\mathbf{y}$, where I use the dot to represent the time derivative and the matrices $X$ and $Y$ have coefficients $X_{i,j}$ and $Y_{i,k}^{j,l}$ in some ordering. If the matrix $X$ has a dimension $N$, then $Y$ has the dimension $N^2$. For time-independent parameters, we can now write the solution $\mathbf{z}_k$, which is the $N$ dimensional vector of two-time correlation functions $\braket{\hat{A}_i(t+\tau)\hat{A}_k(t)}$, as
\begin{equation}
    \mathbf{z}_k(t,\tau)=e^{\tau X}\left[e^{tY}\mathbf{y}(0)\right]_k,
\end{equation}
where the subscript $k$ takes the $k^\mathrm{th}$ $N$-vector corresponding to the set of $\braket{\hat{A}_i(t)\hat{A}_k(t)}$.

The quantum regression theorem approach is particularly useful to solve time integrals of two-time correlation functions for systems where the equations of motion for the first and second order correlation functions are uncoupled, such as the optical Bloch equations described in section \ref{chapter1:cavityQED}. This is because it allows for the separability of functions dependent on $t$ and $\tau$ and hence the integrated correlations can be solved by computing two one-dimensional integrals rather than one two-dimensional integral.

The quantum regression theorem is valid for Markovian systems. However, if the Markovian master equation accurately predicts the evolution of the reduced system state, this does not necessarily imply that the quantum regression theorem will also accurately predict the multi-time correlation functions of the same system. That is, the validation of the quantum regression theorem is stronger evidence for Markovianity than the validation of the Markovian master equation. In this sense, it has been suggested that a good test (or perhaps definition) of Markovian behaviour of a system is its satisfaction of the quantum regression theorem \cite{guarnieri2014quantum}.

\subsection{Non-Hermitian Hamiltonians}
\label{chapter1:nonhermitian}

Moving from the full Liouville-von Neumann system evolution description to a Markovian master equation simplifies a model significantly. In some specific cases, it is possible to further simplify the system dynamics back to an effective Hamiltonian that governs the reduced state evolution via the Schr\"{o}dinger equation. 

To see how this is possible, first consider the cavity-emitter dissipative master equation of the form
\begin{equation}
    \frac{d}{dt}\hat{\rho} = -\frac{i}{\hbar}[\hat{H},\hat{\rho}] + \gamma\mathcal{D}(\hat{\sigma})\hat{\rho} + \kappa\mathcal{D}(\hat{a})\hat{\rho}
\end{equation}
This can be rewritten as \cite{daley2014quantum}
\begin{equation}
\begin{aligned}
    \frac{d}{dt}\hat{\rho} &= -\frac{i}{\hbar}\left[\hat{H},\hat{\rho}\right] - \frac{1}{2}\{\gamma \hat{\sigma}^\dagger\hat{\sigma}+\kappa\hat{a}^\dagger\hat{a},\hat{\rho}\} + \gamma\hat{\sigma}\hat{\rho}\hat{\sigma}^\dagger + \kappa\hat{a}\hat{\rho}\hat{a}^\dagger\\
    &=-\frac{i}{\hbar}\left(\tilde{H}_\text{eff}\hat{\rho}-\hat{\rho}\tilde{H}_\text{eff}^\dagger\right)+\gamma\hat{\sigma}\hat{\rho}\hat{\sigma}^\dagger + \kappa\hat{a}\hat{\rho}\hat{a}^\dagger,
\end{aligned}
\end{equation}
where
\begin{equation}
    \tilde{H}_\text{eff} = \hat{H} -\frac{i\hbar}{2}\left(\gamma \hat{\sigma}^\dagger\hat{\sigma}+\kappa\hat{a}^\dagger\hat{a}\right)
\end{equation}
is an effective non-Hermitian Hamiltonian that captures the amplitude decay of $\hat{\sigma}$ and $\hat{a}$. The remaining terms that cannot be brought into the effective Hamiltonian correspond to the jump operators $\mathcal{J}(\sigd)$ and $\mathcal{J}(\ad)$. These terms cause the system state to jump down by a quantum of energy perhaps after the spontaneous emission of a photon. Hence, they serve to preserve the trace of $\hat{\rho}$ by recycling the population into the ground state(s) after decay. In fact, this rearrangement is the starting point for the quantum trajectories formalism, where the stochastic action of the jump operators are simulated via Monte Carlo type simulations to reproduce the dynamics of the full density operator $\hat{\rho}$.

Due to the absence of the jump operators, the effective Schr\"{o}dinger equation defined by the effective non-Hermitian Hamiltonian $\tilde{H}_\text{eff}$ describes the evolution of a quantum trajectory $\ket{\phi(t)}$ during which no photon is emitted \cite{daley2014quantum}. As a consequence, this evolution does not preserve the normalization condition. That is, we have $p(t)=\braket{\phi(t)|\phi(t)}=\tr{\ketbra{\phi(t)}{\phi(t)}}\leq 1$, which is the time-dependent probability of finding the system in the pure state trajectory $\ket{\phi(t)}$. I will define $\hat{\rho}_{\gamma\kappa}$ where $\tr{\hat{\rho}_{\gamma\kappa}}=1$ as the state of the system if it is not in the state $\ket{\phi(t)}$. Then, the total master equation solution can be written $\hat{\rho}(t)=\ket{\phi(t)}\!\bra{\phi(t)}+(1-p(t))\hat{\rho}_{\gamma\kappa}(t)$, where $\hat{\rho}_{\gamma\kappa}(t)$ is the state of the system at time $t$ given that at least one jump occurred. 

The state $\ket{\phi(t)}$ is usually much easier to compute than $\hat{\rho}(t)$. If $p(t)$ is small enough, then $\ket{\phi(t)}$ approximates the state of the system with high fidelity. In turn, it can be used to estimate the fidelity compared to a desired state $\ket{\psi(t_\mathrm{f})}$ at some final time $t_\mathrm{f}$. In this case, the fidelity is
\begin{equation}
    \euscr{F} = p \euscr{F}_0 + (1-p)\euscr{F}_{\gamma\kappa}
\end{equation}
where $\euscr{F}_0=(1/p)\left|\braket{\phi(t_\mathrm{f})|\psi(t_\mathrm{f})}\right|^2$ is the fidelity provided that no decay occurred, $p=p(t_\mathrm{f})$, and $\euscr{F}_{\gamma\kappa}=\bra{\psi(t_\mathrm{f})}\hat{\rho}_{\gamma\kappa}(t_\mathrm{f})\ket{\psi(t_\mathrm{f})}$ is the potentially non-zero fidelity after a decay event.

When computing the fidelity by solving only the effective non-Hermitian Hamiltonian part of the master equation, we are making the approximation that $\euscr{F}\simeq p\euscr{F}_0$. This approximation is accurate when $p\simeq 1$ or $\mathcal{F}_{\gamma\kappa}\simeq 0$. The precision of this approximation depends on $\euscr{F}_0$ and $\euscr{F}_{\gamma\kappa}$ for a given implementation. If a protocol is interrupted by a decay event, then $\euscr{F}_{\gamma\kappa}< \euscr{F}_0$ and so $\euscr{F}_0$ can properly capture the scaling of $\euscr{F}$ with respect to parameters $\gamma$ and $\kappa$ in the high-fidelity regime. This approach is used in section \ref{chapter4:cavitygates}.

\subsection{Adiabatic elimination}
\label{chapter1:adiabaticelimination}

In section \ref{chapter1:cavityQED}, I touched briefly on one application of adiabatic elimination to correlation functions. The same approach can be applied to amplitudes of states or elements of the density matrix, which can be another powerful tool to reduce complicated system dynamics and obtain analytically tractable and simple solutions that can be easily applied in different situations. In this section, I would like to formalise adiabatic elimination in the general sense. To do this, I will follow the succinct definition of adiabatic elimination given in Ref.~\cite{sanz2016beyond}.

Suppose we have a system of first-order linear differential equations $\dot{\mathbf{z}}=Z\mathbf{z}$. The vector $\mathbf{z}$ can be a quantum state $\ket{\psi}$ corresponding to $Z=-i\hat{H}/\hbar$, it can be the density vector $\kett{\rho}$ corresponding to $Z=\tilde{\mathcal{L}}$, or it can be a vector of correlation functions as in sections \ref{chapter1:cavityQED} and \ref{chapter1:quantumregression}. Now, suppose we can partition the vector space into two subspaces $\mathbb{X}$ and $\mathbb{Y}$, where $\mathbb{X}$ contains the initial state. Let $\Pi_x$ and $\Pi_y=1-\Pi_x$ be complementary projection matrices onto these two subspaces, respectively. Then we can write our original set of differential equations as two coupled sets of first-order differential equations
\begin{equation}
\begin{aligned}
    \dot{\mathbf{x}}&=\Pi_xZ\Pi_x\mathbf{x}+\Pi_xZ\Pi_y\mathbf{y}\\
    \dot{\mathbf{y}}&=\Pi_yZ\Pi_x\mathbf{x}+\Pi_yZ\Pi_y\mathbf{y}\\
\end{aligned}
\end{equation}
where $\Pi_x\mathbf{z}=\mathbf{x}\in\mathbb{X}$ and $\Pi_y\mathbf{z}=\mathbf{y}\in\mathbb{Y}$. We now make the assumption that the eigenvalues of $\Pi_yZ\Pi_y$ are far separated from those of $\Pi_xZ\Pi_x$ so that the coupling between subspaces is weak. Then let $\dot{\mathbf{y}}\simeq0$ so that $\Pi_yZ\Pi_x\mathbf{x}\simeq-\Pi_yZ\Pi_y\mathbf{y}$. Substituting this into the equation for $\dot{\mathbf{x}}$ gives the effective equation of motion for the vector $\mathbf{x}$ in the subspace $\mathbb{X}$
\begin{equation}
    \dot{\mathbf{x}}=\left[\Pi_xZ\Pi_x-\Pi_xZ\Pi_y\left(\Pi_yZ\Pi_y\right)^{-1}\Pi_yZ\Pi_x\right]\mathbf{x}=\tilde{Z}\mathbf{x},
\end{equation}
provided that $\Pi_yZ\Pi_y$ is invertible. Hence, $\tilde{Z}$ is the effective Hamiltonian, Liouville superoperator, or generator of the correlation dynamics, of the subspace $\mathbb{X}$ undergoing adiabatic evolution with respect to the subspace $\mathbb{Y}$.

The validity of adiabatic elimination is not always clear and usually requires justification based on the norm of the correction $\tilde{Z}-\Pi_xZ\Pi_x$ to the homogeneous evolution of $\mathbf{x}$. In this thesis, I will provide comparisons to the full solutions computed numerically to validate analytic solutions obtained using adiabatic elimination.

\subsection{Perturbation theory}
\label{chapter1:perturbationtheory}

Suppose we can separate the Liouville superoperator $\mathcal{L}$ into a part that is easily diagonalized $\mathcal{L}_0$ and a perturbation $\epsilon\mathcal{L}_\epsilon$ where the eigenvalues of $\epsilon\mathcal{L}_\epsilon$ are much smaller than those of $\mathcal{L}_0$. Then, we can write the master equation as
\begin{equation}
    \frac{d}{dt}\hat{\rho}(t)=\mathcal{L}_0(t)\hat{\rho}(t)+\epsilon\mathcal{L}_\epsilon(t)\hat{\rho}(t).
\end{equation}
From here, we can write the integral form of the density operator solution (see appendix \ref{AppendixA:variationofparameters}) as
\begin{equation}
\label{chapter1eq:variationparsol}
    \hat{\rho}(t) = \mathcal{U}_0(t,t_0)\hat{\rho}(t_0) + \epsilon\int_{t_0}^t\mathcal{U}_0(t,t^\prime)\mathcal{L}_\epsilon(t^\prime)\hat{\rho}(t^\prime)dt^\prime
\end{equation}
for initial condition $\hat{\rho}(t_0)$ and where $\mathcal{U}_0$ is the propagation superoperator corresponding to the generator $\mathcal{L}_0$. If $\epsilon$ is small, we can use the zero-order `homogeneous' solution to approximate the integrand of the first-order correction: $\hat{\rho}(t^\prime)\simeq\mathcal{U}_0(t^\prime,t_0)\hat{\rho}(t_0)$. This gives a solution $\hat{\rho}(t)\simeq\left(\mathcal{U}_0(t,t_0)+\mathcal{U}_1(t,t_0)\right)\hat{\rho}(t_0)$ to first order in $\epsilon$ where
\begin{equation}
    \mathcal{U}_1(t,t_0)=\epsilon\int_{t_0}^t\mathcal{U}_0(t,t^\prime)\mathcal{L}_\epsilon(t^\prime)\mathcal{U}_0(t^\prime,t_0)dt^\prime.
\end{equation}
To obtain higher-order solutions, we can repeatedly substitute Eq.~(\ref{chapter1eq:variationparsol}) into itself in a Picard iterative fashion to obtain the series solution for $\mathcal{U}$, provided that it converges:
\vspace{-1mm}
\begin{equation}
    \mathcal{U}(t,t_0)=\sum_{n=0}^\infty\mathcal{U}_n(t,t_0),
\end{equation}
where for $n>0$
\begin{equation}
    \mathcal{U}_n(t,t_0)=\epsilon\int_{t_0}^t\mathcal{U}_{0}(t,t^\prime)\mathcal{L}_\epsilon(t^\prime)\mathcal{U}_{n-1}(t^\prime,t_0)dt^\prime.
\end{equation}
Then, any order of solution can be computed by truncating the summation:
\begin{equation}
    \hat{\rho}(t)=\sum_{n=0}^N\mathcal{U}_n(t,t_0)\hat{\rho}(t_0)+\euscr{O}\left(\epsilon^{N+1}\right).
\end{equation}
This approach can also be applied to a Schr\"{o}dinger equation, where it is called the Dyson series, and to any first-order linear differential equation, such as the optical Bloch equations.

Physically, the perturbation order $n$ considers all possible times between $t_0$ and $t$ that the superoperator $\mathcal{L}_\epsilon$ can be applied to the system state $n$ times. Moving from order $n-1$ to order $n$ requires the integration of one additional operator $\mathcal{L}_\epsilon$ applied sometime between $t_0$ and $t$. In the following chapter, I will discuss the case where $\mathcal{L}_\epsilon$ is a jump superoperator so that the order $n$ corresponds to the number of jumps that the system experienced during an interval of time. 
\chapter{Photon number decomposition}
\label{chapter2}

One goal of quantum optics theory is to understand and compute the complicated photon statistics of non-classical light, such as photon counting distributions and detection waiting times. This led to the development of the quantum trajectories approach for solving quantum optical master equations \cite{zoller1987quantum,carmichael,daley2014quantum}. In the last chapter, I introduced the Markovian master equation, non-Hermitian Hamiltonians, and perturbation theory. In this chapter, I will outline how these concepts are connected to the quantum trajectories formalism and show that they provide a useful approach for analyzing optically active quantum systems that may experience excess decoherence.

A quantum trajectory is a particular pure-state evolution of a quantum system that occurs with a certain probability \cite{daley2014quantum}. The total mixed state evolution of the reduced quantum system is then seen to be a composition of all possible quantum trajectories, in the same sense as the definition of the density operator (see section \ref{chapter1:quantumsystems}). In section \ref{chapter1:nonhermitian}, I introduced the effective non-Hermitian Hamiltonian that describes the reduced state evolution of the quantum trajectory corresponding to no photon emission. This was used to estimate the system state for the case where the chance for emission is very small. I also highlighted that the terms neglected in the master equation to obtain the effective Hamiltonian correspond to the jump superoperators that would otherwise cause the system to jump to the ground state after the potential emission of a photon. Instead of neglecting these terms, we can treat them perturbatively using the approach described in section \ref{chapter1:perturbationtheory}. This leads to a series expansion for the density operator in the number of jumps that occurred during the system evolution.

The perturbative expansion of a master equation described above is one example of a master equation \emph{unravelling} \cite{carmichael,breuer2002theory}. There are infinitely many possible unravellings that describe the master equation dynamics \cite{carmichael}, each corresponding to a different choice of perturbation. In this thesis, I will refer to a master equation unravelling in the number of jumps as a photon number decomposition, since it can provide information on the state of the emitter or cavity-emitter system conditioned on the number of photons emitted or detected.

Master equation unravellings are the foundation for the quantum trajectories method \cite{carmichael,wiseman1994quantum,daley2014quantum} and are usually applied to reduce the computational complexity of a master equation. This is done by solving an effective Schr\"{o}dinger equation (as in section \ref{chapter1:nonhermitian}) or by numerically solving a stochastic equation of the open quantum system \cite{zoller1987quantum,carmichael1989photoelectron,browne2003robust,dodonov2005microscopic,barrett2005efficient,zhang2018quantum,zhang2019heralded}. However, the underlying concept of quantum trajectories is not numerical in nature and can be analytically useful in certain cases even when it does not reduce the complexity of the computation.

Recently, the concept of a photon number decomposition that I will describe in this chapter has been used in part to compute photon statistics for complicated emitter dynamics \cite{hanschke2018quantum,zhang2018quantum}. It is also connected to scattering matrix theory \cite{trivedi2018few} and the solution for a quantum emitter coupled to a waveguide using the technique of coarse graining time \cite{fischer2018scattering}. Furthermore, the decomposition is intrinsically related to continuous measurement and quantum feedback theories \cite{wiseman1994quantum,wiseman2009quantum}, which can describe active or coherent feedback schemes \cite{zhang2019heralded,crowder2020quantum}. As I will show in this thesis, the photon number decomposition approach is also a useful tool to study the temporal properties of the waveguide photonic state (section \ref{chapter3:photonicstate}) and analyze photon counting post-selection schemes (section \ref{chapter4:entanglementgeneration}) under the effects of excess emitter decoherence.

The results in this chapter, such as conditional correlation functions, may be formally derived beginning from the total light-matter Hamiltonian as in \cite{fischer2018particle,fischer2018scattering}. However, in section \ref{chapter2:conditionalpropagationsuperoperators}, I will instead present the decomposition beginning from a Markovian master equation using the principles introduced in the previous chapter. This will hopefully give a concise and intuitive picture of the methods I employ in this thesis. In section \ref{chapter2:photoncountingmeasurements}, I will then use the decomposition to describe an imperfect photon counting measurement of the quantum system, which is applied in section \ref{chapter4:entanglementgeneration}. In sections \ref{chapter2:conditionalcorrelations} and \ref{chapter2:temporaldensityfunctions}, I demonstrate that the photon number decomposition can allow for a partial reconstruction of the photonic temporal density function of the collection waveguide that accounts for excess decoherence of the emitter. These results are then applied in section \ref{chapter3:photonicstate}. Finally, in section \ref{chapter2:photonstatistics}, I show that the photon number decomposition of the reduced system dynamics directly satisfies the expected photon statistics relations for time integrated intensity correlation functions of the waveguide mode. 

Sections \ref{chapter2:conditionalpropagationsuperoperators} and \ref{chapter2:photoncountingmeasurements} are published in Ref.~\ref{wein2020entanglement}, although I have expanded some of the descriptions. The remaining material in this chapter is original research that I have not published.

\section{Conditional propagation superoperators}
\label{chapter2:conditionalpropagationsuperoperators}

We begin with the general Markovian master equation of the form
\begin{equation}
\label{meq}
    \frac{d}{dt}\hat{\rho}(t) = \mathcal{L}\hat{\rho}(t),
\end{equation}
where $\mathcal{L}$ is the Liouville superoperator that contains all the reversible and irreversible dynamics of the open system, which may contain one or more emitters or emitter-cavity systems. To decompose the master equation dynamics into evolution conditioned on single-photon emission or detection, we can rearrange the master equation in a way similar to the derivation of the non-Hermitian Hamiltonian in section \ref{chapter1:nonhermitian}:
\begin{equation}
\label{meqC}
    \frac{d}{dt}\hat{\rho}(t)=\mathcal{L}_0\hat{\rho}(t) +\sum_{i=1}^N\mathcal{J}_{i}\hat{\rho}(t),
\end{equation}
where $\mathcal{L}_0=\mathcal{L}-\sum_i^N\mathcal{J}_i$ is the Liouville superoperator of the state evolution that lacks jumps accounted for by all $\mathcal{J}_i$. Each jump superoperator $\mathcal{J}_i\hat{\rho}=\hat{C}_i\hat{\rho}\hat{C}_i^\dagger$ is given by the collapse operator $\hat{C}_i$ describing the effect of a photon emission or detection event on the system \cite{carmichael}. These collapse operators are in units of square root rate.

For a single reduced system in the Markovian regime described by the standard form of the master equation (see section \ref{chapter1:markovian}), the collapse operators are given by the source field \cite{carmichael1989photoelectron,kiraz2004quantum,barrett2005efficient}. For a two-level emitter, the source field is $\hat{C}=\sqrt{\gamma\eta_\mathrm{r}}\sigd$, where the coefficient $\gamma\eta_\mathrm{r}$ is the emitter Einstein A coefficient \cite{carmichael}. For a cavity coupled to a one-dimensional waveguide, we have $\hat{C}=\sqrt{\kappa\eta_\mathrm{c}}\ad$. Note that if there is non-radiative decay $\eta_\mathrm{r}<1$ or imperfect waveguide collection efficiency $\eta_\mathrm{c}<1$, then $\mathcal{L}_0$ still contains jump dynamics and cannot, in general, be reduced to an effective non-Hermitian Hamiltonian as in section \ref{chapter1:nonhermitian}. This illustrates the fact that an emitted photon always implies that the system jumped but a jump does not necessarily imply a photon was emitted. In the language of photon counting measurements, which is the topic of the next section, $\eta_\mathrm{c}<1$ simply implies that not all cavity photons were detected. Hence, the absence of a photon does not mean that the system followed a pure state trajectory.

By choosing a collapse operator $\hat{C}=\sqrt{\kappa\eta_\mathrm{c}}\ad$, we would be decomposing the dynamics of the reduced system in terms of the jump statistics that give rise to the emitted photonic state at position $x=0$ of the waveguide. This is illustrated by the input-output relation $\bd-\bd_0=\sqrt{\kappa\eta_\mathrm{c}}\ad$. In the quantum-optical regime where we assume that the input mode $\bd_0$ is in the vacuum state, a photon detected at position $x=0$ implies that the system jumped by the collapse operator $\hat{C}$. Hence, the input-output relations give us a natural way to choose our decomposition to represent the number of jumps giving rise to the photonic state at different points along the waveguide, which can capture additional photon losses or beam splitter operations applied to the source fields from multiple emitters. The input-output relations also imply that the photon number decomposition is related to a particular indirect measurement of the reduced system performed by measuring the emitted photonic state.

In general, I will refer to a reduced system composed of one or more emitters or emitter-cavity systems as the source. For a source being monitored by $N$ single-photon detectors, we can write $\mathcal{J}_i\hat{\rho}=\hat{d}_i\hat{\rho}\hat{d}_i^\dagger$ where $\hat{d}_i$ is the source field at the $i^\mathrm{th}$ single-photon detector. If the detected photonic state is fully described by the source, meaning that we are not performing quadrature measurements using an auxiliary laser, this field operator is given by operators of the source. Interestingly, quadrature measurements can be modeled using master equation unravellings and proportionality relations provided that the quantum dynamics of the local oscillator are explicitly included in the master equation \cite{carmichael}. 

From Eq.~(\ref{meqC}), we can now expand the solution $\hat{\rho}(t)$ perturbatively around jump superoperators $\mathcal{J}_i$ \cite{carmichael,horoshko1998multimode}. Using the method described in section \ref{chapter1:perturbationtheory} (see also appendix \ref{AppendixA:variationofparameters}), the density operator solution $\hat{\rho}(t)$ can be decomposed into a set of conditional states dependent on the cumulative photon count $n_i$ in the $i^\mathrm{th}$ mode of $N$ modes using the Liouville-Neumann series
\begin{equation}
    \hat{\rho}(t) = \sum_{i=1}^N\sum_{n_i=0}^\infty \hat{\rho}_{(n_1,n_2,\cdots,n_N)}(t) = \sum_{\mathbf{n}\in\mathbb{N}_0^N}\hat{\rho}_\mathbf{n}(t),
\end{equation}
where $\mathbf{n}=\sum_i^N n_i\mathbf{e}_i$, and $\mathbf{e}_i$ is the $i^\mathrm{th}$ natural basis vector in the $N$-dimensional space of non-negative integers $\mathbb{N}_0^N$ that labels the number of photons at each detector. The conditional state $\hat{\rho}_\mathbf{n}(t)$ is the unnormalized density matrix $\hat{\rho}_{\mathbf{n}}(t) = \mathcal{U}_\mathbf{n}(t,t_0)\hat{\rho}(t_0)$, where $\mathcal{U}_\mathbf{n}$ is the conditional propagation superoperator given recursively by
\begin{equation}
\label{conditionalpropagator}
    \mathcal{U}_\mathbf{n}(t,t_0) = \sum_{i=1}^N\int_{t_0}^t\mathcal{U}_0(t,t^\prime)\mathcal{J}_i\mathcal{U}_{\mathbf{n}-\mathbf{e}_i}(t^\prime,t_0)dt^\prime,
\end{equation}
and $\mathcal{U}_\mathbf{0}$ is the propagation superoperator of the equation $d\hat{\rho}(t)/dt = \mathcal{L}_0\hat{\rho}(t)$. For convenience, we define $\mathcal{U}_\mathbf{n}=0$ if $\mathbf{n}\notin \mathbb{N}_0^N$. This recursive relation is the superoperator equivalent to the stochastic description given by Eq.~(6.36) in Ref.~\cite{breuer2002theory}.

The conditional state $\hat{\rho}_\mathbf{n}(t)$ occurs with the probability $\prb{\mathbf{n}}= \text{Tr}[\hat{\rho}_\mathbf{n}(t)]$, where $\sum_{\mathbf{n}}\prb{\mathbf{n}}=1$. It is dependent on both the initial state and the evolution time. The conditional state corresponding to $\mathbf{n}$ can be interpreted as the mixed-state analog of a quantum trajectory, which has been referred to as a quantum corridor \cite{mensky1993continuous}, as it can be seen as a collection of quantum trajectories that each satisfies a particular property. In this case, that property is the number of photons $n_i$ in each mode corresponding to $\hat{d}_i$.

The photon number decomposition implemented in this way is an exact description of the original master equation dynamics because the total propagation superoperator of Eq.~(\ref{meq}) is given by $\mathcal{U}(t,t_0) = \sum_{\mathbf{n}}\mathcal{U}_\mathbf{n}(t,t_0).$ As a consequence, this decomposition provides access to the state of the source after post selecting based on the number of emitted or detected photons (depending on $\mathcal{J}$). It is also very important to recognize that this post selection does not provide a measurement back-action on the source dynamics. It only selects out a particular quantum corridor through which the source evolved, corresponding to the fact that the underlying master equation is unchanged.

Since the total propagation superoperator $\mathcal{U}$ has the semi-group property that $\mathcal{U}(t,t_0) = \mathcal{U}(t,t^\prime)\mathcal{U}(t^\prime,t_0)$, we can also discuss conditional states of the source for detection during a window of time $(t^\prime,t)$ with duration $T=t^\prime-t$ between an initial time $t_0$ and a final time $t_\text{f}$. The conditional propagator for this window is given by
\begin{equation}
\label{windowpropagator}
    \mathcal{W}_\mathbf{n}(t_\text{f},t^\prime,t,t_0) = \mathcal{U}(t_\text{f},t^\prime)\mathcal{U}_\mathbf{n}(t^\prime,t)\mathcal{U}(t,t_0),
\end{equation}
where $t_0\leq t\leq t^\prime\leq t_\mathrm{f}$ and $\mathcal{U}=\sum_{\mathbf{n}}\mathcal{W}_\mathbf{n}$. The conditional state given by $\mathcal{W}_\mathbf{n}$ does not distinguish between trajectories that give rise to different numbers of photons outside of the window of time. In the next section, I discuss how these conditional propagators are related to a measurement of the source and how they can be used to define an imperfect gated photon counting measurement where the detectors may falsely indicate the number of detected photons due to classical noise.

\section{Photon counting measurements}
\label{chapter2:photoncountingmeasurements}

For a gated detector operating at a distance $L_\text{d}$ from the source, the measurement depends on the state of the source at the retarded time $r(t) = t - L_\text{d}/c$, where $c$ is the transmission speed of light. The gated detector begins at time $t_\text{d}$ where $r(t_\text{d})\geq t_0$ and remains open for duration $T_\text{d}=t_\text{d}^\prime-t_\text{d}=r(t_\text{d}^\prime)-r(t_\text{d})$. The conditional state $\hat{\rho}_\mathbf{n}$ at time $t_\text{f}$ after a retarded detection window is
\begin{equation}
\label{equ:dmat_dis}
\begin{aligned}
    \hat{\rho}_\mathbf{n}(t_\text{f}) &= \mathcal{W}_\mathbf{n}(t_\text{f},r(t_\text{d}^\prime),r(t_\text{d}),t_0)\hat{\rho}(t_0),
\end{aligned}
\end{equation}
where $t_\text{f}-t_0=T_\text{d}+2L_\text{d}/c$ would be the minimum protocol time after a two-way classical communication. The purpose of describing this measurement delay is to place a limit on the protocol time, which becomes important in section \ref{chapter4:entanglementgeneration}. It is not used to describe active feedback, measurement back-action, or relativistic consequences. These are not considered in this thesis. 

Suppose that we use number-resolving single-photon detectors that are ideal so that they always project the waveguide mode at the detector onto the photon-number basis. In this idealized scenario, the conditional state $\hat{\rho}_\mathbf{n}$ is the state of the source after measuring photon numbers $\mathbf{n}$. That is, the photon number decomposition is equivalent to a selective indirect measurement of the source whereby the photonic waveguide is an ideal quantum probe that provides no measurement back-action \cite{breuer2002theory}, analogous to a PVM applied to the waveguide (recall section \ref{chapter1:measurements}). In this context, the conditional propagation superoperator $\mathcal{U}_\mathbf{n}$ is known as the operation of a generalized measurement scheme \cite{breuer2002theory}.

To account for classical detector imperfections such as inefficiency, dark counts, and number resolving limitations, we can remix \cite{breuer2002theory} the conditional states to form a non-selective measurement. That is, we can write the imperfect measured state as a linear combination of perfect measured states weighted by conditional probabilities. This is the analogous case to applying a POVM to the waveguide (recall section \ref{chapter1:measurements}). Although detector inefficiency could be applied during this remixing step, it is much more convenient to model detector inefficiency using a beam-splitter loss model, whereby the waveguide losses are artificially increased just before each detector. To do this, we simply modify the source fields by $\hat{d}_i\rightarrow\sqrt{\eta_\mathrm{d_i}}\hat{d}_i$. Note that this does not yet reduce the selective measurement to a non-selective measurement because of our choice of measurement basis. For the remaining two imperfections, we can perform a remixing.

Let $\mathbb{M}$ be the space of all measurement outcomes from the set of $N$ single-photon detectors. Then, the source state after measuring outcome $\mathbf{m}\in\mathbb{M}$ is given by the classical remixing of the conditional states
\begin{equation}
\label{eq:POVM}
\begin{aligned}
    \hat{\varrho}_\mathbf{m}(t_\text{f}) =\sum_{\mathbf{n}\in\mathbb{N}_0^N} \prb{\mathbf{m}|\mathbf{n}}\hat{\rho}_\mathbf{n}(t_\text{f}),
\end{aligned}
\end{equation}
where $\prb{\mathbf{m}|\mathbf{n}}$ is the probability for measurement outcome $\mathbf{m}$ given the state $\hat{\rho}_\mathbf{n}$. Note that we distinguish between the conditional state of the source $\hat{\rho}_\mathbf{n}$, where $\mathbf{n}$ denotes the true photon distribution (after losses), and the state after the measurement $\hat{\varrho}_\mathbf{m}$, where $\mathbf{m}$ includes dark counts and/or number resolution limitations. Naturally, we also require that $\sum_{\mathbf{m}\in\mathbb{M}} \prb{\mathbf{m}|\mathbf{n}}= 1$ for all $\mathbf{n}\in\mathbb{N}_0^N$.

By assuming that the detectors are identical and independent, we can simplify the conditional probability to $\prb{\mathbf{m}|\mathbf{n}} \equiv \prod_i^N P_\text{d}(m_i|n_i)$. For fast photon-number-resolving detectors (PNRDs) that can count all photons arriving during the gate duration $T_\text{d}$, we have $\mathbb{M}=\mathbb{N}_0^N$. Then $P_\text{d}(m|n)$ is given by all possible combinations of dark counts such that $n$ can appear to be $m$:
\begin{equation}
\label{eq:PNRD}
    P_\text{PNRD}(m|n)=\sum_{k=0}^\infty\delta_{m,k+n}p_{\mathrm{d},k}(T_\text{d},\gamma_\mathrm{d}),
\end{equation}
where $\delta$ is the Kronecker delta and $p_{\mathrm{d},k}$ characterizes the dark count distribution. For Poissonian noise, we have that $p_{\mathrm{d},k}(T_\mathrm{d},\gamma_\mathrm{d})=\gamma_\mathrm{d}^k T_\mathrm{d}^ke^{-\gamma_\mathrm{d} T_\mathrm{d}}/k!$, where $\gamma_\mathrm{d}$ is the detector dark count rate.

For a bin detector (BD) that simply indicates the presence of one or more photons arriving during the gate duration, then $\mathbb{M}=\Sigma^N$ is the set of binary vectors of length $N$ and $P_\text{d}(m|n)$ is given by
\begin{equation}
\label{eq:BD}
\begin{aligned}
    \hspace{-2mm}P_\text{BD}(m|n) &= \sum_{q=0}^\infty\delta_{m,\text{sgn}(q)}P_\text{PNRD}(q|n)\\
    &=\!\delta_{m,\text{sgn}(n)}p_{\mathrm{d},0}(T_\text{d},\gamma_\mathrm{d})\!+\!\delta_{m,1}\!\left(1\!-\!p_{\mathrm{d},0}(T_\text{d},\gamma_\mathrm{d})\right),
\end{aligned}
\end{equation}
where $\text{sgn}:\mathbb{N}_0\to\Sigma$ is the signum function.

The PNRD and BD models are appropriate for many different detector types \cite{hadfield2009single} (see also Ref.~\ref{wein2016bellmeasurement}). For example, single-photon avalanche photodiodes (APDs) and superconducting nanowire single-photon detectors (SNSPDs) \cite{natarajan2012superconducting} can be modeled by BDs while transition edge sensors (TESs) \cite{rosenberg2005noise,lita2008counting} or pixel arrays are ideally considered to be PNRDs.

\section{Conditional correlation functions}
\label{chapter2:conditionalcorrelations}

Using the conditional propagation superoperators $\mathcal{U}_\mathbf{n}$, we can also perform a decomposition for time-ordered correlation functions. I will demonstrate this by giving an explicit example for the two-time correlation function $G^{(1)}$ for a decomposition of a single cavity mode $\ad$ of a source. I will also give the decomposition for $G^{(2)}$. However, the idea can applied to a two-level emitter, be extended to multi-mode decompositions, and other multi-time correlations.

In section \ref{chapter2:conditionalpropagationsuperoperators}, I introduced the decomposition in terms of jump operators $\mathcal{J}$ corresponding to the source field. For a single mode decomposition at $x=0$, we have $\hat{C}=\sqrt{\kappa\eta_\mathrm{c}}\ad$ so that $\mathcal{J}\hat{\rho}=\hat{C}\hat{\rho}\hat{C}^\dagger$. Let $\mathcal{S}$ and $\mathcal{R}$ be the superoperators corresponding to $\mathcal{S}\hat{\rho}=\ad\hat{\rho}$ and $\mathcal{R}\hat{\rho}=\hat{\rho}\au$, respectively. Then we have that $\mathcal{S}\mathcal{R}=\mathcal{R}\mathcal{S}$ and $\mathcal{J} = \kappa\eta_\mathrm{c}\mathcal{S}\mathcal{R}$. From section (\ref{chapter1:correlations}), we have
\begin{equation}
    G^{(1)}(t^\prime,t)=\braket{\au(t^\prime)\ad(t)} = \tr{\mathcal{R}\mathcal{U}(t^\prime,t)\mathcal{S}\mathcal{U}(t,t_0)\hat{\rho}(t_0)}
\end{equation}
where $t^\prime\geq t\geq t_0$. By substituting the photon number decomposition for $\mathcal{U}$, we have
\begin{equation}
    \braket{\au(t^\prime)\ad(t)}=\sum_{l,m}\tr{\mathcal{R}\mathcal{U}_l(t^\prime,t)\mathcal{S}\mathcal{U}_m(t,t_0)\hat{\rho}(t_0)}.
\end{equation}
In order to obtain a correlation function that probes the corridor corresponding to a specific outcome $n$, we need to consider the decomposition of evolution after $t^\prime$ as well. To do this, we can note that $\mathcal{U}=\sum_k\mathcal{U}_k$ is trace preserving because it is the propagation superoperator corresponding to a master equation. Hence, we can substitute it after $\mathcal{R}$ and obtain
 \begin{equation}
 \begin{aligned}
    \braket{\au(t^\prime)\ad(t)}
    &=\sum_{k,l,m}\tr{\mathcal{U}_k(t_\mathrm{f},t^\prime)\mathcal{R}\mathcal{U}_l(t^\prime,t)\mathcal{S}\mathcal{U}_m(t^\prime,t)\hat{\rho}(t_0)},\\
\end{aligned}
\end{equation}
which also requires that $t_\mathrm{f}\geq t^\prime$. We can now count how many $n$ total jumps occur between time $t_0$ and $t_\mathrm{f}$. To do so, we add the condition from each propagation superoperator plus the combined action of $\mathcal{R}$ and $\mathcal{S}$ to get $n=k+l+m+1$.

By rearranging the summation so that $k+l+m+1=n$, and taking $t_f\rightarrow\infty$, we have our final result
\begin{equation}
    \braket{\au(t^\prime)\ad(t)}=
    \sum_{n}\braket{\au(t^\prime)\ad(t)}_n
\end{equation}
where
\begin{equation}
\label{chapter2eq:conditionalcorrelationsamptwotime}
    \braket{\au(t^\prime)\ad(t)}_n = \sum_{k,l,m}\lim_{t_\mathrm{f}\rightarrow\infty}\tr{\mathcal{U}_k(t_\mathrm{f},t^\prime)\mathcal{R}\mathcal{U}_l(t^\prime,t)\mathcal{S}\mathcal{U}_m(t,t_0)\hat{\rho}(t_0)}\delta_{n,k+l+m+1}
\end{equation}
is the correlation function conditioned on $n$ jumps $\mathcal{J}$ between an initial state $\hat{\rho}(t_0)$ until time $t_f\rightarrow\infty$. Note that this conditional correlation is only well-defined if the master equation satisfies $\lim_{t\rightarrow\infty}\mathcal{J}\hat{\rho}(t)=0$ so that the source eventually reaches a dark state. Otherwise, the correlation could be defined for a finite time $t_\mathrm{f}$. The necessity for adding $1$ to $k+l+m$ to account for the combined action of $\mathcal{R}$ and $\mathcal{S}$ becomes apparent when noting that we require that $\braket{\au(t^\prime)\ad(t)}_0=0$ but that the case for $k+l+m=0$ can give a nonzero result in general. Hence, $k=l=m=0$ should correspond to the first non-zero term of the expansion, which is the correlation over the single-jump corridor $\braket{\au(t^\prime)\ad(t)}_1$. If this argument is not convincing, we could take $n$ to be an arbitrary definition and leave its interpretation as `number of jumps' to be justified in the following section. At this point, the validity of Eq.~(\ref{chapter2eq:conditionalcorrelationsamptwotime}) does not depend on our interpretation of $n$.

This decomposition can also be used to compute any time-ordered correlation function. For example, the second-order correlation $G^{(2)}$ is
\begin{equation}
\begin{aligned}
    \braket{\au(t)\au(t^\prime)\ad(t^\prime)\ad(t)} &=\tr{\mathcal{RS}\mathcal{U}(t^\prime,t)\mathcal{RS}\mathcal{U}(t,t_0)\hat{\rho}(t_0)}\\
    &=\sum_n\braket{\au(t)\au(t^\prime)\ad(t^\prime)\ad(t)}_n
\end{aligned}
\end{equation}
where
\begin{equation}
\label{chapter2eq:correlationg2n}
\begin{aligned}
  \braket{\au(t)\au(t^\prime)\ad(t^\prime)\ad(t)}_n&=\sum_{k,l,m}\lim_{t_f\rightarrow\infty} \tr{\mathcal{U}_k(t_\mathrm{f},t^\prime)\mathcal{RS}\mathcal{U}_l(t^\prime,t)\mathcal{RS}\mathcal{U}_m(t,t_0)\hat{\rho}(t_0)}\delta_{k+l+m+2,n}.
\end{aligned}
\end{equation}
Similar to above, I am adding $2$ to $k+l+m$ to account for the additional two jumps $\mathcal{R}\mathcal{S}$ within the dynamics probed by $\braket{~~}_n$. As before, the association of $n=k+l+m+2$ to the number of jumps becomes more apparent in the following section where I will show that this additional counting is necessary so that $\braket{~~}_n$ defined in Eq.~(\ref{chapter2eq:correlationg2n}) is consistent with that of Eq.~(\ref{chapter2eq:conditionalcorrelationsamptwotime}) when reconstructing the state of the waveguide within photon number subspaces.

\section{Temporal density functions}
\label{chapter2:temporaldensityfunctions}

Suppose that $\mathcal{L}$ satisfies $\lim_{t\rightarrow\infty}\mathcal{J}\hat{\rho}(t)=0$ so that the source reaches a dark state. Suppose also that the total system operates in the quantum optical regime where the Fourier transform of plane wave states allows for a photonic state description of the waveguide in the continuous temporal basis of modes $\bd(t)$ \cite{fischer2018scattering}. Then, if the waveguide state is pure, it can be written as $\ket{\Psi}=\sum_n\ket{\psi_n}$ where
\begin{equation}
\label{chapter2eq:temporalwavefunctions}
    \ket{\psi_n} = \sqrt{p_n}\int f_n(\mathbf{t})\prod_{k=1}^n\bu(t_k)\ket{0}\mathbf{dt}
\end{equation}
for $n\geq 1$ and $\ket{\psi_0}=\sqrt{p_0}\ket{0}$. The values $0\leq p_n\leq 1$ are the photon-number probabilities of the waveguide photonic state where $\sum_n p_n=1$. The complex temporal wavefunctions $f_n(\mathbf{t})$ describing the state of $n$ photons are normalized so that $\int |f_n(\mathbf{t})|^2\mathbf{dt}=1$, where $\mathbf{t}=(t_1,\cdots, t_n)$. I will assume the time-ordered convention where the integration is taken so that $t_i\leq t_{i+1}$, or equivalently, over the entire real space $\mathbb{R}^n$ provided that we restrict $f_n(\mathbf{t})=0$ if $\mathbf{t}$ is not time ordered. An alternative convention would be to take $f_n$ to be symmetric upon the exchange of two times, which differs from the time-ordered convention by a normalization factor of $n!$.

In general, if the source suffers from excess decoherence, the waveguide may not be in a pure state. By extending Eq.~(\ref{chapter2eq:temporalwavefunctions}) to the mixed-state picture, we have the waveguide density operator $\hat{\varrho} = \sum_{n,m}\hat{\varrho}_{n,m}$ where
\begin{equation}
    \hat{\varrho}_{n,m} = \sqrt{p_np_m}\iint\xi_{n,m}(\mathbf{t},\mathbf{t}^\prime)\left[\prod_{k=1}^n\hat{b}^\dagger(t_k)\right]\ket{0}\bra{0}\left[\prod_{l=1}^m\hat{b}(t_l^\prime)\right]\mathbf{dt}\mathbf{dt}^\prime,
\end{equation}
for $n,m\geq 1$ and
\begin{equation}
\begin{aligned}
\hat{\varrho}_{0,0} &= p_0\ket{0}\bra{0}\\
\hat{\varrho}_{1,0} &= \sqrt{p_0p_1}\int\xi_{1,0}(t_1)\hat{b}^\dagger(t_1)\ket{0}\bra{0}dt_1\\
\hat{\varrho}_{n,m} &= \hat{\varrho}_{m,n}^\dagger,\\
\end{aligned}
\end{equation}
where the temporal density functions $\xi_{n}=\xi_{n,n}$ are normalized such that $\text{Tr}\left[\hat{\varrho}_{n,n}\right] = p_n$ and the coherence functions $\zeta_{n,m}=\xi_{n,m}$ ($n\neq m$) are bounded by $|\zeta_{n,m}(\mathbf{t},\mathbf{t}^\prime)|^2\leq\xi_n(\mathbf{t},\mathbf{t})\xi_m(\mathbf{t}^\prime,\mathbf{t}^\prime)$.
By the Hermitian property, we have that $\xi_{n,m}^*(\mathbf{t},\mathbf{t}^\prime)=\xi_{n,m}(\mathbf{t}^\prime,\mathbf{t})$. In addition, we assume the same time ordering convention as for the pure state description where the integration is taken over $t_i\leq t_{i+1}$ and $t_{i}^\prime\leq t_{i+1}^\prime$. We can again extend the integration over the entire $\mathbb{R}^{n+m}$ space by restricting $\xi_{n,m}$ appropriately so that $\xi_{n,m}(\mathbf{t},\mathbf{t}^\prime)=0$ if $\mathbf{t}$ or $\mathbf{t}^\prime$ is not time ordered. For a pure state, we have $\xi_{n,m}(\mathbf{t},\mathbf{t}^\prime)=f_n(\mathbf{t})f_m^*(\mathbf{t}^\prime)$ for $n,m\geq 1$ and $\zeta_{1,0}(t)=\zeta_{0,1}^*(t)=f_1(t)$.

As a consequence of the input-output relation $\bd=\sqrt{\kappa\eta_\mathrm{c}}\ad+\bd_o$, we have that normally-ordered, time-ordered correlations of the waveguide mode can be written in terms of the conditional correlations of the source
\begin{equation}
    \braket{\bu(t^\prime)\bd(t)}=\kappa\eta_\mathrm{c}\sum_n\braket{\au(t^\prime)\ad(t)}_n,
\end{equation}
where $t^\prime\geq t$. This naturally defines a corresponding decomposition of the waveguide correlation functions by association $\braket{\bu(t^\prime)\bd(t)}_n=\kappa\eta_\mathrm{c}\braket{\au(t^\prime)\ad(t)}_n$. If $n$ indeed corresponds to the number of jumps giving rise to $n$ photons in the waveguide, then we expect that $\braket{~~}_n$ is the expectation value of the waveguide substate $\hat{\varrho}_{n,n}$. A proof of this claim would require a derivation from the total system Hamiltonian \cite{fischer2018scattering}. However, we can show that this association is perfectly consistent with all the properties of the photonic state and other known results in the literature. To this end, let us assume that $\braket{~~}_n$ is the expectation value for the state $\hat{\varrho}$ after being projected onto the $n$-photon subspace. This assumption should rely on all the same assumptions used to arrive at the Markovian master equation and input-output relation for a vacuum input.

Using this notion, we have a clear definition of the emitted photon number probabilities in terms of the source dynamics:
\begin{equation}
\begin{aligned}
    p_n &= \tr{\hat{\varrho}_{n,n}}=\braket{\hat{I}}_n\\
    &=\lim_{t_\mathrm{f}\rightarrow\infty}\tr{\mathcal{U}_n(t_\mathrm{f},t_0)\hat{\rho}(t_0)},
\end{aligned}
\end{equation}
which is equivalent to the Mandel counting formula \cite{fischer2018particle}. This derivation can be seen as an analytic application of quantum trajectories simulations. In this case, all of the quantum trajectories (or quantum corridors in the mixed-state picture) are buried inside the conditional propagator $\mathcal{U}_n$ and $t_\mathrm{f}\rightarrow\infty$ implies we simulate for much longer than the source lifetime.

The waveguide field correlations in the photon number subspace are directly related to the temporal density functions:
\begin{equation}
\label{chapter2eq:firstordercorr}
\begin{aligned}
    \braket{\bu(t^\prime)\bd(t)}_0 &= 0\\
    \braket{\bu(t^\prime)\bd(t)}_1 &= p_1\xi_1(t,t^\prime)\\
    \braket{\bu(t^\prime)\bd(t)}_2 &= p_2\int\left[\xi_2(t,t^{\prime\prime},t^\prime,t^{\prime\prime})+\xi_2(t,t^{\prime\prime},t^{\prime\prime},t^\prime)+\xi_2(t^{\prime\prime},t,t^{\prime\prime},t^\prime)\right]dt^{\prime\prime}\\
    &\!~~\vdots
\end{aligned}
\end{equation}
Where I have applied the time-ordering convention so that the term $\xi_2(t^{\prime\prime},t,t^\prime,t^{\prime\prime})$ does not appear. From these relations, we can then compute the single-photon temporal density function $\xi_1(t,t^\prime)$ from the master equation dynamics by
\begin{equation}
\label{chapter2eq:onephotondensity}
\begin{aligned}
    \xi_1(t,t^\prime) &= \frac{\kappa\eta_\mathrm{c}}{p_1}\braket{\hat{a}^\dagger(t^\prime)\hat{a}(t)}_1\\
    &=\frac{\kappa\eta_\mathrm{c}}{p_1}\tr{\mathcal{U}_0(\infty,t^\prime)\mathcal{R}_0(t^\prime,t)\mathcal{S}_0(t,t_0)\hat{\rho}(t_0)}
\end{aligned}
\end{equation}
for $t^\prime\geq t$ and $\xi_1(t^\prime,t)=\xi_1^*(t,t^\prime)$. In this expression, I have introduced the shorthand $\mathcal{U}_0(\infty,t)=\lim_{t_\mathrm{f}\rightarrow\infty}\mathcal{U}_0(t_\mathrm{f},t)$ and the Heisenberg-like notations $\mathcal{R}_n=\mathcal{R}\mathcal{U}_n$ and  $\mathcal{S}_n=\mathcal{S}\mathcal{U}_n$. The result for the single-photon temporal density function in Eq.~(\ref{chapter2eq:onephotondensity}) is also given in Ref.~\cite{fischer2018particle} (up to a difference in normalization convention) as derived from the full light-matter interaction for an emitter coupled to a waveguide. This justifies defining the index of the decomposition in the previous section such that $n=1$ corresponds to a single jump resulting in a photon in the waveguide. We can also show that the extension to $n=2$ is self-consistent with the two-photon temporal density function.

The correlation function $\braket{\bu(t^\prime)\bd(t)}_2$ computed using Eq.~(\ref{chapter2eq:conditionalcorrelationsamptwotime}) gives three components for when $t^\prime\geq t$. These components correspond to the cases where conditional evolution $\mathcal{U}_1$ occurs before $t$, in between $t$ and $t^\prime$, and after $t^\prime$, respectively. Expanding $\mathcal{U}_1$ in terms of $\mathcal{U}_0$ and $\mathcal{J}=\kappa\eta_\mathrm{c}\mathcal{S}\mathcal{R}$ gives
\begin{equation}
\label{chapter2eq:twophotonfirstorderdynamics}
\begin{aligned}
    \braket{\bu(t^\prime)\bd(t)}_2&=\kappa\eta_\mathrm{c}\int_{t^\prime}^{\infty}\tr{\mathcal{U}_0(\infty,t^{\prime\prime})\mathcal{J}_0(t^{\prime\prime},t^{\prime})\mathcal{R}_0(t^\prime,t)\mathcal{S}_0(t,t_0)\hat{\rho}(t_0)}dt^{\prime\prime}\\
    &+\kappa\eta_\mathrm{c}\int_{t}^{t^\prime}\tr{\mathcal{U}_0(\infty,t^\prime)\mathcal{R}_0(t^\prime,t^{\prime\prime})\mathcal{J}_0(t^{\prime\prime},t)\mathcal{S}_0(t,t_0)\hat{\rho}(t_0)}dt^{\prime\prime}\\
    &+\kappa\eta_\mathrm{c}\int_{t_0}^t\tr{\mathcal{U}_0(\infty,t^{\prime})\mathcal{R}_0(t^\prime,t)\mathcal{S}_0(t,t^{\prime\prime})\mathcal{J}_0(t^{\prime\prime},t_0)\hat{\rho}(t_0)}dt^{\prime\prime},
\end{aligned}
\end{equation}
where $\mathcal{J}_0(t^\prime,t)=\mathcal{J}\mathcal{U}_0(t^\prime,t)$. In this form, we can identify the similarity between this expression written in terms of the source dynamics and the expression in terms of temporal density functions in Eq.~(\ref{chapter2eq:firstordercorr}). When restricting $t^\prime\geq t$, we can notice that indeed $\xi_2(t,t^{\prime\prime},t^\prime,t^{\prime\prime})$ is nonzero only for $t^\prime\leq t^{\prime\prime}$, $\xi_2(t,t^{\prime\prime},t^{\prime\prime},t^\prime)$ is nonzero only for $t\leq t^{\prime\prime}\leq t^\prime$, and $\xi_2(t^{\prime\prime},t,t^{\prime\prime},t^\prime)$ is nonzero only for $t^{\prime\prime}\leq t$. These three components have clear physical interpretations. Given that two photons were emitted between $t_0$ and $\infty$, the first term describes the temporal coherence within the first photon emitted, the second term describes the temporal entanglement between the two photons, and the last term describes the temporal coherence within the second photon. The integrands of Eq.~(\ref{chapter2eq:twophotonfirstorderdynamics}) hint at the form to solve for the full two-photon temporal density function $\xi_2$ in terms of the source dynamics. We can solve for $\xi_2$ using the second-order conditional correlation for two photons and then show that it is indeed consistent with Eq.~(\ref{chapter2eq:twophotonfirstorderdynamics}).

The second-order four-time amplitude correlation for the two-photon component is
\begin{equation}
\begin{aligned}
    \braket{\bu(t_1^\prime)\bu(t_2^\prime)\bd(t_2)\bd(t_1)}_2 &=p_2\xi_2(t_1,t_2,t_1^\prime,t_2^\prime),
\end{aligned}
\end{equation}
where I have applied the time-ordered convention where $\xi_2(\mathbf{t},\mathbf{t}^\prime)=0$ if $\mathbf{t}=(t_1,t_2)$ or $\mathbf{t}^\prime=(t_1^\prime,t_2^\prime)$ is not time ordered. In terms of the source dynamics, we can expand the total two-photon temporal density function into
\begin{equation}
\begin{aligned}
\label{chapter2eq:twophotondensity}
    &\xi_2(t_1,t_2,t_1^\prime,t_2^\prime) =\\
    &\frac{\kappa^2\eta_\mathrm{c}^2}{p_2}\left\{\begin{aligned} &\tr{\mathcal{U}_0(\infty,t_2^\prime)\mathcal{R}_0(t_2^\prime,t_2)\mathcal{S}_0(t_2,t_1^\prime)\mathcal{R}_0(t_1^\prime,t_1)\mathcal{S}_0(t_1,t_0)\hat{\rho}(t_0)}&&t_1\leq t_1^\prime\leq t_2\leq t_2^\prime\\
    &\tr{\mathcal{U}_0(\infty,t_2^\prime)\mathcal{R}_0(t_2^\prime,t_2)\mathcal{S}_0(t_2,t_1)\mathcal{S}_0(t_1,t_1^\prime)\mathcal{R}_0(t_1^\prime,t_0)\hat{\rho}(t_0)}&&t_1^\prime\leq t_1\leq t_2\leq t_2^\prime\\
    &\tr{\mathcal{U}_0(\infty,t_2^\prime)\mathcal{R}_0(t_2^\prime,t_1^\prime)\mathcal{R}_0(t_1^\prime,t_2)\mathcal{S}_0(t_2,t_1)\mathcal{S}_0(t_1,t_0)\hat{\rho}(t_0)}&&t_1\leq t_2\leq t_1^\prime\leq t_2^\prime\\
    \end{aligned}
    \right.
\end{aligned}
\end{equation}
and $\xi_2(t_1^\prime,t_2^\prime,t_1,t_2)=\xi^*_2(t_1,t_2,t_1^\prime,t_2^\prime)$. In its general form, we can now see that the expression for the case $t_1\leq t_1^\prime\leq t_2\leq t_2^\prime$ applied to $\braket{\bu(t^\prime)\bd(t)}_2$ in Eq.~(\ref{chapter2eq:firstordercorr}) indeed recovers Eq.~(\ref{chapter2eq:twophotonfirstorderdynamics}). This consistency then justifies defining $n=2$ as the lowest-order nonzero term for correlations of quantities involving two jumps.

It may be possible to extend this method to compute the number coherence functions $\zeta_{n,m}$ by looking at correlations such as $\braket{\bd(t)}_n\propto\zeta_{n+1,n}(t)$ and $\braket{\bd(t_2)\bd(t_1)}_n\propto\zeta_{n+2,n}(t_1,t_2)$ under the same decomposition. However, the association is not as intuitive because of our choice to decompose the master equation in terms of the photon number in the waveguide. To my knowledge, unlike $\xi_1$, there are no results in the literature with which to compare such an approach when including excess decoherence, such as pure dephasing. Alternatively, we could unravel the master equation in the context of a physical phase measurement setup, such as a homodyne measurement \cite{carmichael}. As I explore in section \ref{chapter3:photonicstate}, it is also possible to obtain some information about number coherence using a self-homodyne setup, where two copies of the same photonic state are interfered in exactly the same spirit as the standard indistinguishability measurements discussed in section \ref{chapter1:indistinguishability}.

\section{Relation to photon statistics}
\label{chapter2:photonstatistics}

Using the photon number decomposition, we can derive the well-known photon statistics relations for brightness $\mu$ and $g^{(2)}$ from the perspective of the source rather than the field. In addition to providing these relations for use later in this thesis, it will also help show that the photon number decomposition is consistent with the properties expected of the emitted photonic state.

To derive these relations, we will need to use an important property of the conditional propagation superoperators. That is, that the propagator does not depend on the order of jumps because all jumps are identical:
\begin{equation}
\begin{aligned}
\int_{t_0}^{t_f}\mathcal{U}_k(t_\mathrm{f},t)\mathcal{J}\mathcal{U}_l(t,t_0)dt = \mathcal{U}_{k+l+1}(t_\mathrm{f},t_0).
\end{aligned}
\end{equation}
I provide a detailed proof of this property in appendix \ref{appendix:klorder} for the sake of interest, because I was not able to find it in the literature, although I imagine this is a known result in perturbation or measurement theory. It is also a very physically intuitive result because we would expect that the dynamics of the reduced system is unchanged when exchanging the order of any two jumps.

The integrated brightness of a pulsed source is defined in section \ref{chapter1:brightness} as $\mu=\int N(t)dt$ where $N(t)=\braket{\bu(t)\bd(t)}$. Using the invariance of order property and the input-output relations for a vacuum input, we can see that
\begin{equation}
\begin{aligned}
    \mu &= \kappa\eta_\mathrm{c}\lim_{t_\mathrm{f}\rightarrow\infty}\int_{t_0}^{t_\mathrm{f}}\braket{\hat{a}^\dagger(t)\hat{a}(t)}dt\\
    &=\sum_{n,k,l}\lim_{t_\mathrm{f}\rightarrow\infty}\tr{\int_{t_0}^{t_\mathrm{f}}\mathcal{U}_k(t_\mathrm{f},t)\mathcal{J}\mathcal{U}_l(t,t_0)dt\hat{\rho}(t_0)}\delta_{k+l+1,n}\\
    &=\sum_{n}n\lim_{t_\mathrm{f}\rightarrow\infty}\tr{\hat{\rho}_n(t_\mathrm{f})}.
\end{aligned}
\end{equation}
Finally, knowing that $p_n=\lim_{t_\mathrm{f}\rightarrow\infty}\tr{\hat{\rho}_n(t_\mathrm{f})}$, we recover the expected result for the average photon number of the pulsed photonic state: $\mu = \sum_n np_n$.

The pulsed $g^{(2)}$ can be computed in a similar fashion:
\begin{equation}
\begin{aligned}
    \frac{\mu^2}{2}g^{(2)} &= \kappa^2\eta_\mathrm{c}^2\lim_{t_\mathrm{f}\rightarrow\infty}\int_{t_0}^{t_\mathrm{f}}\int_{t}^{t_\mathrm{f}}\braket{\hat{a}^\dagger(t)\hat{a}^\dagger(t^\prime)\hat{a}(t^\prime)\hat{a}(t)}dt^\prime dt\\
    &=\sum_{k,l,m,n}\lim_{t_\mathrm{f}\rightarrow\infty} \tr{\int_{t_0}^{t_\mathrm{f}}\left(\int_{t}^{t_\mathrm{f}}\mathcal{U}_k(t_\mathrm{f},t^\prime)\mathcal{J}\mathcal{U}_l(t^\prime,t)dt^\prime\right)\mathcal{J}\mathcal{U}_m(t,t_0) dt \hat{\rho}(t_0)}\delta_{n,k+l+m+2}\\
    &=\sum_{n}\frac{n(n-1)}{2}\lim_{t_\mathrm{f}\rightarrow\infty} \tr{ \hat{\rho}_{n}(t_0)}.
\end{aligned}
\end{equation}
Hence, we obtain the well-known \cite{fischer2018scattering} expression for the pulsed integrated intensity correlation
\begin{equation}
g^{(2)} = \frac{\sum_n n(n-1)p_n}{\left(\sum_n np_n\right)^2}  
\end{equation}
as expected. These results confirm that the conditional expectation $\braket{~~}_n$ correctly preserves the total photon number and that it is capable of reconstructing the photonic state within photon number subspaces that are consistent with both the photon statistics and correlations predicted by the master equation. Although I will not demonstrate this, we could generalize this derivation for $g^{(m)}$ where $m\geq 2$ to confirm that
\begin{equation}
    g^{(m)} = \frac{m!}{\mu^m}\sum_n\binom{n}{m}p_n.
\end{equation}

To summarize, the results presented in this chapter are direct consequences of assuming that the source is modeled by a Markovian master equation that satisfies an effective proportionality $\bd\propto \ad$ with the waveguide mode. Physically, we can imagine that these two assumptions together imply that the state of the waveguide is a chronicle of the state of the source over time. By treating the waveguide as a quantum probe for the source state (section \ref{chapter2:photoncountingmeasurements}), we can find the state of the source conditioned on the number of emitted or detected photons. Conversely, by analyzing the conditional correlations of the source using this same indirect measurement (section \ref{chapter2:conditionalcorrelations}), we can also reconstruct the corresponding state of the waveguide (section \ref{chapter2:temporaldensityfunctions}) and describe the resulting photon counting statistics in terms of the source dynamics (section \ref{chapter2:photonstatistics}).
\chapter{Single photons}
\label{chapter3}

Single photons are one of the prime physical systems used to demonstrate quantum phenomena theoretically, experimentally, and pedagogically. Their presence is practically ubiquitous in quantum physics. Understanding their properties and behaviour is key to developing fundamental and technological advancements. Single photons can be used to demonstrate particle anti-bunching, single-particle interference, and the Hong-Ou-Mandel (HOM) effect \cite{hong1987measurement}. As discussed in section \ref{chapter1:applications}, they are key ingredients for quantum communication and quantum sensing. Many proposals for universal quantum computers also rely on photonic states built from single photons \cite{knill2001scheme,kok2007linear}.

In this chapter, I explore three main topics related to single photons. In section \ref{chapter3:roomtemperature}, I discuss the feasibility of developing a deterministic solid-state source of indistinguishable single photons that can operate at room temperature. Section \ref{chapter3:HOM} then expands on section \ref{chapter1:indistinguishability} by going into more detail regarding HOM interference, which is the standard method for quantifying single-photon indistinguishability. Finally, in section \ref{chapter3:photonicstate}, I demonstrate how the photon number decomposition can be used to explore the time dynamics of pulsed photonic states emitted by a purely-dephased two-level system.

The content of section \ref{chapter3:roomtemperature}, with the exception of subsection \ref{chapter3:generalizedpurcell}, is published in Ref.~\ref{wein2018plasmonics}. This content is presented in its published form with only minor changes made for clarification and consistency. Because this paper was published three years before this thesis, I provide a brief preface after this chapter outline to give an updated context.

Section \ref{chapter3:HOM} is based on projects that were completed in collaboration with the group of Prof. Pascale Senellart. The analysis of the effect of multi-photon events on indistinguishability is published in the experimental paper given by Ref.~\ref{ollivier2020g2hom}. This project also spurred the unpublished content of subsection \ref{chapter3:pulsedemitterHOM}, which explores the same problem for a specific emitter model. Subsection \ref{chapter3:selfhomodyne} is also unpublished material that was inspired by Refs. \ref{ollivier2020g2hom} and \ref{loredo2020deterministic}. Some of these results may be published in combination with an ongoing project related to photon number coherence with the Senellart group.

The final section \ref{chapter3:photonicstate} of this chapter deals more explicitly with the time dynamics and properties of an emitted photonic state from a pulsed two-level system that may experience pure dephasing. Although I focus on pure dephasing, the methods can be applied to other Markovian master equation models with excess decoherence, at least numerically if not analytically. The majority of the content in this section is not published, although it is related to Refs. \ref{ollivier2020g2hom} and \ref{loredo2020deterministic}. Some of the results in subsection \ref{chapter3:photonicbellstates} do appear in the supplementary of Ref.~\ref{loredo2020deterministic}.

\emph{Preface for Ref.~\ref{wein2018plasmonics}.}---The primary goal of the project presented in section \ref{chapter3:roomtemperature} was to motivate the plasmonics community to focus on developing coherent light-matter interactions for use in quantum information processing applications. However, although this paper has seen success in terms of citations since its publication in 2018 and the topic continues to be relevant \cite{palstra2019hybrid}, the model used in this paper is outdated and neglects some important effects. I will now outline these issues in the following three paragraphs. 

First, the QNM master equation, developed by Ref.~\cite{franke2019quantization} and discussed in section \ref{chapter1:qnmmastereq}, supersedes the simpler dissipative Jaynes-Cummings model and multi-pole Markovian quenching model \cite{peyskens2017} that is used in section \ref{chapter3:roomtemperature}. The latter model is only valid for specific cavity geometries. Instead, the QNM approach generally describes a dissipative cavity in terms of modes that each account for both radiative and non-radiative components, giving rise to inter-dependent coupling strengths and dissipative rates.

Second, section \ref{chapter3:roomtemperature} only explores limitations due to photon emission, and hence makes a crucial assumption that the emitter is excited on a timescale faster than the Purcell-enhanced lifetime, which itself must be sufficiently short at room temperature to overcome dephasing. Otherwise, detrimental re-excitation of the emitter can degrade the emission single-photon purity and HOM visibility. For QDs, this amounts to requiring pulses on the femtosecond timescale \cite{hughes2019theory}, which can then cause additional problems such as reducing the excitation pulse filtering efficiency \ref{ollivier2020g2hom} or incidentally exciting QD higher-energy biexciton states. However, as we mention in Ref.~\ref{wein2018plasmonics}, a QD is unlikely to ever be successful as a room temperature single-photon source even using fast pulses due to having a very broad emission zero-phonon line and phonon sideband at room temperature, which fundamentally limits indistinguishability even with a Purcell enhancement \cite{smith2017,grange2017reducing}.

Third, for solid-state emitters, excitation pulses can cause additional power-dependent dephasing due to phonon interactions \cite{nazir2016modelling}, which is not captured by the constant phenomenological pure dephasing model used in this thesis. In addition, non-Markovian behaviour becomes important for pulses on the order of the phonon bath memory timescale. This effect has been well-studied for quantum dots \cite{mccutcheon2010quantum}, and the lower bound is around 5-10 ps at low temperature. However, to my knowledge, this effect has not been thoroughly investigated for atomic-scale defects such as rare-earth ions and color centers in diamond. These latter systems generally have longer lifetimes and more spectrally-separated phonon sidebands than quantum dots. Hence, some of these systems could remain in a Markovian regime even under ultrafast optical excitation. As a motivating case, ultrafast coherent optical control using pulses as short as 1 ps has been demonstrated with negatively-charged silicon vacancy centers in diamond \cite{becker2016}, albeit at low temperature. Moreover, ultrafast excitation that is faster than the bath memory time can decouple the phonon bath, allowing for efficient excitation that should be quite robust even at room temperature \cite{vagov2007nonmonotonic,kaldewey2017demonstrating}.

Given its shortcomings, I still believe that Ref.~\ref{wein2018plasmonics} is an insightful study that contains valid conclusions. First, it provides a good analytic approach for estimating the emission-limited indistinguishability for a cavity-emitter system in the critical regime that includes a correction due to unfiltered phonon sideband emission. Second, we find that room temperature solid-state indistinguishable single photon sources should still be a feasible goal for the plasmonics community, but that accomplishing this goal will necessarily require emitters more narrow than QDs or NV centers. Third, plasmonic antennas or cavities with low $Q$ factors cannot filter the large phonon sidebands of solid-state emitters at room temperature. Therefore, a mid-$Q$ hybrid plasmonic-dielectric cavity combined with a relatively narrow emitter and ultrafast excitation is the most promising route.

\section{Single-photon sources at room temperature}
\label{chapter3:roomtemperature}

While there is substantial excitement about quantum technology and quantum information processing (QIP), many practical applications are still held back by the fact that critical components are restricted to operating at cryogenic temperatures. Trying to overcome the thermal restrictions of quantum devices also tests fundamental questions about the physical regimes in which quantum processes can exist and be manipulated.

Indistinguishable single-photon sources (SPSs) are basic components of numerous different optical QIP implementations. As discussed earlier in this thesis, they are required for tasks such as linear-optical quantum computing \cite{kok2007linear} and boson sampling \cite{spring2012, broome2013}. In addition, an efficient indistinguishable photon source can be used to construct quantum repeaters and would assist in the development of a quantum internet \cite{gisin2002quantum,sangouard2007, kimble2008internet, simon2017internet}.

The most common way to generate indistinguishable photons at room temperature is by heralded spontaneous parametric down-conversion (SPDC) \cite{kwiat1995}. This technique has seen pioneering success in quantum research, yet its probabilistic nature limits its range of applications, although this limitation could in principle be addressed by multiplexing many SPDC sources \cite{joshi2017,kaneda2019high}.

In contrast, individual quantum emitters not only promise near-deterministic single-photon emission at room temperature, but moreover the emitted photon could be entangled with coherent solid-state spins \cite{balasubramanian2009}. This would allow many QIP applications to be performed at room temperature, including optically-mediated entanglement of distant spins in solids \cite{bernien2013heralded}.

Low-temperature indistinguishable SPSs have been achieved \cite{santori2002} and are becoming more efficient \cite{ding2016demand, somaschi2016}. Ref.~\cite{grange2015cavity} showed that relatively inefficient indistinguishable SPSs can be realized beyond the low-temperature regime using weakly-coupled narrow-bandwidth micro-cavities. These results were applied to extend quantum dot indistinguishable SPS operation up to 20 K \cite{grange2017reducing}. Distinguishable SPSs have also been demonstrated at room temperature using solid-state quantum emitters \cite{aharonovich2016}. In particular, recently, plasmonic cavities have been proposed to enhance emission rates for distinguishable SPSs in spite of high losses \cite{bozhevolnyi2016,bozhevolnyi2017}.

Achieving indistinguishable photon emission from a solid-state emitter is a difficult task, especially at higher temperatures. The main problem is that optical transitions in solid-state materials experience rapid phonon-induced dephasing that homogeneously broadens the zero-phonon line (ZPL) at room temperature \cite{davies1974, fu2009observation}. This dephasing reduces the degree of indistinguishability between emitted photons \cite{kiraz2004quantum, grange2015cavity}. In addition, phonon-assisted optical transitions can produce a phonon sideband (PSB) that reduces indistinguishability and consequently must be filtered, which sacrifices efficiency. If the PSB spectrum overlaps with the desired ZPL emission, it cannot be entirely filtered and hence fundamentally limits indistinguishability \cite{iles2017phonon, smith2017}. Furthermore, when using plasmonic materials to enhance emission, plasmon-induced quenching poses another detrimental effect that affects both efficiency and indistinguishability.  A very recent analysis on a single spherical metallic nanoparticle with the goal of producing an on-chip room temperature single-photon source \cite{peyskens2017} found that simultaneous high efficiency and high indistinguishability are difficult to achieve with integrated plasmonics.

Here we show that achieving high efficiency and high indistinguishability simultaneously at room temperature should be possible by using ultrasmall mode volume cavities and by operating in the critical regime (see section \ref{chapter1:cavityregimes}). We argue that such systems are within reach of current technology, and we
provide theoretical guidelines for designing successful plasmonic cavities for this purpose.

A common approach to improving indistinguishability and efficiency simultaneously is to place a quantum emitter inside a cavity.  In addition to suppressing off-resonant PSB emission, the cavity reduces the emitter lifetime through the Purcell effect \cite{purcell1946} and allows the photon to be emitted before the emitter coherence is destroyed by interactions with the phonon bath. At room temperature, this fast emitter dephasing is very difficult to overcome, requiring Purcell factors exceeding $10^4$ for most emitters \cite{grange2015cavity}. Obtaining a large Purcell factor can be accomplished by either increasing the cavity quality factor or by decreasing the effective mode volume. However, for highly dissipative emitters, increasing the quality factor too high prolongs the interaction between the cavity photon and the emitter, which causes emitter dephasing. As a result, it is necessary to use cavities with mode volumes far below the diffraction limit.

Such ultrasmall mode volume cavities have seen significant development over the last decade and, in particular, the last few years \cite{robinson2005,seo2009,kuttge2009,russell2010,kang2011,hu2016,choi2017,gurlek2018manipulation,santhosh2016,chikkaraddy2016,peng2017enhancing}. Many of these cavities utilize plasmonic materials to concentrate the electromagnetic field along a material interface in the form of plasmon-polaritons \cite{seo2009,kuttge2009,russell2010,kang2011,santhosh2016,chikkaraddy2016,gurlek2018manipulation,peng2017enhancing}. There are also interesting proposals for pure-dielectric ultrasmall cavities \cite{robinson2005,hu2016,choi2017} and plasmonic-Fabry-P\'{e}rot hybrid cavities \cite{gurlek2018manipulation,peng2017enhancing}. However, the application of ultrasmall mode volume cavities to quantum information processing is still relatively unexplored.

The interaction between a quantum emitter and the phonon bath will broaden the emitter's ZPL and could also create a PSB. Recently, the effect of the PSB on the indistinguishability of single-photon sources was studied \cite{smith2017,iles2017phonon}. It was found that the indistinguishability is limited by the fraction of PSB not filtered by the cavity. This limitation can be highly detrimental at room temperature where the ZPL is very broad, which necessitates a broad cavity that might not filter the PSB. Therefore, an ideal emitter for an efficient room-temperature indistinguishable SPS should have a small PSB that is spectrally well-separated from its ZPL. For such an emitter, pure dephasing and possible quenching effects will be the primary limitation. Consequently, we first consider a Markovian system that neglects non-Markovian PSB effects in order to describe the most ideal parameter regime for a good (Markovian) emitter. Then we estimate the correction induced by a non-zero PSB using the results of Iles-Smith \emph{et al.} \cite{smith2017}.

\subsection{Indistinguishability in the critical regime}
\label{chapter3:indistinguishabilityquenching}

We now describe our proposed approach in detail. We begin with the interaction Hamiltonian for a driven two-level system coupled to a resonant cavity $\hat{H} = \hbar g(\sigd\au+\sigu\ad)+(\hbar\Omega/2)(\sigd+\sigu)$. The driving term of the Hamiltonian significantly complicates the derivation for indistinguishability. However, the single-photon purity of emission from a two-level system depends on how the system is excited. Slow excitation allows for multi-photon emission, which increases $g^{(2)}$. For a high single-photon purity, excitation of the emitter will require ultrafast optical control, which has been demonstrated for defects in diamond \cite{bassett2014,becker2016}. The common practice, which we also adopt, is to assume $\Omega\gg g$ during excitation so that the system is effectively instantaneously prepared in the excited state \cite{grange2015cavity,peyskens2017}, implying $g^{(2)}\simeq 0$. We then explore the dominant emission dynamics as governed by the system when $\Omega=0$. This allows us to make use of the single-excitation approximation and decouple the optical Bloch equations (see section \ref{chapter1:cavityQED}). However, for large Purcell factors, the excitation timescale must still be shorter than the Purcell-enhanced lifetime.

We describe the dissipative dynamics of the system using the Markovian master equation governed by the Liouville superoperator $\mathcal{L} = -(i/\hbar)\mathcal{H} + \kappa\mathcal{D}(\ad) + \gamma^\star\mathcal{D}(\sigu\sigd)+\gamma\mathcal{D}(\sigd)$. Please note that the definition of $\gamma^\star$ differs by a factor of 2 in this entire section \ref{chapter3:roomtemperature} and appendix \ref{AppendixB} compared to the rest of this thesis. This choice follows the convention in \cite{grange2015cavity} so that $\Gamma=\gamma+\gamma^\star$ is the FWHM of the ZPL rather than $\gamma+2\gamma^\star$. We take the cavity linewidth $\kappa=\kappa_{r}+\kappa_\text{nr}$ to have a radiative part $\kappa_\text{r}$ and non-radiative part $\kappa_\text{nr}$. The cavity quality factor is $Q = \omega/\kappa$ and we define the cavity quantum efficiency as $\eta_\text{c}=\kappa_\text{r}/\kappa$. Similarly, the natural decay rate $\gamma=\gamma_\text{r}+\gamma_\text{nr}$ also has a radiative part $\gamma_\text{r}$ and a non-radiative part $\gamma_\text{nr}$.

We can now state the requirements for high efficiency and indistinguishability. As discussed in section \ref{chapter1:cavityregimes}, the rate of population transfer from the emitter to the cavity must exceed the total emitter decoherence rate $R>\Gamma$. In addition, the photon must escape the cavity before it is dephased by the emitter: $\kappa>\Gamma$. In this section, we define $R=4g^2/\kappa$, which is consistent with the more general definition given in section \ref{chapter1:cavityQED} only in the critical regime where $\Gamma\ll \kappa$ and when the cavity and emitter are resonant so that $\Delta=0$. In this way, $R$ is directly related to the uninhibited Purcell factor by $F_\mathrm{p}=R/\gamma_\mathrm{r}$.

To quantify the efficiency of the system, we use the cavity efficiency $\mu =\eta_\mathrm{c}\beta$ where $\beta=\kappa\int_0^\infty\braket{\au(t)\ad(t)}dt$ \cite{grange2015cavity}. The derivation is given in appendix \ref{AppendixB} by inverting the optical bloch equations, and the result is $\beta=R\kappa/(R(\gamma+\kappa)+\gamma(\kappa+\Gamma))$.
In our analysis, we only explicitly compute $\beta$---the efficiency of population dissipated through the cavity mode---hereafter referred to as the intrinsic cavity efficiency. The total collected radiative quantum efficiency is given by $\mu=\beta\eta_\text{c}$. The efficiency of an SPS in the presence of plasmonic materials has been studied \cite{bozhevolnyi2016,bozhevolnyi2017}. However, for many quantum information applications, the efficiency is not the only relevant quantity, but the indistinguishability of the emitted photons is also essential. 

The quantity that we use for indistinguishability is derived from the probability that two photons emitted from the same source interfere and bunch at a beamsplitter \cite{kiraz2004quantum, grange2015cavity} (see also section \ref{chapter1:indistinguishability}). An ideal indistinguishable SPS should have near-unity indistinguishability $I$ and efficiency $\beta$. To this end, we focus on maximizing the indistinguishability-efficiency product $I\beta$.

Using the quantum regression theorem \cite{gardiner2004}, we derived an expression for indistinguishability valid in the critical regime for arbitrary $\gamma$. As we will show, this allows the expression to capture possible effects of plasmon quenching. The details of the derivation and the full solution valid for arbitrary $\gamma$ are given in appendix \ref{AppendixB}. Here we only write the expression in the case that $\gamma<\gamma^\star<\kappa$, which arises when quenching is weak:
\begin{equation}
\label{ind}
I=\frac{R^2\kappa^2\left(1+I_1\right)}{(R+\gamma)(\kappa+\gamma)(R+\Gamma)(\kappa+\Gamma)\beta^2},
\end{equation}
where $I_1=(\gamma^\star/\kappa)(6\kappa-R)/(3\kappa+4R)$. This expression is accurate in the critical regime to first order in $\gamma^\star/\kappa$.

When there are no plasmon quenching effects and when the system is far within the critical regime boundaries, we have that $\gamma\simeq 0$ and $I_1\simeq 0$; hence, \mbox{$I=R\kappa(R+\gamma^\star)^{-1}(\kappa+\gamma^\star)^{-1}$} and $\beta=1$ for perfect excitation. From this, it can be seen that $I\beta$ is maximized when $R=\kappa$ for a given cavity coupling rate $g$ (see Fig.~\ref{quenchingcouplingind}~(a)). This implies $2g=\kappa$, which is also the strong-coupling boundary in the limit that $\kappa\gg\Gamma$.

\subsection{A Markovian quenching model}
\label{chapter2:emitterquenching}

\begin{figure}[t]
\centering
\hspace{-45mm}(a)\hspace{55mm}(b)\\
\includegraphics[trim={5mm 5mm 0 0},scale=0.6]{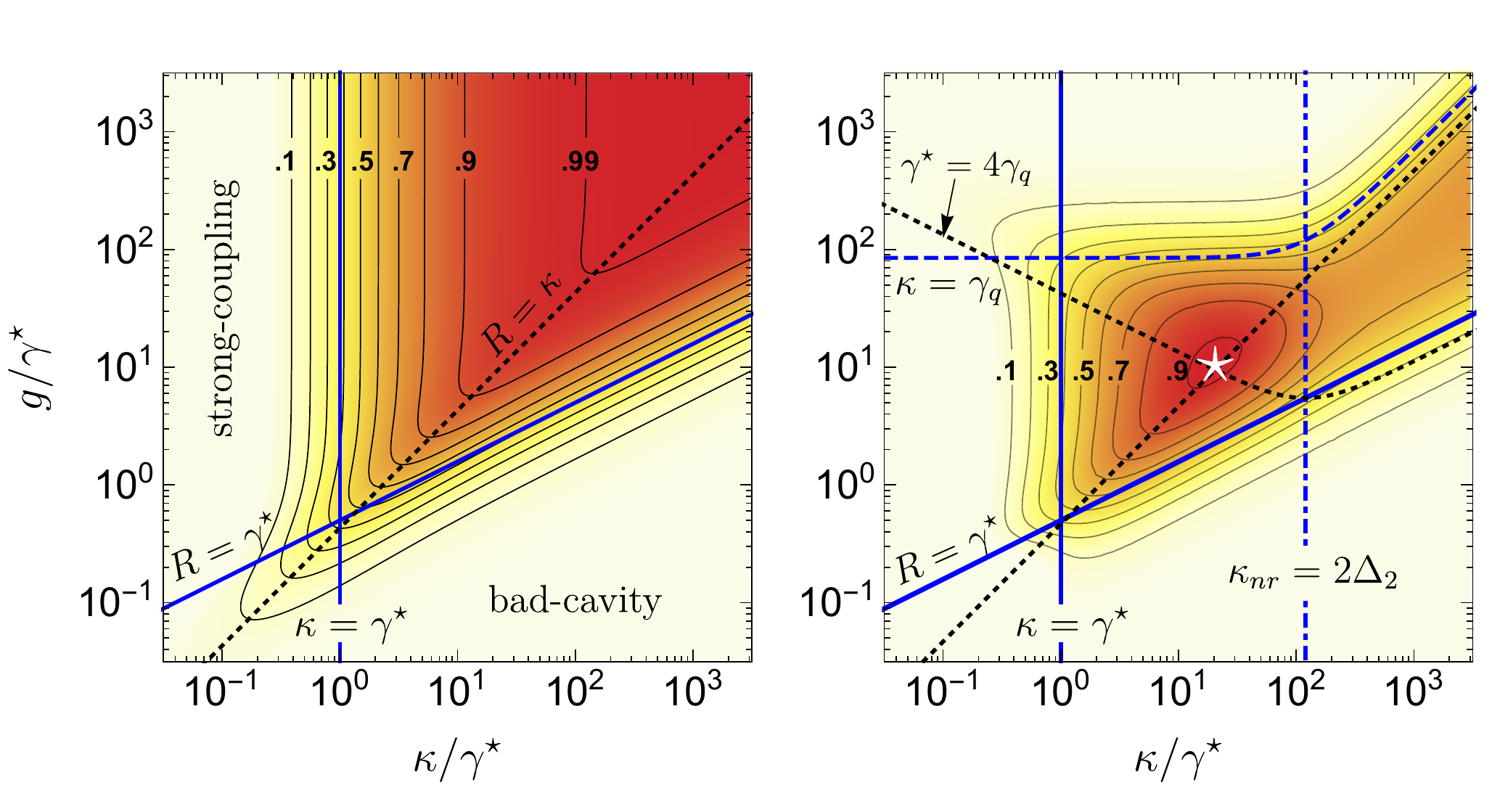}
\caption[Product of the indistinguishability $I$ and intrinsic cavity efficiency $\beta$ for an emitter-cavity system in the critical regime.]{\small \textbf{Product of indistinguishability $I$ and intrinsic cavity efficiency $\beta$ for an emitter-cavity system in the critical regime} with cavity coupling $g$, emitter dephasing rate $\gamma^\star\simeq \Gamma$, and cavity linewidth $\kappa$ with non-radiative portion $\kappa_\text{nr}=(1-\eta_\mathrm{c})\kappa$ where $\eta_\mathrm{c}$ is the bare cavity radiative efficiency. (a) The case without quenching ($\gamma_\text{q}$=0). $I\beta$ plotted along with the critical regime boundaries (blue solid lines) and the boundary between the strong-coupling and bad-cavity regimes, $R=\kappa$ (black dotted line). (b) The case with quenching ($\gamma_\text{q}\neq 0$). $I\beta$ for a simple one-mode example, $\gamma_\text{q}=g_2^2\kappa_\text{nr}/(\Delta_2^2+(\kappa_\text{nr}/2)^2)$, to illustrate the different regimes and boundaries. Here we use $g_2=g/2$, $\Delta_2=30\gamma^\star$, and $\eta_\mathrm{c}=0.5$, corresponding to $\Delta_\text{q}=60\gamma^\star$. The maximum of $I\beta=0.92$ occurs at the intersection between $\gamma^\star=4\gamma_\text{q}$ and $R=\kappa$ (white star). The regime above the $\kappa=\gamma_\text{q}$ line is dominated by quenching. The vertical dashed line divides the mode-detuned case to the left and the nearly resonant case to the right.}
\label{quenchingcouplingind}
\end{figure}

For a plasmonic cavity, $g$ is the coupling rate to the dominant radiating mode; however, the emitter will also couple to higher-order modes \cite{delga2014,bozhevolnyi2017,peyskens2017}. These higher-order modes contribute to the enhanced decay rate of the emitter but are predominantly non-radiative and hence quench the emission.

We treat the coupling of the emitter to higher-order modes using the Markovian approximation \cite{peyskens2017}. In this case, the emitter decay rate $\gamma$ is increased by the quenching rate $\gamma_\text{q}$ so that $\gamma$ becomes $\gamma=\gamma_\text{r}+\gamma_\text{nr}+\gamma_\text{q}$. The quenching contribution $\gamma_\text{q}$ can be described by the plasmon spectral density, which is approximated by a sum of Lorentzian functions in the quasistatic limit \cite{delga2014}:
\vspace{-2mm}
\begin{equation}
\label{quenching}
\gamma_\text{q} = \sum_{l=2}^\infty \frac{g_l^2\kappa_\text{nr}}{\Delta_l^2+(\kappa_\text{nr}/2)^2},
\vspace{-2mm}
\end{equation}
where $\Delta_l>0$ is the detuning of the respective mode from the emitter, $g_l=k_l g$ is the coupling rate where $k_l$ is approximately independent of $g$, and we assume that each mode has the same non-radiative rate $\kappa_\text{nr}=\kappa(1-\eta_\mathrm{c})$ as the dominant mode \cite{peyskens2017}. This Markovian approximation is justified when the coupling rate to each individual higher-order mode is not too strong---when \mbox{$g_l^2/(\Delta_l^2+(\kappa_\text{nr}/2)^2)<1$}. Since we only require the system to achieve the optimal relation $R=\kappa$ in the critical regime, the system is only on the brink of strong coupling with the resonant dominant mode. Hence, we expect that the detuned higher-order modes should not display any significant strong coupling. In the critical regime, the quenching rate dominates $\gamma_\text{q}>\gamma_\text{r}+\gamma_\text{nr}$ so that $\gamma\simeq\gamma_\text{q}$.

To attain large $I\beta$, the dissipation through the cavity must be faster than the quenching rate, implying: $4g^2/\kappa> \gamma_\text{q}$ and $\kappa> \gamma_\text{q}$. There are two upper-bound cases to consider for $\gamma_\text{q}$. In the limit that the higher-order modes are near-resonant with the emitter ($4\Delta_l^2< \kappa_\text{nr}^2$), we have $\gamma_\text{q}\simeq (4g^2/\kappa_\text{nr})\sum_{l=2}^\infty k_l^2$. Then $4g^2/\kappa>\gamma_\text{q}$ implies $\sum_{l=2}^\infty k_l^2< (1-\eta_\mathrm{c})$. This condition opposes high radiative efficiency and it is very difficult to satisfy when there are many modes, making it unsuitable. On the other hand, if the modes are detuned from the emitter \mbox{($4\Delta_l^2> \kappa_\text{nr}^2$)}, we have $\gamma_\text{q}\simeq g^2\kappa_\text{nr}/\Delta_\text{q}^2$ where we define $1/\Delta_\text{q}^2 = \sum_{l=2}^\infty k_l^2/\Delta_l^2$ for simplicity. In this case, $4g^2/\kappa,\kappa> \gamma_\text{q}$ implies that we require $\kappa,2g< 2\Delta_\text{q}(1-\eta_\mathrm{c})^{-1/2}$ to achieve large $I\beta$. Applying the condition $\kappa>\gamma^\star$ that is required to reach the critical regime, we find the main condition: $\gamma^\star(1-\eta_\mathrm{c})^{1/2} <2\Delta_\text{q}$.

The quantity $\Delta_\text{q}$ can be seen as an effective detuning parameter that describes the severity of quenching for a plasmonic-emitter system. The value of $\Delta_\text{q}$ depends on the geometry of the cavity and the position of the emitter relative to the cavity. For example, $\Delta_\text{q}$ for a single spherical metallic nanoparticle is dependent on the ratio $d/r$ of the distance between the emitter and the particle surface $d$ and particle radius $r$ \cite{delga2014}. For a silver sphere, $\Delta_\text{q}$ can range from $2\pi\times 6$ THz for $d/r=0.05$ to $2\pi\times 117$ THz for $d/r=2$. In this example, the limit $\gamma_\text{q}\simeq g^2\kappa_\text{nr}/\Delta_\text{q}^2$ is also a good approximation when $Q>5$.

In the absence of quenching, $I\beta$ can be increased arbitrarily by increasing both $g$ and $\kappa$ while following $R=\kappa$ to maximize the Purcell enhancement (see Fig.~\ref{quenchingcouplingind}~(a)). However, in the presence of quenching, $I\beta$ decreases when $\kappa,2g>2\Delta_\text{q}(1-\eta_\mathrm{c})^{-1/2}$. This restricts $I\beta$ to a maximum value for a given $\Delta_\text{q}$ and $\gamma^\star$. We analytically maximized $I\beta$ in the mode-detuned case for small $\gamma^\star/\Delta_\text{q}$. The values of $g$ and $\kappa$ that maximize $I\beta$ were found to be $\kappa_\text{max}\simeq 2g_\text{max}\simeq [\Delta_\text{q}^2\gamma^\star/(1-\eta_\mathrm{c})]^{1/3}$. From this solution, we see that decreasing $\gamma^\star/\Delta_\text{q}$ or increasing $\eta_\mathrm{c}$ will increase the maximum possible $I\beta$. We also notice that, at the $I\beta$ maximum, $\gamma_\text{q}\simeq \gamma^\star/4$ is independent of $\Delta_\text{q}$ and $\eta_\mathrm{c}$. Hence, the maximum occurs roughly at the intersection between $2g=\kappa$ and $4\gamma_\text{q}=\gamma^\star$, see Fig.~\ref{quenchingcouplingind}~(b).

\subsection{Phonon sideband corrections}

As we have shown, decreasing $\gamma^\star/\Delta_\text{q}$ can increase the maximum attainable $I\beta$. This heavily favors narrow linewidth emitters. An emitter with a smaller $\gamma^\star/\gamma_\text{r}$ ratio is also advantageous because the Purcell factor required to reach the maximum (at $2g=\kappa$ and $\gamma_q=4\gamma^\star$) is not as large. Furthermore, a good emitter for this application must have a small PSB that has a small overlap with the cavity spectrum \cite{smith2017, iles2017phonon}. This is necessary because any photons emitted through the PSB into the cavity will decrease the indistinguishability and any photons emitted from the PSB directly will decrease the intrinsic cavity efficiency.

The reduction of indistinguishability and intrinsic cavity efficiency due to PSB non-Markovian effects can be approximated in the cavity-emitter weak-coupling regime by \cite{smith2017}:
\begin{equation}
\label{Icorrect}
I = I_0\left[\frac{B^2}{B^2+F(1-B^2)}\right]^2
\end{equation}
and
\begin{equation}
\label{betacorrect}
\beta = \beta_0\frac{B^2+F(1-B^2)}{1-\beta_0(1-F)(1-B^2)},
\end{equation}
where $I_0$ and $\beta_0$ are the values computed in the Markovian approximation; $B$ is the Franck-Condon factor, and $F$ is the fraction of the PSB not filtered by the cavity, which depends on the cavity linewidth $\kappa$ and the exact PSB spectrum for an emitter. In writing Eq.~(\ref{Icorrect}), we also assume that the additional interaction between the PSB and ZPL via the cavity does not significantly alter the phonon-induced pure-dephasing rate $\gamma^\star$, which is a reasonable assumption in the regime of interest and at room temperature where $\gamma^\star$ is already quite large (see appendix \ref{AppendixB} for a more detailed discussion).

In general, $F$ increases with cavity linewidth. In the broad-cavity limit where no PSB is filtered ($F\rightarrow 1$), the $I\beta$ product is limited to $I\beta=I_0\beta_0B^4$, where $B^4$ can be interpreted as the probability that both of the photons being interfered at a beamsplitter were emitted from the ZPL.

\subsection{Proposal for silicon vacancy centers}

The above restrictions on ZPL and PSB do not favor popular emitters such as quantum dots (QDs) and nitrogen-vacancy (NV$^-$) centers in diamond, both of which have significant PSBs at room temperature and generally broad ZPLs. However, a promising candidate is the negatively-charged silicon-vacancy (SiV$ ^{-}$) center in diamond. A nanodiamond SiV$ ^{-}$ center can have $\gamma^\star$ as small as $2\pi\times 380$ GHz \cite{neu2011} and $1/\gamma_\text{r}\simeq 8.3$ ns ($1/\gamma=0.58$ ns with $\gamma_\text{r}/\gamma=0.07$) \cite{benedikter2017}. They can also have a very small PSB, with a ZPL emission proportion (Debye-Waller factor) up to $B^2=\text{DW}=0.88$ \cite{neu2011}. The SiV$^-$ PSB is also spectrally well-separated from the ZPL, allowing most of the PSB emission to be filtered by the cavity.

For a nanodiamond SiV$^-$ center with $\text{DW}=0.88$, the PSB corrections from the previous section implies $I\beta\simeq 0.77I_0\beta_0$ whereas an NV$^-$ center with $\text{DW}\simeq 0.03$ \cite{santori2010} implies $I\beta\simeq 0.0009I_0\beta_0$. A cavity can improve this by allowing $F<1$, provided that the PSB is spectrally well-separated from the cavity resonance.

A promising plasmonic cavity design for room-temperature applications is the plasmonic bowtie antenna, which was used to demonstrate vacuum Rabi splitting with single QDs at room temperature \cite{santhosh2016}. Unfortunately, the close proximity of the emitter to the bowtie structure makes it difficult to achieve a large $\Delta_\text{q}$ and $\eta_\mathrm{c}$, causing the system to be dominated by quenching. Moreover, a bowtie antenna has a very broad resonance ($Q\simeq 7$), which is a poor filter for any PSB emission. These problems can be solved by placing the bowtie inside a detuned Fabry-P\'{e}rot cavity \cite{gurlek2018manipulation}. This hybrid approach promises to alleviate quenching effects, improve cavity quantum efficiency, and increase the cavity quality factor. A simulated Fabry-P\'{e}rot-bowtie hybrid cavity shows a Purcell factor (as determined from local density of states (LDOS) enhancement) up to $R/\gamma_\text{r}=2.7\times 10^5$ with a $Q$ as high as $10^3$ and near-unity radiative efficiency ($\beta\eta_\mathrm{c}\simeq 0.95$). In addition, high collection efficiencies should be possible with these systems, e.g. 81\%\cite{gurlek2018manipulation}. Furthermore, by tuning the parameters of a hybrid cavity, it should be possible to optimize the system to produce a large $I\beta$ using the guidelines derived above. 

To estimate the $I_0\beta_0\eta_\mathrm{c}$ achievable using a nanodiamond SiV$^-$ center inside a hybrid cavity, we used the spectrum of sample 5 from Neu \emph{et al.} \cite{neu2011} along with the predictions for cavity properties from Gurlek \emph{et al.} \cite{gurlek2018manipulation}. For the emitter, we used resonance frequency $\omega=2\pi\times 405$ THz, $1/\gamma_\text{r}=8.3$ ns \cite{benedikter2017}, and $\gamma^\star=2\pi\times500$ GHz \cite{neu2011}. For the cavity, we used $R/\gamma_\text{r}=2.7\times 10^5$, $Q=60$, and $\beta_0\eta_\mathrm{c}=0.95$. With these parameters, the Markovian approximation gives $I_0\beta_0\eta_\mathrm{c}=0.86$ ($I_0=0.90$, $\beta_0=0.98$) for \mbox{$\Delta_\text{q}=2\pi\times 5$ THz}.

\begin{figure}[p]
\centering
\includegraphics[trim={6mm 5mm 0mm 30mm},scale=0.55]{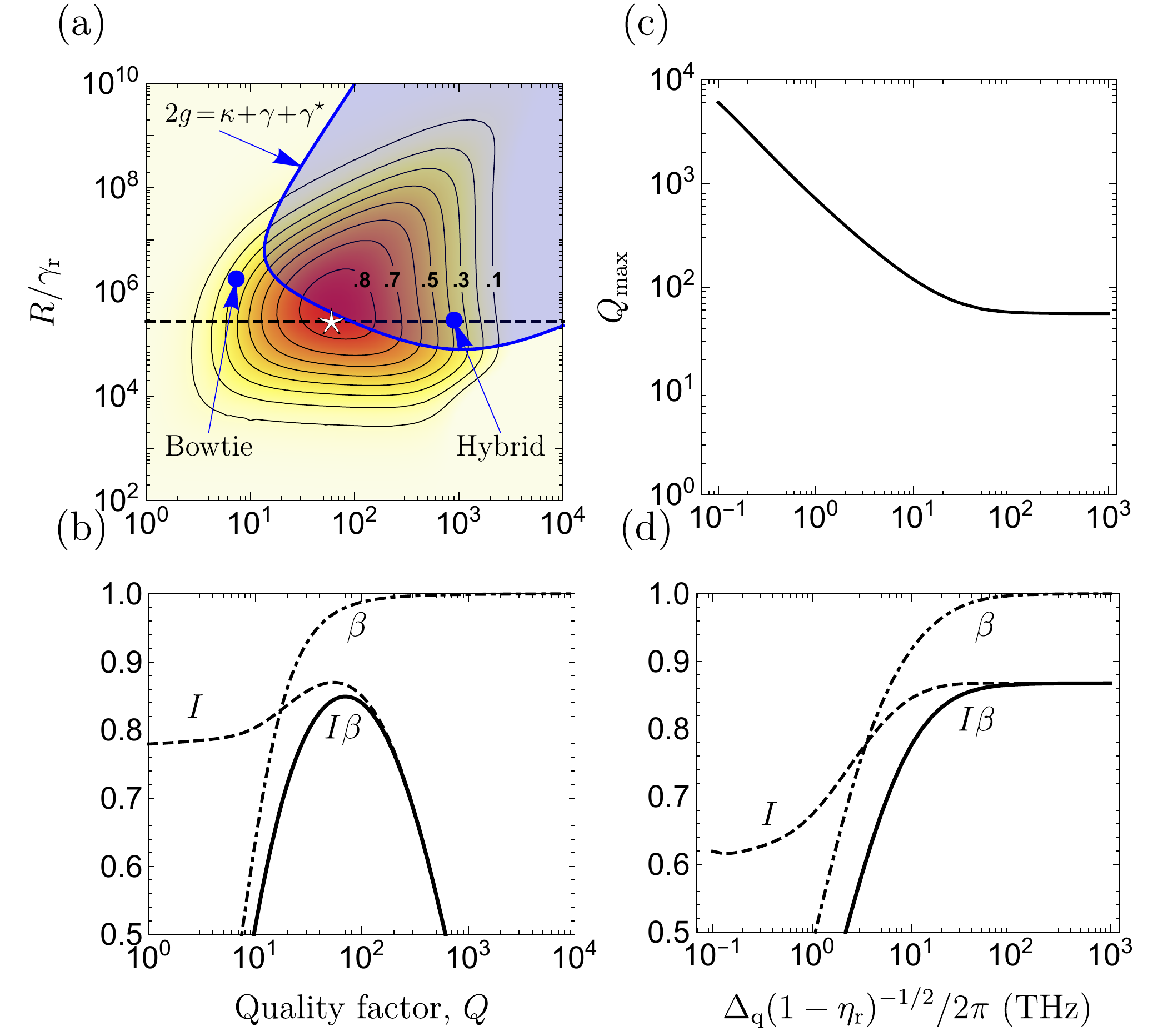}
\caption[Indistinguishability and efficiency estimations for a nanodiamond negatively-charged silicon vacancy (SiV$^-$) center enhanced by an ultrasmall mode volume cavity.]{\small \textbf{Indistinguishability and efficiency estimations for a nanodiamond negatively-charged silicon vacancy (SiV$^-$) center enhanced by an ultrasmall mode volume cavity} when taking into account the effects of quenching and a non-zero phonon sideband (PSB). The effect of the SiV$^-$ PSB on indistinguishability $I$ and intrinsic cavity efficiency $\beta$ is estimated using an expression valid in the weak-coupling regime \cite{smith2017}. This small correction could be inaccurate in the strong-coupling regime (blue shaded region in (a)) where the zero-phonon line (ZPL) begins to display a vacuum Rabi splitting. Parameters used for the SiV$^-$ center are as follows: resonance frequency $\omega=2\pi\times 405$ THz, radiative lifetime $1/\gamma_\text{r}=8.3$ ns \cite{benedikter2017}, phonon-induced pure-dephasing rate $\gamma^\star=2\pi\times 500$ GHz \cite{neu2011}, and Debye-Waller factor $\text{DW}=0.88$ \cite{neu2011}. See appendix \ref{AppendixB} for the PSB spectrum used to calculate the correction.
(a) $I\beta$ plotted in the critical regime and in the mode-detuned case with $\Delta_\text{q}(1-\eta_\mathrm{c})^{-1/2} = 2\pi\times 30$ THz, where $\Delta_\text{q}$ is the effective detuning parameter for higher-order non-radiative plasmon modes and $\eta_\mathrm{c}$ is the bare cavity quantum efficiency. Here $R=4g^2/\kappa$ where $g$ is the cavity coupling rate and $\kappa$ is the bare cavity linewidth. The blue dots represent the plasmonic bowtie \cite{santhosh2016} and a plasmonic-Fabry-P\'{e}rot hybrid cavity \cite{gurlek2018manipulation}. For the bowtie, $R/\gamma_\text{r}=1.7\times 10^6$ is determined from $g=60$ meV, $Q=7.3$ \cite{santhosh2016}, and $1/\gamma_\text{r}=20$ ns \cite{brokmann2004}. The dashed line marks the $R/\gamma_\text{r}=2.7\times 10^5$ expected for the hybrid cavity (at $Q=986$) from the enhancement of the local density of states (LDOS) \cite{gurlek2018manipulation}. The white star shows the proposed single-photon source. (b) Cross-section for the dashed line in (a). The indistinguishability reaches the value of $I=\text{DW}^2\simeq 0.77$ in the limit $Q\rightarrow 0$ where the PSB is not filtered. (c) Quality factor $Q_\text{max}$ required to reach the maximum $I\beta$ for a given $\Delta_\text{q}$ and $R/\gamma_\text{r}=2.7\times 10^5$. (d) $I\beta$ expected at $Q_\text{max}$. $I$ is limited by both ZPL broadening and PSB emission in the limit that quenching is negligible.}
\label{sivplots}
\end{figure}

To approximate the effect of the PSB, we estimated the PSB spectrum using a sum of Lorentzian functions to match the measured spectrum and Debye-Waller factor $B^2=\text{DW}=0.88$ of sample 5 \cite{neu2011} (see appendix \ref{AppendixB} for the expression). For a cavity quality factor of $Q=60$, the fraction of PSB not filtered by the cavity is $F=0.15$. This leads to a correction of $I\simeq 0.96I_0$ and $\beta\simeq 0.997\beta_0$. Hence the estimation becomes $I\beta\eta_\mathrm{c}=0.83$ ($I=0.87$, $\beta=0.97$). This correction is accurate in the weak-coupling regime \mbox{$2g<\kappa$} \cite{smith2017}. For the parameters used in this estimation we have $2g/\kappa\simeq 0.8$. See Fig.~\ref{sivplots} for estimations using other $Q$ and $\Delta_\text{q}$ values.

For applications in quantum information processing, it is necessary to have near-unity indistinguishability for high fidelity of quantum operations. This can be accomplished by spectrally filtering the broad emission from the hybrid cavity, which sacrifices efficiency. For an approximately-Markovian source, the $I\beta$ product after spectral filtering cannot exceed the $I\beta$ product from the source \cite{grange2015cavity}; hence, at best it remains constant. By assuming an outcoupling of $0.81$ \cite{gurlek2018manipulation} and an initial value of $I\beta\eta_\mathrm{c}=0.83$, this source could be capable of providing near-unity indistinguishability with a single-photon emission efficiency as high as $0.67$. This would be comparable to state-of-the-art semi-conductor sources that operate at low temperature \cite{ding2016demand, somaschi2016}.

Although we only discussed the SiV$^-$ center as a candidate emitter, there are other emitters that have potential. A few other diamond defects might have narrow homogeneous linewidths at room temperature, such as the N3, H2, H1b, and H1c defects \cite{stoneham2009b}. There is also evidence that the infrared transition of the NV$^-$ center is weakly coupled to phonons \cite{rogers2008, doherty2014, rogers2015} and could have a very narrow homogeneous linewidth with a relatively small spectrally-separated PSB \cite{rogers2008}. In addition, the neutral silicon vacancy (SiV$^0$) center shows a promising combination of optical and spin properties \cite{green2017, rose2018observation}. Its symmetry and electronic configuration allow it to exhibit stable optical properties and a long spin coherence time, combining the best aspects of the SiV$^-$ and NV$^-$ centers, respectively. Moreover, it has a reported Debye-Waller factor of $\text{DW}>0.9$---a single-defect measurement limited only by the noise floor \cite{rose2018observation}. If the spin properties of the SiV$^0$ center can be made comparable to those of the NV$^-$ center, it could function (in combination with a hybrid plasmonic cavity) as a spin-photon interface for a room-temperature solid-state quantum network.

An interesting alternative approach could be to use proposed pure-dielectric ultrasmall cavities \cite{robinson2005,hu2016,choi2017}. These cavities should maintain high quality factors ($Q\simeq 10^6$) with mode volumes as small as $7\times 10^{-5}\lambda^3$ providing unprecedented Purcell factors without being affected by plasmon quenching \cite{choi2017}. These high-Q ultrasmall cavities would need to be used with very narrow emitters, since the cavity must decay faster than the pure-dephasing rate ($\kappa>\gamma^\star$). In particular, single rare-earth ions are known for having very narrow but dim lines \cite{perrot2013}. These quantum emitters can contain very phonon-resistant transitions (such as the $^5$D$_0\rightarrow ^7$F$_0$ transition in the europium(III) ion \cite{binnemans2015}) that might remain quite narrow at room temperature and exhibit a very small phonon sideband. Such a pure-dielectric ultrasmall cavity could also be useful with broader emitters if its cavity quality factor could be lowered to \mbox{$\sim 10^2$--$10^3$} without sacrificing an increase in the effective mode volume.

We have shown that highly efficient solid-state room-temperature indistinguishable SPSs should be within reach using ultrasmall mode volume cavities, and we have described the most promising regime of operation. For cavities containing plasmonic materials, this regime exists only for emitters that are narrow and bright enough to allow a large effective Purcell enhancement without being dominated by plasmon quenching. In addition, it exists only for emitters with a small PSB that is spectrally well-separated from its ZPL. In particular, a nanodiamond SiV$^-$ defect combined with a hybrid plasmonic-Fabry-P\'{e}rot cavity appears to be exceptionally promising. Room-temperature indistinguishable SPSs would be a significant advance for quantum technology, while also helping to answer fundamental questions about the physical regimes in which quantum phenomena can be observed \cite{plumhof2014, kumar2016}.

\subsection{Generalized Purcell factor}
\label{chapter3:generalizedpurcell}

The above four sections were published in Ref.~\ref{wein2018plasmonics}. Here, I would like to make a few brief comments to connect some of the results in \ref{wein2018plasmonics} to the content presented in section \ref{chapter1:cavityQED}. First, the expression for intrinsic cavity efficiency, or brightness, presented in section \ref{chapter3:indistinguishabilityquenching} can be written in terms of the more general definition of cavity-emitter population transfer rate $R$ given in section \ref{chapter1:cavityQED}. This is
\begin{equation}
    \beta = \frac{R\kappa}{\gamma\kappa + R(\gamma+\kappa)},
\end{equation}
which holds for all regimes of the purely-dephased Jaynes-Cummings model. This result was also derived using a retarded Green's function approach in Ref.~\cite{grange2015cavity}. Second, the brightness for a two-level system with a Purcell factor $F_\mathrm{p}$ is also defined for all parameter regimes as $\beta=F_\mathrm{p}/(F_\mathrm{p}+\eta_\mathrm{r}^{-1})$, where $\eta_\mathrm{r}=\gamma_\mathrm{r}/\gamma$ is the radiative efficiency of the emitter. By equating these results, we can write a generalized Purcell factor that holds in all the previously discussed regimes, including the strong-coupling regime:
\begin{equation}
\label{chapter3eq:purcellsaturationfactor}
    F_\mathrm{p}=\frac{R\kappa}{\gamma_\mathrm{r}(\kappa+R)}.
\end{equation}
This can be written in terms of cavity quality factor $Q$ and $V$ as in Eq.~(\ref{chapter1eq:purcellfactor}) but with a slightly modified inhibition factor $F_\text{inh}$ that now exactly takes into account the Purcell saturation in the strong coupling regime:
\begin{equation}
    F_\text{inh}=\frac{R\kappa}{\gamma C(\kappa+R)},
\end{equation}
where $C=4g^2/\kappa\gamma$ is the uninhibited cavity cooperativity and
\begin{equation}
\label{chapter3eq:trueRfactor}
    R=\frac{4g^2(\kappa+\Gamma)}{(\kappa+\Gamma)^2+4\Delta^2}
\end{equation}
depends on $\Delta$ and $g$, as well as $\gamma^\star$ through $\Gamma$. Taking into account both the saturation and pure dephasing inhibition of the Purcell factor in the critical regime, we can now see that the timescale for excitation must be smaller than $(\gamma+F_\mathrm{p}\gamma_\mathrm{r})^{-1}\simeq (F_\mathrm{p}\gamma_\mathrm{r})^{-1}$ to avoid significant re-excitation. For the scenario described in Fig.~\ref{quenchingcouplingind} for the SiV$^-$ center, we have $(F_\mathrm{p}\gamma_\mathrm{r})^{-1}\simeq 60$~fs, which is almost two orders of magnitude faster than what has been demonstrated for solid-state systems \cite{liu2018high}\ref{ollivier2020g2hom} and may only be achievable for multi-level systems using chirped pulse engineering \cite{kumar2013ultrafast,kaldewey2017demonstrating}.

Third, because this generalized Purcell factor holds in all regimes, we can use it to estimate the indistinguishability of photons in the critical regime. Knowing that, for a purely dephased emitter without a cavity, the indistinguishability is $I=\gamma/\Gamma$, we can apply the generalized Purcell factor to get
\begin{equation}
    I \simeq \frac{F_p\gamma_\mathrm{r}+\gamma}{F_p\gamma_\mathrm{r}+\Gamma} = \frac{\gamma(\kappa+R)+R\kappa}{\Gamma(\kappa+R)+R\kappa},
\end{equation}
which turns out to be a decent estimation. However, it underestimates the indistinguishability near the critical point $R=\kappa=\Gamma$ when moving towards the cavity funneling regime and along the edges of the critical regime compared to the more accurate first-order expression derived in section \ref{chapter2:emitterquenching}.

\section{Hong-Ou-Mandel interference}
\label{chapter3:HOM}

In section \ref{chapter1:indistinguishability}, I introduced the quantities of mean wavepacket overlap and indistinguishability for pulsed systems using Hong-Ou-Mandel (HOM) interference of single-photon states. In this section, I will discuss the relationship between indistinguishability, mean wavepacket overlap, and the integrated intensity correlation for two specific scenarios. The first scenario occurs when one photon of the two-photon component causing a non-zero $g^{(2)}$ is approximately unentangled with the desired single-photon signal. This `separable noise model' is highly applicable to state-of-the-art sources, as demonstrated in the experimental papers of Refs. \ref{ollivier2020g2hom} and \ref{ollivier2020reproducibility}. For the second scenario, I will further analyze this relationship for states produced by a pulsed purely-dephased two-level emitter. I will also introduce the notion that a HOM indistinguishability measurement is actually just a special case of a self-homodyne measurement. The broader picture of self-homodyne measurements expands the photonic state properties that can be probed using an identical optical setup to an indistinguishability measurement. This includes quantifying the quantum properties of photonic states that carry photon number coherence or have two-mode entanglement, which is a topic explored in section \ref{chapter3:photonicstate} related to Ref.~\ref{loredo2020deterministic}.

\subsection{Separable noise model}
\label{chapter3:separablenoisemodel}

\begin{figure}
    \centering
    \hspace{-75mm}(a)\hspace{65mm}(b)
    \includegraphics[width=\textwidth]{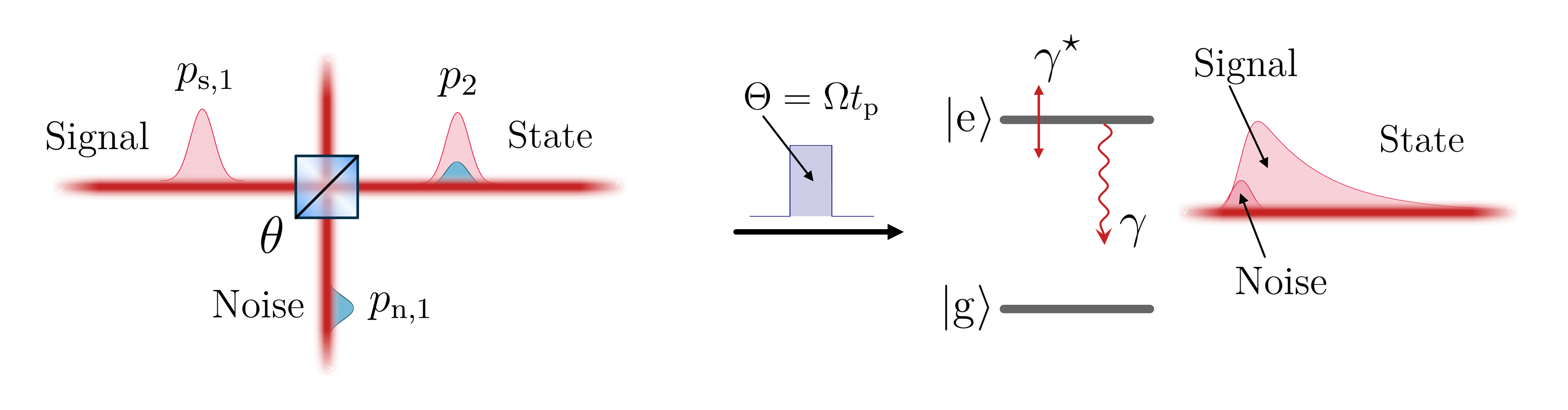}
    \caption[Two specific models for an imperfect single-photon source.]{\small\textbf{Two specific models for an imperfect single-photon source.} (a) The separable noise model where the nonzero two-photon component is composed of a perfect single photon `signal' combined with a `noise' single photon. The state is constructed using a beam-splitter interaction with a variable ratio controlled by the parameter $\theta$. In addition, both the signal and the noise may not be perfectly efficient ($p_{\mathrm{n},1},p_{\mathrm{s},1}\leq 1$). (b) A single-photon produced by the pulsed excitation of a purely-dephased two-level emitter. The pulse length $t_\mathrm{p}$ compared to the system lifetime $1/\gamma$ determines the amount of multi-photon contribution to the desired single-photon state.}
    \label{chapter3:specificmodels}
\end{figure}

We now develop a theoretical model to describe the visibility of HOM interference for a specific type of imperfect single photon. This imperfect single photon source is modeled by adding a small amount of `noise' to an ideal `signal' single photon using a beam splitter interaction, as shown in Fig.~\ref{chapter3:specificmodels}~(a). Here, we assume that the noise is separable and exhibits no entanglement with the single photon. For simplicity, we also model the noise by another single photon, which should be valid in the limit of a weak noise field or, equivalently, when $g^{(2)}$ is small. The approach and results in this section are derived from the material presented in the supplementary of \ref{ollivier2020g2hom}.
	
The initial state of the signal ($\mathrm{s}$) and noise ($\mathrm{n}$) is given by $\hat{\varrho} =\hat{\varrho}_\mathrm{s} \otimes \hat{\varrho}_\mathrm{n}  $ where $\hat{\varrho}_i=p_{i,0}\ket{0}\bra{0}+p_{i,1}\hat{\varrho}_{i,1}$ and $    \hat{\varrho}_{i,1} = \iint \xi_i(t,t^\prime) \bu _i (t)\ket{0} \bra{0} \bd _i (t^\prime) dt dt^\prime$,
where $\xi_i(t,t^\prime)$ is the normalized single-photon temporal density wavefunction for $i \in \{ \mathrm{s},\mathrm{n}\} $ (recall section \ref{chapter2:temporaldensityfunctions}). The imperfect photon in the output transmission mode of the beam splitter $\bd$ is obtained by tracing out the reflected loss mode after applying the beam splitter relation, so that $\hat{\varrho} = \mathrm{Tr}_\text{loss}(\hat{\varrho}_\mathrm{s} \otimes \hat{\varrho}_\mathrm{n}) $, where 
\begin{equation}
  \begin{pmatrix} \bd(t) \\ \bd_\text{loss}(t) \end{pmatrix} =  \begin{pmatrix} \cos\theta & -\sin\theta \\ \sin\theta & \cos \theta \end{pmatrix} \begin{pmatrix} \bd_\mathrm{s}(t) \\ \bd_\mathrm{n}(t) \end{pmatrix}.
\end{equation}

The total state of the imperfect single photon can then be written as $\hat{\varrho} = p_{0}\ket{0}\bra{0} + p_{1}\hat{\varrho}_{1} +p_{2}\hat{\varrho}_{2} $. From this, we can write $g^{(2)}$ and $\mu$ using the photon number probability relations $\mu = p_1 + 2p_2$ and $g^{(2)}=2p_2/\mu^2$ derived in section \ref{chapter2:photonstatistics}. For the average photon number, we have $\mu = p_{\mathrm{s},1}\cos^2\theta+p_{\mathrm{n},1}\sin^2\theta$ and $g^{(2)}$ is given by $ \mu^2g^{(2)} = 2p_{\mathrm{s},1}p_{\mathrm{n},1}(1+M_\mathrm{sn})\cos^2\theta\sin^2\theta$,
where
\begin{equation}
\label{chapter2eq:meanwavepacketoverlapdifferent}
M_\mathrm{sn} = \iint\xi_\mathrm{s} (t,t^\prime) \xi_\mathrm{n}^*(t,t^\prime)dt dt^\prime    
\end{equation}
is the mean wavepacket overlap of the single photon and noise, and $\theta$ along with $p_{\mathrm{n},1}$ quantifies the amount of noise. Note that $M_\mathrm{sn}$ is a real value as a consequence of the Hermitian property of $\xi_\mathrm{s}$ and $\xi_\mathrm{n}$. However, one could restrict $t\leq t^\prime$ and compute the integral over $2\text{Re}(\xi_\mathrm{s} (t,t^\prime) \xi_\mathrm{n}^*(t,t^\prime))$ to obtain the same result. This case is useful when evaluating $\xi$ using the time-ordered convention described in section \ref{chapter2:temporaldensityfunctions}.

The total mean wavepacket overlap of the imperfect photon with itself is given by
\begin{equation}
\label{separableoverlap}
\begin{aligned}
\mu^2M &= \iint\left|\braket{\bu(t)\bd(t^\prime)}\right|^2dt^\prime dt\\
&= p_{\mathrm{s},1}^2I\cos^4\theta + p_{\mathrm{n},1}^2 M_\mathrm{n}\sin^4\theta + 2p_{\mathrm{s},1}p_{\mathrm{n},1}M_\mathrm{sn}\cos^2\theta\sin^2\theta,
\end{aligned}
\end{equation}
where $M_\mathrm{s}=\iint\left|\xi_\mathrm{s}(t,t^\prime)\right|
    ^2 dtdt^\prime = \text{Tr}\left[\hat{\varrho}_{\mathrm{s},1}^2\right]=I$
quantifies the intrinsic single-photon indistinguishability or single-photon trace purity \cite{fischer2018particle, trivedi2020generation} and $M_\mathrm{n}=\iint |\xi_\mathrm{n}(t,t^\prime)|^2dtdt^\prime$ is the indistinguishability of the noise photon with itself. It is worth emphasizing that Eq.~(\ref{separableoverlap}) holds only when the states of modes $\bd_\mathrm{s}$ and $\bd_\mathrm{n}$ are uncorrelated and carry no number coherence, which is the case for $\varrho$ as defined above.

We can now reparametrize the expressions for $\mu$, $g^{(2)}$, and $M$ by defining an effective noise parameter $\vartheta$ so that $\cos^2\!\vartheta=(p_{\mathrm{s},1}\cos^2\theta)/\mu$ and $\sin^2\!\vartheta=(p_{\mathrm{n},1}\sin^2\theta)/\mu$. The fact that this reparametrization exists stems from the independence of $M$ and $g^{(2)}$ from photon loss. It also implies that the fundamental quantity affecting the photon statistics of this imperfect single photon model is $\vartheta$, which depends on both the beam splitter angle $\theta$ and the relative input intensities through $p_{\mathrm{s},1}$ and $p_{\mathrm{n},1}$. Knowing that $V_\text{HOM}=M-g^{(2)}$ (see section \ref{chapter1:indistinguishability}), the visibility and $g^{(2)}$ in terms of the noise parameter $\vartheta$ are
\begin{equation}
\begin{aligned}
    V_\text{HOM}(\vartheta) &= I\cos^4\vartheta+M_\mathrm{n}\sin^4\vartheta -2\cos^2\vartheta\sin^2\vartheta\\
    g^{(2)}(\vartheta) &= 2(1+M_\mathrm{sn})\cos^2\vartheta\sin^2\vartheta.
\end{aligned}
\end{equation}

From an experimental point of view, we are interested in the relationship between $V_\text{HOM}$ and $g^{(2)}$ in the limit that $g^{(2)}$ is small. To this end, let us analyze the slope and intercept of the parametric curve formed by $\{g^{(2)}(\vartheta),V_\text{HOM}(\vartheta)\}$. The solution for the intercept is clear since $g^{(2)}(\vartheta)=0$ implies $\vartheta=0$ and $V_\text{HOM}(0) = I$. To solve for the slope at small $\vartheta$, we have
\begin{equation}
\begin{aligned}
    \lim_{\vartheta\rightarrow 0}\frac{dV_\text{HOM}(\vartheta)}{dg^{(2)}(\vartheta)}
    &= -\frac{1+I}{1+M_\mathrm{sn}}.
\end{aligned}
\end{equation}
Hence, the HOM visibility for small $g^{(2)}$ is given by
\begin{equation}
\label{vhomRT}
    V_\text{HOM} = I-\left(\frac{1+I}{1+M_\mathrm{sn}}\right)g^{(2)}.
\end{equation}
This implies that sources with identical noise where $M_\mathrm{sn}=I$ (i.e. some heralded sources from SPDC) should show a higher HOM visibility compared to sources with distinguishable noise ($M_\mathrm{sn}=0$) for the same $g^{(2)}$ and single-photon indistinguishability $I$. Note that for identical noise ($M_\mathrm{sn}=I$), we recover the usual estimate for single-photon indistinguishability $I=V_\text{HOM}+g^{(2)}=M$ from the measured values. For the distinguishable noise case ($M_\mathrm{sn}=0$) we obtain the correction $I=(V_\text{HOM}+g^{(2)})/(1-g^{(2)})=M/(1-g^{(2)})$, which is larger by a factor of $(1-g^{(2)})^{-1}$ compared to the identical noise case.

The relationship between $V_\text{HOM}$ and $g^{(2)}$ derived in this section has been confirmed experimentally in a collaboration with the group of Prof. Pascale Senellart for the cases $M_\mathrm{sn}=I$ and $M_\mathrm{sn}=0$ \ref{ollivier2020g2hom}. In that work, we also found that pulsed single-photon sources based on semi-conductor QDs, in both trion and exciton configurations, follow very closely to the $M_\mathrm{sn}= 0$ case when $g^{(2)}$ is below 0.1. Furthermore, in Ref.~\ref{ollivier2020reproducibility}, we applied Eq.~(\ref{vhomRT}) using $M_\text{sn}=0$ to 15 different QD sources with varying $g^{(2)}$ and $V_\text{HOM}$ values. The resulting 15 estimates for $I$ showed a smaller deviation than that of $M=V_\text{HOM}-g^{(2)}$, suggesting that the estimated value of $I$ is indeed less dependent on $g^{(2)}$ than $M$ is. These experimental results are particularly interesting because `noise' from re-excitation processes should $\emph{not}$, in general, be separable from the single photon signal because of entanglement due to emission time jitter. Hence, one might expect these sources to violate the primary assumption made to derive Eq.~(\ref{vhomRT}). In the next section, I will look deeper into this relationship for a pulsed two-level system to discuss under which regimes the separable noise model is actually valid for single-photon sources where the nonzero $g^{(2)}$ arises from re-excitation.

\subsection{Pulsed two-level emitter}
\label{chapter3:pulsedemitterHOM}

In this section, I consider the relationship between visibility and $g^{(2)}$ for the specific case where the nonzero $g^{(2)}$ is caused only by re-excitation of the source (see Fig.~\ref{chapter3:specificmodels}~(b)). I model the emission using a driven two-level system that may experience pure dephasing. Then, I compute the HOM visibility and $g^{(2)}$ using a Markovian master equation. The results shown in this section hold in the far-field of the dipole radiation from a two-level emitter where the proportionality relationship $\bd\propto\sigd$ is valid \cite{kiraz2004quantum}. If the two-level system is inside a cavity, as described in section \ref{chapter1:cavityQED}, we must be sufficiently far into the bad-cavity regime where adiabatic elimination of the cavity mode is valid (recall section \ref{chapter1:1datom}). Otherwise, this effective two-level model neglects important emitter-cavity non-Markovian effects such as Rabi oscillations and the drive-dependent Purcell factor \cite{gustin2018pulsed}. For this two-level model, the correlation functions of the waveguide can be computed using the proportionality relation $\bd=\sqrt{\eta_\mathrm{r}\gamma}\sigd$ where $\eta_\mathrm{r}=\eta_\mathrm{c}F_\mathrm{p}/(1+F_\mathrm{p})$ now depends on the Purcell factor of the cavity. However, the important quantities computed in this section are loss-independent and so I will consider $\eta_\mathrm{r}=1$ so that $\bd=\sqrt{\gamma}\sigd$ for simplicity.

I begin by writing the Liouville superoperator for a purely-dephased classically-driven two-level system:
\begin{equation}
\label{chapter3:liouvilleTLS}
    \mathcal{L}=-\frac{i}{\hbar}\mathcal{H}(t)+\gamma\mathcal{D}(\sigd)+2\gamma^\star\mathcal{D}(\sigd^\dagger\sigd),
\end{equation}
where $\hat{H}(t)=(\hbar\Omega(t)/2)(\sigd+\sigd^\dagger)$ and $\mathcal{H}\hat{\rho}=[\hat{H},\hat{\rho}]$. It is then possible to evaluate the total intensity $\mu$, mean wavepacket overlap $M$, and intensity correlation $g^{(2)}$ for any pulse shape $\Omega(t)$ using time-dependent integration methods. However, the quantities $M$ and $g^{(2)}$ require the evaluation of two-time correlation functions, which can be very time-consuming to compute when exploring a very large parameter set. Alternatively, we can assume that the pulse is square in shape and divide the Hamiltonian into two separate time-independent parts. For a square pulse with temporal width $t_\mathrm{p}$ and Rabi frequency $\Omega$ beginning at $t=0$, we can write the propagation superoperator as 
\begin{equation}
\label{chapter3eq:piecewisetimeind}
    \mathcal{U}(t^\prime,t)=\left\{
    \begin{aligned}
    &\mathcal{U}_\mathrm{p}(t^\prime,t)&&t^\prime\leq t_\mathrm{p}\\
    &\mathcal{U}_\mathrm{d}(t^\prime,t_\mathrm{p})\mathcal{U}_\mathrm{p}(t_\mathrm{p},t)&&t\leq t_\mathrm{p}\leq t^\prime\\
    &\mathcal{U}_\mathrm{d}(t^\prime,t)&&t>t_\mathrm{p}\\
    \end{aligned}\right.,
\end{equation}
where $\mathcal{U}_\mathrm{p}(t^\prime,t)=e^{(t^\prime-t)\mathcal{L}_\mathrm{p}}$ is the propagation superoperator when the pulse is on, corresponding to $\mathcal{L}_\mathrm{p}=\mathcal{L}(\Omega(t)=\Omega)$, and $\mathcal{U}_\mathrm{d}(t^\prime,t)=e^{(t^\prime-t)\mathcal{L}_\mathrm{d}}$ is the propagation superoperator when only decay is occurring, corresponding to $\mathcal{L}_\mathrm{d}=\mathcal{L}(\Omega(t)= 0)$.

For the piecewise evolution, the two-time integrals for $M$ and $g^{(2)}$ can be divided into three parts: (1) $0\leq t\leq t_\mathrm{p}$ and $t\leq t^\prime \leq t_\mathrm{p}$, (2) $0\leq t \leq t_\mathrm{p}$ and $t_\mathrm{p} < t^\prime <\infty$, and (3) $t_\mathrm{p}< t<\infty$ and $t \leq t^\prime < \infty$. For example, $M$ can be evaluated by
\begin{equation}
\begin{aligned}
\label{chapter3eq:piecewisetwotime}
        \frac{\mu^2}{2\gamma^2}M &=\int_0^{t_\mathrm{p}}\int_t^{t_\mathrm{p}}\left|\braket{\hat{\sigma}^\dagger(t^\prime)\hat{\sigma}(t)}_\mathrm{p}\right|^2dt^\prime dt+
        \int_0^{t_\mathrm{p}}\int_{t_\mathrm{p}}^{\infty}\left|\braket{\hat{\sigma}^\dagger(t^\prime)\hat{\sigma}(t)}_\mathrm{dp}\right|^2dt^\prime dt\\
        &\hspace{5mm}+
      \int_{t_\mathrm{p}}^{\infty}\int_t^\infty\left|\braket{\hat{\sigma}^\dagger(t^\prime)\hat{\sigma}(t)}_\mathrm{d}\right|^2dt^\prime dt,
\end{aligned}
\end{equation}
where using the superoperators $\mathcal{S}\hat{\rho}=\sigd\hat{\rho}$ and $\mathcal{R}\hat{\rho}=\hat{\rho}\sigu$ we have
\begin{equation}
\begin{aligned}
    \braket{\sigu(t^\prime)\sigd(t)}_\mathrm{p} &= \tr{\mathcal{R}\mathcal{U}_\mathrm{p}(t^\prime,t)\mathcal{S}\mathcal{U}_\mathrm{p}(t,0)\hat{\rho}(0)}\\
    \braket{\sigu(t^\prime)\sigd(t)}_\mathrm{dp} &= \tr{\mathcal{R}\mathcal{U}_\mathrm{d}(t^\prime,t_\mathrm{p})\mathcal{U}_\mathrm{p}(t_\mathrm{p},t)\mathcal{S}\mathcal{U}_\mathrm{p}(t,0)\hat{\rho}(0)}\\
    \braket{\sigu(t^\prime)\sigd(t)}_\mathrm{d} &= \tr{\mathcal{R}\mathcal{U}_\mathrm{d}(t^\prime,t)\mathcal{S}\mathcal{U}_\mathrm{d}(t,t_\mathrm{p})\mathcal{U}_\mathrm{p}(t_\mathrm{p},0)\hat{\rho}(0)}.
\end{aligned}
\end{equation}

\begin{figure}
    \centering
    \hspace{-5cm}(a)\hspace{7cm}(b)\\
    \includegraphics[scale=0.25]{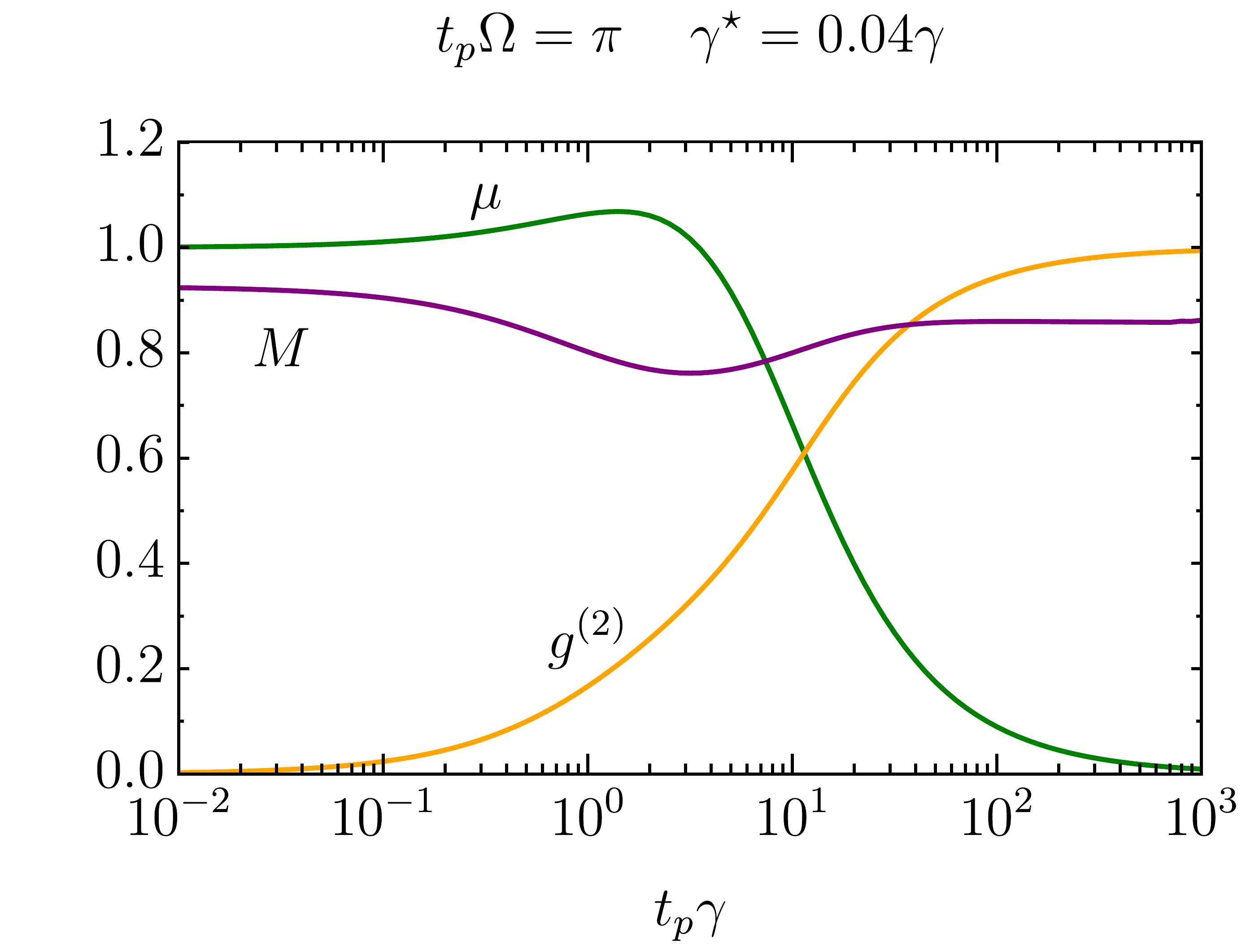}
    \includegraphics[scale=0.25]{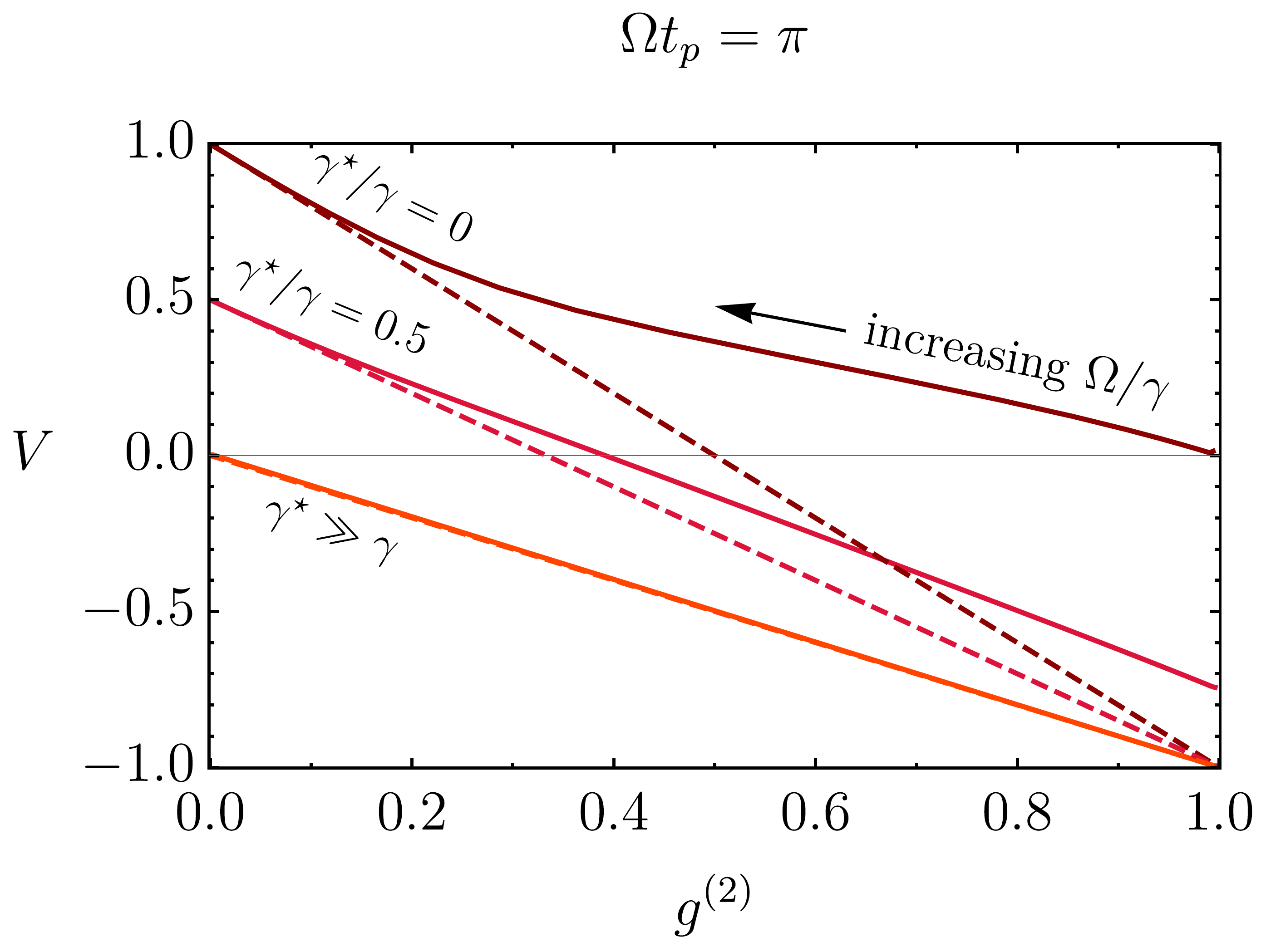}
    \caption[Single-photon figures of merit for a purely-dephased two-level emitter driven by a square $\pi$ pulse.]{\small\textbf{Single-photon figures of merit for a purely-dephased two-level emitter driven by a square $\pi$ pulse.} (a) Single-photon quality quantified by the mean photon number $\mu$, second-order correlation $g^{(2)}$, and mean wavepacket overlap $M$ computed using the proportionality relation $\bd=\sqrt{\gamma}\sigd$. (b) The Hong-Ou-Mandel visibility $V=V_\text{HOM}=M-g^{(2)}$ plotted against $g^{(2)}$ for different pure dephasing values $\gamma^\star$. The dashed lines indicate the slope in the limit of small $g^{(2)}$ and correspond to the $V_\text{HOM} = I - (1+I)g^{(2)}$ result from the previous section.}
    \label{fig:pipulse_pipulse}
\end{figure}

\begin{figure}
    \centering
    \includegraphics[scale=0.3]{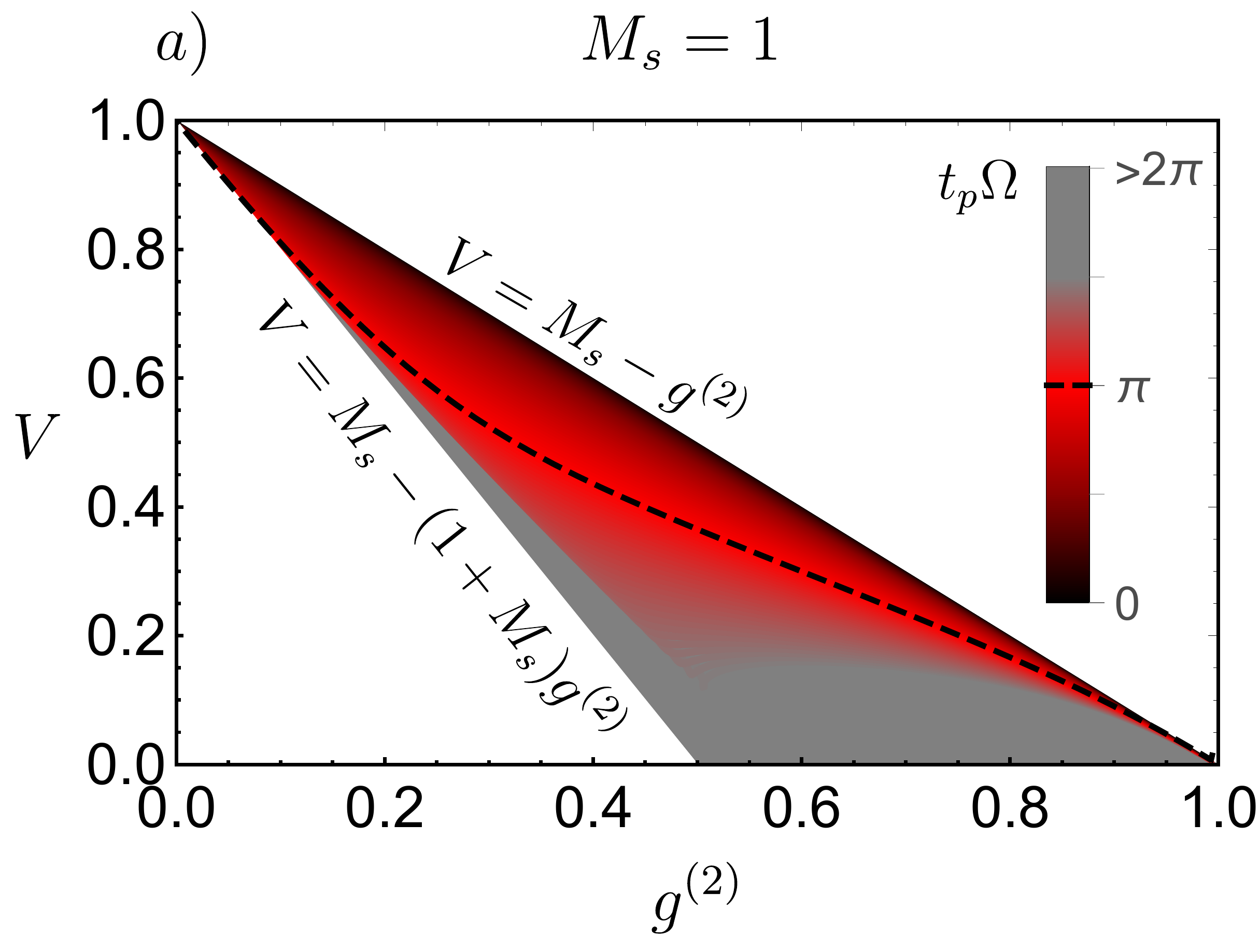}
    \includegraphics[scale=0.3]{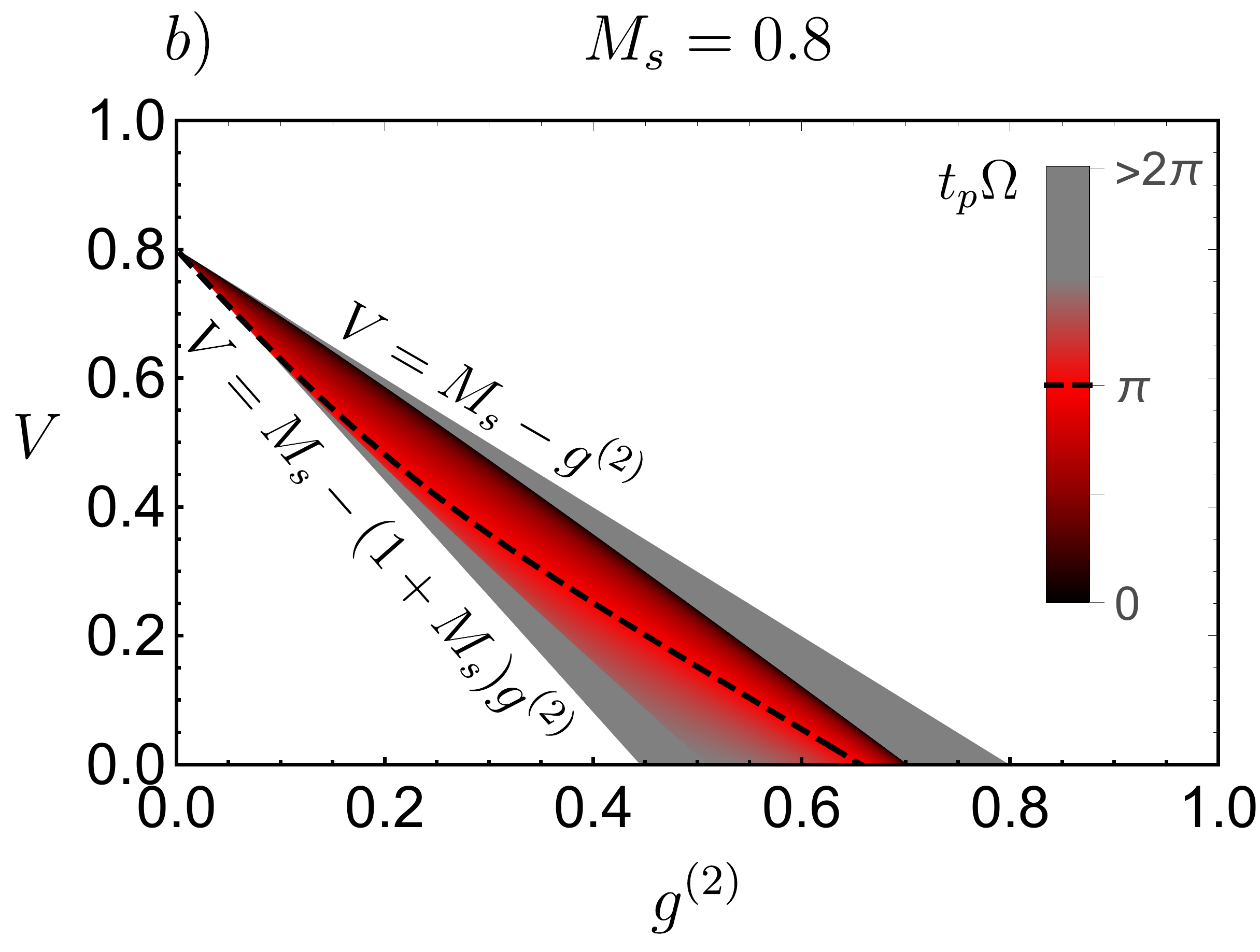}
    \caption[The relationship between Hong-Ou-Mandel visibility and the integrated intensity correlation for a purely-dephased two-level emitter.]{\small\textbf{The relationship between Hong-Ou-Mandel visibility and the integrated intensity correlation for a purely-dephased two-level emitter.} Numerically exact computed positive values of $V=V_\text{HOM}=M-g^{(2)}$ and $g^{(2)}$ reachable using a square pulse with width $t_\mathrm{p}$ and Rabi frequency $\Omega$ for a) no pure dephasing and b) pure dephasing giving $M_s=I=0.8$ corresponding to $\gamma^\star=\gamma/8$. For small $g^{(2)}$, pulses near an area of $\pi$ approach the lower bound of $V_\text{HOM}$ while pulses near an area of $0$ approach the upper bound. Points corresponding to smaller pulse areas are plotted on top of those corresponding to larger pulse areas if there is overlap.}
    \label{fig:g2HOM_tls}
\end{figure}

By evaluating the dynamics in this piecewise time-independent way, it is possible to evaluate exact analytic expressions for $\mu$ and $g^{(2)}$, although the expressions are quite large and do not provide much physical insight. It may also be possible to evaluate an exact expression for $M$ but the squared integrand complicates the solution. Regardless, we are still able to evaluate exact analytic expressions for the two-time correlation functions used to compute $M$, which can drastically reduce the computation time required to solve $M$ numerically. Fig.~\ref{fig:g2HOM_tls}~(a) shows $\mu$, $g^{(2)}$, and $M$ for a two-level system with a pure dephasing rate of $\gamma^\star=0.04\gamma$ driven by a $\pi$ pulse where $\Theta = t_\mathrm{p}\Omega = \pi$.
With the piecewise evaluation approach, we can also efficiently analyze the shape of the $\{g^{(2)},V_\text{HOM}\}$ curve when varying different pulse parameters. For the $\pi$ pulse  case in the limit that $t_\mathrm{p} \rightarrow 0$, the slope of the $\{g^{(2)},V_\text{HOM}\}$ curve can indeed be seen to approach $-(1+I)$ where $I=\gamma/\Gamma$ is the value of $M$ at $g^{(2)}=0$ (see Fig.~\ref{fig:pipulse_pipulse}). This corresponds to the distinguishable noise scenario in the previous section, as was found experimentally. However, if $\Theta$ is not fixed, the shape and slope of the parameteric curve $\{g^{(2)},V_\text{HOM}\}$ actually depends on which parameter is being varied, $\Omega$ or $t_\mathrm{p}$. Interestingly, regardless of the parameters $\Omega$ and $t_\mathrm{p}$, we can find that the value of $V_\mathrm{HOM}$ is always bounded between the straight lines $V_\text{HOM} = I - (1+I)g^{(2)}$ and $V_\text{HOM} = I - g^{(2)}$ (see Fig.~\ref{fig:g2HOM_tls}), just as we found from the separable noise model. Thus we can postulate that
\begin{equation}
    V_\text{HOM}+g^{(2)} \leq I \leq \frac{V_\text{HOM}+g^{(2)}}{1-g^{(2)}}
\end{equation}
is true for any pulse shape applied to a driven emitter following a two-level model. This result is quite useful since it allows one to estimate both a lower bound and an upper bound on $I$ by measuring $V_\text{HOM}$ and $g^{(2)}$ only once.

Numerically computing the $\{g^{(2)},V_\text{HOM}\}$ curve gives a good prediction and a useful relation, but it doesn't really give much physical insight into the nature of the emitted wavepacket. Ultimately, it would be nice to understand how the properties of the re-excited photon affects the HOM visibility. Let us begin that endeavour now. 

Recall from section \ref{chapter2:temporaldensityfunctions} that the emitted photonic state density matrix $\hat{\varrho}$ can be decomposed into the photon number subspaces by $\hat{\varrho} = \sum_{n,m}\hat{\varrho}_{n,m}$ where $\hat{\varrho}_{n}=\hat{\varrho}_{n,n}$ is the unnormalized state given that $n$ photons were emitted and $\hat{\varrho}_{n,m}$ quantifies possible coherence between photon number states. The probability of emitting $n$ photons after one pulse is given by $p_n = \text{Tr}\left[\hat{\varrho}_{n}\right]$. In this notation, the $g^{(2)}$ and $\mu$ can be written in terms of $p_n$ using the standard photon statistics relations (see again section \ref{chapter2:photonstatistics}). Similarly, we can decompose the mean wavepacket overlap using conditional correlation functions of the waveguide (see sections \ref{chapter2:conditionalcorrelations} and \ref{chapter2:temporaldensityfunctions}):
\begin{equation}
\begin{aligned}
\label{chapter3eq:mwpodecomp}
    M &= \frac{1}{\mu^2}\sum_{j,k}\iint\braket{\bu(t^\prime)\bd(t)}_j\braket{\bu(t^\prime)\bd(t)}_k^*dt^\prime dt\\
    &=\frac{1}{\mu^2}\sum_{j,k}\mu_j\mu_k M_{j,k},\\
\end{aligned}
\end{equation}
where $\mu_n = n p_n$ and
\begin{equation}
    M_{j,k}= \frac{1}{\mu_j\mu_k}\iint\text{Re}\left[\braket{\bu(t^\prime)\bd(t)}_j\braket{\bu(t^\prime)\bd(t)}_k^*\right]dt^\prime dt,
\end{equation}
is the normalized mean wavepacket overlap between density operators $\hat{\varrho}_k$ and $\hat{\varrho}_j$ in different photon-number subspaces. Here, I have used the property $M_{j,k}=M_{k,j}^*$ to define the intensity normalized overlap $M_{j,k}$ as a real valued quantity.

\begin{figure}
    \centering
    \includegraphics[scale=0.35]{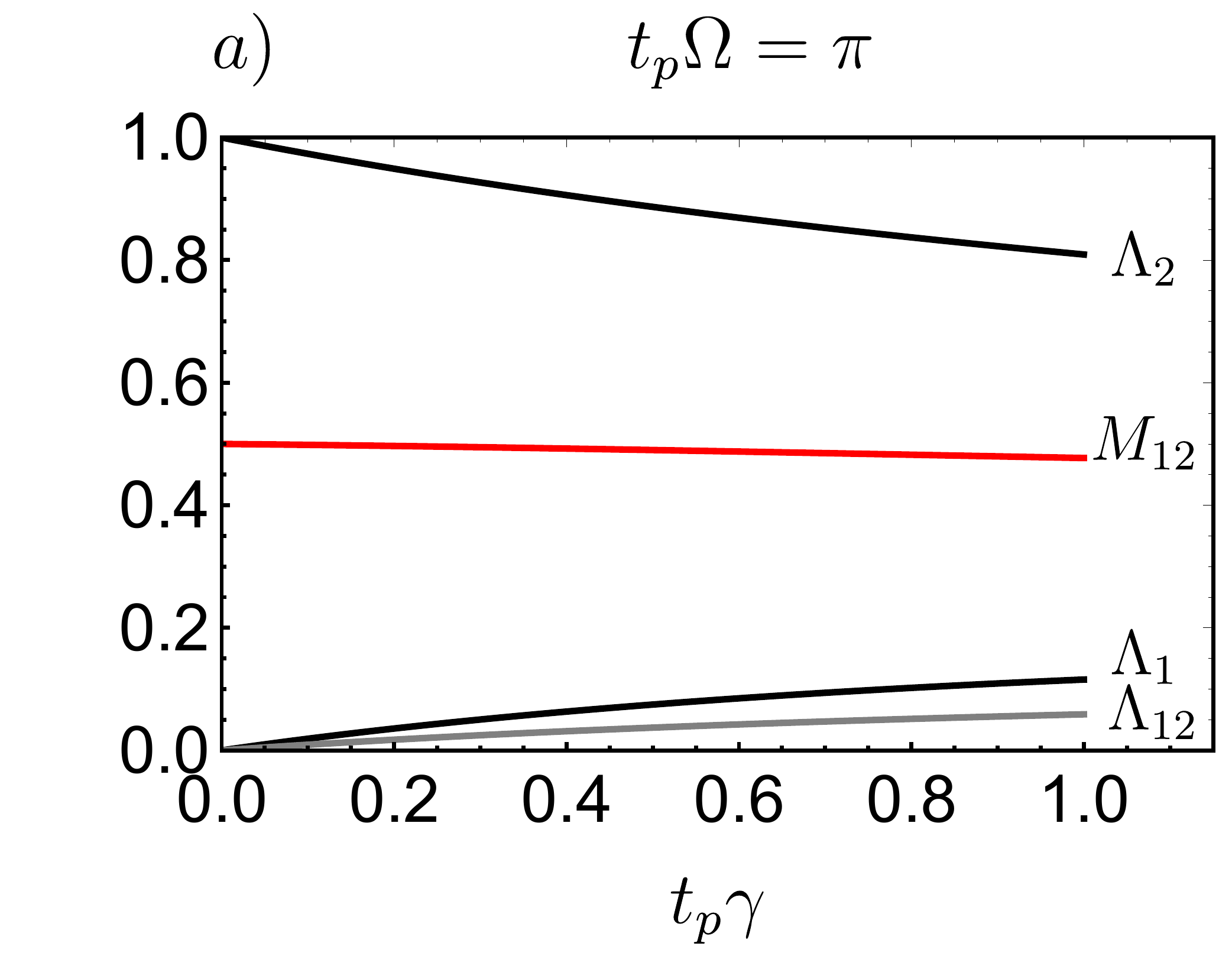}
    \includegraphics[scale=0.35]{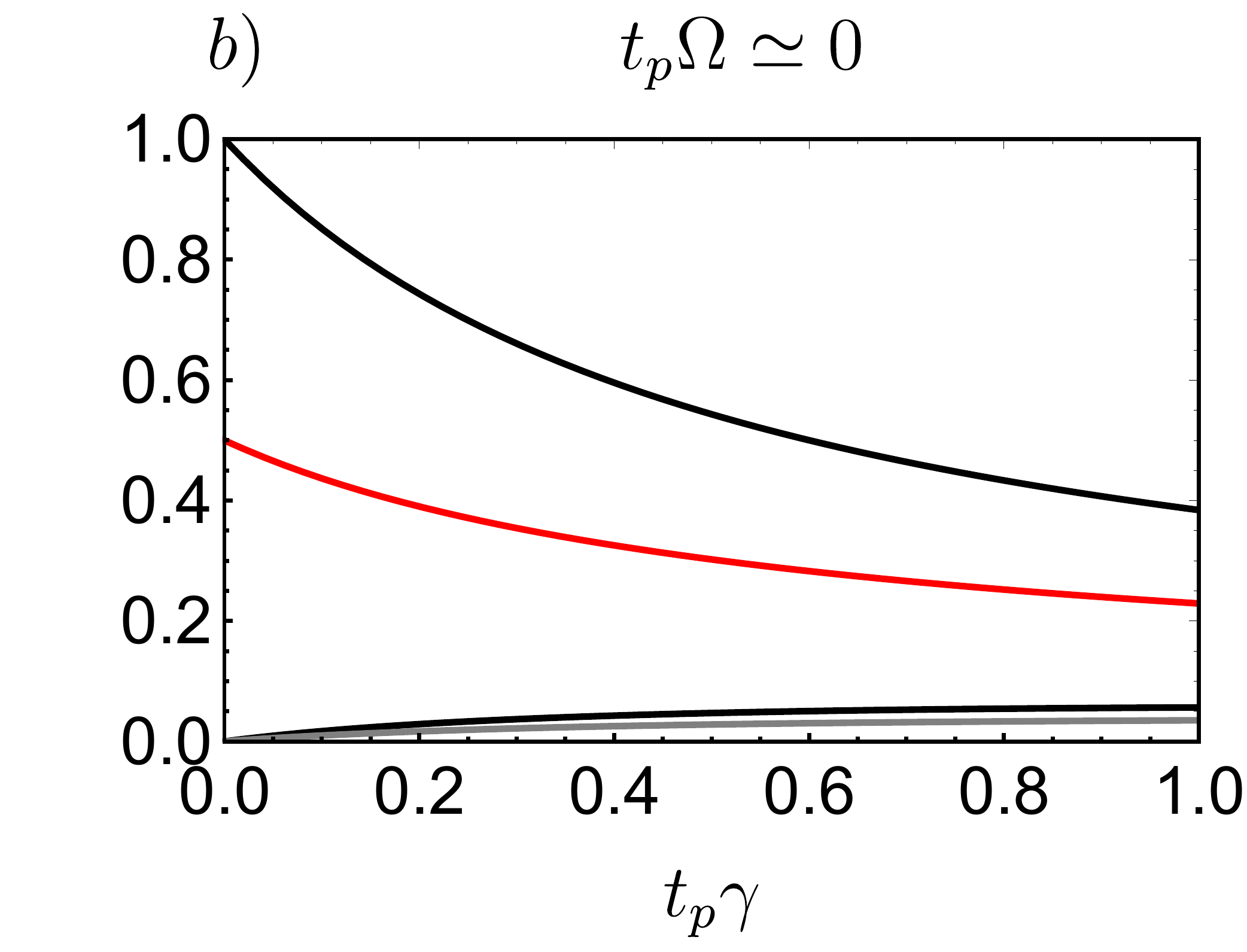}
    \caption[Constituent mean wavepacket overlap integrals illustrating the effect of re-excitation on the single-photon indistinguishability from a purely-dephased emitter.]{\small\textbf{Constituent mean wavepacket overlap integrals illustrating the effect of re-excitation on the single-photon indistinguishability from a purely-dephased emitter.} Components of $M_{1,2} = (\Lambda_1+\Lambda_2+\Lambda_{12})/2$ computed from the temporal density functions when $\gamma^\star=0$ at (a) $\pi$-pulse excitation and (b) in the weak-driving limit. Regardless of $t_\mathrm{p}\Omega$, $M_{1,2}$ converges to $1/2$ from below in the limit that $t_\mathrm{p}\gamma\rightarrow 0$. The curves in this panel correspond to those in panel (a) and follow the same ordering of $\Lambda_{12}\leq\Lambda_{1}\leq M_{1,2}\leq \Lambda_{2}$.}
    \label{fig:m12}
\end{figure}

We can now analyze the overlap in terms of the one- and two-photon temporal density functions introduced in section \ref{chapter2:temporaldensityfunctions}. Using this formalism, we can compute relevant $M_{j,k}$ using the time-ordered normalization convention by
\begin{equation}
\begin{aligned}
    M_{1,1} &= I = \iint\left|\xi_1(t,t^\prime)\right|^2dtdt^\prime\\
    M_{1,2} &=\frac{1}{2}\left(\Lambda_{1}+\Lambda_{2}+\Lambda_{12}\right)\\
\end{aligned}
\end{equation}
and 
\begin{equation}
\label{chapter3eq:lambdaintegrals}
\begin{aligned}
    \Lambda_1 &= \iiint\text{Re}\left[\xi_1(t,t^\prime)\xi_2^*(t,t^{\prime\prime},t^\prime,t^{\prime\prime})\right]dtdt^\prime dt^{\prime\prime}\\
    \Lambda_2 &= \iiint\text{Re}\left[\xi_1(t,t^\prime)\xi_2^*(t^{\prime\prime},t,t^{\prime\prime},t^\prime)\right]dtdt^\prime dt^{\prime\prime}\\
    \Lambda_{12} &= \iiint\text{Re}\left[\xi_1(t,t^\prime)\xi_2^*(t,t^{\prime\prime},t^{\prime\prime},t^\prime)\right]dtdt^\prime dt^{\prime\prime},
\end{aligned}
\end{equation}
where $\Lambda_{1}$ quantifies the overlap of the single photon and the first photon of the two-photon component, $\Lambda_{2}$ quantifies the overlap between the single photon and the second photon of the two-photon component, and $\Lambda_{12}$ is related to how much entanglement exists between the noise and the desired single photon. We can compute these integrals exactly for the square pulse case by applying a photon number decomposition to $\mathcal{U}_\mathrm{p}$ and $\mathcal{U}_\mathrm{d}$ for $\mathcal{J}=\gamma\mathcal{R}\mathcal{S}$ (see section \ref{chapter2:conditionalpropagationsuperoperators}) and then applying Eqs.~(\ref{chapter2eq:onephotondensity}) and (\ref{chapter2eq:twophotondensity}) using the two-level emitter proportionality relation $\bd=\sqrt{\gamma}\sigd$ instead of the cavity-waveguide relation $\bd=\sqrt{\kappa\eta_\mathrm{c}}\ad$. In Fig.~\ref{fig:m12}, we can see that $\Lambda_1$ and $\Lambda_{12}$ approach zero as $t_p$ decreases, regardless of $\Theta$. In addition $\Lambda_2$ approaches 1 when $\gamma^\star=0$. I will discuss the one- and two-photon density functions in much more detail in section \ref{chapter3:photonicstate}. For now, let us use this numerical result to motivate a useful approximation.

The result in Fig.~\ref{fig:m12} is intuitive in the limit $t_\mathrm{p}\gamma\rightarrow 0$ because the first photon of the two-photon component $\xi_2$ $\emph{must}$ have been emitted during the pulse if the emitter was re-excited. Hence, if the pulse is short, the first photon must have a very small temporal overlap with the single-photon component, which is emitted primarily after the pulse (hence $\Lambda_1\rightarrow 0$). The second photon of the two-photon component must also have very little temporal entanglement with the first photon when the pulse is short (hence $\Lambda_{12}\rightarrow 0$), simply because they have very different timescales. Finally, the second photon emitted in the two-photon component must have a very good overlap with the single-photon component, since both are emitted primarily after the pulse and have similar intensity profiles (hence $\Lambda_2\rightarrow 1$).

We can formalize this separability approximation in the fast-pulse limit as $\xi_2(t_1,t_2,t_1^\prime,t_2^\prime)\simeq \xi_\mathrm{n}(t_1,t_1^\prime)\xi_1(t_2,t_2^\prime)$, where $\mathrm{n}$ symbolizes the noise photon (the first photon emitted from $\xi_2$). Then, applying this to Eq.~(\ref{chapter3eq:lambdaintegrals}), we find that $\Lambda_1\simeq 0$, $\Lambda_2\simeq I$, and $\Lambda_{12}\simeq0$, which implies $M_{1,2} = I/2$. Applying this approximation to $M$ gives
\begin{equation}
\begin{aligned}
    M &\simeq \frac{\mu_1^2I+\mu_1\mu_2I+\mu_2^2M_{2,2}}{\mu^2}\\
    &\simeq I\left(1 - p_1g^{(2)}\right).
\end{aligned}
\end{equation}
Finally, in the fast-pulse limit of a driven two-level system, we know that $p_1 \simeq \sin^2\left(\Theta/2\right)$. Thus, for a driven two-level system in the fast-pulse limit and when $g^{(2)}$ is small, we have
\begin{equation}
    V_\text{HOM} = I - \left(1 + I\sin^2\left(\Theta/2\right)\right)g^{(2)}.
\end{equation}
For a $\pi$-pulse, this becomes $V_\text{HOM} = I - (1+I)g^{(2)}$ and for very low power we have $V_\text{HOM} = I - g^{(2)}$. These limits agree with the numerical simulations in Fig.~\ref{fig:g2HOM_tls} and, in particular, when $g^{(2)}$ is small. However, this expression is not accurate for $\Theta$ around $2n\pi$ for integers $n\geq 1$, because $g^{(2)}$ will not be small for these pulse areas \cite{fischer2017signatures}. This high-$g^{(2)}$ regime is within the gray area in Fig.~\ref{fig:g2HOM_tls}.

The results presented in this section are only valid for HOM interference measurements that are normalized by the total squared intensity using an interferometer that is completely phase averaged during the timescale of the measurement. This is because, as I will show in the next section, a photonic state with photon number coherence can alter the HOM visibility if it is normalized by the uncorrelated peaks or if the interferometer has a stable phase on the timescale of the measurement. This latter situation allows a HOM setup to perform a self-homodyne measurement of one- and two-photon coherence.

\subsection{Self-homodyne measurements}
\label{chapter3:selfhomodyne}

Hong-Ou-Mandel interference of single photons produced by one emitter can be seen as a special case of a self-homodyne measurement, where two copies of the same photonic state are interfered. Instead of using a local oscillator (LO) as a phase reference, as is the case for classical homodyne detection, we can instead define one of the two interfering states as the reference state. The phase of this reference state is primarily defined by the LO used to excite the emitter(s). Hence, it can serve in place of the LO in the phase measurement. However, in addition to phase fluctuations of the LO, the phase of the self-homodyne reference state is affected by phase fluctuations due to the source environment, such as pure dephasing and spectral diffusion. This makes self-homodyne measurements useful for characterizing the purity of a photonic state, such as indistinguishability \cite{somaschi2016} and photon-number purity \cite{loredo2019generation}.

When performing HOM interference in this way, the individual detector rates can be affected by the phase of the interferometer in a way analogous to classical homodyne detection, provided that the input photonic states carry coherence in the number basis. For a balanced beam splitter with input modes $\bd_1$ and $\bd_2$, the output modes $\bd_3$ and $\bd_4$ can be described by the relation
\begin{equation}
\label{bsrelation}
    \begin{pmatrix}
    \bd_3(t)\\
    \bd_4(t)
    \end{pmatrix}
    =
    \frac{1}{\sqrt{2}}\begin{pmatrix}
    1&-e^{-i\phi}\\
    e^{i\phi}&1
    \end{pmatrix}
    \begin{pmatrix}
    \bd_1(t)\\
    \bd_2(t)
    \end{pmatrix}.
\end{equation}
Here, we take $\phi$ to be the relative phase difference imparted on the two otherwise identical photonic input states. The intensities received at each detector, denoted $\mu^+=\int N_3(t)dt$ and $\mu^-=\int N_4(t)dt$, will fluctuate in opposition by $\mu^{\pm}=\mu[1\pm \lambda^{(1)}\cos(\phi)]$ where $\lambda^{(1)}=\mu^{-1}\!\int\left|\braket{\bd(t)}\right|^2dt$ is the integrated squared magnitude of the one-photon coherence and $\mu=\mu^++\mu^-$. Thus, we can define the pulsed self-homodyne signal to be
\begin{equation}
\label{chapter3eq:Vshdef}
    V_\text{SH} = \frac{\mu^+-\mu^-}{\mu^++\mu^-} = \lambda^{(1)}\cos(\phi).
\end{equation}

The coincident events of the detectors at the output are determined from the two-time intensity correlation between the output fields $ G_\text{HOM}^{(2)}(t,t^\prime)=\braket{\bu_3(t)\bu_4(t^\prime)\bd_4(t^\prime)\bd_3(t)}$. Using the same approach as in section \ref{chapter1:indistinguishability}, but keeping terms arising from coherence, we obtain
\begin{equation}
\begin{aligned}
    2G_\text{HOM}^{(2)}(t,t^\prime)&=
  N(t)N(t^\prime)+G^{(2)}(t,t^\prime)-\left|G^{(1)}(t,t^\prime)\right|^2\\
  &\hspace{5mm}+2\Lambda^{(1,2)}_-(t,t^\prime)\cos(\phi)-\left|\Lambda^{(2)}(t,t^\prime)\right|^2\cos(2\phi),
\end{aligned}
\end{equation}
where the new term $\Lambda^{(2)}(t,t^\prime)=\braket{\bd(t^\prime)\bd(t)}$ captures the two-photon coherence and $\Lambda_-^{(1,2)}(t,t^\prime)=\text{Re}\left[\braket{\bd(t^\prime)}\braket{\bu(t^\prime)\bu(t)\bd(t)}-\braket{\bd(t)}\braket{\bu(t)\bu(t^\prime)\bd(t^\prime)}\right]$ is a time antisymmetric term that is nonzero only if there is one-photon coherence and specifically coherence between the 1 and 2 photon subspaces. For emission from a two-level system or a cavity-emitter device operating in the Purcell regime, the term $\Lambda^{(1,2)}_-$ is nonzero only during the coherent excitation of the system. Because of this, it could serve as a way to measure the time dynamics of the re-excitation processes, although I do not explore that idea in this thesis.

After integrating $G^{(2)}_\text{HOM}$ over the photonic state and normalizing by the total intensity $\mu=\int N(t)dt$, we get
\begin{equation}
\begin{aligned}
    V_\text{HOM}=M-g^{(2)}-2\lambda^{(1,2)}_-\cos(\phi)+\lambda^{(2)}\cos(2\phi),
\end{aligned}
\end{equation}
where we now have two new quantities affecting the HOM visibility: the integrated two-photon coherence $\lambda^{(2)}=\mu^{-2}\!\iint \left|\Lambda^{(2)}(t,t^\prime)\right|^2 dt^\prime dt$ and the integrated asymmetry component $\lambda^{(1,2)}_-=\mu^{-2}\!\iint \left|\Lambda^{(1,2)}_-(t,t^\prime)\right|^2 dt^\prime dt$. I call this last term the asymmetry component because, when looking at the parameteric curve $\{V_\text{SH}(\phi),V_\text{HOM}(\phi)\}$, we would expect a quadratic relationship centered about $V_\text{SH}=1/2$ when $\lambda_-^{(1,2)}=0$ and $\lambda^{(2)}\neq 0$ \ref{loredo2020deterministic}. If the time integral is not symmetric over $t$ and $t^\prime$, the presence of a very small $\lambda_-^{(1,2)}$ component will add a linear dependence of $V_\text{HOM}$ on $V_\text{SH}$, skewing this quadratic relationship and making the parametric curve asymmetric about $V_\text{SH}=1/2$. By measuring $V_\text{SH}$ and $V_\text{HOM}$ simultaneously while varying $\phi$, one can extract a lot of information about an input photonic state \cite{loredo2019generation}\ref{loredo2020deterministic}.

It is very important to note that I have defined all of the above HOM quantities normalized with respect to $\mu^2$ and not the total uncorrelated coincidence counts, even though the latter normalization is standard for single-photon source characterization \ref{ollivier2020g2hom}. The purpose for choosing to normalize by the intensity is to avoid conflating important quantities that we are attempting to extract from the measurement. If the measured photonic state carries one-photon coherence, then the uncorrelated coincidence counts will depend on $\braket{\bd(t)}$ \cite{gustin2018pulsed} and hence a calculated HOM visibility normalized by the uncorrelated coincidence counts will depend on $V_\text{SH}$. Then, it would be necessary to correct for nonzero $\lambda^{(1)}$ in order to obtain $M$, $\lambda^{(2)}$, and $\lambda^{(1,2)}_-$.

To see this normalization problem explicitly, consider $G^{(2)}_\text{HOM}(t,t^\prime+mT_\mathrm{p})$ where $T_\mathrm{p}$ is the delay between successive photonic states. Here we must choose $m\geq 2$ so that we do not probe the first sideband peak, which can still contain correlations due to the Mach-Zehnder setup (see the supplementary of \ref{ollivier2020g2hom}), but we must also not choose $m$ so large that potential emitter blinking introduces correlations \cite{hilaire2020deterministic}. For this uncorrelated peak, we have that $G^{(2)}_\text{HOM}(t,t^\prime+mT_\mathrm{p})=\braket{\bu_3(t)\bd_3(t)}\braket{\bu_4(t^\prime)\bd_4(t^\prime)}=N_3(t)N_4(t^\prime)$, where I have applied the periodic condition of our source to eliminate $T_\mathrm{p}$. By integrating over the entire uncorrelated peak, we obtain $g^{(2)}_\text{unc.}=\mu^+\mu^-=\mu^2(1-V_\text{SH}^2)$. Hence, the uncorrelated peak underestimates $\mu^2$ depending on the amount of one-photon coherence and the interferometer phase. It is critical to note that, even if the interferometer phase $\phi$ is averaged during data collection, so long as the phase is stable on the timescale of $mT_\mathrm{p}$ then $g^{(2)}_\text{unc.}=\mu^2(1-(\lambda^{(1)})^2/2)$ still underestimates $\mu^2$. If $g^{(2)}_\text{HOM}$ is normalized with respect to $g^{(2)}_\text{unc.}$, then the important quantities such as $M$ will be over-estimated by the factor $(1-V_\text{SH}^2)^{-1}$.

If the two states being measured are produced by the same emitter, the phase coherence probed by the self-homodyne measurement is susceptible to all phase fluctuations occurring on the timescale of the delay $T_\mathrm{p}$ between the reference state and the measured state. This case arises when measuring indistinguishability, as was discussed in the previous section. If the two states are produced by independent (but identical) emitters, then the self-homodyne measurement is susceptible to all emitter dephasing mechanisms on all timescales. This case arises when performing spin-spin entanglement via which-path erasure, which will be discussed in section \ref{chapter4:entanglementgeneration}.

\section{Photonic state from a purely-dephased emitter}
\label{chapter3:photonicstate}

In this section, I will further explore the one- and two-photon temporal density functions that were discussed in the previous section to justify the separability approximation. I will then extend the idea of a photon number decomposition to analyze self-homodyne measurements of photonic states to gain insight into the amount of coherence between photon number subspaces. Finally, I will discuss and analyze how a two-level emitter can be manipulated to generate photon-number Bell states encoded in discrete time bins.

\subsection{Temporal coherence and photon statistics}
\label{chapter3:temporalcoherenceandstatistics}

\begin{figure}
    \centering
    \includegraphics[width=0.32\textwidth]{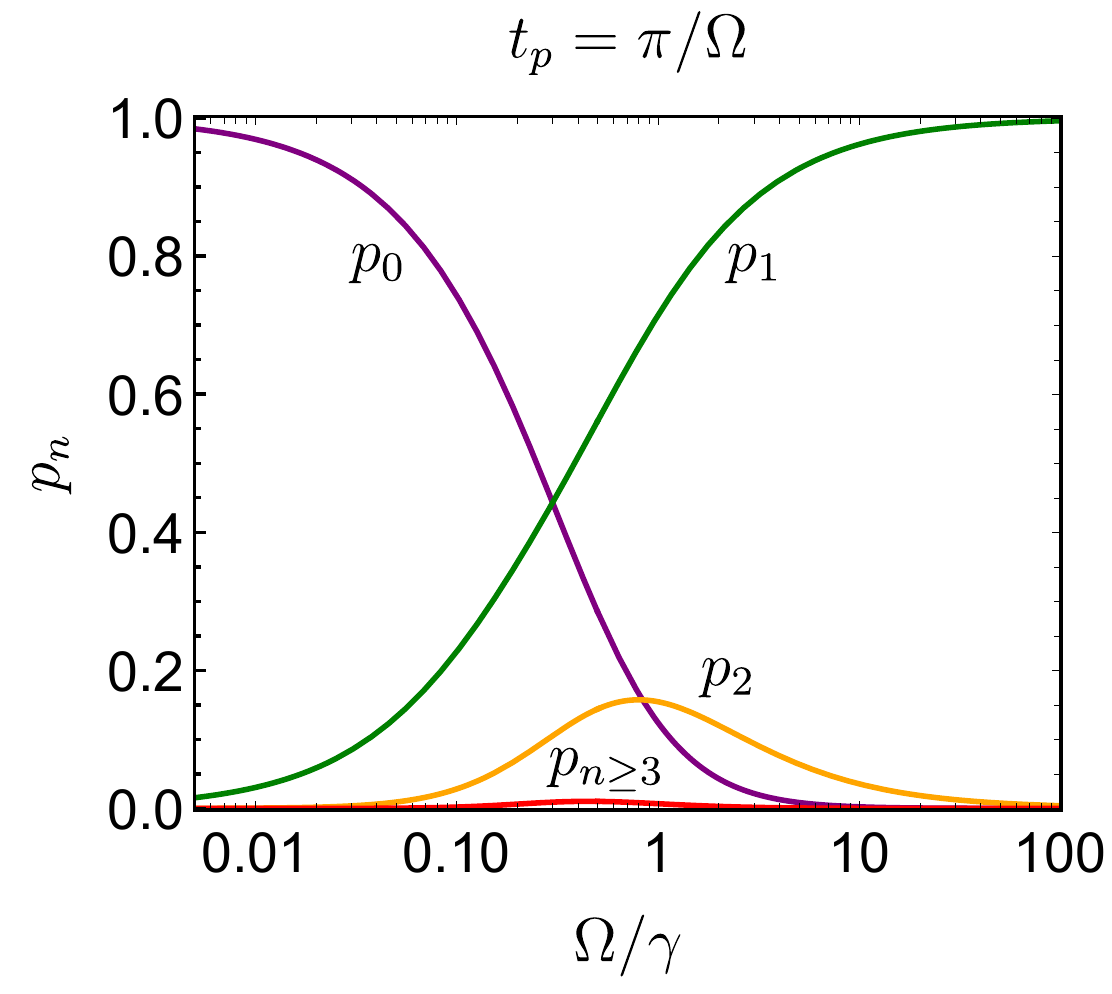}
    \includegraphics[width=0.32\textwidth]{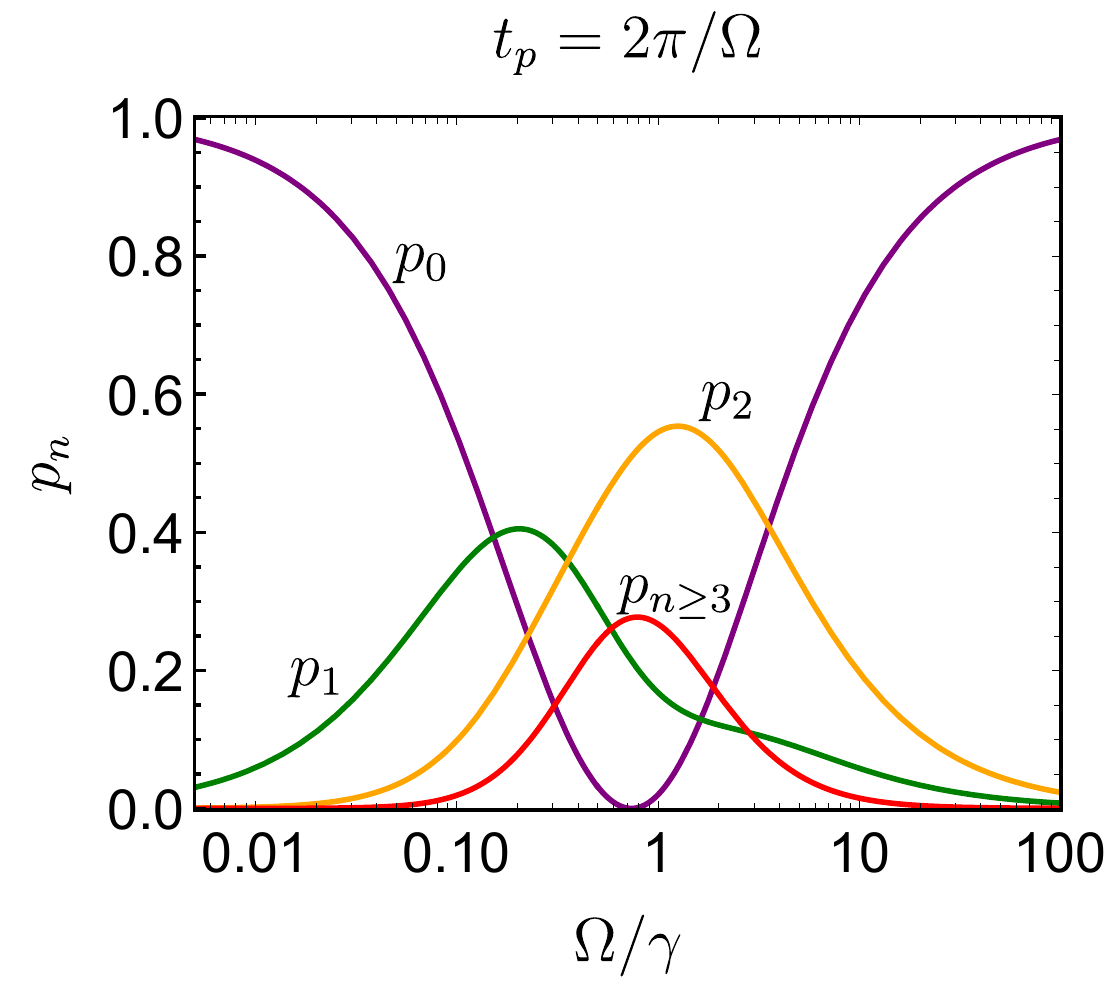}
    \includegraphics[width=0.32\textwidth]{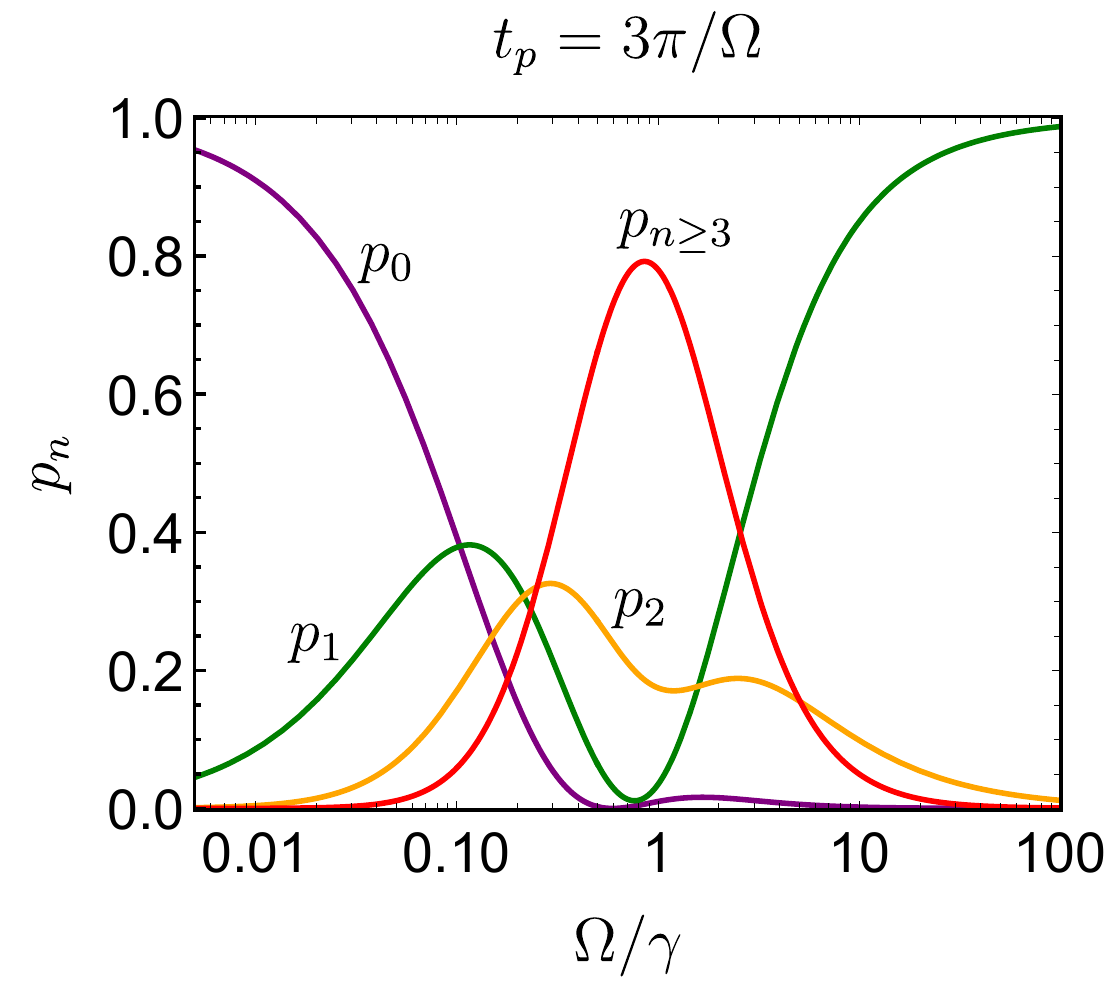}
    \caption[Photon emission statistics for pulsed excitation of a two-level system.]{\small\textbf{Photon emission statistics for pulsed excitation of a two-level system} for square pulse area $\Theta=\pi$, $2\pi$, and $3\pi$ from left to right, respectively. Chosen parameters are $\Delta=0$, $\eta_\mathrm{r}=1$, and $\gamma^\star=0$. The curve labeled $p_{n\geq 3}$ (red curve) is computed by $1-p_0-p_1-p_2$.}
    \label{fig:pns}
\end{figure}

Let us again consider a two-level system experiencing some pure dephasing that is excited by a square pulse in the semi-classical approximation. The Liouville superoperator for this system is given by Eq.~(\ref{chapter3:liouvilleTLS}). We can now perform a photon number decomposition using the jump superoperator $\mathcal{J}\hat{\rho}=\eta_\mathrm{r}\gamma\sigd\hat{\rho}\sigu$ to obtain the conditional propagation superoperators $\mathcal{U}_n$ that are of the order $\eta_\mathrm{r}^n$. Furthermore, we can break the evolution into piecewise time-independent parts using Eq.~(\ref{chapter3eq:piecewisetimeind}). Using the conditional propagation superoperators along with the expressions given in section \ref{chapter2:temporaldensityfunctions}, it is possible to compute the photon number probabilities $p_n$ and temporal density functions of photon number subspaces $\xi_n$ by correctly choosing the integration limits for each $\mathcal{U}_n$, similar to Eq.~(\ref{chapter3eq:piecewisetwotime}). With this method, we can obtain analytically exact expressions by evaluating the conditional propagation superoperators in the Fock-Liouville space (see section \ref{chapter1:fockliouville}). When $\gamma^\star\neq 0$, analytic solutions become very difficult to simplify, but the problem can still be numerically computed in an efficient way. If we choose $\gamma^\star=0$ and $\eta_\mathrm{r}=1$, then for $n=0$ we can reproduce the known result \cite{fischer2018scattering} for the vacuum probability for a pulsed two-level system
\begin{equation}
    p_0 = \frac{1}{4\tilde{\Omega}^2}\left[2\tilde{\Omega}\cos\left(\frac{t_\mathrm{p}\tilde{\Omega}}{2}\right)+\gamma\sin\left(\frac{t_\mathrm{p}\tilde{\Omega}}{2}\right)\right]^2e^{t_\mathrm{p}\gamma/2},
\end{equation}
where $2\tilde{\Omega}=\sqrt{4\Omega^2-\gamma^2}$. For $n\geq 1$ the exact expressions are complicated, but in the approximation that $t_\mathrm{p}\ll 1/\gamma$, we can find that the photon number decomposition of the master equation gives the solutions
\begin{equation}
\begin{aligned}
    p_0 &\simeq \cos^2\left(\frac{\Omega}{2}t_\mathrm{p}\right)e^{-\gamma t_\mathrm{p}/2}\\
    p_1 &\simeq 
    \sin^2\left(\frac{\Omega}{2}t_\mathrm{p}\right)e^{-\gamma t_\mathrm{p}/2}\\
    p_2 &\simeq \frac{\gamma t_\mathrm{p}}{8}\left(2+\cos(\Omega t_\mathrm{p})\right)e^{-\gamma t_\mathrm{p}/2}.
\end{aligned}
\end{equation}
See Fig.~\ref{fig:pns} for plots of the exact $p_{n<3}$ for an emitter driven at $\Theta=\pi$, $2\pi$, and $3\pi$ when $\gamma^\star=0$. The above expressions for $p_n$ are not novel, and can also be derived from a Hamiltonian emitter-waveguide system \cite{fischer2018scattering}, which (for the case of pure states) reduces to the photon number decomposition I have presented. This photon decomposition approach has also been used in Ref.~\cite{hanschke2018quantum} to compute photon number probabilities for quantum dot sources.

The full analytic expressions for $\xi_1$ and $\xi_2$ that include $\gamma^\star$ are quite complicated and do not have a form that is physically insightful. However, if $\gamma^\star=0$, we find that the solutions for $\xi_1$ and $\xi_2$ using Eqs.~(\ref{chapter2eq:onephotondensity}) and (\ref{chapter2eq:twophotondensity}) can be factored into amplitudes $\xi_1(t,t^\prime) = f_1(t)f_1^*(t^\prime)$ and $\xi_2(t_1,t_2,t_1^\prime,t_2^\prime)=f_2(t_1,t_2)f_2^*(t_1^\prime,t_2^\prime)$. The solutions for the amplitudes $f_1$ and $f_2$ obtained from $\xi_1$ and $\xi_2$, respectively, written in the frame rotating at the emitter frequency are
\begin{equation}
    f_1(t)=\frac{\sqrt{\gamma}e^{i\phi_1}}{\sqrt{p_1}}\left\{
    \begin{aligned}
        &\frac{\Omega}{2\tilde{\Omega}^2}\sin\left(\frac{\tilde{\Omega}}{2}t\right)\left[2\tilde{\Omega}\cos\left(\frac{\tilde{\Omega}}{2}(t_\mathrm{p}-t)\right)+\gamma\sin\left(\frac{\tilde{\Omega}}{2}(t_\mathrm{p}-t)\right)\right]e^{-\gamma t_\mathrm{p}/4}&& t<t_\mathrm{p}\\
        &\frac{\Omega}{\tilde{\Omega}}\sin\left(\frac{\tilde{\Omega}}{2}t_\mathrm{p}\right)e^{-\gamma(2t-t_\mathrm{p})/4}&& t\geq t_\mathrm{p}\\
    \end{aligned}
    \right.
\end{equation}
and
\begin{equation}
\begin{aligned}
  f_2(t_1,t_2)=\frac{\sqrt{p_1\gamma} e^{i\phi_2}}{\sqrt{p_2}}
    \left\{
    \begin{aligned}
        &\frac{\Omega}{\tilde{\Omega}}
        \sin\left(\frac{\tilde{\Omega}}{2}t_1\right)\csc\left(\frac{\tilde{\Omega}}{2}t_2\right)
        \sin\left(\frac{\tilde{\Omega}}{2}(t_2-t_1)\right)f_1(t_2)&& t_1\leq t_2 \leq t_\mathrm{p}\\
        &\frac{\Omega}{\tilde{\Omega}}\sin\left(\frac{\tilde{\Omega}}{2}t_1\right)\csc\left(\frac{\tilde{\Omega}}{2}t_\mathrm{p}\right)\sin\left(\frac{\tilde{\Omega}}{2}(t_\mathrm{p}-t_1)\right)f_1(t_2)&& t_1 \leq t_\mathrm{p} < t_2\\
        &0&& \text{otherwise}
    \end{aligned}
    \right.
\end{aligned}
\end{equation}
where I have restricted $t_1\leq t_2$ by our time ordering convention.  This factorization is not the same as the separability approximation discussed in section \ref{chapter3:pulsedemitterHOM}. In that case, we were separating the two-photon density function $\xi_2$ into two one-photon density functions $\xi_\mathrm{n}$ and $\xi_1$. In this case, we are factoring the two-photon density function $\xi_2$ into two two-photon amplitudes $f_2$ and $f_2^*$. The former case is valid when the two photons are not entangled, but they may be temporally impure. The latter case is valid when the total two-photon component is temporally pure, but there may still be entanglement. As I will discuss below, there can be situations where both of these cases apply so that the two photons described by $\xi_2$ are both unentangled and temporally pure.

Before moving on, I would like to note that the above solutions for the one and two-photon temporal wavefunctions are also derived in section 4.1 of Ref.~\cite{fischer2018scattering} by explicit consideration of the full emitter-waveguide system undergoing reversible evolution. However, the equivalent solutions (up to a difference in normalization convention) presented here are lacking information about the relative phase between $f_1$ and $f_2$. This is because we cannot obtain $\phi_1$ and $\phi_2$ when deriving the amplitudes by factoring $\xi_1$ and $\xi_2$. To obtain this phase information, it is necessary to compute $\zeta_{n,m}$. However, often only $\xi_n$ are relevant for a particular analysis, and the derivation using conditional correlations to compute density functions $\xi_n$ instead of wavefunctions $f_n$ can account for excess emitter decoherence and can be applied to a wide variety of emitter systems in a straightforward way.

I have chosen to write $f_2$ in terms of $f_1$ to illustrate some important points about $f_2$. First, we can see that $f_2$ is implicitly dependent on $t_2$ through $f_1(t_2)$ when $t_1\leq t_\mathrm{p}\leq t_2$. Hence, it can be separated into $f_2(t_1,t_2)=f_\mathrm{n}(t_1)f_1(t_2)$, which is the pure-state analog to the separability assumption made in section \ref{chapter3:pulsedemitterHOM} for $\xi_2$. This separability is not true when $t_1\leq t_2\leq t_\mathrm{p}$, because $f_2$ is dependent on $t_2$ explicitly. This implies that the two photons are entangled in time. Second, when $t_\mathrm{p}$ is small compared to $1/\gamma$, the part of $f_2$ corresponding to $t_1\leq t_2\leq t_\mathrm{p}$ has a negligible contribution. Therefore, when $t_\mathrm{p}\ll 1/\gamma$ we have $f_2(t_1,t_2)\simeq f_\mathrm{n}(t_1)f_1(t_2)$ and $\tilde{\Omega}\simeq\Omega$, where the `noise' photon due to re-excitation has a temporal amplitude of
\begin{equation}
    f_\mathrm{n}(t) = \frac{\sqrt{p_1\gamma}e^{i
    \phi_2}}{\sqrt{p_2}}\sin\left(\frac{\Omega}{2}t\right)\csc\left(\frac{\Omega}{2}t_\mathrm{p}\right)\sin\left(\frac{\Omega}{2}(t_\mathrm{p}-t)\right)
\end{equation}
for $t\leq t_p$ and $f_\mathrm{n}(t)=0$ otherwise. This separability of `noise' from `signal' agrees with the observations made in section \ref{chapter3:pulsedemitterHOM} and the assumptions of the derivation in section \ref{chapter3:separablenoisemodel}.

The result that $\xi_1$ and $\xi_2$ correspond to temporally pure quantum states when $\gamma^\star=0$ may seem physically obvious, but it is not mathematically obvious in the context of the photon number decomposition. When solving the evolution from the total emitter-waveguide Hamiltonian as in Ref.~\cite{fischer2018scattering}, the temporal amplitudes $f_1$ and $f_2$ are solved naturally by applying a unitary transformation to the total system state. Once the emitter has decayed to the ground state, the waveguide mode is left in a pure quantum state. However, using the photon number decomposition, the entire derivation was performed in the mixed state picture of the reduced system evolution. I find it interesting that the dynamics of just the emitter, which experiences decoherence during emission (recall section \ref{chapter1:spontaneousemission}), can still allow us to compute the temporally-pure state of the waveguide after emission.

\begin{figure}[t]
    \centering
    \hspace{-36mm}(a)\hspace{48mm}(b)\\\vspace{3mm}
    \hspace{2mm}\includegraphics[width=0.295\textwidth]{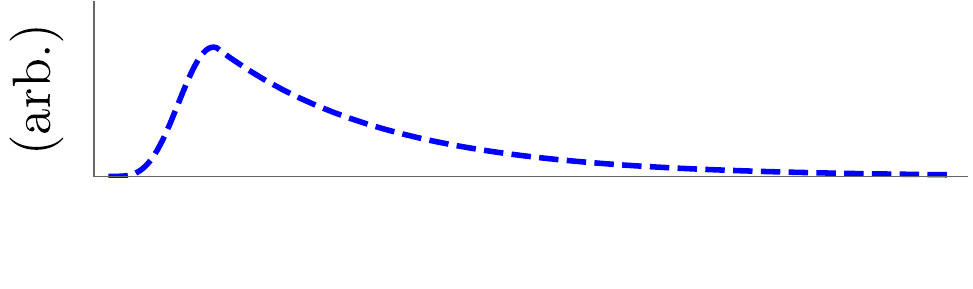}\hspace{4mm}
    \includegraphics[width=0.295\textwidth]{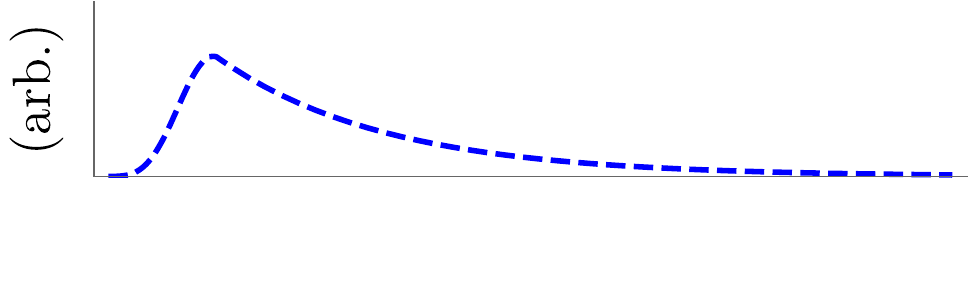}\\\vspace{-5mm}
    \includegraphics[width=0.32\textwidth]{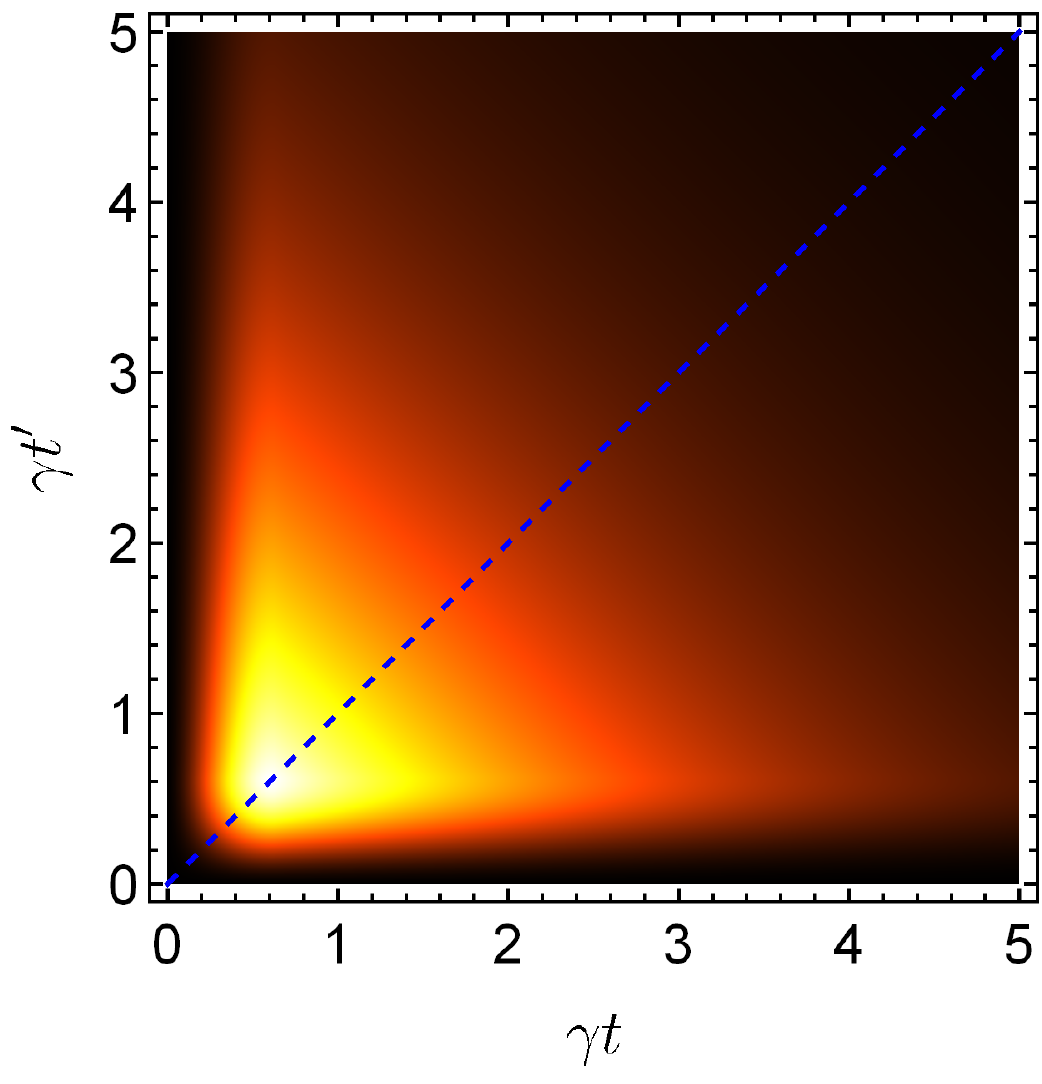}
    \includegraphics[width=0.32\textwidth]{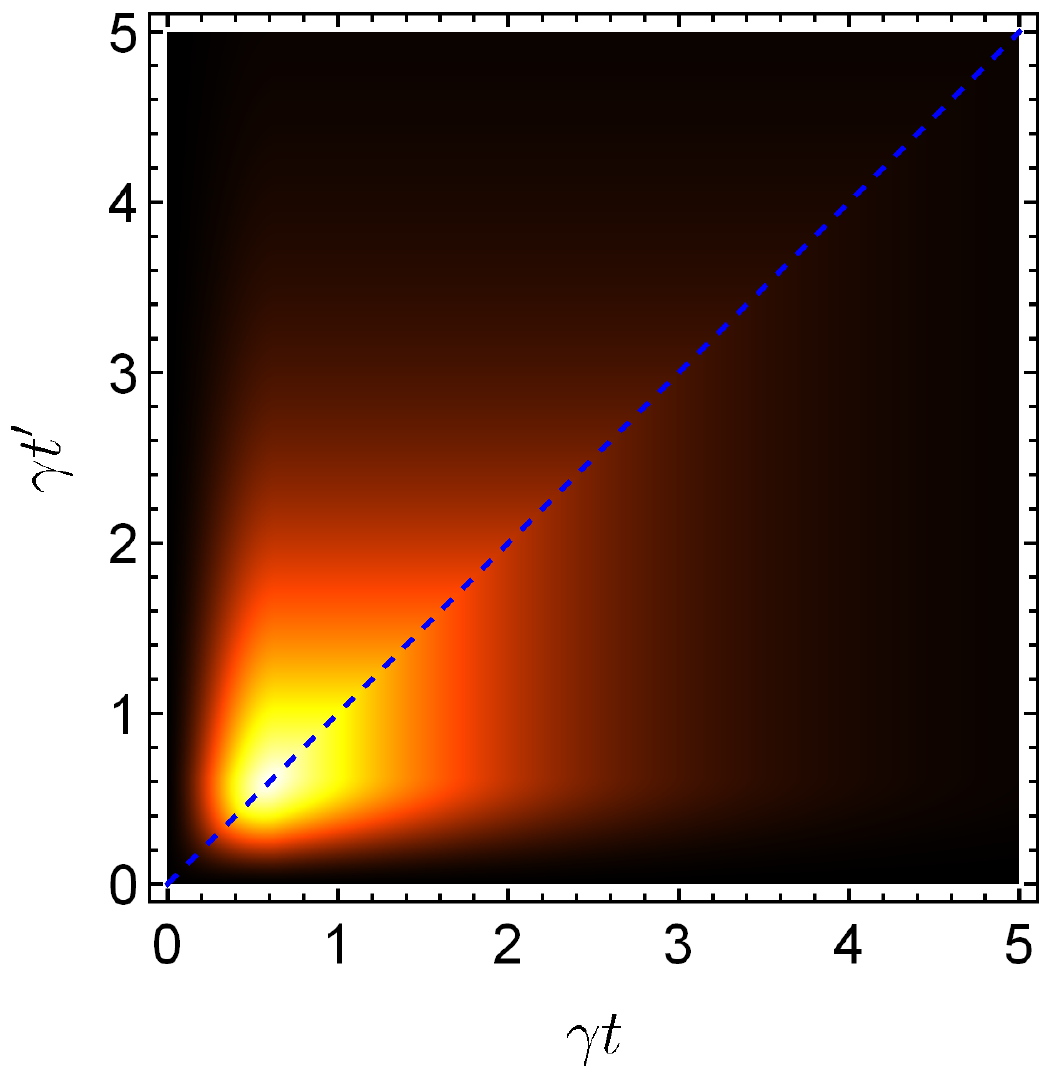}
    \caption[Single-photon temporal density functions for a two-level system coherently driven with a square $\pi$ pulse.]{\small\textbf{Single-photon temporal density function $\xi_1(t,t^\prime)$ for a two-level system  coherently driven with a square $\pi$ pulse} (a) without pure dephasing and (b) with pure dephasing. The dashed blue line shows the single-photon intensity $\xi_1(t,t)=N(t)$ corresponding to the diagonal of the density function. Chosen parameters are $\Omega/\gamma=5$, $\eta_\mathrm{r}=1$, and $\gamma^\star=\gamma/2$ if not zero.}
    \label{fig:singlephoton}
\end{figure}

Let us now turn to the case with a nonzero pure dephasing rate. By evaluating $\xi_1$, we can see that the temporal coherence, $\xi_1(t_1,t_2)$ when $t_1\neq t_2$, is suppressed by pure dephasing (see Fig.~\ref{fig:singlephoton}). However, the intensity $\xi_1(t,t)$ remains mostly unchanged for $\gamma^\star<\Omega$. Any small change in the intensity is due to the damping of the coherent driving.

\begin{figure}
    \centering
    \hspace{-36mm}(a)\hspace{48mm}(b)\\\vspace{3mm}
    \hspace{2mm}\includegraphics[width=0.295\textwidth]{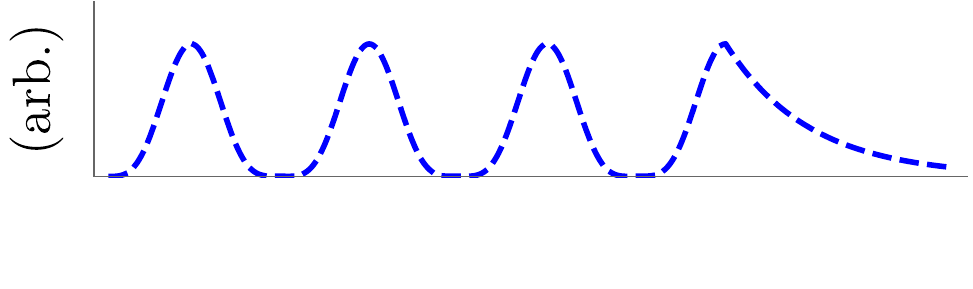}\hspace{4mm}
    \includegraphics[width=0.295\textwidth]{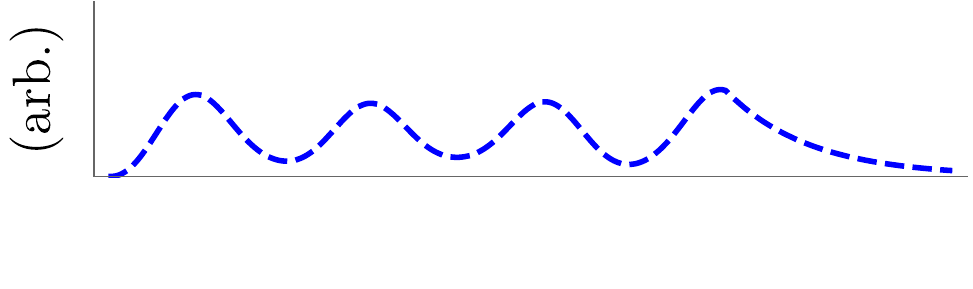}\\\vspace{-5mm}
    \includegraphics[width=0.32\textwidth]{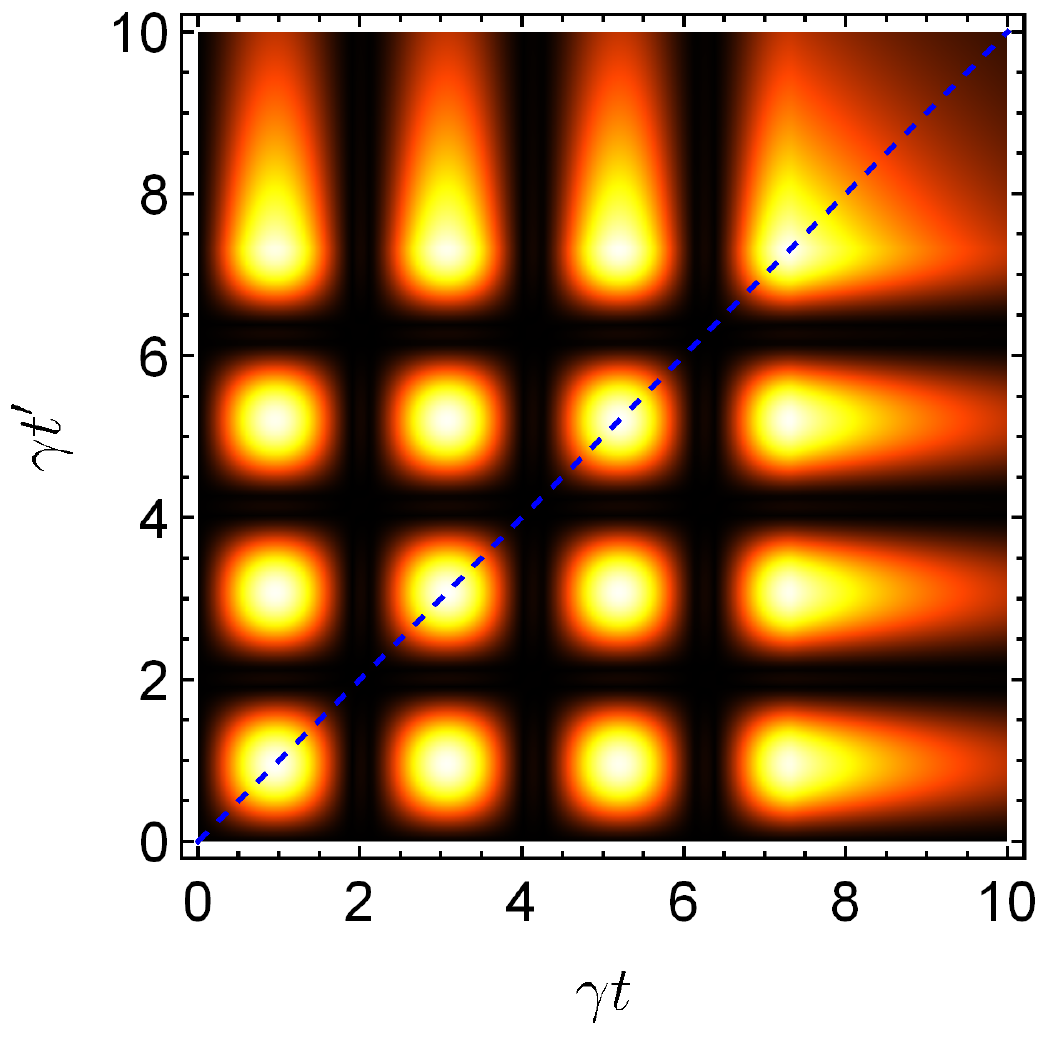}
    \includegraphics[width=0.32\textwidth]{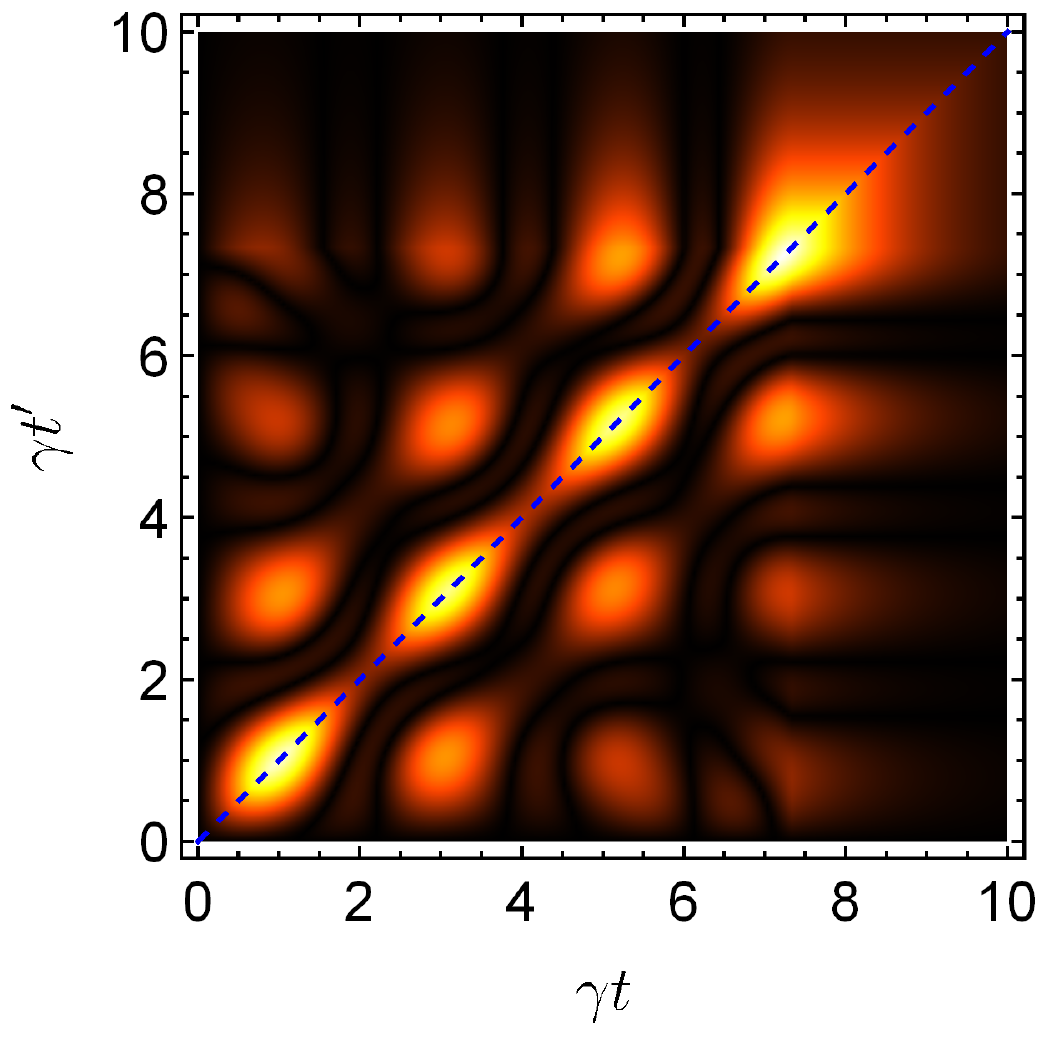}
    \caption[Single-photon temporal density functions for a two-level system coherently driven by a square $7\pi$ pulse.]{\small\textbf{Single-photon temporal density function $\xi_1(t,t^\prime)$ for a two-level system coherently driven by a square $7\pi$ pulse} (a) without pure dephasing and (b) with pure dephasing. The dashed blue line shows the single-photon intensity $\xi_1(t,t)=N(t)$ corresponding to the diagonal of the density function. Chosen parameters are $\Omega/\gamma=3$, $\eta_\mathrm{r}=1$, and $\gamma^\star=\gamma/2$ for panel (b).}
    \label{fig:7piphoton}
\end{figure}

The photon number decomposition approach is not only useful for analyzing single-photon generation, it is also a powerful tool if one needs to access and analyze a particular photon-number subspace of the full photonic state, even if there are potentially many photons emitted. As an interesting example, consider a $7\pi$ pulse for $\Omega=3\gamma$. In this case, we expect significant multi-photon processes, yet we can easily extract the temporal density function of just the single-photon subspace. In Fig.~\ref{fig:7piphoton}, we can see quite a large effect of pure dephasing on both the temporal coherence and the intensity profile of the single photon component for a $7\pi$ pulse. The shape of this temporal density function suggests that the single-photon subspace is in an entangled state of four well-defined time-bin modes---taking a form similar to $(\ket{1000}+\ket{0100}+\ket{0010}+\ket{0001})/2$. Adding pure dephasing significantly reduces the temporal coherence of this state, with the most degradation affecting the coherence between time bins that are far-separated in time (coherence further from the diagonal). It is also important to note that I have shown only the result when neglecting losses so that $\eta_\mathrm{r}=1$. If there are multiphoton processes, the single-photon component will be altered as the higher-photon number subspaces decay into the single-photon subspace. With this method, it is also very straightforward to include losses by decreasing $\eta_\mathrm{r}$.

\begin{figure}
    \centering
    \hspace{-5mm}(a)\hspace{45mm}(b)\hspace{45mm}(c)\hspace{25mm}(d)\\
    \includegraphics[width=0.3\textwidth]{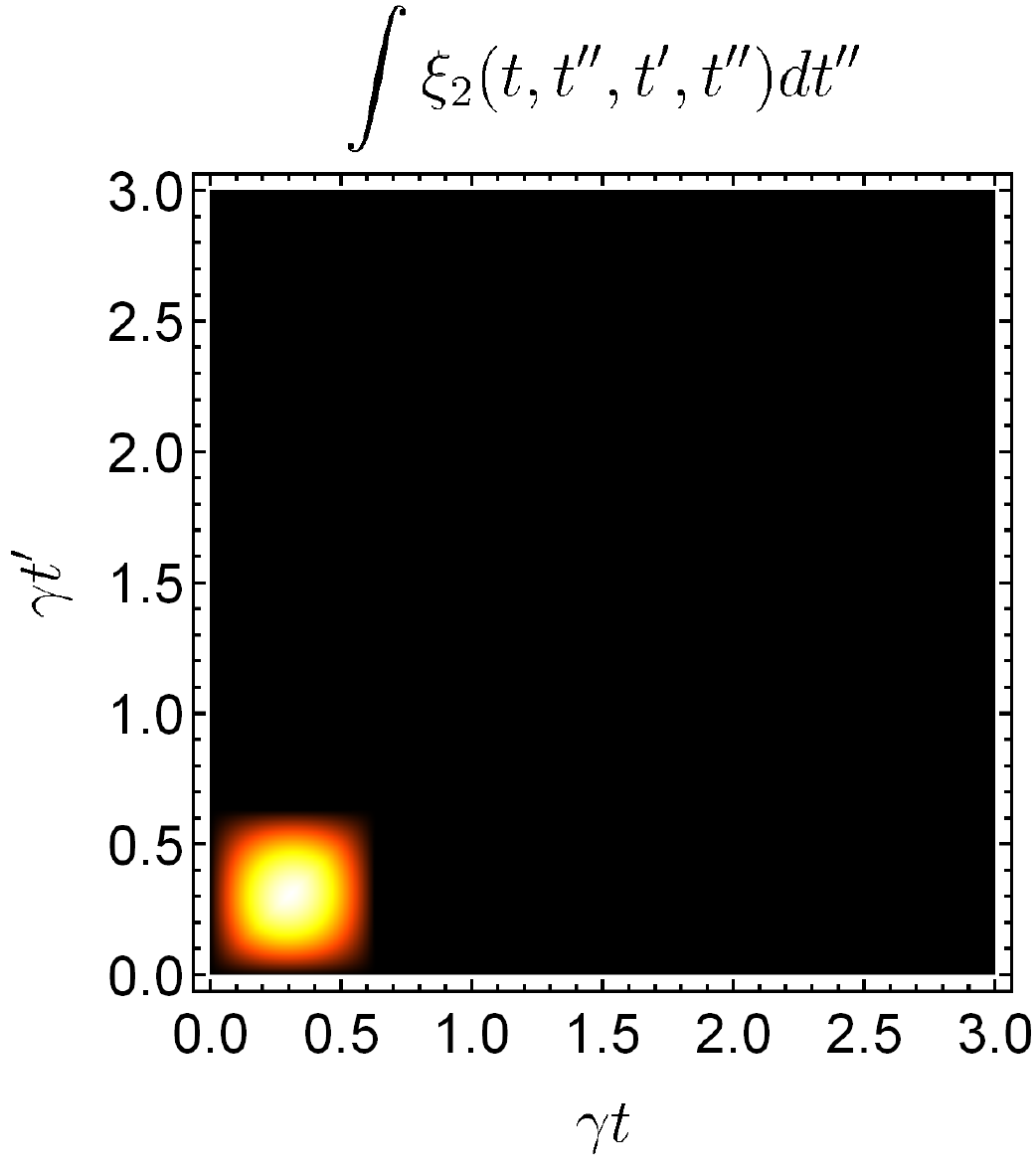}
    \includegraphics[width=0.3\textwidth]{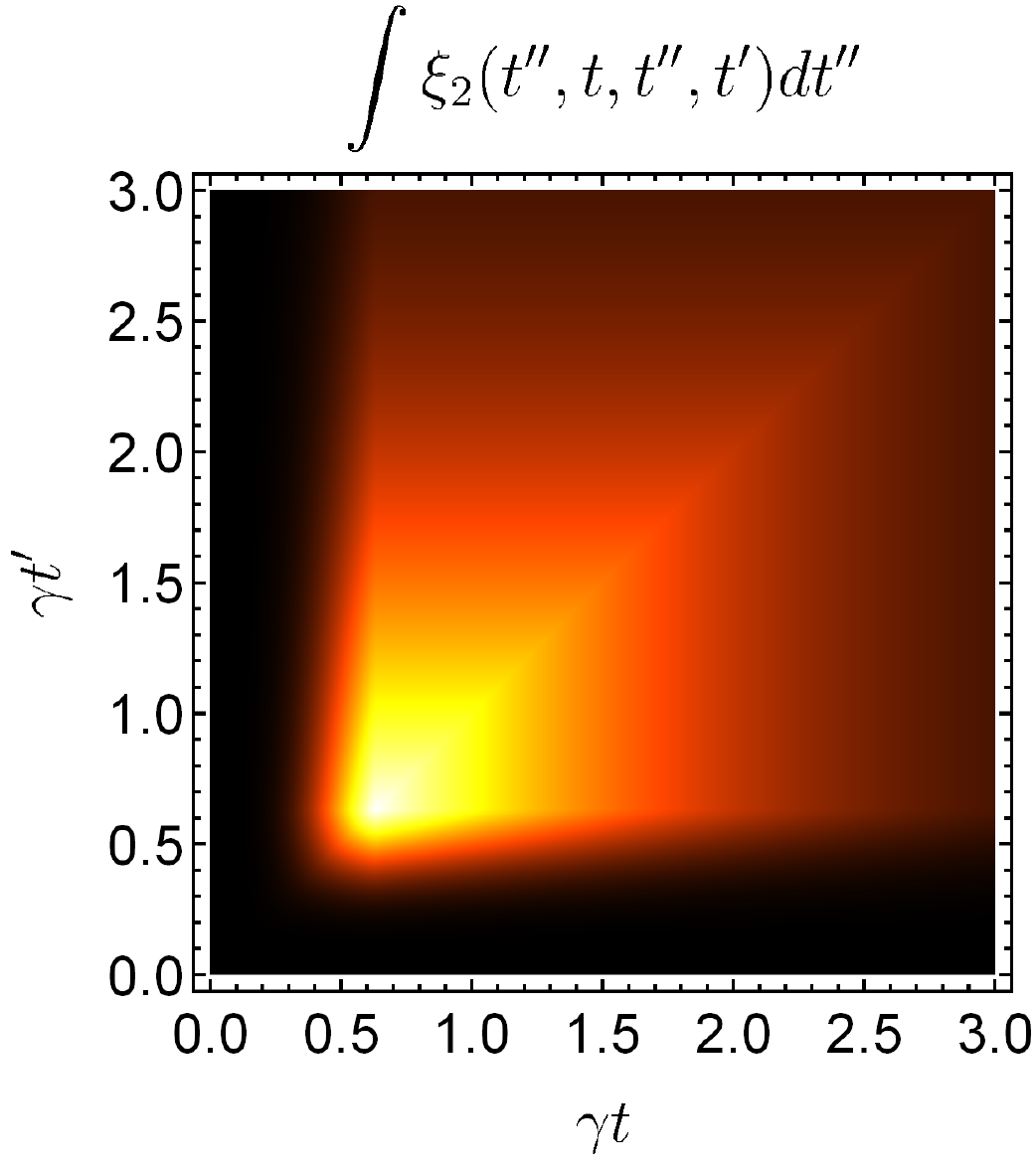}
    \includegraphics[width=0.157\textwidth]{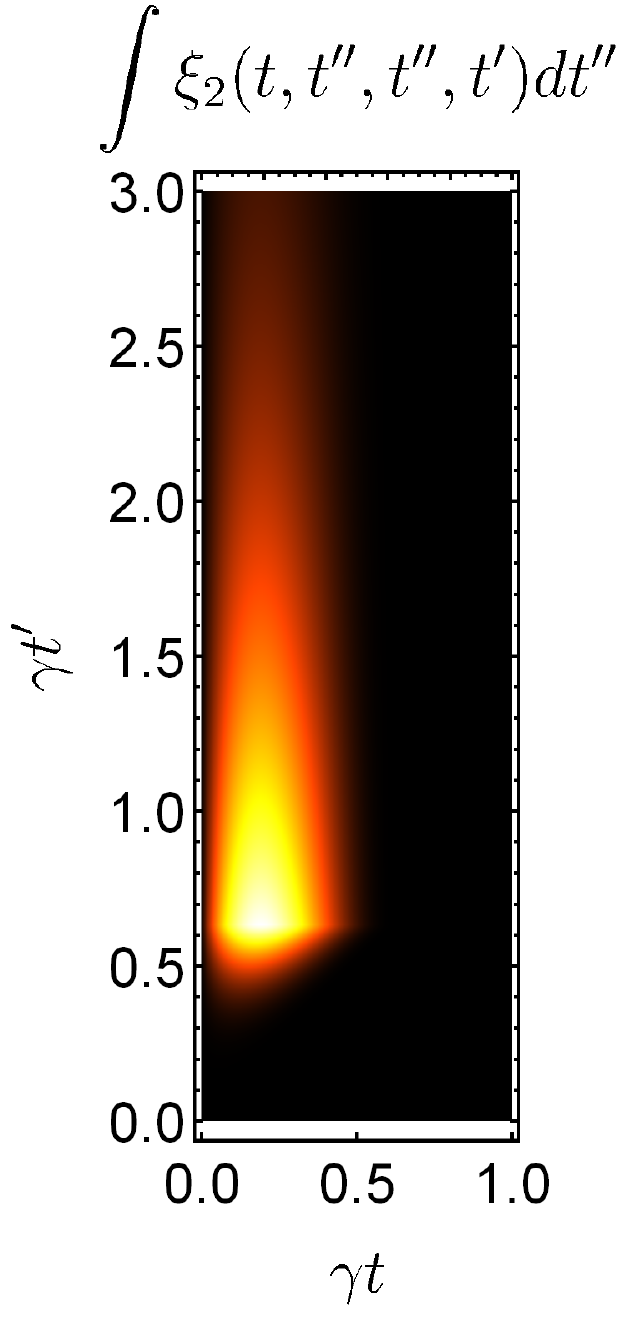}
    \includegraphics[width=0.14\textwidth]{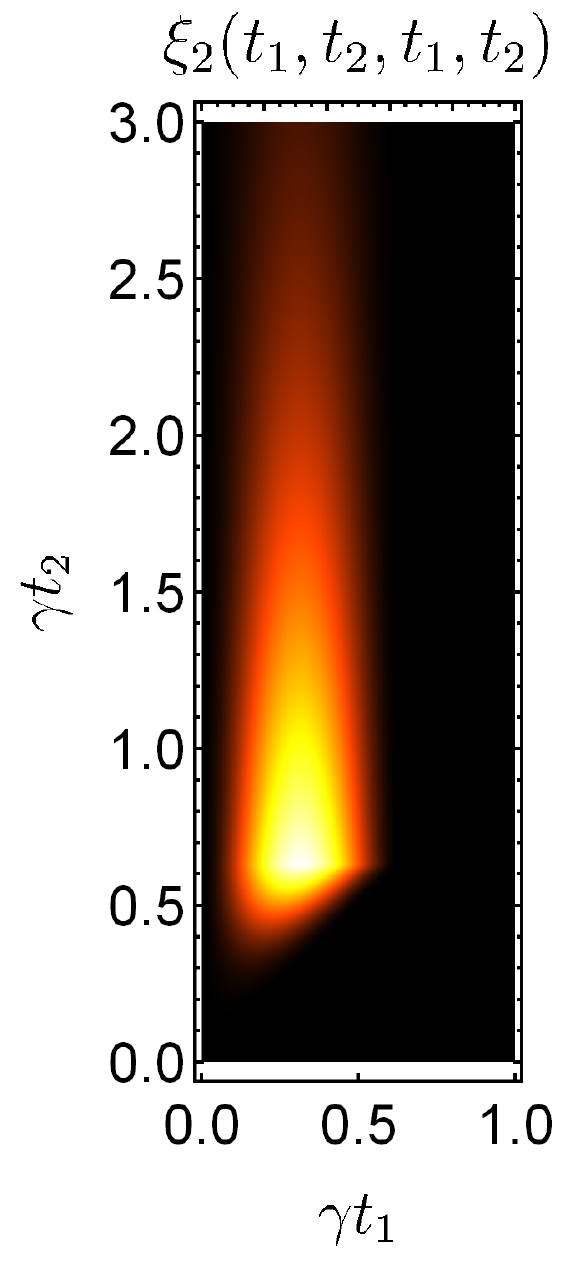}
    \caption[Two-photon amplitude correlation components and intensity correlation for a purely-dephased two-level emitter coherently driven by a square $\pi$ pulse.]{\small\textbf{Two-photon amplitude correlation components and intensity correlation for a purely-dephased two-level emitter coherently driven by a square $\pi$ pulse.} (a) The amplitude correlation of the first photon emitted of the two-photon component. (b) The amplitude correlation of the second photon emitted of the two-photon component. (c) The amplitude correlation between the first and second photons emitted of the two-photon component. (d) The intensity correlation function $G^{(2)}(t_1,t_2)=\xi_2(t_1,t_2,t_1,t_2)$ of the two photon component. Parameters chosen are $\gamma^\star=\gamma/2$, $\eta_\mathrm{r}=1$, and $\Omega=5\gamma$.}
    \label{fig:twophoton}
\end{figure}

Unlike the single-photon density function, the two-photon density function $\xi_2$ cannot be easily visualized. This is because it is a four-dimensional function of time. It is more enlightening to analyze the two photon intensity correlation $\xi_2(t_1,t_2,t_1,t_2)$ and the three components composing $\braket{\bu(t^\prime)\bd(t)}_2$ (see Eq.~(\ref{chapter2eq:firstordercorr})). Fig.~\ref{fig:twophoton} shows these four quantities for the two-photon component produced by a finite-width $\pi$-pulse. From this, we can clearly see that the first photon emitted is very temporally separated from the second. Also, we can see that the second photon has a shape quite similar to the single-photon component illustrated in Fig.~\ref{fig:singlephoton}.

\begin{figure}[t]
    \centering
    \hspace{-58mm}(a)\hspace{73mm}(b)\\
    \includegraphics[width=0.49\textwidth]{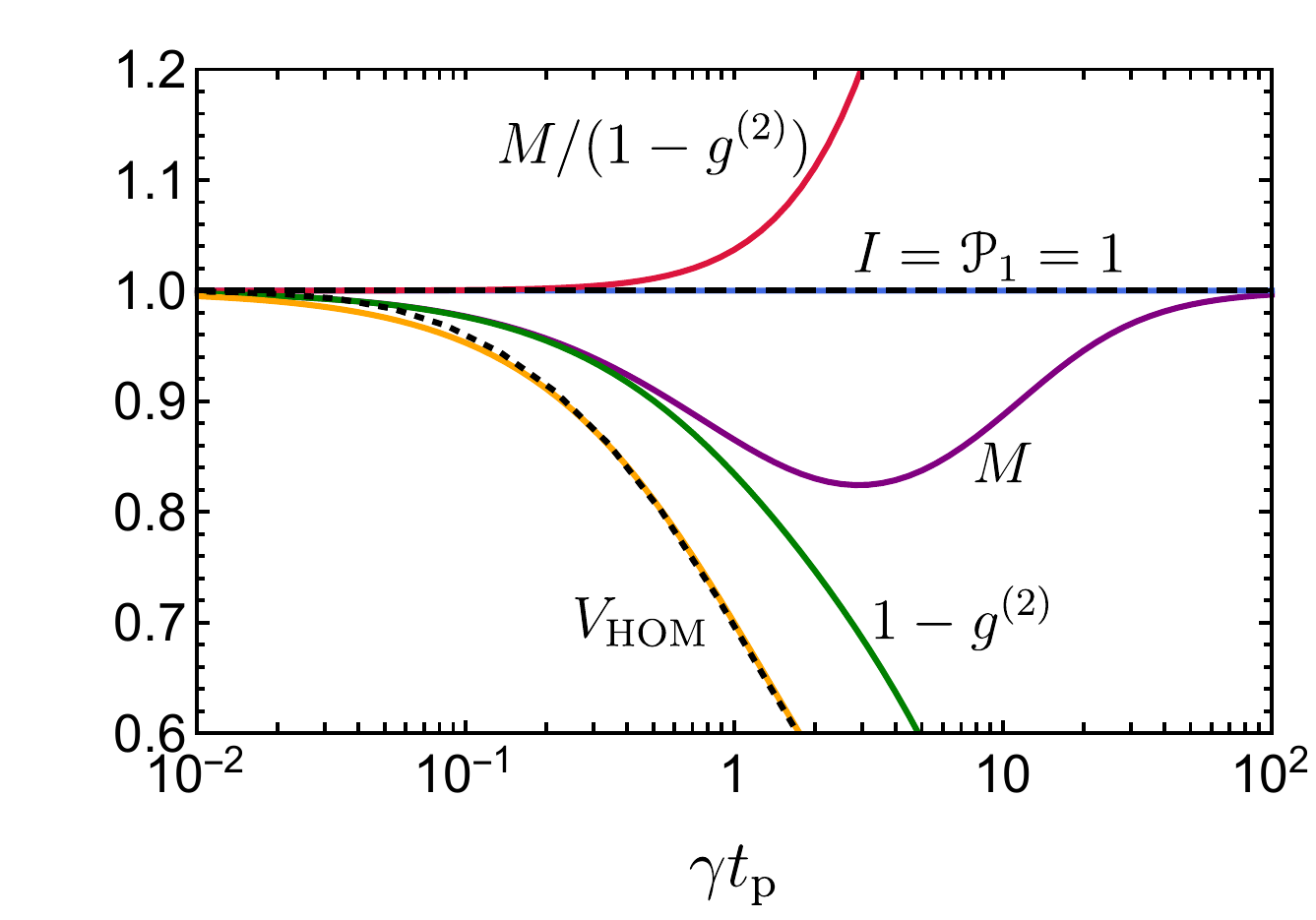}
    \includegraphics[width=0.49\textwidth]{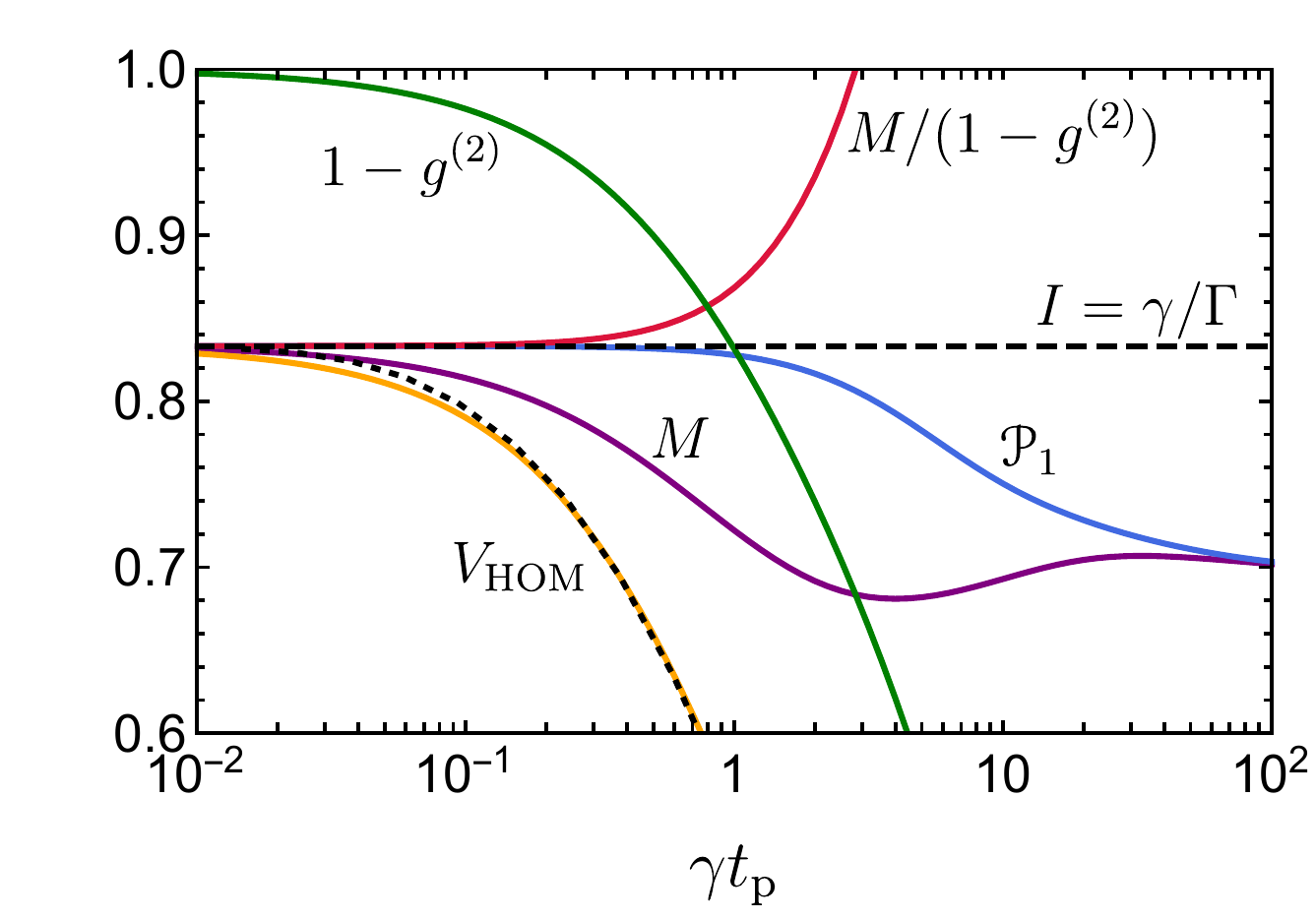}
    \caption[A comparison of measures for the indistinguishability of photons emitted from a pulsed two-level emitter experiencing emitter pure dephasing.]{\small\textbf{A comparison of measures for the indistinguishability of photons emitted from a pulsed two-level emitter experiencing emitter pure dephasing.} (a) The intensity-normalized HOM visibility $V_\text{HOM}$, mean wavepacket overlap $M$, single-photon purity $1-g^{(2)}$, and single-photon trace purity $\euscr{P}_1$ plotted for different $\pi$-pulse widths $t_\mathrm{p}$ relative to the spontaneous emission rate $\gamma$. The quantity $M/(1-g^{(2)})$ is also shown. This was derived in section \ref{chapter3:HOM} as an upper-bound approximation to $I$ and $\euscr{P}_1$ in the limit that $\gamma t_\mathrm{p}\rightarrow 0$. (b) The same quantities as in panel (a) but for a nonzero emitter pure dephasing rate $\gamma^\star=0.1\gamma$. Note that the vertical axis scale is different in panel (b) than in panel (a). In both panels, the black dotted line illustrates $V_\text{HOM}$ as computed from the full pulsed cavity-emitter model for $\hbar \gamma_\mathrm{r}=0.5\mu$eV, $\hbar g=17\mu$eV, and $\hbar\kappa=400\mu$eV corresponding to QD devices used in Refs.~\ref{ollivier2020reproducibility} and \ref{ollivier2020g2hom} that fall well within the bad-cavity regime giving $R/\kappa\simeq 0.007$ for $\gamma=\gamma^\prime=\gamma_\mathrm{r}(1+F_\mathrm{p})\simeq3.4\mu$eV. Note that an effective $\gamma^\star$ for these QD devices at $~7$K is usually estimated to be $\gamma^\star\simeq 0.04\gamma^\prime$, which falls between the cases of panels (a) and (b).}
    \label{Chapter3fig:g2homcompare}
\end{figure}

Using $\xi_1$, we are finally in a good position to check the validity of the estimation for $I$ derived using the separable noise model and also compare it to other quantities describing HOM interference. Fig.~\ref{Chapter3fig:g2homcompare} shows that, regardless of dephasing or pulse width, both the indistinguishability $I$ and the trace purity of the single-photon component $\euscr{P}_1=\iint|\xi_1(t,t^\prime)|^2dtdt^\prime$ at the source are bounded between the mean wavepacket overlap $M$ and the curve $M/(1-g^{(2)})$. In addition, the curve $M/(1-g^{(2)})$ is a much more accurate (upper-bound) estimate of both of these quantities when $\gamma t_\mathrm{p}<1$. Furthermore, we can see that the indistinguishability as computed by $I=\gamma/\Gamma$ and the trace purity of the single-photon component are only truly equal in the fast-pulse regime where $g^{(2)}\rightarrow 0$.

\begin{figure}[t]
    \centering
    \includegraphics[width=0.45\textwidth]{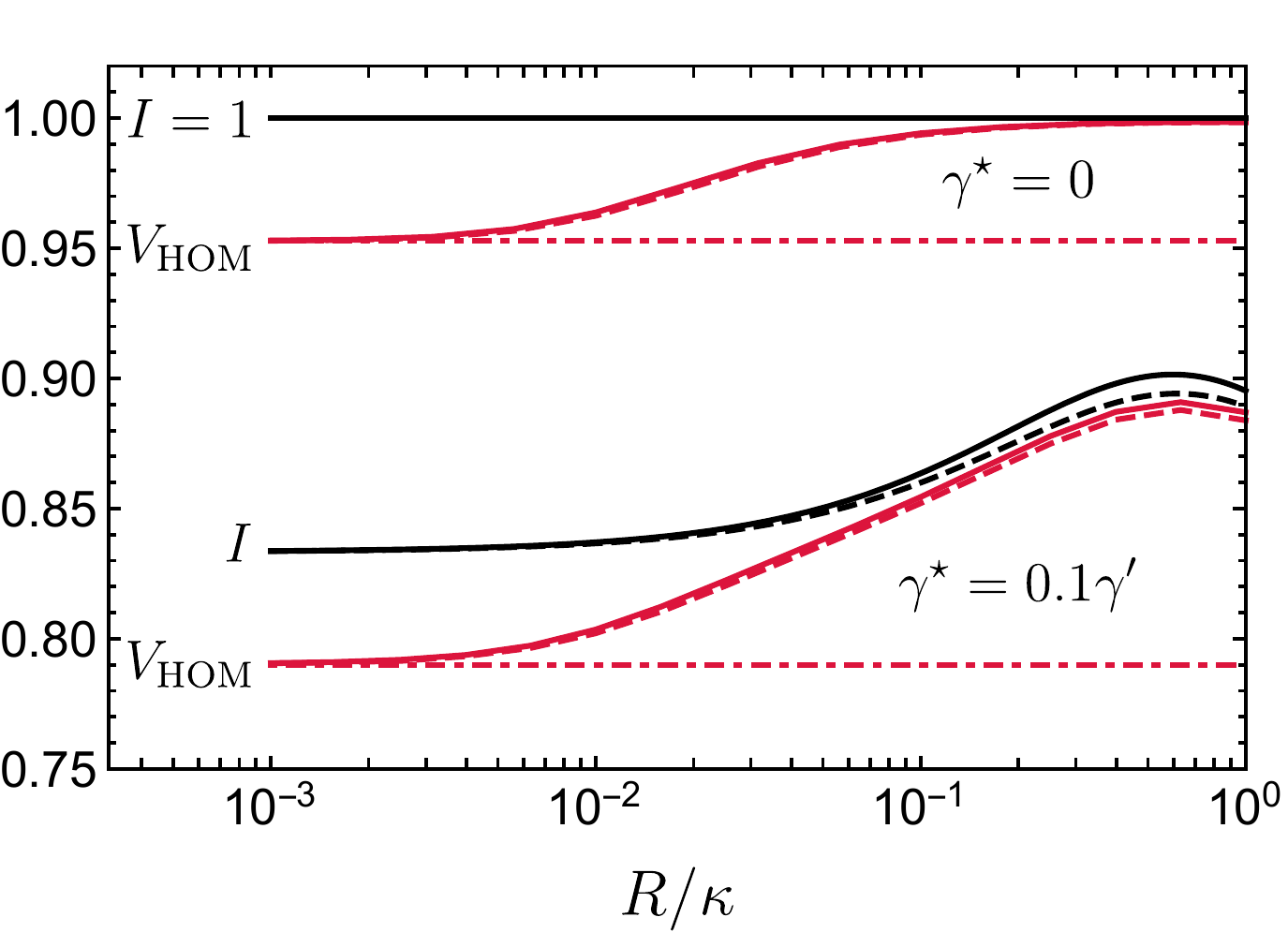}
    \caption[A comparison of the HOM visibility from a driven two-level emitter and from a driven cavity-emitter source in the bad-cavity regime.]{\small\textbf{A comparison of the HOM visibility from a driven two-level emitter and from a driven cavity-emitter source in the bad-cavity regime.} The HOM visibility $V_\text{HOM}$ is plotted for a square $\pi$ pulse of width $\gamma^\prime t_\mathrm{p}=0.1$ where $\gamma^\prime=(1+F_\mathrm{p})\gamma$ and $F_\mathrm{p}$ is given by Eq.~(\ref{chapter3eq:purcellsaturationfactor}) that captures the Purcell saturation in the critical regime. To allow for a comparison of the approximation accuracy for different Purcell factors when dephasing is nonzero, I have scaled the dephasing rate $\gamma^\star$ proportional to the Purcell-enhanced decay rate. For a given $R/\kappa$ and $F_\mathrm{p}$ of either 10 (solid curves) or 1000 (dashed curves), the values of $g$ and $\kappa$ are determined relative to $\gamma$, which would then fix the timescale if specified. The cavity-emitter non-Markovian behaviour during pulsed excitation suppresses $g^{(2)}$ when the system is not very far into the bad-cavity regime where $R/\kappa \ll 1$. This brings $V_\text{HOM}$ (red curves) from the bad-cavity limit value (red dot-dashed horizontal lines) nearly up to the emission-limited indistinguishability $I$ (black curves) computed using Eq.~(\ref{ind}). Although I do not show the mean wavepacket overlap $M$, it is necessarily bounded between $I$ and $V_\text{HOM}$. In addition, $(V_\text{HOM}+g^{(2)})/(1-g^{(2)})$ would very closely approximate $I$ in this scenario.}
    \label{fig:badcavitycompare}
\end{figure}

In Fig.~\ref{Chapter3fig:g2homcompare}, I also show an example of the HOM visibility computed using the cavity-emitter model described in section \ref{chapter1:cavityQED} compared to the cavity-enhanced 1-dimensional atom model described in section \ref{chapter1:1datom}. For a system operating very far into the bad-cavity regime as in Refs. \ref{ollivier2020reproducibility} and \ref{ollivier2020g2hom}, the two-level approximation used extensively in this section is quite accurate. However, adiabatic elimination of the cavity mode cannot capture the brief suppression of the Purcell factor during excitation \cite{gustin2018pulsed}. This transient non-Markovian behaviour can increase $V_\text{HOM}$ by suppressing re-excitation for pulses shorter than the Purcell-enhanced lifetime even when $R/\kappa< 1$. To illustrate this point more clearly, in Fig.~\ref{fig:badcavitycompare} I have plotted $V_\text{HOM}$ as a function of the bad-cavity parameter $R/\kappa$ for two fixed Purcell factors $F_\mathrm{p}=10$ and $F_\mathrm{p}=1000$ where $F_\mathrm{p}$ is given by Eq.~(\ref{chapter3eq:purcellsaturationfactor}). For this plot, I have chosen to fix the excitation pulse so that $\gamma^\prime t_\mathrm{p}=0.1$, where $\gamma^\prime=\gamma(1+F_\mathrm{p})$ is the Purcell-enhanced decay rate. Thus, $0.1/\gamma^\prime$ determines the pulse timescale, which may or may not be realistic for a given implementation. To relate the results in this section to the results of section \ref{chapter3:roomtemperature} that neglect non-zero $g^{(2)}$, I have also plotted the upper-bound value of $I$ given by Eq.~(\ref{ind}) that was derived by considering only the emission dynamics of a cavity-emitter system. Note that $R$ and $\gamma^\star$ are both defined slightly differently in Eq.~(\ref{ind}), which comes from Ref.~\ref{wein2018plasmonics}, compared to Eq.~(\ref{chapter3eq:purcellsaturationfactor}), which uses the standard convention in this thesis. From this result, we can see that the two-level emitter model is accurate in the bad-cavity regime when $R/\kappa<0.01$ (roughly implying $2g/\kappa<0.1$) and that the accuracy is nearly independent of the inhibited Purcell factor for a given $R/\kappa$. In addition, we can see that the HOM visibility is truly dominated by the emission dynamics in the critical regime where $R/\kappa=1$ due to the suppression of $g^{(2)}$ if the excitation pulse is much faster than $1/\gamma^\prime$.

All of the values for $g^{(2)}$ computed in this thesis assume that the collected emission is perfectly filtered from the resonant excitation pulse. For example, this can be accomplished using a cross-polarization setup. However, often the excitation pulse is applied to the emitter through another mode of the same cavity that is used to collect the emission. If the polarization of excitation is not ideally matched to the correct cavity mode, the cavity can induce a birefringence effect \ref{ollivier2020reproducibility} where some of the excitation pulse is collected along with the emission from the emitter, leading to an increased $g^{(2)}$. This effect becomes worse with increasing excitation power, leading to an increased $g^{(2)}$ for short pulses \ref{ollivier2020g2hom} that may partially cancel any gain due to potential non-Markovian suppression of the Purcell effect. Cross-polarization also inherently limits the efficiency of the source due to necessarily filtering approximately half of the desired emission. Therefore, excitation techniques that allow for near-perfect filtering of the excitation pulse without sacrificing collection efficiency are highly desired. This can be accomplished by using two-photon excitation \cite{hanschke2018quantum,gustin2020efficient}, phonon-assisted off-resonant excitation at low temperature \ref{thomas2020LAphonons}\cite{quilter2015phonon,manson2016polaron,cosacchi2019emission,gustin2020efficient}, or cavities with highly non-degenerate orthogonally-polarized modes \cite{wang2019towards}.

\subsection{Number coherence}
\label{chapter3:numbercoherence}

In the previous section, I discussed how we can access the temporal coherence within photon number subspaces of a photonic state produced by a driven emitter using the photon number decomposition. In this section, I will explore one approach to gaining information about the number coherence functions $\zeta_{n,m}$ of the photonic state.

Master equation unravellings have already been used to study classical homodyne measurements \cite{carmichael}, which can give access to the homodyne photon statistics and phase information about a photonic state. However, in this section, I will demonstrate that applying the photon number decomposition to a pulsed self-homodyne measurement (see section \ref{chapter3:selfhomodyne}) can also give us access to some information about the number coherence between distinct photon number subspaces. In particular, the coherence function $\zeta_{1,0}$. This technique corresponds to a physically realistic setup, and this setup was recently used to quantify the magnitude of the coherence produced by a coherently driven artificial atom \cite{loredo2019generation}. The work of Ref.~\cite{loredo2019generation} inspired the material in this section and motivated a related project on photon number entanglement \ref{loredo2020deterministic}, which I will discuss in the sections following this one.

I will begin by studying a slightly simplified case compared to the previous section by first assuming that the emitter is prepared in a pure superposition state so that we are only analyzing the emission dynamics for when $g^{(2)}=0$. In this case, our Liouville superoperator is $\mathcal{L}_\mathrm{s}=-(i/\hbar)\mathcal{H}+\gamma\mathcal{D}(\sigd)+2\gamma^\star\mathcal{D}(\sigu\sigd)$ and $\hat{H}=\hbar\omega\sigu\sigd$. Let us now consider the initial state $\hat{\rho}(0)=\ketbra{\psi(0)}{\psi(0)}$, where $\ket{\psi(0)}=\cos(\vartheta)\ket{\text{g}}+\sin(\vartheta)\ket{\text{e}}$ and $\vartheta=\Theta/2$ is half the pulse area of the perfect state preparation pulse. This allows our emitter to be prepared with some amount of coherence that will then be transferred to our photonic state.

For perfect state preparation, we can have at most one photon emitted from the system and so the photonic state $\hat{\varrho}$ is truncated at $n\leq 1$. As a consequence, any potential photon number coherence must be contained between the $0$- and $1$-photon subspaces. Thus, from the input-output relation $\bd-\bd_0=\sqrt{\gamma\eta}\sigd$, we must have $\sqrt{p_0p_1}\zeta_{1,0}(t)=\sqrt{\gamma\eta}\braket{\sigd(t)}$ since all higher-order coherence functions vanish: $\zeta_{n+1,n}(t)=0$ for $n>0$. This simplified scenario allows us to obtain $\zeta_{1,0}(t)$ using a standard approach, which will then allow us to verify that the self-homodyne photon number decomposition approach reproduces this known solution.

For the simplified scenario described above, it is straightforward to solve for $\zeta_{1,0}(t)$ without the photon number decomposition. To do so, we can first recognize that $p_n=0$ for $n>1$ and so we must have $p_1=\eta\sin^2(\vartheta)$ and $p_0=1-p_1$. We can then apply the full propagation superoperator $\mathcal{U}(t,0)$ to compute $\braket{\sigd(t)}=\text{Tr}\left(S\mathcal{U}(t,0)\hat{\rho}(0)\right)$ and obtain
\begin{equation}
\label{chapte3eq:zeta1Simple}
    \zeta_{1,0}(t)=\sqrt{\gamma F_\eta}e^{-\Gamma t/2-i\omega t},
\end{equation} 
where $F_\eta=\cos^2(\vartheta)/(1-\eta\sin^2(\vartheta))$ is a factor that captures decoherence due to photon losses. This factor is important in the context of quantum repeaters \cite{rozpkedek2019near} and it arises again in section \ref{chapter4:entanglementgeneration}. 

Using $\braket{\sigd(t)}$, we can also compute the self-homodyne signal $V_\text{SH}$ from Eq.~(\ref{chapter3eq:Vshdef}):
\begin{equation}
    V_\text{SH} = \lambda^{(1)}\cos(\phi)=I\cos^2(\vartheta)\cos(\phi),
\end{equation}
where $I=\gamma/\Gamma$ and the self-homodyne signal amplitude $\lambda^{(1)}=I\cos^2(\vartheta)$ is the total integrated one-photon coherence. We can see that the magnitude of the number coherence is limited by pure dephasing in the same way as the HOM visibility. To see this explicitly, we can compute the one-photon temporal density function for this model in a similar way to $\zeta_{1,0}$ using $p_1\xi_1(t,t^\prime)=\eta\gamma\braket{\sigu(t^\prime)\sigd(t)}$, which gives
\begin{equation}
\label{chapter3eq:singlephotondensityTLS}
    \xi_1(t,t^\prime)=\gamma e^{-\gamma(t+t^\prime)/2 -i\omega(t-t^\prime)-\gamma^\star|t-t^\prime|}.
\end{equation}
From this, we derive the indistinguishability $\iint|\xi(t,t^\prime)|^2dtdt^\prime=I$, which in this scenario is equivalent to the HOM visibility. Knowing $\zeta_{1,0}$, $\xi_1$, $p_1=\eta\sin^2\vartheta$, and $p_0=1-p_1$, we have now reconstructed the source field photonic density operator $\hat{\varrho}$ of the waveguide arising from the initial state $\hat{\rho}(0)$ of the emitter.

If more than one photon can be emitted, then the amplitude $\braket{\sigd(t)}$ and correlation function $\braket{\sigu(t^\prime)\sigd(t)}$ will have contributions from multiple photon number subspaces. Hence, the approach illustrated above will not allow us to reconstruct the coherence between specific photon-number subspaces. In section \ref{chapter2:conditionalcorrelations}, I illustrated how to decompose $\braket{\sigu(t^\prime)\sigd(t)}$ into different photon number subspaces. Let us now consider a self-homodyne measurement to decompose $\braket{\sigd(t)}$ in a similar way.

Consider the system describing the emission of two identical photonic states. One photonic state, which I will denote using the subscript $\mathrm{r}$, provides the reference frame for our phase measurement. The other, which I will denote using $\mathrm{s}$, is the photonic state that we would like to measure. These two states could be produced by two identical emitters, or by a single emitter sequentially excited, so long as the two photonic states are initially uncorrelated. For conceptual simplicity, let us describe the situation using two identical emitters. The reduced system evolution of two identical and independent emitters can be described by the Liouville superoperator $\mathcal{L}=\mathcal{L}_\mathrm{r}\otimes\mathcal{I}+\mathcal{I}\otimes\mathcal{L}_\mathrm{s}$ where $\mathcal{L}_\mathrm{r}=\mathcal{L}_\mathrm{s}$ is given as above.

I will now outline the general idea of the self-homodyne decomposition. Suppose that the waveguide modes collecting emission from these two emitters are interfered at a balanced beam splitter with the relative phase $\phi$. Then, the measurement of a single photon at either detector will induce a joint measurement of the emitters described by the jump superoperators $\mathcal{J}_\pm\hat{\rho}=(\eta\gamma/2)(\sigd_\mathrm{r}\pm\sigd_\mathrm{s}e^{i\phi})\hat{\rho}(\sigu_\mathrm{r}\pm\sigu_\mathrm{s}e^{-i\phi})$ \cite{carmichael}, where $\sigma_\mathrm{r}=\sigd\otimes\hat{I}$ and $\sigd_\mathrm{s}=\hat{I}\otimes\sigd$. Here, I assume that both inputs of the beam splitter have the same intensity and both detectors have the same detection efficiency $\eta_\mathrm{d}$ so that $\eta=\eta_\mathrm{r}\eta_\mathrm{t}\eta_\mathrm{d}$, where $\eta_\mathrm{t}$ the transmission efficiency. Using these jump superoperators, we have that $\mathcal{L}_\mathbf{0}=\mathcal{L}-\mathcal{J}_+-\mathcal{J}_-$ describes the reduced system evolution conditioned on no detection at either detector of our self-homodyne setup. We can then decompose the reduced system dynamics conditioned on the individual photon numbers at each detector.

Since we have at most two photons emitted in our idealized scenario (at most one from each emitter), the full propagation superoperator is decomposed into 7 conditional propagators. The vacuum case $\mathcal{U}_\mathbf{0}(t,t_0)$ is computed from $\mathcal{L}_\mathbf{0}$ in the usual way. Then, the two one-photon propagators are given by
\begin{equation}
    \mathcal{U}_{\pm}(t,t_0)=\int_{t_0}^t\mathcal{U}_\mathbf{0}(t,t^\prime)\mathcal{J}_\pm\mathcal{U}_\mathbf{0}(t^\prime,t_0)dt^\prime,
\end{equation}
and the four two-photon cases are
\begin{equation}
\begin{aligned}
    \mathcal{U}_{\pm\pm}(t,t_0)&=\int_{t_0}^t\mathcal{U}_\mathbf{0}(t,t^\prime)\mathcal{J}_\pm\mathcal{U}_\pm(t^\prime,t_0)dt^\prime\\
    \mathcal{U}_{\pm\mp}(t,t_0)&=\int_{t_0}^t\mathcal{U}_\mathbf{0}(t,t^\prime)\mathcal{J}_\mp\mathcal{U}_\pm(t^\prime,t_0)dt^\prime.
\end{aligned}
\end{equation}
Here, I am distinguishing between the detection order, but this may not always be the case in practice. 

Before moving on to solve $\zeta_{1,0}$ with this approach, let us take a look at the detection probabilities of the self-homodyne measurement. These are given by
\begin{equation}
\begin{aligned}
    \prb{0} &= \lim_{t_\mathrm{f}\rightarrow \infty}\tr{\mathcal{U}_\mathbf{0}(t_\mathrm{f},t_0)\hat{\rho}(t_0)} = \left(1-\eta\sin^2(\vartheta)\right)^2=p_0^2\\
    \prb{\pm} &=\lim_{t_\mathrm{f}\rightarrow \infty}\tr{\mathcal{U}_\pm(t_\mathrm{f},t_0)\hat{\rho}(t_0)} = p_1\left(p_0\pm I\cos^2(\vartheta)\cos(\phi)\right)\\
    \prb{\pm\pm} &= \lim_{t_\mathrm{f}\rightarrow \infty}\tr{\mathcal{U}_{\pm\pm}(t_\mathrm{f},t_0)\hat{\rho}(t_0)} = \frac{1}{4}(1+I)\eta^2\sin^4(\vartheta)=\frac{1}{4}(1+I)p_1^2\\
    \prb{\pm\mp} &= \lim_{t_\mathrm{f}\rightarrow \infty}\tr{\mathcal{U}_{\pm\mp}(t_\mathrm{f},t_0)\hat{\rho}(t_0)} = \frac{1}{4}(1-I)\eta^2\sin^4(\vartheta)=\frac{1}{4}(1-I)p_1^2.
\end{aligned}
\end{equation}
The detection probabilities sum to 1, as expected, which verifies the completeness of the decomposition. It is also clear from these probabilities that the self-homodyne decomposition approach captures the degradation of the HOM visibility due to dephasing, since the indistinguishability $0\leq I\leq1$ dictates the photon bunching probability.

Using the above detection probabilities, we can compute the normalized self-homodyne signal within the single-photon subspace of the measured joint photonic state
\begin{equation}
    V_{\text{SH},1}=\frac{\prb+-\prb-}{\prb++\prb-} = I\cos(\phi)F_\eta\geq V_\text{SH}.
\end{equation}
It is important to note that $V_{\text{SH},1}$ is normalized by the probability of measuring the joint photonic state in the single-photon subspace whereas the full signal $V_\text{SH}$ is normalized by the total intensity. Hence, even though the only nonzero signal arises from the single-photon subspace, $V_{\text{SH},1}$ is not necessarily equal to the total signal $V_\text{SH}$. This result illustrates two important points about self-homodyne measurements and photon-number post-selection. In general, post-selecting on photon number resolved detection produces loss-dependent quantities. However, post-selection can increase the self-homodyne signal and hence purify the coherence, which is evident from $V_{\text{SH},1}\geq V_\text{SH}$. This inequality saturates in the limit that $\eta\sin^2(\vartheta)\ll 1$, which is valid when losses are high or driving is weak so that multi-photon events at the detectors are rare.

Let us now turn back to the original goal of finding $\zeta_{1,0}$. In the same spirit as section \ref{chapter2:temporaldensityfunctions}, we can compare the general form of the photonic state density operator of the joint waveguide system to the correlations of our two-emitter system. Doing this, we can see that \begin{equation}
\label{chapter3eq:vsh1exp}
    \braket{\mathcal{V}_\text{SH}(t)}_1=p_{1,0}\text{Re}\left(\zeta_{\mathrm{s},1,0}(t)\zeta^*_{\mathrm{r},1,0}(t)\right)\equiv p_{0,1}\left(\zeta_{\mathrm{r},1,0}(t)\right)^2\cos(\phi),
\end{equation}
where $\mathcal{V}_\text{SH}=\mathcal{J}_+-\mathcal{J}_-$ is the self-homodyne signal superoperator, $p_{1,0}=\prb{+}+\prb{-}=2p_1p_0$ is the probability of detecting one photon, and $\braket{~~}_1$ is the expectation over the single-photon subspace of the joint photonic state. The coherence function $\zeta_{\mathrm{r},1,0}$ is \emph{defined} to be real because we choose it to be the phase reference for the self-homodyne measurement. Also, since we have two identical photonic states, $\zeta_{\mathrm{s},1,0}(t)=\zeta_{\mathrm{r},1,0}(t)e^{i\phi}$. Eq.~(\ref{chapter3eq:vsh1exp}) is analogous to classical homodyne detection where, instead of $\zeta_{\mathrm{r},1,0}$, we would have a classical coherent state amplitude. Computing this expression for our simple scenario gives
\begin{equation}
\label{chapter3eq:condevolVSH}
\begin{aligned}
    \braket{\mathcal{V}_\text{SH}(t)}_1&=\tr{\mathcal{U}_\mathbf{0}(\infty,t)\mathcal{V}_\text{SH}\mathcal{U}_\mathbf{0}(t,t_0)\hat{\rho}(t_0)}\\
    &=\frac{\gamma}{2} e^{-\Gamma t}\cos(\phi)\sin^2(\Theta).
\end{aligned}
\end{equation}
Finally, combining Eqs.~(\ref{chapter3eq:vsh1exp}) and (\ref{chapter3eq:condevolVSH}), we find
\begin{equation}
    \zeta_{\mathrm{s},1,0}(t)= \sqrt{\gamma F_\eta}e^{-\Gamma t/2}e^{i\phi}
\end{equation}
in the rotating frame of the reference state. By adding back the rotating frame $e^{-i\omega t}$ and taking $\phi=0$, we recover the solution given in Eq.~(\ref{chapte3eq:zeta1Simple}).

Of course, the power of this self-homodyne photon number decomposition is not fully illustrated with this simple example since we are able to derive the same result in a much simpler way at the beginning of the section. However, the agreement between these two approaches verifies that the self-homodyne measurement post-selected in the single-photon subspace extracts the coherence between the vacuum and the single-photon component. It also allows us to access the coherence functions between individual photon-number subspaces even when the emitter dynamics includes both multi-photon processes and excess decoherence for a wide variety of emitter models. Furthermore, it corresponds to a physically realistic optical setup that is already commonly used to characterize the indistinguishability of single photons.

\begin{figure}
    \centering
    \hspace{-67mm}(a)\hspace{75mm}(b)\\
    \includegraphics[width=0.49\textwidth,trim=0 2 0 0,clip]{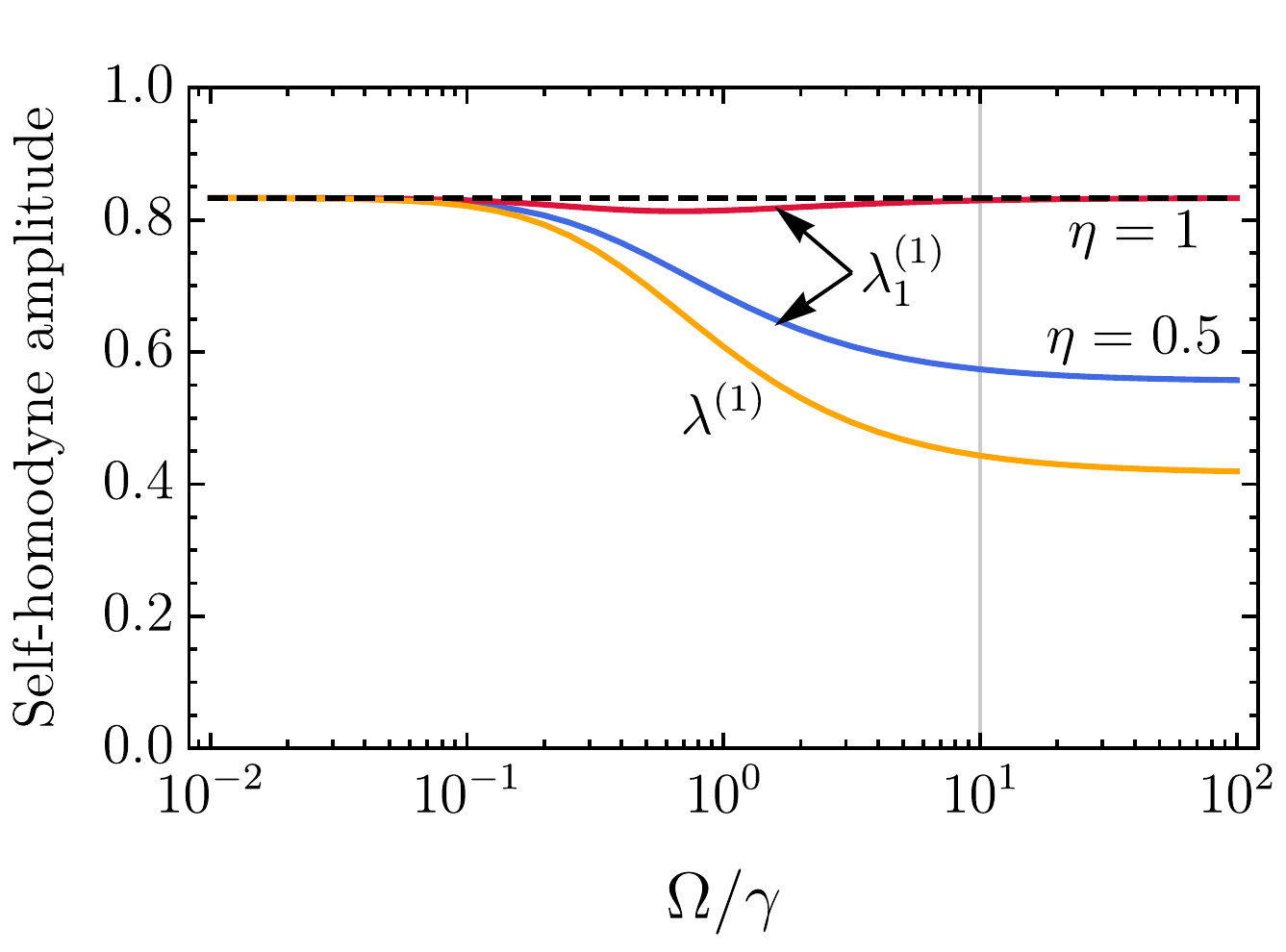}
    \includegraphics[width=0.49\textwidth,trim=0 0 0 1,clip]{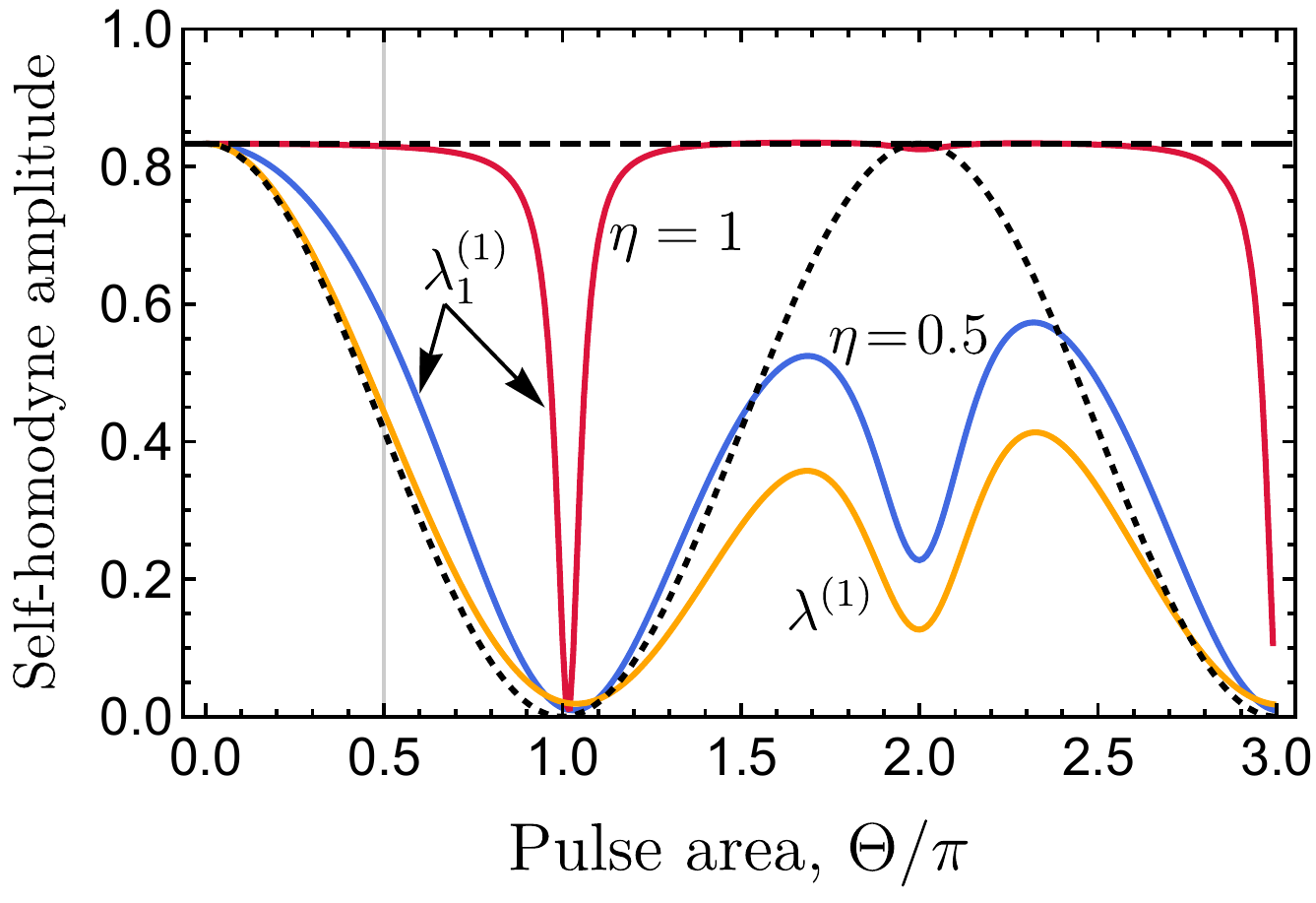}
    \caption[The self-homodyne signal amplitude for a purely-dephased emitter driven by coherent pulses.]{\small\textbf{The self-homodyne signal amplitude for a purely-dephased emitter driven by coherent pulses.} (a) The total self-homodyne signal amplitude $\lambda^{(1)}$ (orange) and the signal amplitude in the single-photon subspace $\lambda^{(1)}_1$ (red for $\eta=1$ and blue for $\eta=0.5$) as a function of Rabi frequency $\Omega$ for a fixed pulse area of $\Theta=\pi/2$. (b) The self-homodyne signal amplitudes for different pulse areas while fixing $t_\mathrm{p}\gamma=\pi/20$ and varying $\Omega$. The dotted black sinusoidal line indicates the curve $\lambda^{(1)}=I\cos^2(\Theta/2)$ that is expected without re-excitation processes. The vertical gray lines indicate the values consistent in both plots. For both plots, I have fixed $\gamma^\star=0.1\gamma$ corresponding to $I=0.833$, which is indicated by the horizontal black dashed line.}
    \label{chapter3fig:lambda1ssquarepulse}
\end{figure}

As an illustrative example of the self-homodyne photon number decomposition approach, I have computed the first-order integrated coherence of the self-homodyne signal amplitude post-selected in the single-photon subspace $\lambda^{(1)}_1=V_{\text{SH},1}/\cos(\phi)$ for a purely-dephased two-level emitter driven with a finite square pulse (see Fig.~\ref{chapter3fig:lambda1ssquarepulse}). I have plotted this value alongside the total self-homodyne signal amplitude $\lambda^{(1)}$ for two different measurement efficiency values. Notably, there is a drastic suppression of the amplitude $\lambda^{(1)}$ at $\Theta=2\pi$ which is not captured when neglecting re-excitation processes. This suppression is alleviated when post-selecting in the single-photon subspace provided that loss is not significant. From this result, we can see that the amount of coherence in the lossy regime is small unless the excitation power is weak (either $\Omega\ll\gamma$ or $\Theta\ll\pi$). This is a well-known conclusion and the reason why single-photon repeater protocols require weak excitation \cite{duan2001long}. I will discuss more about this consequence in section \ref{chapter4:entanglementgeneration} in the context of heralded spin-spin entanglement generation. Before getting to that discussion, I would like to briefly touch on how self-homodyne measurements can also be used to characterize photon number Bell states encoded in time.

\subsection{Single-photon entanglement}
\label{chapter3:photonicbellstates}

It is well-known that a single photon can give rise to multi-partite entanglement \cite{van2005single}. The simplest example is achieved by separating a single photon into multiple spatial paths using beam splitters. A balanced beam splitter takes a single-photon state $\ket{1}$ and divides it into two equal spatial modes, where it becomes a single-particle Bell state of the form $\ket{\psi^+}=(\ket{10}+\ket{01})/\sqrt{2}$. This operation can be repeated using multiple beam splitters, giving rise to a single excitation encoded in many spatial modes. The multi-partite single-particle entanglement carried by a single photon also becomes evident when mapped onto an ensemble of atoms using a frequency comb atomic memory \cite{zarkeshian2017entanglement}, where it generates a Dicke state encoded in frequency.

As we have seen so far, the time dynamics of a single-photon wavepacket emitted by a two-level system into a waveguide can be described using a temporal wavefunction $f_1(t)$. In the more general case, the photonic state can be described by a temporal density function $\xi_1(t,t^\prime)$ where $\xi_1(t,t^\prime)=f_1(t)f_1^*(t^\prime)$ for a pure state. Having access to the time-dependent dynamics of the single-photon wavepacket allows us to probe deeper quantum properties of a single photon, such as the amount of single particle entanglement encoded in time.

Consider the ideal single-photon state in the waveguide
\begin{equation}
    \ket{1} = \int f_1(t)\bu(t)\ket{0}dt.
\end{equation}
Suppose now that we have a detector that can resolve the time dynamics of this system. We can then divide our single photon into two time bins: an early photon which arrives before time $T$ since the state preparation, and a late photon that arrives after time $T$. I will refer to $T$ as the time bin threshold. Let us denote the early photon arriving during the first interval $\Delta S=(-\infty,T)$ using the subscript $\mathrm{s}$ and the late photon arriving during the interval $\Delta T=(T,\infty)$ using the subscript $\mathrm{t}$. If our detector is perfectly capable of resolving these two time bins, we can write our single photon as a bipartite state
\begin{equation}
    \ket{1} = \sqrt{\mu_\mathrm{s}}\su\ket{0}+\sqrt{\mu_\mathrm{t}}\tu\ket{0}.
\end{equation}
The corresponding second-quantized time-bin modes are given by
\begin{equation}
\begin{aligned}
\label{chapter3eq:timebinmodes}
    \su &= \frac{1}{\sqrt{\mu_\mathrm{s}}}\int_{\Delta S}f_1(t)\bu(t)dt\\
    \tu &= \frac{1}{\sqrt{\mu_\mathrm{t}}}\int_{\Delta T} f_1(t)\bu(t)dt,
\end{aligned}
\end{equation}
where $\mu_\mathrm{s}=\int_{\Delta S}|f_1(t)|^2dt$ and $\mu_\mathrm{t}=\int_{\Delta T}|f_1(t)|^2dt$.

Suppose now that we have an imperfect photonic state $\hat{\varrho}$. We can then compute the density matrix elements in the second-quantized picture $\varrho_{nmkl}=\braket{nm|\hat{\varrho}|kl}$, $\ket{nm}=(\su)^n(\tu)^m\ket{0}/\sqrt{n!m!}$ for our ideal time bin modes. For simplicity, I use the abbreviated notation $\varrho_{1010}=\varrho_\mathrm{ss}$, $\varrho_{0101}=\varrho_\mathrm{tt}$, and $\varrho_{1001}=\varrho_\mathrm{st}$. Using this second quantized picture, it is then straightforward to compute the entanglement fidelity and concurrence (see sections \ref{chapter1:fidelity} and \ref{chapter1:concurrence}) of the imperfect single photon in the two-qubit photon-number subspace of $\{\ket{00},\ket{10},\ket{01},\ket{11}\}$.

The entanglement carried by a single photon can be seen to be rooted in the process of spontaneous emission. Recall from section \ref{chapter1:spontaneousemission} that, during spontaneous emission, the emitter briefly becomes entangled with its environment. This manifests as a dip in the purity of the emitter state. For example, consider a two level system in its excited state $\ket{\text{e}}$. As time passes, it decays to its ground state by spontaneously emitting a photon into the waveguide through a \emph{coherent} interaction with the continuum of modes. At time $T$ after state preparation, the joint emitter-waveguide system is in the state $\alpha\ket{\text{e}}\ket{0}+\beta\ket{\text{g}}\ket{1}$, where $\ket{0}$ and $\ket{1}$ represent the state of the time-bin mode $\sd$. After complete emission, the state becomes $\ket{\text{g}}\left(\beta\ket{10}+\alpha\ket{01}\right)=\ket{\mathrm{g}}(\beta\su+\alpha\tu)\ket{0}$, where the emitter is now separable leaving a single-photon entangled state in time.

\begin{figure}
    \centering
    \hspace{-58mm}(a)\hspace{75mm}(b)\\
    \includegraphics[width=0.48\textwidth]{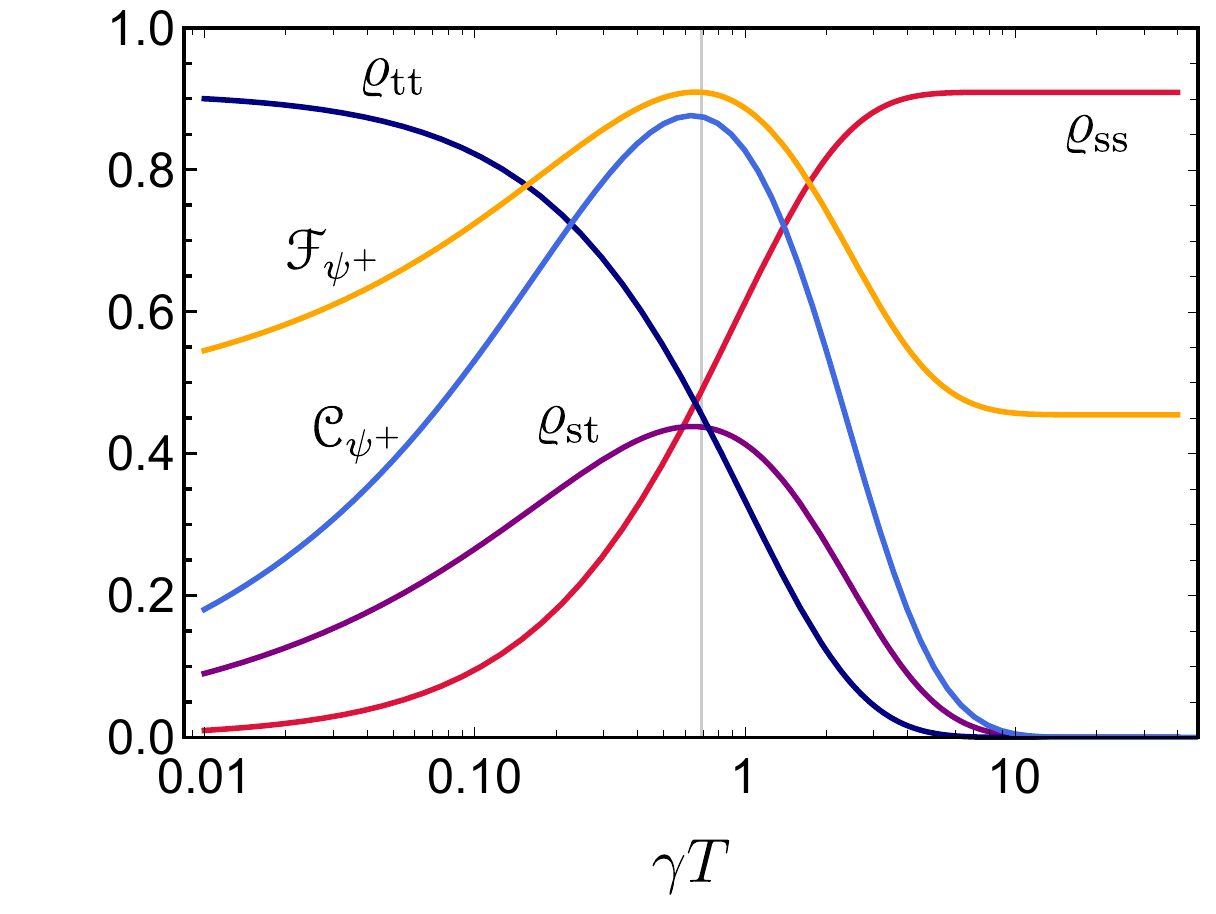}
    \includegraphics[width=0.44\textwidth]{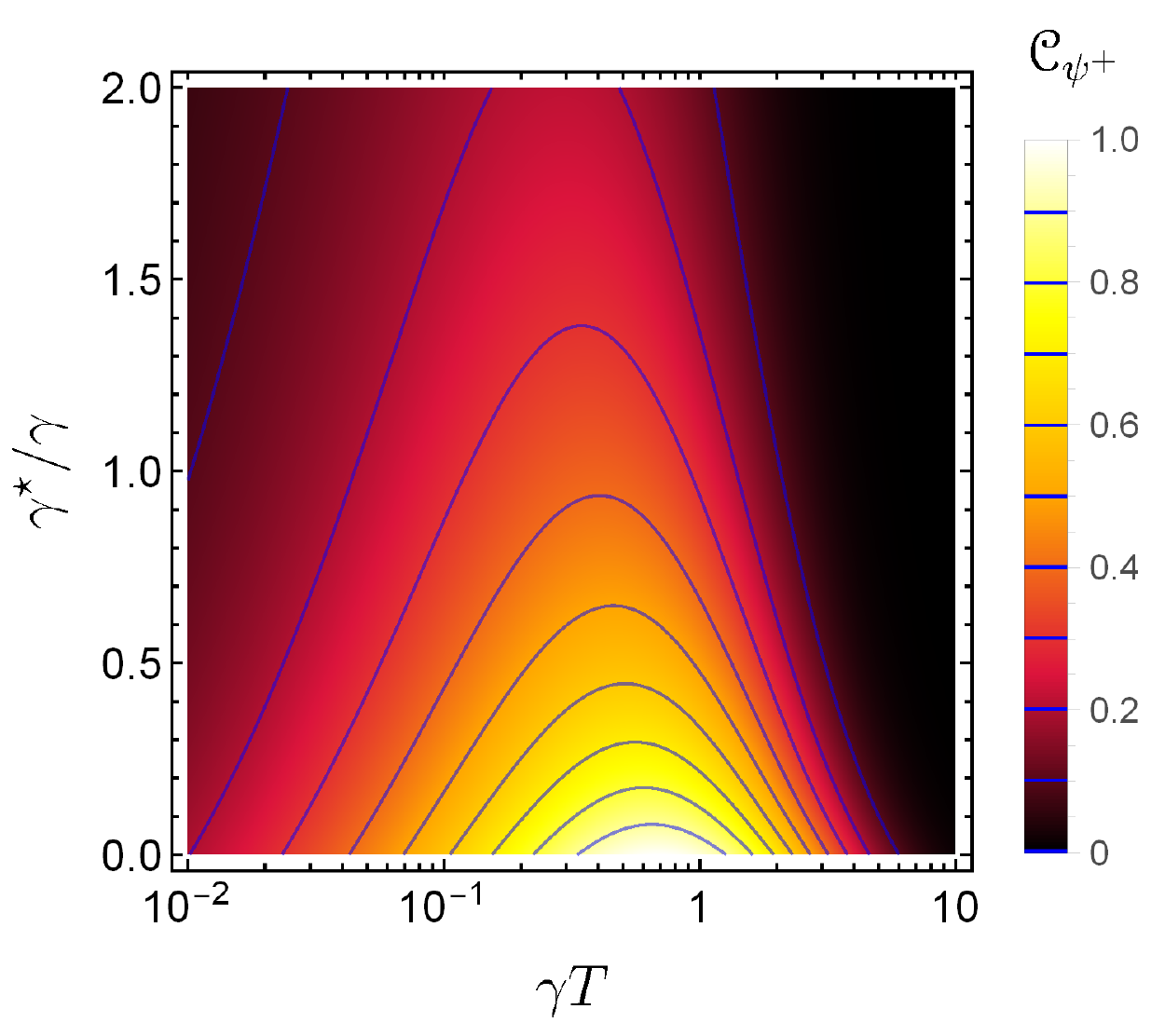}
    \caption[Bipartite time-bin decomposition of a single photon: density matrix elements, Bell-state fidelity, and concurrence.]{\small\textbf{Bipartite time-bin decomposition of a single photon: density matrix elements, Bell-state fidelity $\euscr{F}_{\psi^+}$, and concurrence $\euscr{C}_{\psi^+}$.} (a) The density matrix elements along with Bell-state fidelity and concurrence in the second-quantized time-bin basis defined by the threshold $T$. For this panel, I have used $\gamma^\star=0.1\gamma$. The vertical line shows the half-life $\gamma T=\ln(2)$. (b) The concurrence as a function of time bin threshold $T$ and pure dephasing rate $\gamma^\star$.}
    \label{chapter3fig:psiplusFormal}
\end{figure}

Following the example above, let us consider the impure single photon produced by a perfectly excited purely-dephased two-level emitter, as described by $\xi_1$ in Eq.~(\ref{chapter3eq:singlephotondensityTLS}). If the emitter has no pure dephasing, we expect that $f_1(t)=\sqrt{\gamma}e^{-\gamma t/2-i\omega t}$. Using $\xi_1$ and $f_1$, we can then compute the entanglement figures of merit for $\varrho$. For this scenario, Fig.~\ref{chapter3fig:psiplusFormal} shows the density matrix elements in the bipartite time bin basis along with the computed Bell-state fidelity and entanglement concurrence as a function of time bin threshold $T$. The fidelity compared to the ideal Bell state $\ket{\psi^+}=(\su+\tu)\ket{0}/\sqrt{2}$ and the concurrence are maximized near when the time bin threshold is chosen to be at the half-life of the emitter $T=\ln{2}/\gamma$. However, the presence of pure dephasing shifts this ideal choice of threshold to smaller values.

\begin{figure}
    \centering
    \hspace{-58mm}(a)\hspace{75mm}(b)\\
    \includegraphics[width=0.49\textwidth]{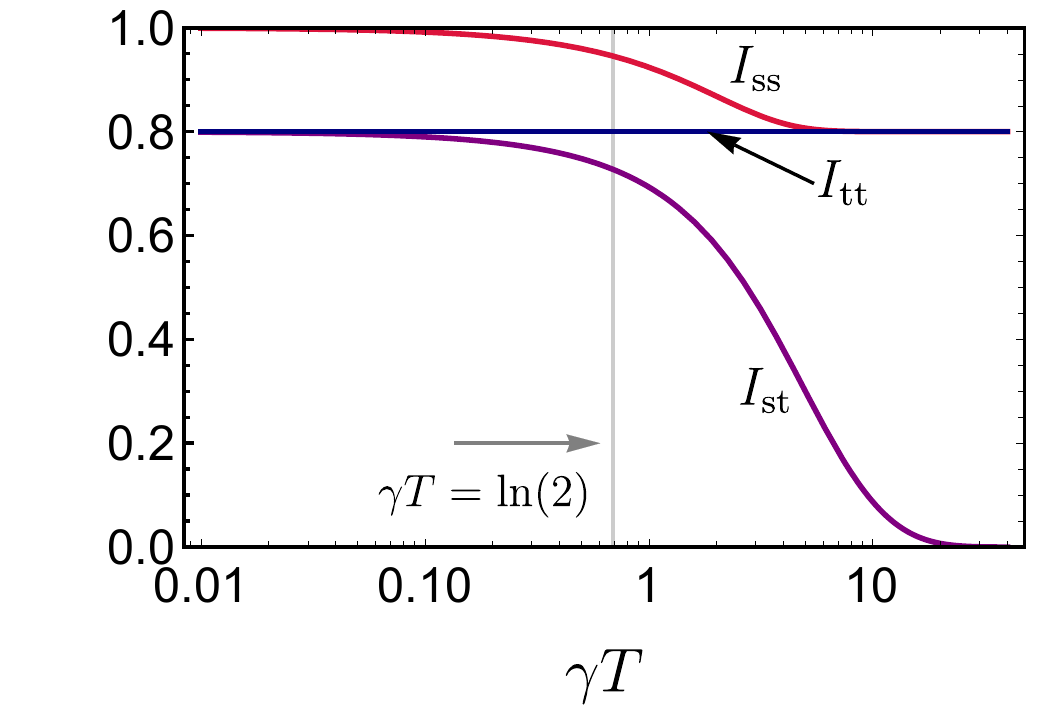}
    \includegraphics[width=0.49\textwidth]{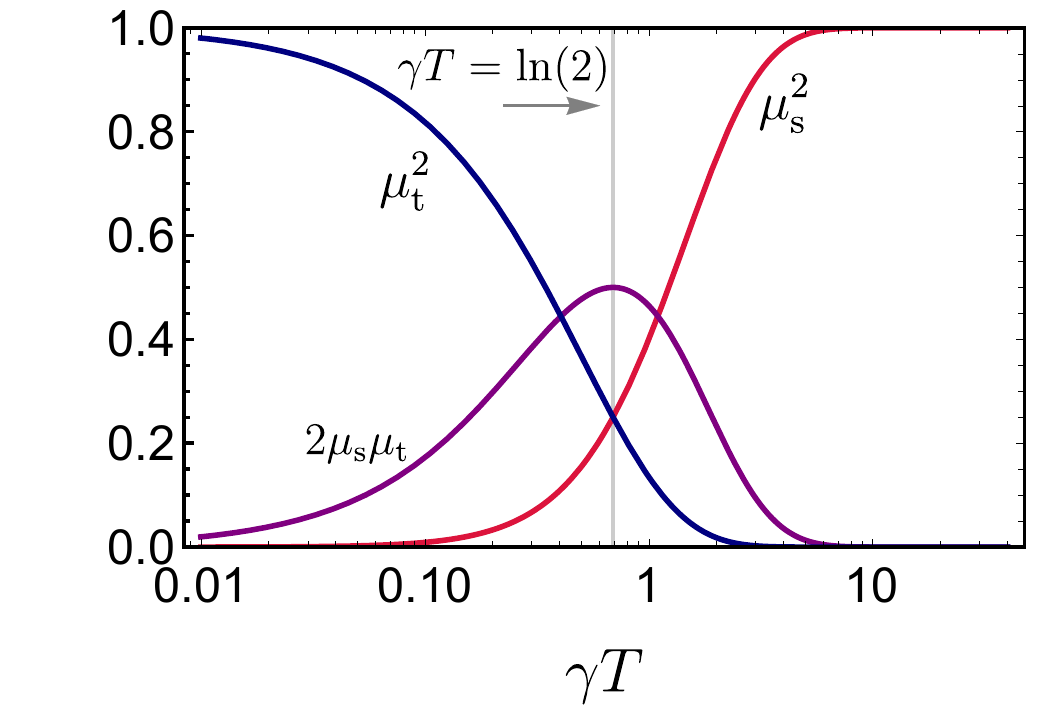}
    \caption[Bipartite time-bin decomposition of the single-photon indistinguishability.]{\small\textbf{Bipartite time-bin decomposition of the single-photon indistinguishability.} (a) The three components of the bipartite indistinguishability $I_\mathrm{ss}$, $I_\mathrm{tt}$ and $I_\mathrm{st}$ composing the total indistinguishability $I=0.8$. (b) The weights of contribution for each of the three components of indistinguishability in panel (a) normalized to the squared intensity $\mu^2$. The vertical line corresponds to the half-life of the emitter decay.}
    \label{chapter3fig:psiplus}
\end{figure}

Let us now discuss the properties of photons within each time bin for an impure single-photon state. From $\xi_1$, we can compute the total indistinguishability $I=\iint|\xi_1(t,t^\prime)|^2dt=\gamma/\Gamma$ as before. The indistinguishability $I$ can then be divided into its bipartite components for modes $\sd$ and $\td$ by:
\begin{equation}
    I = \frac{\mu_\mathrm{s}^2I_\mathrm{ss}+2\mu_\mathrm{s}\mu_\mathrm{t}I_\mathrm{st}+\mu_\mathrm{t}^2I_\mathrm{tt}}{\mu^2}
\end{equation}
where $\mu =\mu_\mathrm{s}+\mu_\mathrm{t}$ is the total intensity. The constituent parts of indistinguishability for post-selected intervals $\Delta S$ and $\Delta T$ are also explicitly related to the total indistinguishability $I$ of the single-photon:
\begin{equation}
\label{chapter3eq:Idecompopsiplus}
\begin{aligned}
    I_\mathrm{ss} &= \frac{1}{\mu_\mathrm{s}^2}\int_{\Delta S}\int_{\Delta S}\left|\xi_1(t,t^\prime)\right|^2dtdt^\prime=\frac{I}{\beta^4}\left[1-\frac{\alpha^4-2I\alpha^{2/I}}{1-2I}\right]\\
    I_\mathrm{st} &=\frac{1}{\mu_\mathrm{s}\mu_\mathrm{t}}\int_{\Delta S}\int_{\Delta T} \left|\xi_1(t,t^\prime)\right|^2dtdt^\prime=\frac{I^2}{\alpha^2\beta^2}\left[\frac{\alpha^4-\alpha^{2/I}}{1-2I}\right]\\
    I_\mathrm{tt} &= \frac{1}{\mu_\mathrm{t}^2}\int_{\Delta T}\int_{\Delta T}\left|\xi_1(t,t^\prime)\right|^2dtdt^\prime = I,
\end{aligned}
\end{equation}
where $\mu_\mathrm{s}=\int_{\Delta S} \xi_1(t,t)dt=\beta^2=1-e^{-\gamma T}$ and $\mu_\mathrm{t}=\int_{\Delta T}\xi_1(t,t)dt=\alpha^2=1-\beta^2$. Fig.~\ref{chapter3fig:psiplus} shows this time bin decomposition of indistinguishability in panel~(a) and the weights of their contribution to the total indistinguishability in panel (b).

The bipartite time bin decomposition illustrates how temporal post-selection of photons that are emitted early can increase the indistinguishability $I_\mathrm{ss}$ at the cost of a reduced efficiency $\mu_\mathrm{s}< 1$. This fast temporal post-selection eliminates the tail of the photon's exponential profile, creating a sharper single-photon pulse in time. The temporal truncation can alternatively be seen as a spectral broadening of the post-selected photon so that it begins to overcome the incoherent broadening induced by dephasing. However, the indistinguishability $I_\mathrm{tt}$ for post-selected photons that are emitted late remains constant regardless of $T$. This is because rejecting early photons does not alter the temporal shape of late photons. That is, the tail always looks like the tail no matter how much we remove from the beginning of the profile. Hence, in this case, the spectrum is not broadened and the indistinguishability is not increased.

\begin{figure}
    \centering
    \hspace{-58mm}(a)\hspace{75mm}(b)\\
    \includegraphics[width=0.49\textwidth,trim=50 0 50 0,clip]{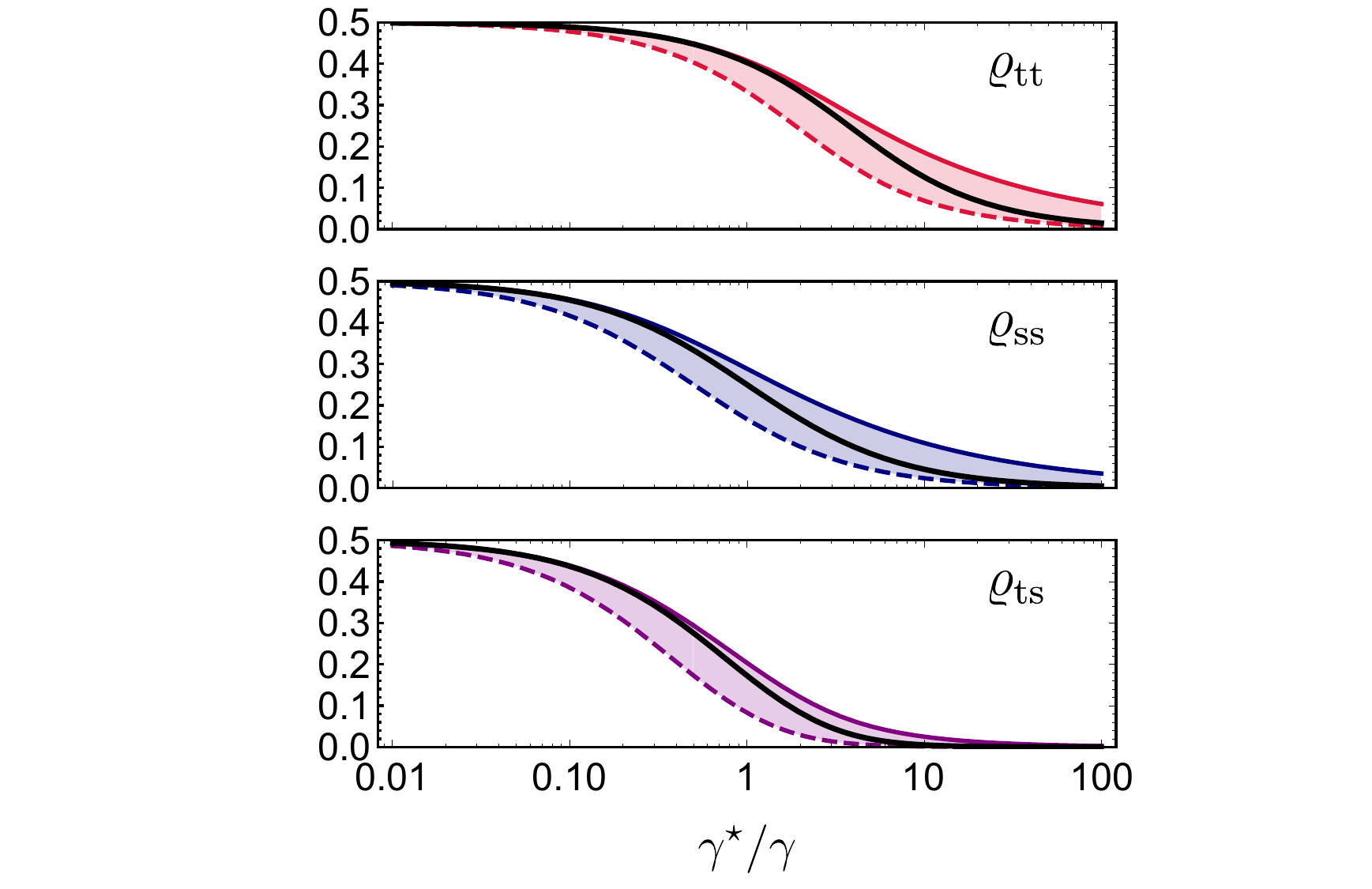}
    \includegraphics[width=0.49\textwidth,trim=10 0 10 0,clip]{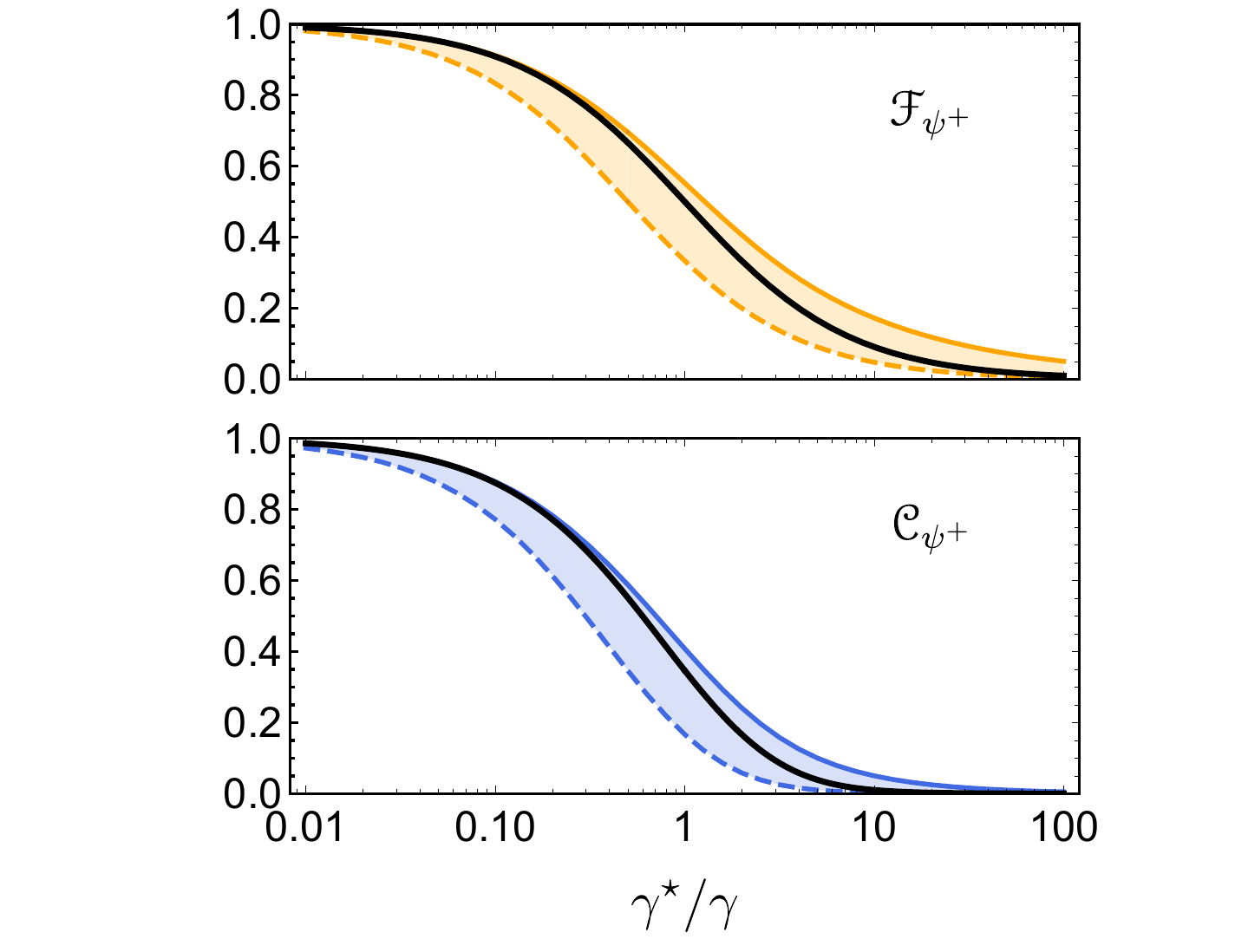}
    \caption[A comparison of the bipartite time bin density matrix elements and their estimates from indistinguishability measurements.]{\small\textbf{A comparison of the bipartite time bin density matrix elements and their estimates from indistinguishability measurements.} (a) The density matrix elements in the second-quantized basis (solid black curves) compared to the upper and lower estimates using bipartite indistinguishability measurements (colored regions) for $\gamma T=\ln(2)$. (b) The same comparison but for fidelity $\euscr{F}_{\psi^+}$ and concurrence $\euscr{C}_{\psi^+}$ computed using the values in panel (a). }
    \label{chapter3fid:psipluscompare}
\end{figure}

The value of $I_\mathrm{st}$ quantifies the amount of two-mode entanglement that the single-particle state carries. It is also much easier to measure than $\varrho_\mathrm{st}$ in practice because $\varrho_\mathrm{st}$ necessarily requires knowledge about the ideal mode amplitude $f_1$. It is then natural to ask how closely the post-selected indistinguishability measurements can estimate the density matrix elements in the second-quantized picture of the ideal time bin modes. To this end, we can find a good estimate by noting that the post-selected values of $I$ are very similar to the overlaps needed to obtain the density matrix elements, except that they account for decoherence twice compared to the elements of $\varrho$ and are normalized differently with respect to intensity. From this, we know that a lower bound on the density matrix elements must be $I_\mathrm{ab}\sqrt{\mu_a\mu_b}\leq\varrho_{ab}$. However, we can do better than this if we attempt to correct for this double-counting of decoherence by taking a square root of the weighted component of $I$. That is, we can also guess that $\varrho_{ab}\sqrt{\mu_a\mu_b I_{ab}}$ for $\{a,b\}\in\{\mathrm{t},\mathrm{s}\}$. Fig.~\ref{chapter3fid:psipluscompare} shows a comparison between these two estimated elements using $I_{ab}$ compared to the true values $\varrho_{ab}$ along with the implied fidelity and concurrence values. From this, we can see that the latter approximation will overestimate the values of the density matrix elements, but it is a much more reasonable approximation for highly-indistinguishable photons $(\gamma^\star<\gamma)$ than the lower-bound approximation.

The single-particle photon number Bell state $\ket{\psi^+}$ and its analysis can be seen as a time-resolved perspective on indistinguishability measurements. However, the underlying concept of using spontaneous emission as a resource to generate entanglement can be extended to more novel situations. I will explore this extension in the following section.

\subsection{Two-photon coherence}

If we can create the $\ket{\psi^+}$ Bell state by exploiting the entangling nature of spontaneous emission, it is natural to ask whether the other three Bell states can also be generated in this way. Interestingly, the answer seems to be yes! The state $\ket{\psi^-}$ can be produced by performing a rapid relative phase flip between state $\ket{\text{e}}$ and $\ket{\text{g}}$ at the half-life of emission, essentially flipping the phase of the dipole half-way through emitting the photon. The states $\ket{\phi^\pm}$ can be generated by applying a second excitation pulse at time $T$ after the first pulse to flip the population of the two-level system. This latter case can be explained by noting that a perfect $\pi$ pulse takes $\alpha\ket{\text{e}}\ket{0}+\beta\ket{\text{g}}\ket{1}$ to $\alpha\ket{\text{g}}\ket{0}+\beta\ket{\text{e}}\ket{1}$. This leaves the photonic modes in the state $\alpha\ket{00}+\beta\ket{11}=(\alpha + \beta\su\tu)\ket{0}$.

The photon number $\ket{\psi^\pm}$ Bell states are very different from the $\ket{\phi^\pm}$ Bell states. Although both carry the same average energy and have the same intensity profiles, they have very different quantum properties. The $\ket{\psi^\pm}$ states are single-photon states that carry significant temporal coherence whereas the $\ket{\phi^\pm}$ states have intensity correlations and carry two-photon coherence. Because of this, they require very different techniques to characterize. We can no longer rely on indistinguishability measurements to understand $\ket{\phi^\pm}$. In this section, I will analyze the state $\ket{\phi^+}$ generated by an imperfect emitter and discuss how to quantify the amount of entanglement using a self-homodyne setup.

To observe the intensity correlations caused by the state $\ket{11}$ within $\ket{\phi^+}$, we can use time-resolved $g^{(2)}$ measurements to look for coincidence counts between time bin modes. In addition, as I alluded to in section \ref{chapter3:selfhomodyne}, by simultaneously monitoring the self-homodyne signal and the HOM interference visibility as a function of the relative phase, it is possible to observe and quantify the effect of two-photon coherence between the vacuum state $\ket{00}$ and the two-photon state $\ket{11}$. This technique was employed in the experimental paper of Ref.~\ref{loredo2020deterministic} to characterize $\ket{\phi^+}$ produced by a coherently driven artificial atom using a self-homodyne setup.

Following the approach in the previous section, let us compute the photonic state $\hat{\varrho}$ corresponding to $\ket{\phi^+}$ generated from a purely-dephased two-level system. Let us again assume that we are using perfect excitation pulses so that we can neglect re-excitation. In this case, the photonic state is truncated at $n\leq 2$ photons and so the density matrix elements that we are interested in can be directly related to correlations of the reduced system operators without considering a photon-number decomposition. Let $\mathcal{S}$, $\mathcal{R}$, $\mathcal{J}=\mathcal{S}\mathcal{R}$ and $\mathcal{X}$ be the superoperators defined by $\mathcal{S}\hat{\rho}=\sigd\hat{\rho}$, $\mathcal{R}\hat{\rho}=\hat{\rho}\sigu$, and $\mathcal{X}\hat{\rho}=\sigd_x\hat{\rho}\sigd_x$. The superoperator $\mathcal{X}$ is the action of an ideal $\pi$-pulse, which we take to arrive at time $\tau$ after the initial state preparation. In addition, let us assume we start in the excited state $\hat{\rho}(0)=\ketbra{\text{e}}{\text{e}}$. Then, we can compute all the necessary time-dependent expectation values and correlation functions of our system in the standard way (see section \ref{chapter1:correlations}).

First, we can verify that the first-order coherence $\Lambda^{(1)}(t)=\braket{\sigd(t)}$ vanishes for all time because we are only ever using perfect $\pi$-pulses, leaving no coherence:
\begin{equation}
    \Lambda^{(1)}(t)=\left\{\begin{array}{ll}
    \sqrt{\gamma}~\!\tr{\mathcal{S}\hat{\rho}(t)}=0&0\leq t  \leq \tau\\
    \sqrt{\gamma}~\!\tr{\mathcal{S}\mathcal{U}(t,\tau)\mathcal{X}\hat{\rho}(\tau)}=0&\hphantom{0\leq~\!}t> \tau\\
    \end{array}\right.
\end{equation}
Second, we can show that the system produces no single-photon temporal coherence between time bins, but that each time bin itself can contain a single photon:
\begin{equation}
    G^{(1)}(t,t^\prime)=\left\{\begin{array}{ll}
    \gamma~\!\tr{\mathcal{R}\mathcal{U}(t^\prime,t)\mathcal{S}\hat{\rho}(t)}=\xi_1(t,t^\prime)&0\leq t \leq t^\prime\leq \tau\\
     \gamma~\!\tr{\mathcal{R}\mathcal{U}(t^\prime,\tau)\mathcal{X}\mathcal{U}(\tau,t)\mathcal{S}\hat{\rho}(t)}=0&0\leq t \leq \tau< t^\prime\\
    \gamma~\!\tr{\mathcal{R}\mathcal{U}(t^\prime,t)\mathcal{S}\mathcal{U}(t,\tau)\mathcal{X}\hat{\rho}(\tau)}=(\beta^2/\alpha^2)\xi_1(t,t^\prime)&0\leq \tau < t\leq t^\prime\\
    \end{array}\right.
\end{equation}
where now $\alpha^2=e^{-\gamma\tau}$ and $\beta^2=1-\alpha^2$ depend on the pulse separation $\tau$. Third, we can show that there are intensity correlations $G^{(2)}(t,t^\prime)$ between the two time bins, but not within each time bin:
\begin{equation}
    G^{(2)}(t,t^\prime)=\left\{\begin{array}{ll}
    \gamma^2~\!\tr{\mathcal{J}\mathcal{U}(t^\prime,t)\mathcal{J}\hat{\rho}(t)}=0&0\leq t \leq t^\prime\leq \tau\\
     \gamma^2~\!\tr{\mathcal{J}\mathcal{U}(t,\tau)\mathcal{X}\mathcal{U}(\tau,t)\mathcal{J}\hat{\rho}(t)}=\gamma^2e^{-(t+t^\prime-\tau)\gamma}&0\leq t \leq \tau< t^\prime\\
    \gamma^2~\!\tr{\mathcal{J}\mathcal{U}(t^\prime,t)\mathcal{J}\mathcal{U}(t,\tau)\mathcal{X}\hat{\rho}(\tau)}=0&0\leq \tau < t\leq t^\prime\\
    \end{array}\right.
\end{equation}
Finally, we can show that there is nonzero two-photon coherence $\Lambda^{(2)}(t,t^\prime)=\braket{\bd(t)\bd(t^\prime)}$ between different time bins, and the two-photon coherence is equal to the temporal density function of a single photon:
\begin{equation}
    \Lambda^{(2)}(t,t^\prime)=\left\{\begin{array}{ll}
    \gamma~\!\tr{\mathcal{S}\mathcal{U}(t^\prime,t)\mathcal{S}\hat{\rho}(t)}=0&0\leq t \leq t^\prime\leq \tau\\
     \gamma~\!\tr{\mathcal{S}\mathcal{U}(t,\tau)\mathcal{X}\mathcal{U}(\tau,t)\mathcal{S}\hat{\rho}(t)}=\xi_1(t,t^\prime)&0\leq t \leq \tau< t^\prime\\
    \gamma~\!\tr{\mathcal{S}\mathcal{U}(t^\prime,t)\mathcal{S}\mathcal{U}(t,\tau)\mathcal{X}\hat{\rho}(\tau)}=0&0\leq \tau < t\leq t^\prime\\
    \end{array}\right.
\end{equation}
Using these correlation functions, we can solve for $\mu$ and $g^{(2)}$ by
\begin{equation}
\begin{aligned}
    \mu &= \int_0^\infty G^{(1)}(t,t)dt = \int_{0}^\tau\xi_1(t,t)dt+\frac{\beta^2}{\alpha^2}\int_\tau^\infty\xi_1(t,t)dt
    =2\beta^2
\end{aligned}
\end{equation}
and
\begin{equation}
\begin{aligned}
    g^{(2)}&=\frac{2}{\mu^2}\int_0^\infty\int_t^\infty G^{(2)}(t,t^\prime)dt^\prime dt=\frac{\gamma^2}{2\beta^4}\int_0^\tau\int_\tau^\infty e^{-(t+t^\prime-\tau)\gamma}dt^\prime dt
    =\frac{1}{2\beta^2}.
\end{aligned}
\end{equation}
Using the photon statistics relations (see section \ref{chapter2:photonstatistics}), and knowing $p_{n\geq3}=0$, we have $\mu=p_1+2p_2=2\beta^2$ and $g^{(2)}=2p_2/\mu^2=\mu^{-1}$. This implies $p_1=0$ and $p_2=\beta^2$, leaving $p_0=1-\beta^2=\alpha^2$. Hence, the single-photon component vanishes for all pulse separations, leaving only the two-photon component and the vacuum component.

The last thing we need is the two-photon density function $\xi_2(t_1,t_2,t_1^\prime,t_2^\prime)$. This can be solved using the form of Eq.~(\ref{chapter2eq:twophotondensity}), although we can use $\mathcal{U}$ in place of $\mathcal{U}_0$ since we know that all components for $n\geq 3$ vanish. With this approach, we can verify that the only non-vanishing component corresponds to $t_1,t_1^\prime\leq \tau\leq t_2,t_2^\prime$ and is given by the separable function
\begin{equation}
    \xi_2(t_1,t_2,t_1^\prime,t_2^\prime)=\frac{1}{\alpha^2\beta^2}\left\{\begin{array}{ll}
    \xi_1(t_1,t_1^\prime)\xi_1(t_2,t_2^\prime)&t_1,t_1^\prime\leq \tau\leq t_2,t_2^\prime\\
    0&\text{otherwise}
    \end{array}
    \right.
\end{equation}
Note that the division by $\alpha^2\beta^2=p_0p_2$ is necessary so that $\xi_2$ is normalized. We now have enough information to write down the photonic density operator of our non-ideal $\ket{\phi^+}$ state:
\begin{equation}
\begin{aligned}
    \hat{\varrho} = p_0\ketbra{0}{0} &+ \frac{p_2}{\alpha^2\beta^2}\int_0^\tau\int_0^\tau \xi_1(t_1,t_1^\prime)\int_\tau^\infty\int_\tau^\infty\xi_1(t_2,t_2^\prime)\bu(t_2)\bu(t_1)\ketbra{0}{0}\bd(t_1^\prime)\bd(t_2^\prime)dt_1dt_1^\prime dt_2dt_2^\prime\\
    &+\int_0^\tau\int_\tau^\infty\xi_1(t_1,t_2)\bu(t_2)\bu(t_1)\ketbra{0}{0}dt_1dt_2 +\text{h.c.}
\end{aligned}
\end{equation}

\begin{figure}
    \centering
    \hspace{-58mm}(a)\hspace{75mm}(b)\\
    \includegraphics[width=0.475\textwidth]{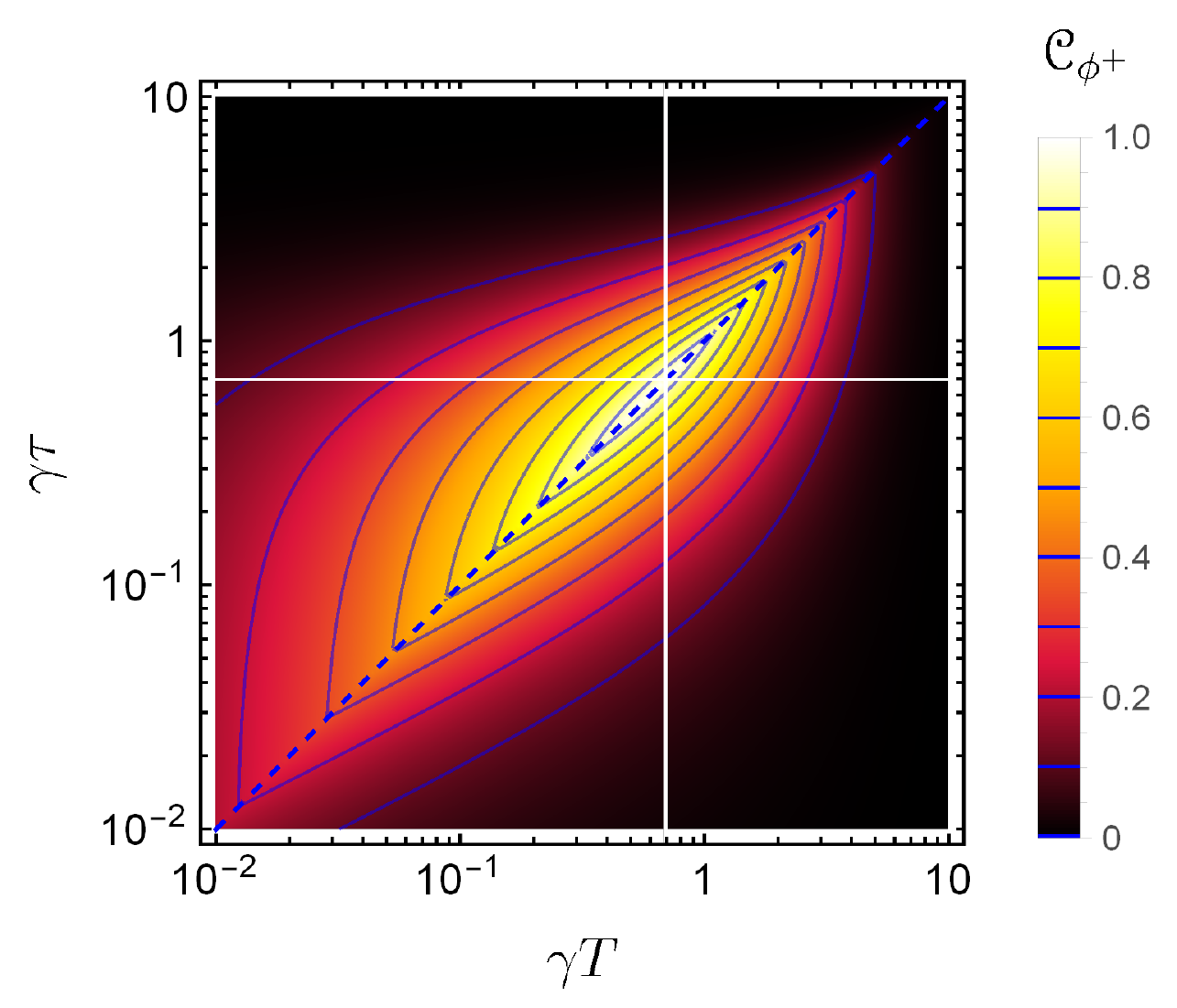}
    \includegraphics[width=0.45\textwidth]{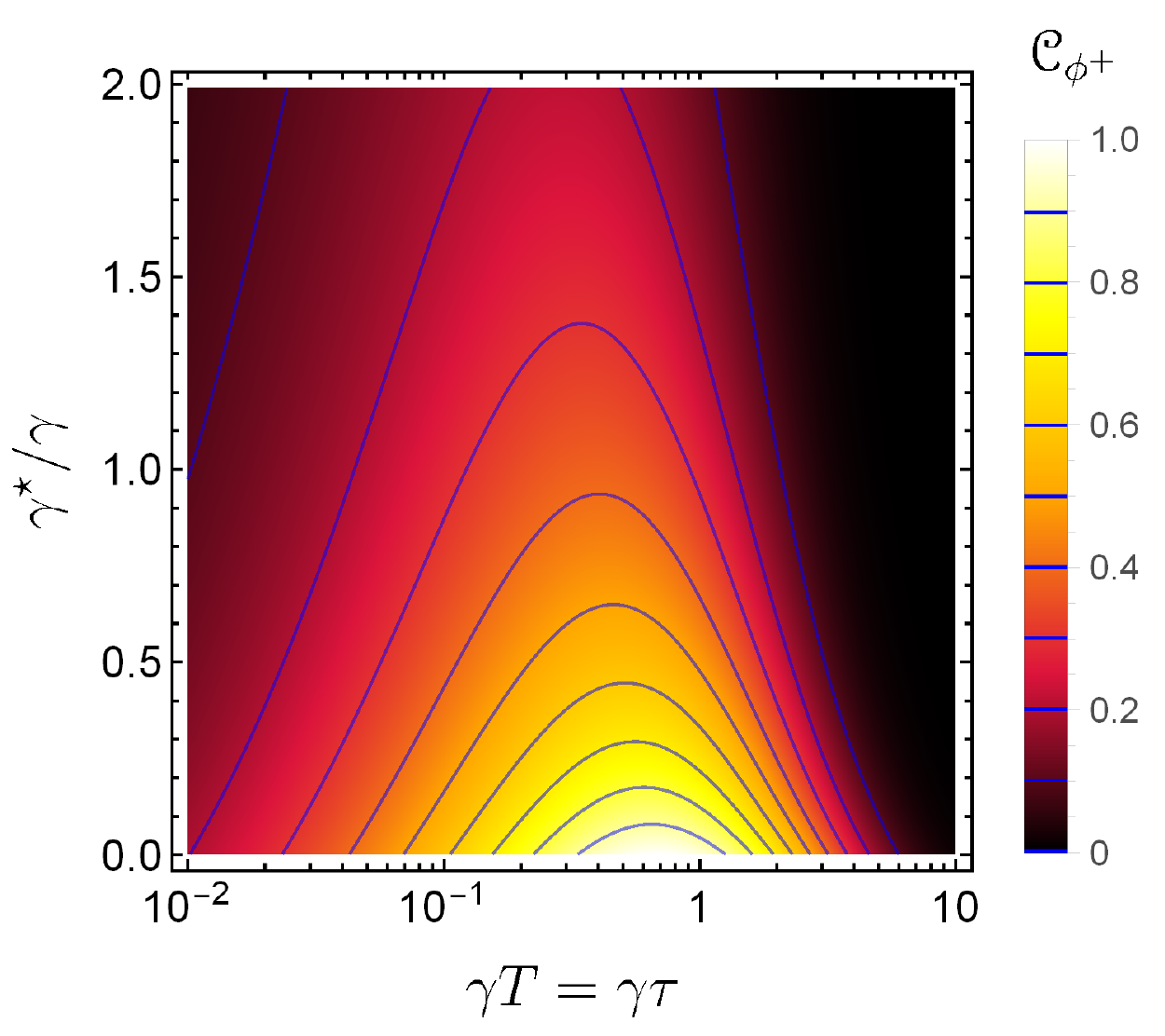}
    \caption[The concurrence of a $\ket{\phi^+}$ photonic Bell state generated by a pulsed two-level emitter experiencing pure dephasing.]{\small\textbf{The concurrence of a $\ket{\phi^+}$ photonic Bell state generated by a pulsed two-level emitter experiencing pure dephasing.} (a) The concurrence $\euscr{C}_{\phi^+}$ as a function of time bin threshold $T$ and pulse separation $\tau$ for $\gamma^\star=0$. The white lines show the $\gamma\tau=\gamma T=\ln(2)$ half-life condition. (b) The concurrence along the $T=\tau$ condition (diagonal dashed blue line in panel (a) as a function of pure dephasing $\gamma^\star$.}
    \label{chapter3fig:phiplusconcurrence}
\end{figure}

Up until this point, I have not discussed the definition for our time bin measurement. We should expect that the best option to maximize entanglement is $T=\tau$, as this is the natural choice to capture all of the two-photon coherence. By computing the density matrix elements in the second-quantized basis, we can compute the concurrence in the $\{\ket{00},\ket{10},\ket{01},\ket{11}\}$ subspace as before. Now, our solution is a function of both the time bin threshold $T$ and pulse separation $\tau$ (see Fig.~\ref{chapter3fig:phiplusconcurrence}~(a)). We can see that the concurrence for a given pulse separation $\tau$ is maximized by choosing $T=\tau$, as expected. Along this condition, we can also look at how dephasing degrades the concurrence to find that it is very similar to the single-particle $\ket{\psi^+}$ case (see Fig.~\ref{chapter3fig:phiplusconcurrence}~(b)).

When choosing $T=\tau$, we can show that the density matrix elements of the $\varrho_{1111}=\braket{11|\hat{\varrho}|11}$ and $\varrho_{0011}=\braket{00|\hat{\varrho}|11}$ are related to $\varrho_\mathrm{ss}$, $\varrho_\mathrm{tt}$, and $\varrho_\mathrm{st}$ from the previous section. By performing the overlaps using time bin modes Eq.~(\ref{chapter3eq:timebinmodes}), we find that $\varrho_{1111}=p_2\varrho_\mathrm{ss}\varrho_{tt}/(\alpha^2\beta^2)$ and $\varrho_{0011}=\varrho_\mathrm{st}$. Also, we have $\varrho_{0000}=\alpha^2=p_0$.

\begin{figure}[t]
    \centering
    \hspace{-58mm}(a)\hspace{75mm}(b)\\
    \includegraphics[width=0.49\textwidth]{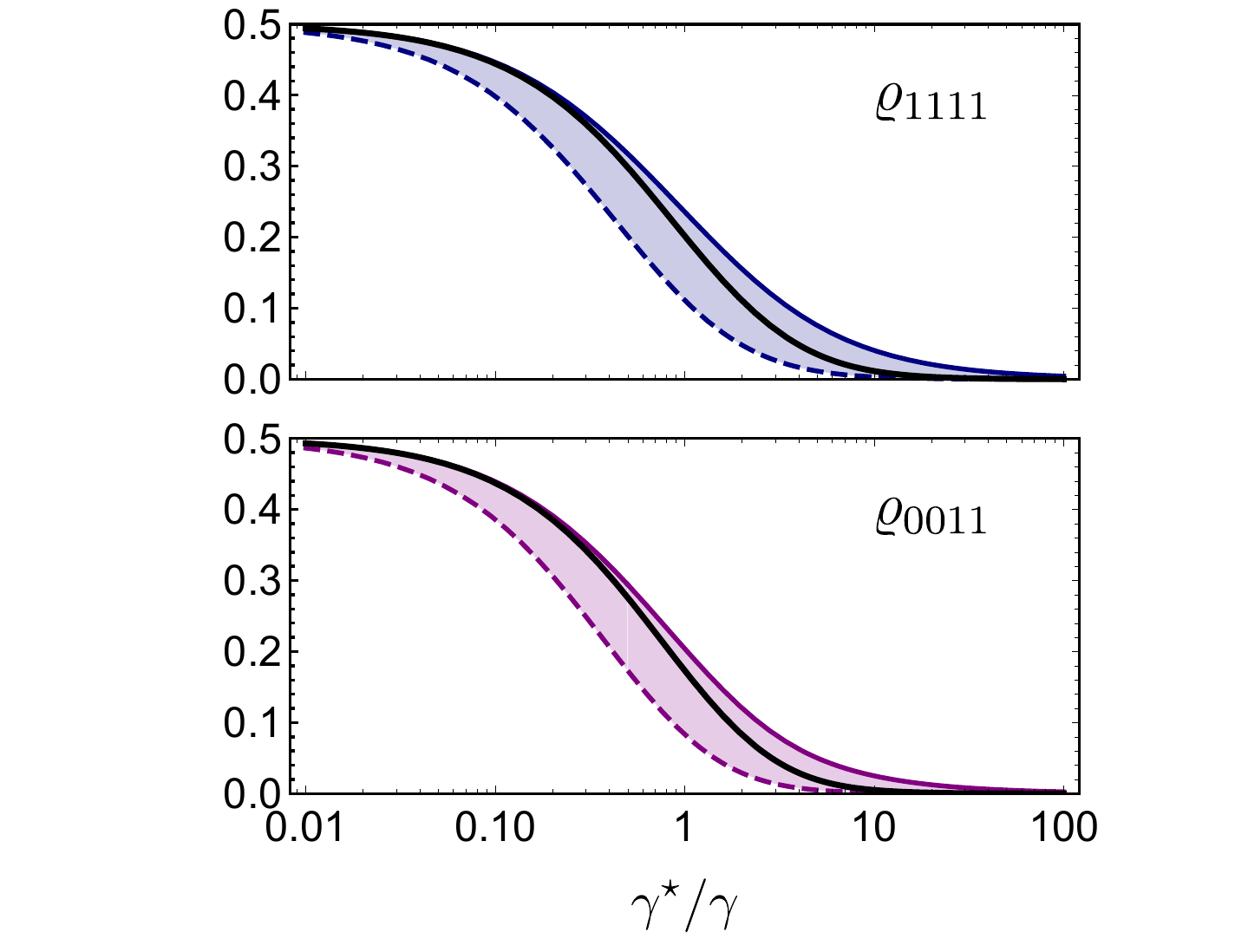}
    \includegraphics[width=0.49\textwidth]{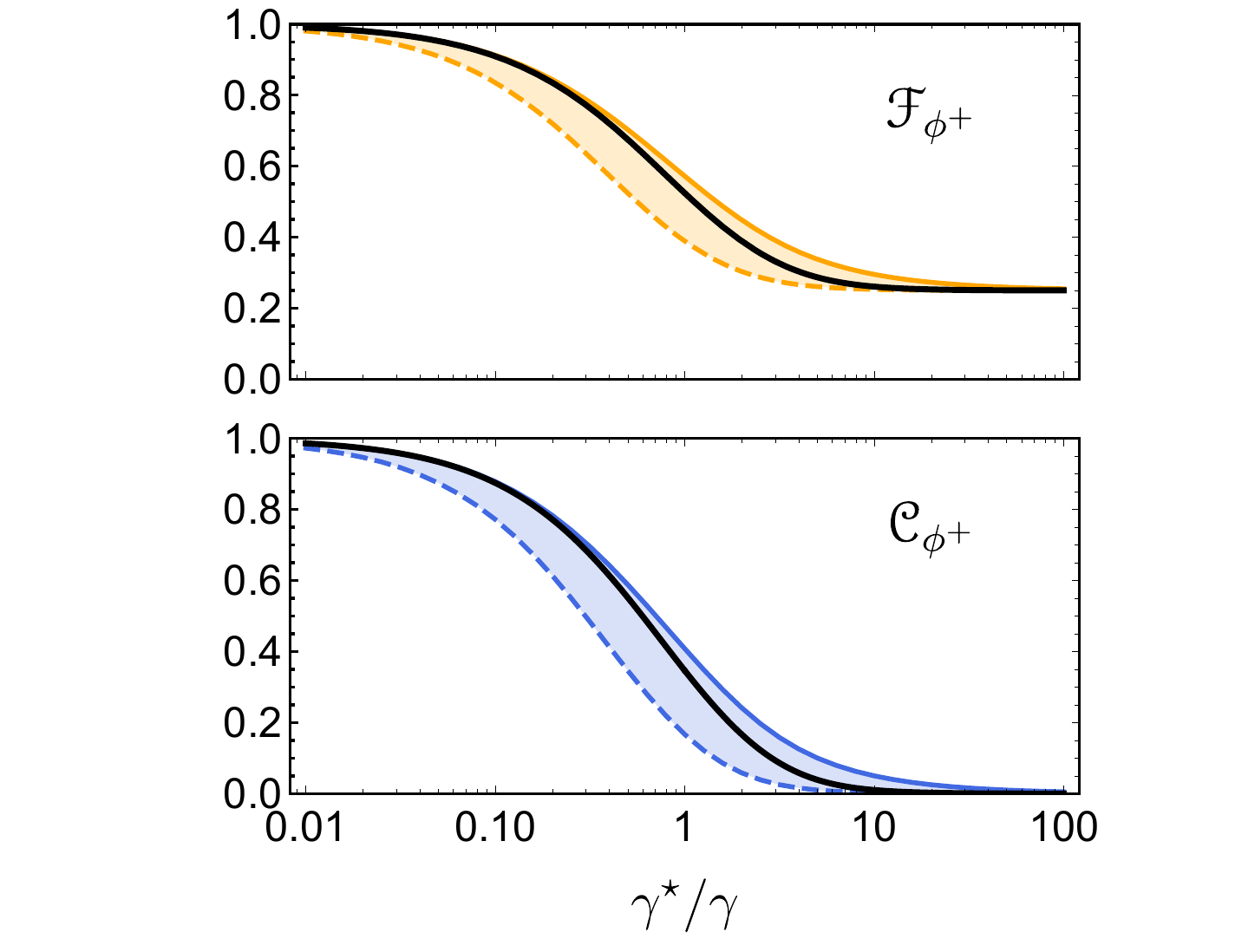}
    \caption[A comparison of the bipartite time bin density matrix elements for a $\ket{\phi^+}$ photon number Bell state and their estimates from self-homodyne measurements.]{\small\textbf{A comparison of the bipartite time bin density matrix elements for a $\ket{\phi^+}$ photon number Bell state and their estimates from self-homodyne measurements.} (a) The density matrix elements in the second-quantized basis (solid black curves) compared to the estimated values using bipartite indistinguishability measurements (blue region, top panel) and bipartite two-photon coherence measurements (purple region, bottom panel) for $\gamma T=\ln(2)$. The third relevant density matrix element $\varrho_{0000}=p_0$ is constant for any given $T$ and $p_0=1/2$ for $T=\ln(2)$. (b) The fidelity and concurrence computed using the values in panel (a). The fidelity approaches $1/4$ for large $\gamma^\star$ due to the constant contribution from $p_0$.}
    \label{chapter3fid:phipluscompare}
\end{figure}

As before, it would be useful to have an alternative, simpler, way to estimate the density matrix elements. This can be done with the help of a time-resolved self-homodyne measurement (recall section \ref{chapter3:selfhomodyne} for self-homodyne measurements), which gives us access to $M_{ab}$ and $\lambda^{(2)}_{ab}$ that can be post-selected into our two time bins $a,b\in\{\mathrm{t},\mathrm{s}\}$ by an appropriate choice of integration limits. For simplicity, let us still focus on the case where $T=\tau$. Using $G^{(1)}$, we can find that $M_\mathrm{st}=0$, $M_\mathrm{ss}=I_\mathrm{ss}$ and $M_\mathrm{tt}=I_\mathrm{tt}$. Furthermore, using $\Lambda^{(2)}$, we can find that $\lambda^{(2)}_\mathrm{ss}=\lambda^{(2)}_\mathrm{tt}=0$ and $\lambda^{(2)}_\mathrm{st}=I_\mathrm{st}$. The values for $I_\mathrm{ss}$, $I_\mathrm{st}$, and $I_\mathrm{tt}$ are given in Eq.~(\ref{chapter3eq:Idecompopsiplus}). From this, and the fact that $\varrho_{0011}=\varrho_\mathrm{st}$, it is clear that applying a second pulse at $\tau=T$ transfers the temporal coherence between $\ket{10}$ and $\ket{01}$ given by $I_\mathrm{st}$ to a two-photon coherence between $\ket{00}$ and $\ket{11}$, as expected. It also implies that, for the condition $\tau=T$, we can estimate the density matrix elements by measuring the integrated two-photon coherence to get $\lambda_\mathrm{st}^{(2)}\sqrt{\mu_\mathrm{s}\mu_\mathrm{t}}\leq\varrho_{0011}\simeq\sqrt{\mu_\mathrm{s}\mu_\mathrm{t}\lambda^{(2)}_\mathrm{st}}$ and the mean wavepacket overlap to get $M_\mathrm{ss}M_\mathrm{tt}\mu_\mathrm{s}\mu_\mathrm{t}/p_2\leq\varrho_{1111}\simeq\mu_\mathrm{s}\mu_\mathrm{t}\sqrt{M_\mathrm{ss}M_\mathrm{tt}}/p_2$ using the time-resolved self-homodyne setup. By comparing these estimates to the full overlaps computed using $f_1$, we can see that they are very accurate (see Fig.~\ref{chapter3fid:phipluscompare}). In addition, the concurrence is very similar to the case for $\ket{\psi^+}$ given in Fig.~\ref{chapter3fid:phipluscompare}, even though the density elements are different. This is a particularity associated with the choice $\tau=T=\ln(2)/\gamma$, which maximizes the entanglement in both cases and when $\gamma^\star$ is small.

The ability to deterministically produce entangled states using only two-level single-photon sources could be very beneficial for quantum information applications. This is because single-photon sources based on two-level system models are well developed \cite{somaschi2016,ding2016demand}. The scheme presented in this section to produce and measure photon-number Bell states can also be extended by applying additional pulses before the emitter has fully decayed. This extension can allow for the deterministic generation of multi-partite W-class entangled states \ref{loredo2020deterministic}, which may have applications in all-optical quantum repeaters or other quantum communication schemes. 

Deterministic generation of entangled photonic states using optically-active defects can also be implemented if the defect contains additional level structure allowing for a quantum memory \cite{schwartz2016deterministic}, such as a ground-state spin qubit. This additional level structure expands the class of entangled states that can be produced. The basic principle takes advantage of the persistent spin-photon entanglement generated between a photonic state produced by the defect and the spin state of the emitter. In the following chapter, I will analyze how this spin-photon entanglement generated by an emitter containing a spin qubit can be used to generate spin-spin entanglement between two remote defects. 
\chapter{Spin-spin entanglement}
\label{chapter4}

Photon-mediated entanglement generation between spins in different quantum defects is important for implementing quantum repeaters \cite{briegel1998quantum, duan2001long,sangouard2009quantum,sangouard2011quantum,rozpkedek2019near} and distributed quantum computing protocols \cite{lim2005repeat, benjamin2009prospects} for quantum networks. Such a quantum network could form the basis for a global quantum internet \cite{kimble2008internet,simon2017internet,wehner2018internet}. The many benefits and applications of quantum networks are detailed in section \ref{chapter1:applications}.

The nodes of a quantum network are composed of clusters of stationary qubits (spins). These qubits serve as a quantum memory to store quantum information. They must also interact locally, enabling quantum logic gates for quantum information processing. To carry quantum information over long distances, nodes of a quantum network must couple to flying qubits (photons). This coupling is mediated by a quantum interface between matter and light that operates in a quantum coherent way to preserve the quantum information \cite{yao2005theory,atature2018material}. Photonic qubits connect nodes together and distribute quantum information over the network. They can propagate using direct optical links over short distances (up to $\sim$500 km). For longer distances (up to $\sim$2000 km), a quantum repeater could be used to overcome transmission loss \cite{briegel1998quantum}. For a truly global network, satellites in low-earth or geostationary orbits might be used \cite{simon2017internet}.

For distances up to about 500 km, photons of telecommunication wavelength ($\sim$ 1550 nm) can be faithfully transmitted directly through readily-available fiber-optic infrastructure \cite{korzh2015provably,valivarthi2016quantum}. By performing a joint Bell-state measurement \ref{wein2016bellmeasurement} on photons emitted from two different nodes, the spin qubits within separated nodes can be projected onto a maximally entangled state \cite{barrett2005efficient}.

Unfortunately, losses of about 0.2 dB/km in optical fibers plague long-distance single-photon transmission. In classical fiber optics communication, this problem is solved by a repeater component that amplifies the signal. However, because of the no-cloning theorem in quantum physics \cite{wootters1982single,dieks1982communication}, it is not possible to faithfully duplicate the quantum state of a single photon. Therefore, to extend single-photon transmission distances, one can use a quantum repeater protocol \cite{briegel1998quantum}.

The basic principle of a quantum repeater is to first generate entanglement between intermediate nodes over shorter ($<$500 km) distances through direct fiber or free-space transmission. These short-range entanglement connections can then be extended over a long distance ($<$5000 km) by performing a local exchange of quantum information between stationary qubits in each intermediate node of repeater, known as an entanglement swapping. Although there are many ensemble-based approaches to quantum repeater memory nodes \cite{sangouard2011quantum}, I will focus on using single spin qubits. In this case, entanglement swapping can be accomplished deterministically by performing a combination of local gates and spin-state measurements \ref{asadi2018repeaters}.

In this chapter, I explore two main types of photon-mediated spin-spin entanglement related to quantum networks for single defects: (1) between two remote optically-active defects that cannot interact locally and (2) between two local defects that can interact via coupling to the same optical cavity mode. These two types of spin-spin entangling processes are directly applicable to the entanglement generation and deterministic entanglement swapping steps, respectively, of quantum repeater schemes using single defects \ref{asadi2018repeaters}\ref{asadi2020cavitygates}.

To achieve remote entanglement between systems that emit visible or near-infrared photons, it is convenient to use pulsed schemes that herald entanglement by the detection of single photons \cite{cabrillo1999creation,barrett2005efficient,simon2003robust,feng2003entangling}. Such schemes have already been implemented using atomic ensembles \cite{chou2005measurement,chou2007functional}, single trapped atoms \cite{hofmann2012heralded} or ions \cite{moehring2007entanglement,slodivcka2013atom}, quantum dots \cite{delteil2016generation,stockill2017phase}, and defects in diamond \cite{bernien2013heralded,hensen2015loophole}. This is usually done by first generating spin-photon entanglement between the spin state of the individual defect and their emission. Then, by performing a probabilistic joint entangling measurement on the emission from each remote system, the joint spin state of the systems is projected onto an entangled state.

For local interactions, if the cavity-emitter system is in the strong-coupling regime, then the vacuum Rabi oscillations can be used to engineer a strong emitter-emitter interaction before dissipation occurs. However, since the strong-coupling regime is not ideal for remote entanglement generation that relies on the Purcell effect to increase fidelity and efficiency, it is attractive to consider deterministic local entanglement schemes that can operate under the same bad-cavity conditions needed for high-fidelity remote entanglement generation. This is possible by engineering an adiabatic cavity-mediated interaction between two defects whereby a photon is never actually created in the cavity mode \ref{asadi2020cavitygates}, and hence decoherence due to the cavity dissipation is suppressed.

As discussed in section \ref{chapter1:solidstateoptialdevices}, solid-state systems are particularly attractive as a quantum technology platform for their scalability, ease of manufacturing, and potential to integrate with classical information processing hardware \cite{awschalom2018quantum,atature2018material}. However, solid-state systems suffer from decoherence (see section \ref{chapter1:deoherence}) that limits the initial amount of generated entanglement between systems as well as their longevity as a quantum memory \cite{benjamin2009prospects}. Spin decoherence can be caused by the interaction of the spin qubit with a surrounding bath of nuclear spins \cite{hanson2007spins} or lattice phonons \cite{golovach2004phonon}. These interactions can randomly flip the spin state of the qubit during or after entanglement generation, or cause a pure-dephasing of the spin coherence. Phonon interactions can also cause homogeneous broadening of the zero-phonon line (ZPL) for solid-state optical transitions \cite{fu2009observation,plakhotnik2015electron}, which degrades the indistinguishability of photons emitted from the quantum system \cite{grange2015cavity} and can degrade the generation of local entanglement that uses superpositions involving the emitter excited state. All these decoherence processes limit the amount of final spin-spin entanglement that can be generated.

The two most critical figures of merit for entanglement generation are efficiency and fidelity (section \ref{chapter1:fidelity}). The efficiency impacts the overall rate of quantum information transfer. For example, the quantum key distribution rate for a repeater protocol is proportional to the entanglement generation efficiency. The fidelity quantifies the quality of entanglement in addition to our knowledge about the state of the system, and can be related to the entanglement concurrence (section \ref{chapter1:concurrence}). High-fidelity entanglement is necessary for many quantum information applications. In addition, purification and error correction protocols require minimum fidelity thresholds to be satisfied \cite{dur1999quantum}.

In section \ref{chapter4:entanglementgeneration}, I will present the work of Ref.~\ref{wein2020entanglement} where we analyze the efficiency and fidelity for three different common remote entanglement generation protocols while taking into account many realistic imperfections including decoherence. These three protocols are respectively implemented via (1) spin-photon number entanglement with a single pulse \cite{cabrillo1999creation}, (2) spin-time bin entanglement with two sequential $\pi$-pulses \cite{barrett2005efficient}, and (3) spin-polarization entanglement using an excited $\Lambda$ system \cite{simon2003robust,feng2003entangling}. Each of these protocols requires fast resonant pulsed excitation of the quantum systems. For each protocol, we derive the spin-spin conditional states for photon counting measurements using the methods introduced in sections \ref{chapter2:conditionalpropagationsuperoperators} and \ref{chapter2:photoncountingmeasurements} to compute expressions for the entanglement figures of merit. We also discuss their relationship to the properties of single-photon emission from the individual quantum systems, such as brightness and mean wavepacket overlap. The results of this section have already been applied to analyze quantum repeaters \ref{asadi2020repeaters}\ref{sharman2020repeaters}\ref{ji2020roomtemperature} and have been included in software for simulating quantum networks \cite{wu2020sequence}.

In section \ref{chapter4:cavitygates}, I will present the analysis of local spin-spin entanglement for one of the three adiabatic photon-mediated approaches discussed in Ref.~\ref{asadi2020cavitygates}. The approach I study considers a controlled phase gate operation between two emitters in a cavity using a virtual photon interaction initiated by exciting one of the two emitters. In my analysis, I will explicitly include the effect of emitter pure dephasing that was not considered in Ref.~\ref{asadi2020cavitygates}. In addition, I will explore the possibility of using a cavity with an optical Fano resonance to greatly enhance the fidelity of this adiabatic phase gate, which perhaps could be implemented using the same Fabry-P\'{e}rot-plasmonic hybrid cavities discussed in section \ref{chapter3:roomtemperature}.

\section{Heralded entanglement generation}
\label{chapter4:entanglementgeneration}

To analyze heralded entanglement generation, we apply a photon number decomposition to compute the entanglement generation efficiency and fidelity of the final spin-spin entanglement conditioned on the number of detected photons. This approach uses a Liouville-Neumann series \cite{carmichael,horoshko1998multimode, brun2000continuous} to decompose the master equation dynamics into a set of propagation superoperators that describe the spin state evolution conditioned on the cumulative detector photon count during a window of time (see chapter \ref{chapter2}).

In section \ref{chapter4:remotesystems}, I will introduce the system model describing two remote optically active defects that contain a spin qubit. This will include a summary of all the imperfections included in the entanglement generation model. Sections \ref{chapter4:numberentanglement} through \ref{chapter4:polarizationentanglement} present the analytical and numerical results for each of the three analyzed protocols. Finally, section \ref{chapter4:entanglementcomparison} presents a comparison of the protocols as a function of loss and distance between the remote spin qubits.

\subsection{System}
\label{chapter4:remotesystems}

\begin{figure}
    \centering
    \includegraphics[width=\textwidth,trim=0 75 0 140,clip]{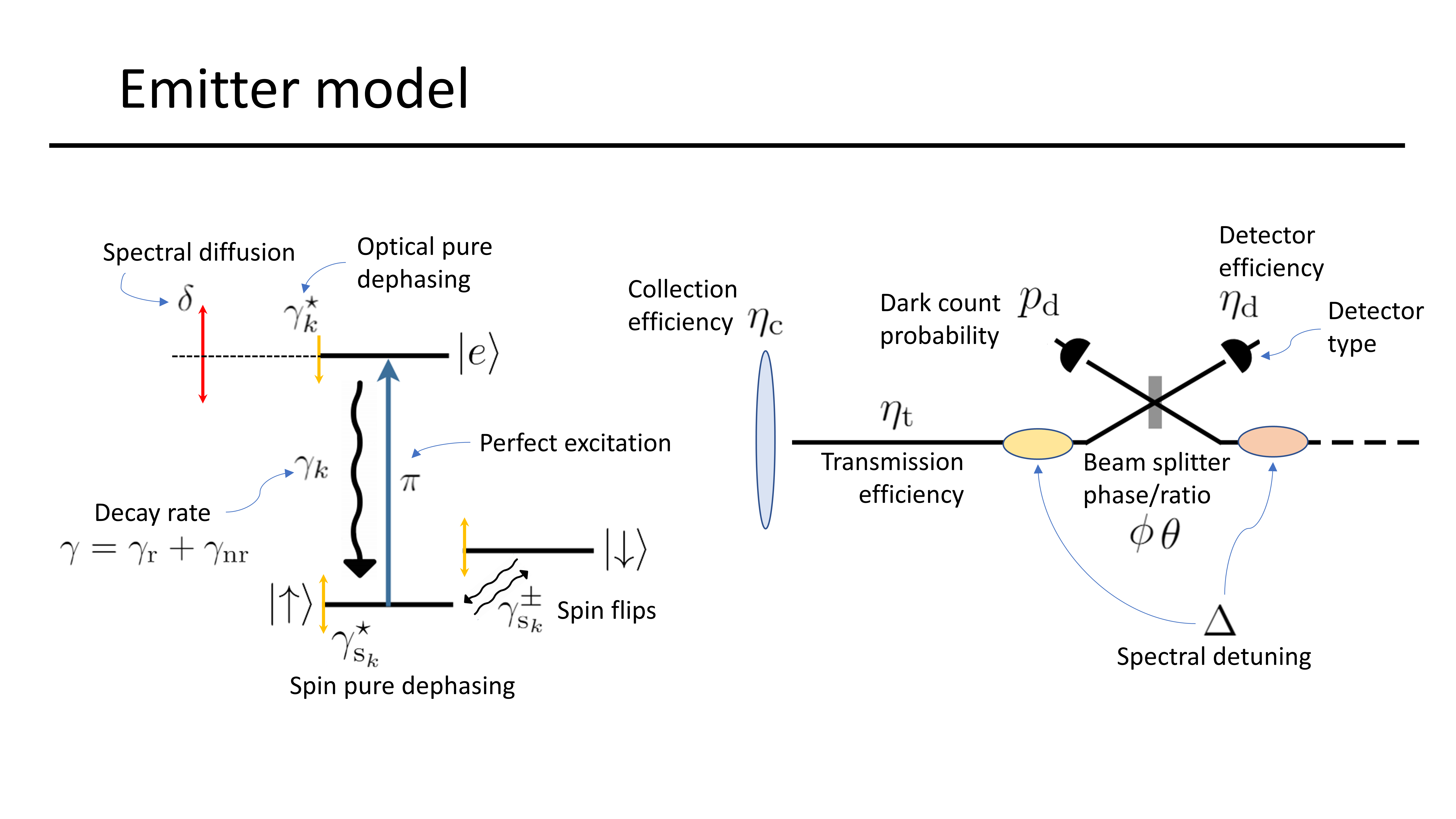}
    \caption[Diagram of the system model used to analyze remote spin-spin entanglement.]{\small\textbf{Diagram of the system model used to analyze remote spin-spin entanglement.} The three level system shown here illustrates the model used for both defects $k=1$ and $k=2$ that emit photons to be interfered at the beam splitter. These two defects may have different parameters, indicated by the subscripts $k$, and a spectral detuning $\Delta$. The defects are subject to decoherence processes such as pure dephasing and spectral diffusion. The spin qubits of the two defects are projected onto an entangled state by performing an imperfect joint measurement of the photons emitted by each defect. The measurement imperfections include transmission loss, detector inefficiency, detector dark counts, and number resolving limitations.}
    \label{chapter4fig:remotesystemdiagram}
\end{figure}

We consider five main imperfections in the entanglement generation process: (1) decoherence, (2) spectral diffusion, (3) photon loss, (4) phase errors, and (5) dark counts. Solid-state systems may suffer from mechanisms that degrade the spin coherence and the coherence of emitted photons. These mechanisms are usually strongly dependent on temperature \cite{fu2009observation,plakhotnik2015electron,grange2017reducing}. These systems can also experience spectral diffusion, which can inhibit the indistinguishability of emitted photons \cite{loredo2016scalable,thoma2016exploring,reimer2016overcoming}. In addition, photon losses due to non-radiative pathways or collection/transmission inefficiency can affect the protocol figures of merit; and in some cases, protocols can moreover be susceptible to initialization and propagation phase errors. Finally, the detectors may have a non-negligible dark count rate \cite{hadfield2009single}.

\emph{Decoherence.---}For each protocol, we consider that a transition with decay rate $\gamma$ is subject to a pure dephasing rate $\gamma^\star$ that degrades the coherence of photons emitted by the system \cite{grange2015cavity}. This dephasing can be seen as a fluctuation of the transition energy on a timescale much faster than its decay rate, and it affects the indistinguishability between photons emitted by the systems. We separate the decay rate of the transition into a radiative component $\gamma_\text{r}$ and a non-radiative component $\gamma_\text{nr}$ so that $\gamma=\gamma_\text{r}+\gamma_\text{nr}$. In addition, we consider that the spin qubits experience incoherent spin flip excitation (decay) at the rate $\gamma_\text{s}^+=1/T_1^+$ ($\gamma_\text{s}^-=1/T_1^-$) and a pure dephasing at the rate $\gamma_\mathrm{s}^\star=1/T_2^\star$ for a total spin decoherence rate of $1/T_2 = 1/T_2^\star+1/2T_1^++1/2T_1^-$.

\emph{Spectral diffusion.---}In contrast to pure dephasing, spectral diffusion is a fluctuation of the transition energy on a timescale much slower than its decay rate. In many solid-state systems, this fluctuation can shift the emitted photon frequency over time by more than its linewidth \cite{thoma2016exploring,reimer2016overcoming}. This degrades the mean wavepacket overlap of photons emitted by the same source at different times \cite{loredo2016scalable,thoma2016exploring,reimer2016overcoming}. Hence, this fluctuation also significantly degrades interference between fields from different sources. We account for spectral diffusion by averaging entanglement figures of merit over a Gaussian distribution \mbox{$h(\omega_k-\overline{\omega}_k,\delta_k) =(\delta_k\sqrt{2\pi})^{-1}e^{-(\omega_k-\overline{\omega}_k)^2/2\delta_k^2}$} for each emitter frequency $\omega_k$ with an average value of $\overline{\omega}_k$ and a spectral diffusion standard deviation $\delta_k$. For example, for two systems, the entanglement fidelity $\euscr{F}$ becomes \mbox{$\iint h(\omega_1-\overline{\omega}_1,\delta_1)h(\omega_2-\overline{\omega}_2,\delta_2)\euscr{F}d\omega_1 d\omega_2$}.

\emph{Photon loss.---}Losses can occur due to non-radiative transitions at a rate $\gamma_\text{nr}$. We also quantify the imperfect collection fraction $\eta_\text{c}$ of emission and the fraction of photons transmitted to the detectors by $\eta_\text{t}$. In addition, we consider that each detector has a probability $\eta_\text{d}$ of detecting an incident photon. This detector inefficiency can be applied during the measurement step. However, the beam-splitter loss model used to describe detector inefficiency can be mapped to the identical model for transmission loss \ref{wein2016bellmeasurement}. Thus, for convenience, we choose to simulate detector inefficiency as part of the conditional dynamics rather than the measurement itself. This allows us to use the total efficiency parameter $\eta=\eta_\text{c}\eta_\text{t}\eta_\text{d}\gamma_\text{r}/\gamma$.

\emph{Phase errors.---}Phase errors can arise when the individual quantum systems are locally initialized and read out using pulses from a source that does not maintain phase stability over the duration of the protocol. We account for this by considering an initial phase $\varphi$ when a quantum system is initialized in a superposition state. Phase errors can also arise when photons from each source do not accumulate the same propagation phase $\phi$ before interference. If these phases are unstable or left uncorrected, then they may degrade the entanglement fidelity. We account for phase errors by assuming that the phase fluctuates between entanglement generation attempts and then average the fidelity over a random phase with a Gaussian distribution.

\emph{Dark counts.---}A realistic detector may falsely indicate the arrival of a photon or detect a photon that did not originate from a desired emitter \cite{hadfield2009single}. In our study, we assume that each detector is gated for an interval $T_\text{d}$ that begins at time $t_\text{d}$ after the start of the protocol and ends at time $t_\text{d}^\prime=t_\text{d}+T_\text{d}$. We also assume that the dark counts are classical noise described by a Poisson distribution with a rate $\gamma_\mathrm{d}$. Then for a given detector, the probability that $n$ dark counts have occurred during the gate duration $T_\text{d}$ is given by $p_{\mathrm{d},n}(T_\text{d},\gamma_\mathrm{d})=\gamma_\mathrm{d}^nT_\text{d}^ne^{-\gamma_\mathrm{d} T_\text{d}}/n!$.

To capture the defect-specific imperfections, we use a Markovian master equation for two independent three-level systems (see Fig.~\ref{chapter4fig:remotesystemdiagram}). The total Liouville superoperator of the two optically-active defects is $\mathcal{L}=\mathcal{L}_1\otimes\mathcal{I}+\mathcal{I}\otimes\mathcal{L}_2$ where
\begin{equation}
\begin{aligned}
\label{chapter4eq:3lvldefect}
    \mathcal{L}_k =& -\frac{i}{\hbar}\mathcal{H}_k+\sum_j\gamma_{j_k}^-\mathcal{D}(\hat{\sigma}_{j})+\gamma_{j_k}^+\mathcal{D}(\hat{\sigma}_{j}^\dagger)
    +2\gamma_k^\star\mathcal{D}(\hat{\sigma}_{\uparrow}^\dagger\hat{\sigma}_{\uparrow})+ \frac{\gamma_{\mathrm{s}_k}^\star}{2}\mathcal{D}(\hat{\sigma}_z),
\end{aligned}
\end{equation}
for $k\in\{1,2\}$. The system operators are defined in their respective Hilbert spaces as $\hat{\sigma}_{\uparrow}=\ket{\uparrow}\!\bra{e}$, $\hat{\sigma}_{\downarrow}=\ket{\downarrow}\!\bra{e}$, $\hat{\sigma}_\text{s}=\ket{\uparrow}\!\bra{\downarrow}$, and $\hat{\sigma}_{z}=\ket{\uparrow}\!\bra{\uparrow}-\ket{\downarrow}\!\bra{\downarrow}$. The rate $\gamma_{j_k}^-$ ($\gamma_{j_k}^+$) is the total incoherent decay (excitation) rate across the transition associated with $\hat{\sigma}_j$ where $j\in\{\uparrow,\downarrow,\text{s}\}$, $\gamma_k^\star$ is the emitter pure dephasing rate, and $\gamma^\star_{\mathrm{s}_k}$ is the spin pure dephasing rate. The three-level system Hamiltonian is $\hat{H}_k = \hbar\omega_{\uparrow_k}\hat{\sigma}_{\uparrow}^\dagger \hat{\sigma}_{\uparrow} + \hbar\omega_{\text{s}_k}\hat{\sigma}_\text{s}^\dagger\hat{\sigma}_\text{s}\!$ where $\omega_{\uparrow_k}$ is the separation between $\ket{\uparrow}$ and $\ket{\mathrm{e}}$, $\omega_{\text{s}_k}$ is the separation between $\ket{\uparrow}$ and $\ket{\downarrow}$.

The model used in this section is the three-level extension of the model used in sections \ref{chapter3:HOM} and \ref{chapter3:photonicstate}. Therefore, if the defects are placed inside a cavity, it is necessary that the system is sufficiently far into the bad-cavity regime such that adiabatic elimination is valid (recall section \ref{chapter1:1datom}). Furthermore, in this section we assume that a coherent state preparation is achieved on a timescale much faster than the total system decay rate. Thus, the results in this section neglect the degrading effects of nonzero $g^{(2)}$. That said, all of the methods used in this section can be directly applied to the full cavity-emitter source model described in section \ref{chapter1:cavityQED} to capture cavity-emitter non-Markovian effects and realistic state preparation pulses.

\subsection{Spin-photon number entanglement}
\label{chapter4:numberentanglement}

\begin{figure}
\centering
\includegraphics[width=0.7\textwidth]{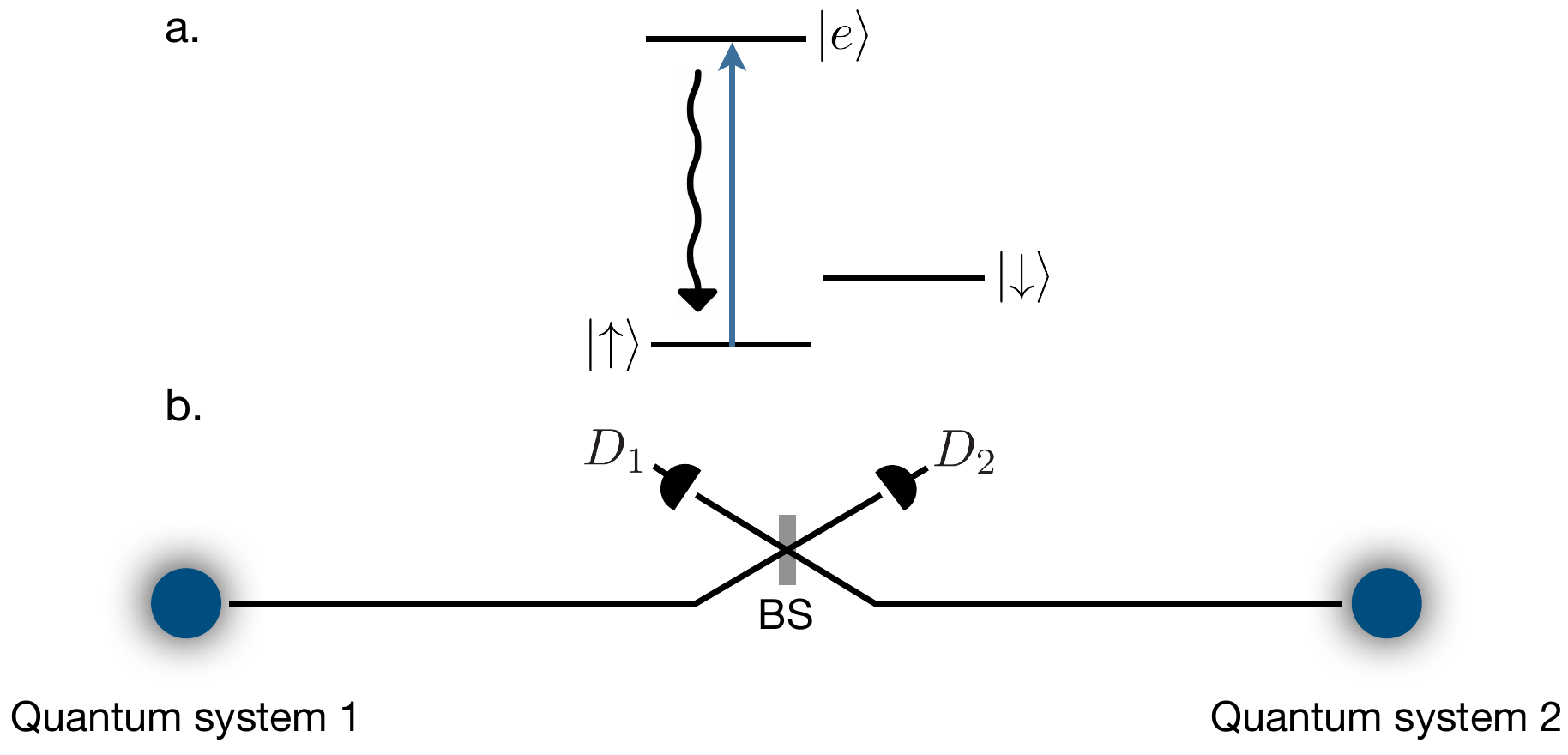}
\caption[Diagram illustrating remote entanglement generation between pulsed L-type systems.]{\small\textbf{Diagram illustrating remote entanglement generation between pulsed L-type systems.} (a) An L-type system as used in protocols $\mathsf{N}$ and $\mathsf{T}$; a ground state doublet $\ket{\uparrow}$ and $\ket{\downarrow}$ with one optically excited state $\ket{\mathrm{e}}$ that decays back to its initial ground state $\ket{\uparrow}$. (b) A diagram illustrating the fields from each quantum system interfering at a central beam splitter (BS) that has its output ports monitored by single-photon detectors $\text{D}_1$ and $\text{D}_2$.}\label{fig:Lsystem}
\end{figure}

Consider the scheme where two spatially separated L-type systems are entangled by heralding a single-photon emission after erasing the which-path information using a beam splitter (see Fig.~\ref{fig:Lsystem}~(a)). This scheme is similar to the scheme used in the DLCZ repeater protocol to generate entanglement between spatially separated quantum memories \cite{duan2001long}. However, by using fast resonant pulses, the quantum system requires only one optical transition. The scheme generates spin-spin entanglement by using spin-photon number entanglement \cite{rozpkedek2019near}. For this reason, we will denote it as protocol $\mathsf{N}$. For this protocol, we assume that the excited state $\ket{\mathrm{e}}$ can only decay to spin state $\ket{\uparrow}$. That is, we assume that $\gamma_{\downarrow_k}^-\ll \gamma_{\uparrow_k}^-=\gamma_k$, where $k\in\{1,2\}$ indexes the system.

\emph{Protocol description.---}Each system is first prepared in the state $\ket{\downarrow}$. Then a microwave pulse resonant with the $\ket{\downarrow}\longleftrightarrow\ket{\uparrow}$ transition with a pulse area of $\Theta=2\vartheta$ and phase $\varphi_k$ brings the spin qubit to the state $\cos(\vartheta)\ket{\downarrow}+\sin(\vartheta)e^{i\varphi_k}\ket{\uparrow}$. After an optical $\pi$-pulse is applied to excite $\ket{\uparrow}$, each system is left in a superposition of ground and excited states. The excited state then decays back to $\ket{\uparrow}$ and the system emits a photon with a probability $\sin^2(\vartheta)$. By perfectly interfering the fields from two quantum systems (see Fig.~\ref{fig:Lsystem}~(b)), the which-path information is erased and a single detection event will herald one of the Bell states $\ket{\psi^\pm}=(\ket{\uparrow\downarrow}\pm\ket{\downarrow\uparrow})/\sqrt{2}$. 

To show protocol $\mathsf{N}$ in detail, consider the simpler case where $\vartheta=\pi/4$ and $\varphi_k=0$. Then the total state of the quantum systems before decay is
\begin{equation}
\label{eq:singlephoton_initialstate}
    \ket{\psi(t_0)}=\frac{1}{2}(\ket{\downarrow}+\ket{\mathrm{e}})_1\otimes(\ket{\downarrow}+\ket{\mathrm{e}})_2.
\end{equation}
After decay, $\ket{\mathrm{e}}\rightarrow\ket{\uparrow}$, each system is in a spin-photon number entangled state $\ket{\psi}_k=(\ket{\downarrow}\ket{0}+\ket{\uparrow}\ket{1})_k/\sqrt{2}$, where $\ket{0}$ is the vacuum state and $\ket{1}$ is the single-photon state of emission mode. After interfering photons $\ket{1}_1$ and $\ket{1}_2$ at a beam splitter, the state before detection is
\begin{equation}
\label{eq:singlephoton_beamsplitter}
    \frac{1}{2}\left(\ket{\downarrow\downarrow}\ket{ 00}+\ket{\psi^+}\ket{01}+\ket{\psi^-}\ket{10}+\ket{\uparrow\uparrow}\ket{\psi^-_{2002}}
\right)_\text{s,p},\\
\end{equation}
where $\ket{n_1n_2}_\text{p}$ is the state with $n_1$ ($n_2$) photons in the mode of detector $\text{D}_1$ ($\text{D}_2$), $\ket{\psi^\pm}_\text{s}\!=\!\left(\ket{\uparrow\downarrow}\pm\ket{\downarrow\uparrow}\right)/\sqrt{2}$ are spin Bell states, and $\ket{\psi^-_{2002}}_\text{p}=(\ket{20}-\ket{02})/\sqrt{2}$ is a two-photon NOON state. Hence, a single photon at either detector heralds a maximally entangled spin state with a phase determined by which detector received the photon.

The maximum efficiency of the above scheme is 50\%, which is the Bell analyzer efficiency of a single beam splitter \cite{calsamiglia2001maximum}\ref{wein2016bellmeasurement}. However, any amount of photon loss will cause errors due to states $\ket{20}$ and $\ket{02}$ contributing to single-photon measurement outcomes. If $\vartheta$ is small enough, then the probability for both quantum systems to emit photons becomes much less than the probability that only one system emits a photon. Thus, to combat errors due to multi-photon events, the parameter $\vartheta$ can be reduced to improve fidelity at the cost of efficiency \cite{rozpkedek2019near}. This trade-off also improves the protocol fidelity for BD type detectors (see section \ref{chapter2:photoncountingmeasurements} for the definition of BD and PNRD detector models).

\emph{Conditional states.---}In the far field limit or for a cavity-emitter system in the Purcell regime, the source field component collected from a quantum emitter dipole by the waveguide is described by the proportionality relation $\hat{b}_k=\hat{\sigma}_{\uparrow_k}\sqrt{\eta_{\text{c}_k}\gamma_{\text{r}_k}}$ \cite{carmichael,kiraz2004quantum,fischer2016dynamical}. After transmission losses and a propagation phase we have $\hat{b}_k\rightarrow\hat{b}_k\sqrt{\eta_{\text{t}_k}}e^{-i\phi_k}$ where $\phi_k=L_k\omega_{\uparrow_k}/v_\text{p}$, $L_k$ is the propagation distance, and $v_\text{p}$ is the phase velocity. Then, the collected fields are interfered at a beam splitter so that the fields $\hat{d}_1$ and $\hat{d}_2$ at detectors D$_1$ and D$_2$, respectively, are
\begin{equation}
\label{beamsplittereq}
    \begin{pmatrix}
    \hat{d}_1\\
    \hat{d}_2\\
    \end{pmatrix}
    =
    \hat{R}(\theta)    \begin{pmatrix}
    \hat{b}_1\sqrt{\eta_{\text{t}_1}}e^{-i\phi_1}\\
    \hat{b}_2\sqrt{\eta_{\text{t}_2}}e^{-i\phi_2}\\
    \end{pmatrix},
\end{equation}
where $\hat{R}(\theta)$ is the $2\times 2$ rotation unitary matrix. Using a beam-splitter model for detector inefficiency, the effective detected field is $\hat{d}_i\rightarrow\hat{d}_i\sqrt{\eta_{\text{d}_i}}$. In this case, the jump superoperator corresponding to the effective detected field at the $i^\text{th}$ detector is $\mathcal{J}_i\hat{\rho} = \eta_{\text{d}_i}\hat{d}_i\hat{\rho}\hat{d}_i^\dagger$, where to simplify calculations we will take $\eta_{\text{d}_i}=\eta_\text{d}$ for each detector.

If we assume that the pulse Rabi frequency is much faster than the rates of dissipation and decoherence, and that the excitation pulses are resonant with the emitters, then we can consider the initial state to be approximated by $\hat{\rho}(t_0)=\ket{\psi(t_0)}\!\bra{\psi(t_0)}_1\otimes\ket{\psi(t_0)}\!\bra{\psi(t_0)}_2$ where $\ket{\psi(t_0)}_k=\cos(\vartheta)\ket{\downarrow}+\sin(\vartheta)e^{i\varphi_k}\ket{\mathrm{e}}$. Under these conditions, the set of conditional states can also be truncated to those $\mathbf{n}$ such that $n_1+n_2\leq 2$ as a consequence of each emitter only being able to emit up to a single photon.

The successful conditional states are associated with the single-photon detection conditions $\mathbf{n}=$ $(1,0)$, and $(0,1)$, which are given by their corresponding conditional propagators $\mathcal{W}_\mathbf{n}$ (see section \ref{chapter2:conditionalpropagationsuperoperators}). For convenience, we notate these vectors by 10, and 01, respectively. For example, the outcome associated with $\ket{\psi^+}$ is $\hat{\rho}_{01}(t_\text{f})=\mathcal{W}_{01}(t_\text{f},t_\text{d}^\prime,t_\text{d},t_0)\hat{\rho}(t_0)$ where $\mathcal{W}_{01}(t_\text{f},t_\text{d}^\prime,t_\text{d},t_0)=\mathcal{U}(t_\text{f},t_\text{d}^\prime)\mathcal{U}_{01}(t_\text{d}^\prime,t_\text{d})\mathcal{U}(t_\text{d},t_0)$ and
\begin{equation}
\label{singlephotonconditionalstates}
\begin{aligned}
    \mathcal{U}_{01}(t_\text{d}^\prime,t_\text{d})
    = \int_{t_\text{d}}^{t_\text{d}^\prime}\mathcal{U}_\mathbf{0}(t_\text{d}^\prime,t)\mathcal{J}_2\mathcal{U}_\mathbf{0}(t,t_\text{d})dt.
\end{aligned}
\end{equation}
Likewise, the outcome $\hat{\rho}_{10}$ associated with $\ket{\psi^-}$ is given by $\mathcal{W}_{10}$, which is determined using Eq.~(\ref{singlephotonconditionalstates}) but with $\mathcal{J}_1$ in place of $\mathcal{J}_2$.
The states corresponding to the remaining relevant conditions $\mathbf{n}=$ (2,0), (1,1), and (0,2) are similarly obtained from Eqs.~(\ref{conditionalpropagator}) and (\ref{windowpropagator}).

\begin{figure}
    \centering
    \includegraphics[width=0.55\textwidth]{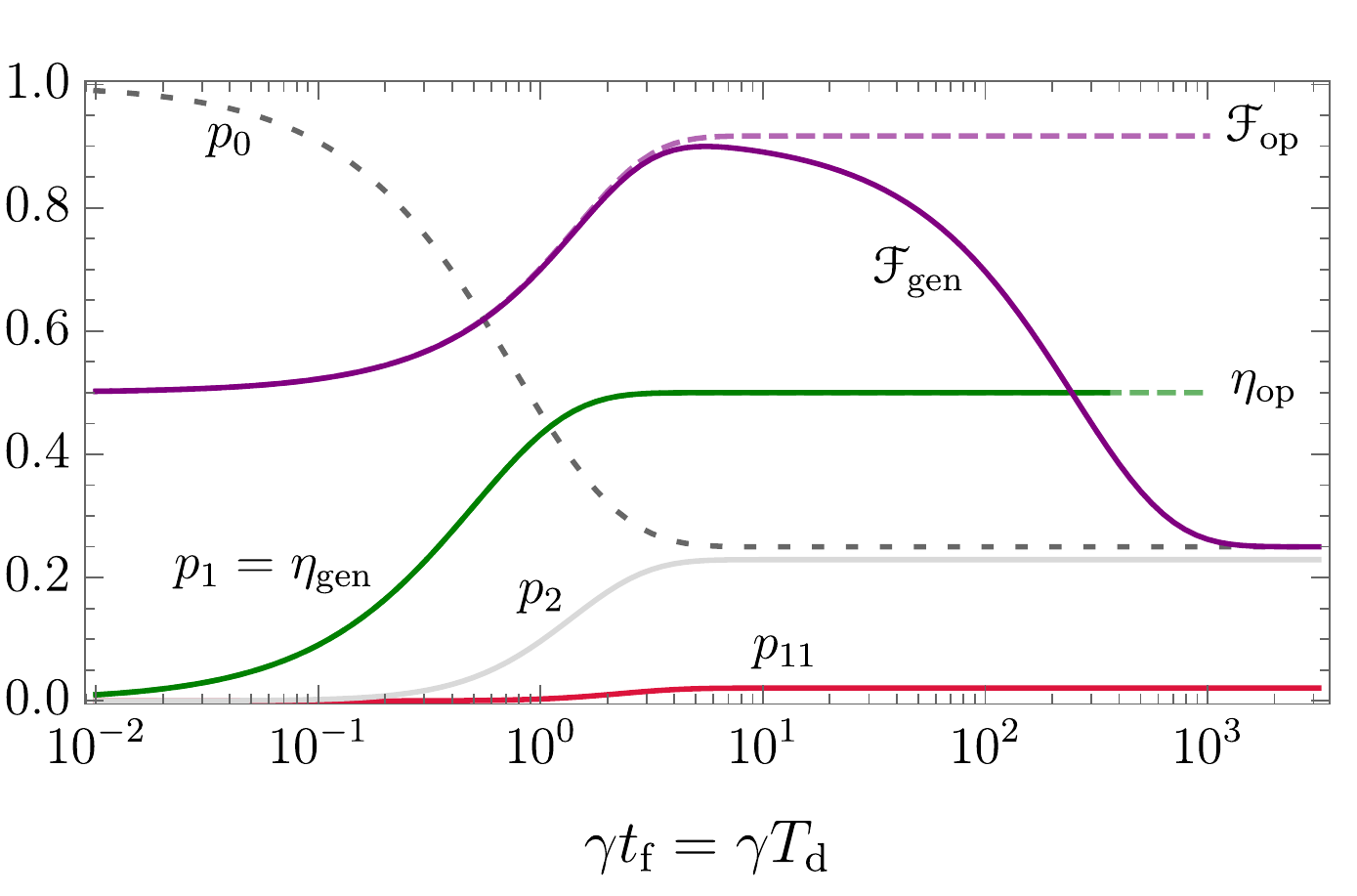}
    \caption[The time dynamics of remote entanglement generation via spin-photon number entanglement.]{\small\textbf{The time dynamics of remote entanglement generation via spin-photon number entanglement: protocol $\mathsf{N}$.} Entanglement generation fidelity $\euscr{F}_\mathrm{gen}$ and efficiency $\eta_\text{gen}$ as a function of protocol time $t_\text{f}$ for an initial state $\hat{\rho}(t_0)=\ket{\psi(t_0)}\!\bra{\psi(t_0)}$ where $\ket{\psi(t_0)}=(1/2)(\ket{\downarrow}+\ket{\mathrm{e}})^{\otimes 2}$ and where there is no loss and noiseless local photon-number resolving detectors. The asymptotic dashed lines indicate the limits on fidelity $\euscr{F}_\mathrm{op}$ and efficiency $\eta_\text{op}$ reached when spin decoherence is neglected and when the detection window encompasses the entire photon lifetime. The detection window is set to be equal to the entire protocol duration: $t_\text{d}=t_0=0$ and $t_\text{d}^\prime=t_\text{f}=T_\text{d}$. The gray lines show the probabilities for no photon emission from either system $p_0=\text{Tr}[\hat{\rho}_\mathbf{0}]$ and for photon bunching $p_2=p_{20}+p_{02}=\text{Tr}[\hat{\rho}_{20}+\hat{\rho}_{02}]$. The red line indicates coincident counts $p_{11}=\text{Tr}[\hat{\rho}_{11}]$ caused by imperfect HOM interference. Parameters chosen: $\gamma_k=\gamma$, $\gamma_k^\star=0.1\gamma$, and $\gamma_{\text{s}_k}^\pm=\gamma_{\mathrm{s}_k}^\star=0.001\gamma$ for $k\in\{1,2\}$.}
    \label{fig:onephotonTimeDynamics}
\end{figure}

\emph{Measurement duration.---}The spin entanglement is generated at the moment a single photon from one of the emitters is detected by one of the two detectors. However, a subsequent detection of another photon will destroy this entanglement \cite{martin2019single}. As a consequence, unless the probability for two-photon events is very small, the detection duration must be long enough to ensure that only one photon was emitted. Hence, the fidelity can be very low for small $T_\text{d}$. On the other hand, for a detection window much longer than the lifetime, the fidelity becomes limited by spin decoherence processes. Fig.~\ref{fig:onephotonTimeDynamics} shows the entanglement generation fidelity and efficiency as the detection window duration is increased, illustrating the peak in fidelity when the duration is on the order of the lifetime. To show this qualitative behaviour, we have chosen parameters in the regime $\gamma>\gamma^\star\gg\gamma_\mathrm{s}^\star$ to represent a solid-state system that could potentially serve as a quantum communication node. The two-photon probabilities for photon bunching $(p_2)$ and coincident counts ($p_{11}$) are also illustrated. Note that the coincident counts are nonzero due to the emitter pure dephasing that degrades the HOM interference \cite{hong1987measurement} between photons from different sources.

Under the condition where the systems experience negligible spin decoherence on the timescale of the lifetime of the emitter, we can analytically solve the conditional states as a function of detection window duration. This can then be used to estimate simple figures of merit for the quality of entanglement based only on the optical properties of the emitters. These limits are illustrated by the asymptotes of the dashed lines in Fig.~\ref{fig:onephotonTimeDynamics}. The analytic solutions can also then be used to estimate the fidelity under the effects of additional imperfections such as noisy detectors and spectral diffusion using the methods outlined in sections \ref{chapter2:photoncountingmeasurements} and \ref{chapter4:remotesystems}, respectively.

Suppose that the detection window begins at $t_\text{d}$ such that $r(t_\text{d})=t_0$ and ends at $t_\text{d}^\prime$ such that $r(t_\text{d}^\prime)=T_\text{d}+t_0$. Using the appropriate conditional propagators $\mathcal{W}_\mathbf{n}$, we compute the final (unnormalized) spin-spin conditional states in the rotating frame of the spin qubits after time $t_\text{f}\gg1/\gamma_k$. In this limit of time, neither quantum system remains in $\ket{\mathrm{e}}$.

The conditional spin state of the quantum systems given that both detector modes do not contain a photon from an emitter is
\begin{equation}
\begin{aligned}
    \hat{\rho}_\mathbf{0} &= \frac{1}{4}\sin^2(2\vartheta)\left((1-\overline{\beta}_1)\ket{\uparrow\downarrow}\!\bra{\uparrow\downarrow}+(1-\overline{\beta}_2)\ket{\downarrow\uparrow}\!\bra{\downarrow\uparrow}\right)\\ &\hspace{5mm}+(1-\overline{\beta}_1)(1-\overline{\beta}_2)\sin^4(\vartheta)\ket{\uparrow\uparrow}\!\bra{\uparrow\uparrow}+\cos^4(\vartheta)\ket{\downarrow\downarrow}\!\bra{\downarrow\downarrow},\\
\end{aligned}
\end{equation}
where $\overline{\beta}_k=\eta_\text{d}\beta_k/\gamma_k=\eta_k\left(1-e^{-T_\text{d}\gamma_k}\right)$ is the rate-normalized brightness and $\eta_k=\eta_\text{d}\eta_{\text{t}_k}\eta_{\text{c}_k}\gamma_{\text{r}_k}/\gamma_k$ is the total single-photon efficiency of each quantum system.

The single-photon conditioned states of the quantum system are
\begin{equation}
\begin{aligned}
    \hat{\rho}_\text{s}^\pm= &\rho_{\uparrow\uparrow}^\pm\ket{\uparrow\uparrow}\!\bra{\uparrow\uparrow}+\rho_{\uparrow\downarrow}^\pm\ket{\uparrow\downarrow}\!\bra{\uparrow\downarrow}+\rho_{\downarrow\uparrow}^\pm\ket{\downarrow\uparrow}\!\bra{\downarrow\uparrow}
    +\left(\rho_\text{c}^\pm\ket{\downarrow\uparrow}\!\bra{\uparrow\downarrow}+\text{h.c.}\right),
\end{aligned}
\end{equation}
where
\begin{equation}
\begin{aligned}
\label{P1condsol}
    \rho_{\uparrow\uparrow}^\pm &= \frac{1}{2}\!\left(\overline{\beta}_1\!+\!\overline{\beta}_2\!-\!2\overline{\beta}_1\overline{\beta}_2\pm(\overline{\beta}_1\!-\!\overline{\beta}_2)\cos(2\theta)\right)\!\sin^4\!\left(\vartheta\right)\\
    \rho_{\downarrow\uparrow}^\pm &=\frac{\overline{\beta}_1}{8}(1\pm\cos(2\theta))\sin^2(2\vartheta)\\
    \rho_{\uparrow\downarrow}^\pm &= \frac{\overline{\beta}_2}{8}(1\mp\cos(2\theta))\sin^2(2\vartheta)\\
    \rho_\text{c}^\pm \!&= \pm\frac{\tilde{C}(T_\text{d})}{8}\sqrt{\eta_1\eta_2}\sin(2\theta)\sin^2(2\vartheta)\\
    \tilde{C}(T_\text{d})&=\frac{2\sqrt{\gamma_
1\gamma_2}}{\Gamma_1+\Gamma_2+2i\Delta}\!\left(1-e^{-\frac{1}{2}T_\text{d}(\Gamma_1+\Gamma_2+2i\Delta)}\right)\!e^{i(\varphi+\phi)},
\end{aligned}
\end{equation}
and where $\Gamma_k = \gamma_k+2\gamma_k^\star$ is the FWHM of the emission ZPL for system $k$, $\Delta=\omega_{\uparrow_1}-\omega_{\uparrow_2}$ is the spectral detuning, $\varphi=\varphi_1-\varphi_2$ is the relative initialization phase, and $\phi=\phi_1-\phi_2$ is the relative propagation phase. The sign of $\hat{\rho}_\text{s}^\pm$ is given by which detector received the photon: $\hat{\rho}_{10}=\hat{\rho}_\text{s}^-$ and $\hat{\rho}_{01}=\hat{\rho}_\text{s}^+$. The quantity $\tilde{C}$ is a complex value related to the spin-spin coherence and entanglement concurrence, not to be confused with the cavity cooperativity $C$ used in other sections of this thesis. Note that $\rho_{\downarrow\downarrow}=0$ because after either system emits one photon, the system is guaranteed to not be in $\ket{\downarrow\downarrow}$ due to our assumptions about the allowed emitter transitions and under the approximation of a negligible spin flip rate.

The individual two-photon conditioned states $\hat{\rho}_{20}$, $\hat{\rho}_{11}$, and $\hat{\rho}_{02}$ are all proportional to $\ket{\uparrow\uparrow}\!\bra{\uparrow\uparrow}$, as expected. However, their trace has a complicated dependency on $T_\text{d}$ due to the HOM effect. Regardless, their sum can be easily simplified to the intuitive result
\begin{equation}
    \hat{\rho}_{20}+\hat{\rho}_{11}+\hat{\rho}_{02}=\overline{\beta}_1\overline{\beta}_2\sin^4(\vartheta)\ket{\uparrow\uparrow}\!\bra{\uparrow\uparrow}.
\end{equation}
Using all these conditional states, we can also verify that
\begin{equation}
\begin{aligned}
    \hat{\rho}_\mathbf{0}+\hat{\rho}_\text{s}^++\hat{\rho}_\text{s}^-+\hat{\rho}_{20}+\hat{\rho}_{11}+\hat{\rho}_{02}&=\sin^4(\vartheta)\ket{\uparrow\uparrow}\!\bra{\uparrow\uparrow}+\cos^4(\vartheta)\ket{\downarrow\downarrow}\!\bra{\downarrow\downarrow}\\&\hspace{5mm}+\frac{1}{4}\sin^2(2\vartheta)\left(\ket{\uparrow\downarrow}\!\bra{\uparrow\downarrow}+\ket{\downarrow\uparrow}\!\bra{\downarrow\uparrow}\right)\\
\end{aligned}
\end{equation}
is indeed the solution $\hat{\rho}(t)$ of the total master equation in the limit $t\gg1/\gamma_k$ and when spin decoherence is neglected. This confirms the completeness of the photon-number decomposition.

The states $\hat{\rho}_\text{s}^\pm$ correspond to the expected Bell states $\ket{\psi^\pm}$ and so the average entanglement fidelity (recall section \ref{chapter1:fidelity}) is
\begin{equation}
    \euscr{F}_\text{gen} = \frac{1}{\eta_\text{gen}}\left(\bra{\psi^+}\hat{\varrho}_\text{s}^+\ket{\psi^+}+\bra{\psi^-}\hat{\varrho}_\text{s}^-\ket{\psi^-}\right),
\end{equation}
where $\eta_\text{gen} = \text{Tr}\left[\hat{\varrho}_\text{s}^+\right]+\text{Tr}\left[\hat{\varrho}_\text{s}^-\right]$ is the entanglement generation efficiency and $\hat{\varrho}_\text{s}^\pm$ is the state after measurement computed from the conditional state $\hat{\rho}_\text{s}^\pm$.

\emph{Optical limits.---}Suppose that the interference is balanced so that $\theta=\pi/4$ and $\eta_1=\eta_2=\eta$. Also, suppose that the protocol is phase corrected so that $\varphi+\phi=0$ (see section \ref{chapter4:entanglementcomparison} for a discussion on phase errors). Then in the limit that $T_\text{d}\gg 1/\gamma_k$ we have $\overline{\beta}_1\rightarrow \eta$, $\overline{\beta}_2\rightarrow \eta$, and
\begin{equation}
\label{ctilde}
    \tilde{C}\rightarrow\frac{2\sqrt{\gamma_1\gamma_2}}{\Gamma_1+\Gamma_2+2i\Delta}.
\end{equation}
If we also assume that the measurement is performed by ideal noiseless PNRDs, then $\hat{\varrho}_\text{s}^\pm=\hat{\rho}_\text{s}^\pm$ and under these conditions---which we refer to as the optical limit---the corresponding entanglement generation fidelity $\euscr{F}_\mathrm{op}$ gives an estimate of the fidelity determined only by the optical properties of the emitters. In principle, this estimate could be exceeded using spectral or temporal post selection of photons, consequently sacrificing efficiency.

In the optical limit, the fidelity for protocol $\mathsf{N}$ becomes 
\begin{equation}
\label{Nfidelity}
\begin{aligned}
     \euscr{F}_\text{op}&=\frac{1}{2}\left(1+\text{Re}(\tilde{C})\right)F_\eta(\vartheta)\\
\end{aligned}
\end{equation}
with concurrence $\euscr{C}_\text{op}=|\tilde{C}|F_\eta(\vartheta)$, where the loss compensation factor is $F_\eta(\vartheta)=\cos^2(\vartheta)/(1\!-\!\eta\sin^2(\vartheta))$ that arose also in the context of self-homodyne measurements of number coherence (recall section \ref{chapter3:numbercoherence}). The efficiency becomes \mbox{$\eta_\text{op}=(\eta/2)\sin^2(2\vartheta)/F_\eta(\vartheta)$} and the two-photon conditioned states reduce to
\begin{equation}
\label{2photoncondstates}
\begin{aligned}
    \hat{\rho}_{20} &= \hat{\rho}_{20} = \frac{1}{4}\left(1+M_{12}\right)\eta^2\sin^4(\vartheta)\ket{\uparrow\uparrow}\!\bra{\uparrow\uparrow}\\
    \hat{\rho}_{11} &= \frac{1}{2}\left(1-M_{12}\right)\eta^2\sin^4(\vartheta)\ket{\uparrow\uparrow}\!\bra{\uparrow\uparrow},
\end{aligned}
 \end{equation}
where
\begin{equation}
\label{meanwavepacketoverlap}
    M_{12} = M_\gamma\frac{(\Gamma_1+\Gamma_2)(\gamma_1+\gamma_2)}{(\Gamma_1+\Gamma_2)^2+4\Delta^2}\leq \sqrt{M_1M_2},
\end{equation}
is the mean wavepacket overlap between photons from defects $k=1$ and $k=2$, $M_k=\gamma_k/\Gamma_k$ is the individual system indistinguishability from Eq.~(\ref{ch1eq:mwpo}), and $M_\gamma=4\gamma_1\gamma_2/(\gamma_1+\gamma_2)^2\geq M_{12}$ quantifies the temporal profile mismatch. We emphasize that Eq.~(\ref{2photoncondstates}) and $M_{12}$ in Eq.~(\ref{meanwavepacketoverlap}) were solved using the methods of chapter \ref{chapter2} and not by computing the mean wavepacket overlap of two different input photons as in Eq.~(\ref{chapter2eq:meanwavepacketoverlapdifferent}). However, we have verified that solving the mean wavepacket overlap indeed gives the same result as Eq.~(\ref{meanwavepacketoverlap}), which confirms that the photon statistics of the HOM interference are independent of whether the calculation is performed from the perspective of the source or the field.

For a given $F_\eta(\vartheta)$, the fidelity and concurrence are limited by the spectral and temporal properties of the individual emitters. In particular, we can identify that $\euscr{C}_\text{op}^2\leq M_{12}F^2_\eta(\vartheta)$. Hence for protocol $\mathsf{N}$, the square root of the mean wavepacket overlap $M_{12}$ gives an upper bound on the entanglement generation concurrence, which itself can be used to determine an upper bound on the entanglement generation fidelity by $\euscr{F}_\text{op} \leq\left(1+\sqrt{M_{12}}\right)F_\eta(\vartheta)$. On the other hand, we have that $\text{Re}(\tilde{C})=M_{12}/\sqrt{M_\gamma}\geq M_{12}$. Hence the optical limit of fidelity $\euscr{F}_\text{op}$ for protocol $\mathsf{N}$ is bounded by
\begin{equation}
    \frac{1}{2}\left(1+M_{12}\right)\leq\frac{\euscr{F}_\text{op}}{F_\eta(\vartheta)}\leq\frac{1}{2}\left(1+\sqrt{M_{12}}\right).
\end{equation}

\emph{Detector noise and number resolution.---}The entangled spin-spin state after a single-photon measurement by a PNRD with non-negligible noise is $\hat{\varrho}_\text{s}^\pm = p_\mathrm{d,1}^2\hat{\rho}_\text{s}^\pm+ p_\mathrm{d,0}p_\mathrm{d,1}\hat{\rho}_\mathbf{0}$, where $p_{\mathrm{d},n}(T_\text{d},\gamma_\mathrm{d})$ is the probability to have $n$ dark counts within the detection window $T_\text{d}$. For $T_\text{d}\gg1/\gamma_k$, we can write
\begin{equation}
\label{fgennoise}
    \euscr{F}_\text{gen} = \frac{1}{\eta_\text{gen}}\left(p_{\mathrm{d},0}^2\euscr{F}_\text{op}\eta_\text{op}+\frac{1}{2}p_{\mathrm{d},0}p_{\mathrm{d},1}(1-\eta)\sin^2(2\vartheta)\right),
    \vspace{-1mm}
\end{equation}
where
\vspace{-1mm}
\begin{equation}
    \eta_\text{gen} = p_{\mathrm{d},0}^2\eta_\text{op} + 2p_{\mathrm{d},0}p_{\mathrm{d},1}\left(1-\eta\sin^2(\vartheta)\right)^2
\end{equation}
is the total efficiency. For a measurement by a BD, the state after heralding is given by
\begin{equation}
\label{bdnoise}
\begin{aligned}
    \hat{\varrho}_\text{s}^-&=p_{\mathrm{d},0}(\hat{\rho}_\text{s}^-+\hat{\rho}_{20}) + p_{\mathrm{d},0}(1-p_{\mathrm{d},0})\hat{\rho}_\mathbf{0}\\
    \hat{\varrho}_\text{s}^+&=p_{\mathrm{d},0}(\hat{\rho}_\text{s}^++\hat{\rho}_{02}) + p_{\mathrm{d},0}(1-p_{\mathrm{d},0})\hat{\rho}_\mathbf{0},
\end{aligned}
\end{equation}
which can be used to compute the fidelity and efficiency in the same way as for the PNRD case.

In the absence of detector noise and for a given $\eta$, $\euscr{F}_\text{op}$ can be maximized by increasing $F_\eta$ arbitrarily close to 1 by taking $\vartheta\rightarrow 0$ and sacrificing efficiency. However, detector noise places an additional constraint on the fidelity due to the presence of a finite noise floor. This gives rise to an optimal $\vartheta\neq 0$ that maximizes fidelity. In the regime where $1-p_{\mathrm{d},0}\simeq p_{\mathrm{d},1}\ll\eta$, we find that Eq.~(\ref{fgennoise}) for the PNRD case is maximized when $\vartheta\simeq[p_{\mathrm{d},1}/(\eta(1-\eta))]^{1/4}$. Note that this estimate is also only accurate for $\eta<p_{\mathrm{d},0}\simeq 1$ as evidently $\vartheta=\pi/4$ is the optimal choice for $\eta=1$ when using a PNRD. As for the BD case, the optimal $\vartheta$ depends on $M_{12}$ due to the contribution from two-photon events. For $M_{12}\simeq 1$ we use the conditional states in Eq.~(\ref{bdnoise}) to find that $\vartheta\simeq [2p_{\mathrm{d},1}/(\eta(2-\eta))]^{1/4}$ maximizes the fidelity. When $\eta=1$ this optimal choice becomes $\vartheta\simeq(2p_{\mathrm{d},1})^{1/4}$. In the regime of quantum communication where $\eta\ll1$, two-photon detections are suppressed due to losses and so the PNRD and BD models give equivalent results.

\subsection{Spin-time bin entanglement}
\label{chapter4:timeentanglement}

For the second protocol (denoted by $\mathsf{T}$), which uses spin-time bin entanglement, we focus on the extension of protocol $\mathsf{N}$ where two successive photons herald entanglement between L-type systems (see Fig.~\ref{fig:Lsystem}). This protocol is also referred to as the Barrett-Kok scheme \cite{barrett2005efficient}, which was utilized to demonstrate the first loophole-free Bell inequality violation \cite{hensen2015loophole}.

\emph{Protocol description.---}Each system is first prepared in the maximal superposition state $(\ket{\uparrow}+\ket{\downarrow})/\sqrt{2}$. Then a resonant  $\pi$ pulse excites the $\ket{\uparrow}$ states at $t_0$, giving Eq.~(\ref{eq:singlephoton_initialstate}). Following protocol $\mathsf{N}$, we could obtain the entangled state $\ket{\psi^{\pm}}$ by post-selecting on a single photon. However, to eliminate the errors caused by both systems emitting photons after the first pulse, we can flip the spin state of both systems and re-excite $\ket{\uparrow}$ some time $t_x-t_0$ after the first pulse. If the quantum systems emit only one photon either before or after the second pulse, then they are each in a spin-time bin entangled state $\ket{\psi}_k=(\ket{\downarrow}\ket{\text{early}}+\ket{\uparrow}\ket{\text{late}})/\sqrt{2}$, where $\ket{\text{early}}$ and $\ket{\text{late}}$ represent the presence of a photon in the early  and late time bin modes, respectively. The joint state $\ket{\psi}_1\otimes\ket{\psi}_2$ can be written in the Bell basis of the spin and photon states
\begin{equation}
\label{bellbasis}
\frac{1}{2}\left(\ket{\psi^+}\ket{\psi^+} -\ket{\psi^-}\ket{\psi^-} +\ket{\phi^+}\ket{\phi^+} -\ket{\phi^-}\ket{\phi^-} \right)_\text{s,p},
\end{equation} 
where $\ket{\psi^{\pm}}_\text{s}$ is as before and $\ket{\phi^\pm}_\text{s}=\frac{1}{\sqrt{2}}(\ket{\uparrow\uparrow}\pm\ket{\downarrow\downarrow})$. The Bell states $\ket{\psi^{\pm}}_\text{p}$ and $\ket{\phi^{\pm}}_\text{p}$ are similarly defined using time bin states $\ket{\text{early}}$ and $\ket{\text{late}}$. Interfering the joint state at a beam splitter performs a partial Bell-state measurement (BSM), allowing the identification of $\ket{\psi^+}_\text{p}$ and $\ket{\psi^-}_\text{p}$ from $\ket{\phi^\pm}_\text{p}$. This projects the spin state onto either $\ket{\psi^+}_\text{s}$ or $\ket{\psi^-}_\text{s}$ with a 50\% total probability.

In the absence of spin-flipping decoherence, and with perfect spin-flipping operations, neither quantum system can emit a photon in both the early and late time bins. Thus, when neglecting detector dark counts, the entanglement fidelity is independent of photon losses and the protocol does not suffer from an inherent efficiency-fidelity trade-off. The ramification is that a photon from each emitter must be transmitted to the beam splitter, which reduces the overall protocol efficiency.

\emph{Conditional states.---}
Let $\mathcal{X}(t_\text{x}^\prime,t_\text{x})$ be the superoperator propagator that performs a spin-flip and re-excitation of both systems beginning at time $t_\text{x}$ and concluding at time $t_\text{x}^\prime$. The conditional state $\hat{\rho}_{\mathbf{n}_l,\mathbf{n}_\text{e}}$ given photon counts $\mathbf{n}_\text{e}$ in the early detection window and $\mathbf{n}_l$ in the late detection window is $\hat{\rho}_{\mathbf{n}_l,\mathbf{n}_\text{e}} =\mathcal{W}_{\mathbf{n}_l}\mathcal{X}\mathcal{W}_{\mathbf{n}_\text{e}}\hat{\rho}$, where $\hat{\rho}$ is the state after the first excitation and $\mathcal{W}_{\mathbf{n}_l}$ ($\mathcal{W}_{\mathbf{n}_\text{e}}$) is the conditional propagator for the late (early) time bin detection window dependent on the time ordering $\mathcal{W}_{\mathbf{n}_l}(t_\text{f},r(t_l^\prime),r(t_l),t_\text{x}^\prime)$ ($\mathcal{W}_{\mathbf{n}_\text{e}}(t_\text{x},r(t_\text{e}^\prime),r(t_\text{e}),t_0)$). The detector window duration for the early and late bins are $T_\text{e} = t_\text{e}^\prime-t_\text{e}$ and $T_l=t_l^\prime-t_l$, respectively. 

There are four measurement outcomes that may indicate successful entanglement: $\{\mathbf{n}_l,\mathbf{n}_\text{e}\}=\{(1,0),(1,0)\}$, $\{(1,0),(0,1)\}$, $\{(0,1),(1,0)\}$, and $\{(0,1),(0,1)\}$. For notation convenience, we concatenate the sets of vectors. For example, $\hat{\rho}_{(1,0),(0,1)}=\hat{\rho}_{1001}$. The conditional states are then given by the appropriate conditional propagators $\mathcal{W}_\mathbf{n}$. Using the case 1001 as an example, we have $\hat{\rho}_{1001} = \mathcal{W}_{10}\mathcal{X}\mathcal{W}_{01}\hat{\rho}$, where $\mathcal{W}_{10}$ and $\mathcal{W}_{01}$ are the same as in protocol $\mathsf{N}$. The remaining conditional states can be similarly expressed in terms of $\mathcal{U}_\mathbf{0}$, $\mathcal{J}$, and $\mathcal{X}$ superoperators, although for brevity we do not display them.

\emph{Measurement duration.---}For protocol $\mathsf{T}$, there are two detection windows beginning at $t_{\text{e}}$ and $t_{\text{l}}$ with duration $T_\text{e}$ and $T_\text{l}$, respectively. Suppose that the detection window is continuous between the pulses. Then we have $t_\text{e}=t_0$, $t_\text{e}^\prime=t_\text{x}$. Also, if the spin-flip and re-excitation is much faster than other system dynamics so that $t_\text{x}^\prime-t_\text{x}\simeq 0$, then we have $t_\text{e}^\prime=t_\text{l}=t_\text{x}$. To simplify the problem, we also make the time bins equal in duration so that $T_\text{e}=T_\text{l}=T_\text{d}$. 

\begin{figure}[t]
    \centering
    \includegraphics[width=0.55\textwidth]{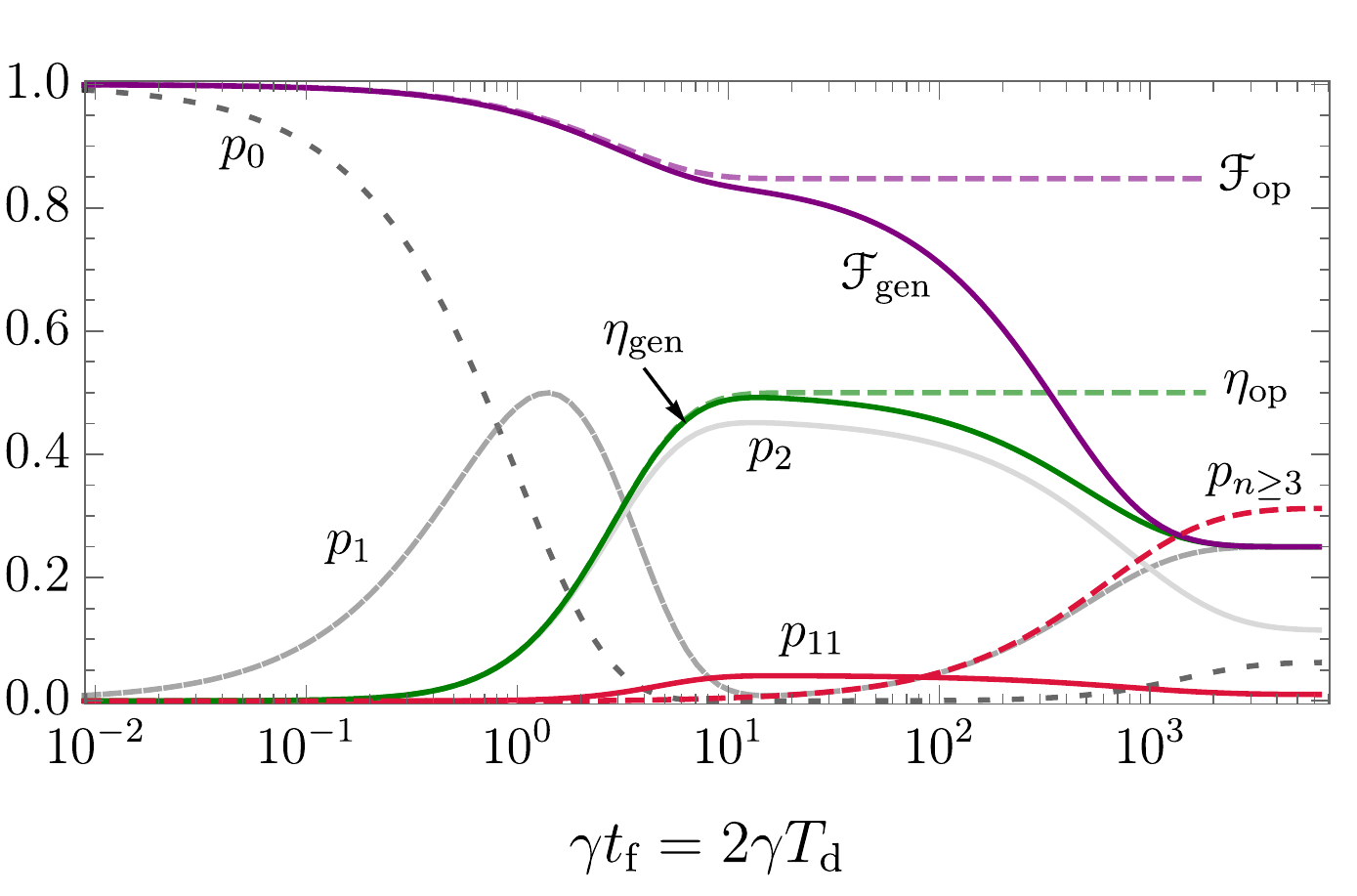}
    \caption[The time dynamics of remote entanglement generation via spin-time bin entanglement.]{\small\textbf{The time dynamics of remote entanglement generation via spin-photon number entanglement: protocol $\mathsf{T}$.} Entanglement generation fidelity $\euscr{F}_\mathrm{gen}$ and efficiency $\eta_\text{gen}$ as a function of protocol time $t_\text{f}$ for an initial state $\hat{\rho}(t_0)=\ket{\psi(t_0)}\!\bra{\psi(t_0)}$ where $\ket{\psi(t_0)}=(1/2)(\ket{\downarrow}+\ket{\mathrm{e}})^{\otimes 2}$ and where there is no loss and noiseless local photon-number resolving detectors. The asymptotic dashed lines indicate the limits of fidelity $\euscr{F}_\mathrm{op}$ and efficiency $\eta_\text{op}$ reached when spin decoherence is neglected. The detection window for each time bin is set to be equal to half the protocol duration, which begins after the first system excitation: $t_\text{d}=t_0=0$, $t_\text{d}^\prime=t_\text{x}=T_\text{d}$, and $t_\text{f}=2T_\text{d}$. The gray lines show the probabilities for no photon emission from either system $p_0$, for single-photon detection events $p_1$, and for photon bunching events $p_2$. The red solid line shows coincident counts $p_{11}$ caused by imperfect HOM interference and the red dashed line indicates the probability $p_{n\geq 3}$ for 3 or more photons to be emitted as a consequence of spin relaxation between the pulses. Parameters chosen: $\gamma_k=\gamma$, $\gamma_k^\star=0.1\gamma$, and $\gamma_{\text{s}_k}^\pm=\gamma_{\mathrm{s}_k}^\star=0.001\gamma$ for $k\in\{1,2\}$.}
    \label{fig:BK-timedynamics}
\end{figure}

If the detection windows do not encompass the entire photon lifetime, then a high fidelity can be attained because after heralding by two-photon events, both systems will be in the ground state with a high probability. When this is the case, the detection window post selects photons that were emitted early compared to the total lifetime (see Fig.~\ref{fig:BK-timedynamics}). This again demonstrates how fast detector gate times can potentially purify photon indistinguishability and increase the overall spin-spin entanglement fidelity (recall section \ref{chapter3:photonicbellstates}). Consequently, the efficiency in this regime is very low. Note that this type of temporal post selection can also be applied to protocol $\mathsf{N}$ provided that $\vartheta$ is very small.

When the time bin duration is on the order of the emission lifetime, the fidelity briefly plateaus at the optical limit where the non-zero coincidence count probability $p_{11}$ indicates imperfect interference. In this regime, the efficiency approaches the ideal Bell-analyzer efficiency of 50$\%$. However, if the duration is much longer than the optical lifetime, then spin flips occurring between the excitation pulses increase the probability to have three or more photons emitted during the protocol, which reduces the efficiency and fidelity to their thermal limits of $0.25$.

For brevity we do not show the full conditional state solutions for protocol $\mathsf{T}$. However, due to the symmetry of this protocol and its close relationship with protocol $\mathsf{N}$, the fidelity for $\theta=\pi/4$ when neglecting spin decoherence and detector noise takes the simple form
\begin{equation}
\begin{aligned}
\label{BKfidelity}
    \euscr{F}_\text{gen} = \frac{1}{2}\left(1 + \frac{\eta_1\eta_2}{2\eta_\text{gen}}|\tilde{C}(T_\text{d})|^2\right),
\end{aligned}
\end{equation}
where $\eta_\text{gen} =\overline{\beta}_1\overline{\beta}_2/2$ is the efficiency, $\tilde{C}(T_\text{d})$ is given by Eq.~(\ref{P1condsol}), and $\overline{\beta}_k$ is the same as in protocol $\mathsf{N}$. This expression accounts for emitter pure dephasing through $\tilde{C}(T_\text{d})$ and can also be averaged for a fluctuating detuning $\Delta$ to capture spectral diffusion.

\emph{Optical limits.---}In the limit that $T_{\text{d}}\gg1/\gamma_k$ we have $\overline{\beta}_k\rightarrow\eta_k$ and $\tilde{C}$ again reduces to Eq.~(\ref{ctilde}). Then the optical limits of efficiency and fidelity are $\eta_\text{op} = \eta^2/2$ and $\euscr{F}_\text{op}=(1+|\tilde{C}|^2)/2$, respectively, for $\eta_k=\eta$. The corresponding concurrence is simply $\euscr{C}_\text{op}=|\tilde{C}|^2$.

As with protocol $\mathsf{N}$, the fidelity and concurrence can be related to the mean wavepacket overlap $M_{12}$ by noting
\begin{equation}
    \euscr{C}_\text{op} = M_{12}\left(\frac{\gamma_1+\gamma_2}{\Gamma_1+\Gamma_2}\right)\leq M_{12}.
\end{equation}
On the other hand, it can be shown that $\euscr{F}_\text{op}\geq M_{12}$. Hence the optical limit of fidelity for $\mathsf{T}$ is bounded by
\begin{equation}
    M_{12}\leq \euscr{F}_\text{op}\leq\frac{1}{2}\left(1+M_{12}\right).
\end{equation}
We note that the upper bound result presented here has also been derived in the supplementary of Ref.~\cite{bernien2013heralded} using arguments from interference visibility. However, since pure dephasing degrades both the interference visibility and the coherence of the emitter states, the final spin-spin fidelity must be less than or equal the value computed from interference visibility alone.

\emph{Detector noise and number resolution.---}Because of detector dark counts, it is possible that zero or single-photon conditioned states appear to give successful measurements. After taking detector noise into consideration with PNRDs as described in subsection~\ref{chapter2:photoncountingmeasurements} we have, for example,
\begin{equation}
\label{BK-PNRD}
\begin{aligned}
\hspace{-1mm}\hat{\varrho}_{1001}= p_{\mathrm{d},0}^4\hat{\rho}_{1001}
    + p_{\mathrm{d},0}^3p_{\mathrm{d},1}\left(\hat{\rho}_{1000}\!+\!\hat{\rho}_{0001}\right)+p_{\mathrm{d},0}^2p_{\mathrm{d},1}^2\hat{\rho}_\mathbf{0}.\\
\end{aligned}
\end{equation}

In the absence of detector noise, only conditional states corresponding to three or more total detected photons will cause errors when using BDs with protocol $\mathsf{T}$. This only occurs if the probability for a spin flip in between the pulses is non-negligible and photon loss is not too low. Conditional states where two photons arrive at one detector can combine with a single dark count at another detector to cause errors. However, for reasonably high photon losses or a reasonably low spin flip probability, both of these contributions to errors are negligible compared to other sources. Hence, Eq.~(\ref{BK-PNRD}) also well-approximates the measured state for BDs in this regime. This illustrates the robustness of protocol $\mathsf{T}$ against losses.

\subsection{Spin-polarization entanglement}
\label{chapter4:polarizationentanglement}

\begin{figure}
\centering
\includegraphics[width=0.7\textwidth]{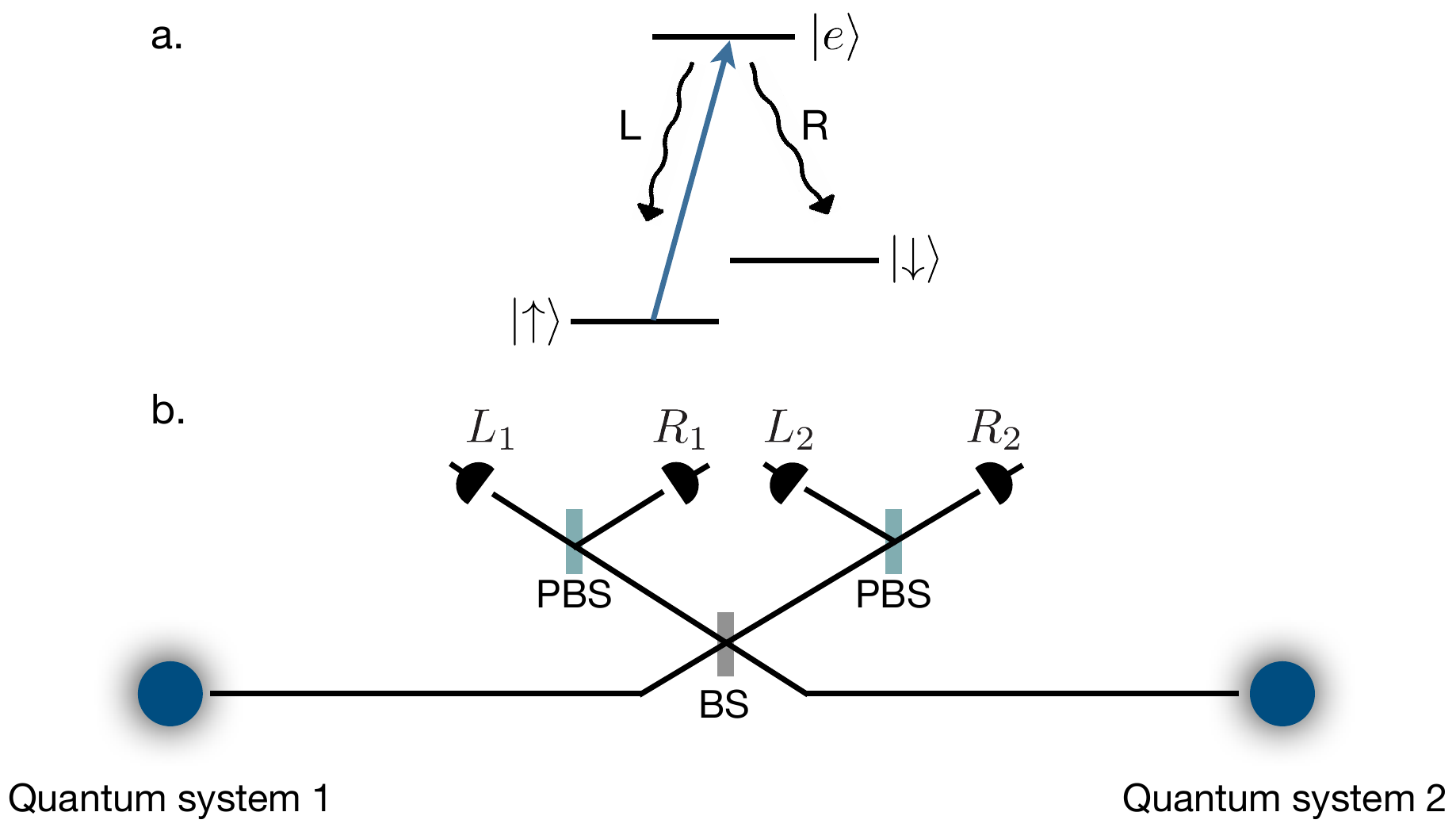}
\caption[Diagram illustrating remote entanglement generation between pulsed $\Lambda$-type systems.]{\small\textbf{Diagram illustrating remote entanglement generation between pulsed $\Lambda$-type systems.} (a) A $\Lambda$-type system as used in protocol $\mathsf{P}$; an excited state $\ket{\mathrm{e}}$ can decay to either ground state $\ket{\uparrow}$ or $\ket{\downarrow}$. (b) Entanglement generation via a polarization Bell-state measurement. Placing $\lambda/2$ and $\lambda/4$ waveplates before each polarizing beam splitter (PBS) can control the measurement basis.}\label{fig:polarization}
\end{figure}

We now look at the third protocol (denoted by $\mathsf{P}$), which is based on spin-spin entanglement generation via spin-polarization entanglement. For this scheme, we analyze a $\Lambda$-type system where a single excited state $\ket{\mathrm{e}}$ can decay to either $\ket{\uparrow}$ or $\ket{\downarrow}$, emitting photons of orthogonal polarization depending on the transition (see Fig.~\ref{fig:polarization}).

\emph{Protocol description.---} Initially, we prepare each of the quantum systems in one of the two ground states. Then, using a short $\pi$-pulse, each system is brought to the excited state. This excited state will then decay to one of the ground states while emitting a photon. 

To illustrate this more clearly, suppose the probability is equal to decay to either ground state. Then the state of the qubit and the emitted photon for each system is $\ket{\psi}_k=\frac{1}{\sqrt{2}}(\ket{\uparrow}\ket{\text{L}}+\ket{\downarrow}\ket{\text{R}})$,
where $\ket{\text{L}}$  and $\ket{\text{R}}$ denote the left and right circular polarization modes of the photon. The joint state of both systems can then be written in the Bell basis for the spin and photon as Eq.~(\ref{bellbasis}), where the polarization modes replace the time bin modes of protocol $\mathsf{T}$.

To perform a BSM, we require a beam splitter (BS) and a polarizing beam splitter (PBS) at each output port of the BS. Then, we place detectors L$_1$ and R$_1$ (L$_2$ and R$_2$) on the left (right) output port of the BS, as shown in Fig.~\ref{fig:polarization}. For $\ket{\psi^+}_{p}$ ($\ket{\psi^-}_{p}$) photon bunching (anti-bunching) happens on the BS. Therefore, considering a perfect interference of the fields, a coincidence in detectors (L$_1$, R$_1$) or (L$_2$, R$_2$) will project the photon state onto the entangled state $\ket{\psi^+}_\text{p}$ and a coincidence in detectors (L$_1$, R$_2$) or (L$_2$, R$_1$) results in the entangled state $\ket{\psi^-}_\text{p}$ \cite{mattle1996dense}. This projects the state of the qubits onto the corresponding spin Bell state. As in protocol $\mathsf{T}$, this setup is not able to distinguish $\ket{\phi^+}_\text{p}$ and $\ket{\phi^-}_\text{p}$ since photon bunching happens for both of these cases. However, with the addition of a source of local auxiliary polarization-entangled photon states, the Bell analyzer success rate could be increased to $75\%$ \cite{grice2011arbitrarily}\ref{wein2016bellmeasurement}.

\emph{Conditional states.---}We can describe the source field collected from each transition by $\hat{b}_{\uparrow_k}= \hat{\sigma}_{\uparrow_k}\sqrt{\eta_{c_k}\gamma_{\text{r}\uparrow_k}}$ and $\hat{b}_{\downarrow_k}=\hat{\sigma} _{\downarrow_k}\sqrt{\eta_{c_k}\gamma_{\text{r}\downarrow_k}}$ where $k$ denotes the quantum system 1 and 2 and $\text{r}$ indicates the radiative decay rate. Considering the transmission loss and the beam splitter, we can compute the L-polarized fields $(\hat{d}_1,~\hat{d}_2)$ at detectors L$_1$ and L$_2$ and the R-polarized fields $(\hat{d}_3,~\hat{d}_4)$ at detectors R$_1$ and R$_2$ using Eq.~(\ref{beamsplittereq}). The associated jump superoperators are then $\mathcal{J}_{i}\hat{\rho} = \eta_\text{d}\hat{d}_i\hat{\rho}\hat{d}_i^\dagger$.

Similar to protocol $\mathsf{T}$, the conditions for a successful protocol are $\mathbf{n} = (1,0,1,0)$, $(1,0,0,1)$, $(0,1,1,0)$, and  $(0,1,0,1)$ where the vectors notate the photon count at the detectors in the order (L$_1$, L$_2$, R$_1$, R$_2$). Like with the previous protocol, we simplify the notation by concatenating the vector elements. In contrast to protocol $\mathsf{T}$, the conditional states for protocol $\mathsf{P}$ are true two-photon events rather than sequential one-photon events. These two-photon conditioned states are computed from their corresponding two-photon conditioned propagators. For example, $\hat{\rho}_{1001}(t_\text{f}) = \mathcal{W}_{1001}(t_\text{f},t_\text{d}^\prime,t_\text{d},t_0)\hat{\rho}(t_0)$ where $\mathcal{W}_{1001}=\mathcal{U}(t_\text{f},t_\text{d}^\prime)\mathcal{U}_{1001}(t_\text{d}^\prime,t_\text{d})\mathcal{U}(t_\text{d},t_0)$ is computed using
\begin{equation}
\begin{aligned}
\label{twophotonpropagator}
    \mathcal{U}_{1001}(t_\text{d}^\prime,t_\text{d})\!&=\!\!\int_{t_\text{d}}^{t_\text{d}^\prime}\!\!\!\int_{t_\text{d}}^{t^{\prime\prime}}\!\!\!\!\mathcal{U}_\mathbf{0}(t_\text{d}^\prime,\!t^{\prime\prime})\mathcal{J}_1\mathcal{U}_\mathbf{0}(t^{\prime\prime}\!,\!t^{\prime})\mathcal{J}_4\mathcal{U}_\mathbf{0}(t^\prime\!\!,t_\text{d})dt^\prime\! dt^{\prime\prime}\\
    &+
    \!\!\int_{t_\text{d}}^{t_\text{d}^\prime}\!\!\!\int_{t_\text{d}}^{t^{\prime\prime}}\!\!\!\!\mathcal{U}_\mathbf{0}(t_\text{d}^\prime,\!t^{\prime\prime})\mathcal{J}_4\mathcal{U}_\mathbf{0}(t^{\prime\prime}\!,\!t^{\prime})\mathcal{J}_1\mathcal{U}_\mathbf{0}(t^\prime\!\!,t_\text{d})dt^\prime\!dt^{\prime\prime}.
\end{aligned}
\end{equation}
Note that by how we defined a photon counting measurement in this thesis, we are not tracking the arrival time within the detection window. Thus $\mathcal{U}_{1001}$ does not discriminate between the cases where L$_1$ clicks before R$_2$ and cases where R$_2$ clicks before L$_1$. This is illustrated in Eq.~(\ref{twophotonpropagator}) as a consequence of the summation in Eq.~(\ref{conditionalpropagator}). Such a restriction can be lifted if the detectors have sufficient time resolution capabilities.

\begin{figure}[t]
    \centering
    \includegraphics[width=0.55\textwidth]{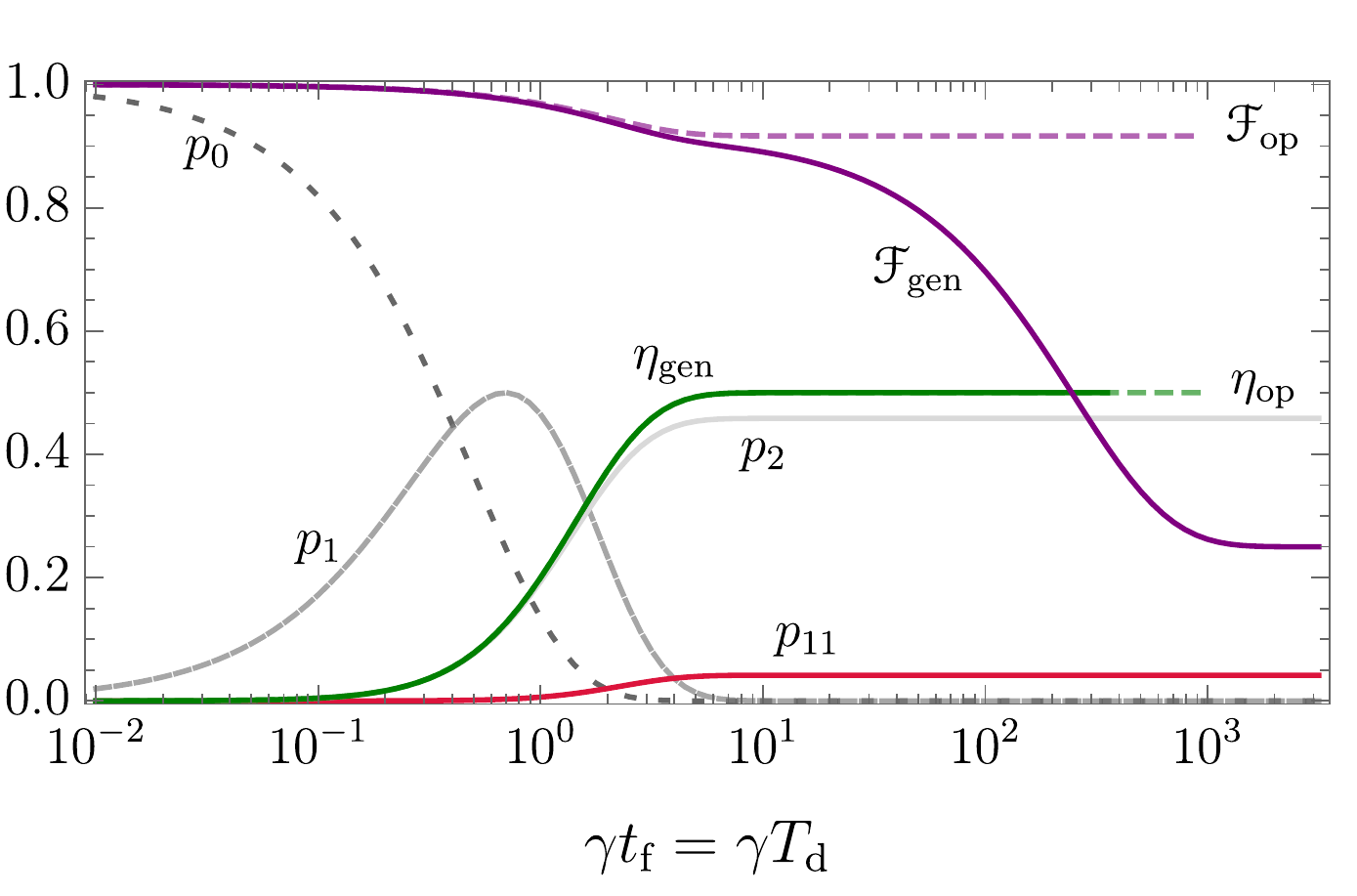}
    \caption[The time dynamics of remote entanglement generation via spin-polarization entanglement.]{\small\textbf{The time dynamics of remote entanglement generation via spin-polarization entanglement: protocol $\mathsf{P}$.} Entanglement generation fidelity $\euscr{F}_\mathrm{gen}$ and efficiency $\eta_\text{gen}$ as a function of protocol time $t_\text{f}$ for an initial state $\hat{\rho}(t_0)=\ket{\mathrm{ee}}\!\bra{\mathrm{ee}}$ when there is no loss and noiseless local photon-number resolving detectors. The asymptotic dashed lines indicate the limits of fidelity $\euscr{F}_\mathrm{op}$ and efficiency $\eta_\text{op}$ reached when spin decoherence is neglected and when the detection window encompasses the entire photon lifetime. The detection window is set to be equal to the entire protocol duration: $t_\text{d}=t_0=0$ and $t_\text{d}^\prime=t_\text{f}=T_\text{d}$. The gray lines show the probabilities for no photon emission from either system $p_0$, for single-photon events $p_1$ where only one photon is detected, and for photon bunching where two photons arrive at one detector $p_2$. The red line indicates the probability for coincident counts $p_{11}$ caused by imperfect HOM interference. Parameters chosen: $\gamma_{j_k}=\gamma/2$ for $j\in\{\uparrow,\downarrow\}$ and $k\in\{1,2\}$ so that $\gamma_k=\gamma_{\uparrow_k}+\gamma_{\downarrow_k}=\gamma$, $\gamma_k^\star=0.1\gamma$; and $\gamma_{\text{s}_k}^\pm=\gamma_{\mathrm{s}_k}^\star=0.001\gamma$.}
    \label{fig:P3timedynamics}
    \vspace{-1mm}
\end{figure}

\emph{Measurement duration.---} The time dynamics of protocol $\mathsf{P}$ shows features in common with both protocols $\mathsf{N}$ and $\mathsf{T}$. Like $\mathsf{T}$, it is a two-photon scheme and so the fidelity is high for small $T_\text{d}$ compared to the system optical lifetimes $1/\gamma_k$. However, like $\mathsf{N}$, $\mathsf{P}$ only requires a single excitation of each system. Thus the efficiency is unaffected by spin flip processes when $T_\text{d}$ is much larger than the lifetime (see Fig.~\ref{fig:P3timedynamics}).

Although it is possible to derive analytic expressions for protocol $\mathsf{P}$ for arbitrary measurement duration when neglecting spin decoherence, they do not provide new physical insight. For brevity, we only show analytic results in the optical limit to compare with protocols $\mathsf{N}$ and $\mathsf{T}$.

\emph{Optical limits.---}
Consider the case where spin decoherence is negligible, the measurement window encompasses the lifetime $T_\text{d}\gg 1/\gamma_{j_k}$, and the interference is balanced so that $\theta=\pi/4$ and $\eta_{j_k}=\eta$ for $j\in\{\uparrow,\downarrow\}$ and $k\in\{1,2\}$. Then the entanglement generation efficiency is given by $\eta_\text{op}=\eta^2/2$ and the fidelity becomes
\begin{equation}
\euscr{F}_\text{op}=\frac{1}{2}\left(1+\text{Re}\!\left(\frac{\tilde{C}_\uparrow^*\tilde{C}_\downarrow}{\tilde{M}_{\gamma^\star}}\right)\right),
\end{equation}
\vspace{-1mm}
where
\begin{equation}
\begin{aligned}
    \tilde{C}_j &= \frac{2\sqrt{\gamma_1\gamma_2}}{\Gamma_1+\Gamma_2+2i\Delta_j}\\
    \tilde{M}_{\gamma^\star}&=\frac{\gamma_1+\gamma_2-i(\Delta_\uparrow-\Delta_\downarrow)}{\Gamma_1+\Gamma_2-i(\Delta_\uparrow-\Delta_\downarrow)},
\end{aligned}
\end{equation}
and where $\gamma_k=\gamma_{\uparrow_k}+\gamma_{\downarrow_k}$ is the total decay rate of the $k^\text{th}$ system, $\Gamma_k=\gamma_k+2\gamma_k^\star$ is the total optical decoherence rate, $\Delta_\uparrow$ and $\Delta_\downarrow$ are the optical detunings between the left and right circularly polarized transitions (respectively) of the systems. Similar to protocols $\mathsf{N}$ and $\mathsf{T}$, the factor $\tilde{C}_j$ quantifies the coherence for the which-path erasure of photons from the $j$ transitions at the beam splitter, which depends only on the total decay rates relative to the detuning and dephasing. We attribute the factor $1/\tilde{M}_{\gamma^\star}$ to the gain in fidelity due to the systems being initialized in the excited state, compared to protocols $\mathsf{N}$ and $\mathsf{T}$ where the systems are initialized in a superposition state and are directly affected by emitter pure dephasing. 

The fidelity is bounded from above by the mean wavepacket overlaps $M_{12\uparrow}$ and $M_{12\downarrow}$ of photons from each transition: \mbox{$\euscr{F}_\text{op}\leq (1+\sqrt{M_{12\uparrow}M_{12\downarrow}})/2$}. Interestingly, this inequality can be saturated if $\Delta_\uparrow-\Delta_\downarrow$ is much smaller than $\gamma_1+\gamma_2$, implying that the systems have nearly identical spin splittings compared to the system decay rate $\gamma_k$. Then we have $\tilde{C}_\uparrow\simeq\tilde{C}_\downarrow$ and $\tilde{M}_{\gamma^\star}\simeq (\gamma_1+\gamma_2)/(\Gamma_1+\Gamma_2)$. In this case, the fidelity becomes $\euscr{F}_\text{op}=(1+M_{12})/2$ where $\Delta = \Delta_\uparrow=\Delta_\downarrow$.

We note that the fidelity of protocol $\mathsf{P}$ does not depend on the ratio of the decay rates to each ground state. Rather, it only depends on the total decay rate $\gamma_k$ of each system, which dictates the photon temporal profile. However, it is still necessary to balance the input intensity at the beam splitter, which may require artificially reducing $\eta_{j_k}$ if some transitions are brighter than others, consequently reducing $\eta$ and the overall efficiency.

\emph{Detector noise and number resolution.---} Because protocol $\mathsf{P}$ is a two-photon heralded scheme, it behaves almost identically to protocol $\mathsf{T}$ in terms of robustness against photon loss and detector noise. This means that the final measured states $\hat{\varrho}_\mathbf{n}$ can be determined using the form of Eq.~(\ref{BK-PNRD}). However, unlike protocol $\mathsf{T}$, protocol $\mathsf{P}$ is quite robust against non-number resolving detectors even when spin flips occur on the order of the emission timescale. This is because spin flips cannot directly affect the photon statistics of protocol $\mathsf{P}$ and so the chance to have more than 1 photon arriving at a given detector remains very small.

\subsection{Protocol comparison}
\label{chapter4:entanglementcomparison}

In this section, we compile the results for each protocol and compare their optical limits of fidelity with respect to each other and the mean wavepacket overlap. We also show the performance of each protocol when including photon loss, detector noise, and spin decoherence.

\begin{figure}
    \centering
    \includegraphics[width=1\textwidth,trim={6mm 4mm 14mm 4mm},clip]{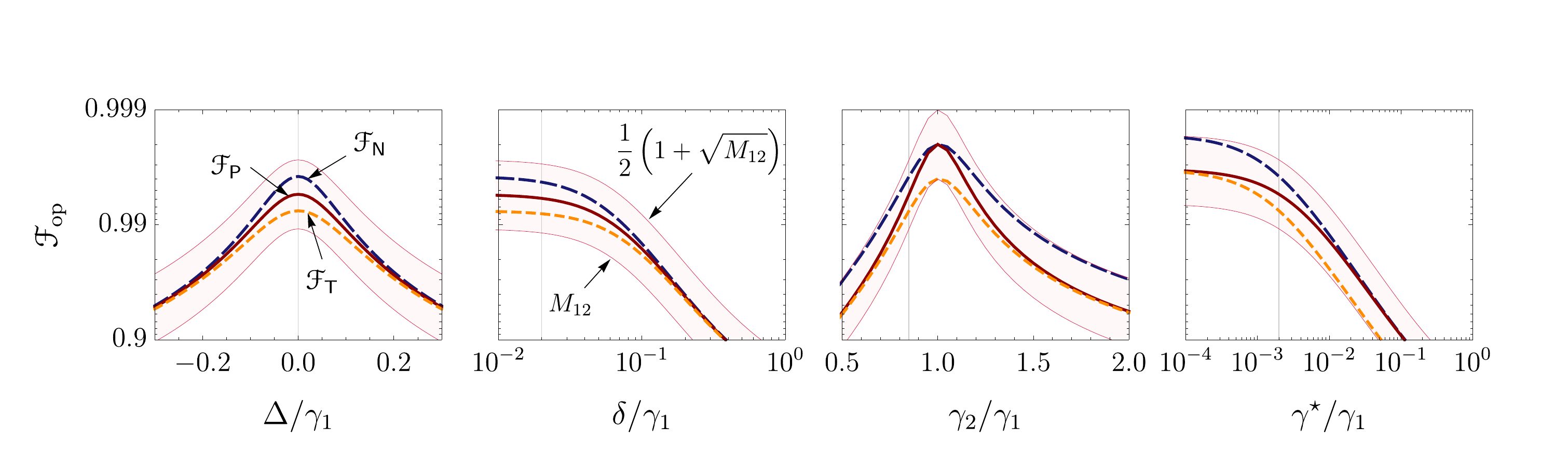}
    \caption[A comparison of the optical limits to fidelity for remote entanglement generation protocols.]{\small\textbf{A comparison of the optical limits to fidelity for remote entanglement generation protocols.} Fidelity in the optical limit $\euscr{F}_\mathrm{op}$ for each protocol as affected by emitter temporal profile mismatch $\gamma_2/\gamma_1$, pure dephasing $\gamma^\star$, spectral detuning $\Delta$, and spectral diffusion standard deviation $\delta$. These limits are attained when the detection window encompasses the entire photon lifetime and when spin decoherence and loss are negligible. The long dashed blue line represents the fidelity limit for protocol $\mathsf{N}$ when $F_\eta\rightarrow 1$, the short dashed orange line represents protocol $\mathsf{T}$, and the solid red line represents protocol $\mathsf{P}$. The thin red lines and shaded region represent values bounded by the mean wavepacket overlap $M_{12}$ of photons from each source: $M_{12}\leq \euscr{F}_\mathrm{op}\leq (1+\sqrt{M_{12}})/2$. The labeling and order of all lines are the same across all four panels. The thin vertical gray lines in each plot show the fixed values used for each of the other plots. Parameters used unless otherwise stated: $\gamma_2=0.85\gamma_1$, $\gamma^\star_1=\gamma^\star_2=0.002\gamma_1$, $\Delta=0$, and $\delta=0.02\gamma_1$. For protocol $\mathsf{P}$, we also assume that $\Delta_\uparrow=\Delta_\downarrow=\Delta$.}
    \label{fig:comparison-opticallimits}
    \vspace{-4mm}
\end{figure}

\emph{Optical limits}.---In section \ref{chapter4:numberentanglement}, the maximum fidelity achievable for the spin-photon number entanglement scheme (protocol $\mathsf{N}$) in the limit that $T_\text{d}\gg1/\gamma_k$, $F_\eta(\vartheta)\rightarrow 1$, and $\varphi+\phi=0$ was found to be $\euscr{F}_\text{$\mathsf{N}$} = (1+\text{Re}(\tilde{C}))/2$ corresponding to a concurrence $\euscr{C}_\text{$\mathsf{N}$}=|\tilde{C}|$. The fidelity in section \ref{chapter4:timeentanglement} for the scheme using time-bin entanglement (protocol $\mathsf{T}$) was found to be $\euscr{F}_\text{$\mathsf{T}$} = (1+|\tilde{C}|^2)/2$ corresponding to a concurrence $\euscr{C}_\text{$\mathsf{T}$}=\euscr{C}_\text{$\mathsf{N}$}^2$. In section \ref{chapter4:polarizationentanglement}, we found the fidelity for the spin-polarization entanglement generation scheme (protocol $\mathsf{P}$) to be $\euscr{F}_\text{$\mathsf{P}$} = (1+M_{12})/2$ when $\Delta_\uparrow\simeq\Delta_\downarrow$. The corresponding concurrence is $\euscr{C}_\text{$\mathsf{P}$}=M_{12}$, where $M_{12}$ is the mean wavepacket overlap of photons from each source. 

Knowing that $|\tilde{C}|^2\leq M_{12}$ (see section \ref{chapter4:timeentanglement}) but also $|\tilde{C}|\geq \text{Re}(\tilde{C})\geq M_{12}$ (see section \ref{chapter4:numberentanglement}), we find that $\euscr{C}_\text{$\mathsf{T}$}\leq \euscr{C}_\text{$\mathsf{P}$}\leq \euscr{C}_\text{$\mathsf{N}$}$. In addition, we have that the order is the same for the fidelity as well: $\euscr{F}_\text{$\mathsf{T}$}\leq \euscr{F}_\text{$\mathsf{P}$}\leq \euscr{F}_\text{$\mathsf{N}$}$. Furthermore, since $\euscr{F}_\text{$\mathsf{N}$}\leq (1+\sqrt{M_{12}})/2$ and $\euscr{F}_\text{$\mathsf{T}$}\geq M_{12}$, the optical limits of fidelity for all three protocols are bounded by
\begin{equation}
    M_{12}\leq \euscr{F}_\text{$\mathsf{T}$}\leq \euscr{F}_\text{$\mathsf{P}$}\leq \euscr{F}_\text{$\mathsf{N}$}\leq \frac{1}{2}\left(1+\sqrt{M_{12}}\right).
\end{equation}

From Fig.~\ref{fig:comparison-opticallimits}, we can see that protocols $\mathsf{N}$ and $\mathsf{T}$ have parallel behaviour in terms of dephasing and temporal overlap due to the fact that $\mathsf{T}$ can be seen as two applications of $\mathsf{N}$. However, protocol $\mathsf{P}$ is implemented with a single pulse on each system like $\mathsf{N}$ but it is still a two-photon scheme like $\mathsf{T}$. Hence, it matches the fidelity of $\mathsf{N}$ or $\mathsf{T}$ in different scenarios.

The dominance of protocol $\mathsf{N}$ in the ideal case is expected because the two-photon schemes can naively be seen as two single-photon schemes applied back-to-back, which would compound the error. Because of this, it is tempting to believe that protocol $\mathsf{N}$ would then also be less susceptible to spectral diffusion. However, protocols $\mathsf{T}$ and $\mathsf{P}$ have a symmetry advantage that protocol $\mathsf{N}$ does not have. In protocol $\mathsf{T}$, the fact that the second photon must come from the opposite side of the beam splitter causes an opposing phase rotation on the entangled spin state. These two phases cancel, leaving only a reduction in the magnitude of the coherence due to nonzero $\Delta$ rather than both a reduction and a phase rotation as seen in protocol $\mathsf{N}$. This is illustrated in Eq.~(\ref{BKfidelity}) where the fidelity depends on $|\tilde{C}|^2$ rather than $\text{Re}(\tilde{C})^2$. A similar symmetry occurs for protocol $\mathsf{P}$, however, the detuning phase is only fully eliminated if $\Delta_\uparrow=\Delta_\downarrow$. Because of these symmetry advantages, a sufficient amount of spectral detuning or spectral diffusion eliminates the fidelity advantage that the single-photon scheme had over the two-photon protocols (see Fig.~\ref{fig:comparison-opticallimits}).

\emph{Phase errors.}---Let us now discuss the impact of the relative initialization phase $\varphi=\varphi_1-\varphi_2$ and propagation phase $\phi=\phi_1-\phi_2$ errors. The initialization phase $\varphi_k$ for each quantum system can independently fluctuate over time causing significant phase errors if the two quantum systems do not share a phase reference. In addition, it may be necessary to stabilize or correct the propagation phase $\phi$ by monitoring the phase fluctuations of the communication channel \cite{minavr2008phase, yu2020entanglement}. Since the propagation phase depends on distance, this propagation phase fluctuation can become severe for large entanglement generation distances.

As discussed in the previous section, protocols $\mathsf{T}$ and $\mathsf{P}$ have a symmetry advantage over protocol $\mathsf{N}$ for the spectral detuning phase. This advantage extends to propagation phase errors and other possible local phase errors such as initialization phase and the relative precession of the two spin qubits. On the other hand, the upper bound on fidelity for protocol $\mathsf{N}$ can be severely degraded by any phase error $\phi$ becoming $(1/2)(1+\text{Re}(\tilde{C}e^{i\phi}))$. If this phase fluctuates in a Gaussian distribution \cite{minavr2008phase} centered around $\phi=0$ with a standard deviation of $\sigma_\phi$, then the fidelity reduces to $\euscr{F}_\textsf{N}=(1/2)(1+\text{Re}(\tilde{C})e^{-\sigma_\phi^2/2})$ where, as in the previous section, we have assumed $F_\eta\rightarrow 1$.

\begin{figure}[t]
    \centering
    \hspace{-5mm}\includegraphics[width=0.5\textwidth]{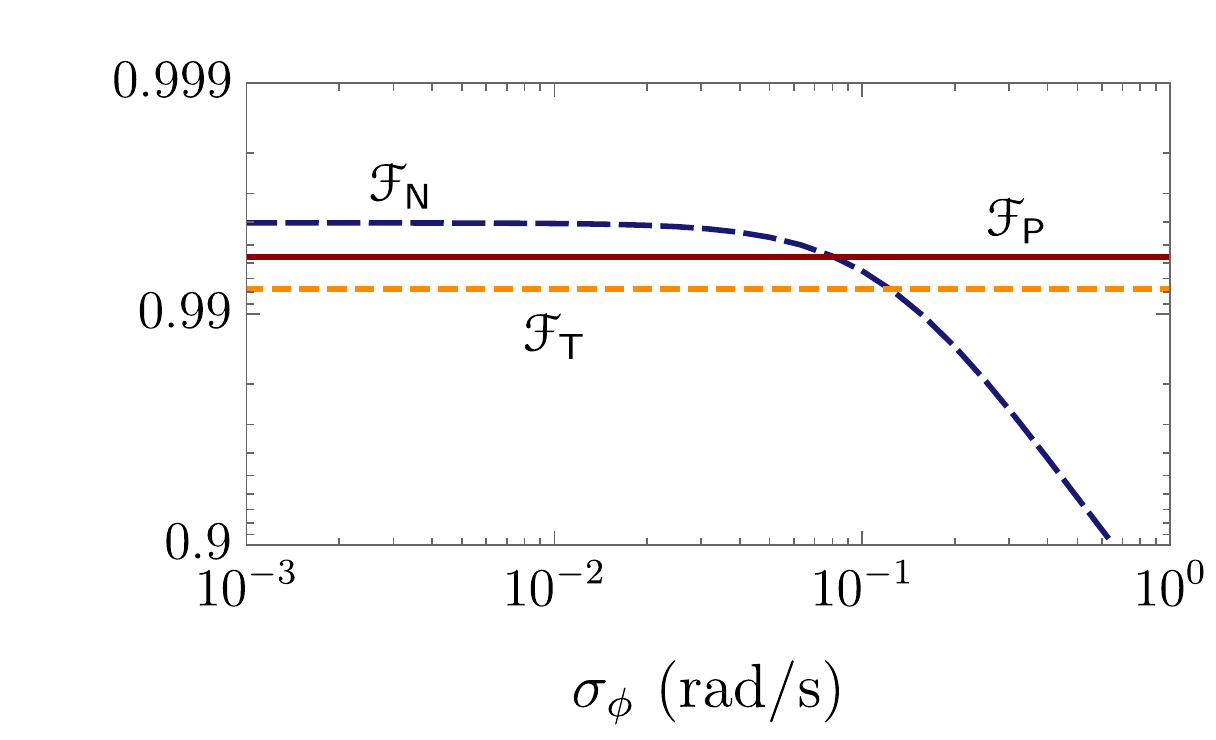}
    \caption[A comparison of phase errors for remote entanglement generation protocols.]{\small\textbf{A comparison of phase errors for remote entanglement generation protocols.} The reduction in entanglement generation fidelity of protocol $\mathsf{N}$ as compared to phase-robust protocols $\mathsf{T}$ and $\mathsf{P}$ for a phase error $\phi$ fluctuating in a Gaussian distribution around $\phi=0$ with a standard deviation of $\sigma_\phi$. For this comparison, we are neglecting spin decoherence, spectral diffusion, detector dark counts, and detector number resolving limitations. Other parameters used: $\gamma_2=0.85\gamma_1$, $\gamma_1^\star=\gamma_2^\star=0.002\gamma_1$, and $\Delta=0.02\gamma_1$. We also assume that $F_\eta\rightarrow 1$ for protocol $\mathsf{N}$ and that $\Delta_\uparrow=\Delta_\downarrow$ for protocol $\mathsf{P}$.}
    \label{fig:comparisonphaseerror}
\end{figure}

For a large enough phase fluctuation $\sigma_\phi$, protocol $\mathsf{N}$ loses its fidelity advantage over the other two protocols (see Fig.~\ref{fig:comparisonphaseerror}). We find that the value for the variance $\sigma_\phi^2$ where $\euscr{F}_\mathsf{N}\leq \euscr{F}_\mathsf{T}$ is $\sigma_\phi^2\geq \ln((\Gamma_1+\Gamma_2)^2/(4\gamma_1\gamma_2))$. Likewise, for $\euscr{F}_\mathsf{N}\leq \euscr{F}_\mathsf{P}$ we would need $\sigma_\phi^2\geq \ln((\gamma_1+\gamma_2)^2/(4\gamma_1\gamma_2))$.

Although protocols $\mathsf{T}$ and $\mathsf{P}$ are very robust against phase errors, they can still be affected under some conditions. If the phase fluctuation occurs on a timescale faster than the separation between pulses for protocol $\mathsf{T}$, then $\euscr{F}_\mathsf{T}$ can be degraded. This could be accounted for in our method by adding different phases for the second detection window when computing the conditional propagators. In addition, significant birefringence in protocol $\mathsf{P}$, quantified by $\omega_{\text{s}_k}$, can cause a small degradation of $\euscr{F}_\mathsf{P}$ due to propagation phase errors. However, since $\omega_{\text{s}_k}\ll\omega_k$, this effect is orders of magnitude smaller than the degradation experienced by protocol $\mathsf{N}$.

\emph{Loss and distance.}---Let us now compare the fidelity and efficiency of all three protocols while taking into account all imperfections aside from spectral diffusion and phase errors, which were discussed above.

\begin{figure}[p]
    \centering
    \hspace{-55mm}(a)\hspace{75mm}(b)\\
    $\hspace{10.mm}
    \overbrace{\hphantom{h\hspace{30.8mm}h}}^\text{
    \normalsize QComm.}\overbrace{\hphantom{h\hspace{22mm}h}}^\text{\normalsize DQC}\hspace{18.1mm}
    \overbrace{\hphantom{h\hspace{26mm}h}}^{\mycom{
    \text{\footnotesize Optically}}{
    \text{\footnotesize Limited}}
    }\overbrace{\hphantom{h\hspace{15.8mm}h}}^{\mycom{
    \text{\footnotesize Spin}}{
    \text{\footnotesize Limited}}
    }\hspace{-0.3mm}\overbrace{\hphantom{h\hspace{7.mm}h}}^{\mycom{
    \text{\footnotesize Noise}}{
    \text{\footnotesize Limited}}
    }$\\\vspace{-4mm}
    \hspace{0.3mm}\includegraphics[width=0.4592\textwidth]{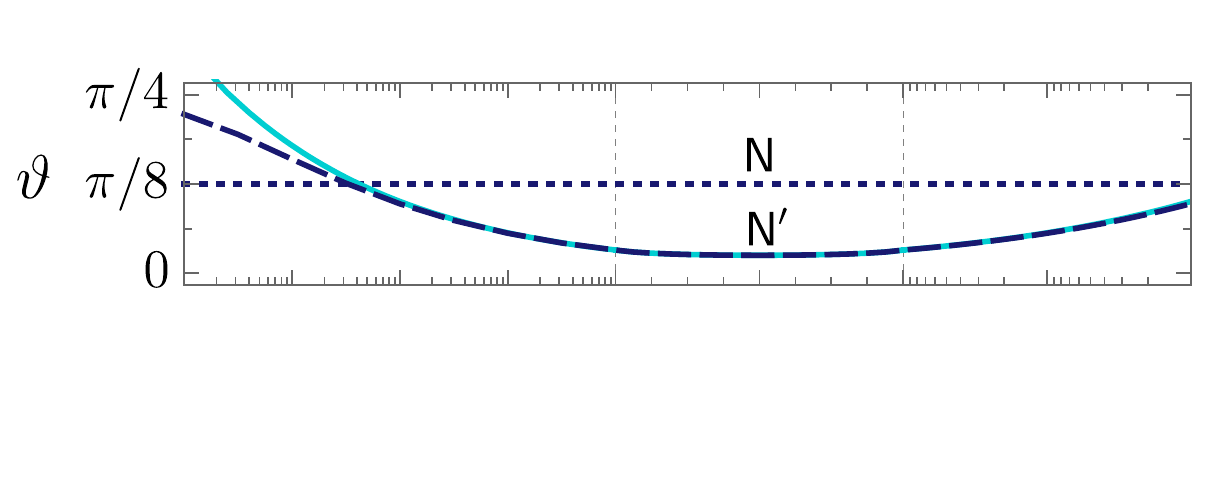}\hspace{5.2mm}
    \includegraphics[width=0.4592\textwidth]{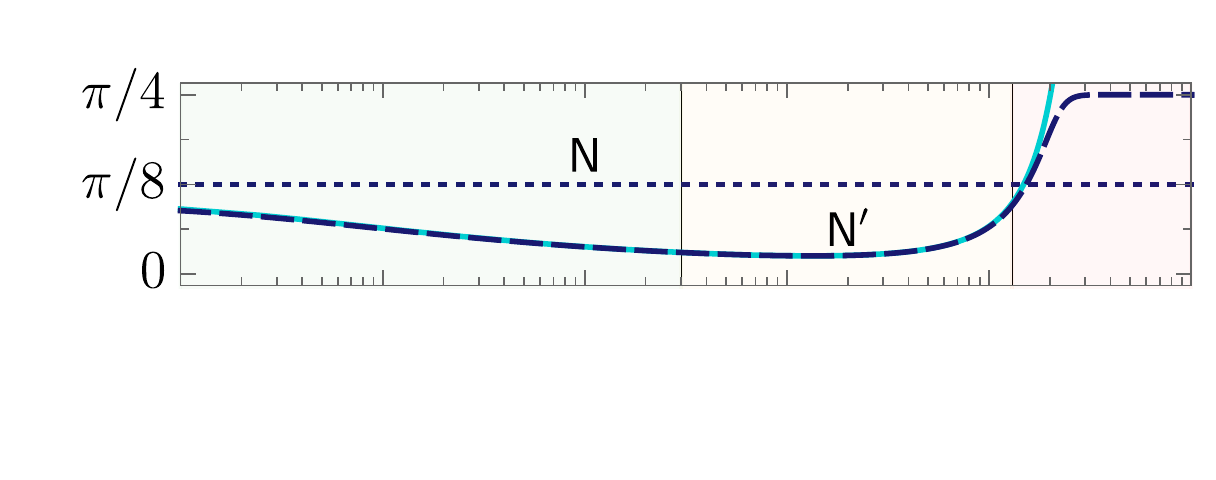}\\\vspace{-13mm}
    \hspace{-3.3mm}\includegraphics[width=0.479472\textwidth]{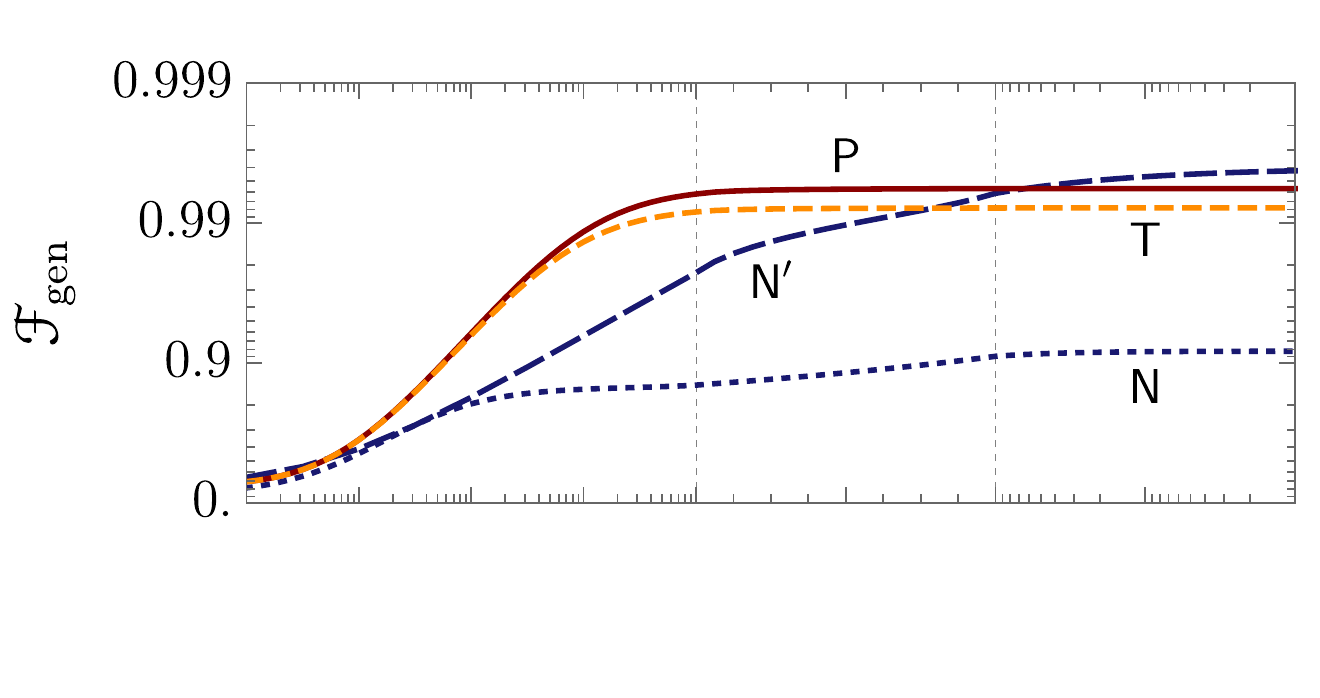}\hspace{1.7mm}
    \includegraphics[width=0.47936\textwidth]{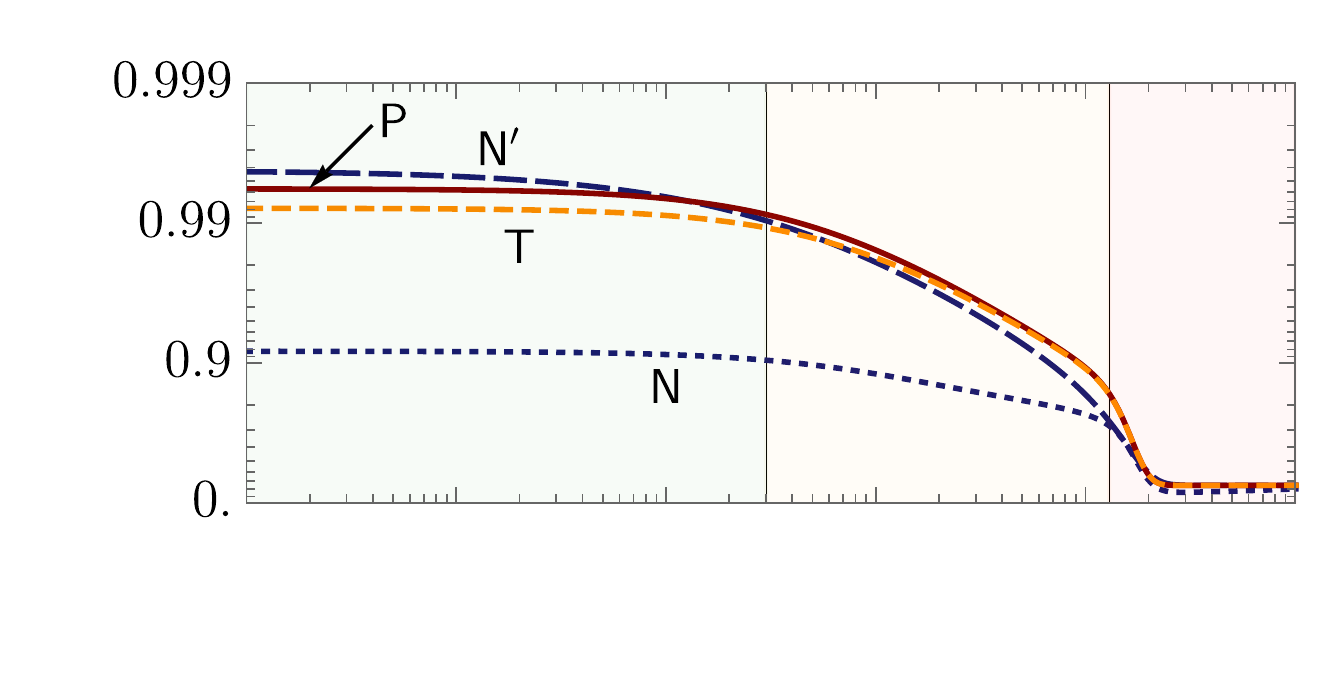}\\\vspace{-13mm}
    \hspace{-1.7mm}\includegraphics[width=0.49392\textwidth]{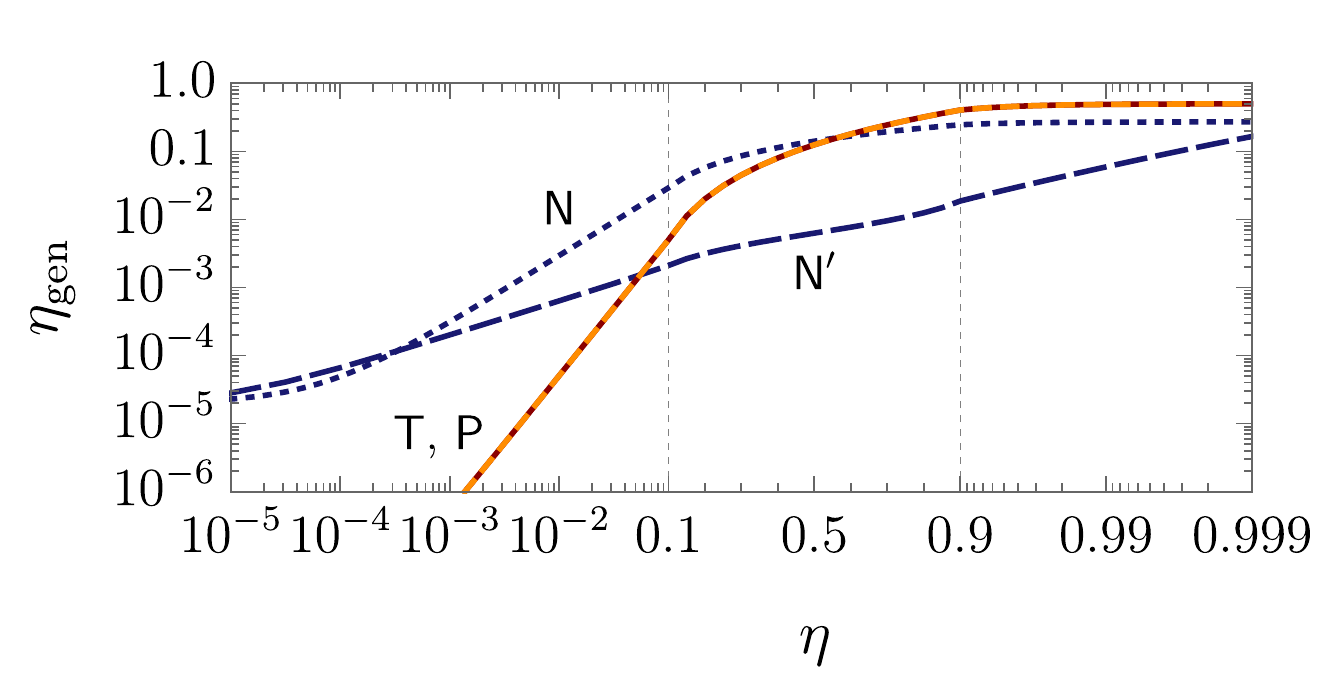}\hspace{-0.1mm}
    \includegraphics[width=0.4816\textwidth]{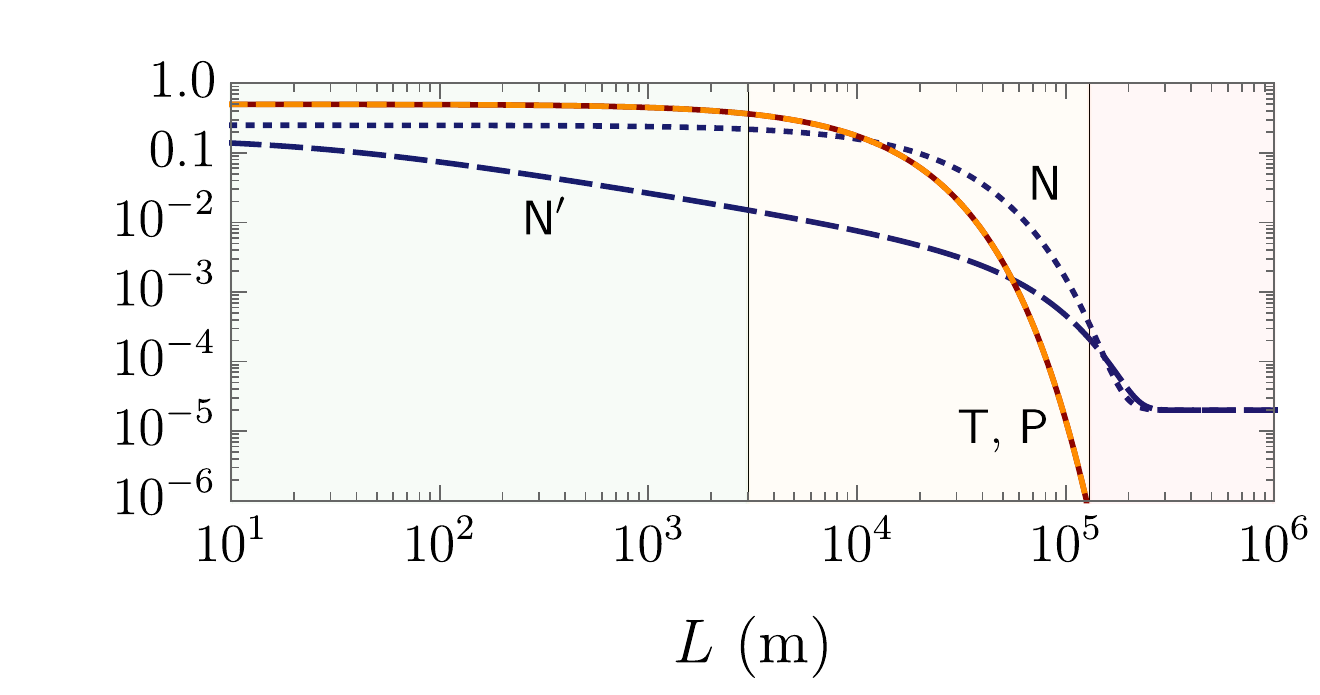}
    \caption[A comparison of loss and distance for remote entanglement generation protocols.]{\textbf{A comparison of loss and distance for remote entanglement generation protocols.} (a) A split-scale plot of the fidelity $\euscr{F}_\mathrm{gen}$ and efficiency $\eta_\text{gen}$ as a function of total single-photon efficiency $\eta$ for entanglement generation via spin-photon number entanglement ($\mathsf{N}$), spin-time bin entanglement ($\mathsf{T}$), and spin-polarization entanglement ($\mathsf{P}$) using local detectors with a finite detector dark count probability of $1-p_{\mathrm{d},0}=10^{-5}\simeq p_{\mathrm{d},1}$ per detection window of duration $T_\text{d}$. (b) The effect of distance $L$ between the quantum systems on the fidelity and efficiency for each protocol in panel (a) when taking into account spin decoherence in addition to photon loss and detector noise. An increased distance affects both the total protocol duration $t_\text{f}=N_\text{w} T_\text{d}+L/c$ and the single-photon efficiency $\eta=\eta_0 10^{-L/2L_\text{att}}$, where $c=2\times 10^{8}$~m/s is the speed of transmission in the fibre, $N_\text{w}=1$ for $\mathsf{N}$ and $\mathsf{P}$, $N_\text{w}=2$ for $\mathsf{T}$, and $L_\text{att}=22$~km is the fibre attenuation length. (a, b) The dotted blue line represents the protocol $\mathsf{N}$ using non-number resolving detectors (BD) and with a fixed $\vartheta=\pi/8$, corresponding to a probability of 1/4 for a source to emit a photon. The long-dashed blue line indicated by $\mathsf{N}^\prime$ illustrates the noise-limited maximum possible fidelity for $\mathsf{N}$ using photon-number resolving detectors (PNRD). The top panel shows the corresponding numerically optimized $\vartheta$ and the solid light blue line indicates the analytic approximation for the PNRD model: $\vartheta^4=(1-p_{\mathrm{d},0})/(\eta(1-\eta))$. The short-dashed orange line and solid red line corresponding to $\mathsf{T}$ and $\mathsf{P}$, respectively, are visually unaffected when accounting for non-number resolving detectors. Other parameters used: $T_\text{d}=5/\gamma_1$, $\gamma_2=0.85\gamma_1$, $\gamma^\star_k = 0.002\gamma_1$, $\Delta=0.02\gamma_1$, $\delta_k=0$, $\gamma_{\text{s}_k}^\pm=0.5\times 10^{-6}\gamma_1$, and $\gamma_{\mathrm{s}_k}^\star=10^{-6}\gamma_1$ for  $k\in\{1,2\}$. For $\mathsf{P}$, we also assume that $\Delta_\uparrow=\Delta_\downarrow=\Delta$. For panel (b) we set $\eta_0=0.999$ for continuity with panel (a) and choose $\gamma_1=10^{8}$~Hz.}
    \label{fig:comparisonLossDistance}
\end{figure}

When including losses and detector noise, the two-photon protocols distinguish themselves significantly from the single-photon protocol. Although less flexible, $\mathsf{T}$ and $\mathsf{P}$ are more robust in terms of fidelity than $\mathsf{N}$ (see Fig.~\ref{fig:comparisonLossDistance}~(a)). However, using PNRDs, $\mathsf{N}$ can exceed $\mathsf{T}$ and $\mathsf{P}$ in terms of fidelity in the regime of distributed quantum computing (DQC) where errors may be dominated by optical imperfections. It can also exceed $\mathsf{T}$ and $\mathsf{P}$ in terms of efficiency in the loss regime of quantum communication (QComm). This latter advantage can come at a significant cost to fidelity if there is significant detector noise, even after optimizing $\vartheta$ to minimize the errors caused by both systems emitting a photon.

To simulate each protocol's performance over distance, it is necessary take the classical communication time into account as shown in Eq.~(\ref{equ:dmat_dis}), such that the measurement takes place at the retarded time $r(t) = t - L/2c$, where $L=2L_\text{d}$ is the total distance between the systems. The final protocol time also cannot be less than  $t_\text{f} = N_\text{w} T_\text{d} + L/c$, where $N_\text{w}$ is the number of detection windows; $N_\text{w}=1$ for $\mathsf{N}$ and $\mathsf{P}$, $N_\text{w}=2$ for $\mathsf{T}$. This delay caused by the classical communication time can cause a degradation of the entanglement generation fidelity due to spin decoherence.

To compare the protocols, we have selected a set of parameters that best illustrate their differences while also remaining relevant to realistic systems. We have chosen an optical lifetime of 10 ns, with a spin $T_1^\pm$ time of 20 ms and a spin $T_2^\star$ of 10 ms typical of a nitrogen-vacancy center in diamond \cite{fu2009observation,bernien2013heralded}. However, we have chosen an optimistic pure dephasing rate of $0.2$ MHz corresponding to nearly Fourier-transform limited lines, which for many systems would likely require some cavity enhancement or spectral filtering to achieve. In Fig.~\ref{fig:comparisonLossDistance}~(b), we set $\eta_0=0.999$ for $L=0$ to illustrate the distance-limited values. In practice, $\eta_0$ is much lower due to other inefficiencies such as collection losses. This may include filtering losses as a consequence of suppressing the excitation laser or phonon sideband emission.

Some differences in fidelity between the protocols seen in Fig.~\ref{fig:comparisonLossDistance}~(a) are washed out by spin decoherence when the distance approaches or exceeds the fibre attenuation length $L_\text{att}=22$~km. However, the differences in efficiency scaling remain apparent, with $\mathsf{N}$ having the potential to exceed the efficiency of $\mathsf{T}$ and $\mathsf{P}$ by a couple orders of magnitude for long-distance entanglement generation, although with a modest fidelity for our chosen parameter set.

\emph{Conclusions.}---In this section, we have demonstrated a powerful and intuitive approach based on conditional propagator superoperators to analytically and numerically compute figures of merit for single-photon heralded entanglement generation protocols subject to dephasing. Our method relies on concepts from quantum trajectories and is apt given its resurgence in related techniques for analyzing emitted field states \cite{fischer2018particle,fischer2018scattering,hanschke2018quantum}. Our approach includes a multitude of realistic imperfections that must be considered when developing a platform for quantum information processing based on solid-state emitters.

We have provided simple relations to estimate the fidelity and efficiency for three popular entanglement generation protocols. These results are directly useful for developing future proposals for system-specific applications and may also help guide the experimental development of solid-state emitters for quantum information processing. Furthermore, we have used our results to compare these three protocols in order to reveal their strengths and weaknesses in detail.

Although the analysis in this section focused on a simplified three-level model for the quantum systems, our approach can be applied to more complicated systems, such as those in the critical cavity coupling regime \ref{wein2018plasmonics}, spin-optomechanical hybrid systems \ref{ghobadi2019cryogenfree}, or perhaps emitters in unconventional hybrid cavities \cite{gurlek2018manipulation,franke2019quantization}. It may also prove to be a powerful tool to analyze other photon counting applications when exposed to decoherence process such as novel single-photon interference phenomena \cite{loredo2019generation} or deterministic entanglement generation using feedback \cite{martin2019single}. Moreover, by extending the decomposition and measurements to include detector temporal resolution, the methods presented in this paper may provide a foundation to analyze the effects of decoherence on photon time-tagging heralded measurements.

\section{Deterministic entanglement generation}
\label{chapter4:cavitygates}

In addition to remote entanglement generation, quantum networks and related information processing applications also require local gates between stationary qubits. For quantum repeaters, this is necessary to perform entanglement swapping between two quantum memories within a repeater node \cite{sangouard2011quantum}. The entanglement swapping step can, in principle, be performed by applying a local version of one of the remote entanglement protocols described in the previous section. However, the efficiency of that approach would be limited to at most 50\% due to the reliance on a passive linear-optical Bell state measurement. Active feedback \cite{martin2019single} or auxiliary photons \ref{wein2016bellmeasurement} could be used to increase this efficiency above 50\%. Alternatively, if the two stationary qubits can interact directly and in a quantum coherent way, it is possible to perform deterministic two-qubit gates that do not rely on post-selection or active feedback. This can significantly increase the entanglement distribution rates for quantum repeater protocols \ref{asadi2018repeaters}.

Different types of interactions have been proposed to mediate local two-qubit gates between solid-state defects. For example, one could utilize phonon-mediated couplings \cite{lemonde2018phonon}
or electric \cite{ohlsson2002quantum}\ref{asadi2018repeaters} or magnetic \cite{you2000quantum} dipole-dipole interactions. However, in many cases, the single defects may already require an optical cavity to enhance the emission rate in order to improve the entanglement generation efficiency and fidelity. Hence, it is natural to explore the option of using the same optical cavity to mediate the local interaction needed for entanglement swapping. 

The idea of using photon-mediated local interactions to build two-qubit gates has been proposed in many forms \cite{duan2005robust,feng2002quantum,majer2007coupling,sillanpaa2007coherent}. One major advantage of photon mediated interactions over direct electric or magnetic dipole-dipole interactions is that the two defects do not need to be in close proximity. This makes fabrication and addressability less of a challenge. Optical photon-mediated interactions also have an advantage over phonon-related approaches because the higher-frequency photon modes are less susceptible to environmental noise than lower-frequency phonon modes, especially at non-cryogenic temperatures.

Defects coupled to a cavity operating in the strong cavity coupling regime can be used to engineer two-qubit gates, and strong cavity-coupling regime has been observed for some optical systems such as quantum dots \cite{SC-QD1,SC-QD2,SC-QD3}. However, it is a generally difficult task to achieve strong coupling with single optically-active solid-state defects. It is particularly difficult using defects that have weaker dipole magnitudes than quantum dots, such as rare earth ions and color centers in diamond. In addition, the strong-coupling regime is not the most ideal regime for implementing remote entanglement generation due to the Purcell saturation and vacuum Rabi splitting in the emission spectrum. Hence, in this section, I take a look at using a cavity system operating in the bad-cavity regime to mediate local interactions between two defects within the same cavity.

The content in this section is related to Ref.~\ref{asadi2020cavitygates} where we analyzed three different approaches to engineering a two-qubit gate between defects coupled to a cavity operating in the bad-cavity regime. The first approach that we studied is a probablistic gate performed by scattering a single photon off of the cavity, as proposed in Refs.~\cite{duan2005robust,welte2018photon}. The second approach is an effective adiabatic dipole-dipole interaction induced by a virtual photon exchange between the defects, which has already been explored for microwave QED systems \cite{blais2004cavity,majer2007coupling}. The required interaction has also been observed for cavity-coupled defects in diamond \cite{evans2018photon}. The third approach that we explored was a modification of the second approach where weak continuous driving of the defect optical transitions induces a direct spin-spin interaction using a Raman transition. This latter Raman-assisted approach is based on the proposal of Ref.~\cite{feng2002quantum}, although we modified the implementation to avoid a flaw in the original proposal.

In this section \ref{chapter4:cavitygates}, I will primarily discuss the second approach analyzed in Ref.~\ref{asadi2020cavitygates} based on a simple virtual photon exchange interaction. However, I will include a more detailed discussion on the effects of pure dephasing in section \ref{chapter4:simplevirtual} that was not included in Ref.~\ref{asadi2020cavitygates}. This discussion about pure dephasing is also related to some results briefly presented in Ref.~\ref{asadi2020repeaters}, which used the results of Ref.~\ref{asadi2020cavitygates} in a repeater analysis. I will comment on this overlap when it arises. In section \ref{chapter4:subradiantvirtual}, I will also extend the analysis of the virtual photon exchange to include a second cavity mode that may interfere with the first, as described by the quasi-normal mode (QNM) model introduced in section \ref{chapter1:qnmmastereq}. To my knowledge, this analysis provides original results that are not present in the literature. 

\subsection{System}
 
The most general case of system model studied in this section is that of two three-level defects described in section \ref{chapter4:remotesystems} coupled to a cavity described by a summation of interfering QNMs as described in section \ref{chapter1:qnmmastereq}. However, I will only use up to two QNMs in my analysis and I will assume that the defect-QNM interactions are independent, which may be violated under certain conditions; for example, if the two dipoles experience a non-negligible electric dipole-dipole interaction. Each defect system is composed of an optically active transition that is coupled to the cavity and a ground state spin qubit. In addition, I assume that the states of each defect can be individually prepared and manipulated using microwave or optical fields.

To capture the defect-specific imperfections, we use the same Markovian master equation for two three-level systems discussed in section \ref{chapter4:remotesystems}, with an additional cavity coupling term $\mathcal{L}=\mathcal{L}_1\otimes\mathcal{I}+\mathcal{I}\otimes\mathcal{L}_2+\mathcal{L}_\mathrm{c}$. The emitter superoperators $\mathcal{L}_k$ are given by Eq.~(\ref{chapter4eq:3lvldefect}) and the cavity-mediated coupling superoperator is given in section \ref{chapter1:qnmmastereq} as
 \begin{equation}
    \mathcal{L}_\text{c} = -\frac{i}{\hbar}\mathcal{H}_c +\sum_{ij}2\chi_{ij}^-\mathcal{D}(\ad_i,\ad_j),
\end{equation}
where
\begin{equation}
    \hat{H}_c = \sum_{ij}\hbar\chi^{+}_{ij}\au_i\ad_j+\sum_{ik}\hbar g_{ik}\sigd_{\uparrow_k}\au_i+\hbar g_{ik}^*\sigu_{\uparrow_k}\ad_i.
\end{equation}
This general system model takes into account the following processes: (1) spin pure dephasing, (2) incoherent spin decay and excitation, (3) emitter pure dephasing, (4) incoherent optical decay and excitation, (5) independent emitter-QNM dipole coupling, (6) QNM interference, (7) QNM dissipation, and (8) independent semi-classical driving of defect optical and spin transitions. However, following the trend in this thesis, I assume that all state preparation and retrieval pulses are perfect in order to obtain upper-bound estimates on the quality of the protocol. In practice, the finite speed and non-unity fidelity of pulsed manipulation of the system will degrade the final protocol fidelity.

To be transparent, this generalized model is composed under some strong assumptions in addition to those made to reach a Markovian master equation. Most importantly, as I mentioned before, the emitter-QNM dipole coupling is assumed to be independent for each defect, which is likely violated under a proper derivation of the light-matter coupling for more than one dipole from first principles. In addition, the semi-classical driving of the defects may incidentally excite the surrounding cavity environment, causing complicated dynamics not captured by this model. Therefore, the analysis presented in this section should be seen as an idealized scenario and the results for fidelity must be taken as an upper bound on what may be possible with a realistic implementation of this model. I will briefly discuss possible implementations in section \ref{chapter5:outlook}.

\subsection{Simple virtual photon exchange}
\label{chapter4:simplevirtual}

The basic principle of the simple virtual photon exchange interaction is to use the cavity to mediate an adiabatic interaction between the defect excited states. This type of interaction may best be described as a long-range dipole-dipole interaction that is enhanced by the confinement of the electromagnetic field, which increases the local density of states between the two dipoles. In the bad-cavity regime, however, the Purcell effect rapidly dissipates any photons that would otherwise mediate the interaction between the two dipoles. This issue can be solved by recognizing that the Purcell effect falls off as $1/\Delta^2$ (see section \ref{chapter1:cavityQED}) while the mediated interaction only falls off as $1/\Delta$ \ref{asadi2020cavitygates}, where $\Delta$ is the detuning between the cavity resonance and the two dipole resonances. Therefore, by detuning the cavity away from the resonant dipoles, one can suppress the Purcell effect to the point where the long-range interaction between the dipoles dominates the dynamics. That is, the rate of population exchange $R$ between the defects and the cavity can be made much smaller than the dipole-dipole interaction rate $|\lambda|$. This implies that, if the cavity is initially in the vacuum state, it will remain approximately in the vacuum state during the entire protocol. Since this long-range interaction causes an efficient coherent exchange of energy between the two defects, yet no photon is ever produced in the cavity, it is said that the dipoles exchange a virtual, interaction mediating, photon.

To put this into a more mathematical context, consider the simplified case where there is no dissipation or decoherence. Then, suppose we prepare our system by exciting one defect so that the initial state includes some non-zero amplitude of the state $\ket{\mathrm{e}\!\uparrow\! 0}_{12\mathrm{c}}$. If we assume that the transitions \mbox{$\ket{\uparrow}_k\longleftrightarrow\ket{\mathrm{e}}_k$} are each coupled to the cavity by the rate $g$, then the state $\ket{\uparrow\uparrow\!1}$ will couple to both our initial state and the state $\ket{\uparrow \!\mathrm{e}~0}_{12\mathrm{c}}$, which is resonant with our initial state. The Hamiltonian for such a simplified case is given by the two-emitter independent Jaynes-Cummings interaction $\hat{H}=\hbar\Delta\au\ad+\sum_k\hbar g(\sigd_{\uparrow_k}\au+\sigu_{\uparrow_k}\ad)$ in the frame rotating with the resonant dipoles. If $\ket{\uparrow\uparrow\!1}$ is far-detuned so that $\Delta\gg g$ we can apply adiabatic elimination (see section \ref{chapter1:adiabaticelimination}) to eliminate the mediating cavity state $\ket{1}_\mathrm{c}$ from the dynamics. This results in the effective interaction Hamiltonian $\hat{H}_\text{eff}=\hbar (g^2/\Delta)(\sigd_{\uparrow_1}\sigu_{\uparrow_2}+\sigu_{\uparrow_1}\sigd_{\uparrow_2})$ in the subspace $\{\ket{\mathrm{e}\!\uparrow\! 0},\ket{\uparrow \!\mathrm{e}~0}\}$. By letting the initial state evolve for time $T_\text{gate}=\pi\Delta/g^2$, the system will return back to the initial state but with an additional relative $\pi$ phase compared to the other product states of the two defects, which have not experienced the exchange interaction. Hence, the virtual photon exchange allows one to implement a deterministic entangling controlled phase gate between the two defects. By de-exciting the initially-excited defect after time $T_\text{gate}$, the interaction is halted and the state is retrieved from the interacting subspace, transferring the entanglement to the spin qubits.

Let us now discuss how decoherence can affect this two-qubit gate. As mentioned before, the interaction is successful only if the effective dipole-dipole coupling rate $g^2/\Delta$ is larger than $R$. However, we must also ensure that $g^2/\Delta$ is larger than the bare decay rates of the individual defects $\gamma$. This is because $\Delta$ cannot be increased too much or the defect excited states will decay before the gate is complete. Furthermore, the phase gate relies on the maintenance of coherence between the two excited states $\ket{\mathrm{e}\!\uparrow\! 0}$ and $\ket{\uparrow \!\mathrm{e}~0}$ and also between these excited states and the ground state spin qubits. Hence, any emitter dephasing $\gamma^\star$ will degrade the entanglement fidelity, in addition to causing a damping effect on the effective interaction rate $g^2/\Delta$. Finally, any decoherence processes affecting the spin qubits, represented by rates $\gamma^\star_\mathrm{s}$ and $\gamma^\pm_\mathrm{s}$, must occur on a timescale slower than $T_\text{gate}$. To summarize, the condition for a successful gate is $g^2/\Delta\gg R,\gamma,\gamma^\star,\gamma^\pm_\mathrm{s},\gamma_\mathrm{s}^\star$. In the remainder of the analysis, I will assume that spin decoherence, represented by parameters $\gamma^\pm_\mathrm{s}$ and $\gamma_\mathrm{s}^\star$, are negligible on the timescale of the gate.

Before discussing more details about how emitter pure dephasing degrades the gate, let us consider just the optical dissipation. In Ref.~\ref{asadi2020cavitygates}, we analyzed the simple virtual photon exchange mediated by a single cavity mode. This model can be recovered from the general QNM master equation described in the previous section by taking $\chi_{11}^+=\omega_\mathrm{c}$, $\chi_{11}^-=\kappa/2$, $g_{11}=g_{12}=g$, and $\chi^\pm_{ij}=g_{jk}=0$ otherwise. Then, I will assume the optical transitions are resonant $\omega_{\uparrow_1}=\omega_{\uparrow_2}$ so that $\Delta=\omega_{\uparrow_k}-\omega_\mathrm{c}$ is the cavity-emitter detuning for both emitters. In Ref.~\ref{asadi2020cavitygates}, we considered the case where the optical transitions may be slightly detuned and may also have different cavity coupling rates, but I will not discuss those subtleties here. 

To obtain an analytic expression for the entanglement fidelity that includes dissipative phenomena, we first apply a non-Hermitian approximation to the Liouville superoperator as described in section \ref{chapter1:nonhermitian}. This gives a non-Hermitian Hamiltonian that describes the evolution of the quantum trajectory conditioned on no photon emission from either defect or the cavity mode. Then, making the adiabatic approximation for $\Delta\gg g$, we eliminate the cavity mode to obtain an effective non-Hermitian Hamiltonian using the methods described in section \ref{chapter1:adiabaticelimination}. The adiabatic non-Hermitian Hamiltonian for the single-excited system subspace is
\begin{equation}
\label{chapter4eq:nonhermitaeham}
    \tilde{H}_\text{AE}=\hbar\lambda(\sigu_{\uparrow_1}\sigd_{\uparrow_2}+\sigd_{\uparrow_1}\sigu_{\uparrow_2})-\frac{i\hbar}{2}(\gamma+R)\left(\sigu_{\uparrow_1}\sigd_{\uparrow_1}+\sigu_{\uparrow_2}\sigd_{\uparrow_2}\right)
\end{equation}
where $\lambda=-2g^2/(2\Delta-i\kappa)$ is the effective (non-Hermitian) dipole-dipole coupling rate and $R=4g^2\kappa/(\kappa^2+4\Delta^2)$ is the Purcell enhancement. The ideal gate time is given by $T_\text{gate}=\pi/|\lambda|=\pi\left(2g^2/\sqrt{\kappa^2+4\Delta^2}\right)^{-1}$, which is slightly damped by the cavity dissipation. However, since high fidelity is only obtained when $\Delta\gg \kappa$, it is reasonable to ignore this damping and choose $T_\text{gate}=\pi g^2/\Delta$. From this effective non-Hermitian dipole-dipole dynamics, it is then straightforward to solve for the phase gate fidelity that includes dissipative effects. 

In this section, I will define the phase gate fidelity $\euscr{F}_\text{gate}$ as the fidelity of the final state after applying the gate to the initially unentangled state after perfect excitation of defect 1: $\ket{\psi(0)}=(1/2)(\ket{\mathrm{e}}+\ket{\downarrow})_1\otimes(\ket{\uparrow}+\ket{\downarrow})_2\otimes\ket{0}_\mathrm{c}$. In this case, a perfect phase gate would provide the expected final entangled state $\ket{\psi_\mathrm{f}}=(1/2)(\ket{\mathrm{e}\!\downarrow}-\ket{\mathrm{e}\! \uparrow}+\ket{\downarrow \downarrow}+\ket{\downarrow \uparrow})_{12}\otimes\ket{0}_\mathrm{c}$. Using this definition along with Eq.~(\ref{chapter4eq:nonhermitaeham}), we obtain
\begin{equation}
\label{chapter4eq:fidgatenonHerm}
    \euscr{F}_\text{gate}=\frac{1}{4}\left[1+e^{-(\gamma+R)t/2}\left|\sin\!\left(\!\frac{t\lambda}{2}\!\right)\right|^2\right]^2.
\end{equation}
The product states $\ket{\downarrow\downarrow}$ and $\ket{\downarrow\uparrow}$ of the initial state $\ket{\psi(0)}$ experience no evolution under Eq.~(\ref{chapter4eq:nonhermitaeham}) and so they contribute 1/4 to the fidelity by default. Only the relative phase and magnitudes of the excited states $\ket{\mathrm{e}\!\downarrow}$ and $\ket{\mathrm{e}\! \uparrow}$ contribute to the degradation of the fidelity, which is captured by the second term within the brackets of Eq.~(\ref{chapter4eq:fidgatenonHerm}). Note also that this fidelity expression may not give the expected $\sin^2$ behaviour in time due to the non-Hermitian coupling term $\lambda$, which is complex in general. 

By setting the evolution time to the chosen value of $t=T_\text{gate}=\pi\Delta/g^2$ we have $t\lambda\simeq \pi$ if $\Delta\gg\kappa$. This latter condition holds in the far-adiabatic regime. In this regime, only the exponential term remains in the fidelity. Thus, the fidelity is limited only by the dissipative rate in the high-fidelity, far-adiabatic, regime. Writing the fidelity in terms of the uninhibited cavity cooperativity $C=4g^2/\kappa\gamma$, we obtain
\begin{equation}
\begin{aligned}
\label{chapter4:sphtexchfid1}
    \euscr{F}_\text{gate}  &\simeq\frac{1}{4}\left(1+e^{-2\pi\Delta/C\kappa-\pi\kappa/2\Delta}\right)^2.
\end{aligned}
\end{equation}
This expression can be maximized to find that $\Delta=(\kappa/2)\sqrt{C}$ is the ideal detuning condition \ref{asadi2020cavitygates}. This condition is identical to requiring that the bare cavity dissipation is equal to the (detuning inhibited) Purcell rate: $\gamma=R\simeq g^2\kappa/\Delta^2$, which illustrates that the fidelity is truly limited by decoherence from the rate trade-off between the cavity dissipation and the defect spontaneous emission. However, the general way to maximize fidelity in this far-adiabatic regime is to minimize the dissipation during the gate time: $T_\text{gate}(R+\gamma)$. It just happens to be the case that $T_\text{gate}R$ and $T_\text{gate}\gamma$ are reciprocally related via $\Delta$, and hence the maximum fidelity is obtained when the two rates are equal. In any case, by substituting this solution back into the fidelity, we can reveal the simple maximum fidelity scaling of $\text{max}\!\left(\euscr{F}_\text{gate}\right)=\left(1+e^{-2\pi/\sqrt{C}}\right)^2/4$, which for large $C$ converges to $1-2\pi/\sqrt{C}$ \ref{asadi2020cavitygates}. This sets a fundamental upper bound on the simple virtual photon exchange control phase gate fidelity for a given cavity cooperativity when only one cavity mode is involved.

Let us now take a look at how pure dephasing degrades the fidelity. In principle, we can perform the non-Hermitian approximation even when pure dephasing is present. But the interpretation in terms of quantum trajectories is not as clear because pure dephasing is an elastic process. That is, it can be seen to cause `jumps' from a state back to that same state. Consider the emitter pure dephasing term $2\gamma^\star\mathcal{D}(\sigu\sigd)$. Expanding this out, we can identify that the pure-dephasing part of the non-Hermitian Hamiltonian is $-i\gamma^\star\sigu\sigd$. Hence, it should have an identical effect as $\gamma/2$ resulting in a simple replacement of $C$ with an inhibited cavity cooperativity $C^\star=4g^2/\kappa\Gamma$; but, as we briefly mentioned in Ref.~\ref{asadi2020repeaters}, this is not actually the case. Recall that, in the context of fidelity, the non-Hermitian approximation makes the assumption that the fidelity obtained when not following the ideal quantum trajectory is much smaller than the fidelity obtained when following the ideal quantum trajectory (see section \ref{chapter1:nonhermitian}). This is always true for dissipation in this scheme because a single jump will remove the system from the ideal subspace causing immediate and total decoherence. However, pure dephasing can never remove the system from the ideal subspace. Hence, dissipation and dephasing do not have an equivalent effect on the state fidelity as they would, for example, on the temporal coherence of an emitted photonic state. For this reason, replacing $C$ with $C^\star$ in the above fidelity expression generally leads to an \emph{underestimation} of the true fidelity.

Instead of using the non-Hermitian approximation to obtain a non-Hermitian Hamiltonian, we can make use of the conditional propagator superoperator $\mathcal{U}_0$ corresponding to no photon emission to keep effects of pure dephasing fully intact. That is, we can consider the dynamics associated with the Liouville superoperator $\mathcal{L}_0=\mathcal{L}-\gamma\mathcal{J}(\sigd_{\uparrow_1})-\gamma\mathcal{J}(\sigd_{\uparrow_2})-\kappa\mathcal{J}(\ad)$, where $\mathcal{J}(\hat{A})\hat{\rho}=\hat{A}\hat{\rho}\hat{A}^\dagger$. Although this does not allow us to reduce the dimension of the problem as before---because the pure dephasing term within $\mathcal{L}_0$ prevents us from writing an effective Hamiltonian---it does still decouple the equations of motion for many of the density matrix elements. By adiabatically eliminating all density matrix elements (coherence and population) corresponding to the cavity mode, we can diagonalize the effective Liouville superoperator and obtain an analytic solution for the fidelity valid in the bad-cavity regime that captures pure dephasing.
 
The exact expression for fidelity obtained by diagonalizing the Liouville superoperator is too large to write out, but by applying the same far-adiabatic regime assumptions used to get Eq.~(\ref{chapter4:sphtexchfid1}), the expression is found to be well-approximated by
\begin{equation}
\label{chapter4:sphtexchfid2}
    \euscr{F}_\text{gate}\simeq\frac{1}{4}+\frac{1}{2}
    e^{-2\pi\Delta/C^\star\kappa-\pi\kappa/2\Delta}+\frac{1}{16}\left(1+3e^{(2\pi\Delta/\kappa)\left(1/C-1/C^\star\right)}
    \right)e^{-4\pi\Delta/C\kappa-\pi\kappa/\Delta},
\end{equation}
Note that by setting $C^\star=C$ (hence $\gamma^\star=0$), we recover Eq.~(\ref{chapter4:sphtexchfid1}) by factorization. This solution explicitly shows that pure dephasing does not affect the fidelity in the same way as dissipation. If it did, $C$ would not appear in the fidelity. Interestingly, it turns out that we can estimate an effective cavity cooperativity $C_\text{eff}$ for this particular scheme. To do so, we can postulate that $C_\text{eff}=4g^2/\kappa(\gamma+c\gamma^\star)=C(1+c(\gamma^\star/\gamma))^{-1}$ for some unknown coefficient $c$. For example, if $c=0$ we get $C_\text{eff}=C$ and if $c=2$ we get $C_\text{eff}=C^\star$. By substituting this in place of $C$ in Eq.~(\ref{chapter4:sphtexchfid1}) and minimizing the difference with Eq.~(\ref{chapter4:sphtexchfid2}) that includes pure dephasing exactly, we can analytically solve for the best value of $c$ to lowest order for when $C\gg 1$. The result is $c=11/8$, which is independent of the relative magnitudes of $\gamma$ and $\gamma^\star$. Actually, it turns out that, without making the additional far-adiabatic assumptions to simplify Eqs.~(\ref{chapter4:sphtexchfid1}) and (\ref{chapter4:sphtexchfid2}), the coefficient can be found to be $c=21/16$, which turns out to be slightly more accurate. In any case, since $c<2$, this implies that pure dephasing has less of an impact on fidelity than would be expected from $C^\star$ alone. I should also note that in Ref.~\ref{asadi2020repeaters}, we mention a value of $c=0.61$, but this was guessed from numerical observations and not analytically derived as was done here.

\begin{figure}[t]
    \centering
    \hspace{-50mm}(a)\hspace{77mm}(b)\\
    \includegraphics[width=0.49\textwidth]{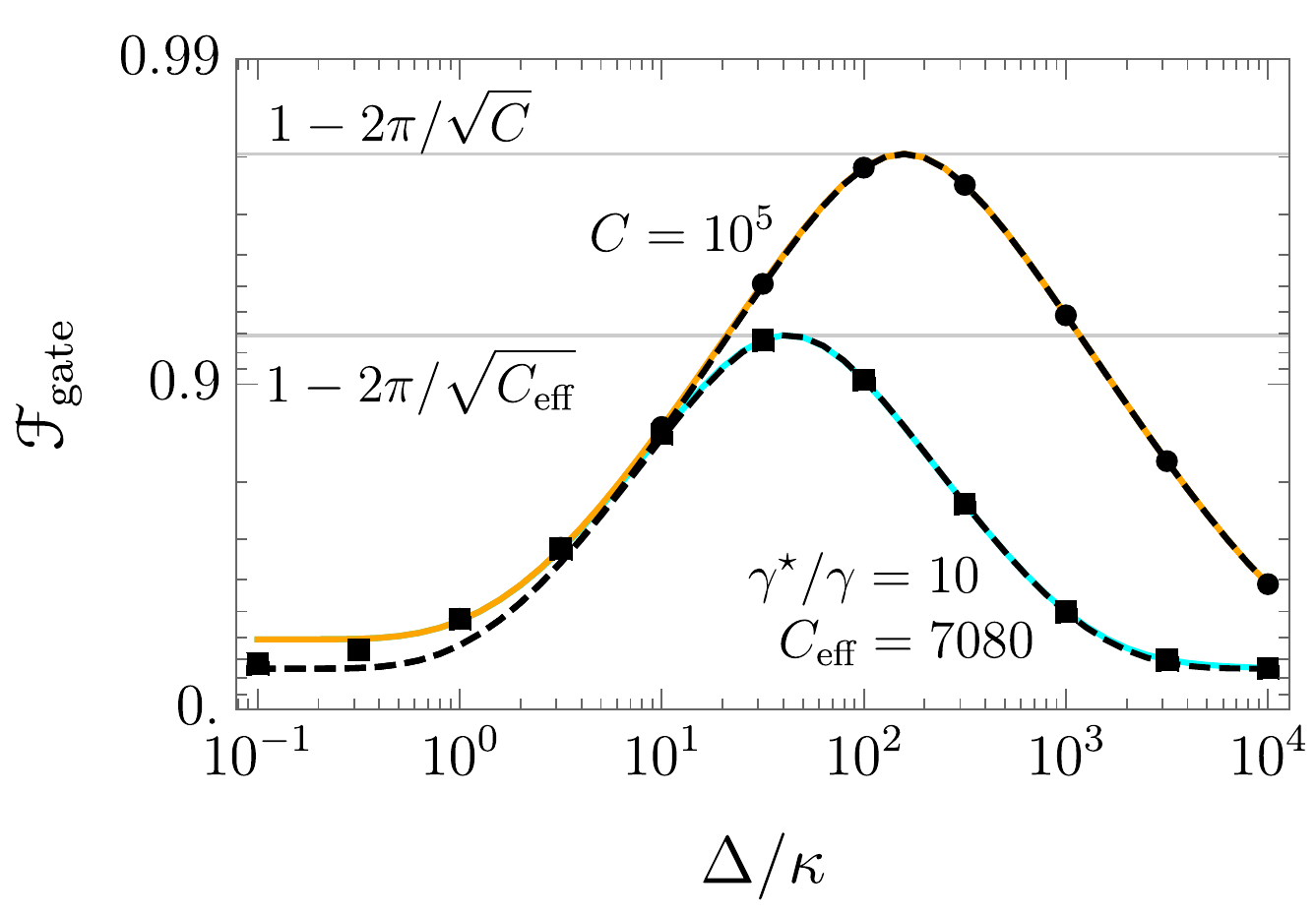}
    \includegraphics[width=0.49\textwidth]{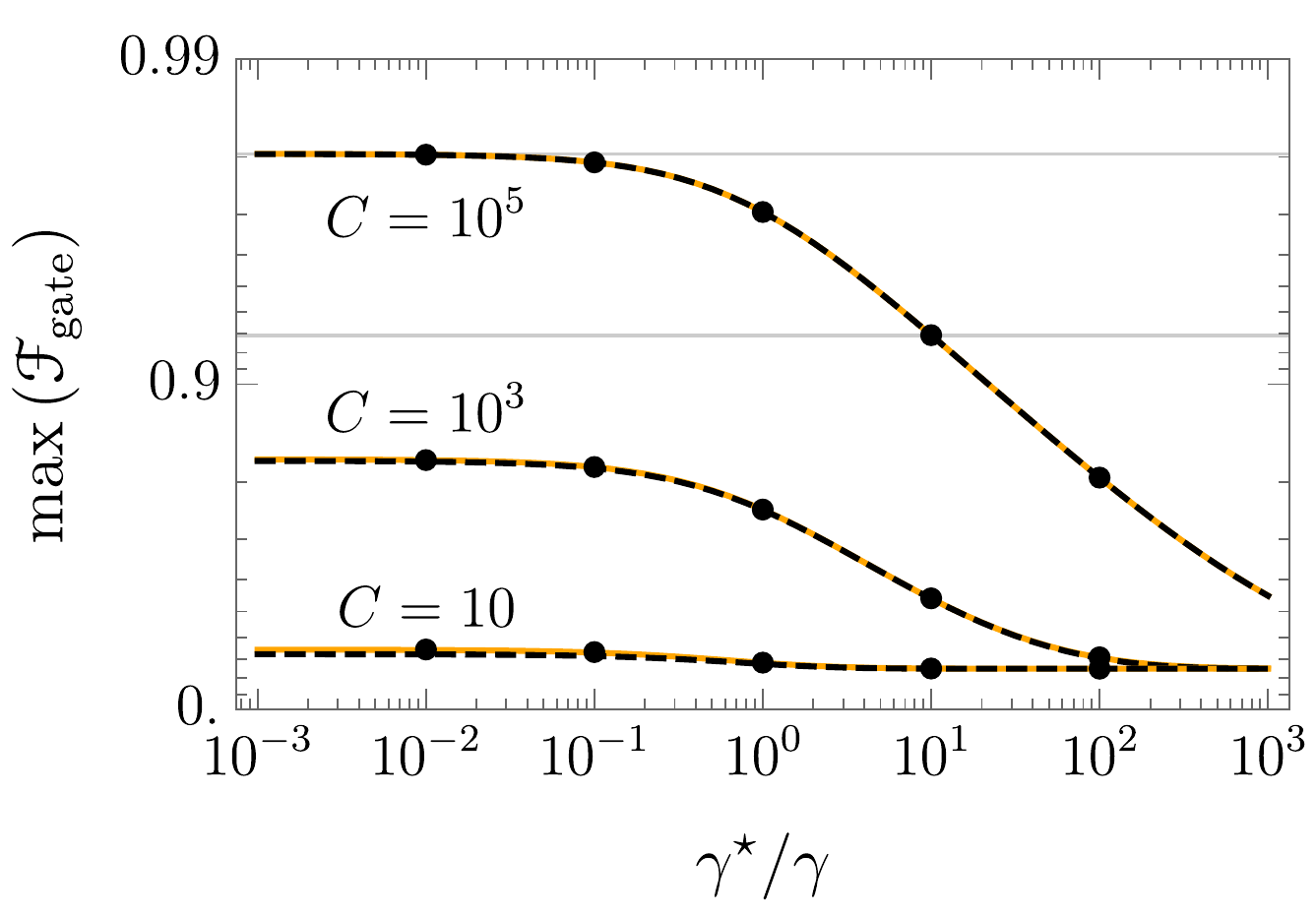}
    \caption[The effect of emitter pure dephasing on the simple photon exchange controlled phase gate fidelity for local entanglement generation.]{\small\textbf{The effect of emitter pure dephasing on the simple photon exchange controlled phase gate fidelity for local entanglement generation.} (a) The phase gate fidelity as a function of cavity detuning $\Delta$ in units of the cavity FWHM linewidth $\kappa$. The black points are numerically exact values computed from simulating the full master equation with perfect state preparation and retrieval pulses. The orange and cyan colored lines indicate the analytic solution obtained from applying adiabatic elimination to the conditional Liouville superoperator using no pure dephasing (orange) and pure dephasing of $\gamma^\star/\gamma=10$ (cyan). The black dashed lines indicate the solution Eq.~(\ref{chapter4:sphtexchfid1}) using an effective cavity cooperativity $C_\text{eff}$. The emitter pure dephasing degrades the fidelity for this scheme by inhibiting the cavity cooperativity by a factor $C_\text{eff}=C(1+c\gamma^\star/\gamma)^{-1}$ where $c=21/16$. The maximum fidelity occurs at $2\Delta=\kappa\sqrt{C_\text{eff}}$ with a value well-approximated by $\text{max}(\euscr{F}_\text{gate})=(1+e^{-2\pi/\sqrt{C}})^2/4\simeq 1-2\pi/\sqrt{C_\text{eff}}$ when $C_\text{eff}\gg 1$. (b) The maximum phase gate fidelity as a function of emitter pure dephasing for three different values of cavity cooperativity. The horizontal gray lines carry over from panel (a).}
    \label{chapter4fig:simpleexchangefidelity}
\end{figure}

Fig.~\ref{chapter4fig:simpleexchangefidelity} shows a comparison of the analytic solution obtained by applying adiabatic elimination to the conditional Liouville superoperator and points computed from a numerical simulation of the full master equation. I have also plotted the fidelity from Eq.~(\ref{chapter4:sphtexchfid1}) computed using the effective cavity cooperativity with $c=21/16$. All these computations assume that the system is in the bad-cavity regime with $\kappa\gg \gamma C$ and $C>1$ (hence $\gamma\ll\kappa)$. For the numerical simulation, I used $\kappa=20g$ and $\gamma=10^{-7}\kappa$ for a cavity cooperativity of $C=10^5$. It is impressive to note that the fidelity computed using the effective cavity cooperativity with $c=21/16$ provides a near flawless estimate of the fidelity in this regime. However, both analytic solutions slightly deviate from the numerical values when $\Delta/\kappa<1$ or $C\simeq 1$, which is expected because this is where adiabatic evolution breaks down.

The main message to take away from this section is that the simple virtual photon exchange interaction using a single cavity mode allows for a maximum fidelity near $1-2\pi/\sqrt{C_\text{eff}}$. This means that, for a high fidelity  such as >0.99, it is necessary to use a system with an effective cooperativity exceeding $10^5$, which is very demanding for solid-state systems that suffer from dephasing. In the next section, I will outline how a cavity with a Fano resonance could surpass this inverse square-root bound, potentially allowing for high fidelity gates using much smaller cavity cooperativity values.

\subsection{Subradiant virtual photon exchange}
\label{chapter4:subradiantvirtual}

As I described in the previous section, aside from pure dephasing, the two main processes hindering the phase gate fidelity are the bare spontaneous emission and the Purcell enhanced emission. These two processes place opposing constraints on the ideal detuning, giving rise to a maximum fidelity at some intermediate detuning value. By introducing a second cavity mode that can interfere with the first, this trade-off can be modified, potentially allowing for regimes that surpass the single-mode case.

A Fano resonance is a spectral phenomenon that occurs when a broad spectral mode resonance interferes with a detuned narrow mode resonance. This interference gives rise to a characteristic asymmetric spectral feature surrounding the narrow mode resonance frequency. This asymmetry occurs because the phase contribution from a detuned broad mode changes monotonically as the frequency is swept over the narrow mode resonance. The abrupt change in phase at the narrow mode resonance then causes the interference between the two modes to be very different on either side of the narrow mode resonance. Thus, the situation can occur where destructive interference occurs on one side of the narrow resonance, forming a `Fano dip' where the modes are out of phase and a constructive interference forms a `Fano peak' where the modes are in phase on the other side.

Fano resonances often arise in the context of energy transfer between two systems, in particular involving plasmonics \cite{nan2016plasmon,hsu2017plasmon}. Fano resonances have also been proposed to enhance single-photon indistinguishability \cite{denning2019quantum} and suppress quenching for hybrid cavities \cite{gurlek2018manipulation}. However, to my knowledge, the quantum properties of a phase gate interaction taking advantage of an optical Fano resonance have not been studied. The methods detailed in the previous section have now set us up to analyze this question using the quantum model recently presented in Ref.~\cite{franke2019quantization}, which I introduced in section \ref{chapter1:qnmmastereq}.

Let us consider two QNMs with complex frequencies $\tilde{\omega}_k=\omega_k-i\kappa_k/2$ coupled to two optically-active three-level systems. For simplicity, I will neglect spin decoherence and emitter pure dephasing for now. In this case, we can use the non-Hermitian approximation to obtain an effective non-Hermitian Hamiltonian. To do so, we must first diagonalize the dissipative part of the Liouville superoperator \cite{franke2020quantized} to obtain a superradiant and a subradiant mode. As explained in section \ref{chapter1:qnmmastereq}, the dissipation for the dual QNM system is given by the $2\times 2$ matrix $\chi^-$ where $\chi_{ij}^-=i(\chi_{ij}-\chi_{ij}^*)/2$, $\chi_{ij}=\sum_k(\mathbf{S}^{-1/2})_{ik}\tilde{\omega}_k(\mathbf{S}^{1/2})_{kj}$, and $\mathbf{S}$ is a positive semi-definite overlap matrix describing the amount of interference between the modes. For simplicity, I will fix the diagonal elements of $\mathbf{S}$ to 1 so that the modes are not renormalized by the interference. However, in practice this renormalization depends strongly on the system geometry and dissipation, ranging from being fairly small \cite{franke2019quantization} to quite significant \cite{franke2020quantized}. I will also consider a general case for the off-diagonal element $S_{12}=S^*_{12}=|S_{12}|e^{i\phi_s}$. The assumption that $S_{12}$ is independent of the diagonal elements may not be true when varying parameters of the system geometry or material composition. Hence, these simplifying assumptions are used to explore what possibilities arise from this model and to motivate further investigation.

In addition to the simplifying assumptions about $\mathbf{S}$ detailed above, we must also ensure that $\mathbf{S}$ cannot allow for a physically impossible amount of interference based on the spectral overlap of the QNMs. This means that we must restrict the eigenvalues of $\chi^-$ to be positive. Applying these additional constraints gives
\begin{equation}
    \left|S_{12}\right| \leq 2\sqrt{\frac{\kappa_1\kappa_2}{4\Delta_\mathrm{c}^2+(\kappa_1+\kappa_2)^2}},
\end{equation}
where $\Delta_\mathrm{c}=\omega_1-\omega_2$ is the detuning between the two QNM resonances. For convenience, I define a new parameter $0\leq s\leq 1$ where $|S_{12}|=2s\sqrt{\kappa_1\kappa_2/(4\Delta_\mathrm{c}^2+(\kappa_1+\kappa_2)^2)}$. Then $s=0$ implies no interference and $s=1$ implies maximum possible interference. Caution must also be taken when considering the case of $s=0$. If the modes still significantly overlap in frequency, then $s=0$ inherently implies that the modes must not overlap significantly in space. This then complicates the notion of a two-mode cavity. For realistic implementations, the matrix $\mathbf{S}$ should be calculated fully from classical electromagnetic simulations  \cite{franke2019quantization,franke2020quantized}.

After diagonalizing $\chi^-$ so that $\sum_{ij}\chi_{ij}^-\mathcal{D}(\ad_i,\ad_j)=\kappa^+\mathcal{D}(\hat{c}_+,\hat{c}_+)+\kappa^-\mathcal{D}(\hat{c}_-,\hat{c}_-)$, we have a superradiant cavity mode $\hat{c}_+$ and a subradiant cavity mode $\hat{c}_-$ corresponding to the exponential dissipative rates $\kappa^+>\kappa^-$ that are the eigenvalues of $\chi^-$. On one hand, if $s=1$, then we have $\kappa^-=0$ and $\kappa^+=\kappa_1+\kappa_2$, which represents the most extreme case of interference. On the other hand, if $s=0$, then we have $\kappa^-=\kappa_2$ and $\kappa^+=\kappa_1$. In this diagonalized dissipative basis, the non-Hermitian approximation can be easily applied by considering the evolution induced by the Liouville superoperator $\mathcal{L}_0=\mathcal{L}-\kappa^+\mathcal{J}(\hat{c}_+)-\kappa^-\mathcal{J}(\hat{c}_-)-\gamma\mathcal{J}(\sigd_{\uparrow_1})-\gamma\mathcal{J}(\sigd_{\uparrow_2})$ conditioned on no photon emission (or absorption) from either cavity mode or bare spontaneous emission from either dipole systems. This approximation gives a lower bound on the phase gate fidelity (see section \ref{chapter1:nonhermitian}). If no other decoherence processes are captured by $\mathcal{L}_0$, such as pure dephasing, we can write an effective non-Hermitian Hamiltonian governing the evolution of the ideal quantum trajectory.
\begin{equation}
    \tilde{H}_\text{eff} = \hat{H}_1+\hat{H}_2+\hat{H}_\mathrm{c}-\frac{i\hbar\gamma}{2}\left(\sigu_{\uparrow_1}\sigd_{\uparrow_1}+\sigu_{\uparrow_2}\sigd_{\uparrow_2}\right)-\frac{i\hbar}{2}\left(\kappa^+\hat{c}_+^\dagger\hat{c}_++\kappa^-\hat{c}_-^\dagger\hat{c}_-\right).
\end{equation}
Note that $\hat{H}_\mathrm{c}$ contains the quantum interaction between the symmetrized modes $\ad_1$ and $\ad_2$ via the rates $\chi_{ij}^+$. Since $\hat{c}_\pm$ are given as a linear combination of $\ad_1$ and $\ad_2$ dictated by the eigenvectors of $\chi^-$, the operators $\hat{c}^\dagger_\pm\hat{c}_\pm$ will alter the QNM interaction in $\hat{H}_\mathrm{c}$. In the symmetrized QNM basis, the non-Hermitian QNM Hamiltonian simply reverts to $\sum_{ij}\hbar\chi_{ij}\au_i\ad_j$. Hence, we have
\begin{equation}
\label{HeffSubradiantvexch}
    \tilde{H}_\text{eff}=\sum_k\hbar\left(\omega_o-\frac{i}{2}\gamma\right)\sigu_{\uparrow_k}\sigd_{\uparrow_k}+\sum_{ij}\hbar\chi_{ij}\au_i\ad_j + \sum_{ik}\hbar\left(g_{ik}\sigd_{\uparrow_k}\au_i+g_{ik}^*\sigu_{\uparrow_k}\ad_i\right).
\end{equation}
By adiabatically eliminating the cavity modes using the methods in section \ref{chapter1:adiabaticelimination}, we can obtain the effective non-Hermitian dipole-dipole Hamiltonian.

The first thing we can look at is the expected Purcell enhancement on the individual dipole systems. Similar to sections \ref{chapter1:cavityQED} and \ref{chapter1:1datom}, we can estimate the dissipative rate in the bad-cavity regime by adiabatically eliminating the cavity modes and looking at the remaining imaginary part of the non-Hermitian Hamiltonian corresponding to one of the dipole excited states. Let $\tilde{H}_\text{AE}$ be the effective non-Hermitian Hamiltonian after the cavity excited state amplitudes are adiabatically eliminated (see section \ref{chapter1:adiabaticelimination}). Then the Purcell rate for an individual dipole is given by $R_k\simeq-2~\text{Im}\!\left[\braket{\mathrm{e}|\tilde{H}_\text{AE}|\mathrm{e}}_k\right]-\gamma$, when the other dipole system is in the state $\ket{\downarrow}$ so as to not be coupled to the cavity. The solution can be computed from applying adiabatic elimination to Eq.~(\ref{HeffSubradiantvexch}) or by expanding the general expression derived for the Purcell rate in Ref.~\cite{franke2019quantization}
\begin{equation}
\label{chapter4eq:purcellratefano}
    R_k = \frac{|\tilde{g}_{1k}|^2\kappa_1}{|\tilde{\Delta}_1|^2} + \frac{|\tilde{g}_{2k}|^2\kappa_2}{|\tilde{\Delta}_2|^2}+2\text{Re}\!\left[\tilde{g}_{1k}\tilde{g}_{2k}^*S_{12}\left(\frac{1}{\tilde{\Delta}_1^*}+\frac{1}{\tilde{\Delta}_2}\right)\right],
\end{equation}
where $\tilde{\Delta}_i=\omega_o-\tilde{\omega}_i$ is the complex detuning between the resonant dipoles and the $i^\mathrm{th}$ cavity mode resonance. Furthermore, we can compute the effective (non-Hermitian) dipole-dipole coupling rates from $\lambda_{12}=\braket{\uparrow\!\mathrm{e}|\tilde{H}_\text{AE}|\mathrm{e}\!\uparrow}$ and $\lambda_{21}=\braket{\mathrm{e}\!\uparrow\!|\tilde{H}_\text{AE}|\!\uparrow\!\mathrm{e}}$, which are
\begin{equation}
\label{chapter4eq:effectivecoupling}
\begin{aligned}
    \lambda_{21} &= -\frac{\tilde{g}_{11}}{\tilde{\Delta}_1}\left(\tilde{g}_{12}^*+S_{12}^*\tilde{g}_{22}^*\right)-\frac{\tilde{g}_{21}}{\tilde{\Delta}_2}\left(\tilde{g}_{22}^*+S_{12}\tilde{g}_{12}^*\right)\\
    \lambda_{12} &= -\frac{\tilde{g}_{12}}{\tilde{\Delta}_1}\left(\tilde{g}_{11}^*+S_{12}^*\tilde{g}_{21}^*\right)-\frac{\tilde{g}_{22}}{\tilde{\Delta}_2}\left(\tilde{g}_{21}^*+S_{12}\tilde{g}_{11}^*\right).
\end{aligned}
\end{equation}
Note that, due to the complex values $S_{12}$ and $\tilde{\Delta}_k$, we do not have $\lambda_{21}=\lambda_{12}^*$ in general. 

\begin{figure}
    \centering
    \hspace{-50mm}(a)\hspace{77mm}(b)\\
    \includegraphics[width=0.49\textwidth]{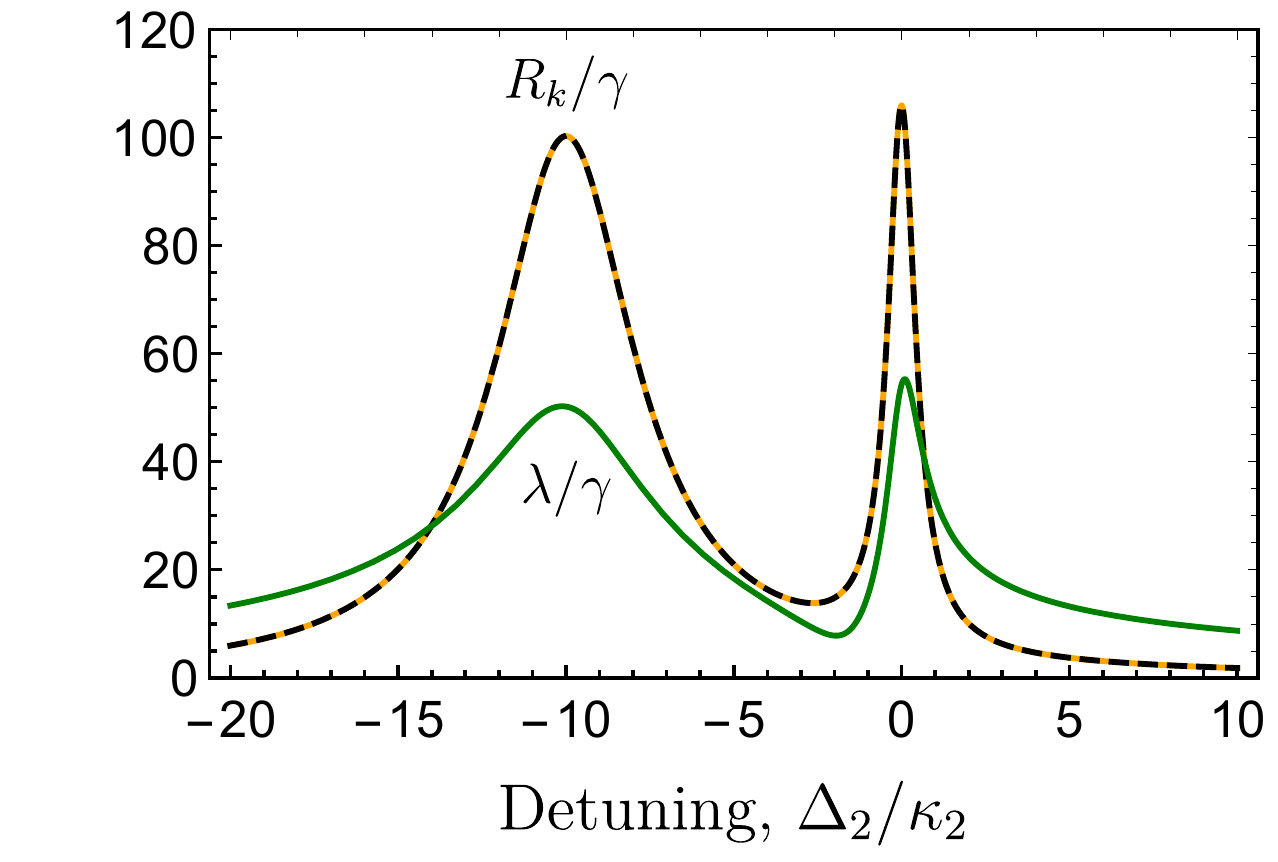}
    \includegraphics[width=0.49\textwidth]{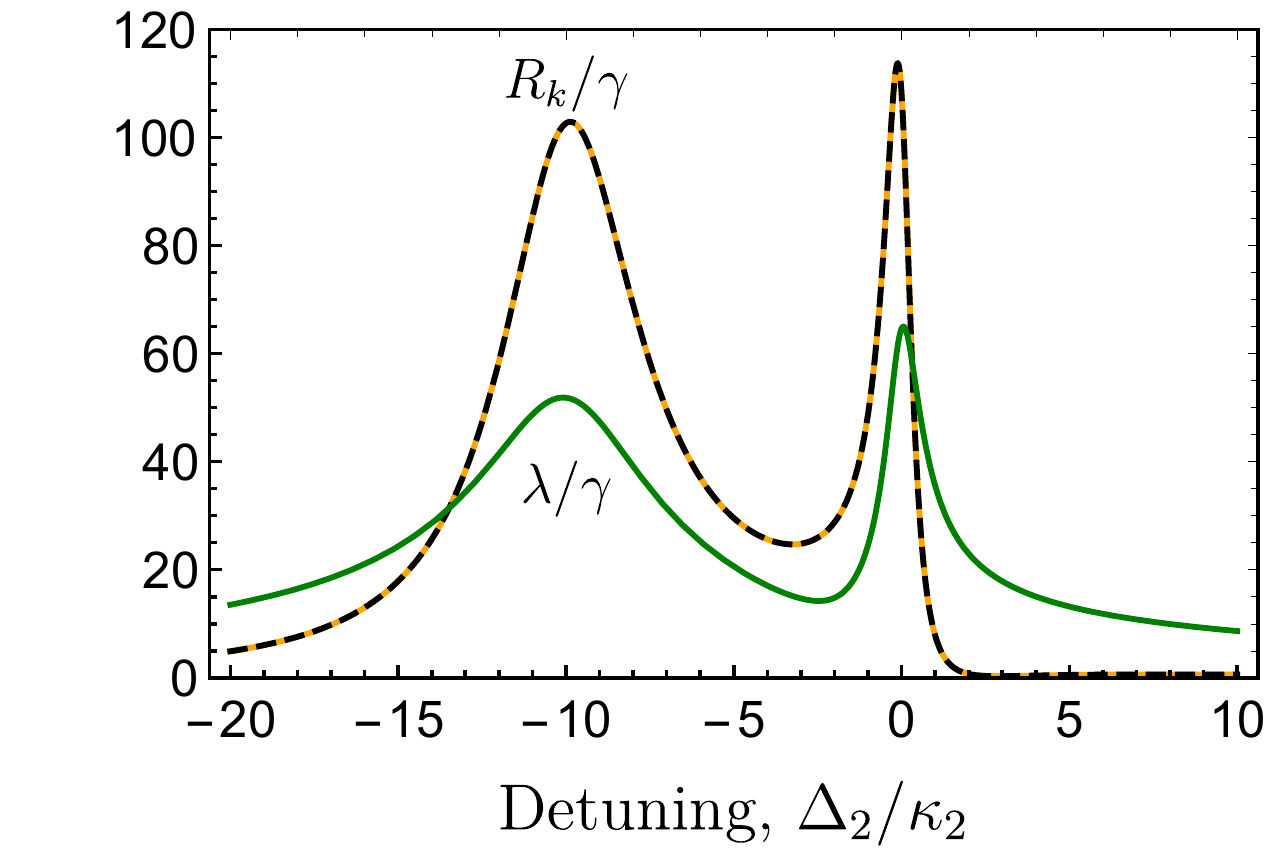}
    \caption[Purcell enhancement spectrum and effective dipole-dipole coupling rate for a dual quasi-normal mode model.]{\small\textbf{Purcell enhancement spectrum and effective dipole-dipole coupling rate for a dual quasi-normal mode model.} (a) Example spectrum for the Purcell enhancement caused by coupling to two QNMs that are not interfering ($s=0$). The orange and black lines show the Purcell decay enhancement $R_k$ given by Eq.~(\ref{chapter4eq:purcellratefano}) of each dipole and the green line shows the effective dipole-dipole coupling rate $\lambda=|\lambda_{12}|=|\lambda_{21}|$ given by Eq.~(\ref{chapter4eq:effectivecoupling}) that is induced by coupling to the cavity modes. (b) The same case as in panel (a) but with maximum interference $s=1$ forming a Fano resonance with a dip appearing to the right of the narrow mode. Notably, the effective dipole-dipole coupling rate remains large even though the Purcell enhancement is suppressed, indicating that an adiabatic coupling can be mediated by the subradiant mode of the cavity. Parameters used: $\gamma=\gamma_1=\gamma_2=10^{-3}\kappa_2$, $\kappa_1=5\kappa_2$, $\Delta_\mathrm{c}=10\kappa_2$, $\phi_s=0$, $\tilde{g}_{11}=\tilde{g}_{12}=i\sqrt{C\kappa_1\gamma/4}$, and $\tilde{g}_{22}=\tilde{g}_{21}=\sqrt{C\kappa_2\gamma/4}$ with $C=100$.}
    \label{chapter4fig:qnmspectra}
\end{figure}

Fig.~\ref{chapter4fig:qnmspectra} shows example spectra for the Purcell enhancement and the effective dipole-dipole coupling with and without interference. It is crucial to note that the characteristics of the spectrum depend strongly on the phases of the dipole-QNM couplings $g_{ik}$, the cavity mode overlap phase $\phi_s$, and magnitude $s$. The values used in Fig.~\ref{chapter4fig:qnmspectra} and throughout this section are chosen without consideration for a given physical implementation. Rather, my intention is to illustrate the limits on what could be physically possible for a system following this model. To that end, let us uncover the ideal operating parameter regime and derive the upper bound cooperativity scaling.

To achieve a high phase gate fidelity using a Fano resonance, it is necessary to engineer the system parameters so that the effective dipole-dipole coupling rates induced by each cavity mode constructively interfere within the Fano dip. When the dipoles are far-detuned from the two modes that are themselves quite far separated, then we have $\Delta_i=\omega_o-\omega_i\gg\kappa_i$ so that $\tilde{\Delta}_i=\Delta_i$ and $|S_{12}|\ll 1$ even if $s=1$. Under these conditions, the effective dipole-dipole coupling rate reduces to $\lambda=\lambda_{12}=\lambda_{21}=\lambda_1+\lambda_2$ where $\lambda_1=-\tilde{g}_{11}\tilde{g}_{12}^*/\Delta_1$ is the contribution from the broad mode and $\lambda_2=-\tilde{g}_{21}\tilde{g}_{22}^*/\Delta_2$ is the contribution from the narrow mode. For constructive quantum interference to occur, it is necessary that we have $\text{Arg}\!\left(\lambda_1\right)=\text{Arg}\!\left(\lambda_2\right)$. We can also identify the ideal gate time in this regime as $T_\text{gate}=\pi/|\lambda|$.

\begin{figure}[t]
    \centering
    \hspace{-54mm}(a)\hspace{72mm}(b)\\
    \hspace{-11.5mm}\includegraphics[width=0.47\textwidth]{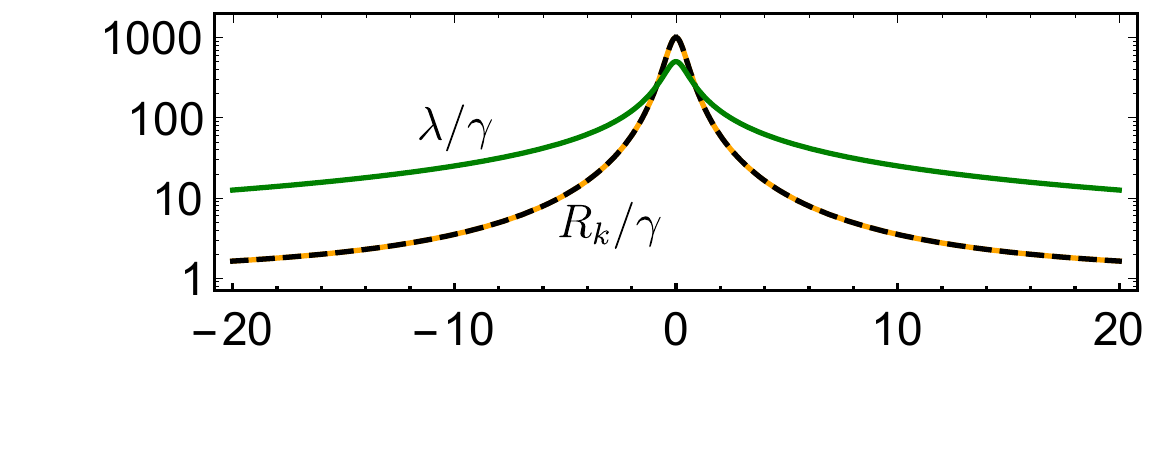}\hspace{-2.5mm}
    \includegraphics[width=0.47\textwidth]{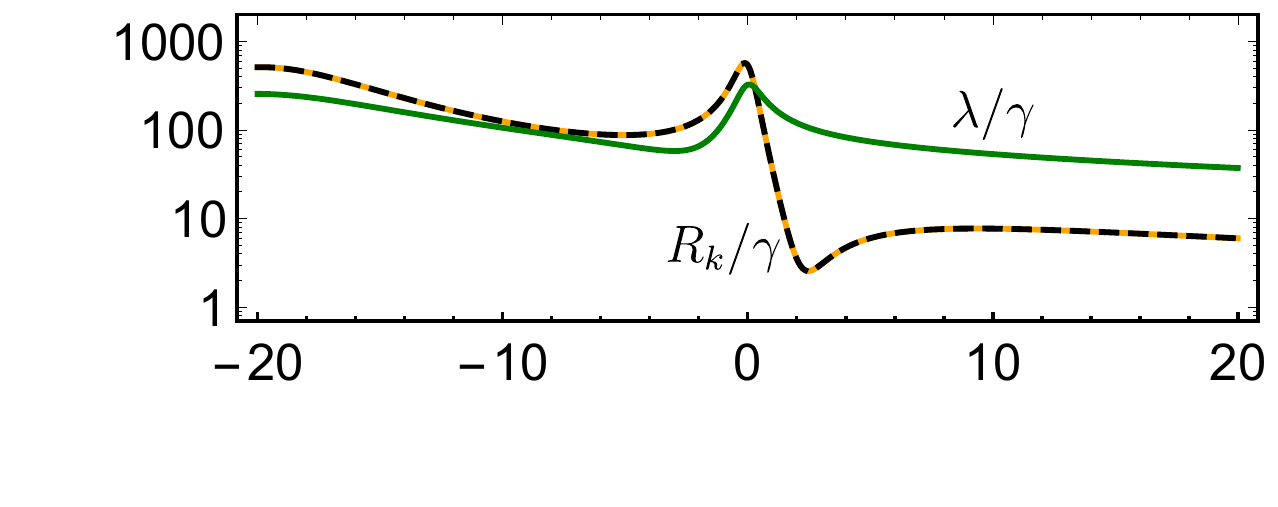}\\\vspace{-8mm}
    \includegraphics[width=0.47\textwidth]{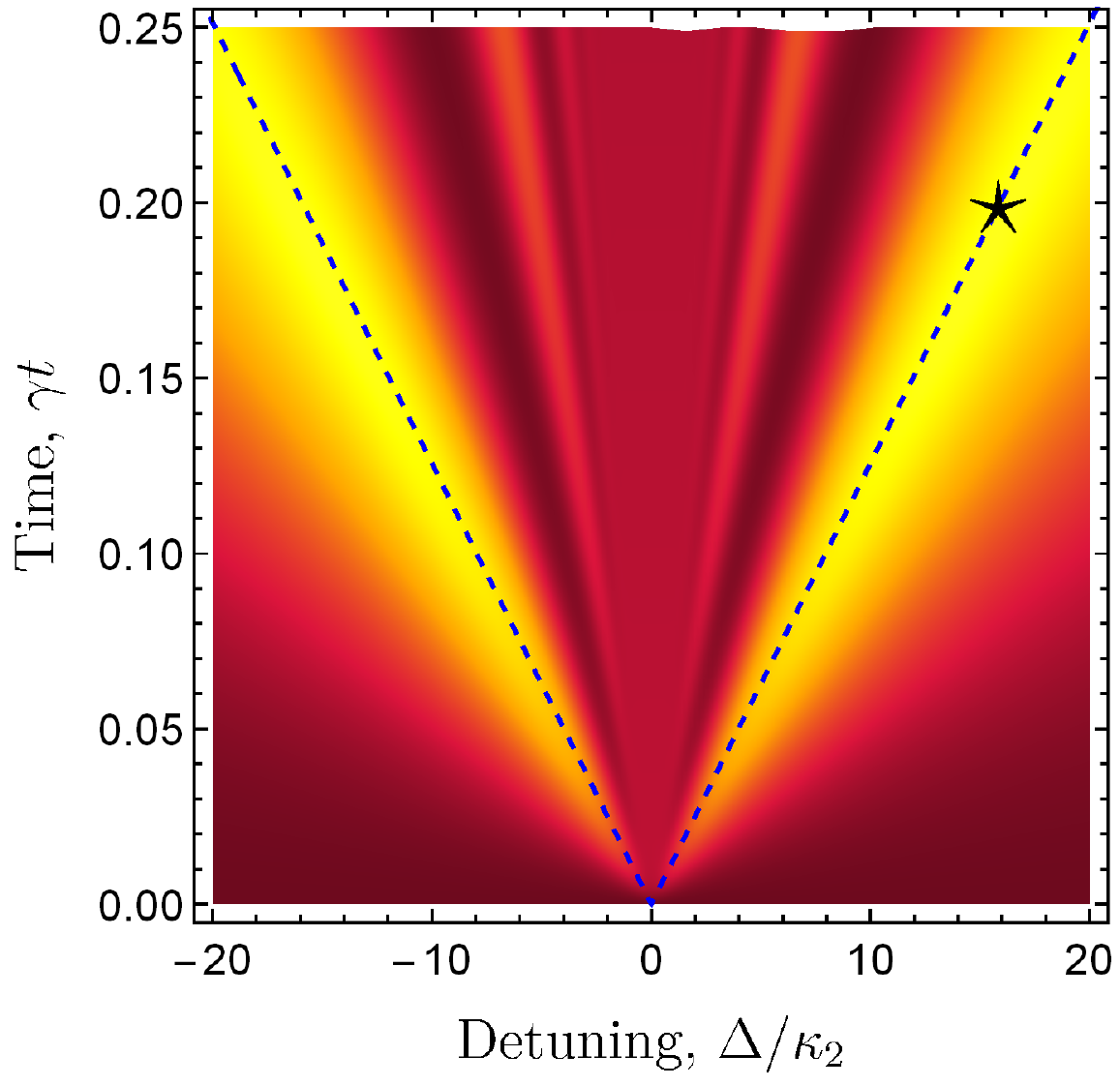}\hspace{2mm}
    \includegraphics[width=0.508\textwidth,trim={20 0 8 0},clip]{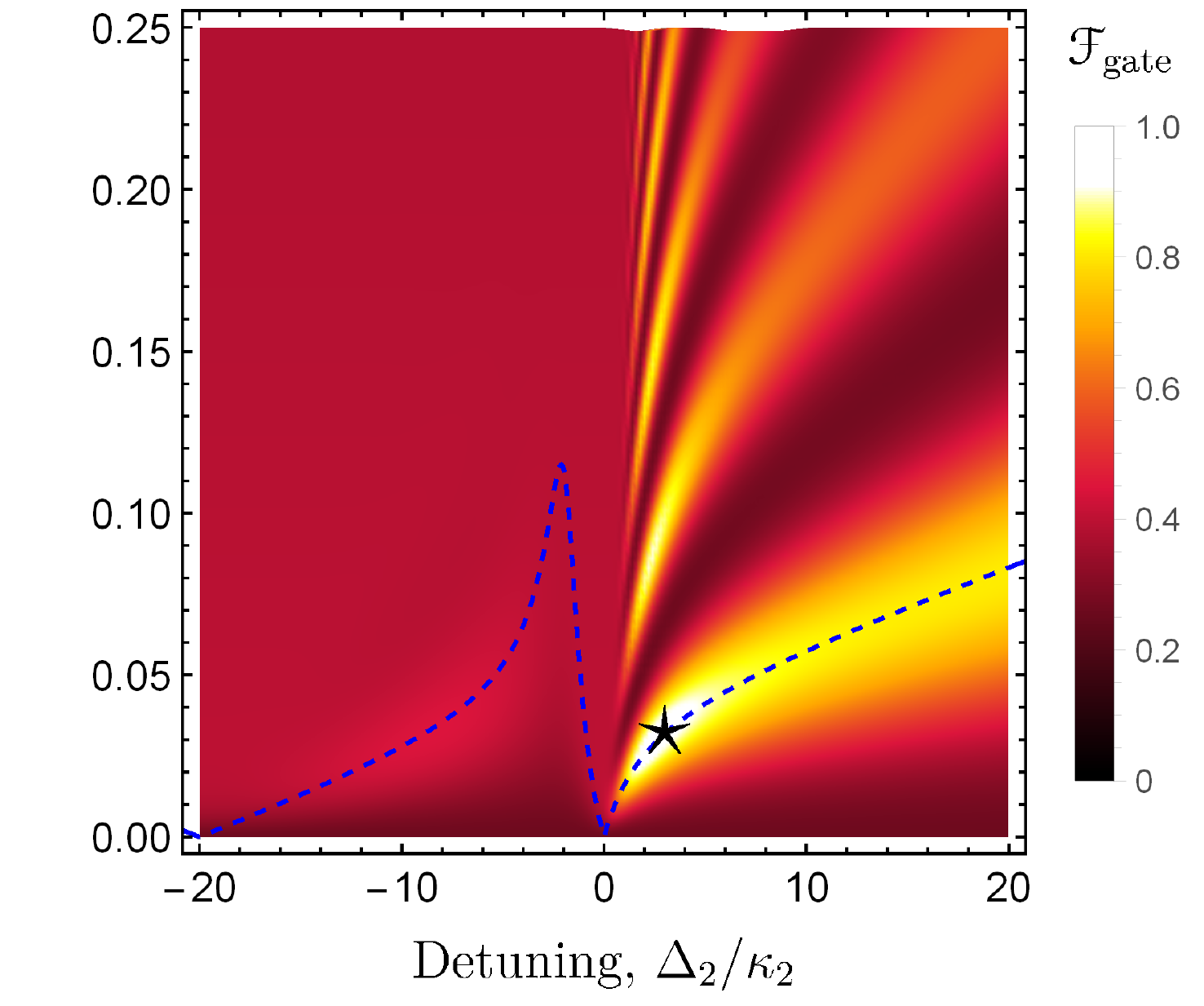}\\\vspace{-3mm}
    \caption[Time dynamics of the virtual photon exchange phase gate fidelity with a Fano resonance enhancement.]{\small\textbf{Time dynamics of the virtual photon exchange phase gate fidelity with a Fano resonance enhancement.} (a) The top panel shows the Purcell enhancement $R_k/\gamma$ and effective dipole-dipole coupling $|\lambda|$ for a single cavity mode with FWHM linewidth $\kappa_2$. The lower panel shows the phase gate fidelity $\euscr{F}_\text{gate}$ as a function of cavity-dipole detuning $\Delta$ and evolution time $t$. The dashed line shows the ideal gate time $T_\text{gate}=\pi|\Delta|/g^2$ and the star indicates the point of maximum fidelity corresponding to $\Delta=\kappa_2\sqrt{C}/2$. (b) The top panel shows the characteristic Fano resonance spectrum and effective dipole-dipole coupling for maximum interference $s=1$. The bottom panel shows the corresponding phase gate fidelity. The maximum fidelity occurs in the Fano dip. The dashed line shows the ideal gate time $T_\text{gate}=\pi/|\lambda|$ where, $\lambda\simeq\lambda_1+\lambda_2$, $\lambda_1=\tilde{g}_{11}\tilde{g}_{12}^*/(\Delta_\mathrm{c}+\Delta_2)$ and $\lambda_2=\tilde{g}_{21}\tilde{g}_{22}^*/\Delta_2$. Parameters used: $\Delta_\mathrm{c}=20\kappa_2$, $\kappa_1=10\kappa_2$, $\gamma=\gamma_1=\gamma_2=\kappa_1 10^{-5}$, $\phi_s=\pi/2$, $\tilde{g}_{11}=\tilde{g}_{12}$, and $\tilde{g}_{21}=\tilde{g}_{22}$. For panel (a): $C=1000$. For panel (b): $C_{ik}=4|\tilde{g}_{ik}|^2/\gamma_i\kappa_k=500$.}
    \label{chapter4fig:singlemodevsafanoFideltiyTime}
\end{figure}

For the phase gate to have an enhancement from the Fano dip, it is necessary that both dipoles experience a Fano dip when they are resonant. From the expression for $R_k$, we can identify that the Fano dip occurs when the third term is negative. Hence, if there is a Fano dip, then for it to occur at the same frequency for both dipoles it is necessary to have $\text{Arg}\!\left(\tilde{g}_{11}\tilde{g}_{21}^*\right)=\text{Arg}\!\left(\tilde{g}_{12}\tilde{g}_{22}^*\right)$. Note that for small $|S_{12}|$, this is also roughly equivalent to the condition $\text{Arg}\!\left(g_{11}g_{21}^*\right)=\text{Arg}\!\left(g_{12}g_{22}^*\right)$. For convenience, let us define the phases of these cavity coupling rates as $\phi_{ik}^g=\text{Arg}\!\left(\tilde{g}_{ik}\right)$ so that the condition becomes $\phi_{11}^g-\phi_{21}^g=\phi_{12}^g-\phi_{22}^g$. 

Going back to the condition $\text{Arg}\!\left(\lambda_1\right)\simeq\text{Arg}\!\left(\lambda_2\right)$, we have three cases: (1) the dipole resonances are blue-detuned from both cavity resonances implying $\Delta_i>0$, (2) the dipole resonances are in between the cavity resonances implying $\Delta_1\Delta_2<0$, and (3) the dipole resonances are red-detuned from both cavity resonances implying $\Delta_i<0$. Cases (1) and (3) imply that we need $\phi_{11}^g-\phi_{12}^g=\phi_{21}^g-\phi_{22}^g$ whereas case (2) implies that we need $\phi_{11}^g-\phi_{12}^g=\phi_{21}^g-\phi_{22}^g+\pi$. By applying the condition $\phi_{11}^g-\phi_{21}^g=\phi_{12}^g-\phi_{22}^g$ required for both dipoles to experience a Fano dip while resonant with each other, we can see that case (2) suggests $\pi=0$, which is a direct contradiction. However, cases (1) and (3) give $\phi_{11}^g+\phi_{22}^g=\phi_{12}^g+\phi_{21}^g$, which can (in principle) be satisfied. Therefore, a high fidelity phase gate is only possible if both dipoles are resonant with each other and either red-detuned from both cavity modes or blue-detuned from both cavity modes, but never in between. Finally, it is still necessary to engineer the system so that a Fano dip actually occurs in the spectrum. From $R_k$, we can see that a good Fano dip would require that the $\text{Re}$ function picks out the most negative term possible. By expanding the third term, we can see that a good estimate of the phase condition showing the best Fano dip is $\phi_s+\phi_{1k}^g-\phi_{2k}^g\simeq\pm\pi/2+2m\pi$ for an integer $m$. This picks out the term(s) $\pm2|\tilde{g}_{1k}\tilde{g}_{2k}S_{12}|(\Delta_1/|\tilde{\Delta}_1|^2-\Delta_2/|\tilde{\Delta}_2|^2)$. Since I have defined $\Delta_1>\Delta_2$, for a Fano dip to occur for case (1) or (3), we must select the $+$ case whereas the $-$ case puts a Fano dip between the mode resonances. All of these conditions are satisfied in the example given in Fig.~\ref{chapter4fig:qnmspectra}, where $\phi_{11}^g=\phi_{12}^g=\pi/2$, $\phi_{22}^g=\phi_{21}^g=0$, and $\phi_s=0$. However, there is no single unique choice. For example, $\phi_{11}^g=\phi_{12}^g=\phi_{22}^g=\phi_{21}^g=0$ and $\phi_s=\pi/2$ also satisfy the conditions.

Using the effective Hamiltonian $\tilde{H}_\text{AE}$, it is very numerically efficient to compute the virtual photon exchange phase gate fidelity. Fig.~\ref{chapter4fig:singlemodevsafanoFideltiyTime} shows the phase gate fidelity as a function of detuning $\Delta$ and evolution time for a Lorentzian resonance compared to a Fano resonance following the phase conditions given above. From this example, it is clear that the Fano resonance dramatically reduces the detuning required to achieve a high phase gate fidelity. This reduces the gate time and increases the fidelity as a consequence. For a reasonably-optimistic cooperativity of $C=1000$, the maximum fidelity for the Lorentzian case is just $\euscr{F}_\text{gate}=0.83$. However, two modes each with $C=500$ in this case can allow for up to $\euscr{F}_\text{gate}=0.96$ while simultaneously reducing the gate time by more than a factor of 6.

\begin{figure}[t]
    \centering
    \hspace{-54mm}(a)\hspace{72mm}(b)\\
    \includegraphics[width=0.49\textwidth]{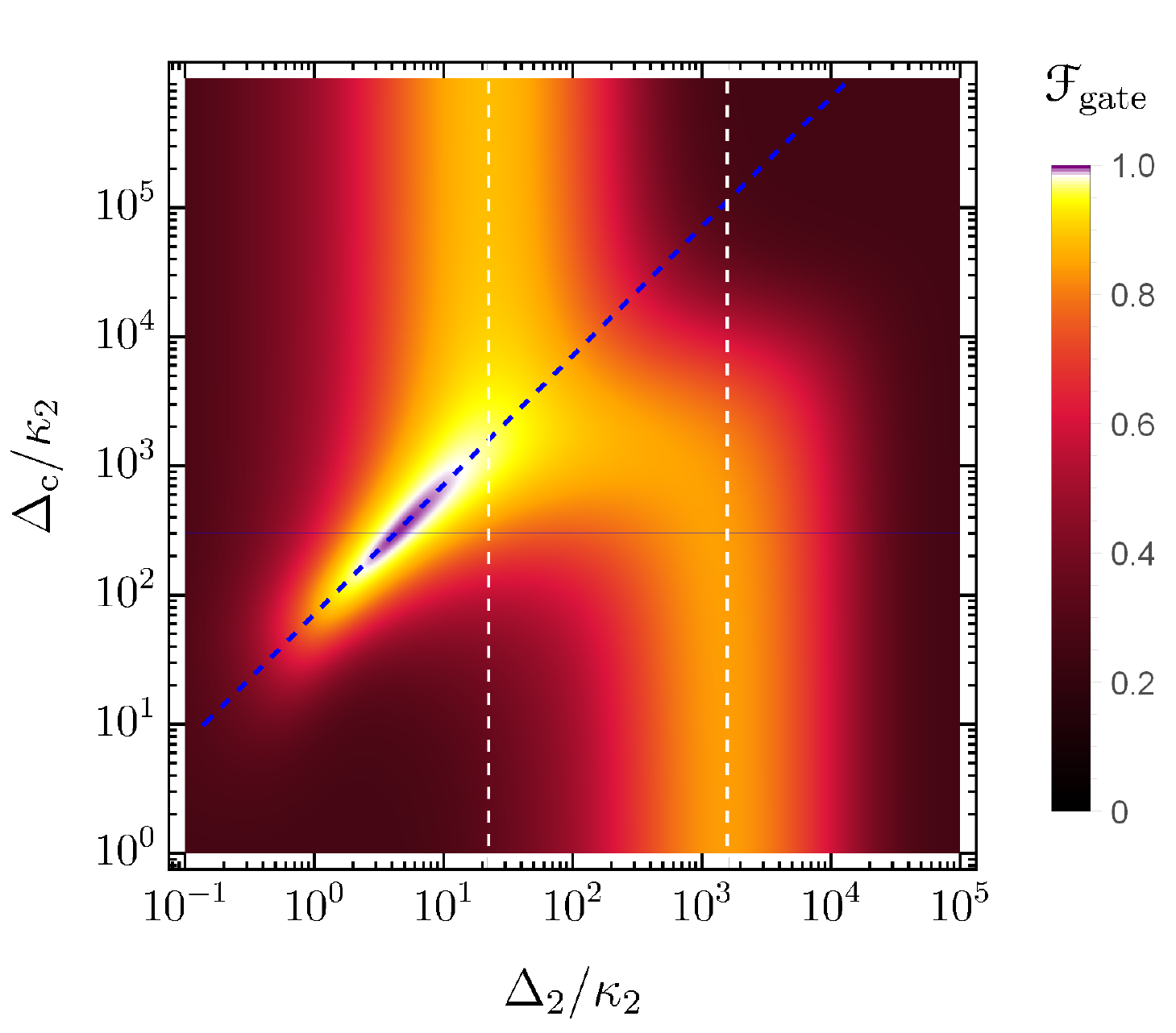}
    \includegraphics[width=0.49\textwidth]{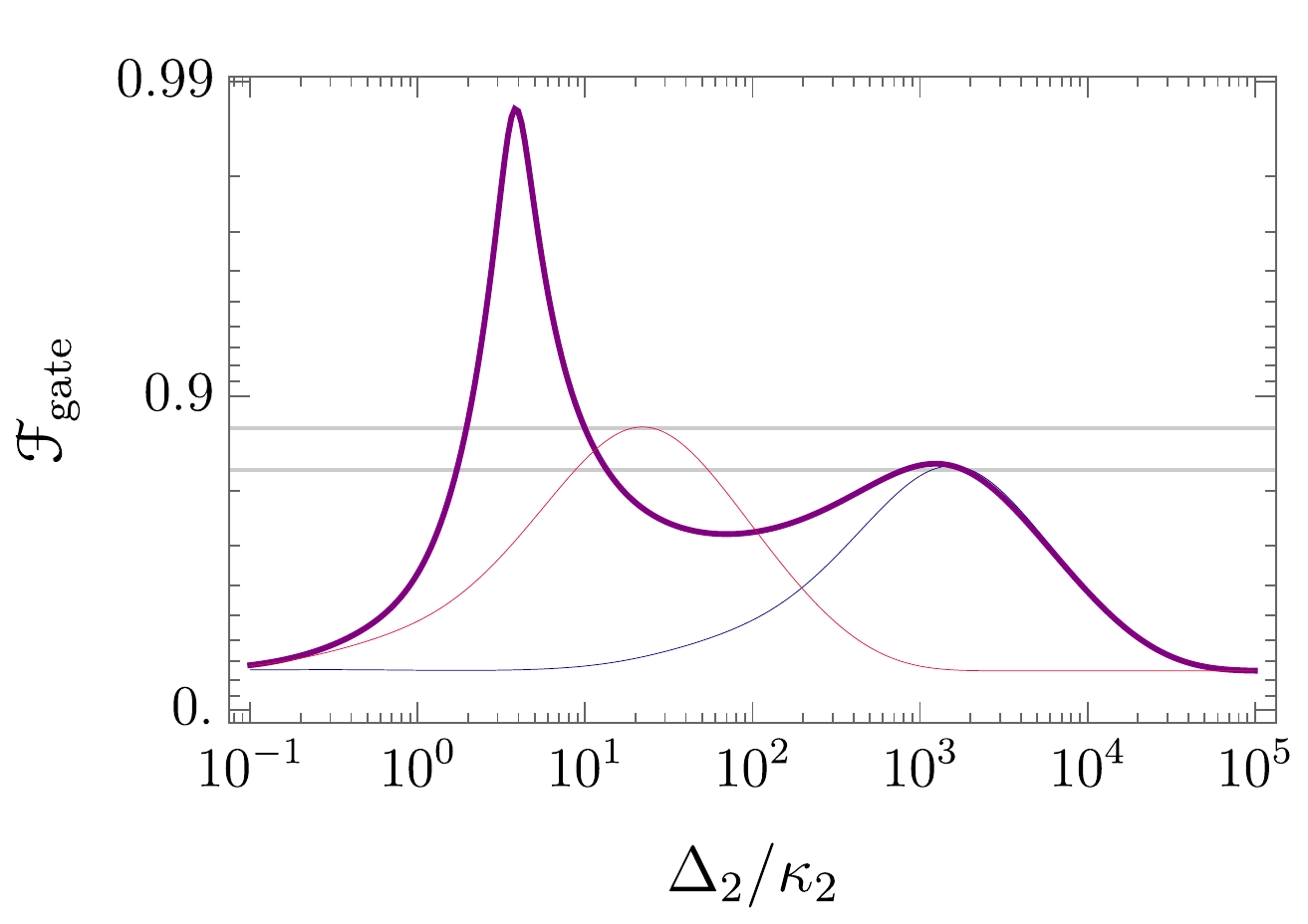}
    \caption[Subradiant virtual photon exchange phase gate fidelity in a perfect Fano dip.]{\small\textbf{Subradiant virtual photon exchange phase gate fidelity in a perfect Fano dip.} (a) The phase gate fidelity for $T_\text{gate}=\pi/|\lambda|$ as a function of the dipole detuning $\Delta_2$ from the narrow cavity mode with a FWHM of $\kappa_2$ and the splitting between the two cavity modes $\Delta_\mathrm{c}=\Delta_2-\Delta_1$. The vertical dashed white lines indicate the ideal detunings for each individual mode given by the condition $\kappa_i\sqrt{C_i}/2$. The dashed blue line indicates the Fano dip condition $\Delta_1/(\kappa_1\sqrt{C_1})=\Delta_2/(\kappa_2\sqrt{C_2})$. (b) Cross-section (purple curve) along the horizontal line from panel (a) superimposed with the fidelity expected from each individual mode (red and blue curves), which are given by the extreme cases of $\Delta_\mathrm{c}$ in panel (a). The horizontal lines indicate the maximum fidelity for each individual mode alone. Parameters used: $\kappa_1=100\kappa_2$, $C_1=1000$, $C_2=2000$, $s=1$, $\tilde{g}_{i1}=\tilde{g}_{i2}>0$, and $\phi_s=\pi/2$.}
    \label{chapter4fig:fanogateatidealTgate}
\end{figure}

Let us now explore the maximum fidelity scaling as a function of cavity cooperativity. To that end, I will assume that the dipole-QNM coupling strengths are equal in magnitude $|\tilde{g}_{i1}|=|\tilde{g}_{i2}|$ so that $R=R_1=R_2$. Then the fidelity is given similar to Eq.~(\ref{chapter4eq:fidgatenonHerm}) by
\begin{equation}
\label{chapter4eq:qnmfidelity}
    \euscr{F}_\text{gate}=\frac{1}{4}\left[1+e^{-T_\text{gate}(R+\gamma)/2}\left|\sin\!\left(\frac{T_\text{gate}\sqrt{\lambda_{12}\lambda_{21}}}{2}\right)\right|^2\right]^2.
\end{equation}
In the regime where the modes do not overlap much in frequency ($|S_{12}|\ll 1$ implying $\Delta_\mathrm{c}\gg\kappa_i$ and/or $\kappa_1\gg\kappa_2$) and the dipoles are far detuned relative to each mode $\Delta_i\gg\kappa_i$ we have $\lambda_{12}\simeq\lambda_{21}\simeq\lambda$. Hence, by choosing $T_\text{gate}=\pi/|\lambda|$ along with satisfaction of the phase conditions mentioned previously, we can find a significant Fano dip suppression of $R$ that increases the fidelity compared to the expected enhancement from each individual mode alone (see Fig.~\ref{chapter4fig:fanogateatidealTgate}). 

Using the same approach as with the single-mode case, we can minimize $T_\text{gate}(R+\gamma)$ to solve for the ideal detuning conditions that achieve a good Fano dip enhancement of the fidelity. By making the assumption that $\kappa_2\ll\kappa_1$ and that $C_i=4\tilde{g}_i^2/\kappa_i\gamma\gg 1$ in the bad-cavity regime for both modes, we can write this term as
\begin{equation}
\label{chapter4eq:qnmgaterate}
    T_\text{gate}(R+\gamma)=4\pi\left(1+\frac{C_1}{1+4r_1^2}+\frac{C_2}{1+4r_2^2}-\frac{4sr_2\sqrt{C_1C_2}}{(1+4r_2^2)\sqrt{1+4r_1^2}}\right)\left(\frac{C_1}{r_1}+\frac{C_2}{r_2}\right)^{-1},
\end{equation}
where $r_1=\Delta_1/\kappa_1$ and $r_2=\Delta_2/\kappa_2$. By observation, we can identify that the Fano dip condition is given when both modes satisfy equal single-mode detuning conditions so that $\Delta_2/(\kappa_2\sqrt{C_2})=\Delta_1/(\kappa_1\sqrt{C_1})$. Under this condition, which corresponds to the dashed blue line in Fig.~\ref{chapter4fig:fanogateatidealTgate}, I minimize Eq.~(\ref{chapter4eq:qnmgaterate}) for $s=1$ to find that $r_1/\sqrt{C_1}=r_2/\sqrt{C_2}\simeq(1/2)(3/C_2)^{1/4}$ maximizes the fidelity. Substituting this solution into the fidelity, under the same ideal conditions, gives our main result
\begin{equation}
\label{chapter4eq:maxfideq}
    \text{max}\!\left(\euscr{F}_\text{gate}\right)\simeq\frac{1}{4}\left(1+e^{-4\pi/\left(3^{3/4}\left(\sqrt{C_1}+\sqrt{C_2}\right)C_2^{1/4}\right)}\right)^2\simeq 1-\frac{4\pi}{3^{3/4}\left(\sqrt{C_1}+\sqrt{C_2}\right)C_2^{1/4}}.
\end{equation}
Here we can identify that the term  $(\sqrt{C_1}+\sqrt{C_2})^{-1}$ in the errors arises from two cavity modes that are split in frequency (two independent single-mode cases). The extra factor $C_2^{-1/4}$ increasing the fidelity is directly due to operating in the Fano dip near the narrow mode. 

As a consequence of the combined action of the two modes plus the Fano dip, the ideal gate time to maximize the fidelity is reduced compared to a gate using a single Lorentzian mode. Consider $T_\text{gate}=\pi/\lambda$ where $|S_{12}|\ll 1$, $\tilde{g}_i=|\tilde{g}_{i1}|=|\tilde{g}_{i2}|$ and where the ideal phase conditions are met. This gives $T_\text{gate}\simeq\pi(\tilde{g}_1^2/\Delta_1+\tilde{g}_2^2/\Delta_2)^{-1}$. Then, in the far adiabatic regime with the ideal detuning conditions $r_1/\sqrt{C_1}=r_2/\sqrt{C_2}\simeq(1/2)(3/C_2)^{1/4}$, the ideal gate time becomes $T_\text{gate}=(2\pi/\gamma)(3/C_2)^{1/4}\left(\sqrt{C_1}+\sqrt{C_2}\right)^{-1}$. Hence, compared to a single mode where the ideal gate time is $2\pi/\gamma\sqrt{C}$, the Fano enhancement can allow for a gate that is multiple times faster primarily due to the additional $C_2^{-1/4}$ scaling from operating in the Fano dip.

For $C=C_1=C_2$, the maximum fidelity for the subradiant virtual photon exchange scheme reduces to $1-2\pi/(3C)^{3/4}$. This shows that the Fano enhancement gives a significant improvement to the cooperativity scaling compared to the $1-2\pi/\sqrt{C}$ scaling for the single mode case. To be fair, we need to recognize that if the two modes were resonant, we would expect the cooperativity of each cavity mode to add giving a scaling of $1-2\pi/\sqrt{C_1+C_2}$. Hence, we should consider the single mode case using $C$ compared to a two-mode case with $C/2$. Even then, the minimum cavity cooperativity needed to possibly reach a fidelity of 0.9 (0.99) for a single mode is 3400 (390000) and for two modes is just 150 (3600). In addition, we can see that the fidelity for the two-mode case benefits more from an increase in $C_2$ than from an equivalent increase in $C_1$.

\begin{figure}[t]
    \centering
    \hspace{-54mm}(a)\hspace{72mm}(b)\\
    \includegraphics[width=0.44\textwidth]{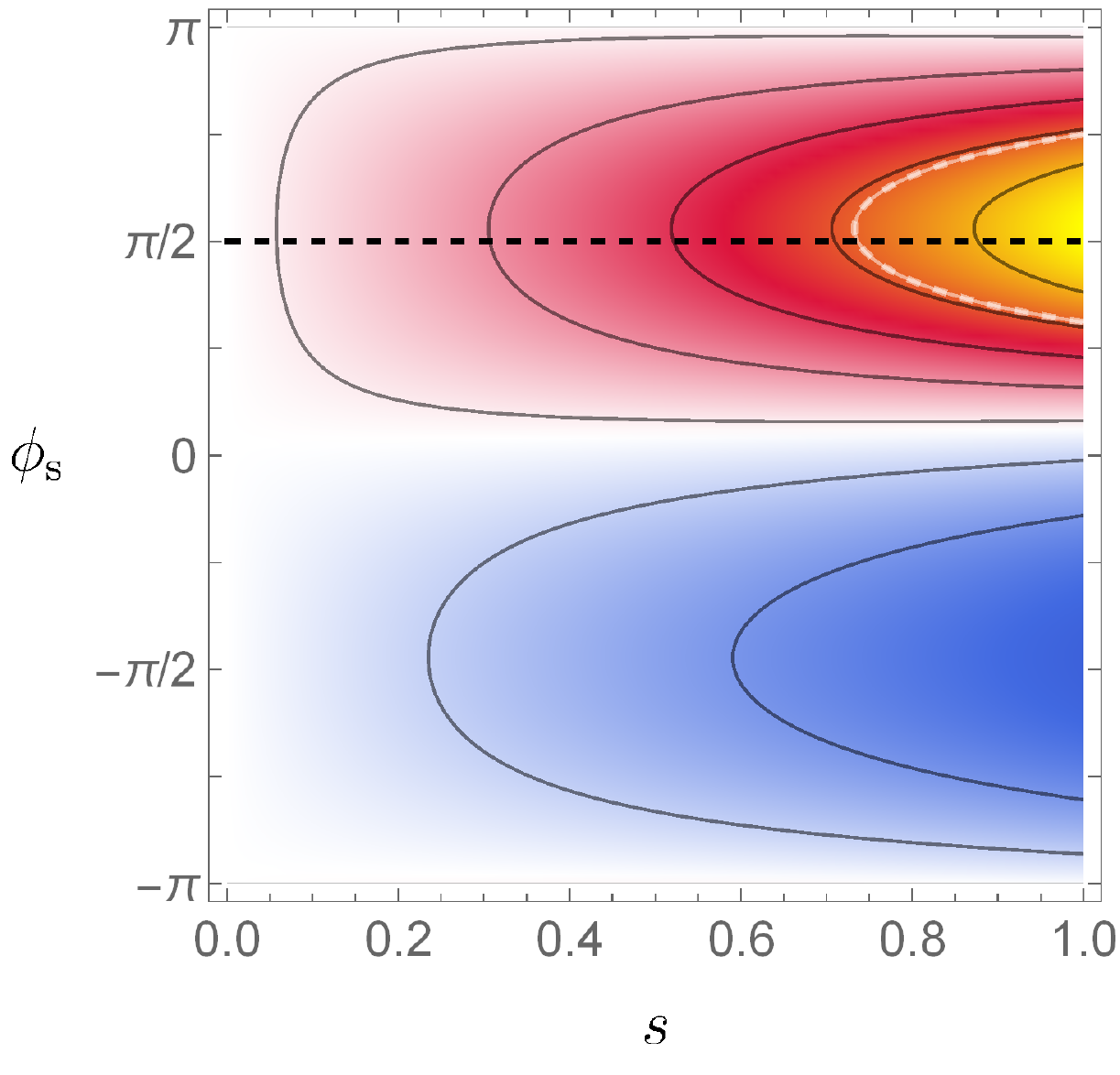}\hspace{5mm}
    \includegraphics[width=0.52\textwidth]{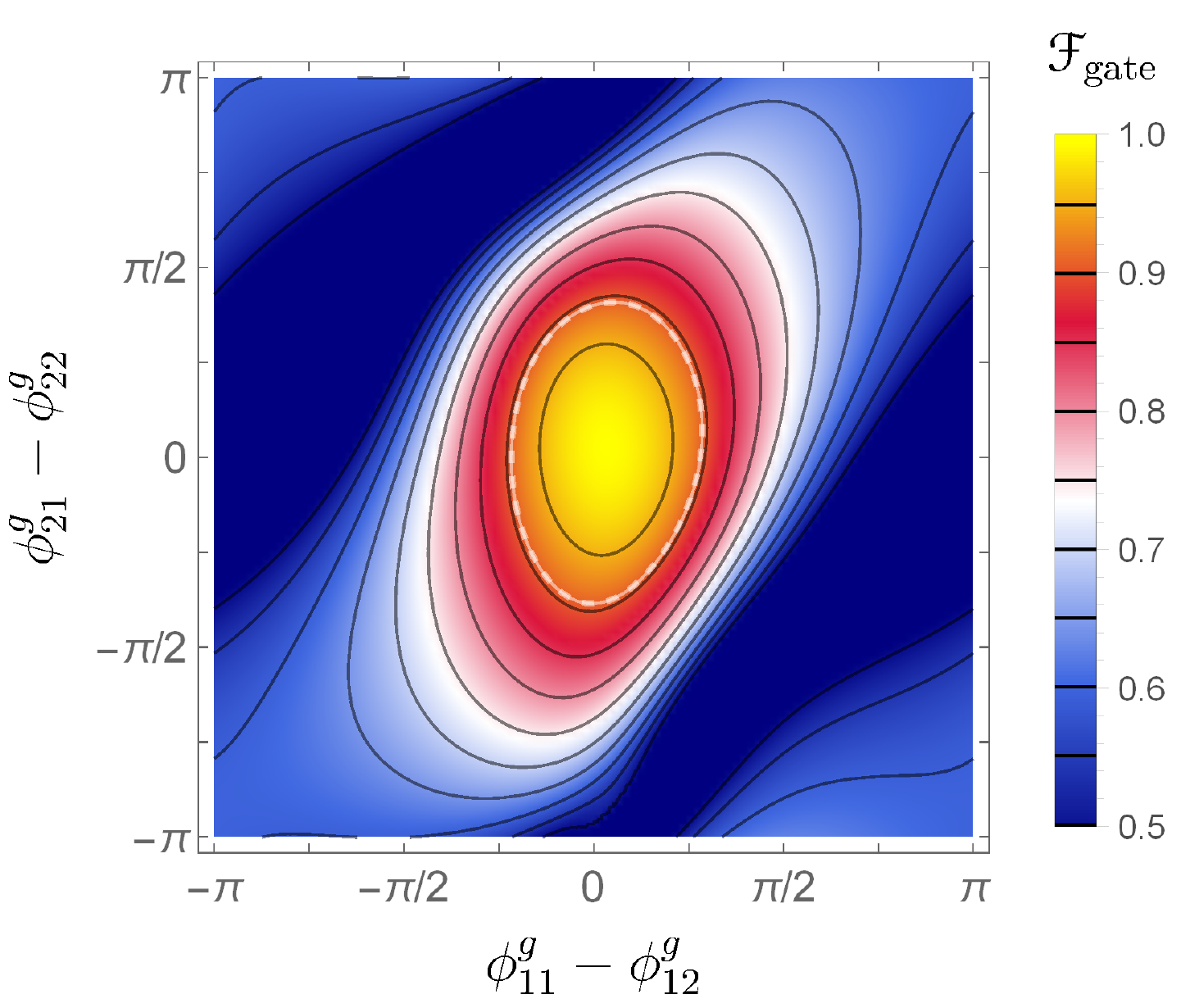}
    \caption[Susceptibility to deviations from the ideal phase conditions for the subradiant virtual photon exchange phase gate fidelity.]{\small\textbf{Susceptibility to deviations from the ideal phase conditions for the subradiant virtual photon exchange phase gate fidelity.} (a) The gate fidelity dependence on the phase and magnitude of the mode overlap parameter $S_{12}$. The white color is set to indicate the fidelity value for no Fano dip enhancement ($s=0$). The dashed white contour indicates the boundary where the fidelity exceeds what is possible using a single mode with twice the cavity cooperativity. In this case, I have set $\phi_{ij}^g=0$ so that $\phi_\mathrm{s}\simeq\pi/2$ indicates the maximum fidelity (horizontal black dashed line). (b) The susceptibility of the fidelity for deviations from the ideal dipole-cavity mode coupling phases. Here, I have set $\phi_\mathrm{s}=\pi/2$ and $s=1$. Parameters used: $\kappa_1=100\kappa_2$, $C=C_1=C_2=2000$, and $2\Delta_i/\kappa_i=(3C)^{1/4}$.}
    \label{chapter4fig:fanogatefidelityphasemaps}
\end{figure}

To summarize, the ideal conditions to achieve maximum fidelity are (1) constructive quantum interference of the dipole-dipole interaction $\phi_{11}^g-\phi_{21}^g=\phi_{12}^g-\phi_{22}^g$, (2) Fano dip arising on the exterior of the two modes $\phi_s+\phi_{1k}^g-\phi_{2k}^g\simeq \pi/2$, (3) maximum allowed mode overlap $s=1$, and (4) Fano dip detuning condition $2\Delta_i/(\kappa_i\sqrt{C_i})=(3/C_2)^{1/4}$. Knowing these conditions, we can now explore how the fidelity is degraded when the phase conditions are not perfectly satisfied. Fig.~\ref{chapter4fig:fanogatefidelityphasemaps} shows how the fidelity depends on the exact values of $s$, $\phi_\mathrm{s}$, and the dipole-cavity coupling phases. From these plots, it is clear that a Fano resonance allows for a significant improvement over what is possible using a Lorentzian mode without requiring extreme precision for the ideal conditions. For $C=C_1=C_2=2000$, there can be an improvement over a single mode with cooperativity $2C$ when $s>0.75$. For $s\simeq 1$, the phases $\phi_s$ and $\phi_{i1}^g-\phi_{i2}^g$ can deviate by up to $\sim~\!\pi/4$ from their ideal values before the Fano resonance is no better than a Lorentzian. This implies that an enhancement could still be possible even with some fabrication imperfections, which would be promising for increasing the yield of high-quality devices.

\begin{figure}[t]
    \centering
    \hspace{-58mm}(a)\hspace{75mm}(b)\\
    \includegraphics[width=0.48\textwidth]{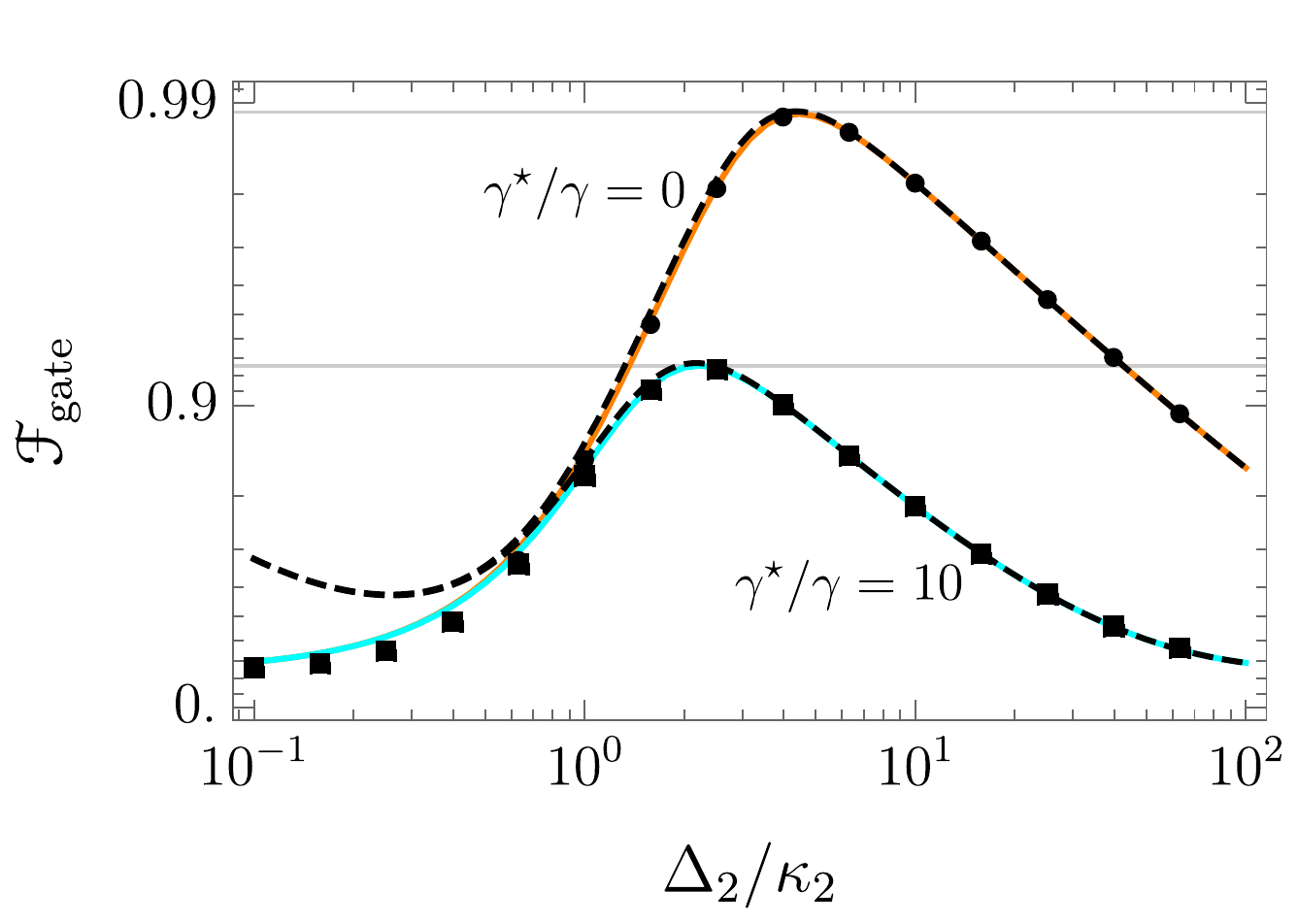}\hspace{5mm}
    \includegraphics[width=0.48\textwidth]{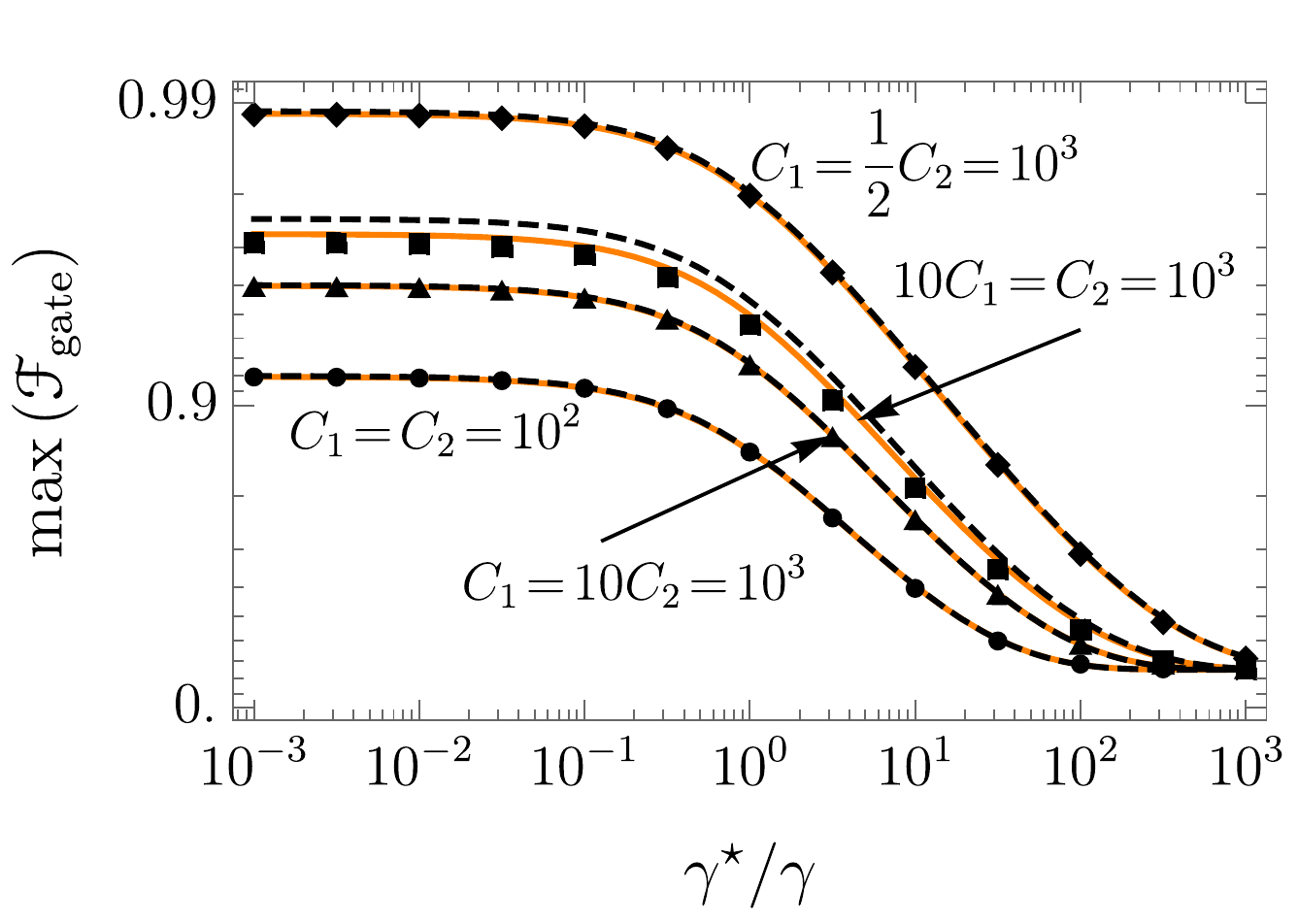}
    \caption[The effect of emitter pure dephasing on the subradiant photon exchange controlled phase gate fidelity using a Fano dip.]{\small\textbf{The effect of emitter pure dephasing on the subradiant photon exchange controlled phase gate fidelity using a Fano dip.} (a) The phase gate fidelity as a function of detuning from the narrow mode resonance $\Delta_2$ with the broad mode detuning set using the Fano dip condition $\Delta_1/(\kappa_1\sqrt{C_1})=\Delta_2/(\kappa_2\sqrt{C_2})$ corresponding to the dashed blue line in Fig.~\ref{chapter4fig:fanogateatidealTgate}~(a). The black points are numerically exact values computed from simulating the full master equation with perfect state preparation and retrieval pulses. The orange and cyan colored curves show the full analytic solution obtained from applying adiabatic elimination to the effective Hamiltonian with no pure dephasing (orange curve) and using the effective cooperativity $C_{\text{eff},i}=C_i(1+(21/16)\gamma^\star/\gamma)^{-1}$ for a pure dephasing of $\gamma^\star=10\gamma$ (cyan curve). The black dashed lines indicate the solution Eq.~(\ref{chapter4eq:qnmfidelity}) assuming $T_\text{gate}\sqrt{\lambda_{12}\lambda_{21}}=\pi$ along with an effective cavity cooperativity to capture pure dephasing. The maximum fidelity occurs at the ideal detuning condition $2\Delta_2=\kappa_2(3C_2)^{1/4}$ with a value given by Eq.~(\ref{chapter4eq:maxfideq}). (b) The phase gate fidelity at the ideal detuning condition as a function of the emitter pure dephasing ratio $\gamma^\star/\gamma$ for four different sets of cavity cooperativity. The points, colored curves, and the dashed curves represent the same approaches as in panel (a). Parameters used: $s=1$, $\phi_s=\pi/2$, and $\phi_{ij}^g=0$. For panel (a): $C_1=C_2/2=1000$, corresponding to the top curve in panel (b).}
    \label{chapter4fig:fanogatefidelityDephasing}
\end{figure}

Let us now consider emitter pure dephasing. Recall that both the simple and subradiant virtual photon exchange schemes reduce to being described by the same effective dipole-dipole Hamiltonian in the ideal regime of operation, albeit with different rates. Also, pure dephasing primarily affects this subspace. Thus, so long as dephasing is small enough to not significantly dampen the dipole-cavity coupling, it is reasonable to expect that the same effective cooperativity derived for the simple scheme will apply for the two-mode case as well. To verify this postulate, we can compare Eq.~(\ref{chapter4eq:qnmfidelity}) to the numerically exact solution from the master equation. Indeed, Fig.~\ref{chapter4fig:fanogatefidelityDephasing} shows that $C_\text{eff,i}=C_i\left(1+(21/16)(\gamma^\star/\gamma)\right)^{-1}$ again provides a very good estimate of the fidelity when there is pure dephasing even if $C_1\neq C_2$. It also illustrates the asymmetric dependence on the cavity cooperativity of each mode, showing a higher fidelity when $C_2>C_1$ than $C_1>C_2$ for the same values of $C_1$ and $C_2$. However, the analytic solution does slightly overestimate the fidelity when $C_2\gg C_1$. The good correspondence between the numerical solution and the analytic solution also helps to verify all of the assumptions made in deriving the analytic solution.

Reducing the cavity cooperativity requirement undoubtedly helps any application of this type of phase gate. However, it is particularly interesting to consider how the Fano enhancement could potentially open up entirely new regimes of operation. For example, plasmonic cavities may allow for an (uninhibited) cavity cooperativity nearing $C\simeq 10^6$ \cite{gurlek2018manipulation,hughes2019theory}. Even with a single mode, this would be enough to reach very high fidelity gates. But, at room temperature where optical defects may experience pure dephasing rates exceeding $\gamma^\star/\gamma=10^4$, the effective phase gate cavity cooperativity will be $C_\text{eff}\simeq 800$. For a single mode with $C=10^6$, this limits the maximum gate fidelity to around 0.78. On the other hand, the Fano dip enhancement with $C=10^6$ may still allow for a gate fidelity of up to $0.98$ even with $\gamma^\star/\gamma=10^4$, which may be sufficient to perform cavity-mediated information processing at or near room temperature.

The results in this section are promising and warrant further investigation for both low and room temperature applications. However, inhibition of the cavity cooperativity due to dephasing is not the only issue that must be overcome for solid-state systems. As discussed in section~\ref{chapter3:roomtemperature}, phonon non-Markovian effects often play a role in determining the ultimate fidelity possible for a realistic implementation of a solid-state device. In addition, it is necessary to excite one of the two defects much faster than the gate time. Luckily, since this scheme operates in the Fano dip where dissipation is suppressed, the excitation problem is much less severe as compared to a single-photon source. That said, for large $C=C_1=C_2$, the ideal gate time $T_\mathrm{gate}=3^{1/4}\pi/\gamma C^{3/4}$ still requires excitation pulses orders of magnitude faster than the bare lifetime $1/\gamma$. When dephasing is included, the ideal gate time is reduced by an additional factor of $(C_\text{eff}/C)^{1/4}$. Therefore, when designing an implementation for a specific system, it is important to consider realistic limitations on the pulse used for state preparation and retrieval.
\chapter{Conclusion}
\label{chapter5}

\section{Summary}
\label{chapter5:summary}

Over the course of this thesis, I have presented the analysis of light-matter interaction models for basic components used to process quantum information with solid-state optical devices. These basic components ranged from single-photon sources to optically-mediated spin-spin entanglement. 

In many cases, the analysis in this thesis relies on material from chapter \ref{chapter2}, where I described a framework based on quantum trajectories to study the state of a quantum optical device when subject to a variety of imperfections. The approach can be used to analytically or numerically decompose a wide variety of emitter Markovian master equations into a set of propagation superoperators conditioned on the emitted photonic state. These operators can then be applied to reconstruct the quantum photonic state of the emission. It can also be used to compute imperfect measurements of the emitter, or joint measurements of multiple emitters. In addition, it is connected to the method of non-Hermitian Hamiltonians, which I used to analyze local entangling gates.

In section \ref{chapter3:roomtemperature}, I discussed how ultra-small mode volume cavities, such as cavities containing plasmonic material, may allow for indistinguishable single-photon sources that operate at room temperature. In particular, we derived an expression for single-photon indistinguishability in the critical cavity coupling regime that is valid for any magnitude of emitter decay rate. With this expression, we studied the optimal regime of operation for when the emitter quenching rate is proportional to the squared magnitude of the cavity coupling rate. Motivated by our results, we proposed that a Fabry-P\'{e}rot-plasmonic hybrid cavity combined with a negatively-charged silicon-vacancy center is a good candidate system to achieve a room temperature indistinguishable single-photon source.

Section \ref{chapter3:HOM} presented an analysis of how multi-photon emission affects the indistinguishability of single-photon sources measured using Hong-Ou-Mandel interference. The main conclusion from this analysis is that the single-photon indistinguishability can be estimated by applying a simple correction using the second-order intensity correlation, and that this correction is accurate for both a general separable noise model and a driven two-level system when the probability for multi-photon emission is small. This section concluded by generalizing the concept of indistinguishability measurements to a self-homodyne measurement setup, which can be used to measure other quantum properties of photonic states such as one- and two-photon number coherence.

In section \ref{chapter3:photonicstate}, I applied the photon number decomposition to analyze the temporal density function of emitted photonic states from a purely dephased emitter. This allowed for a much more detailed perspective on the results obtained in section \ref{chapter3:HOM} on Hong-Ou-Mandel interference. I also showed how to use the photon number decomposition in the context of a self-homodyne measurement to probe the photon number coherence in time between specific photon-number subspaces. This section concluded with an exploration on how a two-level system can be manipulated to deterministically generate photon number entangled states encoded in discretized time bins. This included an analysis of the degrading effect of pure dephasing on the entanglement concurrence for photon number Bell states encoded in time.

In section \ref{chapter4:entanglementgeneration}, I built on the concept of self-homodyne measurements and applied the photon number decomposition to analyze joint photon counting measurements of two remote optically-active defects containing spin qubits. I presented an analysis of three popular photon-heralded entanglement generation protocols based on (1) spin-photon number entanglement, (2) spin-time bin entanglement, and (3) spin-polarization entanglement. These three protocols were found to have different upper bounds on the entanglement generation fidelity. Interestingly, the maximum fidelity values of the three protocols were found to satisfy a specific ordering and were also all bounded by the mean wavepacket overlap of photons from each emitter. The section concluded with a detailed comparison of the three schemes while taking into account photon losses, detector noise, and detector number resolution limitations in addition to optical and spin dephasing.

The discussion of upper bounds on fidelity was continued in section \ref{chapter4:cavitygates} in the context of local entanglement generation between spin qubits in optically-active defects mediated by an optical cavity. In this section, I explored how a cavity-emitter system operating in the bad-cavity regime can allow for high-fidelity entanglement even in the presence of pure emitter dephasing. This was emphasized by deriving a simple relation to estimate the entanglement fidelity from an effective cavity cooperativity. This analysis was then extended to cavities that are described by two interfering quasi-normal modes, which is relevant for cavities containing dissipative material such as plasmonics. This extension revealed that the subradiant mode of two interfering optical modes could mediate an adiabatic interaction between two emitters. By placing the two emitters in a Fano dip feature of the two-mode cavity, the maximum gate fidelity was shown to be dramatically higher than for a single mode cavity, even if the emitters experience severe dephasing.

\section{Outlook}
\label{chapter5:outlook}

This thesis focused heavily on phenomenological pure dephasing as a non-trivial example requiring a master equation model and the superoperator formalism presented in chapter \ref{chapter2}. I would like to emphasize that the methods I have demonstrated can be applied to a wide variety of Markovian master equations, including those with time-dependent parameters. For example, phonon interactions can cause an increased dephasing rate when a solid-state defect is driven by optical pulses. A rough approximation for this phenomenon can be made by taking the pure dephasing rate to be proportional to the square of the driving Rabi frequency \cite{nazir2016modelling}. Multi-level systems with many different dissipative and dephasing channels can also be easily handled in a phenomenological way, allowing for straightforward coarse estimates of figures of merit for complicated quantum information processing protocols. Moving beyond this phenomenological approach, weak-coupling and polaron Markovian master equations \cite{nazir2016modelling,gustin2017influence,gustin2020efficient} may provide a natural extension to account for more detailed electron-phonon interactions for quantum dots. Alternatively, a quantum optical master equation with discrete damped phonon resonances may be more relevant to model atomic defects in diamond \cite{betzholz2014quantum}. Lastly, there may be other interesting applications to analyze that involve the quasi-normal mode Markovian master equation model \cite{franke2019quantization}. For these reasons, I hope that the methods and results presented in this thesis will benefit the study and development of a wide range of quantum devices based on light-matter interactions.

In many ways, the content of this thesis was motivated by a desire to understand and circumvent processes that may restrict solid-state optical devices to cryogenic temperatures. This original motivation is illustrated by the content of section \ref{chapter3:roomtemperature} on the feasibility of room temperature indistinguishable single-photon sources. However, there are still many open questions regarding the feasibility of room-temperature solid-state devices. A promising direction to achieve this goal is to discover or engineer new optically-active solid-state defects that have a reduced electron-phonon coupling at room temperature. Current research in this direction suggests that defects in 2-dimensional materials, such as hexagonal boron-nitride, are promising candidates \cite{li2019purification,dietrich2020solid}. These 2-dimensional materials may also be easily incorporated into small mode volume cavities or layered with plasmonic material to achieve the Purcell enhancements needed to overcome any remaining dephasing or emission inefficiency at room temperature. That said, the thermal limits to cavity devices operating at room temperature are still being explored \cite{panuski2020fundamental}.

An additional approach that could be applied to improve the quality of photonic state sources is to post process the emission. This approach is often only useful if the source quality is already quite high, but it could provide a last push needed to reach quality thresholds required for quantum information processing. The methods and results presented in sections \ref{chapter3:HOM} and \ref{chapter3:photonicstate} provide a detailed picture of the time dynamics and quantum properties of imperfect photonic states. This can be used to optimize post-processing methods such as temporal filtering \cite{abudayyeh2019purification} to minimize the effect of pure dephasing and unwanted multi-photon noise on the desired state. The content of section \ref{chapter3:photonicstate} also provides a basis to study the effect of emitter imperfections on more complicated processes, such as photonic graph state generation for all-optical quantum repeaters \cite{buterakos2017deterministic}.

Imperfect emission also affects the quality of entanglement generation between nodes of quantum networks based on solid-state defects. A natural continuation of the content presented in section \ref{chapter4:entanglementgeneration} would be to include realistic excitation pulses and study the effects of unwanted multi-photon emission on the entanglement generation fidelity. One of the imperfections that we did explore in section \ref{chapter4:entanglementgeneration} was the relative detuning between remote emitters. Reducing this source of error for solid-state defects is quite difficult because optically-active defects are often quite spectrally distinct due to microscopic differences in their respective solid-state environments. If the emitters are close in frequency, it may be possible to use a Stark shift or strain tuning to bring two emitters into resonance \cite{bernien2013heralded,hensen2015loophole,stockill2017phase}. For any remaining detuning, fast time-tagging measurements may still allow for successful spin-spin entanglement generation using spectrally distinct photons \cite{zhao2014entangling}. These time-tagging measurements could also be modeled using a photon number decomposition, which could extend the results in section \ref{chapter4:entanglementgeneration} to compensate for finite emitter detuning.

The phase gate schemes presented in section \ref{chapter4:cavitygates} also require two emitters to be resonant. However, even if the emitters are brought fully into resonance, it may be difficult to individually excite just one of these two dipoles. This is a particularly large problem if the cavity has a very small mode volume, which would potentially require the defects to be closer than the diffraction limit for spatial addressability. As we studied in Ref.~\ref{asadi2020cavitygates} for the simple virtual photon scheme, this problem could be overcome by using a Raman-assisted approach where weak optical driving can allow for a virtual photon exchange interaction without needing the dipoles to be resonant. This approach could also be applied to the subradiant virtual photon exchange scheme in a similar way. Another modification that can improve these phase gate schemes is post-selection. As we studied in Ref.~\ref{asadi2020repeaters}, it is possible to monitor emission from the cavity mode(s) and reject any gate attempts where one or more photon is detected. As a result, the phase gate fidelity could be improved at the cost of some efficiency. The photon number decomposition used in this thesis also provides the framework to analyze this type of post-selection as a function of the monitoring efficiency.

The fidelity limits on a subradiant virtual photon phase gate using a Fano resonance studied in section \ref{chapter4:subradiantvirtual} suggest a significant improvement in the cavity cooperativity scaling versus a single-mode scheme. However, it remains to explore whether the derived ideal parameter regime for the quasi-normal mode model can be physically implemented. Fortunately, one of the strengths of this model is that the parameters of the master equation can be efficiently computed from classical electromagnetic simulations \cite{franke2019quantization}. Guided by the phase, detuning, and mode linewidth conditions presented in section \ref{chapter4:subradiantvirtual}, it may be possible to engineer and optimize a cavity design that can perform fast and high-fidelity optically-mediated gates using a Fano dip. A promising approach is to combine a high-$Q$ Fabry-P\'{e}rot cavity with a low-$Q$ plasmonic resonator. Such hybrid cavities can give good optical Fano resonance spectra \cite{dezfouli2017modal,gurlek2018manipulation}. That said, one challenge will be to engineer a Fano resonance that has a Fano dip arising on the exterior of the two main interfering modes. This type of Fano resonance can occur for plasmonic hybrid cavities \cite{doeleman2020observation} but it seems to be rare, with most studies showing examples with a dip between modes \cite{barth2010nanoassembled,dezfouli2017modal,gurlek2018manipulation,franke2019quantization}. Performing the gate fidelity analysis for more than two modes may also modify the phase conditions for a high-fidelity phase gate and allow for more flexibility to design an optimal cavity.

The development of efficient and robust quantum optical devices stands to make a tremendous impact on the technological progress of humanity. In particular, optically-active solid-state systems will undoubtedly continue pushing the frontier of the ongoing quantum revolution. These quantum systems may also provide the perfect platform to complement and enhance current classical computation and communication technology. This, in turn, could allow for a multitude of unprecedented advances that may fundamentally shape the future of our world. From clean energy production and agricultural practices to medical procedures and cryptography, the potential breadth of quantum technological applications is truly astonishing. Moreover, alongside all of these practical benefits, quantum physics will continue to allow humanity to ask and answer profound questions about the nature of our universe.



\cleardoublepage 
\phantomsection  
\renewcommand*{\bibname}{References}

\addcontentsline{toc}{chapter}{\textbf{References}}

\printbibliography


\appendix
\addcontentsline{toc}{chapter}{APPENDICES}
\chapter{Properties of conditional propagation superoperators}
\label{AppendixA}

In this appendix section, I will present three proofs related to conditional propagation superoperators in the interest of completion. The first proof is a validation that the integral form of the density operator solution satisfies the master equation. Although the form is commonly obtained by the method of variation of parameters, I validate it here in the notation of the thesis for consistency. The second is an interesting extension of the semi-group property to a set of conditional propagation superoperators. The last proof shows the invariance of time ordering for the jump operators composing a conditional propagator. Although I haven't found a proper citation for these last two proofs, I am quite confident that they are not novel.

\section{Integral form of the master equation}
\label{AppendixA:variationofparameters}

The photon number decomposition relies on the solution
\begin{equation}
\label{meqint}
    \hat{\rho}_s(t) = \mathcal{U}_0(t,t_0)\hat{\rho}(t_0) + \int_{t_0}^t\mathcal{U}_0(t,t^\prime)\mathcal{J}\hat{\rho}_s(t^\prime)dt^\prime
\end{equation}
being the solution to the master equation $\dot{\rho}=\mathcal{L}_0\hat{\rho}(t)+\mathcal{J}\hat{\rho}(t)$. I am using the notation $\hat{\rho}_s(t)$ to represent the proposed solution and then I will prove it is equal to the general solution $\hat{\rho}(t)$ for initial condition $\hat{\rho}(t_0)$. The solution $\hat{\rho}_s$ is constructed using the method of variation of parameters (which is equivalent to solving using an integrating factor for first-order differential equations) where the first term is the solution to the equation $\dot{\rho}_0=\mathcal{L}_0\hat{\rho}_0(t)$ and the second term is a particular solution to the total equation $\dot{\rho}=\mathcal{L}\hat{\rho}(t)$ where $\mathcal{L} = \mathcal{L}_0+\mathcal{J}$. Here, I am using $\hat{\rho}_0(t)$ to again avoid confusion with the solution $\hat{\rho}(t)$ to the master equation. To prove that $\hat{\rho}_s(t)$ is the general solution to the master equation $\dot{\rho}=\mathcal{L}\hat{\rho}(t)$, we must first prove some properties of $\mathcal{U}_0$. The propagator $\mathcal{U}_0$ is defined as $\hat{\rho}_0(t) = \mathcal{U}_0(t,t_0)\hat{\rho}_0(t_0)$ where $\dot{\rho}_0 = \mathcal{L}_0\hat{\rho}_0(t)$. The above definition of $\mathcal{U}_0$ implies that it also satisfies the same differential equation
\begin{equation}
    \frac{d\mathcal{U}_0(t,t_0)}{dt} =\mathcal{L}_0\mathcal{U}_0(t,t_0)\hspace{10mm}[\text{I}]
\end{equation}
for any $t_0$. In addition, 
\begin{equation}
    \mathcal{U}_0(t_\mathrm{f},t_0)\hat{\rho}_0(t_0) = \hat{\rho}_0(t_\mathrm{f}) = \mathcal{U}_0(t_\mathrm{f},t)\hat{\rho}_0(t) = \mathcal{U}_0(t_\mathrm{f},t)\mathcal{U}_0(t,t_0)\hat{\rho}_0(t_0)
\end{equation}
implies $\mathcal{U}_0(t_\mathrm{f},t_0) = \mathcal{U}_0(t_\mathrm{f},t)\mathcal{U}_0(t,t_0)$ for any $t$ [II]. Finally, $\mathcal{U}_0(t,t)\hat{\rho}_0(t)=\hat{\rho}_0(t)$ for all $t$ implies $\mathcal{U}_0(t,t)=\mathcal{I}$ [III] for all $t$ where $\mathcal{I}$ is the identity superoperator. In addition to the above three properties, we need to use the product rule [IV] and the fundamental theorem of calculus [V].

Now, let us take the time derivative of the variation of parameters solution
\begin{equation}
\begin{aligned}
    \dot{\rho}_s&=\frac{d}{dt}\left[ \mathcal{U}_0(t,t_0)\hat{\rho}(t_0) + \int_{t_0}^t\mathcal{U}_0(t,t^\prime)\mathcal{J}\hat{\rho}_s(t^\prime)dt^\prime\right]\\
    &=\frac{d\mathcal{U}_0(t,t_0)}{dt}\hat{\rho}(t_0) + \frac{d}{dt}\int_{t_0}^t\mathcal{U}_0(t,t^\prime)\mathcal{J}\hat{\rho}_s(t^\prime)dt^\prime\\
    &= \mathcal{L}_0\mathcal{U}_0(t,t_0)\hat{\rho}(t_0) + \frac{d}{dt}\mathcal{U}_0(t,t^{\prime\prime})\int_{t_0}^t\mathcal{U}_0(t^{\prime\prime},t^\prime)\mathcal{J}\hat{\rho}_s(t^\prime)dt^\prime&\text{by [I] and [II]}\\
    &=\mathcal{L}_0\mathcal{U}_0(t,t_0)\hat{\rho}(t_0) + \frac{d\mathcal{U}_0(t,t^{\prime\prime})}{dt}\int_{t_0}^t\mathcal{U}_0(t^{\prime\prime},t^\prime)\mathcal{J}\hat{\rho}_s(t^\prime)dt^\prime\\
    &\hspace{33mm}+\mathcal{U}_0(t,t^{\prime\prime})\frac{d}{dt}\int_{t_0}^t\mathcal{U}_0(t^{\prime\prime},t^\prime)\mathcal{J}\hat{\rho}_s(t^\prime)dt^\prime&\text{by [IV]}\\
    &=\mathcal{L}_0\mathcal{U}_0(t,t_0)\hat{\rho}(t_0) + \mathcal{L}_0\mathcal{U}_0(t,t^{\prime\prime})\int_{t_0}^t\mathcal{U}_0(t^{\prime\prime},t^\prime)\mathcal{J}\hat{\rho}_s(t^\prime)dt^\prime\\
    &\hspace{33mm}+\mathcal{U}_0(t,t^{\prime\prime})\mathcal{U}_0(t^{\prime\prime},t)\mathcal{J}\hat{\rho}_s(t)&\text{by [I] and [V]}\\
    &=\mathcal{L}_0\left[\mathcal{U}_0(t,t_0)\hat{\rho}(t_0) + \int_{t_0}^t\mathcal{U}_0(t,t^\prime)\mathcal{J}\hat{\rho}_s(t^\prime)dt^\prime\right]+\mathcal{U}_0(t,t)\mathcal{J}\hat{\rho}_s(t)&\text{by [II]}\\
    &=\mathcal{L}_0\hat{\rho}_s(t)+\mathcal{J}\hat{\rho}_s(t)&\text{by [III]}\\
    &=\mathcal{L}\hat{\rho}_s(t).
\end{aligned}
\end{equation}
Hence $\hat{\rho}_s(t)$ is a solution to $\dot{\rho} = \mathcal{L}\hat{\rho}(t)$. Finally, we can confirm that
\begin{equation}
\begin{aligned}
    \hat{\rho}_s(t_0) &= \mathcal{U}_0(t_0,t_0)\hat{\rho}(t_0) + \int_{t_0}^{t_0}\mathcal{U}_0(t_0,t^\prime)\mathcal{J}\hat{\rho}_s(t^\prime)dt^\prime &\\
    &=\mathcal{U}_0(t_0,t_0)\hat{\rho}(t_0) + 0&\\
    &= \hat{\rho}(t_0).&\text{by [III]}
\end{aligned}
\end{equation}
Thus $\hat{\rho}_s(t)$ satisfies the initial condition $\hat{\rho}(t_0)$. Therefore, by the uniqueness theorem for first-order differential equations, we have proven $\hat{\rho}(t)=\hat{\rho}_s(t)$.

\section{Subintervals of time}
The total propagation superoperator can be divided into subintervals
\begin{equation}
\mathcal{U}(t_\mathrm{f},t_0)=\mathcal{U}(t_\mathrm{f},t)\mathcal{U}(t,t_0) = \sum_{k,l}\mathcal{U}_k(t_\mathrm{f},t)\mathcal{U}_l(t,t_0) = \sum_{n}\mathcal{U}_n(t_\mathrm{f},t_0).
\end{equation}
We can show that a similar relation
\begin{equation}
\begin{aligned}
\mathcal{U}_{n}(t_\mathrm{f},t_0)=\sum_{k,l}\mathcal{U}_k(t_\mathrm{f},t)\mathcal{U}_l(t,t_0)\delta_{k+l,n}
\end{aligned}
\end{equation}
holds for the conditional propagation superoperators. This relies on the fact that it holds for the homogeneous case
\begin{equation}
    \mathcal{U}_0(t_\mathrm{f},t_0) = \mathcal{U}_0(t_\mathrm{f},t)\mathcal{U}_0(t,t_0)
\end{equation}
as a consequence of $\mathcal{U}_0$ being a propagator superoperator of a Markovian master equation $\dot{\rho}=\mathcal{L}_0\hat{\rho}(t)$. First, consider the base case for $n=1$
\begin{equation}
\begin{aligned}
    &\mathcal{U}_0(t_\mathrm{f},t)\mathcal{U}_1(t,t_0)+\mathcal{U}_1(t_\mathrm{f},t)\mathcal{U}_0(t,t_0)\\
    &=\int_{t_0}^{t}\mathcal{U}_0(t_\mathrm{f},t)\mathcal{U}_0(t,t^{\prime})\mathcal{J}\mathcal{U}_0(t^{\prime},t_0)dt^{\prime}+\int_{t}^{t_\mathrm{f}}\mathcal{U}_0(t_\mathrm{f},t^{\prime})\mathcal{J}\mathcal{U}_0(t^{\prime},t)\mathcal{U}_0(t,t_0)dt^{\prime}\\
    &=\int_{t_0}^{t}\mathcal{U}_0(t_\mathrm{f},t^{\prime})\mathcal{J}\mathcal{U}_0(t^{\prime},t_0)dt^{\prime}+\int_{t}^{t_\mathrm{f}}\mathcal{U}_0(t_\mathrm{f},t^{\prime})\mathcal{J}\mathcal{U}_0(t^{\prime},t_0)dt^{\prime}\\
    &=\mathcal{U}_1(t,t_0).
\end{aligned}
\end{equation}
Now, assume that for some $n\geq 0$ we have
\begin{equation}
\mathcal{U}_n(t,t_0) = \sum_{k,l}\mathcal{U}_k(t,t^\prime)\mathcal{U}_l(t^\prime,t_0)\delta_{k+l,n}.
\end{equation}
Then this implies
\begin{equation}
\begin{aligned}
\mathcal{U}_{n+1}(t_\mathrm{f},t_0)&=\int_{t^\prime}^{t_\mathrm{f}}\mathcal{U}_0(t_\mathrm{f},t)\mathcal{J}\sum_{k,l}\mathcal{U}_k(t,t^\prime)\mathcal{U}_l(t^\prime,t_0)\delta_{k+l,n}dt+\int_{t_0}^{t^\prime}\mathcal{U}_0(t_\mathrm{f},t)\mathcal{J}\mathcal{U}_n(t,t_0)dt\\
&=\sum_{k,l}\int_{t^\prime}^{t_\mathrm{f}}\mathcal{U}_0(t_\mathrm{f},t)\mathcal{J}\mathcal{U}_k(t,t^\prime)dt\mathcal{U}_l(t^\prime,t_0)\delta_{k+l,n}+\mathcal{U}_0(t_\mathrm{f},t^\prime)\int_{t_0}^{t^\prime}\mathcal{U}_0(t^\prime,t)\mathcal{J}\mathcal{U}_n(t,t_0)dt\\
&=\sum_{k=1,l}\mathcal{U}_{k}(t_\mathrm{f},t^\prime)\mathcal{U}_l(t^\prime,t_0)\delta_{k+l,n+1}+\mathcal{U}_0(t_\mathrm{f},t^\prime)\mathcal{U}_{n+1}(t^\prime,t_0)\\
&=\sum_{k,l}\mathcal{U}_{k}(t_\mathrm{f},t^\prime)\mathcal{U}_l(t^\prime,t_0)\delta_{k+l,n+1}
\end{aligned}
\end{equation}
which completes the proof. As a consequence of this proof, we know that dividing conditional propagators into subintervals of time conserves the total photon number, as expected.

\section{Invariance of order}
\label{appendix:klorder}

The definition of $\mathcal{U}_n$ assumes that all $n-1$ photons are emitted before the integrated time
\begin{equation}
    \mathcal{U}_n(t_\mathrm{f},t_0) = \int_{t_0}^{t_\mathrm{f}}\mathcal{U}_0(t_\mathrm{f},t)\mathcal{J}\mathcal{U}_{n-1}(t,t_0)dt.
\end{equation}
However, we can show that the conditional propagator does not depend on the order of emission. That is, the more general case 
\begin{equation}
\begin{aligned}
\mathcal{U}_{k+l+1}(t_\mathrm{f},t_0)=\int_{t_0}^{t_\mathrm{f}}\mathcal{U}_k(t_\mathrm{f},t)\mathcal{J}\mathcal{U}_l(t,t_0)dt
\end{aligned}
\end{equation}
holds as a consequence. First consider the base case where $k=0$,
\begin{equation}
\begin{aligned}
\mathcal{U}_{l+1}(t_\mathrm{f},t_0)=\int_{t_0}^{t_\mathrm{f}}\mathcal{U}_0(t_\mathrm{f},t)\mathcal{J}\mathcal{U}_{l}(t,t_0)dt &= \int_{t_0}^{t_\mathrm{f}}\int_{t_0}^t\mathcal{U}_0(t_\mathrm{f},t)\mathcal{J}\mathcal{U}_0(t,t^\prime)\mathcal{J}\mathcal{U}_{l-1}(t^\prime,t_0)dt^\prime dt\\
&=\int_{t_0}^{t_\mathrm{f}}\int_{t^\prime}^{t_\mathrm{f}}\mathcal{U}_0(t_\mathrm{f},t)\mathcal{J}\mathcal{U}_0(t,t^\prime)\mathcal{J}\mathcal{U}_{l-1}(t^\prime,t_0)dt dt^\prime\\
&=\int_{t_0}^{t_\mathrm{f}}\mathcal{U}_1(t_\mathrm{f},t^\prime)\mathcal{J}\mathcal{U}_{l-1}(t^\prime,t_0)dt^\prime,
\end{aligned}
\end{equation}
where I have used the change of integration order relation
\begin{equation}
    \int_{t_0}^{t_\mathrm{f}}\int_{t_0}^{t}f(t,t^\prime)dt^\prime dt=\int_{t_0}^{t_\mathrm{f}}\int_{t^\prime}^{t_\mathrm{f}}f(t,t^\prime)dtdt^\prime.
\end{equation}
Now, let us assume that for some $k\geq 0$ we have
\begin{equation}
\begin{aligned}
\int_{t_0}^{t_\mathrm{f}}\mathcal{U}_k(t_\mathrm{f},t)\mathcal{J}\mathcal{U}_{l}(t,t_0)dt &=\int_{t_0}^{t_\mathrm{f}}\mathcal{U}_{k+1}(t_\mathrm{f},t)\mathcal{J}\mathcal{U}_{l-1}(t,t_0)dt.
\end{aligned}
\end{equation}
Then this implies
\begin{equation}
\begin{aligned}
\int_{t_0}^{t_\mathrm{f}}\mathcal{U}_{k+1}(t_\mathrm{f},t)\mathcal{J}\mathcal{U}_{l}(t,t_0)dt &=\int_{t_0}^{t_\mathrm{f}}\int_{t}^{t_\mathrm{f}}\mathcal{U}_0(t_\mathrm{f},t^\prime)\mathcal{J}\mathcal{U}_k(t^\prime,t)dt^\prime\mathcal{J}\mathcal{U}_{l}(t,t_0) dt\\
&=\int_{t_0}^{t_\mathrm{f}}\mathcal{U}_0(t_\mathrm{f},t^\prime)\mathcal{J}\int_{t_0}^{t^\prime}\mathcal{U}_k(t^\prime,t)\mathcal{J}\mathcal{U}_{l}(t,t_0)dtdt^\prime\\
&=\int_{t_0}^{t_\mathrm{f}}\mathcal{U}_0(t_\mathrm{f},t^\prime)\mathcal{J}\int_{t_0}^{t^\prime}\mathcal{U}_{k+1}(t^\prime,t)\mathcal{J}\mathcal{U}_{l-1}(t,t_0)dtdt^\prime\\
&=\int_{t_0}^{t_\mathrm{f}}\int_{t}^{t_\mathrm{f}}\mathcal{U}_0(t_\mathrm{f},t^\prime)\mathcal{J}\mathcal{U}_{k+1}(t^\prime,t)dt^\prime\mathcal{J}\mathcal{U}_{l-1}(t,t_0)dt\\
&=\int_{t_0}^{t_\mathrm{f}}\mathcal{U}_{k+2}(t_\mathrm{f},t)\mathcal{J}\mathcal{U}_{l-1}(t,t_0)dt.
\end{aligned}
\end{equation}
Then for any $k$ and $l$, we have
\begin{equation}
\begin{aligned}
\int_{t_0}^{t_\mathrm{f}}\mathcal{U}_{k}(t_\mathrm{f},t)\mathcal{J}\mathcal{U}_{l}(t,t_0)dt&=\int_{t_0}^{t_\mathrm{f}}\mathcal{U}_{0}(t_\mathrm{f},t)\mathcal{J}\mathcal{U}_{l+k}(t,t_0)dt\\
&=\mathcal{U}_{k+l+1}(t_\mathrm{f},t_0).\\
\end{aligned}
\end{equation}
\chapter{Critical regime and phonon sideband corrections}
\label{AppendixB}

\section{Efficiency}
\label{ssec:efficiency}

To derive the indistinguishability and brightness in the critical coupling cavity QED regime, we begin by writing the optical Bloch equations in the single-excitation regime. By assuming that the cavity and emitter are resonant so that $\Delta=0$, from section \ref{chapter1:cavityQED} we have
\begin{equation}
\frac{d}{dt}\left[\begin{matrix}
\braket{\au}\\
\braket{\sigu}
\end{matrix}\right]
=
A_1
\left[\begin{matrix}
\braket{\au}\\
\braket{\sigu}
\end{matrix}\right],
\end{equation}
and
\begin{equation}
\frac{d}{dt}\left[\begin{matrix}
\braket{\au\ad}\\
\braket{\au\sigd}\\
\braket{\sigu\ad}\\
\braket{\sigu\sigd}\\
\end{matrix}\right]
=
A_2
\left[\begin{matrix}
\braket{\au\ad}\\
\braket{\au\sigd}\\
\braket{\sigu\ad}\\
\braket{\sigu\sigd}\\
\end{matrix}\right],
\end{equation}
where
\begin{equation}
\label{A1}
A_1=-\frac{1}{2}\left[\begin{matrix}
\kappa&-2ig\\
-2ig&\Gamma
\end{matrix}\right],
\end{equation}
and
\begin{equation}
\label{A2}
A_2=-\frac{1}{2}\left[\begin{matrix}
2\kappa&2ig&-2ig&0\\
2ig&\kappa+\Gamma&0&-2ig\\
-2ig&0&\kappa+\Gamma&2ig\\
0&-2ig&2ig&2\gamma\\
\end{matrix}\right].
\end{equation}

The cavity emission efficiency, or brightness, is defined by $\beta = \kappa\int_0^\infty \braket{\hat{a}^\dagger(t)\hat{a}(t)}dt$ in section \ref{chapter1:brightness}. An exact solution to $\beta$ can be derived by integrating the solution to Eq. (\ref{A2}). Assuming instantaneous excitation of the system, we set the initial condition to $\rho(0)=\ket{e}\bra{e}$. This implies that the only nonzero initial condition for the optical Bloch equations is $\braket{\sigu\sigd}=1$. With this initial condition, we have
\begin{equation}
\begin{aligned}
\beta&=\kappa\int_0^\infty\text{exp}\!\left(tA_2\right)_{14}dt = -\kappa \left(A_2^{-1}\right)_{14}\\
&=\frac{4g^2\kappa}{4g^2(\gamma+\kappa)+\gamma\kappa(\kappa+\Gamma)},
\end{aligned}
\end{equation}
where the subscripts denote the element of the matrix. This expression for brightness holds in all parameter regimes for a Markovian system and is equal to the result given in \cite{grange2015cavity}. In this section, as with section \ref{chapter3:roomtemperature}, we have defined $R=4g^2/\kappa$ and $\Gamma=\gamma+\gamma^\star$. Please note that $\gamma^\star$ is defined differently here by a factor of 2 compared to the rest of this thesis. By substituting \mbox{$g=(R\kappa)^{1/2}/2$} into this expression, it can be arranged to the form given in section \ref{chapter3:roomtemperature}. Also, notice that $\gamma\ll\kappa,g,\gamma^\star$ gives $\beta=1$.

\section{Indistinguishability}
\label{ssec:indistinguishability}
We can analytically solve for the indistinguishability in a way similar to that for the efficiency. We are interested in the regime where pure dephasing $\gamma^\star$ is small relative to $g$ and $\kappa$. Since $A_1$ can be easily diagonalized, the evolution given by $A_1$ can be solved exactly by computing $U(t)=\text{exp}\!\left(A_1t\right)$. However, to simplify solving the propagator $W(t)=\text{exp}\!\left(A_2t\right)$, we treat it perturbatively for $\gamma^\star/(\kappa+\gamma)<1$. Interestingly, this condition is satisfied if we only assume $\gamma^\star<\kappa$ and so we need not make assumptions about the relative magnitude of $\gamma$ and $\gamma^\star$ or $g$ and $\gamma^\star$.

Let $A_2^{(0)}=A_2(\gamma^\star=0)$, and let $A_2^{(1)}=A_2-A_2^{(0)}$. Then we can write $W(t)= W^{(0)}+W^{(1)}+\mathcal{O}\left(\gamma^{\star 2}\right)$ where
\begin{equation}
\begin{aligned}
W^{(0)}&=\text{exp}\!\left(A_2^{(0)}t\right),\\
W^{(1)}&=W^{(0)}\int_0^t \text{exp}\!\left(-A_2^{(0)}t^\prime\right)A_2^{(1)}\text{exp}\!\left(A_2^{(0)}t^\prime\right)\text{d}t^\prime,
\end{aligned}
\end{equation}
using the same perturbation approach introduced in section \ref{chapter1:perturbationtheory} but applied to the optical Bloch equations. The indistinguishability from section \ref{chapter1:indistinguishability} is
\begin{equation}
I = \frac{2\kappa^2}{\beta^2}\int_0^\infty\int_0^\infty|\braket{\hat{a}^\dagger(t+\tau)\hat{a}(t)}|^2dtd\tau.
\end{equation}
Using the quantum regression theorem (section \ref{chapter1:quantumregression}), we can write
\begin{equation}
\braket{\hat{a}^\dagger(t+\tau)\hat{a}(t)} = U_{11}(\tau)\braket{\au(t)\ad(t)}+U_{12}(\tau)\braket{\sigu(t)\ad(t)},
\end{equation}
where the subscripts denote the element of the matrix propagator $U$. Taking the initial condition to be $\braket{\sigu(0)\sigd(0)}=1$, we have $\braket{\au(t)\ad(t)} = W_{14}$ and $\braket{\sigu(t)\ad(t)} = W_{34}$. Then the correlation function becomes
\begin{equation}
\braket{\hat{a}^\dagger(t+\tau)\hat{a}(t)} = U_{11}(\tau)W_{14}(t)+U_{12}(\tau)W_{34}(t).
\end{equation}
From this, we can write the indistinguishability as:
\begin{equation}
\begin{aligned}
I\beta^2 &= 2\kappa^2\left[h_1+2\text{Re}(h_2)+h_3
\right],
\end{aligned}
\end{equation}
where
\begin{equation}
\begin{aligned}
h_1&=\int_0^\infty|U_{11}|^2\text{d}\tau\int_0^\infty|W_{14}|^2\text{d}t,\\
h_2&=\int_0^\infty U^\star_{11}U_{12}\text{d}\tau\int_0^\infty W^\star_{14}W_{34}\text{d}t,\\
h_3&=\int_0^\infty|U_{12}|^2\text{d}\tau\int_0^\infty|W_{34}|^2\text{d}t.
\end{aligned}
\end{equation}

This expression can be split into two parts: $I=I^{(0)}+I^{(1)}+\mathcal{O}({\gamma^\star}^2/(\kappa+\gamma)^2)$ where $I^{(0)}$ is computed using terms with $U$ and $W^{(0)}$. With a little help from Mathematica, we found $I^{(0)}$ to be:
\begin{equation}
I^{(0)}\beta^2=\frac{R^2\kappa ^2 \left[3\gamma^\star(2\gamma+3\kappa+\gamma^\star)+\Gamma_1^2\right]}{(R+\gamma)(\kappa+\gamma)(R+\gamma+\gamma^\star)(\kappa+\Gamma) \Gamma_1^2},
\end{equation}
where $\Gamma_1^2=(3 \gamma +\kappa ) (\gamma +3 \kappa )+4 \kappa  R$.

The perturbation $I^{(1)}$ contains the correction required for $I$ to be exact to first-order in $\gamma^\star/(\kappa+\gamma)$. Since we are only interested in the first-order correction in the perturbation of $W(t)$, we only compute the $W$ cross-terms that are first-order in $\gamma^\star/(\kappa+\gamma)$. This means, for example, only computing terms with ${W^{(0)}}^\star_{14}W^{(1)}_{34}\propto\gamma^\star$ but not those with ${W^{(1)}}^\star_{14}W^{(1)}_{34}\propto{\gamma^\star}^2$. However, we keep $U(t)$ terms exact because any expansion of $U(t)$ in $\gamma^\star$ would require the assumption that $\gamma^\star/\gamma<1$, which is not the desired case. For this reason, the result for $I^{(1)}$ for arbitrary $\gamma$ still contains some higher-order $\gamma^\star$ terms:

\begin{equation}
\begin{aligned}
\frac{I^{(1)}}{\gamma^\star I^{(0)}}&=\frac{(R-2 \gamma ) \left[(\gamma +\gamma^\star )^2+\gamma  \kappa \right]}{ \left[3\gamma^\star(2\gamma+3\kappa+\gamma^\star)+\Gamma_1^2\right]\Gamma_2^2}\\
   &-\frac{\gamma^\star(\gamma-\gamma^\star)  (4 \gamma +R)+2 \gamma  (\gamma +R) (2 \gamma +R)}{2 (\gamma +\kappa ) (\gamma +R) \Gamma_2^2}\\
    &-\frac{(\gamma +\kappa ) (8 \gamma +5 R)}{2 (\gamma +R) \Gamma_1^2},
 \end{aligned}
\end{equation}
where $\Gamma_2^2=3\gamma^\star(\gamma -\gamma^\star)+4 \gamma  (\gamma +R)$.

We use this solution to compute estimations and generate plots in section \ref{chapter3:roomtemperature}. The expression can be simplified under the assumption that $\gamma\ll\gamma^\star$, which arises when quenching is weak. In this case, the expression $I=I^{(0)}+I^{(1)}$ can be reduced to the form given by Eq. (\ref{ind}).

\section{Phonon sideband corrections}
\label{ssec:psb}

To estimate the correction to $I\beta$ expected due to the presence of a phonon sideband (PSB) in the SiV$^-$ spectrum, we first estimated the fraction $F$ of PSB not removed by the cavity \cite{smith2017}. In terms of wavelength, this can be defined as:
\begin{equation}
F(Q)=\frac{\int_0^\infty S_\text{cav}(\lambda,Q)\times S_\text{PSB}(\lambda)\text{d}\lambda}{\int_0^\infty S_\text{PSB}(\lambda)\text{d}\lambda},
\end{equation}
where
\begin{equation}
S_\text{cav}(\lambda,Q)=\frac{1}{1+4Q^2\frac{(\lambda-\lambda_0)^2}{\lambda_0^2}},
\end{equation}
and
\begin{equation}
S_\text{PSB}(\lambda) = \sum_i\frac{a_i}{1+\frac{(\lambda-\lambda_0-c_i)^2}{b_i^2}}.
\end{equation}
The coefficients $(a_i,b_i,c_i)$ were determined by visually fitting the total spectrum $S(\lambda) = S_\text{ZPL}(\lambda) + S_\text{PSB}(\lambda)$ to the measured spectrum of samples reported by Neu \emph{et al.} \cite{neu2011}, while also maintaining that:
\begin{equation}
\text{DW} \simeq \frac{\int_0^\infty S_\text{ZPL}(\lambda)\text{d}\lambda}{\int_0^\infty S(\lambda)\text{d}\lambda},
\end{equation}
for the associated Debye-Waller ($\text{DW}$) values reported for that sample. Here we use:
\begin{equation}
S_\text{ZPL}(\lambda) = \frac{1}{1+\frac{(\lambda-\lambda_0)^2}{\delta^2}},
\end{equation}
where $\lambda_0$ is the zero-phonon line (ZPL) resonance of the sample and $\delta$ is the ZPL width in wavelength. We chose to fit the sample with the smallest $\delta$ (sample 3), and the sample with the largest $\text{DW}$ (sample 5). The PSB coefficients that we determined are given in Table \ref{coefTable}.

\begin{figure}[t]
\centering
\includegraphics[trim={6mm 5mm 0 0},scale=0.6]{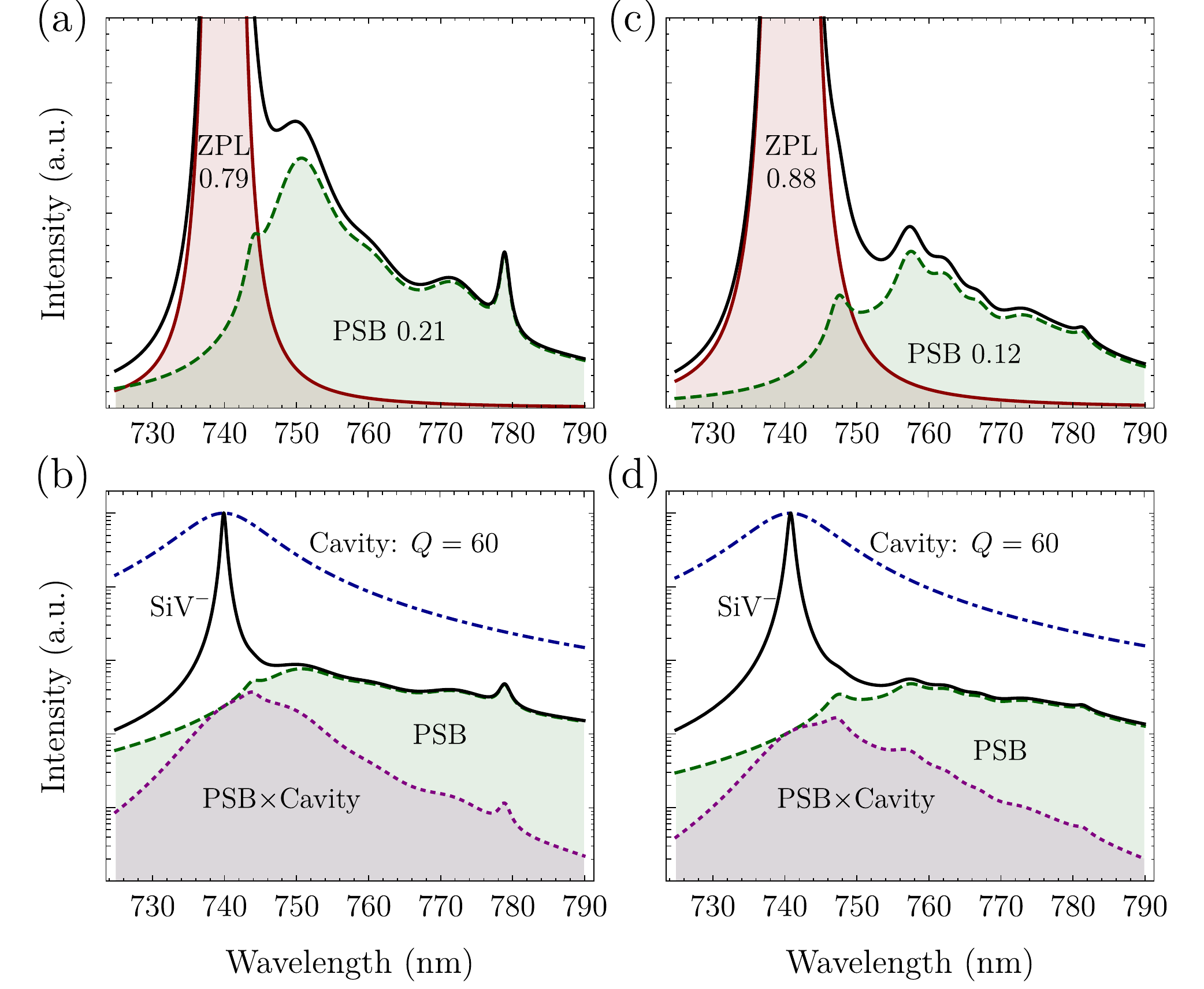}
\vspace{-2mm}
\caption[Estimated nanodiamond negatively-charged silicon-vacancy (SiV$^-$) center emission spectrum.]{\small \textbf{Estimated nanodiamond negatively-charged silicon-vacancy (SiV$^-$) center emission spectrum} of sample 3 (a,b) and sample 5 (c,d) from Neu \emph{et al.} \cite{neu2011}. (b,d) Intensity log-scale plot illustrating the reduction of the PSB due to a cavity with quality factor $Q=60$ on resonance with the zero-phonon line (ZPL) of the sample.}
\label{spectrumplots}
\end{figure}

The estimated spectra for samples 3 and 5 are illustrated in Fig.~\ref{spectrumplots}. The corresponding DW factors were calculated to be $\text{DW}_3=0.791$ and $\text{DW}_5=0.884$, which closely match the measured values of $0.79$ and $0.88$ for samples 3 and 5, respectively. For a cavity with $Q=60$, we estimated the fraction $F$ to be $F_3(60) = 0.19$ and $F_5(60)=0.15$. Using $B^2=\text{DW}$ and the estimated values for $F(Q)$, we applied the PSB corrections using Eqs.~(\ref{Icorrect}) and (\ref{betacorrect}) to our results for indistinguishability $I_0$ and intrinsic cavity efficiency $\beta_0$ under the Markovian approximation.

\begin{table}[t]
\caption[Coefficients for the phonon sideband spectrum used to represent the nanodiamond negatively-charged silicon vacancy centre spectrum.]{Coefficients for $S_\text{PSB}(\lambda)$ used to represent the nanodiamond negatively-charged silicon vacancy centre spectrum of samples 3 and 5 from Neu \emph{et al.} \cite{neu2011}.}
\begin{center}
\mbox{\hspace{5mm}Sample 3\hspace{23mm} Sample 5}
\vspace{-6.5mm}
\end{center}
\centering
\begin{tabular}{c|ccc|ccc}
 & & & & & & \\
i&$a_i\times 10^{3}$&$b_i$ (nm)&$c_i$ (nm) &$a_i\times 10^{3}$&$b_i$ (nm)&$c_i$ (nm) 
\\\hline
1&1.4        &1.3     &4.0      &1.4             &1.7     &7.5\\
2&6.7        &6.5     &10.5     &1.8             &8.0     &11.0\\
3&2.0        &6.0     &20.5     &2.6             &2.9     &17.5\\
4&2.2        &6.0     &32.0     &2.0             &3.3     &22.5\\
5&2.4        &0.9     &39.0     &0.9             &2.5     &27.0\\
6&1.0        &20.0    &47.0     &1.2             &5.0     &33.0\\
7&           &        &         &1.0             &8.0     &39.0\\
8&           &        &         &0.3             &1.1     &41.5\\
9&           &        &         &0.7             &18.0    &49.0\\
\end{tabular}
\label{coefTable}
\end{table}

The narrower linewidth of sample 3 would provide a higher $I_0\beta_0$ than sample 5. However, the larger PSB of sample 3 would restrict the indistinguishability for lower $Q$ values. Hence the maximum $I\beta$ for sample 3 is attained at $Q=100$ rather than $Q=60$. For $Q=100$, $R/\gamma_\text{r}=2.7\times 10^5$, and $\Delta_q(1-\eta_\text{r})^{-1/2}=2\pi\times 30$ THz, sample 3 reaches $I=0.85$ and $\beta=0.99$ giving $I\beta=0.84$, which is comparable to sample 5 at $Q=60$ ($I=0.87$, $\beta=0.97$, $I\beta=0.84$).

\begin{figure}[t]
\vspace{-4mm}
\centering
\includegraphics[trim={0mm 12mm 0 0},scale=0.65]{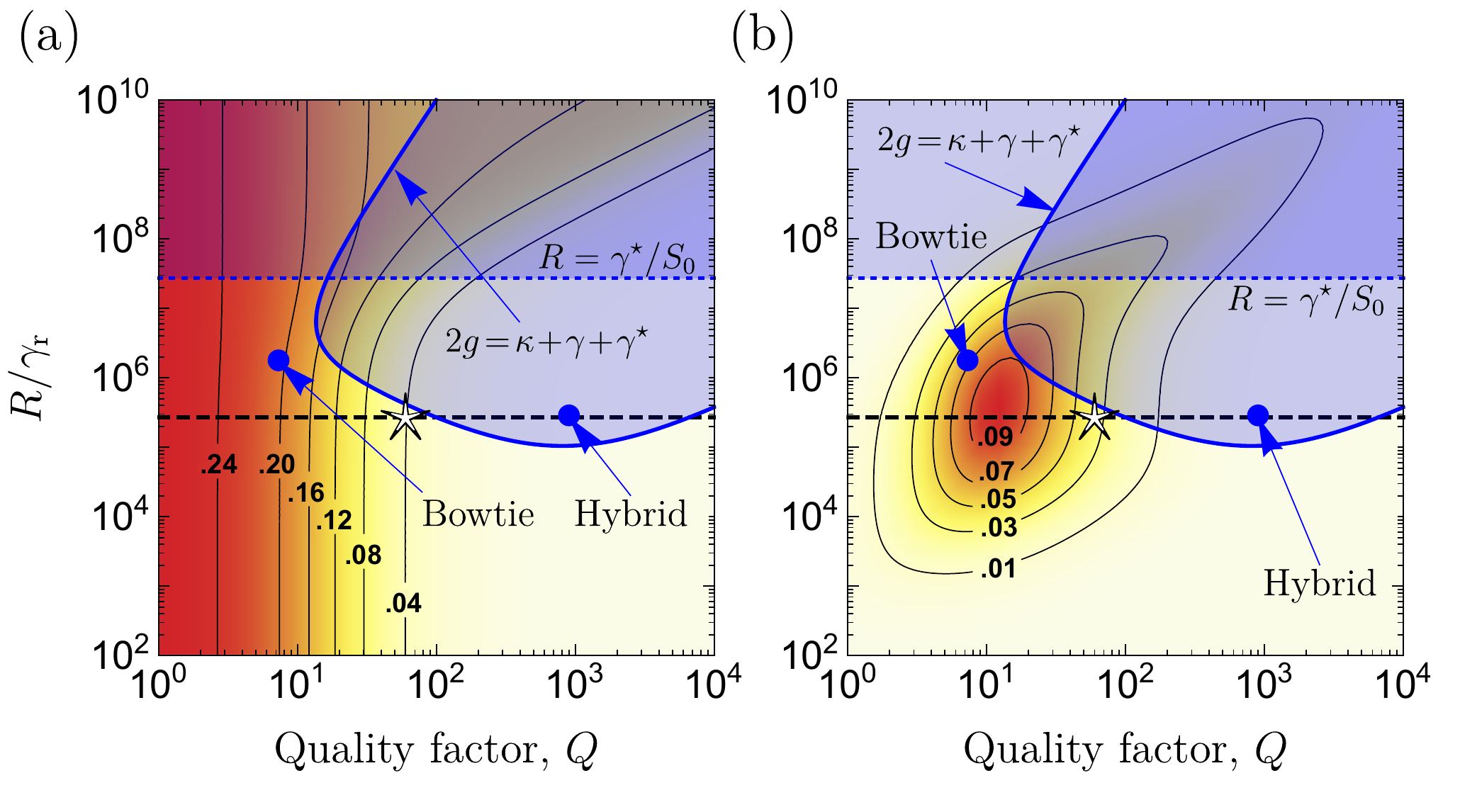}
\caption[Error estimates for the Markovian approximation using a phonon sideband correction.]{\small \textbf{Error estimates for the Markovian approximation using a phonon sideband correction.} (a) The relative error $2(I_0\beta_0-I\beta)/(I_0\beta_0+I\beta)$ for the PSB correction plotted in the critical regime and in the mode-detuned case with $\Delta_\text{q}(1-\eta_\text{r})^{-1/2} = 2\pi\times 30$ THz, where $\Delta_\text{q}$ is the effective detuning parameter for higher-order non-radiative plasmon modes and $\eta_\text{r}$ is the bare cavity quantum efficiency. Here $R=4g^2/\kappa$ where $g$ is the cavity coupling rate and $\kappa$ is the bare cavity linewidth. $I_0\beta_0$ is the derived estimation under the Markovian approximation and $I\beta$ is the value after including the correction due to the phonon sideband (PSB). This small correction could be inaccurate in the strong-coupling regime ($2g>\kappa+\gamma^\star+\gamma$) and when the PSB might begin to enhance the ZPL dephasing rate $R>\gamma^\star/S_0$ (blue shaded regions). Parameters used for the SiV$^-$: $\omega=2\pi\times 405$ THz, $1/\gamma_\text{r}=8.3$ ns \cite{benedikter2017}, $\gamma^\star=2\pi\times 500$ GHz \cite{neu2011}, $\text{DW}=0.88$ \cite{neu2011}, and $S_0=9.6\times 10^{-4}$. See the Appendix text for the PSB spectrum used to calculate the correction. The blue dots represent the plasmonic bowtie \cite{santhosh2016} and a plasmonic-Fabry-P\'{e}rot hybrid cavity \cite{gurlek2018manipulation}. For the bowtie, a $R/\gamma_\text{r}=1.7\times 10^6$ ratio is determined from $g=60$ meV, $Q=7.3$ \cite{santhosh2016}, and $1/\gamma_\text{r}=20$ ns \cite{brokmann2004}. The dashed line marks $R/\gamma_\text{r}=2.7\times 10^5$ expected for the hybrid cavity (at $Q=986$) from the enhancement of the local density of states (LDOS) \cite{gurlek2018manipulation}. The white star shows the proposed single-photon source. (b) Absolute difference $I_0\beta_0-I\beta$ plotted to illustrate the magnitude of the correction.}
\label{errorplots}
\vspace{-6mm}
\end{figure}

The regime in which the Markovian approximation is valid depends strongly on the shape and size of the emitter's PSB. In general, the approximation becomes less accurate as the cavity quality factor is decreased (see Fig.~\ref{errorplots} (a)). In addition, when computing $I_0$ to use in Eq.~(\ref{Icorrect}), we assume that the phonon-induced pure-dephasing rate is not significantly altered by the coupling between the PSB and the ZPL via the cavity mode. A sufficient condition to ensure that this assumption is valid is given when the pure-dephasing rate is much larger than the total effective rate between the PSB and the cavity mode: $\gamma^\star\gg R S_\text{PSB}(\lambda_0)$. In this case, the cavity-PSB interaction can be considered as predominantly a filtering effect. For samples 3 and 5, we can estimate $S_0=S_\text{PSB}(\lambda_0)$ as $2.4\times 10^{-3}$ and $9.6\times 10^{-4}$, respectively. For $R>\gamma^\star/S_0$, it may be necessary to consider the influence of the cavity-PSB interaction on the ZPL dephasing rate.

We also note that we have not considered interactions between the PSB and the higher-order plasmon modes. However, the PSB is predominantly Stokes-shifted to lower energy whereas the higher-order plasmon modes are generally of higher energy than the cavity resonance. Hence, the higher-order plasmon modes should not enhance the PSB and so this additional interaction can be safely neglected in the same regime where quenching does not dominate.

For narrow emitters with a spectrally-separated PSB, such as the nanodiamond SiV$^-$, the $I\beta$ maximum can exist in a parameter regime where the detrimental effects of the PSB are small, leaving the maximum to be primarily limited by Markovian processes (see Fig.~\ref{errorplots} (b)). This leaves a cavity-parameter range where efficient indistinguishable emission should be attainable at room temperature.

\chapter{Copyright permissions}
\label{AppendixC}

In this appendix section, I give the copyright permissions to fully include papers \ref{wein2018plasmonics} and \ref{wein2020entanglement} in this thesis in their published form.

\section{Permissions from journals}

\begin{figure}
    \centering
    \includegraphics[width=\textwidth]{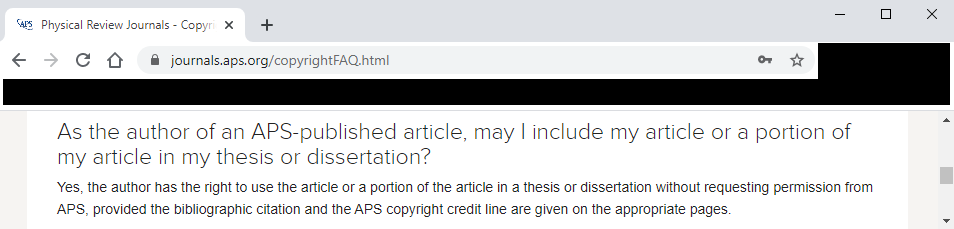}
    \caption{Permission to include APS papers in this thesis.}
    \label{figappdxC:apsrights}
\end{figure}

Paper \ref{wein2018plasmonics} is published in Physical Review B and paper \ref{wein2020entanglement} is published in Physical Review A. Both journals are of the American Physical Society (APS), which grants an author the right to include published papers in their thesis (see Fig. \ref{figappdxC:apsrights}).

\section{Permissions from co-authors}

Each co-author of papers \ref{wein2018plasmonics} and \ref{wein2020entanglement} have given their written permission by email to include these papers in this thesis.

\begin{figure}[p]
    \centering
    \includegraphics[width=\textwidth]{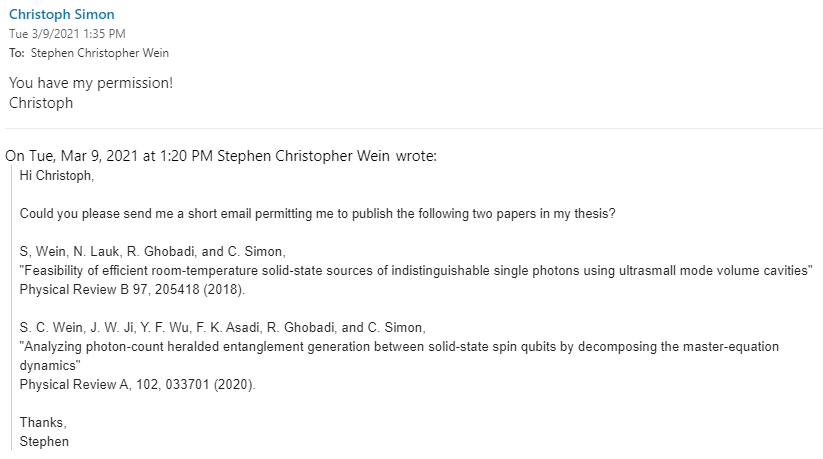}
    \caption{Permission from Christoph Simon to include Ref.~\ref{wein2018plasmonics} and Ref.~\ref{wein2020entanglement} in this thesis.}
    \label{figappdxC:christoph}
\end{figure}

\begin{figure}[p]
    \centering
    \includegraphics[width=\textwidth]{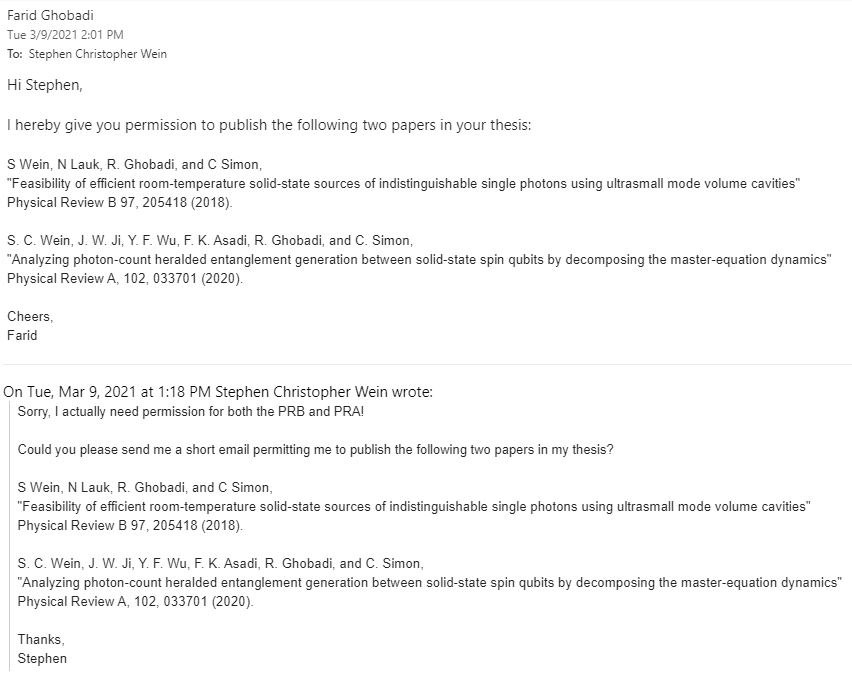}
    \caption{Permission from Roohollah (Farid) Ghobadi to include Ref.~\ref{wein2018plasmonics} and Ref.~\ref{wein2020entanglement} in this thesis.}
    \label{figappdxC:farid}
\end{figure}

\begin{figure}[p]
    \centering
    \includegraphics[width=\textwidth]{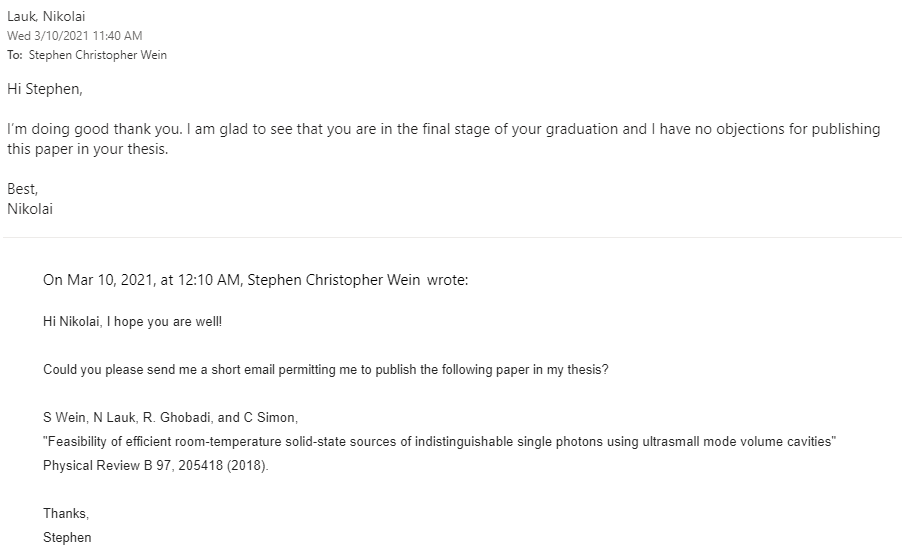}
    \caption{Permission from Nikolai Lauk to include Ref.~\ref{wein2018plasmonics} in this thesis.}
    \label{figappdxC:nikolai}
\end{figure}

\begin{figure}[p]
    \centering
    \includegraphics[width=\textwidth]{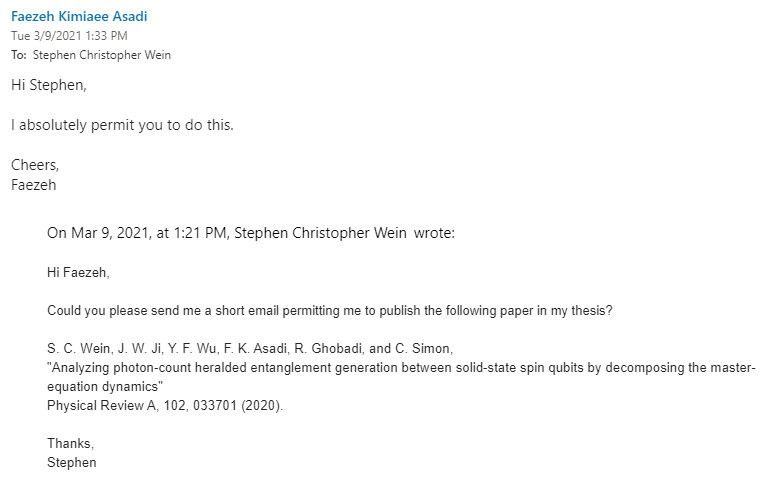}
    \caption{Permission from Faezeh Kimiaee Asadi to include Ref.~\ref{wein2020entanglement} in this thesis.}
    \label{figappdxC:faezeh}
\end{figure}

\begin{figure}[p]
    \centering
    \includegraphics[width=\textwidth]{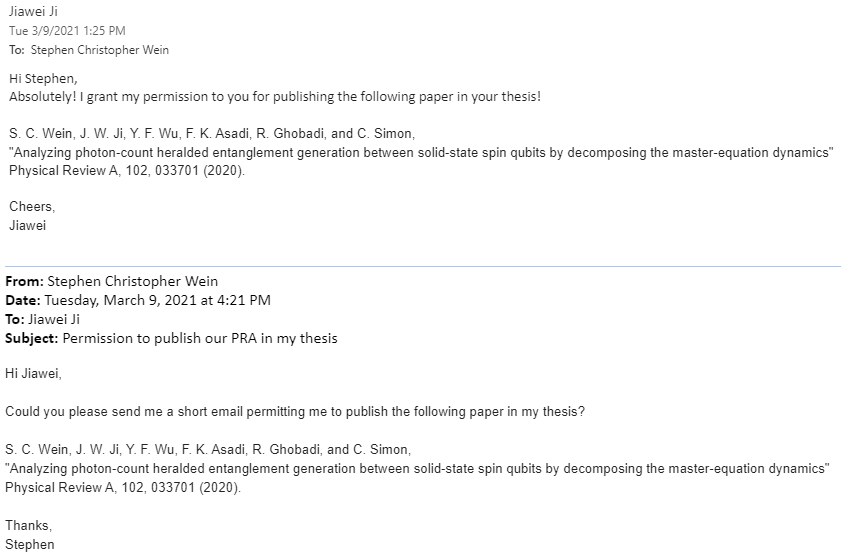}
    \caption{Permission from Jia-Wei Ji to include Ref.~\ref{wein2020entanglement} in this thesis.}
    \label{figappdxC:jiawei}
\end{figure}

\begin{figure}[p]
    \centering
    \includegraphics[width=\textwidth]{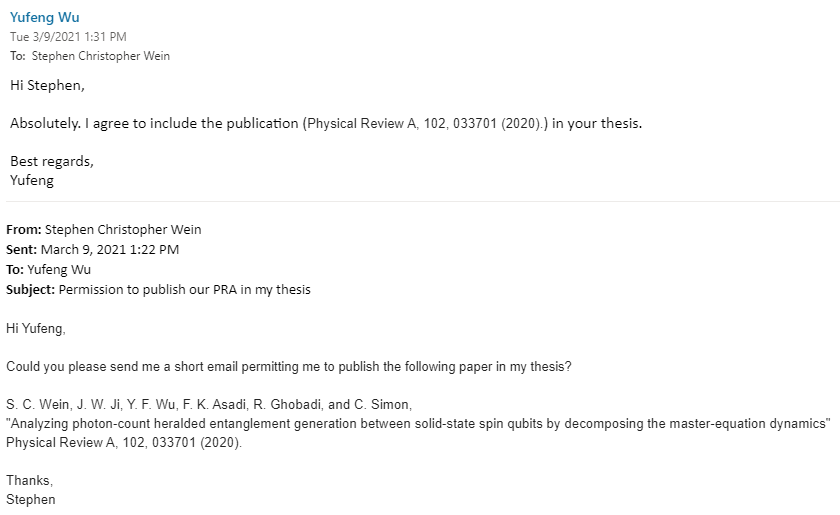}
    \caption{Permission from Yu-Feng Wu to include Ref.~\ref{wein2020entanglement} in this thesis.}
    \label{figappdxC:yufeng}
\end{figure}

\begin{itemize}
    \item Christoph Simon, see Fig.~\ref{figappdxC:christoph}
    \item Roohollah (Farid) Ghobadi, see Fig.~\ref{figappdxC:farid}
    \item Nikolai Lauk, see Fig.~\ref{figappdxC:nikolai}
    \item Faezeh Kimiaee Asadi, see Fig.~\ref{figappdxC:faezeh}
    \item Jia-Wei Ji, see Fig.~\ref{figappdxC:jiawei}
    \item Yu-Feng Wu, see Fig.~\ref{figappdxC:yufeng}
\end{itemize}

\end{document}